\documentclass[11pt,a4paper]{book}

\usepackage{vmargin}
\setpapersize{A4}
\setmarg{40mm}{45mm}{130mm}{210mm}
\usepackage[utf8]{inputenc}

\usepackage[T1]{fontenc}

\usepackage{graphicx}
\usepackage{amstext,amsmath,amssymb}
\usepackage[small]{caption}

\usepackage{fancyhdr}
\renewcommand{\chaptermark}[1]{\markboth{\chaptername \ \thechapter.\ #1}{}}

\fancypagestyle{plain}{
\fancyhead{}

}

 \makeatletter
  \def\cleardoublepage{\clearpage\if@twoside \ifodd\c@page\else%
    \hbox{}
    \thispagestyle{empty}               
       \newpage
    \if@twocolumn\hbox{}\newpage\fi\fi\fi}
\makeatother

\usepackage{siunitx}
\usepackage{units}
\usepackage[version=3]{mhchem}
\usepackage{chapterbib}
\usepackage{etoolbox}
\usepackage{comment}
\usepackage{gensymb}
\patchcmd{\thebibliography}{\chapter*}{\section*}{}{}

\newcommand{\um}[1]{\SI{#1}{\micro\meter}}
\newcommand{\ket}[1]{|#1 \rangle}
\newcommand{\bra}[1]{\langle #1|}
\newcommand{\refeq}[1]{(\ref{#1})}

\usepackage{amsmath}
\DeclareMathOperator{\Tr}{Tr}
\usepackage{url}
\usepackage{relsize}
\newcommand\CC{C\nolinebreak[4]\hspace{-.05em}\raisebox{.4ex}{\relsize{-3}{\textbf{++}}}}

\begin{document}

%
%
%
%
%

\pagenumbering{roman}

\thispagestyle{empty}

\noindent {\Large \bf Colecci\'on de Estudios de F\'{\i}sica}
\qquad \qquad \qquad \qquad \quad {\Large \bf Vol. 124}

\vspace{20mm}

\begin{center}

{\LARGE \bf Coupling quantum circuits \\[2mm] to magnetic molecular qubits \\[2mm]}

\vspace{50mm}

{\Large \bf Mark David Jenkins Sánchez}

\vspace{60mm}
\includegraphics[width=0.6\textwidth]{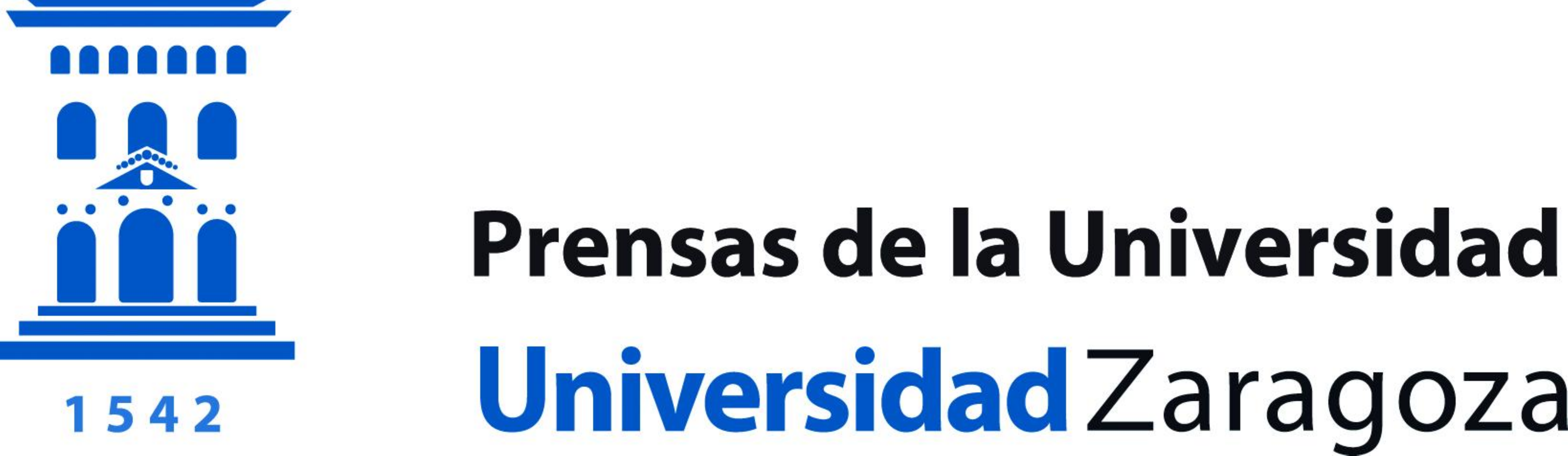}
\end{center}

\newpage
\newpage

\thispagestyle{plain}

\vspace*{15mm}
\begin{quotation}
\noindent {\bf \em Nothing travels faster than the speed of light with the possible exception of bad news, which obeys its own special laws.}\\

\noindent \emph{Douglas Adams}, ``The Hitchhiker's Guide to the Galaxy''\\[3cm]
\begin{flushright}
{\bf \em A mis padres}
\end{flushright}
\end{quotation}


\thispagestyle{empty}\
\mbox{}
\newpage


\parskip 0cm
\tableofcontents 
\markboth{Contents}{Contents}
\parskip 0.2cm

\cleardoublepage
\renewcommand{\chaptermark}[1]{\markboth{#1}{}}
\renewcommand\bibname{Bibliografía}
\chapter*{Introducción}\label{chap:introesp}
\chaptermark{Introducción}

Uno de los principales retos en el campo de la computación cuántica es la fabricación de un ordenador cuántico escalable.  La posibilidad de crear una generalización cuántica de un ordenador clásico fue propuesta por R. Feynman en 1982 \cite{Feynman1982} cuando discutía el problema de la simulación eficiente de sistemas cuánticos.  En un ordenador cuántico, la unidad de información de los ordenadores clásicos, el bit, es sustituido por su análogo cuántico, el qubit.  Mientras un bit clásico solo puede estar en uno de dos estados, 0 o 1, un bit cuántico puede almacenar cualquier superposición cuántica de estos, como por ejemplo $\alpha |1\rangle + \beta |0\rangle$.  La naturaleza cuántica de los qubits los hace componentes ideales para la simulación de sistemas cuánticos reales donde sistemas de computación clásicos experimentarían una ralentización exponencial al aumentar el numero de sistemas interactuantes.  Además, la habilidad de poder manejar estados superposición se presta a una especie de procesamiento paralelo que permite la implementación de algoritmos y procedimientos que serían imposibles con tecnologías clásicas.

Mas de tres décadas han pasado desde la propuesta original de Feynman y mucho trabajo, tanto teórico como aplicado, se ha hecho en el intento de hacer realidad esta propuesta.  Los principales obstáculos que han de ser superados si se ha de construir un ordenador cuántico son lo siguientes

\begin{enumerate}
\item Encontrar sistemas físicos con propiedades adecuadas que les permitan almacenar bits cuánticos.  El sistema en cuestión debe en algún régimen comportarse como un sistema cuántico de dos niveles.  Hay muchos sistemas físicos que de forma natural ya presentan esta característica, como por ejemplo espines nucleares, atómicos o moleculares o sistemas dipolares eléctricos.  Por otra parte, muchos sistemas artificiales pueden ser fabricados con las cualidades deseadas.  Algunos ejemplos propuestos de qubits sintéticos son los conocidos como circuitos cuánticos, como qubits superconductores de flujo o de carga, o sistemas de \emph{dots} cuánticos semiconductores.
\item Encontrar mecanismos para inicializar, leer, escribir y manipular qubits individuales además de acoplar múltiples qubits produciendo entrelazamiento y puertas lógicas cuánticas.  Además de almacenar información, los qubits deben además permitir que esta información sea manipulada de forma efectiva.  Debe ser posible transferir esta información entre diferentes regiones de un sistema a través de lo que se conoce como un bus cuántico.  Uno de los métodos más directos para cumplir estos requisitos es usar radiación electromagnética.  Los fotones son de forma natural objetos cuánticos y proporcionan un medio excelente para transferir información cuántica.
\item Mantener la coherencia del qubit.  Los fenómenos cuánticos son tremendamente frágiles y son muy sensibles a perturbaciones externas.  Incluso pequeñas interacciones con el entorno pueden rápidamente destruir la coherencia cuántica del sistema y producir un comportamiento clásico.  Este factor compite directamente con la facilidad de manipulación del qubit ya que cuanto más aislado tenga que estar el sistema, más difícil será interaccionar coherentemente con él.
\end{enumerate}

En este contexto, tiene especial importancia el estudio de la interacción de fotones individuales con sistemas cuánticos candidatos a ser qubits.  El ejemplo arquetípico para el estudio de esta interacción son los sistemas de electrodinámica cúantica de cavidades (Cavity Quantum Electrodynamics o CQED en inglés)\cite{Raimond2001}.  Éstos consisten en un sistema cuántico de dos niveles acoplado a único modo de la radiación electromagnética, por ejemplo, a fotones almacenados en una cavidad.  La intensidad del acoplo fotón-qubit ha de superar las frecuencias de decoherencia ambos subsistemas.  Este régimen de acoplo fuerte permite que, durante el tiempo de vida del sistema, se pueda realizar un numero suficientemente elevado de operaciones.

Con el objeto de aumentar el acoplo entre los fotones de una cavidad y un qubit, se han investigado distintos tipos de cavidades y qubits que permiten alcanzar el régimen de acoplo fuerte \cite{Raimond2001,Meschede1985,Rempe1987,Thompson1992,Brune1996}.  Uno de los más interesantes es lo que ahora se conoce como electrodinámica cuántica de circuitos (Circuit QED), que consiste en usar tecnologías de circuitos integrados y superconductores para fabricar cavidades y qubits adaptados para aplicaciones en el procesado cuántico de información \cite{Blais2004}.  Uno de los sistemas más utilizados en Circuit QED es el qubit de carga \cite{Makhlin2001} (Cooper pair box) acoplado a través de la componente eléctrica del campo a un resonador coplanar superconductor \cite{Goppl2008}(superconducting coplanar waveguide resonator).  La geometría de una guía de ondas coplanar permite aumentar la densidad de energía electromagnética (y por consiguiente el acoplo) con respecto a una cavidad tridimensional estándar.  Este sistema presenta acoplo fuerte y permite la realización de operaciones sobre el qubit e incluso entrelazar múltiples qubits acoplados a la misma cavidad \cite{Wallraff2004,Majer2007,DiCarlo2009,Dicarlo2010}.

Sistemas candidatos a qubits que se acoplan al campo magnético, como por ejemplo sistemas de espines de estado sólido, también son interesantes por sus posibles aplicaciones para el almacenamiento de información cuántica y para proporcionar una interconexión entre fotones de radio-frecuencias con fotones ópticos \cite{Imamoglu2009,Wesenberg2009,Marcos2010}.  Experimentos realizados en los últimos años demuestran la posibilidad de acoplar coherentemente centros NV y P1 de diamante con circuitos cuánticos, como resonadores superconductores \cite{Schuster2010,Kubo2010,Amsuss2011} o qubits de flujo \cite{Zhu2011}.  Estos defectos en diamante actúan como sistemas de espín 1 y se acoplan colectivamente al circuito proporcionando un aumento en el acoplo proporcional a la raíz cuadrada de número de espines.  El acoplo fuerte se consigue gracias a este factor junto con la alta coherencia de los centros magnéticos (tiempos de coherencia de 1-2 ms a temperatura ambiente).  También se ha demostrado acoplo fuerte magnético, incluso a temperatura ambiente, entre radicales paramagnéticos de espín 1/2 y cavidades de microondas tridimensionales \cite{Chiorescu2010,Miyashita2012}.

El uso qubits de espín abre también la posibilidad a otros sistemas algo más complejos.  Los ``single molecule magnets'' (moleculas iman ó SMMs) \cite{Christou2000,Gatteschi2003,Bartolome2013} son moléculas organometálicas formadas por un núcleo de alto espín rodeado de ligandos orgánicos que, de forma natural, se organizan en cristales moleculares.  El alto valor de su espín puede permitir que el régimen de acoplo fuerte sea más fácil de alcanzar y podría llevar a su aplicación como memorias cuánticas \cite{Imamoglu2009,Wesenberg2009,Marcos2010}. Los SMMs son también interesantes por que permiten manipular químicamente sus propiedades para distintas aplicaciones.  Si nos restringimos solo al campo de la información cuántica, estos sistemas pueden actuar no sólo como qubits individuales \cite{Meier2003,Troiani2005} sino que se pueden diseñar para que una única molécula contenga múltiples qubits débil o fuertemente entrelazados \cite{Timco2009,Candini2010,Aromi2012}, para actuar como puertas lógicas  \cite{Luis2011} o como simuladores cuánticos \cite{Santini2011}.

\begin{figure}[tb]
\centering
\includegraphics[width=0.7\columnwidth]{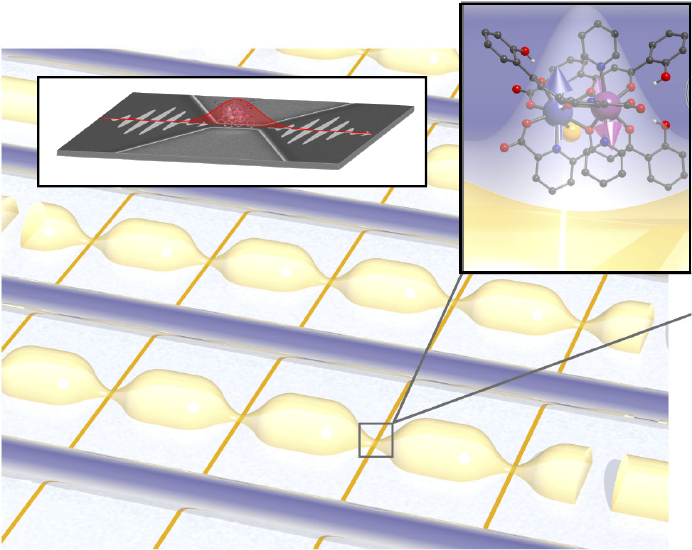}
\caption{Posible arquitectura para un procesador cu\'antico de spin.  Los qubits de spin se colocarían a lo largo de una guia de ondas coplanar superconductora en las posiciones donde el campo magnético es máximo para el modo de la cavidad considerado.  Cada spin se podría sintonizar para estar en resonancia o no mediante un campo magnetico aplicado localmente utilizando un cable o micro-bobina.  Entrelazamiento entre disntintos qubits en una misma cavidad tambien sería viable.}\label{fig:fantasyesp}
\end{figure}

Muchas de las aplicaciones de SMMs más sofisticados requieren acoplar moléculas individuales al campo magnético de un resonador, un reto aún más complicado que el acoplo a cristales macroscópicos.  Si este límite fuera alcanzable, se podrían utilizar circuitos superconductores para manipular y transferir coherentemente información entre qubits de espín, proporcionando una arquitectura adecuada para implementar un procesador cuántico de espines.  Si nos inspiramos en los sistemas de Circuit QED, podemos imaginar construir un sistema similar al representado en la figura \ref{fig:fantasyesp}.  Se colocarían moléculas individuales en las posiciones de la cavidad donde haya máximo campo magnético para el modo deseado de las microondas.  Los espines podrían entonces sintonizarse individualmente utilizando campos magnéticos locales para ponerlos en resonancia con la cavidad e intercambiarían información.

El principal objetivo en esta tesis, es comprobar la viabilidad de propuestas como la descrita en la figura \ref{fig:fantasyesp} y dar los primeros pasos hacia su realización.  Los acoplos alcanzables en la actualidad para espines individuales son todavía demasiado bajos como para superar sus los límites impuestos por sus tiempos de coherencia.  Esto nos deja con dos alternativas para mejorar el acoplo:
\begin{enumerate}
\item Encontrar sistemas de espín con mejores propiedades (tanto acoplo más fuerte como mejor coherencia)
\item Modificar las cavidades para aumentar el campo magnético y, por tanto, el acoplo.
\end{enumerate}

Teniendo esto en cuenta, los objetivos concretos de la tesis son:
\begin{enumerate}
\item Establecer que características deben tener los sistemas de espín para acoplarse fuertemente con un circuito cuántico
\item Estudiar las propiedades de diferentes familias de SMMs y ver si cumplen con las condiciones requeridas para el acoplo fuerte, tanto en forma de cristales moleculares como de moléculas individuales.
\item Diseñar nuevos resonadores de Circuit QED para concentrar el campo de microondas en determinadas regiones, optimizando así la interacción con pequeños grupos de espines o moléculas magnéticas.  Explorar también si estas modificaciones cambian de forma sustancial las propiedades básicas del resonador.
\end{enumerate}

La organización de esta tesis se detalla a continuación:
\begin{itemize}
\item Primeramente, en el capítulo \ref{chap:tech}, presentamos un resumen de las técnicas experimentales principales utilizadas a lo largo de este trabajo.
\item En el capítulo \ref{chap:Theo0} introducimos los conceptos básicos involucrados en la electrodinámica cuántica de cavidades y circuitos y resumimos como pueden ser aplicados a la computación cuántica.  Presentamos una propuesta para un procesador cuántico de espín y discutimos los principales obstáculos para su realización.
\item En el capítulo \ref{chap:Theo1} introducimos las moléculas imán (Single Molecule Magnets, SMMs) como posibles candidatos para ser qubits de espín.  Exploramos su Hamiltoniano de espín y discutimos las cualidades deseables que deben tener para su aplicación como qubits.  Utilizando métodos numéricos simulamos el campo rf generado por un resonador coplanar y estimamos su acoplo a sistemas de SMM conocidos.
\item En el capítulo \ref{chap:SIMs}, se introduce un tipo específico de SMM, los imanes de un solo ion (Single Ion Magnets, SIMs).  Se discuten las propiedades específicas que hacen que esta familia sea atractiva para la computación cuántica y dos ejemplos, \ce{GdW30} y \ce{TbW30}, se estudian en detalle.  Finalmente, los acoplos esperados y condiciones de operación de estos SIMs son comparados con los de SMM convencionales.
\item Después, en el capítulo \ref{chap:SIMs},  discutimos el diseño y fabricación de resonadores coplanares superconductores y damos detalles sobre el rendimiento y características de los resonadores fabricados para este trabajo.  También fabricamos y medimos los efectos de constricciones nanométricas en la línea central de nuestros resonadores y estudiamos sus efectos sobre los parámetros de transmisión.
\item En el capítulo \ref{chap:samp} presentamos medidas de espectroscopía sobre las distintas muestras magnéticas utilizando ambos resonadores y guías de onda coplanares y comparamos las medidas con los valores esperados según los modelos teóricos.  Finalmente comprobamos el rendimiento de una constricción nanométrica a la hora de hacer espectroscopía de muestras micrométricas.
\item Finalmente, el capitulo \ref{chap:concl} se dedica a resumir nuestras conclusiones.
\end{itemize}

\bibliographystyle{h-physrev3}
\bibliography{mybiblio}

\renewcommand{\chaptermark}[1]{\markboth{\chaptername \ \thechapter.\ #1}{}}
\renewcommand\bibname{References}
\cleardoublepage
\pagenumbering{arabic}
\chapter{Introduction}\label{chap:intro}

One of the main challenges in the field of quantum information is the fabrication of a scalable quantum computer.  The possibility to realize a quantum generalization of a classical computing system was proposed in 1982 by R. Feynman \cite{Feynman1982} when addressing the issue of the efficient simulation of quantum systems.  In a quantum computer the classical unit of information, the bit, is substituted by its quantum analogue, the qubit.  While a classical bit can be in one of two states, 0 or 1, a quantum bit can store any quantum superposition state, such as $\alpha |1\rangle + \beta |0\rangle$.  The quantum nature of qubits would make them ideal building blocks for the simulation of real quantum systems where classical systems would experience exponential slowdown when scaling up the number of interacting systems.  Also, the ability to manage superposition states gives rise to a kind of parallel processing allowing the implementation of algorithms and procedures that would be impossible with classical technologies.

Over three decades have passed since Feynman's original proposal and much work has been done, both theoretical and applied, attempting to make this proposal a reality.  The main obstacles that need to be overcome if a quantum computer is to be built can be broadly summarized as follows:
\begin{enumerate}
\item Find physical systems with suitable properties to store quantum bits.  In a general sense, the system in question should be, in some regime, equivalent to a quantum two level system.  There are many physical systems that naturally already present this characteristic, such as nuclear, atomic or molecular spins or electric dipole systems.  On the other hand, many artificial systems can be fabricated with the desired qualities.  Some examples of proposed synthetic qubits are those known as quantum circuits, like superconducting charge or flux qubits, or semiconducting quantum dots.
\item Find mechanisms to initialize, read-out, write and manipulate individual qubits as well as to couple multiple qubits producing entanglement and creating quantum logic gates.  Besides storing information, qubits must also allow this information to be manipulated effectively.  Also, it must be possible to transfer information between different areas of the system through what is known as a quantum bus.  One of the more natural methods to achieve many of these requirements is to use electromagnetic radiation.  Photons are naturally quantum objects and provide an excellent means for transferring quantum information.
\item Maintain the qubit quantum coherence.  Quantum phenomena are extremely fragile and very sensitive to external perturbations.  Even small interactions with the surrounding environment can quickly destroy the quantum coherence of the system and produce a classical behavior.  This factor competes directly with the ease of manipulation of a qubit since the more isolated the system needs to be from the environment, the harder it will be to coherently interact with the system.
\end{enumerate}

With these obstacles in mind, the study of the interaction of individual photons with candidate qubit systems is of special importance.  The archetypal example for this type of interaction are systems based on Cavity Quantum Electrodynamics (CQED) \cite{Raimond2001}.  These consist of a quantum two level system coupled to a single mode of electromagnetic radiation, for instance, the photons stored in a cavity.  For a system of this type to be useful for quantum computing, the photon-qubit coupling intensity must surpass the decoherence rates of both the qubit and the cavity.  This is known as the strong coupling regime and would allow a large number of quantum operations to be performed in the system's lifetime.

With the objective of attaining larger and larger couplings between cavity photons and qubit systems, many different cavity and qubit combinations have been studied, some of which allow the strong coupling regime to be reached. \cite{Meschede1985,Rempe1987,Thompson1992,Brune1996,Raimond2001}.  One of the more promising approaches is the field known as Circuit Quantum Electrodynamics (Circuit QED).  This approach consists in using microwave integrated circuit technologies and superconductors to fabricate cavities and qubits specially adapted to the quantum processing of information \cite{Blais2004}.  One of the more successful systems is the charge qubit \cite{Makhlin2001} (or Cooper pair box) coupled to a superconducting coplanar waveguide resonator through the electric component of the electromagnetic field \cite{Goppl2008}.  The geometry of a coplanar waveguide allows the electromagnetic energy density (and consequently the coupling) to be increased when compared to a three dimensional microwave cavity.  This system presents strong coupling and allows operations to be performed on a qubit as well as establishing coupling between multiple qubits and entanglement through the cavity \cite{Wallraff2004,Majer2007,DiCarlo2009,Dicarlo2010}.

Candidate qubit systems that couple to the magnetic field component, such as solid state spin ensembles, are also interesting for their possible applications for quantum information storage and because they can provide an interconnect between radio frequency and optical photons \cite{Imamoglu2009,Wesenberg2009,Marcos2010}.  Experiments performed in recent years have demonstrated the possibility of coherently coupling ensembles of NV and P1 centers in diamond to superconducting resonators \cite{Schuster2010,Kubo2010,Amsuss2011} or flux qubits \cite{Zhu2011}.  These diamond defects act as $S=1$ spins and collectively couple to the quantum circuit providing a $\sqrt{N}$ enhancement to the coupling, $N$ being the number of spins.  The strong coupling regime is achieved in this case thanks to this enhancement and to the excellent coherence time of these magnetic centers (up to 1-\SI{2}{\milli\second} at room temperature).  Strong magnetic coupling has also been demonstrated between spin $1/2$ paramagnetic radicals and three dimensional microwave cavities \cite{Chiorescu2010,Miyashita2012}.

The use of spin qubits also opens up the possibility of using somewhat more complex magnetic systems.  Many magnetic systems present important qualitative differences from the usual qubit candidates and could provide interesting possibilities for quantum information processing.  In particular, Single Molecule Magnets (SMMs) \cite{Christou2000,Gatteschi2003,Bartolome2013} are a type of organometallic molecule consisting of a high spin magnetic core surrounded by organic ligands that naturally form molecular crystals.  The high spin value should contribute to attain stronger coupling regimes and allow these crystals to be used as quantum memories \cite{Imamoglu2009,Wesenberg2009,Marcos2010}.  SMMs are also interesting because they can be chemically engineered to fulfill many different roles.  Restricting ourselves to quantum information applications, these molecules could not only act as individual qubits \cite{Meier2003,Troiani2005}, but also be designed to embody multiple weakly or strongly coupled qubits \cite{Timco2009,Candini2010,Aromi2012} performing as quantum logic gates \cite{Luis2011} or quantum simulators \cite{Santini2011}.

\begin{figure}[tb]
\centering
\includegraphics[width=0.7\columnwidth]{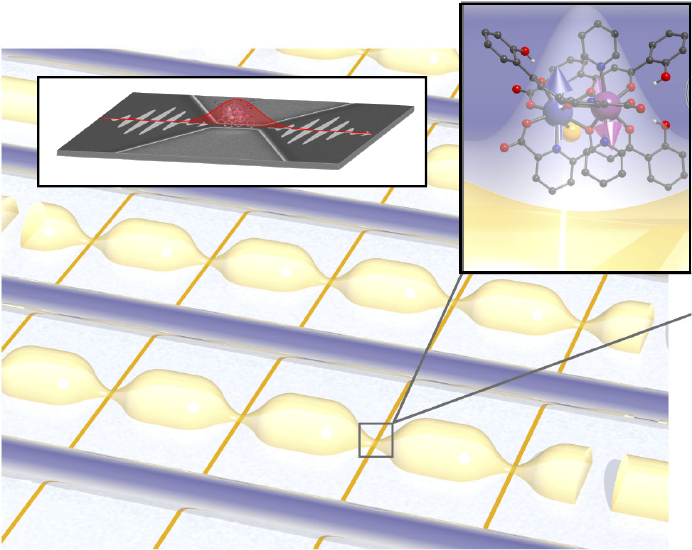}
\caption{Possible architecture for an all-spin quantum processor.  Each spin would be placed along a coplanar waveguide resonator at locations where microwave the magnetic field is maximum for the desired cavity mode.  Each spin could be tuned into and out of resonance using locally applied magnetic fields through wires or microcoils.  Entanglement between multiple qubits in the same cavity would also be viable.}\label{fig:fantasy1}
\end{figure}

Many of these more elaborate applications of SMMs to quantum information require the coupling of individual molecules to a resonator's magnetic field, an even greater challenge than the coupling to macroscopic crystals.  If this limit were achievable, it would be possible to use superconducting circuits to coherently manipulate and transfer information between spin qubits therefore providing an adequate architecture to implement an all-spin quantum processor.  Taking inspiration from Circuit QED systems, we can imagine building a system similar to the one presented in figure \ref{fig:fantasy1}.  Individual molecules would be placed at cavity locations with maximum magnetic field for the desired mode.  Each spin could be tuned into and out of resonance with the cavity using local magnetic fields applied using wires or microcoils thus selectively allowing the transfer of information between molecules and cavity photons.

The main objective of this thesis is to test the viability of proposals like the one presented in figure \ref{fig:fantasy1} and take the first steps towards its realization.  The currently achievable couplings for single spins are still too small to overcome the limits given by their coherence times.  This leaves us two alternatives to improve the coupling:
\begin{enumerate}
\item Find spin systems with better properties, both with better coupling and better coherence.
\item Modify the cavities used to increase the  magnetic field, and therefore, the coupling.
\end{enumerate}

With this in mind, this thesis will address the following points:
\begin{enumerate}
\item Establish what characteristics spin systems should have in order to optimize their coupling to a quantum circuit.
\item Study the properties of different families of SMMs and check if they fulfill the required conditions for strong coupling, both as crystalline ensembles and as individual molecules.
\item Design new resonators for application in Circuit QED schemes to concentrate the microwave magnetic field in small regions.  This would optimize the coupling to small groups of spins or magnetic molecules.  Explore also whether or not these modifications introduce substantial changes to the basic resonator properties.
\end{enumerate}

Following this logical sequence, this thesis is organized as follows:
\begin{itemize}
\item Firstly, in chapter \ref{chap:tech}, we present a review of the main experimental techniques involved in this work.
\item In chapter \ref{chap:Theo0} we introduce the basic concepts involved in Cavity and Circuit QED and review how they can be applied to quantum computing.  We present a proposal for an all-spin quantum information processing and summarize the main obstacles for it.
\item In chapter \ref{chap:Theo1} we introduce Single Molecule Magnets (SMMs) as possible candidates to realize spin qubits.  We explore their spin Hamiltonian and discuss the desirable qualities for their application in quantum computing.  Using numerical methods we simulate the rf fields that are generated by a superconducting coplanar waveguide resonator and estimate its coupling to well known SMM systems.
\item A specific type of SMM is introduced in chapter \ref{chap:SIMs}, namely Single Ion Magnets (SIMs).  The specific properties that make this type of system attractive for quantum computation are discussed and two examples, \ce{GdW30} and \ce{TbW30}, are studied in detail.  Finally, the expected couplings and operating conditions of these SIMs are compared to those of standard SMMs.
\item In chapter \ref{chap:CPWG} we discuss the design and fabrication of superconducting coplanar waveguide resonators and give details on the performance and characteristics the resonators fabricated for this work.  We then fabricate and measure the effects of nanometric constrictions in the center line of our resonators and study the effect they have on the transmission parameters.
\item In chapter \ref{chap:samp} we present spectroscopy measurements of different magnetic samples using both resonators and open transmission lines and compare the measurements to the values predicted using theoretical models.  We test the performance of nanoscale constrictions in resonators when it comes to performing spectroscopy on micro-metric samples.
\item Finally, chapter \ref{chap:concl} is dedicated to summarizing our conclusions.
\end{itemize}

\bibliographystyle{h-physrev3}
\bibliography{mybiblio}

\chapter{Experimental Techniques}\label{chap:tech}

\section{Introduction}

In this chapter we summarize the main experimental techniques and equipment used throughout this thesis.  They can be broadly divided into the following categories:
\begin{itemize}
\item General magnetic sample characterization
\item Lithography and circuit fabrication techniques
\item RF circuit measurement
\item Local scanning probe systems
\item Software and computational techniques
\end{itemize}

\section{General magnetic sample characterization}

\subsection{Electron Paramagnetic Resonance}\label{sec:EPRtech}

Paramagnetic resonance is a form of spectroscopy in which an oscillating magnetic field induces magnetic dipole transitions between energy levels of a system of spins \cite{Pake1973}.  These resonances are usually in the radio frequency (RF) (1 MHz to 1 GHz) or microwave frequency range (1 GHz to 100 GHz).  Electron paramagnetic resonance (or EPR) is restricted to the study of magnetic dipoles of electronic origin, usually in the microwave range.  It is conceptually very similar to nuclear magnetic resonances (NMR) with the difference that NMR works with nuclear spins and, hence, much lower frequencies.  Most materials do not exhibit electronic paramagnetism and hence do not produce an EPR signal (although they generally produce NMR signals).  The electronic shells are usually filled and have no resultant electronic angular momentum or magnetic moment.  This makes EPR less widely used than NMR but it also means that, when its use is possible, it can offer very clear signals and a great deal of information on the sample.

The typical EPR setup consists of a resonant cavity that can be irradiated with a fixed frequency microwave source.  The paramagnetic sample is then placed inside the cavity and can be subjected to an external DC magnetic field, usually applied perpendicular to the cavity microwave field using large electromagnets.  In general, a paramagnetic ion or molecule has discrete energy levels usually characterized by an angular momentum quantum number.  In the absence of magnetic field, the energy levels will be degenerate save for possible zero-field splitting effects given by the relevant spin Hamiltonian.  The application of an external magnetic field further splits the energy levels through the Zeeman effect, making the level separations change as the field changes.  When the level splitting equals the photon energy of the cavity photons, a net absorption is observed.  The exact position and number of the absorption lines gives information on the zero-field Hamiltonian, its gyromagnetic ratios, as well as how the paramagnet interacts with its environment, such as nuclear spins (hyperfine splitting).

\begin{figure}[tbh]
\centering
\includegraphics[width=1.\columnwidth]{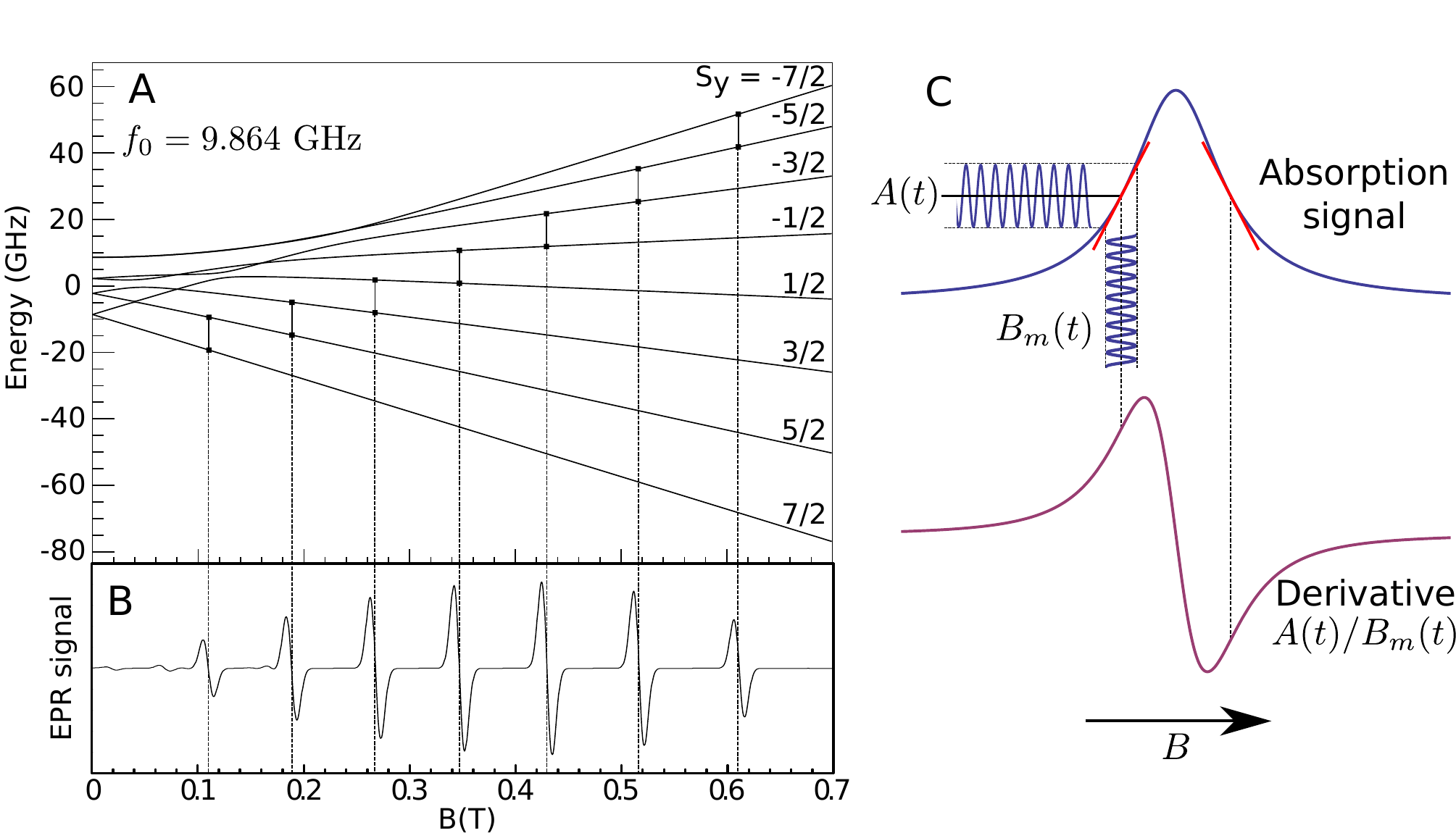}
\caption{Example of a continuous wave (CW) EPR spectrum.  Graph A shows the energy levels of a \ce{GdW30} polioxometalate molecule \cite{AlDamen2009,Cardona-Serra2012} as a function of an applied magnetic field (y-axis direction).  At high fields, the levels approximately correspond to well defined values of $S_y$. Graph B shows the simulated CW EPR spectrum for the same molecule.  The absorption lines coincide with magnetic fields for which there are level separations of 9.864 GHz, the chosen EPR cavity frequency (X-band).  Graph C shows the modulated magnetic field measurement scheme for CW EPR.  The signal recorded is the derivative (or slope) of the absorption curve.}\label{fig:epr}
\end{figure}

In this work we are specifically interested in paramagnetic rare earth ions.  These ions have incomplete $4f$ electronic shells, giving rise to a net atomic angular momentum and magnetic moment.  A typical energy level spectrum is shown in figure \ref{fig:epr} along with a simulated EPR spectrum.  The characteristic shape of the absorption lines in the EPR spectrum is due to the fact that most EPR systems use a small harmonic field modulation.  This allows greater sensitivity by using lock-in detection to reduce noise but produces the derivative instead of the absolute value of the absorption.

The EPR spectrum shown in figure \ref{fig:epr} is an example of a continuous wave EPR experiment.  In this type of measurement, the background DC field is swept slowly while the cavity is constantly irradiated.  A set of small coils apply the field modulation and a spectrum like that shown in figure \ref{fig:epr} is recorded.  In this scheme, the application of the RF field and the measurement are simultaneous.  Unless the field modulation has a very high frequency, this method only provides the splittings of the energy levels of the spin system, but doesn't give us any information of the spin dynamics or relaxation.

Information on the relaxation and coherence of a spin system can be obtained using pulsed EPR methods \cite{Brustolon2009}.  The prototypical experiment consists of measuring the signal emitted by the spin system after a series of microwave pulses.  The measured signal depends on the dynamics of the state the spin system was prepared in by the microwave pulses.  Pulsed EPR signals contain spectral information as well as spin relaxation information that can be extracted by further processing.  During the signal acquisition there are no microwaves applied but the pulse timing and the resonator bandwidth are critical factors.

\begin{figure}[tbh]
\centering
\includegraphics[width=0.75\columnwidth]{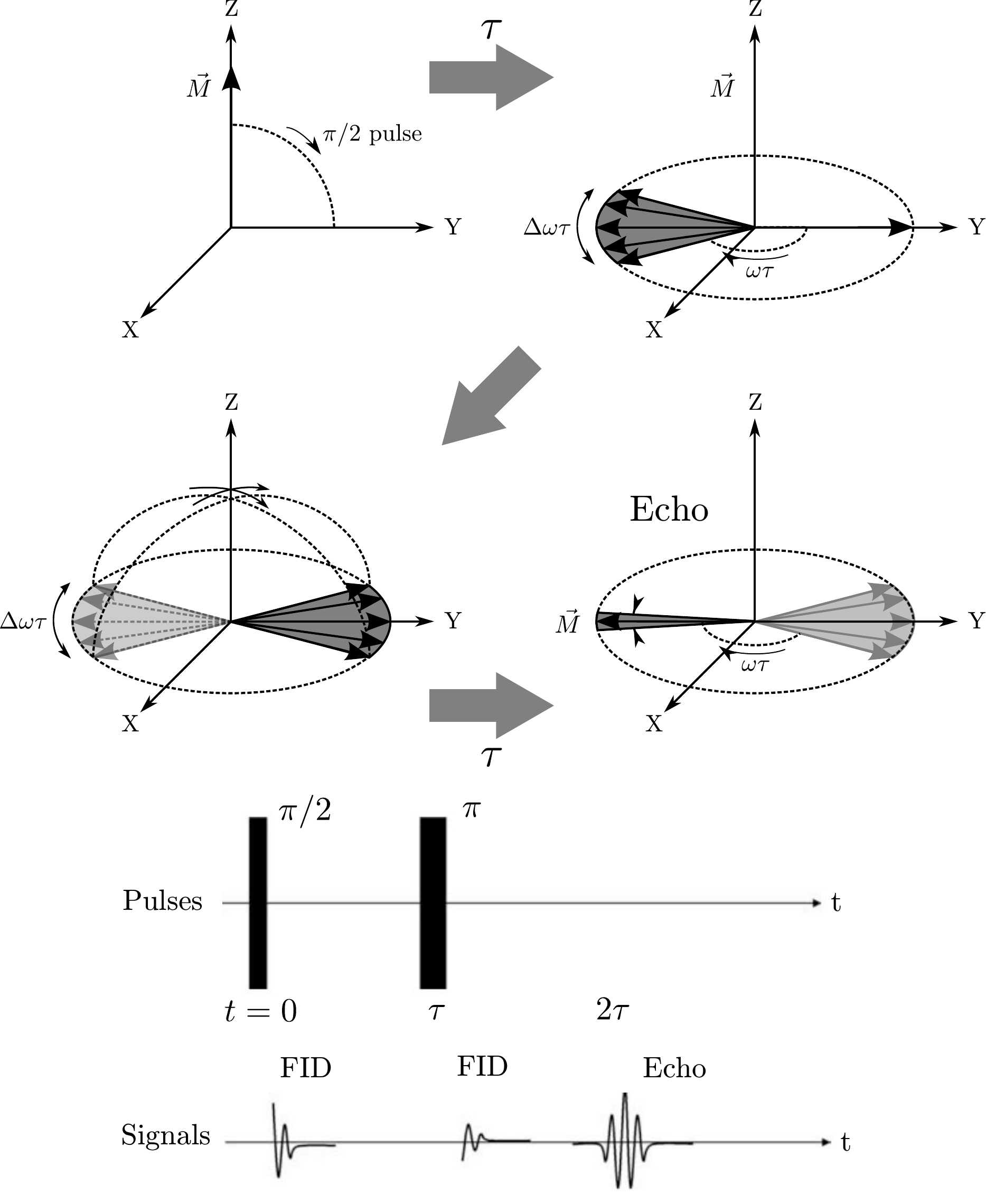}
\caption{Diagram of a spin echo procedure.  Fistly a $\pi/2$ pulse rotates the magnetization $\vec{M}$ onto the XY plane.  After a time $\tau$ the spin package will have spread according to its relaxation dynamics.  Then, a $\pi$ pulse is applied that reverses the spins placing the faster rotating components behind the slower components.  At time $2\tau$ the faster precessing spins catch up with the slower precessing spins generating the spin echo.}\label{fig:pulsedEPR}
\end{figure}

There are many pulse sequences used in pulsed EPR, but one of the most common is the \emph{two-pulse echo}, shown in figure \ref{fig:pulsedEPR}.  For this sequence, the background field of the EPR system is tuned to the transition we wish to study.  At equilibrium, the spin system will have its magnetization aligned with this DC magnetic field which we assume is along the Z direction.  Then, by applying an RF $\pi/2$ pulse, the spins can be rotated onto the XY plane.  The spins then precess at the transition Larmor frequency ($\omega = \omega_{\rm RF} = \gamma B_z$ where $\gamma$ is the gyromagnetic ratio).  The induced magnetization produces low intensity microwaves ($\simeq\si{\nano\watt}$) that can be monitored, but the signal decays (known as free induction decay) as each individual spin precesses at slightly different speeds than the rest.  After a time $\tau$ a $\pi$ pulse is applied that inverts all the spins directions, thus placing the spins with higher precession speeds behind the ones with a slower precession speed.  After a further $\tau$ period, the spin system produces an echo when the faster spins catch up with the slower spins.  The intensity of the echo is studied for different values of $\tau$.  From this measurement, information on the $T_2$ decoherence time can be extracted \cite{Brustolon2009}.  This procedure is schematically represented in figure \ref{fig:pulsedEPR}.

\subsubsection{Elexsys E-580 spectrometer}
\begin{figure}[tbh]
\centering
\includegraphics[width=0.7\columnwidth]{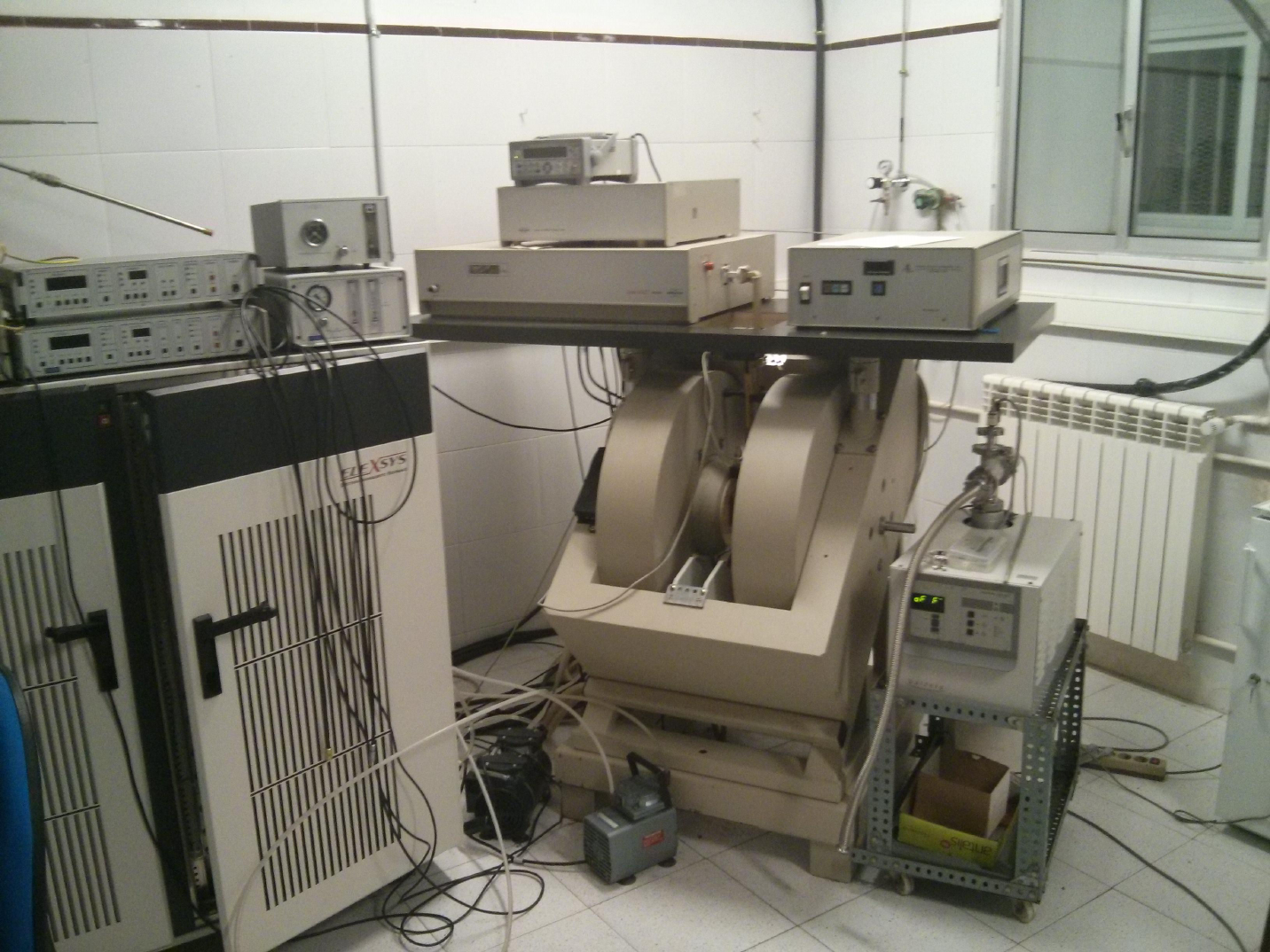}
\caption{Elexsys E-580 EPR spectrometer at the University of Zaragoza}\label{fig:brukerepr}
\end{figure}

The EPR system used for the measurements presented in this work is an Elexsys E-580 spectrometer by Bruker Corporation \cite{bruker} property of the Instituto de Ciencia de Materiales de Aragón (figure \ref{fig:brukerepr}).  It is capable of taking EPR measurements in both X-band (8-10 GHz) and Q-band (35 GHz) and in continuous wave mode and pulsed mode (X-band only).  For anisotropic samples, the microwave cavity can be fitted with a rotating stage mechanism that allows the automatic acquisition of spectra for different orientations.  It is also equipped with a helium flow cryostat that can cool the samples down to 5 K.  The electromagnet used for the DC fields can apply fields of up to 1.4 T.

\subsection{Physical Property Measurement System}\label{sec:PPMS}
The Physical Property Measurement System (PPMS) is a commercial system by Quantum Design \cite{qdppms} that allows the measurement of different physical properties (such as resistivity, heat capacity, magnetic properties, heat transport, etc.) of samples under variable temperature, magnetic field and pressure conditions.  Its architecture allows the use of different sample stages for different applications and allows for the possibility of custom built accessories.  Two PPMS systems are available for use at the Servicio de Medidas Físicas at the Universidad de Zaragoza \cite{smf}.

\begin{figure}[tbh]
\centering
\includegraphics[width=0.75\columnwidth]{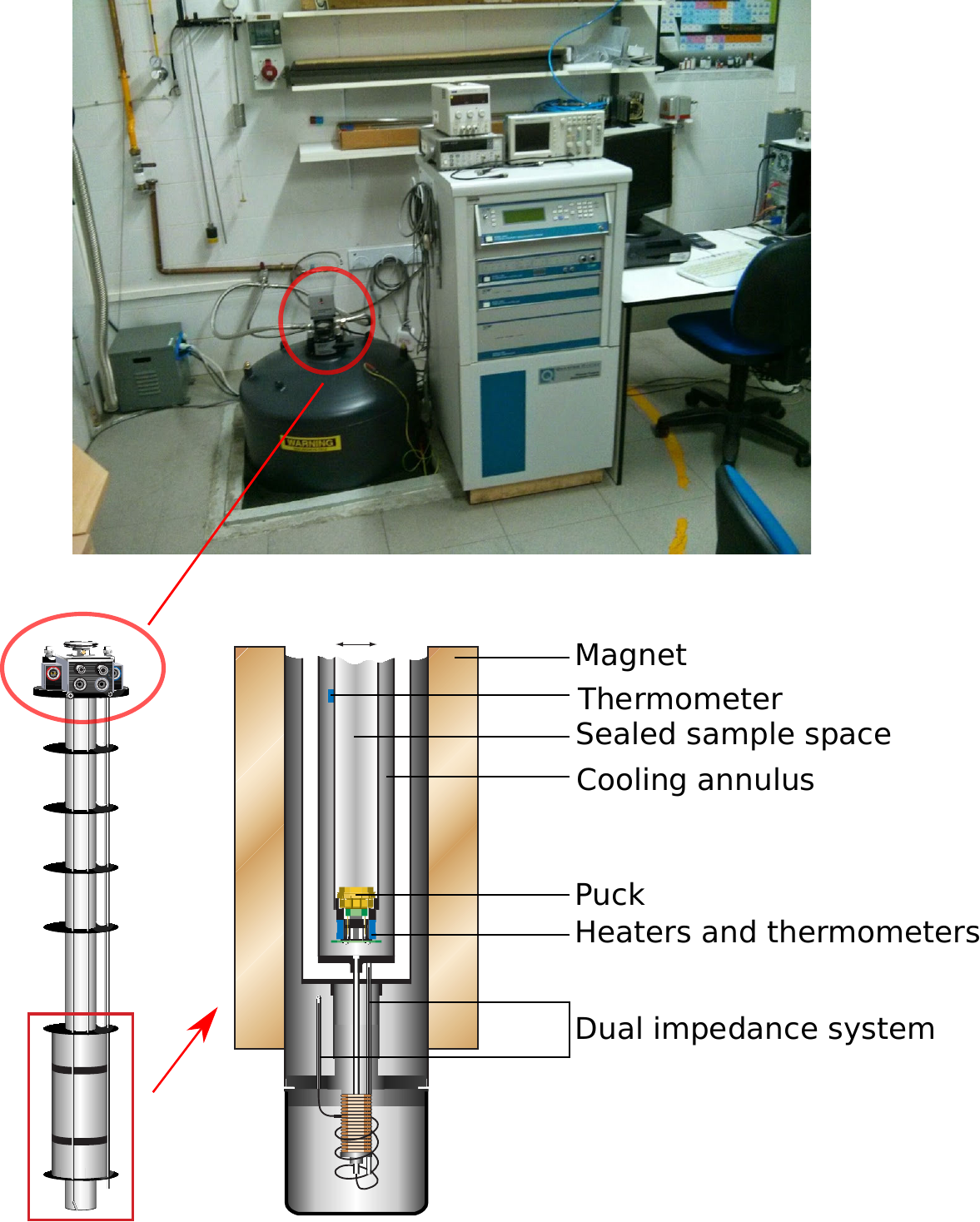}
\caption{PPMS dewar, model 6000 and PPMS probe schematics}\label{fig:ppms}
\end{figure}

The system consists mainly of a liquid helium dewar with a probe inserted into the helium bath. This probe integrates a \ce{^{4}He} cryostat \cite{Pobell2007}, sample chamber and superconducting magnet.  The cryostat allows the sample chamber's temperature to be controlled and varied in the range of \SI{1.9}{\kelvin}-\SI{400}{\kelvin} and allows continuous operation as long as the liquid helium supply is maintained.  The superconducting magnet has a maximum field of 9 or 14 T for our available models. The sample a chamber also has electrical connections at the bottom (figure \ref{fig:ppms}) which are used to interact with the sample and take the measurements.  The different sample holders (or pucks) are made to fit this connection.

The entire system is controlled through an external electronic system (model 6000) that includes all the electronics necessary for the control of the chamber pressure, temperature and field.  Additional modules are installed for each of the measurement options desired.  An external PC and software allow simple programming of automated measurement sequences.

\begin{figure}[tbh]
\centering
\includegraphics[width=0.75\columnwidth]{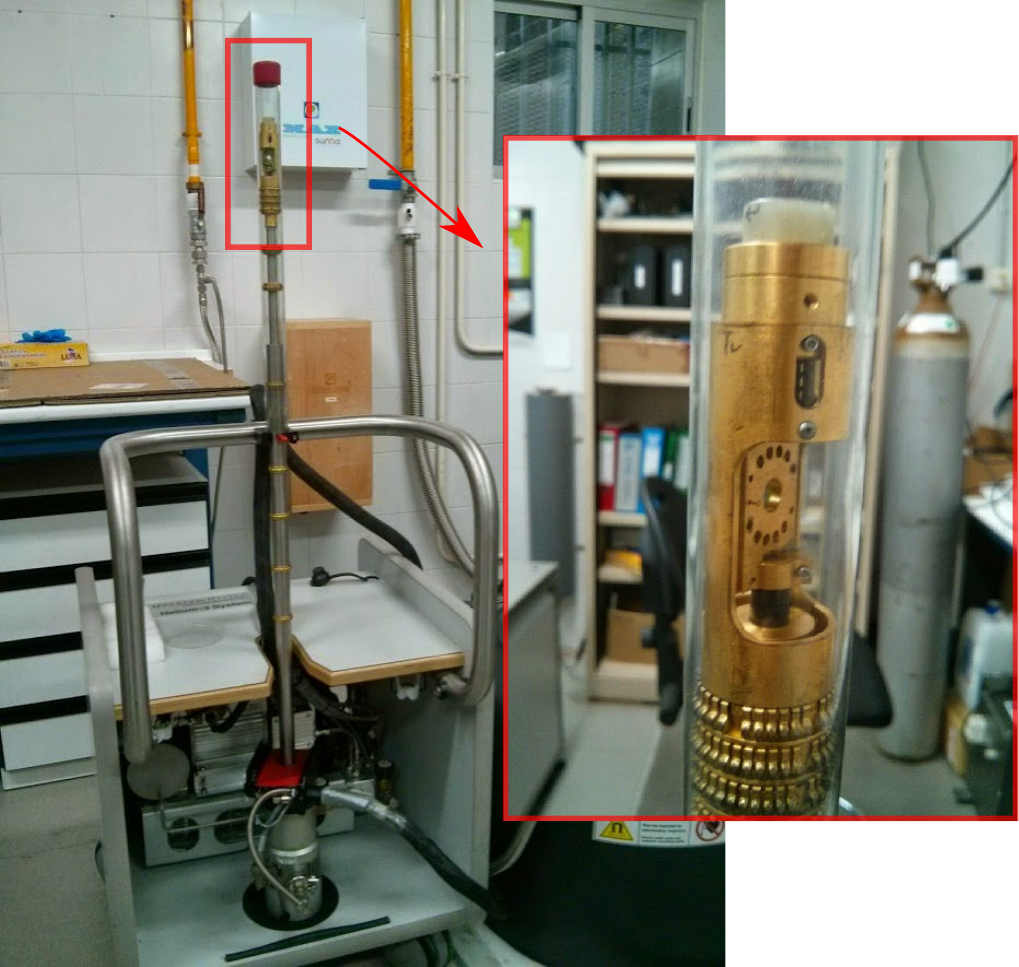}
\caption{\ce{^{3}He} cryostat insert for PPMS system}\label{fig:He3insert}
\end{figure}

One of the options we will commonly use is a \ce{^{3}He} cryostat \cite{Pobell2007} that can be inserted into the PPMS probe to achieve temperatures down to \SI{350}{\milli\kelvin} (figure \ref{fig:He3insert}).  The insert has different electrical connections to samples than the base system and specific pucks must be used.

We now give some additional details for the two measurement types we perform on these systems:  Resistivity and heat capacity.

\subsubsection{DC resistivity measurements on the PPMS system}
DC resistivity measurements are the most basic measurements possible and require no additional options apart from the base PPMS system.  The measurement uses a standard 4 point resistance measurement to determine the resistivity given the sample geometry.  A given current is circulated from the $+I$ to the $-I$ lead and the voltage across the $+V$ to $-V$ leads is measured.  The resistance on the $\pm V$ branch is designed to be much higher than that of the $\pm I$ branch, thus guaranteeing that almost all the current circulates from $+I$ to $-I$ and that there are no contributions of the connecting wires to the potential difference $\Delta V$ on the $\pm V$ branch.  The resistance is then simply $\Delta V/I$.  The resistivity of the sample can be obtained knowing the sample length ($l$) between the $\pm V$ leads and the sample cross section ($S$) through the definition $\rho = R S/l$.

\begin{figure}[tbh]
\centering
\includegraphics[width=0.75\columnwidth]{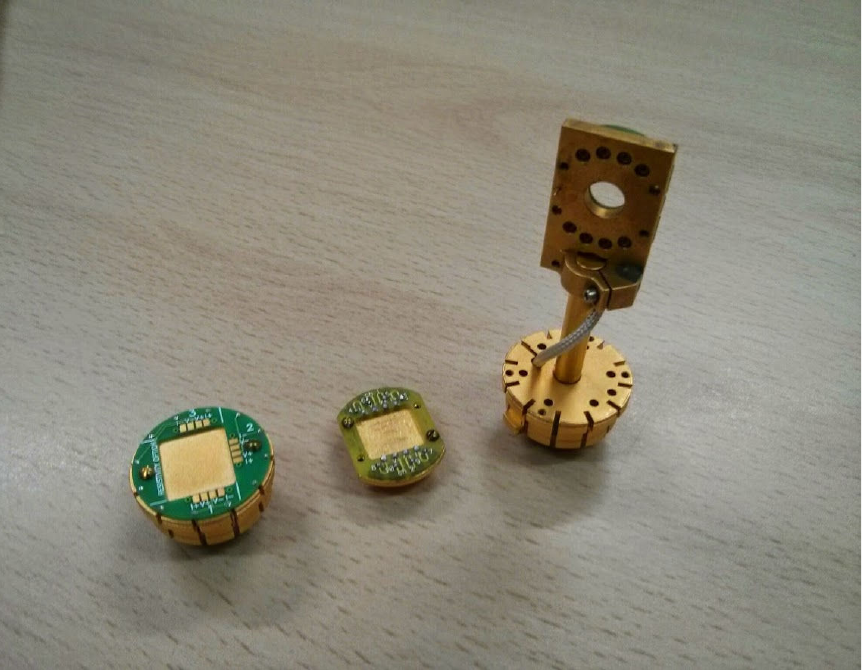}
\\[2mm]
\begin{footnotesize}
\begin{tabular}{|c|c|c|c|}
\hline
Mode & Current range & Voltage range & Frequencies \\
\hline
DC & $\pm 0.01-\SI{5000}{\micro\ampere}$ & $\pm 1-\SI{95}{\milli\volt}$ & DC or 50 Hz \\
\hline
ACT & $\pm \SI{0.02}{\micro\ampere}-\SI{2}{\ampere}$ & $\pm \SI{40}{\micro\volt}-\SI{5}{\volt}$ & $\SI{1}{\hertz}-\SI{1}{\kilo\hertz}$ \\
\hline
\end{tabular}
\end{footnotesize}
\caption{DC resistivity pucks for the PPMS and parameter limits.  From left to right: Standard puck, \ce{^{3}He} puck, in plane field adapter.}\label{fig:DCresist}
\end{figure}

The different resistivity pucks are shown in figure \ref{fig:DCresist}.  The standard puck has electrical contacts for up to three samples simultaneously.  It also allows measurement of Hall voltages on a fifth lead ($V_{\rm hall}$).  Samples are attached or glued to the flat puck surface and usually wire bonded to the contacts.  In the standard configuration the magnetic field is applied perpendicular to the puck surface.  To apply an in plane magnetic field, an adapter can be used to hold a \ce{^{3}He} type puck in the correct position.  The \ce{^{3}He} puck is very similar to the standard puck but is smaller and has only two measurement channels.

The PPMS also includes an optional AC transport module (ACT) can also be used to apply higher frequencies and currents.  It is specifically designed for measuring critical currents and IV curves for superconductors.  The current, voltage and frequency limits for both DC and ACT are given in figure \ref{fig:DCresist}.

\subsubsection{Heat capacity measurements on the PPMS system}\label{sec:cp}

The heat capacity option measures heat capacity at constant pressure \linebreak $C_p=\left(\frac{dQ}{dT}\right)_p$ using a relaxation technique \cite{Stewart1983}.  The heat capacity option controls the heat added and removed from the sample while monitoring the temperature changes.  A constant power is applied to the sample for a fixed time followed by a cooling period of the same duration.   A simple model predicts that the temperature $T$ of the calorimeter block (the calorimeter itself plus the sample) obeys the equation:
\begin{equation}
C_{\rm total} \frac{dT}{dt} = -K_w(T-T_b)+P(t);
\end{equation}
where $C_{\rm total}$ is the total heat capacity of the sample and platform, $K_w$ is the thermal conductance of the supporting wires, $T_b$ is the thermal bath temperature (puck temperature) and $P(t)$ the applied heat power (equal to $P_0$ during heating and $0$ during cooling).  The solution to this equation is given by exponential functions with time constants $\tau$ equal to $C_{\rm total}/K_w$.  Since $K_w$ is known from previous calibration experiments, this data enables the determination of $C_{\rm total}$ and, from it, the sample contribution.  This is the model commonly used by the PPMS and it assumes good thermal contact between the sample and platform.  More complex models with multiple time constants are available for cases when the thermal contact is poor.

\begin{figure}[tbh]
\centering
\includegraphics[width=0.8\columnwidth]{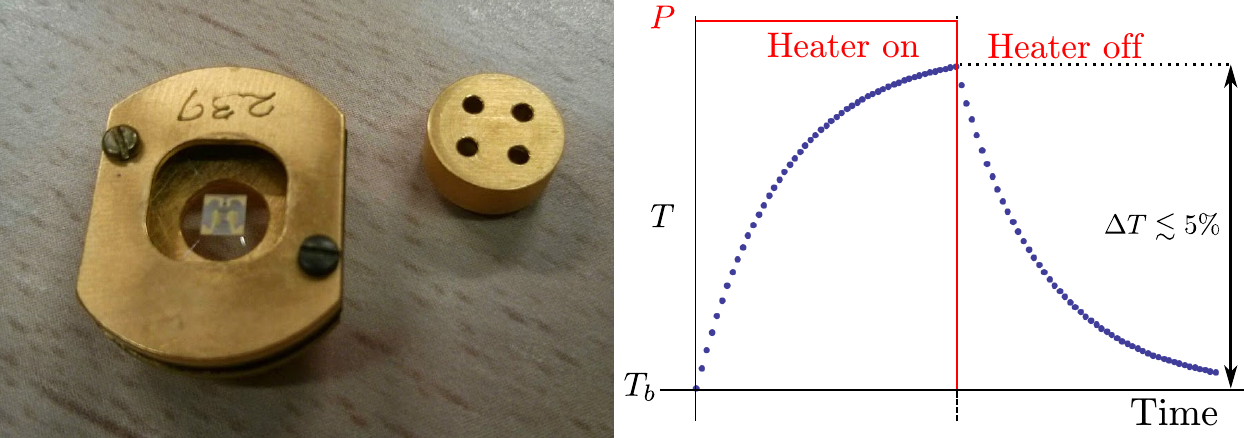}
\caption{Heat capacity puck for \ce{^{3}He} insert and typical temperature response during a measurement}\label{fig:CP}
\end{figure}

The sample is mounted on a small sapphire platform (about \SI{2}{\milli\meter} side) with a platform heater and a thermometer on the bottom side.  The wires connecting the heater and thermometer provide the only structural support to the platform and, since measurements are performed in high vacuum, the dominant thermal connection.  Power pulses are set to induce only small excursions from the bath temperature, of the order of $3-5\%$ of $T_b$ in order to avoid nonlinear effects.  The complete puck adequate for the \ce{^{3}He} system can be seen in figure \ref{fig:CP} along with a typical response curve while measuring.  To correctly evaluate the sample heat capacity, the heat capacity of the empty platform must be calibrated (addenda) as well as any adhesive material used (addenda offset).

The system specifications allow us to measure the heat capacity of relatively flat samples of between 1 and 200 mg mass at temperatures from 0.35 K to room temperature with or without an applied magnetic field.  The sensitivity is higher than adiabatic methods as it relies on the measurement of the time constant $\tau$.  Our samples are usually small crystals or powders that are pressed into flat pellets using a hand press ($\simeq \SI{1}{\milli\meter}$ and $\lesssim \SI{10}{\milli\gram}$).  They are attached to the platform using Apiezon N grease \cite{apiezonN} that provides good thermal contact.

\subsection{Magnetic Property Measurement System}\label{sec:mpms}

\begin{figure}[tbh]
\centering
\includegraphics[width=1.\columnwidth]{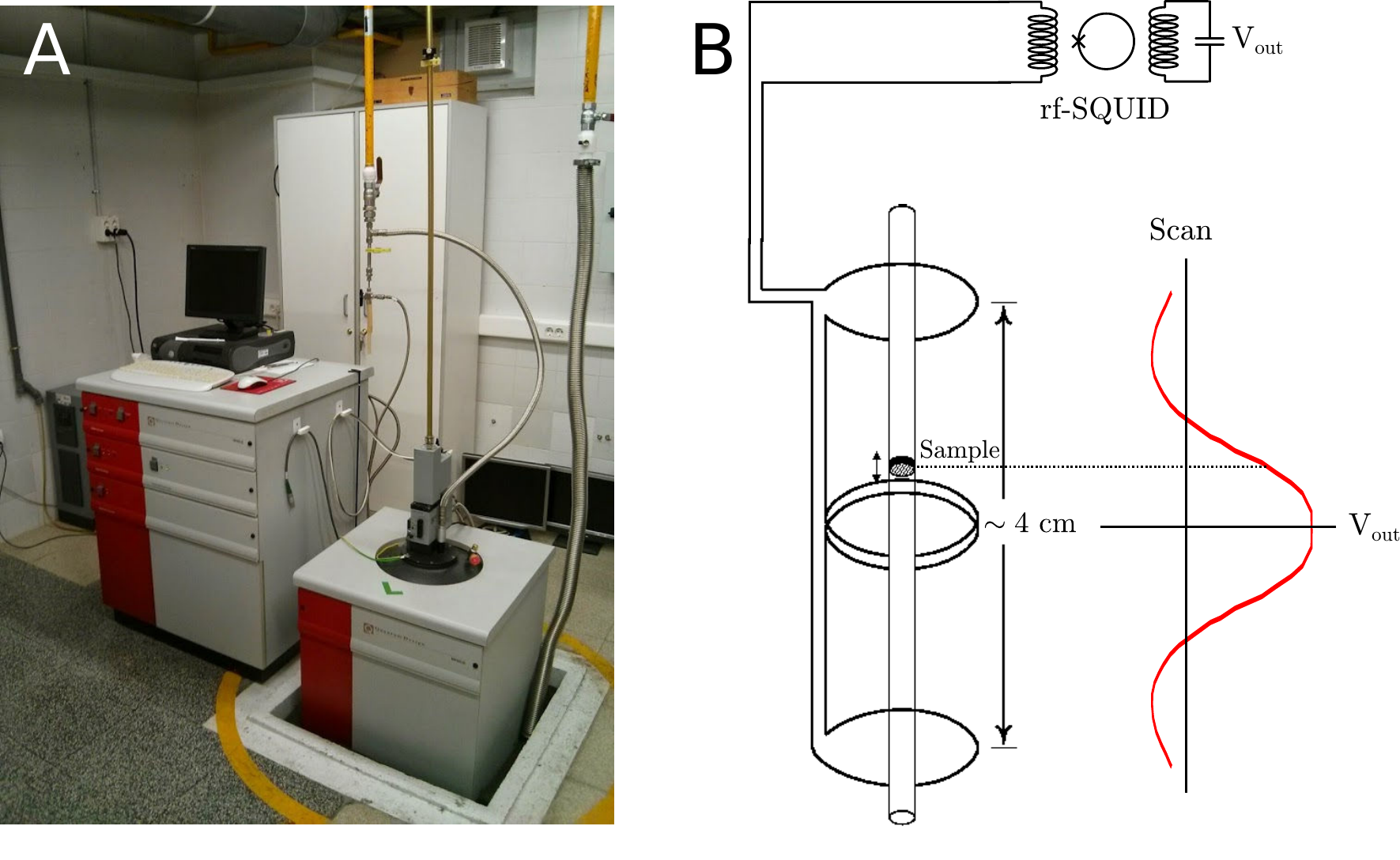}
\caption{A) MPMS-XL system at Servicio de Medidas Físicas (Universidad de Zaragoza). B) Diagram of the superconducting pickup coils of the second degree gradiometer connected to SQUID electronics}\label{fig:MPMS}
\end{figure}

The Magnetic Property Measurement System or MPMS is another commercial system by Quantum design \cite{qdmpms} that is designed specifically for measuring both DC and AC magnetic responses of a material.  The system used is a model MPMS-XL also provided by Servicio de Medidas Físicas (Universidad de Zaragoza) \cite{smf} as in the case of the PPMS system.  It is very similar to the PPMS system (section \ref{sec:PPMS}) except in that the probe contains additional components intended for the magnetic measurements.  It also operates as a \ce{^{4}He} cryostat and can operate at temperatures from $1.8-\SI{300}{\kelvin}$.  Its superconducting magnet can apply static fields of up to \SI{5}{\tesla} and AC magnetic fields of frequencies in the range $\SI{0.1}{\hertz}-\SI{10}{\kilo\hertz}$ and up to \SI{4}{Oe} in amplitude.

\begin{figure}[p]
\centering
\includegraphics[width=1.\columnwidth]{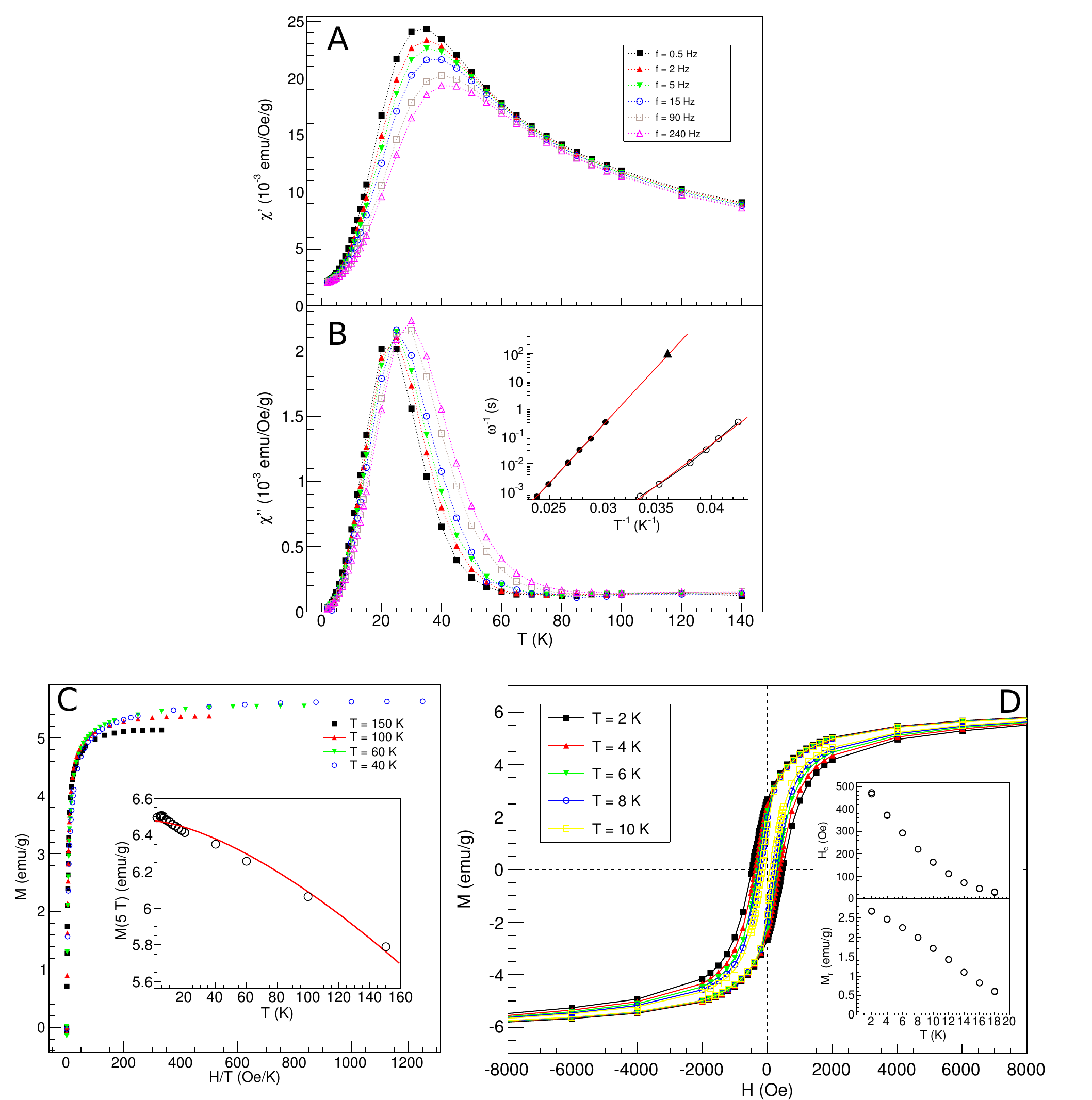}
\caption{AC susceptibility and DC magnetization of polycrystalline magnetoferritin \cite{Moro2014}.  Graphs A and B show the AC susceptibility as a function of temperature and for different AC field frequencies.  The AC field amplitude ($h_{\rm ac}$) used was \SI{4}{Oe}.  The graph A shows the real component ($\chi' = m'/h_{\rm ac}$) and graph B shows the imaginary component ($\chi''=m''/h_{\rm ac}$).  The inset shows a fit of the blocking temperature (maximum value of $\chi'$ for filled circles and $\chi''$ for open circles) as a function of the AC frequency to an Arrhenius Law, $\log{\omega} = A + B/T$ where $A$ and $B$ fitted constants related to the attempt time and the anisotropy energy barrier preventing the flipping of magnetoferritin magnetic moments.  The triangle in the inset is the value for the DC blocking temperature.  Graph C shows the DC magnetization as a function of $H/T$ for different temperatures while the inset shows the saturation magnetization as a function temperature.  Graph D shows hysteresis cycles for a collection of temperatures.  The insets show the cohercive field and remanent magnetization for all the measured hysteresis cycles.}\label{fig:ferritinDC}
\end{figure}

The main component of the MPMS system is the SQUID (Superconducting Quantum Interference Device) magnetometer.  SQUIDs are very sensitive magnetic flux-to-voltage transducers and can be used as low noise current amplifiers \cite{Martinis1985,Muck2010}.  This allows them to be used in the detection of very small magnetic signals by amplifying small currents generated by moving magnetized samples through superconducting coils.  The maximum sensitivity that can be achieved in our system is of \SI{2e-8}{emu}.  A basic schematic of the SQUID magnetometer can be found in figure \ref{fig:MPMS}.  The sample is mounted at the end of a thin rod that is attached to the sample motor.  This allows the sample to be moved vertically through a system of superconducting (NiTi) second-order gradiometer coils, i.e., four coils placed one above the other with the two central coils spooled in the opposite direction as the two external coils.  This configuration suppresses any contribution to the flux from external uniform fields.  If the sample on the other hand has a net magnetization, it will produce a current signal as is passes through each coil.  This current is then amplified using a pickup circuit that incorporates an rf-SQUID.  The voltage signal is proportional to the sample magnetization.  When operating in AC mode, a coil generates an AC magnetic field and a lock-in detection system is used to extract the real and imaginary components of the signal.

As in the PPMS case, the MPMS is controlled through external electronics and computer software that can automate many of the standard measurements and tasks.  Sequences of commands and measurements may be pre-programmed and run unattended.

As an example measurement we present measurements of samples of polycrystalline magnetoferritin done in the MPMS at the University of Zaragoza \cite{Moro2014}.  Ferritin is a protein with the approximate shape of a spherical cage that provides a confined vessel where guest species can enter and react to give a core with a defined shape and size.  In the case of magnetoferritin, the interior is filled with a maghemite core with an average diameter of \SI{5.7}{\nano\meter} for our samples.  Figures \ref{fig:ferritinDC}AB show measurements for an AC magnetic field while figures \ref{fig:ferritinDC}CD show DC magnetization and hysteresis cycles.  The experiments provide information on basic parameters, such as the magnetic moment and its distribution, as well as the spin lattice relaxation time scales.

\subsubsection{Rotating Sample Stage}

\begin{figure}[tbh]
\centering
\includegraphics[width=0.9\columnwidth]{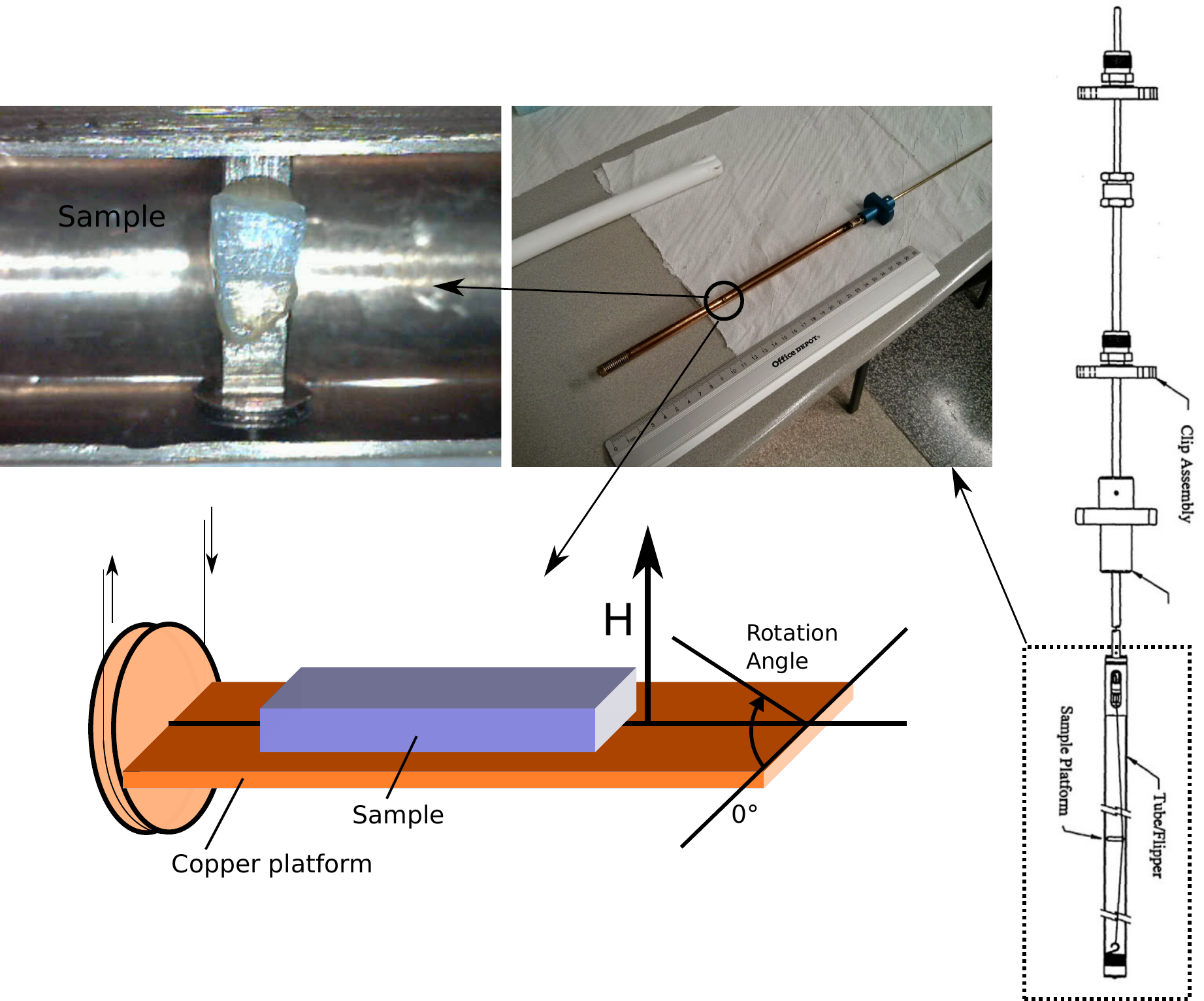}
\caption{Rotating sample stage for MPMS}\label{fig:mpmsrot}
\end{figure}

When measuring the magnetic properties of crystals it is often important to measure the response when the fields are applied along different directions with respect to the sample.  When working with powders or isotropic samples, these are usually placed in capsules inside a plastic tube that can be easily attached to the rod that the MPMS moves through the magnetometer coils (see figure \ref{fig:MPMS}).  The tube is chosen to be uniform and long enough to occupy all 4 coils giving a very low background contribution.  This setup however does not allow us to rotate the sample.  A different stage must be used when we wish to study anisotropic effects.

The rotating sample stage allows us to rotate a sample with respect to an axis perpendicular to the field direction (the magnetic field is vertical and the rotation axis horizontal in the laboratory frame).  It is shown schematically in figure \ref{fig:mpmsrot}.  The stage consists of a small copper platform within a copper tube (oxygen free).  The platform has a pulley attached to a thin wire that allows it to be rotated by motors at the head of the rod in \ang{0.1} steps.  The sample is usually attached to the platform using a thin layer of Apiezon N grease \cite{apiezonN}.

\begin{figure}[tbh]
\centering
\includegraphics[width=0.75\columnwidth]{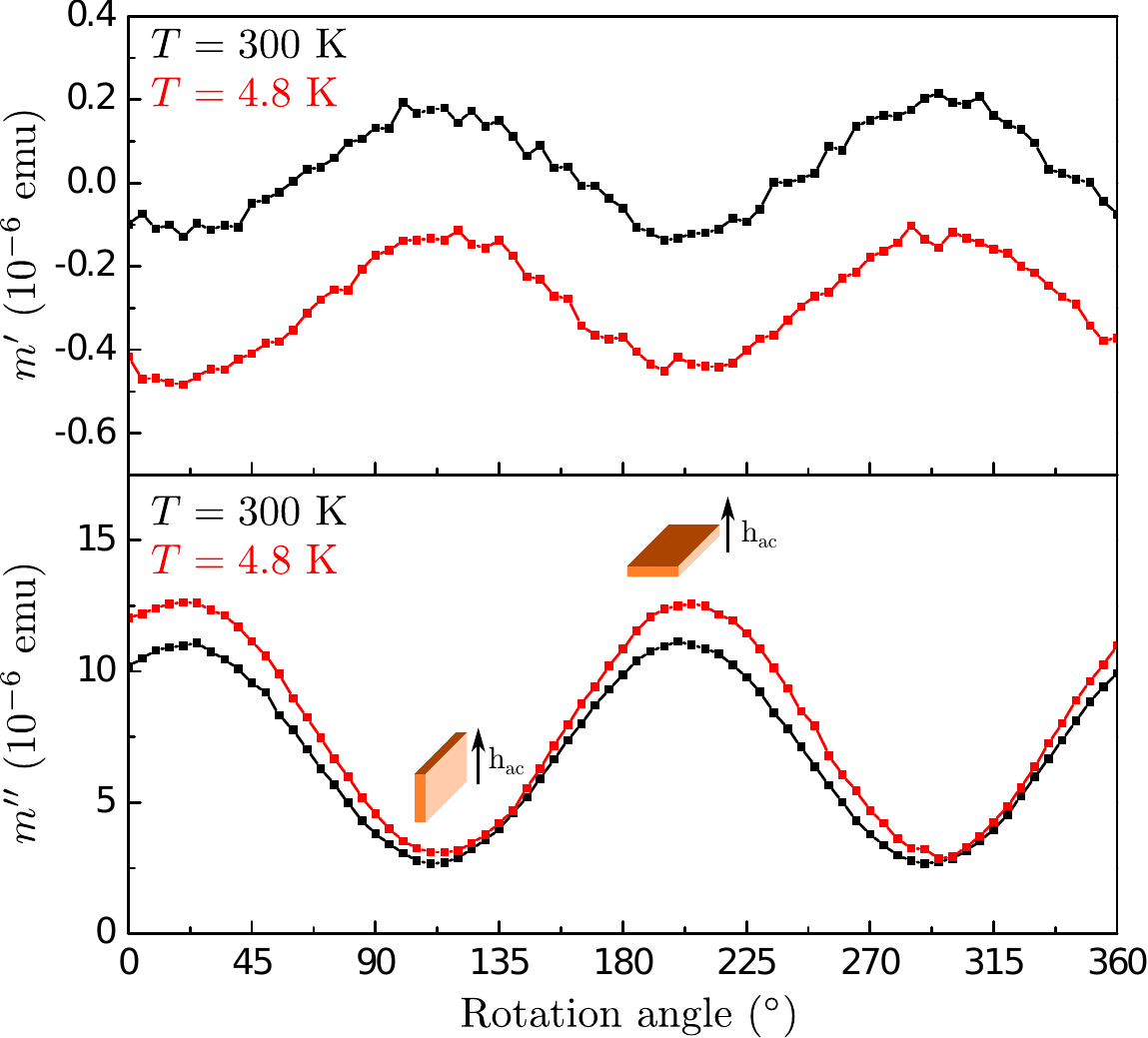}
\caption{Contribution of rotating sample stage to MPMS AC magnetization measurements at 976 Hz and $T=\SI{300}{\kelvin}$ and $T=\SI{4.8}{\kelvin}$.  Top graphs correspond to the real component (in phase) magnetization while the lower graphs correspond to the imaginary (reactive) component.} \label{fig:cuplat}
\end{figure}

The copper platform contributes a relatively small signal to measurements of magnetization.  However, when measuring AC signals at high frequencies (1 kHz and higher), the complex or reactive component of the magnetization has a characteristic contribution from the platform (see figure \ref{fig:cuplat}).  The signal is due to induction currents in the platform and they are maximum when the platform is perpendicular to the applied field.  Assuming our sample has small contributions at these frequencies, this allows us to easily find a reference position from which to measure the rotation angles.

\section{Lithography and circuit fabrication techniques}

\subsection{Ultra-Violet Photolithograpy}\label{sec:liftoff}
The microwave resonators and waveguides we study in later chapters are fabricated mainly through a standard ultra-violet (UV) photolithograpy procedure.  Here we detail the basic procedure and materials that were used in this phase.  The fabrication was done almost entirely using the facilities at the Instituto de Nanociencia de Aragón (Universidad de Zaragoza).

\begin{figure}[tbh]
\centering
\includegraphics[width=0.95\columnwidth]{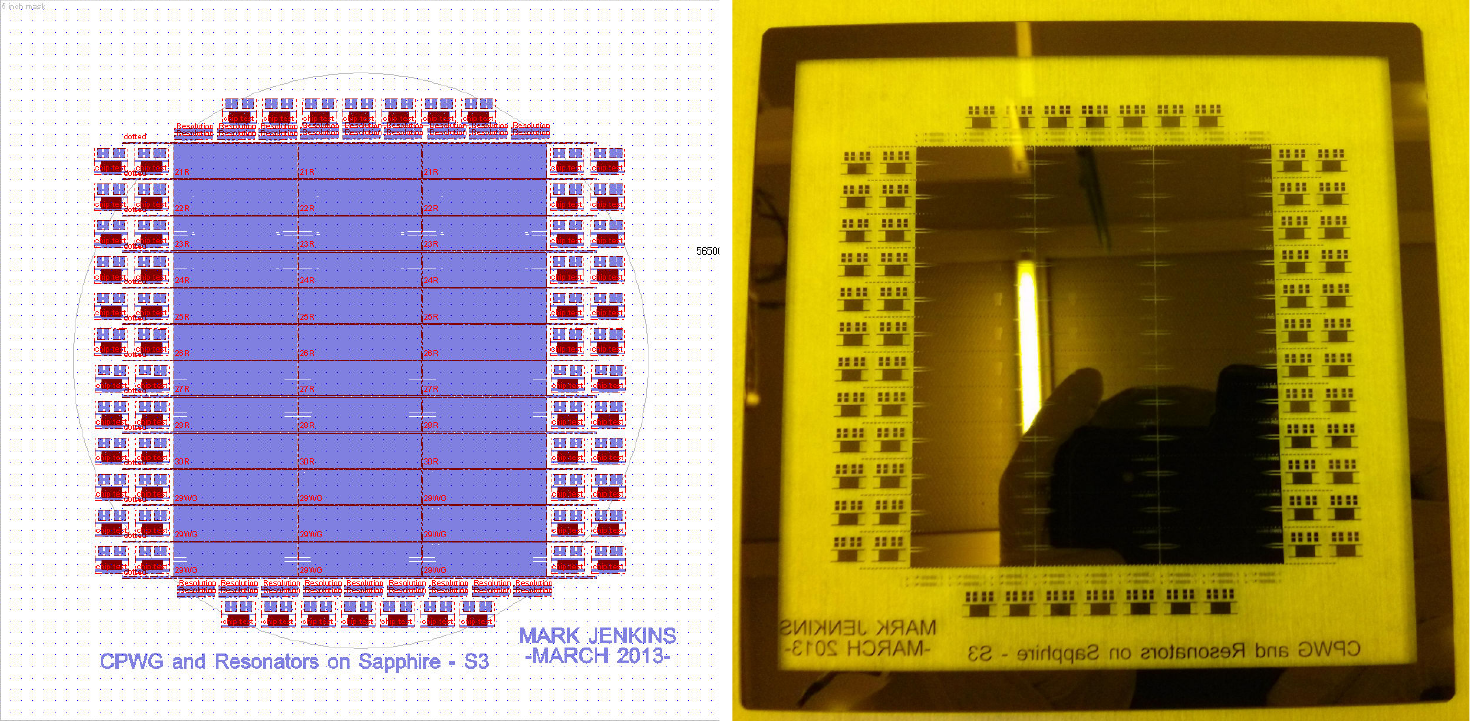}
\caption{UV photolithography mask CAD design and finished mask}\label{fig:masks}
\end{figure}

UV photolithography is a microfabrication technique that allows patterns to be transferred from a mask to a photosensitive resin or photoresist on a substrate.  Our mask patterns were designed using standard CAD software and provided to the company Delta Mask b.v. that produced the finished photomasks.  An example mask is shown in figure \ref{fig:masks}.  Different chemical and physical procedures then allow the pattern to be engraved in some way on the underlying substrate.  The minimum feature size depends on the desired depth of the pattern and on the materials and procedures involved.  However, under optimal conditions, the minimum feature sizes achievable are of the order of \SI{1}{\micro\meter}.

\begin{figure}[tbh]
\centering
\includegraphics[width=0.95\columnwidth]{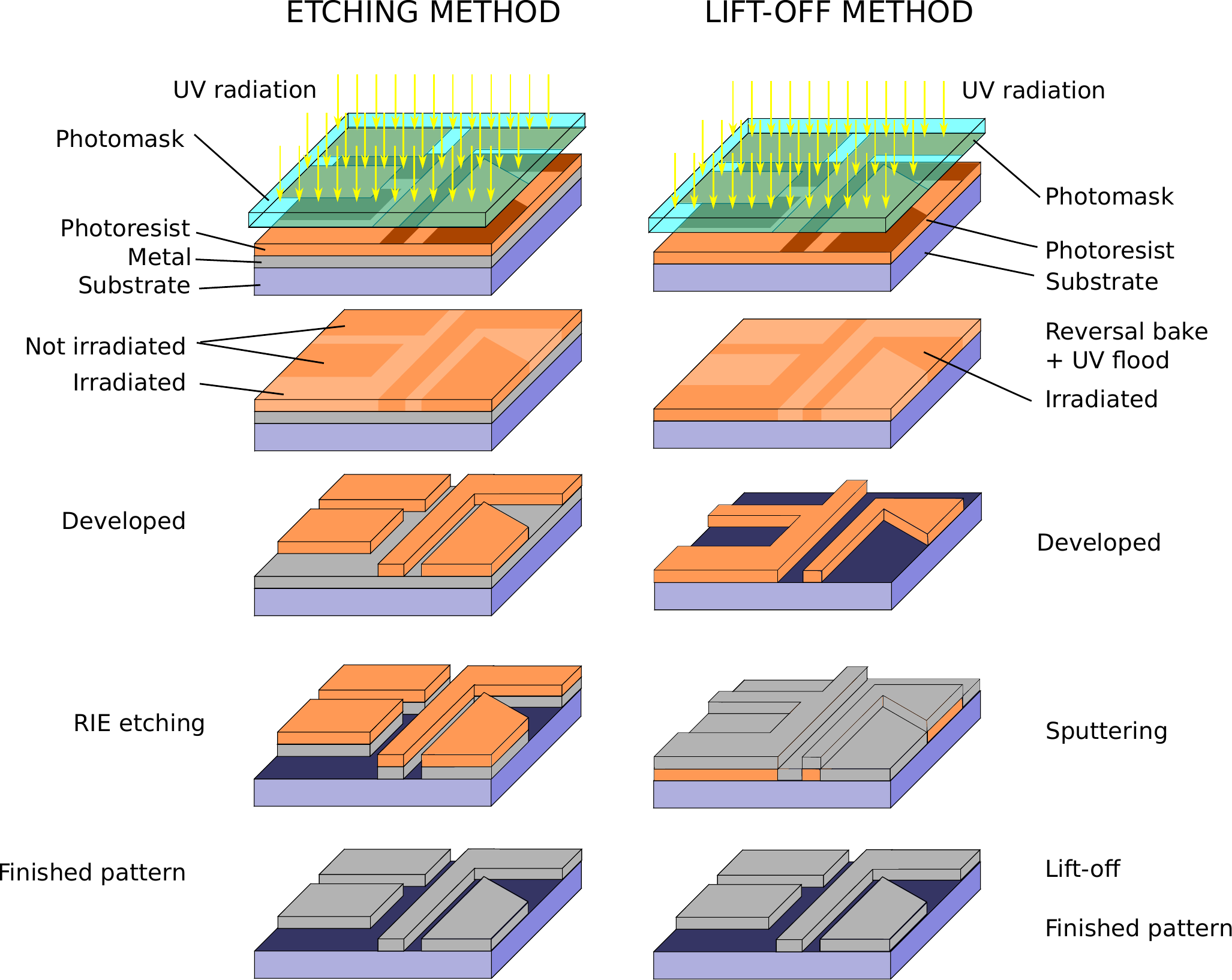}
\caption{Two UV photolithography procedures.}\label{fig:lithoproc}
\end{figure}

As an example we will explain what procedures were used during the fabrication of our devices.  The process is shown schematically in figure \ref{fig:lithoproc}.  Our devices are fabricated on 4 inch in diameter and \SI{500}{\micro\meter} thick C-plane sapphire wafers.  Our objective is to \emph{print} superconducting niobium circuits onto these wafers with a minimum feature size of around \SI{4}{\micro\meter}.

The first step involves covering one side of the sapphire wafer with a 150 nm layer of Nb using ion-beam sputtering (IBS).  Many other metals do not require IBS to create layers and some of our tests used layers of copper and gold which are deposited by evaporation using electron beam physical vapor deposition (EBPVD) (for gold) or even thermal evaporation with a high power resistor (for copper).  In any case, after the metal layer is deposited, the wafer is processed under clean room conditions to avoid dust and impurities from damaging the final patterns.  The following steps are then carried out in sequence.

\begin{itemize}
\item The wafers are thoroughly cleaned using ultrasound baths in acetone, isopropanol and de-ionized (DI) water.  They are then heated at \SI{120}{\celsius} degrees for 10 minutes to remove humidity.
\item After the wafer has cooled, a spin coater is used to apply a thin layer of adhesion promoter (TI prime) and a \SI{2.4}{\micro\meter} layer of photoresist (AZ 6624).
\item The wafer is then \emph{soft} baked at \SI{110}{\celsius} for \SI{50}{\second}
\item It is loaded into the mask aligner for irradiation.  The mask aligner holds the mask and mercury UV lamp necessary to imprint the patterns onto the resin.  The photomask with the circuit patterns is aligned with the resin covered wafer and put into close contact by applying vacuum.  The wafer is then irradiated through the mask for 10-20 seconds (calculated to get a dose of about \SI{154}{\milli\joule\per\centi\meter\squared}).
\item Once imprinted, the wafer is bathed in a 1:1 solution of photoresist developer (AZ developer) and DI water for about 50-\SI{60}{\second} to remove the exposed areas of resin.  The wafer is then inspected under a microscope.  If the patterns are defective in some way, the resin can be removed using ultrasound and acetone an the process can be repeated. 
\item If the patterns are well reproduced, the wafer is \emph{hard} baked at \SI{125}{\celsius} for \SI{2}{\minute} to harden the photoresist.
\item The next step involves etching the exposed niobium using reactive ion etching (RIE).  The wafer is loaded into the RIE chamber and attacked using \ce{SF6} gas preceded by 1 minute of \ce{02} plasma.  The base standard parameters are 200 Watts of RF power, \SI{0.19}{\milli\bar} gas pressure and a 20 sccm gas flux and an attack time of 10 minutes.
\item The remaining resin is finally removed using an acetone bath leaving the finished devices.
\end{itemize}
It is worth noting that the typical procedure to fabricate a photolithography mask is very similar to the procedure detailed here, with the exception of the exposure step.  Instead of illuminating the sample (in the mask case a layer of chromium on fused quartz) through a mask, an electron beam or a laser is used to irradiate the desired areas of the resin.

Instead of RIE etching, lift-off lithography was also used in some cases.  The procedure is also very similar but involves using different resins and a few additional steps.  The resin patterning is done directly on the bare sapphire wafer using a \emph{negative} resin (TI 35 ES).  This resin is first exposed through the same photomask (for a \SI{200}{\milli\joule\per\centi\meter\squared} radiation dose) and then a \emph{reversal} bake (at \SI{130}{\celsius} for \SI{2}{\minute}) is applied.  This makes the exposed areas resistant to to UV radiation and to the photoresist developer.  After it cools, the entire wafer is flooded in UV light (no photomask) which will now affect only the initially unexposed parts.  The wafer is then developed leaving an inverted resin pattern.  Niobium is then sputtered on top of this pattern in a uniform layer.  The niobium resting on the resin covered areas is \emph{lifted off} by bathing the wafer in acetone while the niobium deposited on the bare sapphire areas is maintained leaving the same patterns as in the RIE case.  Both the RIE and the lift-off procedures are schematically represented in figure \ref{fig:lithoproc}.

\begin{figure}[tbh]
\centering
\includegraphics[width=0.85\columnwidth]{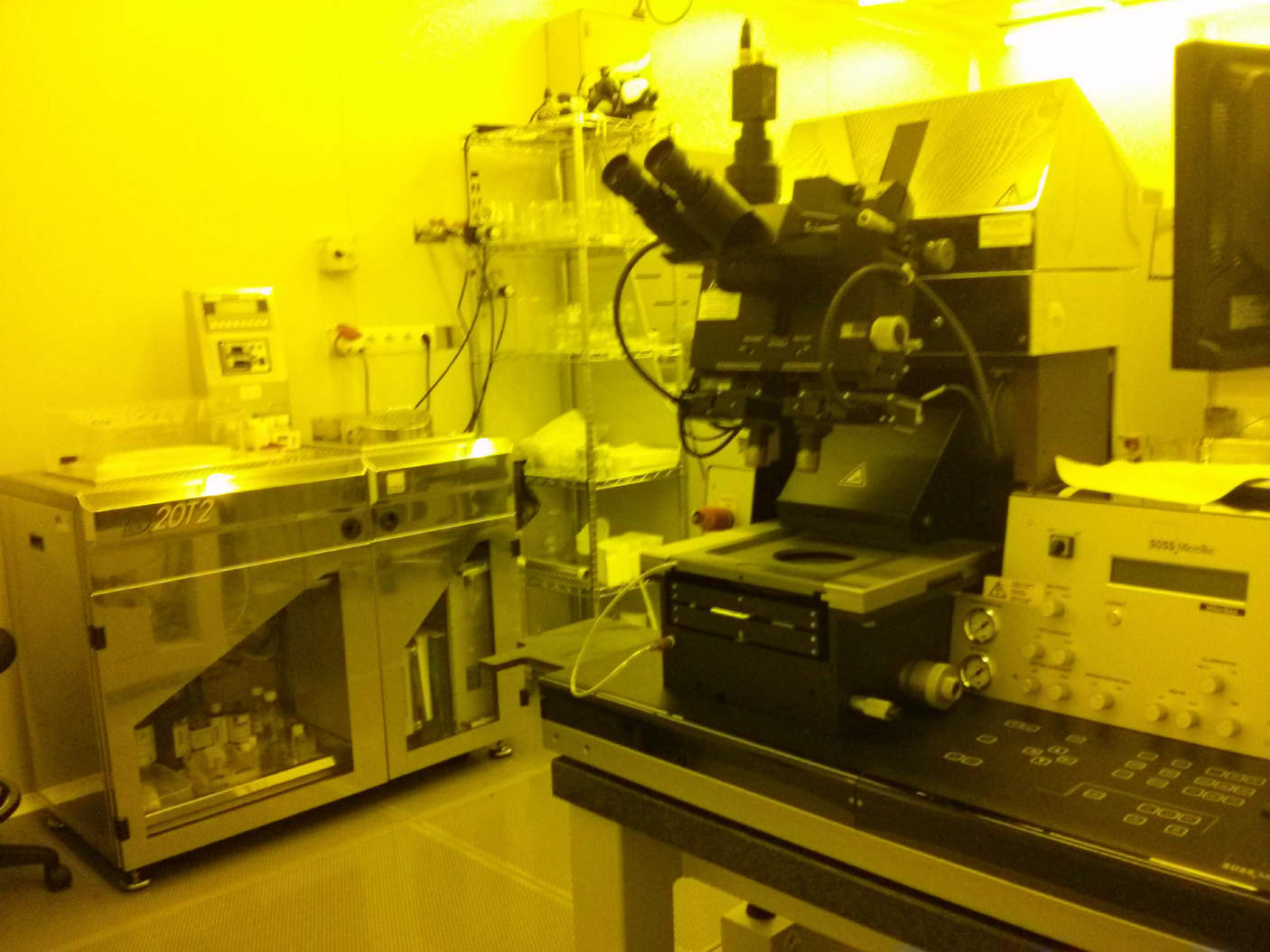}
\caption{UV photolithography clean room with the mask aligner (foreground) and the spin coater and hot plate (background).}\label{fig:cleanroom}
\end{figure}

As a final step in both the RIE and lift off cases, the wafer must be diced to separate the different devices (there are about 30 different devices on every wafer).  This is done using a diamond saw and is usually the lengthiest (and often the most critical) part of the whole procedure.  Given the mechanical properties of sapphire, the cuts must be done very slowly (no more than \SI{2}{\milli\meter\per\minute}) and with an adequate saw.  The wafers easily shatter if cut too quickly.

\subsection{Scanning Electron Microscopy (SEM) and Focused Ion Beam (FIB)}\label{sec:semfib}

Scanning electron microscopy (SEM) and focused ion beam (FIB) are two techniques that we use in the fabrication of nanoscale features in our superconducting circuits.

SEM is a microscopy technique that consists in bombarding a sample with a high energy electron beam ($\sim\si{keV}$).  These electron can be either backscattered or transmitted through the sample and can produce secondary electrons as a result of ionization ($\sim\si{eV}$).  In SEM, these secondary electrons are collected to obtain information on the sample and its topography.  This technique allows very high resolution images (down to 1 nm) and has a large depth of field allowing images with a characteristic 3D appearance.

FIB is similar to SEM but the electrons are replaced with a beam of accelerated ions (in our case \ce{Ga+} at \SI{30}{kV}).  Although imaging is possible with an ion beam, it is inherently destructive to the sample surface.  This makes sample etching the beam's primary use.  Nanoscale (as small as $\sim\SI{10}{nm}$) structures and designs can be etched into a wide variety of substrates and samples.  This same ion beam can also be used to build structures by injecting a precursor gas close to the sample surface in a process known as focused ion beam induced deposition (FIBID)\cite{Utke2008}.  The molecules of a precursor gas are injected close to the sample surface and are adsorbed to the surface.  The incoming ions decompose these molecules and remove the volatile components leaving a deposit composed of non-volatile elements of the precursor gas with implanted ions from the beam.  Materials that can be deposited include tungsten, platinum, cobalt, carbon and gold.  In the specific case of tungsten, the combination the the \ce{Ga+} ions allows the fabrication of superconducting wires \cite{Guillamon2008,Martinez-Perez2009}.

\subsubsection{Helios NanoLab DualBeam}

\begin{figure}[tbh]
\centering
\includegraphics[width=0.6\columnwidth]{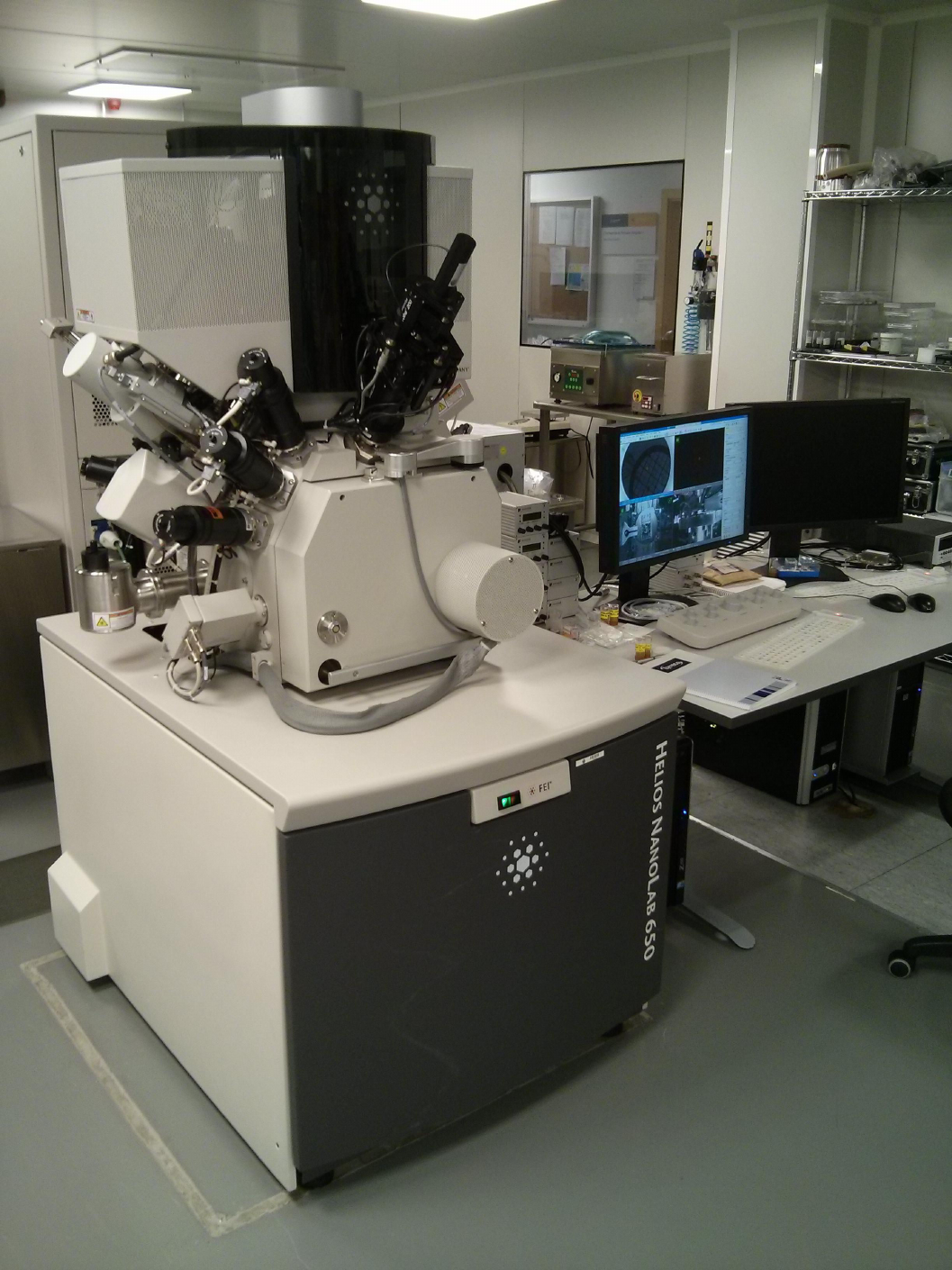}
\caption{Helios 650 system in clean room at Instituto de Nanociencia de Aragón}\label{fig:nanolab}
\end{figure}

The SEM/FIB system we use in this work is a Helios Nanolab DualBeam by FEI \cite{helios}(see figure \ref{fig:nanolab}).  It is provided by the Laboratorio de Microscopías Avanzadas (LMA) \cite{lma} at the Instituto de Nanociencia de Aragón (INA) \cite{ina} clean room.  There are three dualbeam systems (model 650, 600 and a older nova 200 model).  The \emph{DualBeam} designation refers to the fact that these systems combine two columns, one for the electron beam and, at $52^\circ$, another for the \ce{Ga+} ion beam.  This allows the sample to be imaged (by SEM) and processed (with FIB or FIBID) simultaneously.

Our samples are usually loaded into the Nanolab chamber taped onto a sample holder using copper or carbon tape to ensure good electrical contact to ground.  This is necessary to avoid the accumulation of charges in the sample since these charges can deflect the beams, reduce our resolution and induce drift in the images and etched patterns.  After the chamber is at vacuum (approximately \SI{e-6}{mbar}), the electron beam is used to image the sample and look for the area we wish to pattern.  Once it is located, the sample is rotated to align it with the ion column and a brief image is taken of the area.  The patterns to be etched are then overlaid on the image and the beam is swept over the designated areas to create the desired structures.

One of the main parameters to adjust is the ion beam current.  The beam current determines the spot size and the minimum resolvable structures as well as the time required to produce the patterns.  High beam currents (up to \SI{21}{nA}) allow fast (1-10 minutes) patterning or cutting of large areas (up to \SI{100}{\micro\meter}) but reduce the resolution.   The lowest current ($\sim\SI{1}{\pico\ampere}$) has a spot size of about \SI{5}{nm} for etching fine details but requires longer etching times.  Very long etching times can be problematic as sample drift and charging can also limit the resolution of our structures.  The currents and patterns have to therefore be adjusted such that the patterning times do not exceed 10 minutes on average to avoid this issue.  In some cases, several steps at different currents may be necessary to generate all the features.

\section{RF circuit measurement}

\subsection{Vector Network Analyzer}\label{sec:pna}

A network analyzer is an instrument that measures the transmission and reflection properties of electrical networks in the radio-frequency (RF) regime.  The device uses a RF source to send a known wave into the electrical system through transmission lines and measures the reflected and transmitted waves at the device's ports \cite{Agilent2004}.

It is helpful to think of these electrical waves using an optical analogy and to replace the electrical network with a system of lenses.  We can imagine incident light striking the optical system and having part of the wave reflected and part transmitted.  If the lenses are lossy, for example, some of the light will be absorbed by the lenses.  If the system includes mirrors, large amounts of light might also be reflected.  These lenses might also behave differently at different wave frequencies if the materials are dispersive.  These concepts are still valid when working with RF signals except that the electromagnetic energy lies in the RF/microwave range instead of the optical range.  Network analysis is concerned with the accurate measurement of the ratios of the reflected and transmitted signals to the incident signal.

It is important to note that a network analyzer, although similar in some respects, has substantial differences with a spectrum analyzer.  Network analyzers are designed to study the properties of electrical networks and include the signal source in the device.  They measure the response of the system to a known signal with, for example, a specific frequency.  On the other hand, spectrum analyzers are designed to determine the properties of a received signal (signal or carrier level, sidebands, harmonics, phase noise, etc.).  They are usually configured as a single channel receiver without a signal source.

\begin{figure}[htb]
\centering
\includegraphics[width=0.65\columnwidth]{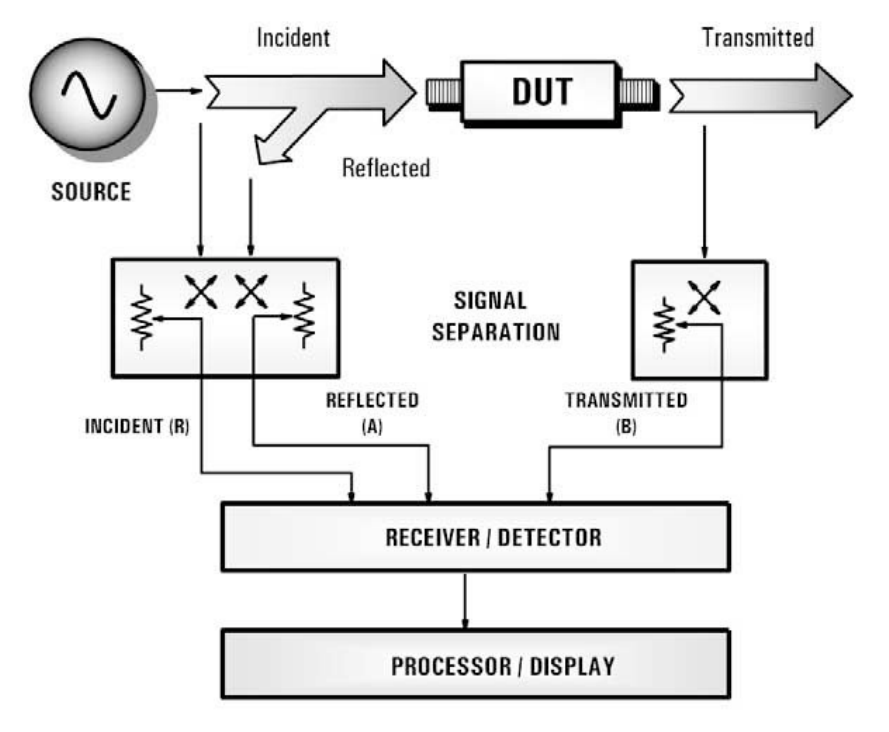}
\caption{Block diagram of a network analyzer \cite{Agilent2004}. }\label{fig:VNAblock}
\end{figure}

Network analyzers come in both scalar and vector models.  A scalar network analyzer only measures the relative amplitude of the reflected and transmitted signals to the source.  This is sufficient in many cases.  However, in other cases the phase difference is also necessary to get a complete characterization of the system.  This information is obtained using a vector network analyzer that measures both the phase difference and amplitude of the signals.

A generalized block diagram of a network analyzer is given in figure \ref{fig:VNAblock}.  The four main components are:
\begin{itemize}
\item A signal source - It provides the stimulus for our system and can usually be programmed to sweep either frequency or power.
\item Signal separation devices - This block is usually called the \emph{test set} and serves two primary functions.  The first is to measure a portion of the input signal to provide a reference for ratioing.  The second is to separate the incident and reflected traveling waves at the input of our electrical device under test (DUT).  Both these tasks are achieved using different components such as splitters, directional couplers and bridges. 
\item Receivers to detect the signals - Can be diode detectors (scalar and broadband) or tuned receivers (magnitude and phase information).  Vector network analyzers use tuned receivers.  These use a local oscillator (LO) to mix the RF down to a lower \emph{intermediate} frequency (IF).  The IF signal is then bandpass filtered which greatly improves sensitivity and dynamic range.  Modern analyzers then use an analog-to-digital converter (ADC) and digital signal processing (DSP) to extract the amplitude and phase data from the IF signal.
\item A processor and display system to show the results - The results are processed and displayed so that they are easy to interpret.  Some systems also allow for automated measurements to be programmed (Programmable Network Analyzer or PNA).
\end{itemize}
Modern commercial network analyzers like the ones used in this thesis have all these components integrated into a single instrument that includes a full operating system and graphical interface.  They can also have 2 or more measurement ports allowing the analysis of multi-port networks.  The excitation can be applied to any of the ports while measuring the response on all of them simultaneously.

In most cases when performing measurements, network analyzers are calibrated using a calibration standard connected at the location of the device under test including all the connecting wires that will be used in the final measurement.  A full two port calibration requires measuring a series of different connections:

\begin{itemize}
\item An \emph{open connection} at each port
\item A \emph{shorted connection} at each port
\item A \emph{matched load} at each port (usually a \SI{50}{\ohm} resistance)
\item A \emph{thru connection} between the two ports
\item An \emph{isolation connection} between the two ports (usually optional)
\end{itemize}

With these measurements, the network analyzer will compensate for effects arising from the connecting wires and measure the response of the DUT only.  On most modern network analyzers these calibrations can be saved and re-used without having to remeasure the calibration standard.  There are also electronic calibration standards that allow some systems to perform automatic calibrations.

\subsection{Rohde \& Schwarz, ZVB14 Vector Network Analyzer}

\begin{figure}[tbh]
\centering
\includegraphics[width=0.65\columnwidth]{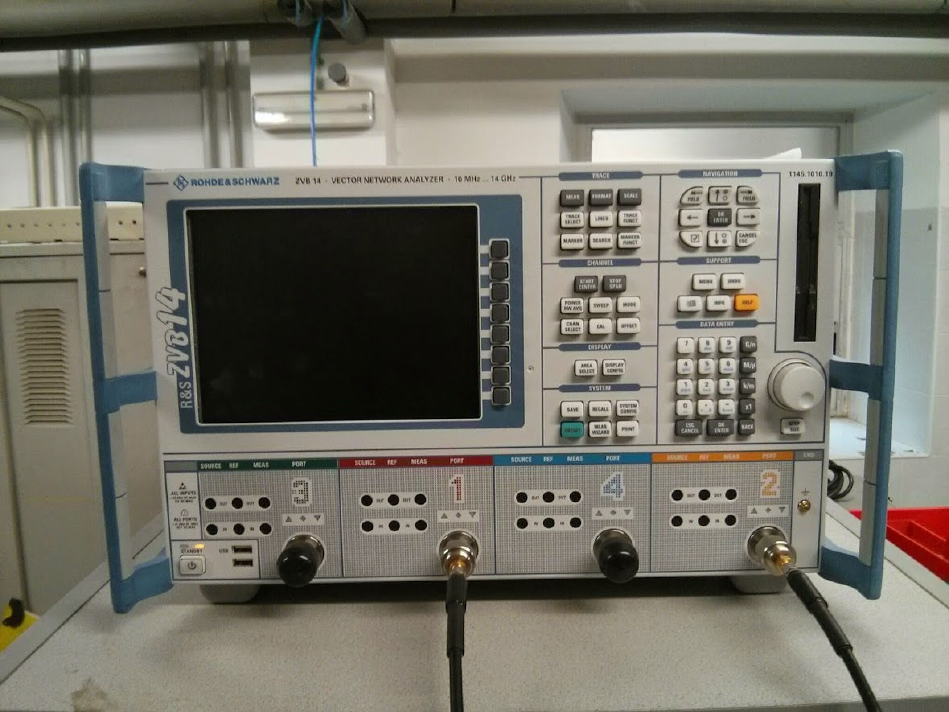}\\[2mm]
\begin{tabular}{|l|r|}
\hline
Number of Ports & 4 \\
\hline
Frequency range & 10 MHz to 14 GHz \\
\hline
Power range & -45 to 13 dBm \\
\hline
Dynamic range (at 10 Hz IF bandwidth) & >120 dB \\
\hline
IF bandwidths & 1 Hz to 500 kHz \\
\hline
Measurement points per trace & 1 to 60001 \\
\hline
Operating system and internal PC & Windows XP \\
\hline
Port connector type & 3.5mm male \\
\hline
\end{tabular}
\caption{Rohde and Schwarz, ZVB14 Vector Network Analyzer}\label{fig:zvb14}
\end{figure}

The specific model used for most measurements in this thesis is a R{\&}S ZVB14 Vector Network Analyzer \cite{Schwarz}.  A photograph and the basic specifications can be seen in figure \ref{fig:zvb14}.  Each of the devices 4 ports has an independent measurement receiver and reference receiver as well as an independent signal generator.  It uses an internal computer to run the measurements and can be programmed remotely either using remote desktop systems or included NI LabView drivers \cite{NationalInstruments2011}.

Our network analyzer ports use \SI{3.5}{\milli\meter} (male) connectors which are compatible with the more widespread SMA connectors.  In measurements performed throughout this work, we use only ports 1 and 2 wired to our system via two 1 meter long coaxial cables (SMA to SMA).  We however do not use any calibration standard to remove these wires from the measurement.  Our actual DUTs (described in chapter \ref{chap:CPWG}) are usually at cryogenic temperatures and have longer wires attached which have much larger losses than these 1 m wires.  We do not have a calibration standard that is adequate for connection at those temperatures.  Therefore our measurements will include the contributions of all these connecting cables.

\subsection{\SI{4}{\kelvin} probe for RF measurements}

\begin{figure}[tbh]
\centering
\includegraphics[width=0.85\columnwidth]{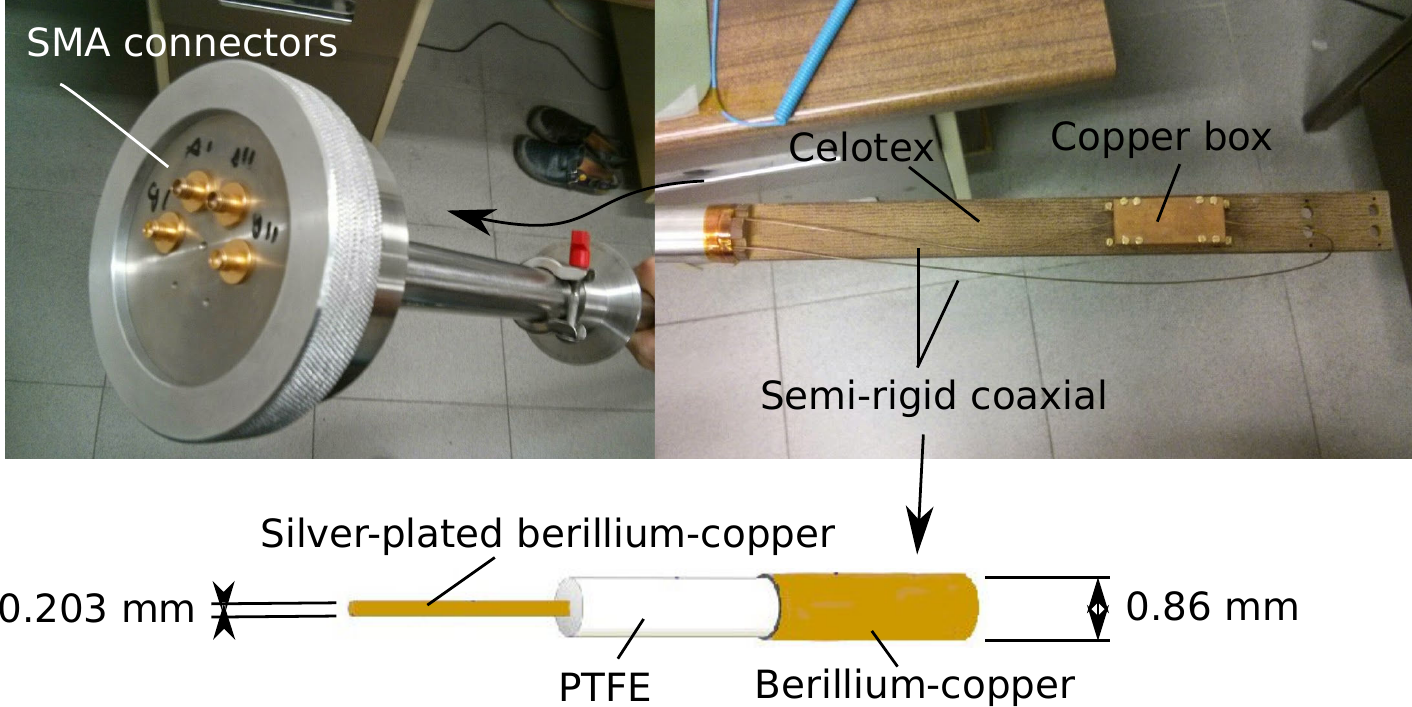}
\caption{\SI{4}{\kelvin} probe for RF measurements and diagram of semi-rigid coaxial cable.  Out of the 4 visible SMA connectors on the probe head, only two are actually wired.}\label{fig:probewires}
\end{figure}

A simple \SI{4}{\kelvin} probe was build to submerge superconducting microwave circuits into liquid helium and measure RF transmission through it.  As we detail in section \ref{sec:testCPWG}, our circuits are mounted on 4 cm long fiberglass holders equipped with SMP \cite{Technologies} connectors.  These holders are screwed onto a copper box that in turn is attached to a celotex board.  This board is at the end of a 2 m steel tube that holds two cryogenic coaxial cables (figure \ref{fig:probewires}).  The cables are model SC-086/50-SB-B by COAX CO. (Japan) \cite{coaxco} and their basic characteristics are shown in figure \ref{fig:probewires}.  The cables join the circuit on the fiberglass holders to bulkhead SMA connectors at the other end of the tube that allow them to be connected to external electronics.  The probe dimensions and flange allow it to be used by directly inserting it in one of our \SI{100}{\liter} liquid helium transport dewars or in the cryostat holding our superconducting vector magnet (section \ref{sec:vecmag}).

\subsection{Vector Magnet}\label{sec:vecmag}
The Low Temperature Lab at the Instituto de Ciencia de Materiales de Aragón is equipped with a superconducting vector magnet manufactured by Oxford Instruments.  This is one of the pieces of equipment used in some EPR-like experiments detailed in chapter \ref{chap:samp}.

\begin{figure}[tbh]
\centering
\includegraphics[width=0.65\columnwidth]{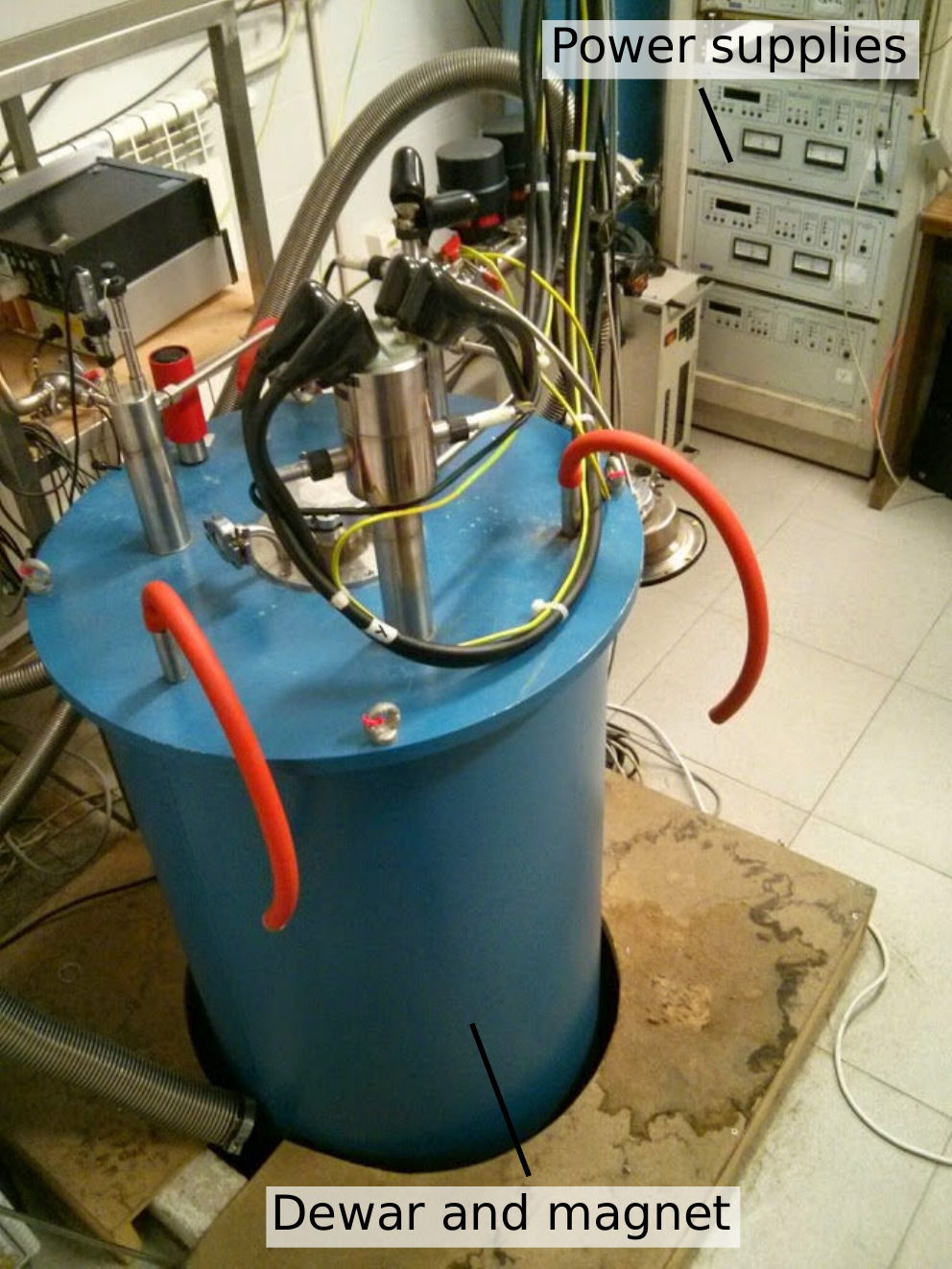}
\caption{Vector magnet and power supplies at the Low Temperature Laboratory (ICMA).}
\end{figure}

Superconducting magnets \cite{OxfordInstruments2007} allow large magnetic fields to be generated in laboratory scale cryostats without the \SI{}{\kilo\watt} or \SI{}{\mega\watt} power supplies necessary for conventional electromagnets.  In most cases, the cost of cooling a superconducting magnet is lower than that of the power required for a resistive electromagnet.

Superconducting magnets consist of a number of coaxial solenoid sections wound using multifilamentary superconducting wire.  One of the main advantages of superconducting magnets is the ability to work in \emph{persistent mode}.  In this mode, the superconducting circuit is closed of using a superconducting switch, keeping the current circulating in a continuous loop.  The power supply can then be switched off leaving the magnet \emph{at field}.  The field then decays very slowly (about 1 part in $10^4$ per hour or lower).

There are three separate superconducting coils that apply magnetic fields in three perpendicular directions.  The largest coil applies field in the vertical (or Z) direction in the laboratory reference frame and can maintain fields of up to $\pm\SI{9}{\tesla}$.  The remaining two coils apply fields in the horizontal laboratory plane (directions X and Y) and can reach up to $\pm\SI{1}{\tesla}$.  Each magnet has an independent power supply that provides the current to maintain the fields in each magnet.  The Z and X magnets use model IPS120-10 power supply while the Y magnet uses an older IPS120-3 model which can not reverse the polarity of the field (only 0 to $+\SI{1}{\tesla}$).

\begin{figure}[tbh]
\centering
\includegraphics[width=0.95\columnwidth]{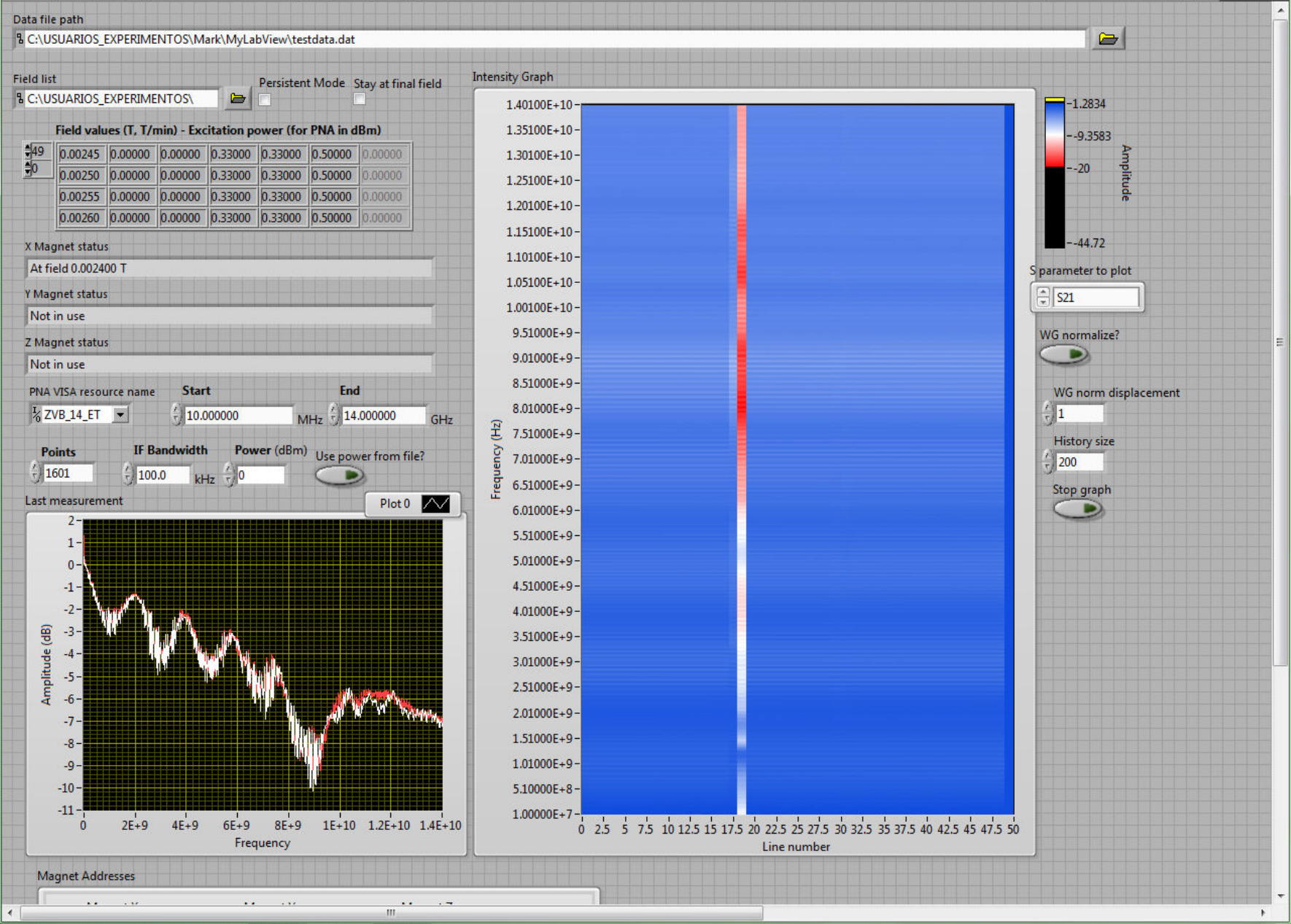}
\caption{Screenshot of Labview program for coordinated control of the superconducting magnets and the vector network analyzer}\label{fig:maglabview}
\end{figure}

The magnet is contained inside a \SI{90}{\liter} capacity liquid helium dewar that keeps the magnet below the superconducting critical temperature.  Any sample that is inserted into the magnet will also be submerged in this liquid helium bath.  The dewar includes a liquid nitrogen cooled radiation shield and an outer vacuum chamber to insulate the liquid helium reservoir and prevent losses.

This magnet system is used in this work for different types of measurements that require external magnetic fields.  In chapter \ref{chap:CPWG} it is used to measure the performance of superconducting coplanar waveguide resonators in the presence of magnetic field and in chapter \ref{chap:samp} it is used in EPR-like experiments to tune level separations into resonance.  The magnets are controlled by a computer through a RS232 interface and an Labview \cite{NationalInstruments2011} driver from Oxford Instruments.  The magnet driver is incorporated into custom Labview programs to coordinate the magnetic field with data acquisition from our network analyzer allowing the automation different measurement sequences (figure \ref{fig:maglabview}).

\section{Local scanning probe systems}

\subsection{Atomic and Magnetic Force Microscopy (AFM/MFM)}\label{sec:afmmfm}

\begin{figure}[tbh]
\centering
\includegraphics[width=0.85\columnwidth]{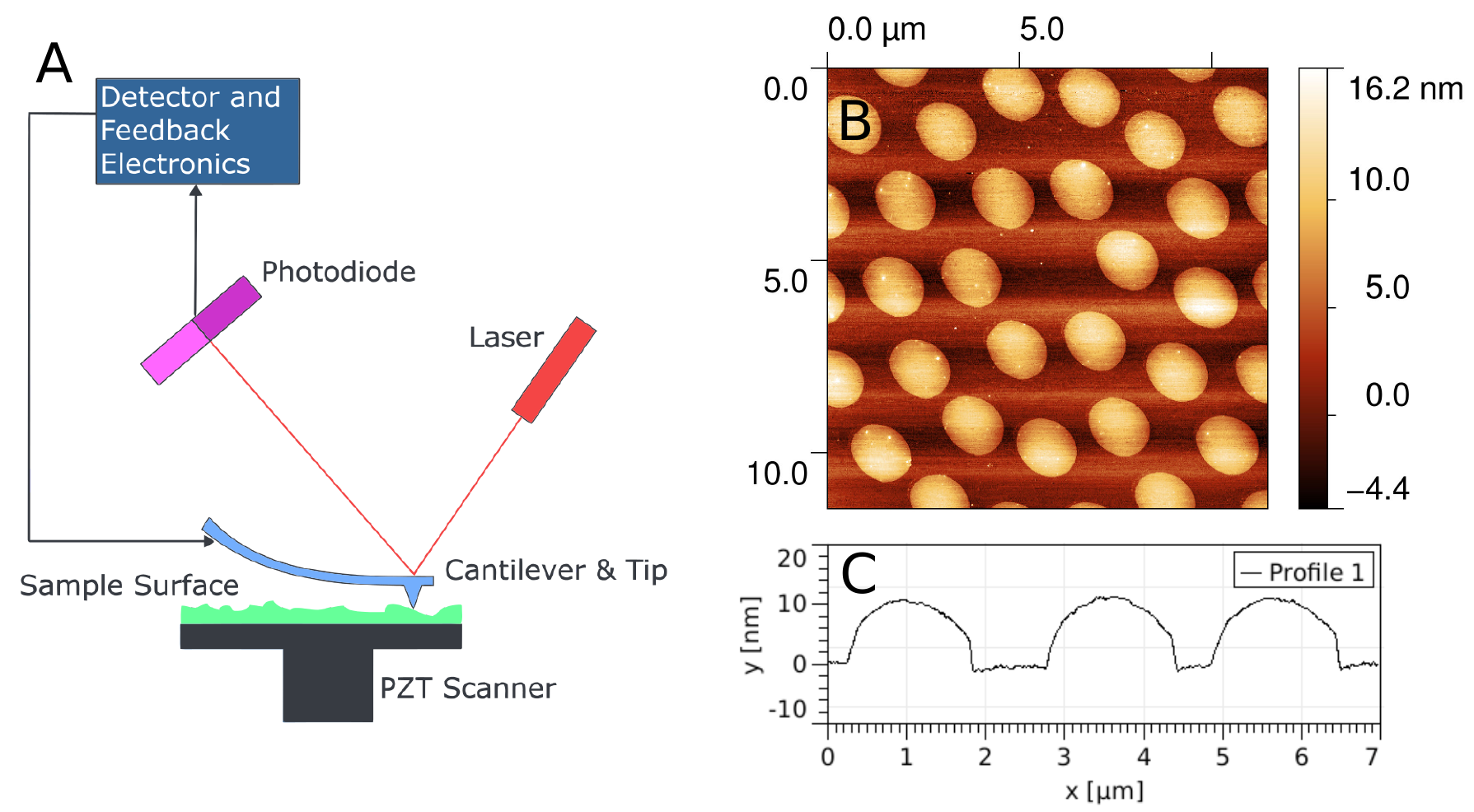}
\caption{Atomic Force Microscopy (AFM).  Graph A shows a schematic of an AFM measurement system.  Graphs B and C show a tapping mode topography measurement of \ce{Gd2}-ac droplets deposited on a \ce{SiO2} substrate by dip-pen nanolithography \cite{Lorusso2013} (see section \ref{sec:DPN})}\label{fig:afm1}
\end{figure}

Atomic force microscopy is a type of scanning probe microscopy that consists in sweeping a sharp tip over a sample surface and detecting changes in its deflection \cite{Binnig1986,Giessibl2003,Jalili2004}.  Typically the tip is at the end of a cantilever and the position of the cantilever is monitored with a laser and photodiode (figure \ref{fig:afm1}).  The sample and tip movements are usually controlled using piezoelectric actuators to precisely control their positions.  AFM has been demonstrated to be able to achieve topographies with atomic resolution.  Although the most common AFM measurements are of sample topographies, other magnitudes can also be measured by coating the tip with different materials sensitive to other interactions.  The tip can be made to register magnetic domains, electric charge, conductivity, etc.

There are several different operating modes for AFM systems and most systems can switch between these modes.
\begin{itemize}
\item Contact mode consists in bringing the tip into direct contact with the sample surface and measuring the deflection of the tip.  As the tip is swept over the surface the deflection changes depending on the topography and a feedback loop adjusts the sample height to keep the deflection constant.  This feedback provides a topography signal which is recorded.  This technique is the most direct way of measuring AFM but is somewhat aggressive for both the tip and sample.  To minimize damage to the tip and sample, the cantilevers used are usually soft in the sense that the have a low elastic constant.
\item Semi-contact or tapping mode consists in actively exciting a vibration resonance in the tip while it is swept over the surface.  The tip is positioned above the surface at such a distance that the tip touches the surface at the end of its oscillation.  The feedback system monitors the vibration amplitude and uses it as the feedback signal to produce the topography.  This method is less aggressive than contact mode since contact only occurs for a fraction of the oscillation and is the most widely used in standard measurements.  Cantilevers have higher elastic constants than in contact mode to get higher quality resonances.  
\item Non-contact or frequency modulated mode is similar to semi-contact mode but the tip is kept out of contact with the surface.  The surface is detected by monitoring changes in the tip vibration phase and resonant frequency.  The resonance is established when the tip is far from the surface and the phase used as a feedback variable.  Deviations from the central value are corrected by shifting the excitation frequency.  This frequency shift is recorded as the tip is swept at constant height over the sample.  This method can obtain the highest resolution and can achieve atomic resolution under adequate conditions.  Since the tip never actually touches the surface, it can be made much sharper than in the other modes where the contact with the surface quickly would wear down the tip.  It can however only be used with relatively flat surfaces since it requires scanning at a constant height and abrupt topography changes can easily crash the tip into the surface.
\end{itemize}

\begin{figure}[tbh]
\centering
\includegraphics[width=0.75\columnwidth]{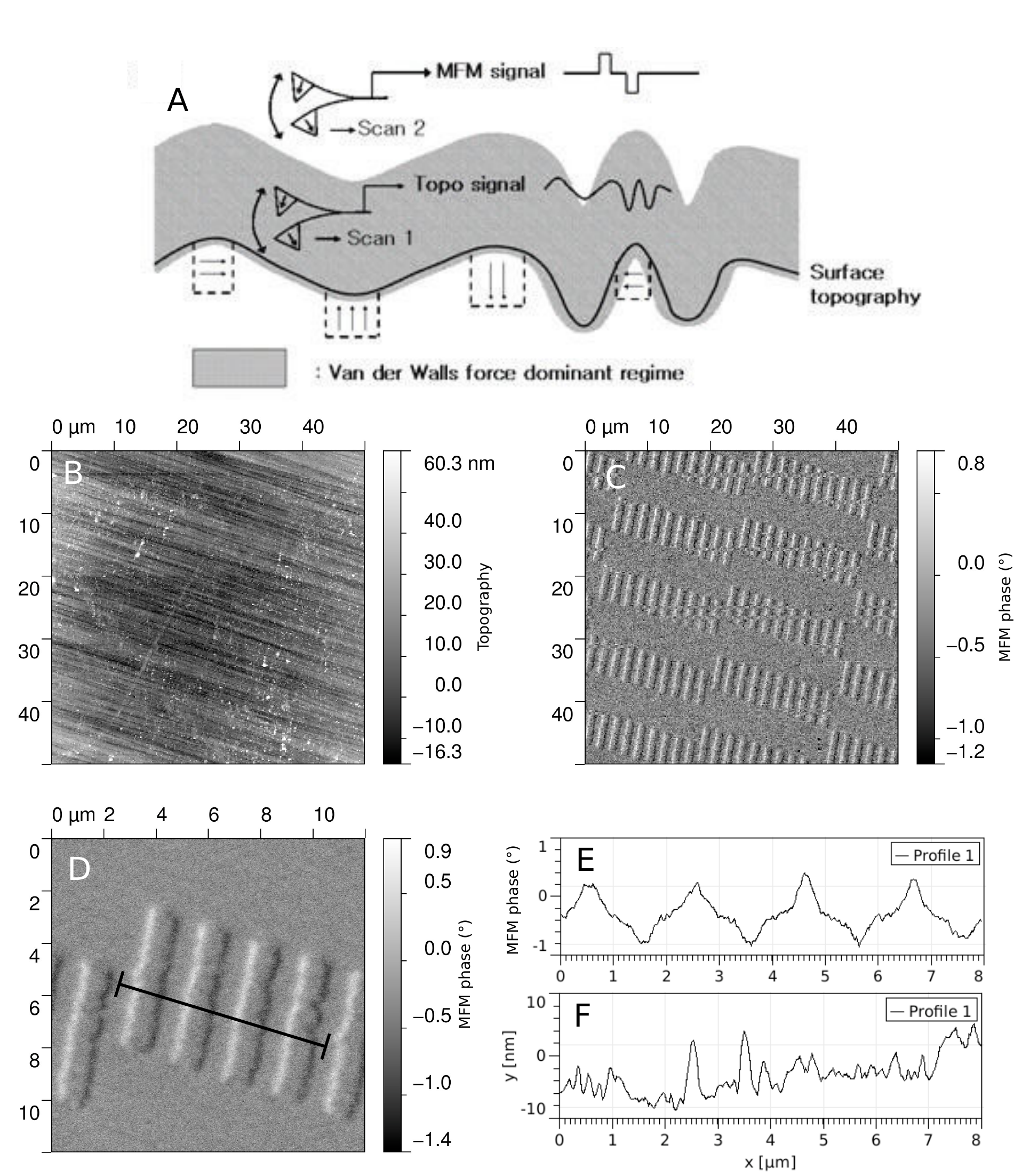}
\caption{Magnetic force microscopy (MFM.  Graph A shows a diagram of a two-pass MFM measurement scheme.  Graph B and C are the topography and MFM passes (100 nm lift) in a two-pass MFM measurement of a hard disk.  Magnetic features can clearly be seen and are not visible on the flat topography (within $\sim \SI{70}{\nano\meter}$ over a region of $50\times\SI{50}{\micro\meter}$).  Graph D shows a $12\times\um{12}$ close up of graph B.  Graphs E and F show the MFM phase and topography of the \um{8} profile marked in graph D.}\label{fig:mfm1}
\end{figure}

In this work we are particularly interested in an AFM technique known as magnetic force microscopy (MFM).  In MFM, the AMF tip is coated in a magnetic material (usually \ce{Co} based compounds) so that the measurement is sensitive to magnetic interactions.  This allows the mapping of magnetic domains and detecting stray field from magnetic regions.  There are two main methods of measuring MFM:
\begin{itemize}
\item The two-pass method involves first taking a topography profile of each of the lines in the image.  In close contact (as in tapping-mode), the tip is only sensitive to the topography since at close range it is much larger than any magnetic contribution.  After this scan the tip is moved up ($\sim\SI{100}{\nano\meter}$) and rescanned over the same line following the measured topography profile.  This allows us to suppress the topography contribution from the final image.  Since magnetic interactions have a much longer range than the interactions due to topography, the second scan height can be adjusted to leave only a magnetic signal and a negligible topography signal.
\item To measure high resolution MFM images, non-contact techniques must be used.  Non-contact MFM is very similar to standard non-contact AFM, but the distance needs to be adjusted to a regime where the topography is invisible and only the long range magnetic contributions are visible.  Again there are limitations how rough the surface can be to get a clean measurement.
\end{itemize}

\subsubsection{NT-MDT AFM system}
\begin{figure}[tbh]
\centering
\includegraphics[width=0.7\columnwidth]{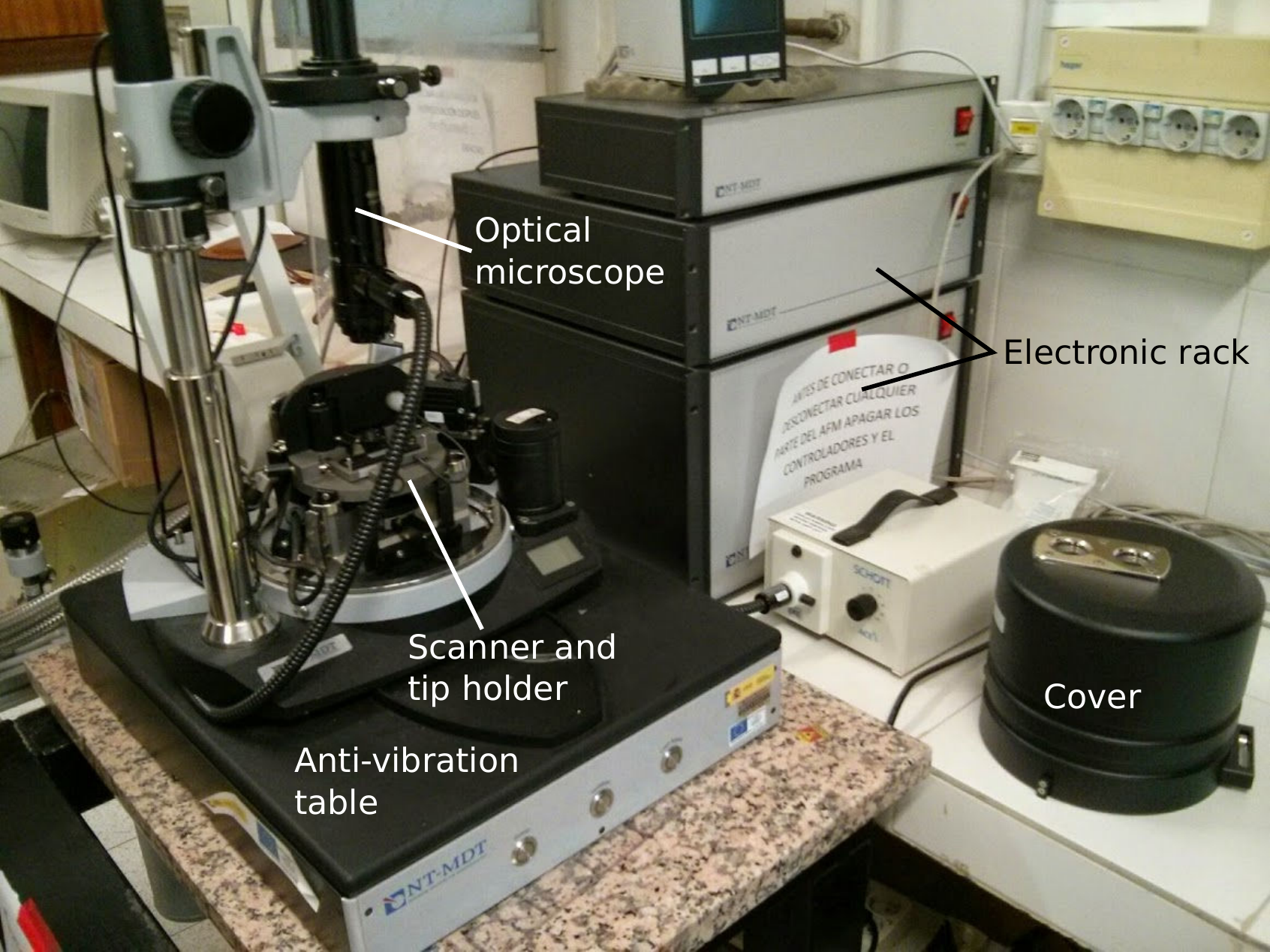}
\caption{NT-MDT NTEGRA system at Instituto de Ciencia de Materiales de Aragón.}\label{fig:ntegra}
\end{figure}

The system used in this work is a general purpose commercial AFM system by NT-MDT \cite{ntmdt} (figure \ref{fig:ntegra}).  The NTEGRA model available at the Instituto de Ciencia de Materiales de Aragón (ICMA) allows many different measurement types.  Apart from standard contact, semi-contact and non-contact modes, it can perform lateral force microscopy, adhesion force imaging, MFM (see figure \ref{fig:mfm1} for an example), electrostatic force microscopy, scanning capacitance microscopy, kelvin probe microscopy, spreading resistance imaging, some basic lithography procedures (force and current) and basic scanning tunneling microscopy.

Different modules and options also allow measurements to be performed in vacuum (\SI{e-2}{\milli\bar}), in a liquid cell, applying an external magnetic field or with temperature control (Room temperature to  \SI{150}{\celsius}).  The available scan heads also allow scanning by sample (usually more stable but samples sizes and weights are limited) or scanning by the tip (usable for larger samples).  Its maximum scan range is 100 by \SI{100}{\micro\meter} and its typical scanner resolution is 0.2 nm in the XY scan plane and 0.05 nm in the Z vertical direction although different combinations of probes, scanners and measurement schemes can improve upon this limit.

\subsection{Dip-Pen Nanolithography}\label{sec:DPN}
A related technique to AFM that is touched upon in this work is Dip-Pen nanolithography \cite{Piner1999}.  On a basic level, it involves using an AFM tip as a pen to transfer a sample to a substrate creating a specific pattern.  Patterns with features in the sub \SI{100}{nm} range can be achieved depending on the sample and substrate \cite{Ginger2004}.

The highest resolution DPN patterns have been achieved using molecular inks that have a specific affinity for a substrate.  These inks are coated onto AFM tips and then brought into contact with the substrate.  The molecules then diffuse through a water meniscus.  Different sizes can be achieved controlling the ambient humidity and dwell time of the tip.  Although high resolutions can be achieved with these inks, they are usually limited to a single type of substrate per molecular ink.

It is also possible to work with liquid inks that allow for a greater variety of samples and substrates.  A solid sample can be dissolved in an appropriate solvent and this ink be transferred with an AFM tip to the substrate.  Depending on the ink properties, the minimum feature sizes are somewhat larger than in the molecular ink case and are of the order of \si{\micro\meter}.  This method also allows bio-molecules such as proteins or DNA as well as inorganic molecules to be patterned.

\begin{figure}[ptbh]
\centering
\includegraphics[width=0.8\columnwidth]{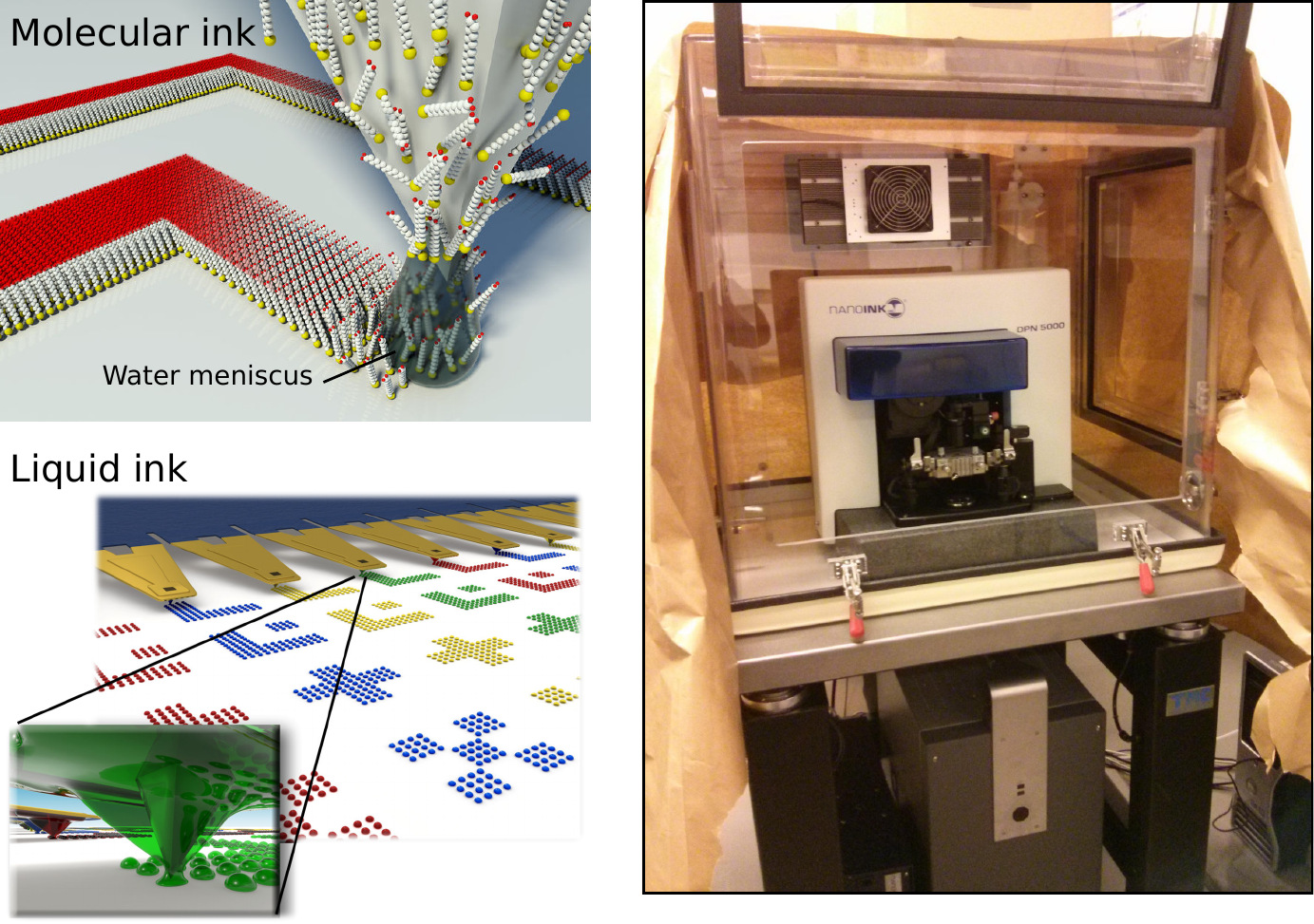}
\caption{Dip-pen nanolithography.  Nanoink DPN 5000 System at Instituto de Nanociencia de Aragón.}\label{fig:dpn}
\end{figure}

\begin{figure}[ptbh]
\centering
\includegraphics[width=1.\columnwidth]{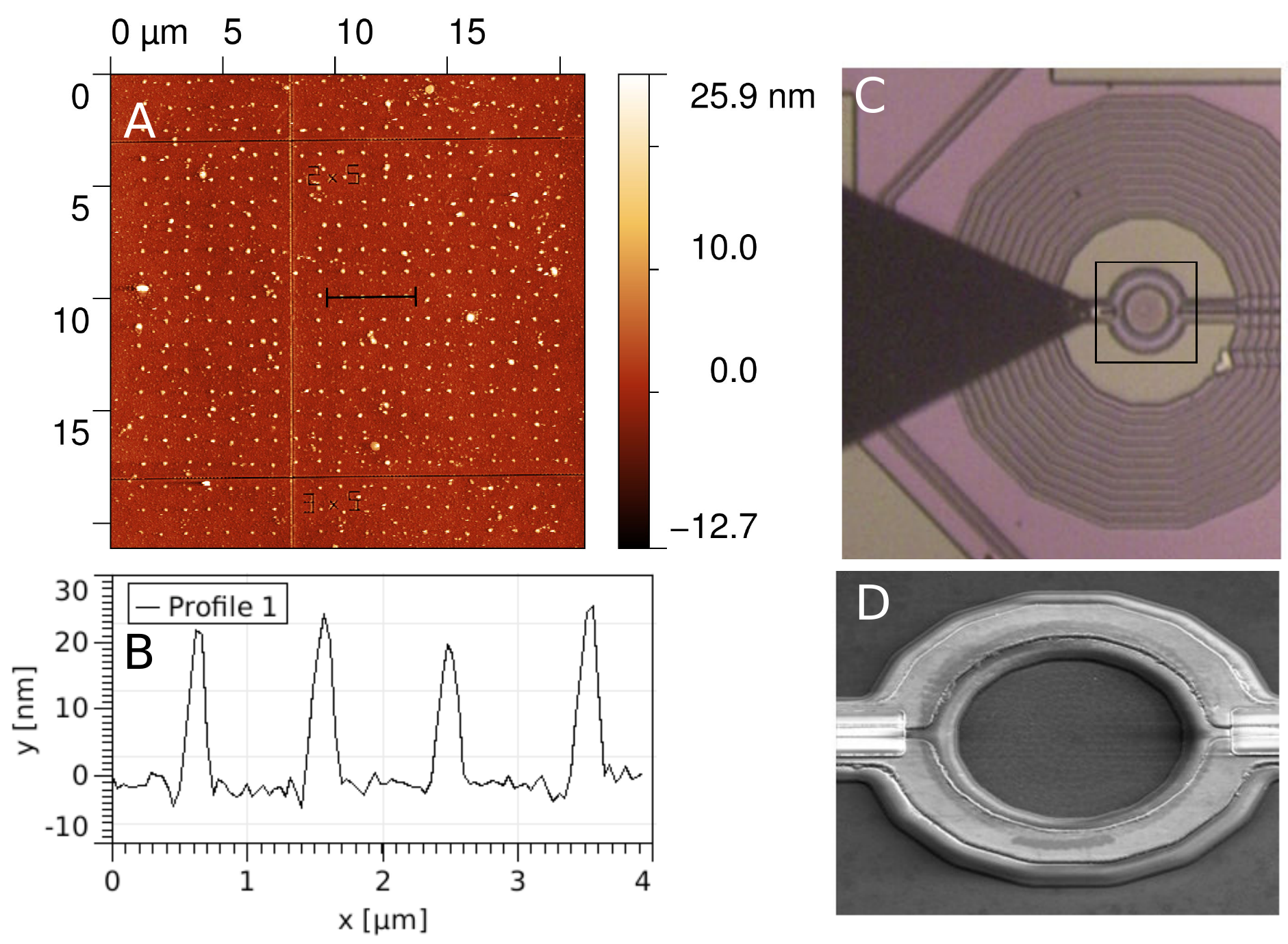}
\caption{Dip-pen nanolithography.  Graphs A and B show an AFM images and profile of droplets of magnetoferritin deposited using DPN.  Graph C shows a \si{\micro}-SQUID sensor immediately after droplets of \ce{Mn12}-benzoate have been deposited inside its \SI{30}{\micro\meter} loop as well as the DPN tip (black triangle).  Graph D shows a SEM image of the \si{\micro}-SQUID and \ce{Mn12} droplets (viewed at a \SI{52}{\degree} angle).}\label{fig:dpn2}
\end{figure}

A couple of examples of DPN depositions using liquid inks are shown in figure \ref{fig:dpn2}.  In one example we show droplets of magnetoferritin on a \ce{SiO2} substrate.  The diameter of a magnetoferritin molecule is of about 10-12 nm so, according to the profile, the droplets consist of about two layers of molecules.  The other example shows a microscope image of \ce{Mn12}-benzoate molecular nanomagnets \cite{Gatteschi2003} droplets just after being deposited by the DPN tip on a \SI{30}{\micro\meter} diameter \si{\micro}-SQUID sensor \cite{Bellido2013}.  As a further example, the sample seen previously in the AFM image in figure \ref{fig:afm1} was also fabricated using DPN \cite{Lorusso2013}.

\section{Software and computational techniques}

\subsection{COMSOL Multiphysics}\label{sec:comsol}
COMSOL Multiphysics \cite{COMSOL2012} is a commercial software package that provides numerical simulations for a wide range of physics problems.  It includes a number of different modules for thermal, mechanical, electrical, fluid flow and chemical simulations.  These different modules can also be cross coupled.  For example, a problem may simultaneously involve fluid flow and heat flow.  The system allows the flow and thermal variables to be coupled and solved as a single problem.

\begin{figure}[tbh]
\centering
\includegraphics[width=1.\columnwidth]{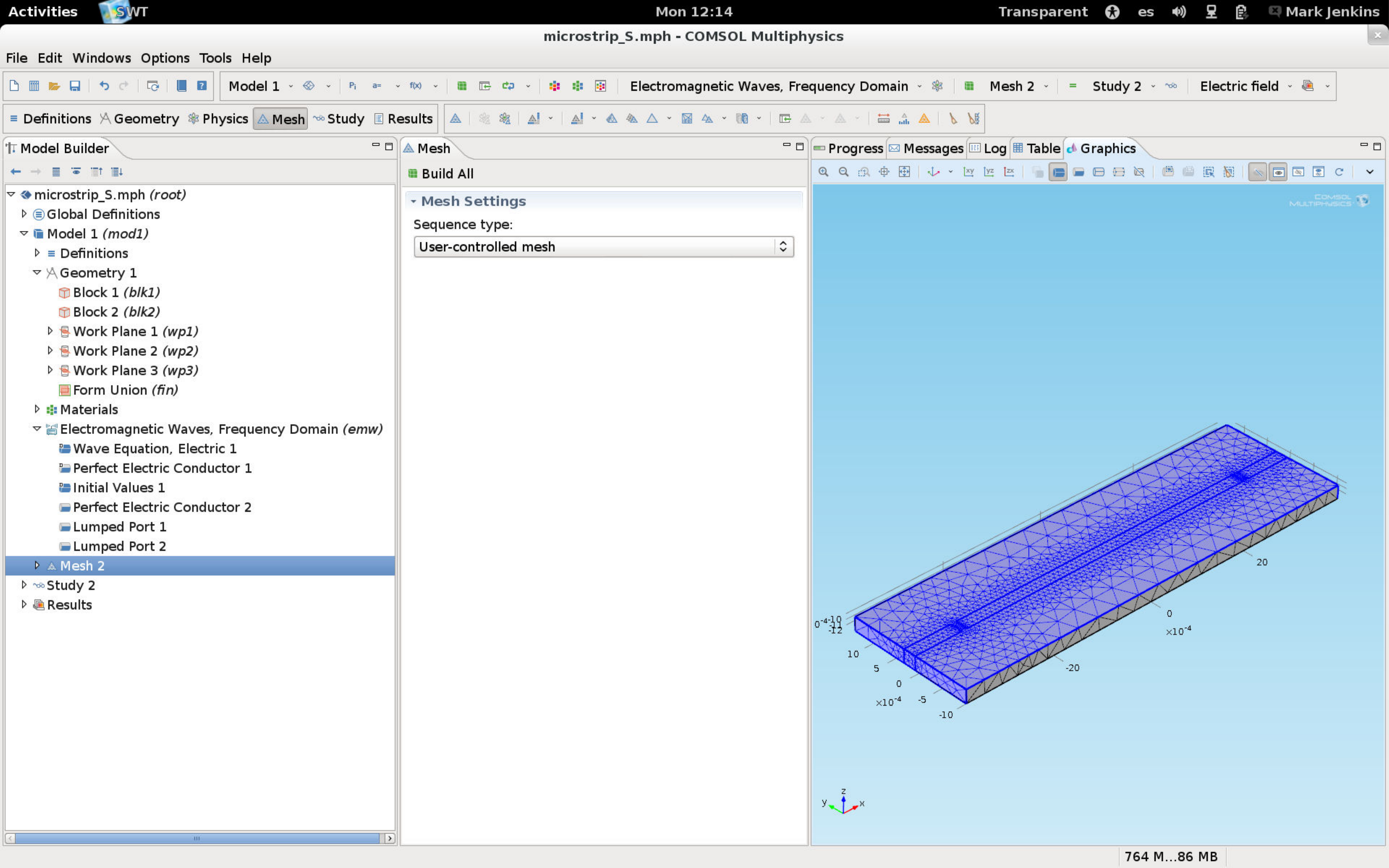}
\caption{Screenshot of Comsol 4.4.  Mesh of a stripline resonators to be simulated using the Electromagnetic Waves module.}\label{fig:comsolline}
\end{figure}

At its heart, COMSOL numerically solves systems of partial differential equations using a finite element methods.  Many common sets of equations are included in the predefined packages, but custom physics are possible.  Different types of elements and solvers can be selected for different types of problems and both stationary and time domain solutions can be found.  A typical simulation begins defining the geometry (1, 2 or 3D) and physical structure of the problem.  Then the physics involved are chosen and applied to the geometry (boundary conditions, applicable domains, etc).  The next step involves meshing the geometry such that the desired features can be resolved.  Finer meshes can resolve more details but of course require additional memory and computation time.  Finally the desired solver is executed and attempts to find a solution.  If the solution converges, the solved variables and other derived variables can then be represented and processed.

The modules used in this work are limited to different electrical and magnetic models.  In particular we use the ``Magnetic and Electric fields'' module (mef), the ``Magnetic fields'' with and without currents modules (mf and mfnc) and the ``Electromagnetic waves'' (emw) module.  All these modules implement different combinations of the Maxwell equations applying different simplifications.

\begin{figure}[tbh]
\centering
\includegraphics[width=0.85\columnwidth]{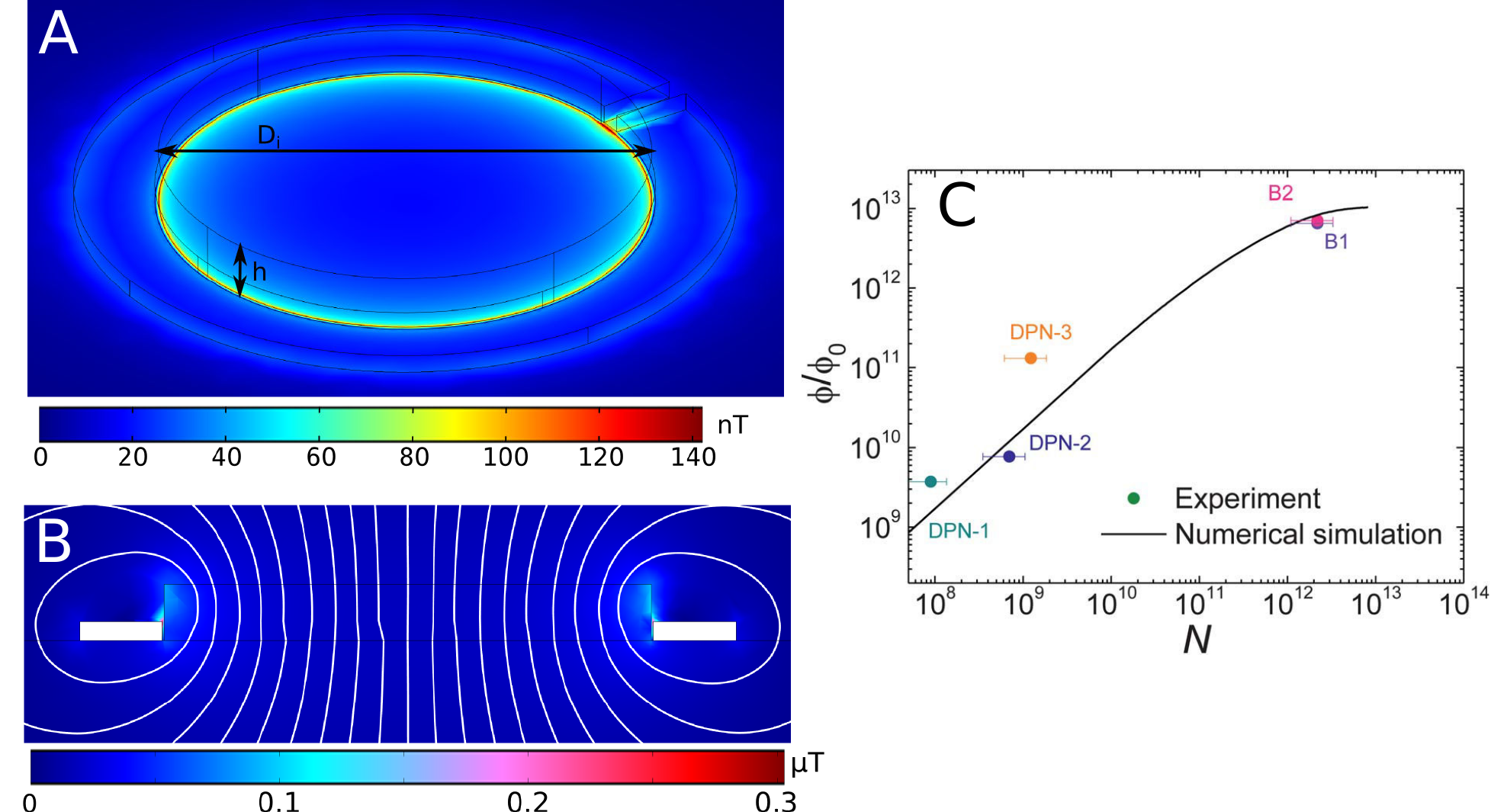}
\caption{Comsol simulation of magnetic flux density through a superconducting (SQUID) loop \cite{Bellido2013}.  The diameter $D_i$ is \SI{26}{\micro\meter}.  The color scale for graphs A and B corresponds to the magnetic flux absolute value.  Graph B shows a cross section of the SQUID loop along a diameter and calculated magnetic field lines.  Graph C shows simulated flux generated by the sample (in units of the flux quanta $\phi_0$) as a function of the number of molecules of \ce{Mn12}-benzoate present.  The points are experimental measurements.}\label{fig:comsolsquid}
\end{figure}

In this thesis work, we have been mostly concerned with the simulation of rf waves propagating via coplanar waveguides and resonators (figure \ref{fig:comsolline} and chapter \ref{chap:CPWG}).  As a further example we show in figure \ref{fig:comsolsquid} results of a Comsol calculation of the magnetic flux density generated by a magnetic layer that fills a SQUID loop \cite{Bellido2013}.  The SQUID loop is meant to model a micro-susceptometer that converts the flux through the loop to voltage.  The SQUID wire is superconducting and can be modeled as a magnetic insulator that does not allow the magnetic field lines to enter the domain and will deform the stray field generated by the magnetized sample.  This simulation is used to estimate the minimum amount of a given magnetic sample and applied DC field necessary to generate sufficient flux through the SQUID to be detected by the SQUID electronics.  It also allows us to approximate the dependence of our signal with the amount and size of the sample since not all areas of the cylinder contribute equally.  Results of these calculations are compared with data measured on layers of \ce{Mn12}-benzoate deposited onto a \si{\micro}-SQUID loop (figure \ref{fig:comsolsquid}C).

\subsection{Other programs and libraries}
Most data processing and analysis has been done using has been done using custom programs written in \CC\ \cite{Stroustrup1986} and compiled using the GNU Compiler Collection (GCC) \cite{gcc}.  Additional libraries were also used to perform standard tasks.  These include:
\begin{itemize}
\item ROOT - An object oriented framework for large scale data analysis developed at CERN \cite{root}.
\item Armadillo - A \CC\ linear algebra library \cite{armadillo}.
\item GNU Scientific Library - A numerical library for C and \CC.  The library provides a wide range of mathematical routines such as random number generators, special functions and least-squares fitting \cite{gsl}.
\end{itemize}

Other commercial and open source software packages used include
\begin{itemize}
\item Wolfram Mathematica (version 7)  \cite{mathematica}
\item Mathworks MatLab R2013a \cite{matlab} and the Easyspin EPR simulation package \cite{Stoll2006}.
\item OriginLab OriginPro 8 \cite{origin}
\item NI LabView 2011 \cite{NationalInstruments2011}
\item LibreCad \cite{librecad}
\item Clewin 3 \cite{clewin}
\end{itemize}

\bibliographystyle{h-physrev3}
\bibliography{mybiblio}

\chapter{Towards a quantum computing architecture with spins}\label{chap:Theo0}

\section{Cavity Quantum Electrodynamics and the Jaynes-Cummings Hamiltonian}\label{sec:CQED}

\begin{figure}[!t]
\centering
\includegraphics[width=0.85\textwidth]{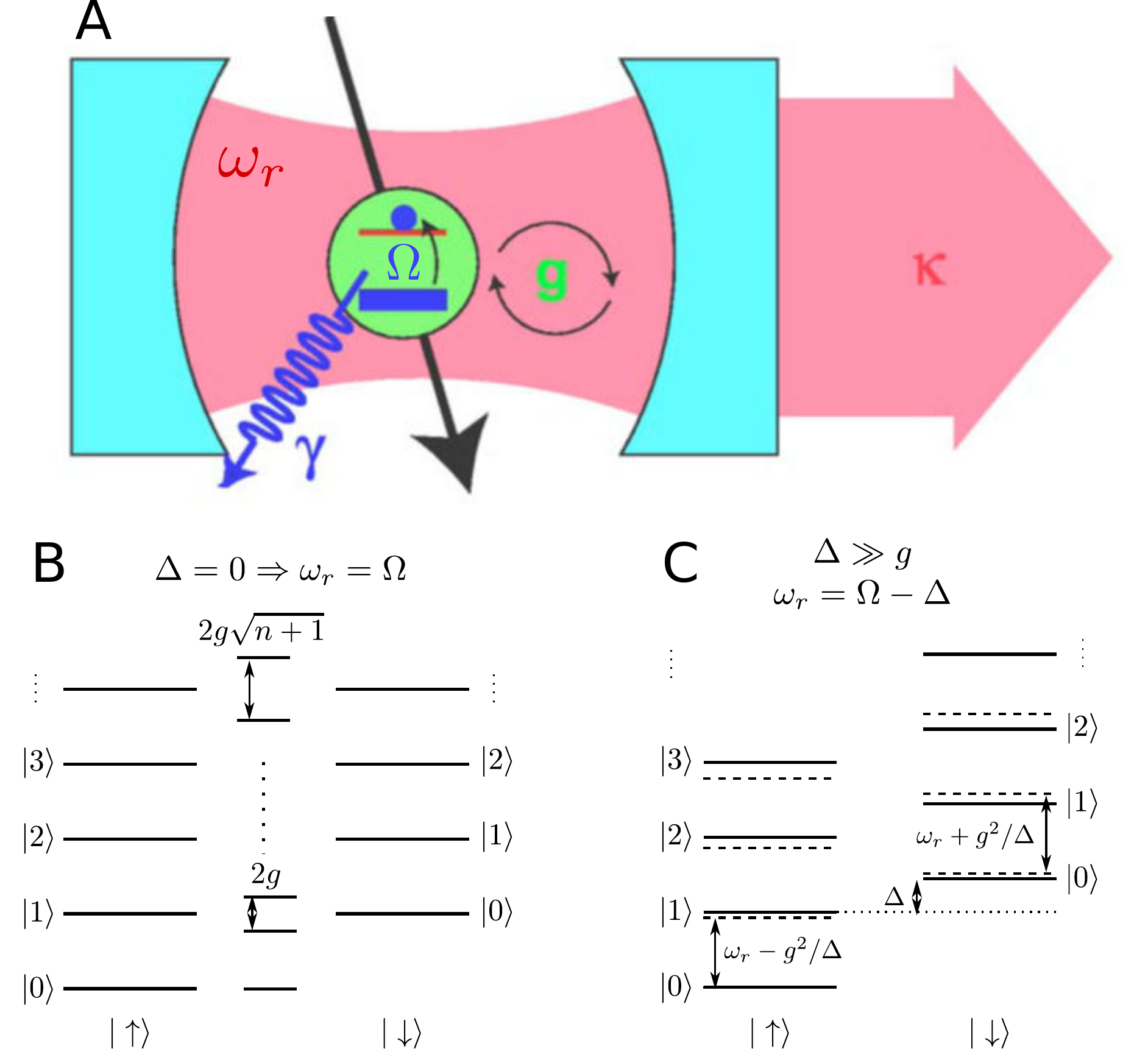}
\caption{A. Basic Cavity QED system and parameters \cite{Blais2004}.  B. Energy level spectrum of the coupled cavity-qubit system in the zero detuning case ($\omega_r = \Omega$).  The states in the left and right columns represent the undressed (or uncoupled) states while the center column shows the dressed states.  C. Energy level spectrum of the coupled cavity-qubit system in the large detuning case ($\Delta = \Omega - \omega_r \gg g$).  The solid lines represent the undressed states while the dotted lines represent the dressed states.  The level spacing is different depending on the state of the qubit.}
\label{fig:CQEDdiag}
\end{figure}

The field of cavity quantum electrodynamics (QED) studies the interaction of photons in resonant cavities with quantum mechanical systems, for example atoms or other systems with a discrete energy level spectrum, that couple either to electric or magnetic fields (or both).  The classical example of a cavity QED system is an optical cavity driven by a laser where the transmission intensity through the cavity is monitored as one drops atoms through it (figure \ref{fig:CQEDdiag}).  Intuitively, changes will be observed when the atom couples to or absorbs radiation from the cavity giving us information on the nature of the atom and its energy levels.  These levels can be tuned with an external electric or magnetic field such that a certain level splitting coincides with the cavity photon energy.  If the coupling is strong enough and the excitation power is such that there is only one photon stored in the cavity, the atom will oscillate coherently between two quantum states in the phenomenon known as Rabi oscillations \cite{Rabi1937}.  The oscillation frequency is known as the Rabi frequency $\Omega_R$ and is proportional to the interaction energy of the atom with the photon electromagnetic field.  If the atom involved can be described in first approximation as a two level quantum system, the Hamiltonian for the composite system is the Jaynes-Cummings Hamiltonian \cite{Jaynes1963}:
\begin{equation}
H = \hbar\omega_r\left(a^\dag a+\frac{1}{2}\right) + \frac{\hbar\Omega}{2}\sigma^z + \hbar g(a^\dag\sigma^-+\sigma^+a)+H_\kappa+H_\gamma, \label{eq:jc}
\end{equation}
where $\omega_r$ is the cavity resonance frequency (typically controlled by the geometry of the cavity and the materials used), $\hbar\Omega$ is the energy difference between the atom energy levels, $g$ is the coupling rate of the two systems (dependent on the strength of the electric or magnetic field at the atom's location as well as the atom's properties such as electric dipole moment or spin), $H_\kappa$ describes the losses of the cavity (decay rate $\kappa = \omega_r/Q$) and $H_\gamma$ contains the decoherence or decay of the atom into other channels (related to the $\textrm{T}_2$ decay time).  The Jaynes-Cummings Hamiltonian assumes that the rotating wave approximation (essentially $g\ll \omega_r,\Omega$ and $\omega_r \sim \Omega$) and the dipole approximation hold \cite{Blais2004,Bina2012}.

In the absence of damping, the Hamiltonian (\ref{eq:jc}) can be readily diagonalized.  We define the detuning parameter as $\Delta = \Omega - \omega_r$, the qubit eigenstates as $\ket{\uparrow}$,$\ket{\downarrow}$, and the cavity eigenstates as $\ket{n}$ for $n=0,1,2,...$.  The energy eigenvalues and eigenstates for the joint system are then:

\begin{eqnarray}
E_{\overline{\pm,n}} & = & (n+1)\hbar\omega_r \pm \frac{\hbar}{2}\sqrt{4g^2(n+1)+\Delta^2} \label{eq:jcenergy}\\
E_{\uparrow,0} & = & -\frac{\hbar\Delta}{2}\label{eq:Eground}\\
\ket{\overline{+,n}} & = & \cos{\theta_n}\ket{\downarrow,n} + \sin{\theta_n}\ket{\uparrow,n+1}\\
\ket{\overline{-,n}} & = & -\sin{\theta_n}\ket{\downarrow,n} + \cos{\theta_n}\ket{\uparrow,n+1}\\
\tan{2\theta_n} & = & \frac{2g\sqrt{n+1}}{\Delta}
\end{eqnarray}
with $\ket{\uparrow,0}$ being the ground state of the system.  These states are represented schematically in figure \ref{fig:CQEDdiag} for both the zero detuning ($\Delta = 0$) and large detuning ($\Delta\gg g$) cases.

If $\Delta = 0$, each pair of states with $n+1$ quanta (n photons in the cavity and a qubit in the excited state $\ket{\downarrow}$ or n+1 photons and the qubit in its ground state $\ket{\uparrow}$) would be degenerate for $g=0$.  Any finite coupling lifts this degeneracy by $2g\sqrt{n+1}$.  In the case of a single excitation, the states are the maximally entangled atom-cavity states $\ket{\overline{\pm,0}}=(\ket{\uparrow,1}+\ket{\downarrow,0})/\sqrt{2}$ separated $2g$ in energy.  This means that a qubit initially in the $\downarrow$ state coupled to an empty cavity will coherently oscillate into a cavity excitation and back at the Rabi frequency $\Omega_R= 2g$.

The large detuning (or dispersive) case corresponds to $\Delta \gg g$ (but still $\Omega \sim \omega_r$).  Expanding to first order, the one excitation space becomes:
\begin{eqnarray}
\ket{-,0} & \sim & \frac{-g}{\Delta}\ket{\downarrow,0} + \ket{\uparrow,1} \nonumber\\
\ket{+,0} & \sim & \ket{\downarrow,0} + \frac{g}{\Delta}\ket{\uparrow,1} \label{eq:statesdispersive}
\end{eqnarray}
The states are approximately the undressed states but have small mixings with other states within the manifolds with the same number of excitations.  Also, as seen in figure \ref{fig:CQEDdiag}C, the energy level separation between states with $n$ and $n+1$ cavity excitations depends on the state of the qubit with the cavity frequency being \emph{pulled} by $\pm \frac{g^2}{\Delta}$ as can be seen expanding expression (\ref{eq:jcenergy}) for the energy eigenvalues.

\subsection{Energy density and limits on the coupling}

\begin{figure}[!t]
\centering
\includegraphics[width=0.75\textwidth]{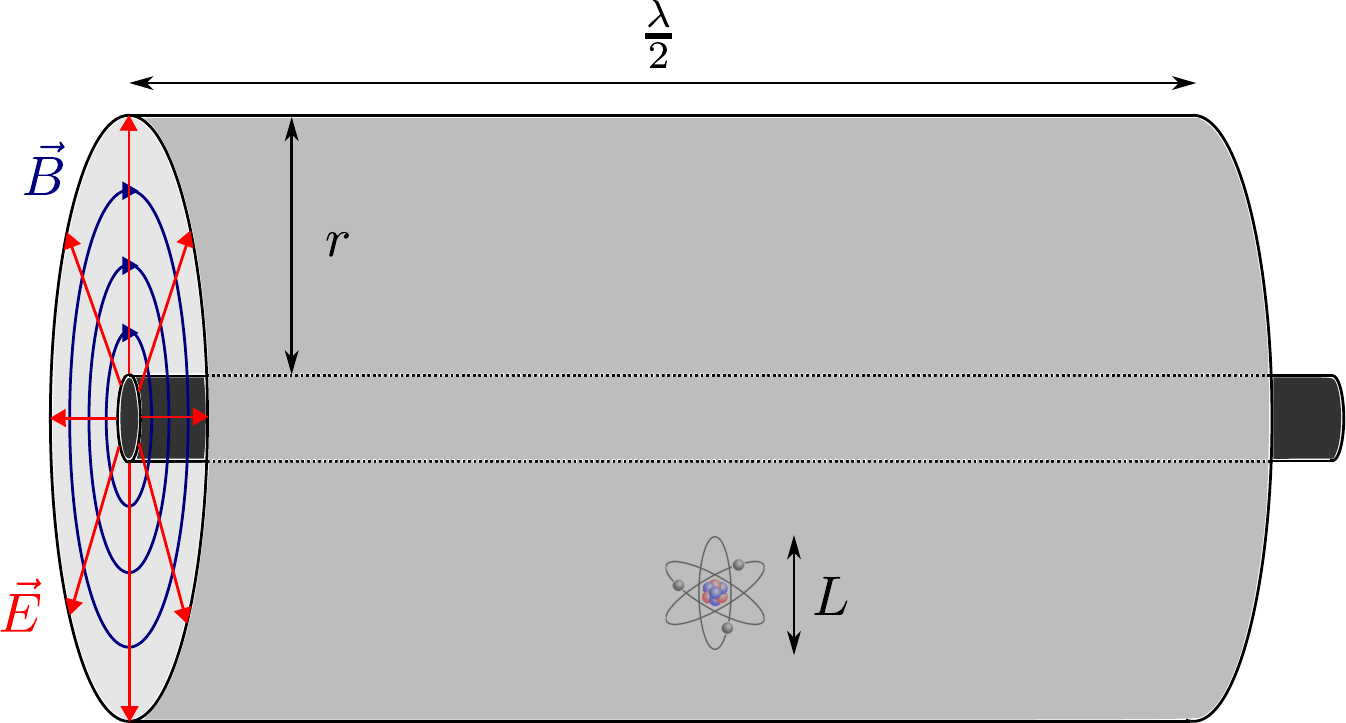}
\caption{Simple cylindrical microwave cavity}
\label{fig:cylindercavity}
\end{figure}

The previous analysis illustrates how the value of $g$ sets the speed at which quantum operations can be performed in this setup.  Therefore, for applications in quantum information processing, the goal is to maximize the coupling $g/\pi$, while keeping it much larger than the two main dephasing rates of the system $g\gg \kappa,\gamma$.  This strong coupling limit allows for a large number of operations to be performed during the lifetime of the quantum state and is necessary for a system to have practical applications in quantum computing.
A simple computation can give us the order of magnitude for the coupling to a 3D cavity of a single electric or magnetic dipole \cite{Schoelkopf2008}.  As a first approximation, the coupling $g$ is given by the field strength at the atom site times its dipole moment:
\begin{equation}
g=\frac{dE_0}{\sqrt{2}\hbar} \; \textrm{(Electric)} \quad g=\frac{\mu B_0}{\sqrt{2}\hbar} \; \textrm{(Magnetic)} \label{eq:g}
\end{equation}
where $E_0$ and $B_0$ are the electric and magnetic field amplitudes.  Given a typical cylindrical cavity geometry (figure \ref{fig:cylindercavity}), we can estimate the average (spatial) field strength remembering that the integral of the field amplitude over the cavity volume is equal to the zero point energy of the cavity \cite{Devoret2007}:
\begin{equation}
\frac{1}{2}\int_V\epsilon |E_0|^2 dV = \frac{1}{2}\int_V \frac{|B_0|^2}{\mu} dV= \frac{\hbar\omega_r}{2}
\end{equation}
Taking the spatial average and setting $\epsilon = \epsilon_0$ and $\mu=\mu_0$ gives:
\begin{equation}
\frac{\hbar\omega_r}{2} = \frac{1}{2}\epsilon_0 E_0^2 V= \frac{1}{2\mu_0} B_0^2 V \label{eq:Erms}
\end{equation}
The cavity dimensions give us $V=\pi r^2 \lambda/2$.  Rearranging terms leads to:
\begin{equation}
E_0 = \frac{1}{r}\sqrt{\frac{\hbar\omega_r^2}{\pi^2\epsilon_0c}}, \quad B_0 = \frac{1}{r}\sqrt{\frac{\hbar\omega_r^2\mu_0}{\pi^2c}}
\end{equation}
These equations show that the field in the cavity (and therefore the coupling $g$) can be increased by reducing the cavity radius or by increasing the operating frequency, both of which have the effect of increasing the energy density.  Since increasing the cavity frequency may not be an option since the atom energy levels may not be suited to the higher frequencies and higher frequencies are also harder to work with, the most direct way to improve the cavity field is to reduce the lateral dimension $r$ of the cavity.  For electric systems, and specially for quantum circuits (cooper pair boxes), the electric dipole can be approximated by $d=eL$ where $e$ is the electron charge and $L$ is the dipole's characteristic lateral dimension.  Introducing this approximation into (\ref{eq:g}) and then using (\ref{eq:Erms}) we get:
\begin{equation}
\frac{g}{\omega_r} = \frac{L}{r}\sqrt{\frac{2\alpha}{\pi}}, \label{eq:glim}
\end{equation}
where $\alpha = e^2/(4\pi\epsilon_0\hbar c)$  is the fine structure constant.  We see then that $g/\omega_r$ depends only on how well the atom ``fills'' the cavity ($L/r$) and on fundamental constants.  This reiterates our previous point that reducing the cavity size to match the system size is a good way to improve the coupling.  For magnetic systems the same principle applies although the magnetic moment is not in general proportional to the system size so the expression for $g$ doesn't reduce to the simple form of equation (\ref{eq:glim}).  However, reducing the cavity size to the system size is not possible in all cases.  For instance, for 3D microwave cavities, the lateral dimension may also be limited by the wavelength of the radiation mode making the development of different cavity geometries necessary to improve the coupling.  This is the approach taken in what has come to be known as circuit QED.

\subsection{Circuit Quantum Electrodynamics}

\begin{figure}[tb]
\centering
\includegraphics[width=0.7\textwidth]{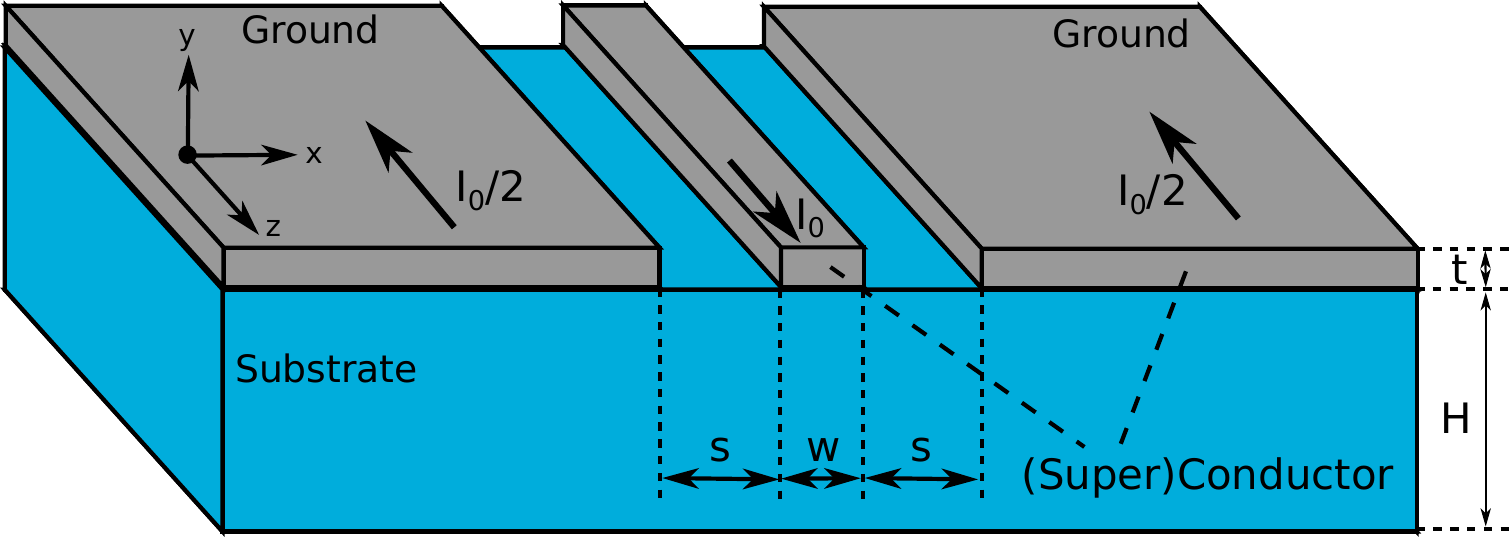}
\caption{Coplanar waveguide cross-section}
\label{fig:CPWGcross}
\end{figure}

In circuit QED, the 3D resonant microwave cavities are replaced by certain types of integrated circuits that are essentially sections of microwave transmission lines.  The prime example of this type of cavity is a coplanar waveguide (CPW) resonator, i.e., a section of a coplanar waveguide capacitively coupled to external feed lines.  A diagram showing the structure of a coplanar waveguide is shown in figure \ref{fig:CPWGcross} as well as the relevant dimensions (see section \ref{sec:basicCPWG} for more details).  For thick substrates the transmission parameters for CPW depend only on the materials used and the ratios of widths of the centerline and gaps \cite{Simons2004}.  This means that the lateral dimension can be made of the order of \SI{}{\micro\metre} using simple photolithograpy techniques while keeping the transmission properties constant.  As a first approximation, these systems may be considered 1D resonators where the resonant frequency is controlled by the length of the transmission line segment.  The photon energy is concentrated in and around the center line, thus the field strengths achieved can be much larger than in typical 3D cavities \cite{Blais2004}.  Furthermore, if superconducting materials such as Nb or Al are used for the circuit, the resistive losses can be suppressed to achieve quality factors of up to $10^6$ \cite{Frunzio2005,Goppl2008} meaning that, in these cases, the losses and dephasing of the system would be limited only by the atom properties.

\section{Early Experiments and applications in Quantum Computing}

Examples of strong coupling to atoms have been achieved previously.  For optical photons, the vacuum Rabi splitting was first observed in 1992 in Caesium atoms \cite{Thompson1992} and, in the microwave regime, strong coupling was also observed using highly excited atomic states known as Rydberg atoms \cite{Meschede1985,Rempe1987,Brune1996,Raimond2001} with transitions of the order of $\Omega = \SI{50}{\giga\hertz}$.  The first instance of a single quantum system coupled to a CPW resonator was a superconducting charge qubit or Cooper-pair box \cite{Wallraff2004}.  The latter can be approximated by a quantum two level system and is fabricated directly in the CPW resonator gap area.  The strong coupling regime in this system has enabled the realization of many other experiments demonstrating single qubit operations \cite{Majer2007}, its use as a quantum bus \cite{DiCarlo2009}, or multiple qubit gates \cite{Dicarlo2010}.  These applications, which form the basis of the most promising architectures for quantum information processing, are succinctly described below.

\subsection{Signatures of strong coupling}
\begin{figure}[tbh]
\centering
\includegraphics[width=0.8\columnwidth]{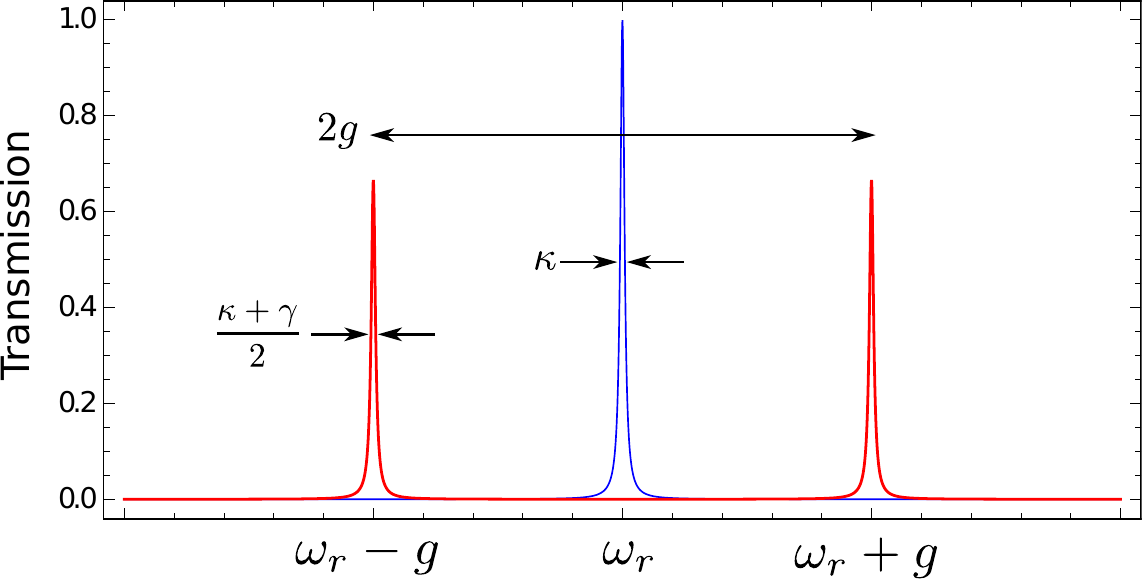}
\caption{Single photon transmission spectrum for a strongly coupled cavity-qubit system with zero detuning.}\label{fig:strongcoupling}
\end{figure}

When a qubit is strongly coupled to a cavity according to the Hamiltonian (\ref{eq:jc}) it produces a specific signature when tuned into resonance (i.e. $\Delta = \Omega - \omega_r = 0$).  Observing figure \ref{fig:CQEDdiag}B, if the cavity is excited with powers such that there is only a single excitation or photon in the cavity, the typical cavity transmission peak of width $\kappa$ splits into two peaks separated $2g$ each of width $(\kappa+\gamma)/2$ (see figure \ref{fig:strongcoupling}) given by the transitions from the ground state $\ket{\uparrow,0}$ to either $\ket{\overline{\pm,1}}$.  Also, the fact that the energy level separation scales as $\sqrt{n+1}$ makes the spectrum anharmonic and can lead to non-linear effects at high drive powers (where $n\gg 1$) \cite{Walls2008}.

\subsection{Quantum Non-Demolition Measurement in Cavity or Circuit QED}

The dispersive or large detuning regime, where $\Delta \gg g$, is in general more interesting for applications in quantum computing as it can allow quantum non-demolition (QND) measurements of the state of the qubit as well as its coherent manipulation \cite{Blais2004}.

\begin{figure}[tbh]
\centering
\includegraphics[width=0.65\columnwidth]{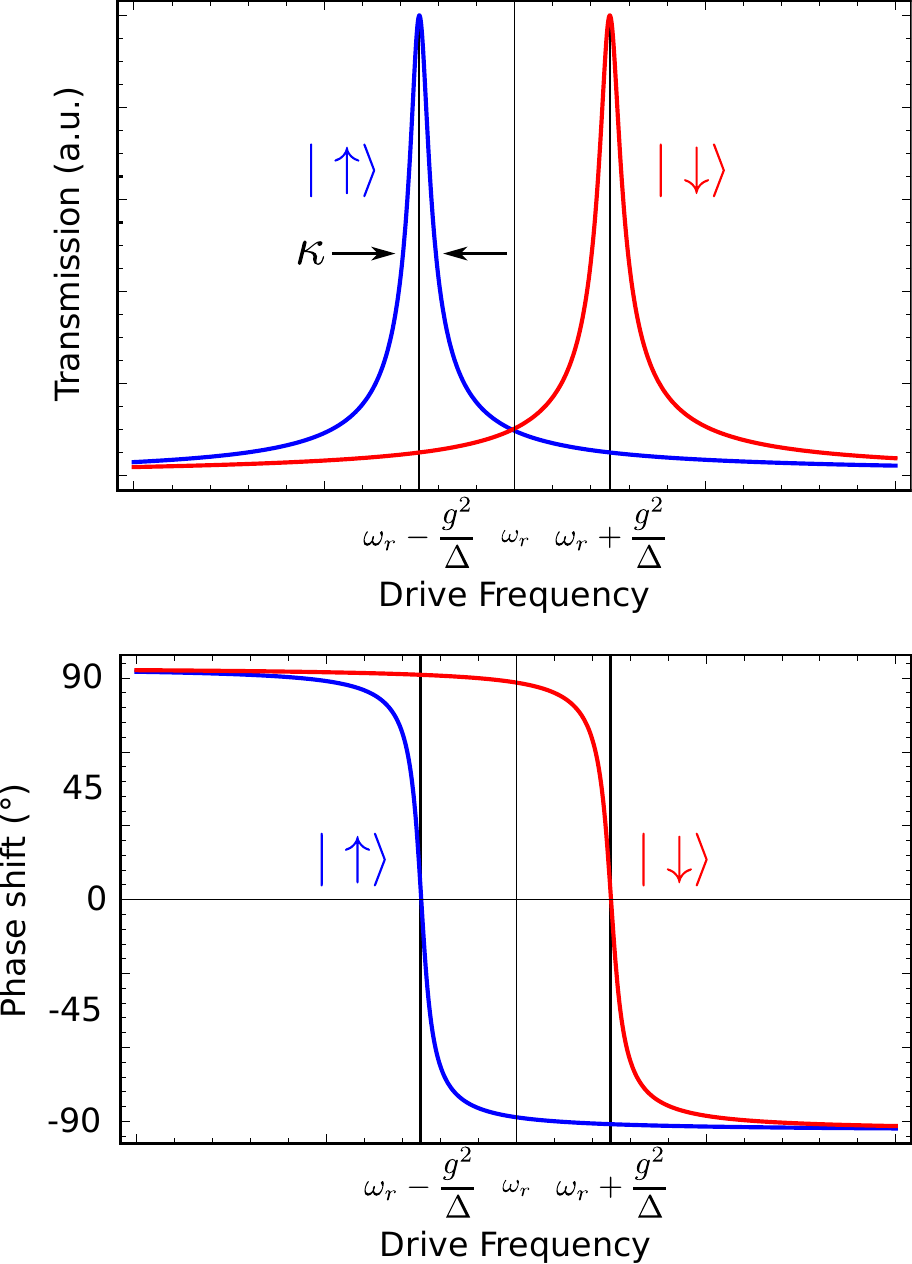}
\caption{Cavity transmission and phase shift in the large detuning or dispersive regime for a strongly coupled cavity-qubit system.  The cavity peak frequency and phase shift depend on the qubit state.}\label{fig:QND}
\end{figure}

On a basic level, the possibility of doing QND measurements of the qubit state is based on the fact that the qubit state \emph{pulls} the cavity frequency by $+\frac{g^2}{\Delta}$ or $-\frac{g^2}{\Delta}$ depending on whether the qubit is in the $\ket{\downarrow}$ or $\ket{\uparrow}$ state.  A driving radiation field of frequency $\omega'$ acting on the qubit-resonator system can be modeled by the following time dependent Hamiltonian \cite{Haroche1992}:
\begin{equation}
H_{\rm drive}(t) = \hbar \epsilon(t)(a^\dagger e^{-i\omega't} + ae^{i\omega't}), \label{eq:drive}
\end{equation}
where $\epsilon$ is the field amplitude.  Observing figure \ref{fig:CQEDdiag}C, we expect that, if the qubit is initially in its ground state $\ket{\uparrow}$, there will be transmission peaks at $\omega_r-\frac{g^2}{\Delta}$ and at $\Omega+\frac{g^2}{\Delta}$.  However, if we compute the transition matrix elements of the drive (in a frame rotating at the drive frequency):
\begin{eqnarray}
&\bra{\uparrow,0}H_{\rm drive}\ket{\overline{-,n}} \sim \epsilon & \\
&\bra{\uparrow,0}H_{\rm drive}\ket{\overline{+,n}} \sim \frac{\epsilon g}{\Delta}. &
\end{eqnarray}
This shows how the peak at $\Omega+g^2/\Delta$ corresponding approximately to a qubit flip (see equations (\ref{eq:statesdispersive})) will be suppressed by the factor $g/\Delta \ll 1$ as compared to the $\omega_r-g^2/\Delta$ peak that does not flip the qubit.  The qubit flip transition from an excited qubit state is similarly suppressed.  The expected transmission spectra for each case are shown in figure \ref{fig:QND}.  Since qubit flips by the driving field are suppressed in either case, this allows us to probe the qubit state by monitoring the cavity transmission at a fixed frequency without altering the state of the qubit.  If we probe, for instance, at the pulled frequency $\omega_r-g^2/\Delta$, we should observe high transmission if the qubit is in the $\ket{\uparrow}$ state but a very small transmission if it is in the $\ket{\downarrow}$ state.  The situation is reversed if we probe at $\omega_r+g^2/\Delta$.  If the phase of the signal is measured, it is also possible to probe the state by probing the cavity at the bare frequency $\omega_r$.  According to figure \ref{fig:QND}  a positive or negative phase shift is observed if the qubit is in the $\ket{\downarrow}$ or $\ket{\uparrow}$ state respectively.  

\subsection{Coherent manipulation in Cavity or Circuit QED}

In the large detuning limit, irradiating at the qubit frequency $\Omega$ can be used to coherently manipulate the qubit.  When irradiating at the cavity frequency $\omega_r$, a sizable photon population is introduced in the cavity. The states of these photons become entangled with the qubit states and can be used to probe the latter.  When irradiating at the qubit frequency, the cavity is only virtually populated (with $\langle n \rangle < 1$) and most of the photons are reflected from the cavity input.  The effect of this driving can be modeled by combining the Hamiltonian (\ref{eq:jc}), (\ref{eq:drive}) and the unitary transformation,
\begin{equation}
U = \exp{\left(\frac{g}{\Delta}(a\sigma^+-a^\dagger \sigma^-)\right)}.
\end{equation}
This leads to the following effective single-qubit Hamiltonian in the frame rotating at the drive frequency $\omega'$:
\begin{eqnarray}
H_{\rm eff} & = & \frac{\hbar}{2}\left(\Omega + 2\frac{g^2}{\Delta}\left(a^\dagger a + \frac{1}{2}\right)-\omega'\right)\sigma^z \nonumber\\ 
&& + \hbar\frac{g\epsilon (t)}{\Delta} \sigma^x + \hbar(\omega_r-\omega')a^\dagger a + \hbar\epsilon(t)(a^\dagger + a).
\end{eqnarray}
If, for instance, we drive at a frequency $\omega' = \Omega + (2n+1)g^2/\Delta$, the first term in the previous equation cancels out and the Hamiltonian generates rotations of the qubit about the $x$ axis with frequency $g\epsilon/\Delta$.  Different drive frequencies produce different types of rotations.  It can be shown that these are sufficient to perform any one-qubit logical operation.  Here we see that the rotation \emph{speed} is proportional to the coupling $g$ and enhancing this value allows for more operations to be performed before the system loses its quantum coherence.

\section{Coupling to magnetic systems: proposal for an all spin quantum processor}\label{sec:theo0smm}

The examples in the previous section deal with systems that couple electrically to the cavity mode.  Systems that couple to the magnetic field, such as solid-state spin ensembles, are also seen as promising media to store quantum information as well as to interconnect radio-frequency and optical photons \cite{Imamoglu2009,Wesenberg2009,Marcos2010}.  Experiments performed in the last few years have shown the feasibility of coherently coupling NV or P1 centres in diamond to either superconducting resonators \cite{Schuster2010,Kubo2010,Amsuss2011} or flux qubits \cite{Zhu2011}.  These diamond defects effectively act as anisotropic spin 1 systems and can be collectively coupled to a CPW resonator providing a $\sqrt{N}$ enhancement to the coupling.  Strong coupling is achieved thanks to this enhancement of the coupling in conjunction with the very low dephasing rate of the magnetic centers (with $T_2\simeq 1$-\SI{2}{\milli\second} at room temperature \cite{Balasubramanian2009}).  Evidences for strong magnetic coupling have also been found, even at room temperature, between spin-$1/2$ paramagnetic radicals and three-dimensional microwave cavities \cite{Chiorescu2010,Abe2011}.  

\begin{figure}[tb]
\centering
\includegraphics[width=0.7\columnwidth]{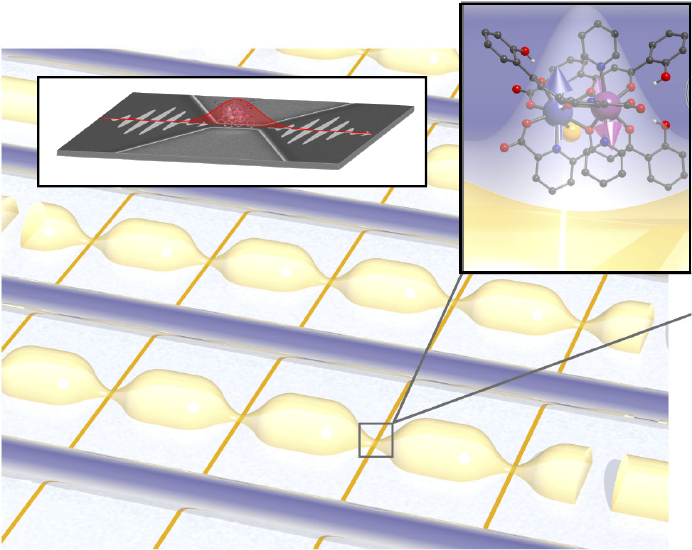}
\caption{Possible architecture of an all-spin quantum processor.  Spin qubits would be placed along a superconducting coplanar waveguide resonator at positions where there is a magnetic field maximum for the cavity mode considered.  Each spin could be tuned into and out of resonance using a wire or microcoil to appliy a local magnetic field.  Entangling between qubits at the different sites would also be feasable.}\label{fig:fantasy}
\end{figure}

Coherently coupling to individual spins is more challenging. Provided this limit can be attained, on-chip superconducting circuits could be used to coherently manipulate and transfer information between spin qubits, thus providing a suitable architecture to implement an all-spin quantum processor \cite{Awschalom2013}.  Taking inspiration from circuit QED systems, we can imagine constructing a quantum computing system such as that shown in figure \ref{fig:fantasy}.  Single spins would be placed at locations along the cavity where the magnetic field is maximum for the desired rf mode.  The spins could then be individually addressed by locally applying magnetic fields to bring them into and out of resonance with the cavity.  This architecture also would allow entangling multiple qubits in the same cavity as described in \cite{DiCarlo2009,Dicarlo2010}.

Our objective throughout this thesis is to test the feasibility of this proposal.  As it has been made clear in the previous sections, the main parameter that needs to optimized is the coupling $g$.  If we assume a typical operating frequency of \SI{10}{\giga\hertz}, a CPW resonator with an effective volume $V=\lambda/2\cdot(\SI{10}{\micro\metre})^2$, a spin system with magnetic moment $\mu=\mu_\textrm{B}$ and use equations (\ref{eq:g}) and (\ref{eq:Erms}), we arrive at a value of $g\simeq \SI{20}{\hertz}$.  This value is far too small to overcome the best dephasing rates known for molecular spin qubits ($\gamma\gtrsim \SI{0.1}{\mega\hertz}$ \cite{Wedge2012}) or even resonators ($\kappa \gtrsim \SI{5}{\kilo\hertz}$ if $Q\sim 10^6$ at 10 GHz).  Observing equation \ref{eq:g}, this limitation leaves us with basically two possibilities to explore if we wish to improve the coupling of single atom or molecule spins and allow their use in quantum computing:

\begin{enumerate}
\item Exploring different spin systems with higher magnetic moments ($\mu$) and transition matrix elements.
\item Increasing the field strength $B_0$ felt by each spin.
\end{enumerate}

In chapter \ref{chap:Theo1} we will address the first approach and explore what desirable qualities a magnetic system should have in order to strongly couple to a quantum circuit.  We will simulate the field distribution in a CPW resonator and investigate what changes in design can be made in order to increase the magnetic field at the spin site.  We also investigate a different family of magnetic materials: single molecule magnets (SMMs) \cite{Christou2000,Gatteschi2003,Bartolome2013}. These are organometallic molecules formed by a high-spin magnetic core surrounded by organic ligands that naturally organize into molecular crystals. In SMMs with strong uniaxial magnetic anisotropy, such as Mn$_{12}$ or Fe$_{8}$, the magnetization shows hysteresis (i.e. magnetic memory) near liquid Helium temperatures \cite{Sessoli1993}. In addition, SMMs show intriguing quantum phenomena such as resonant spin tunneling \cite{Friedman1996,Hernandez1996,Thomas1996,Sangregorio1997} and Berry phase interferences between different tunneling paths \cite{Wernsdorfer1999}.

Later, in chapter \ref{chap:CPWG} we will address the second approach and give more details on CPW resonators and describe a practical method for optimizing their coupling to spin systems.  In broad terms, the optimization can be done by optimizing the positioning of the magnetic molecule by fixing it in a higher field area of the cavity and by engineering the cavity to have higher field intensities.  These results show that improvements in the cavity architecture and in the choice of spin qubit that are within the reach of current state of the art, could eventually make the realization of the spin processor sketched in figure \ref{fig:fantasy} feasible.

\bibliographystyle{h-physrev3}
\bibliography{mybiblio}

\chapter{Theoretical basis for the coupling of Quantum Circuits to Spin Qubits}
\chaptermark{Coupling of Quantum Circuits to Spin Qubits}
\label{chap:Theo1}

\section{Introduction}

As mentioned in section \ref{sec:theo0smm}, Single Molecule Magnets (SMMs) are attractive candidates to act as either spin qubits \cite{Leuenberger2001,Tejada2001,Troiani2005,Affronte2009,Ardavan2009,Stamp2009} or spin-based quantum memories.  Several characteristics make the particularly attractive: the ability to tune their properties, e.g. spin, magnetic anisotropy, resonance frequencies, etc, by chemical design and their high spins (e.g. $S=10$ for both Fe$_{8}$ and Mn$_{12}$), large densities (typically $\sim 10^{20}-10^{21}$ spins/cm$^{3}$), and the fact that, in many SMM crystals, the anisotropy axes of each magnetic centre are aligned parallel to each other, which might enable the attainment of stronger couplings than those previously achieved with other natural spin systems.   In this chapter we study from a theoretical perspective the possibility that strong coupling to single molecules might be achieved in the near future with available technologies and, on the other, what new physics, or new physical regimes, can be expected from the coupling of SMMs crystals to these devices.

The chapter in organized as follows. Section \ref{sec:SMMs} describes the basic features and the spin Hamiltonian of SMMs. A generic framework to calculate the magnetic coupling to electromagnetic rf fields is introduced, and then applied to discuss how such a coupling depends on molecular properties, such as spin and anisotropy, as well as on the intensity and orientation of the external magnetic field. The following section \ref{sec:resonators} gives realistic estimates of the coherent coupling of some well known SMMs to superconducting coplanar resonators as a function of their dimensions and geometries. The final section \ref{sec:conclusions} gives the conclusions for this chapter which provide guidance for the experiments carried out in the course of this thesis work.

\section{Coupling of single molecule magnets to quantum radiation fields}
\label{sec:SMMs}

\subsection{Basic properties and spin Hamiltonian of a SMM}

A magnetic spin with no anisotropy in an external magnetic field has a Hamiltonian defined by the Zeeman interaction:
\begin{equation}
\mathcal{H}= -\vec{\mu}\vec{B} = -g\mu_\textrm{B}\vec{S}\vec{B} \label{eq:zeeman}
\end{equation}
where $\vec{\mu} = g\mu_\textrm{B}\vec{S}/\hbar$ is the magnetic moment of the spin system, g is the g-factor (related to the gyromagnetic ratio $\gamma = g\mu_\textrm{B}/\hbar$) and $\vec{S}=(S_X,S_Y,S_Z)$ are the spin operators (without the $\hbar$ factor).  Since the system is isotropic, at zero field there is no preferred spin direction and there will be $2S+1$ degenerate quantum states which correspond to the possible values of the spin projection along any given direction.  When a field is applied along an arbitrary axis, which for simplicity can be chosen to be the Z axis, the Hamiltonian eigenstates are also eigenstates of the $S_z$ spin operator with their energies split evenly according to their $S_z$ quantum number, $E_m = mg\mu_\textrm{B}B$ with $m=-S,(-S+1),\ldots,(S-1),S$.

\begin{figure}
\includegraphics[width=\textwidth]{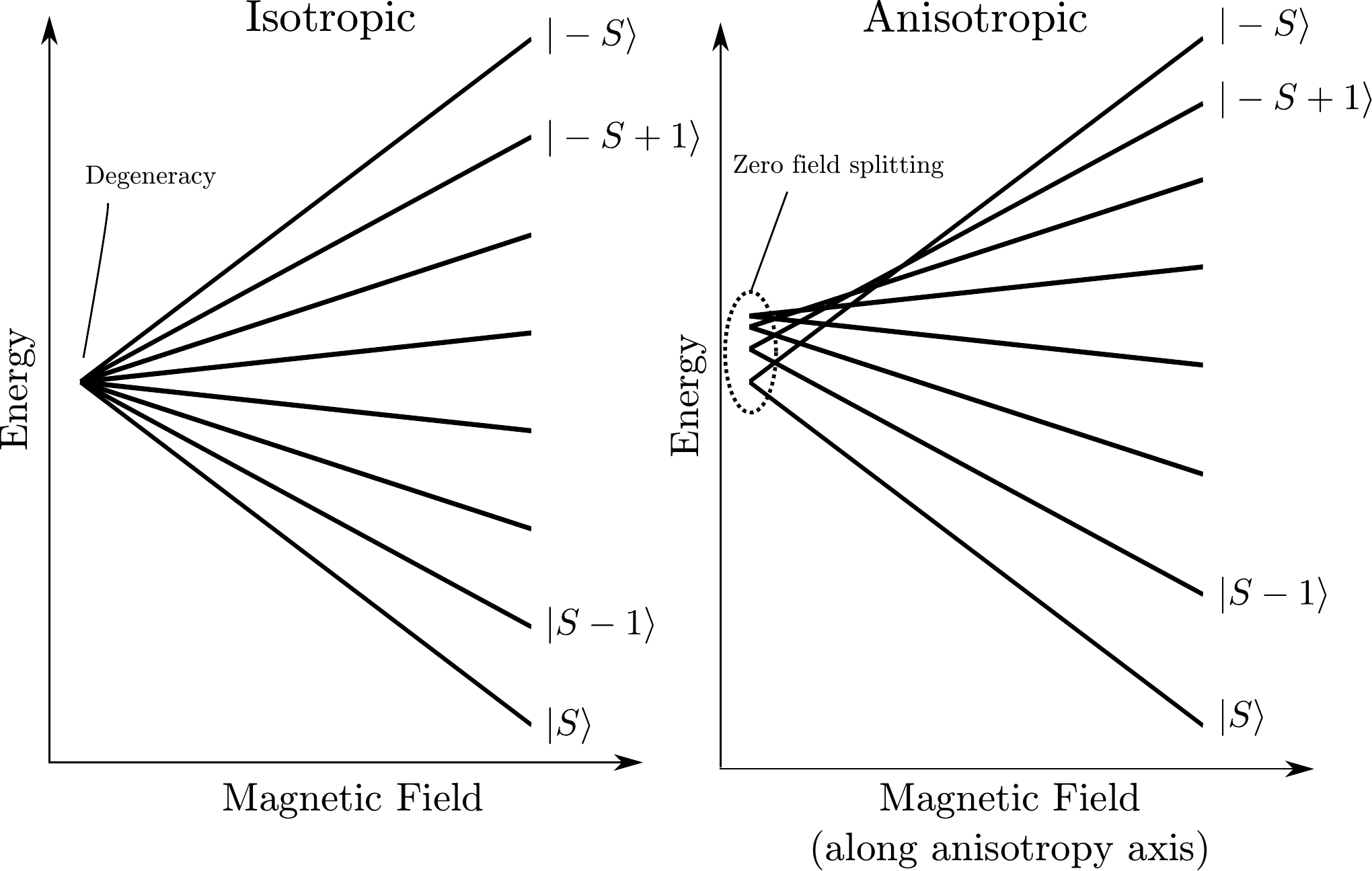}
\caption{Energy levels of an isotropic and an anisotropic spin in a magnetic field.  In the case of the anisotropic spin, there is a single anisotropy axis and the field is applied along this direction.}
\label{fig:anisotropy}
\end{figure}

In the case of a more complex system where anisotropies have important contributions, a zero field splitting term must be added to the Zeeman interaction energy (see figure \ref{fig:anisotropy}).  The zero field Hamiltonian usually has a preferred axis direction (which we commonly identify with the molecular $Z$ axis) and the spin along this axis ($S_z$) approximately determines the eigenstate.  The level spacing is uneven in general and, in addition, mixing among levels is possible (associated to terms containing spin operators transverse to the quantization axis, i.e., $S_{x,y}$).

\begin{figure}
\centering
\includegraphics[width=0.5\textwidth]{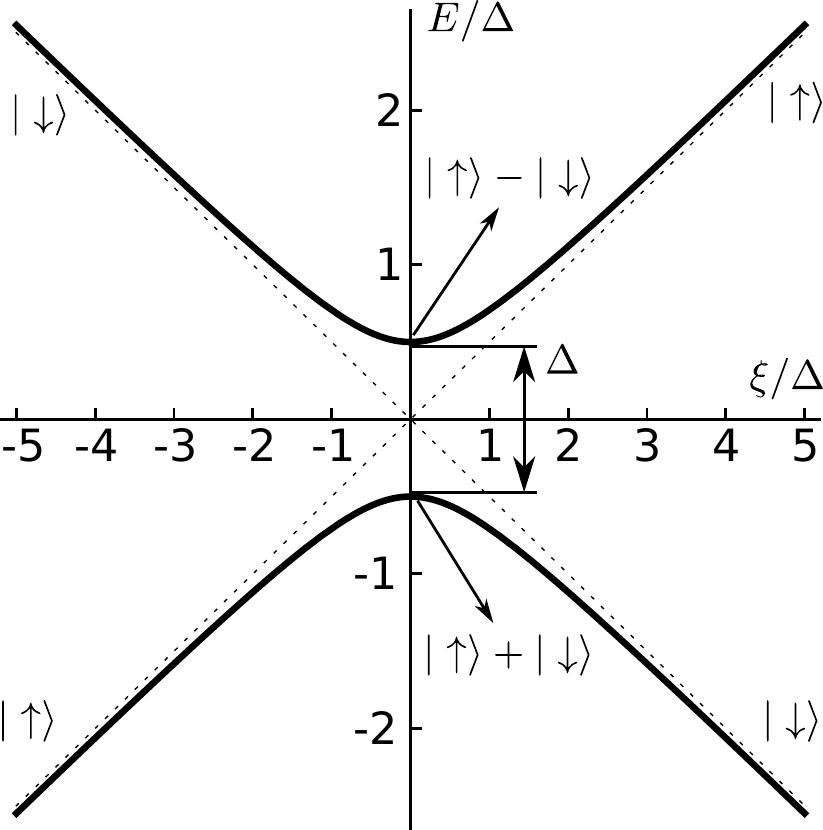}
\caption{Energy levels of an effective spin $1/2$ with a transverse anisotropy.  The graph shows the tunnel splitting ($\Delta$) at low bias ($\xi$).}
\label{fig:splitting}
\end{figure}

To illustrate such mixing effects, we use a very simple effective spin $1/2$ Hamiltonian as an example (figure \ref{fig:splitting}):
\begin{equation}
\mathcal{H} = -\Delta S_x + \xi S_z, \label{eq:tunnel}
\end{equation}
where $\xi$ is a \emph{bias} energy (arising, for example, from an external field applied along Z) and $\Delta$ corresponds to the quantum tunnel splitting.  This system can be solved analytically and the energy eigenvalues and eigenstates are given by
\begin{eqnarray}
E_\pm & = & \pm\frac{1}{2}\sqrt{\Delta^2+\xi^2}, \\
\ket{+} & = & \frac{1}{\sqrt{1+\alpha^2}}\left(\ket{\uparrow}-\alpha\ket{\downarrow}\right), \\
\ket{-} & = & \frac{1}{\sqrt{1+\alpha^2}}\left(\ket{\downarrow}+\alpha\ket{\uparrow}\right),
\end{eqnarray}
where $\alpha = \frac{\sqrt{\Delta^2+\xi^2} - \xi}{\Delta}$ and $\ket{\uparrow}$ and $\ket{\downarrow}$ are the $S_z$ eigenstates.  For $\xi \gg \Delta$, we have $\alpha \simeq 0$ and $\ket{\uparrow},\ket{\downarrow}$ become approximate energy eigenstates with energies $\pm \xi/2$, basically recovering the isotropic case.  However, if $\xi \ll \Delta$, $\alpha \simeq 1$ and the eigenstates are symmetric and antisymmetric superpositions of $\ket{\uparrow},\ket{\downarrow}$ separated in energy by the anisotropy energy or tunnel splitting $\Delta$.  If $\Delta =0$ these two states would simply cross over as the bias was swept through $\xi=0$ and be degenerate at this point.  However, this splitting introduces an \emph{anticrossing} effect when there is a strong tunnel splitting $\Delta$.  This level anticrossing in often seen between nearly degenerate levels of high spin systems with strong quantum tunneling effects \cite{Friedman1996,Hernandez1996,Thomas1996, Sangregorio1997,Wernsdorfer1999,Luis2000, DelBarco2004}.

\begin{figure}
\includegraphics[width=\textwidth]{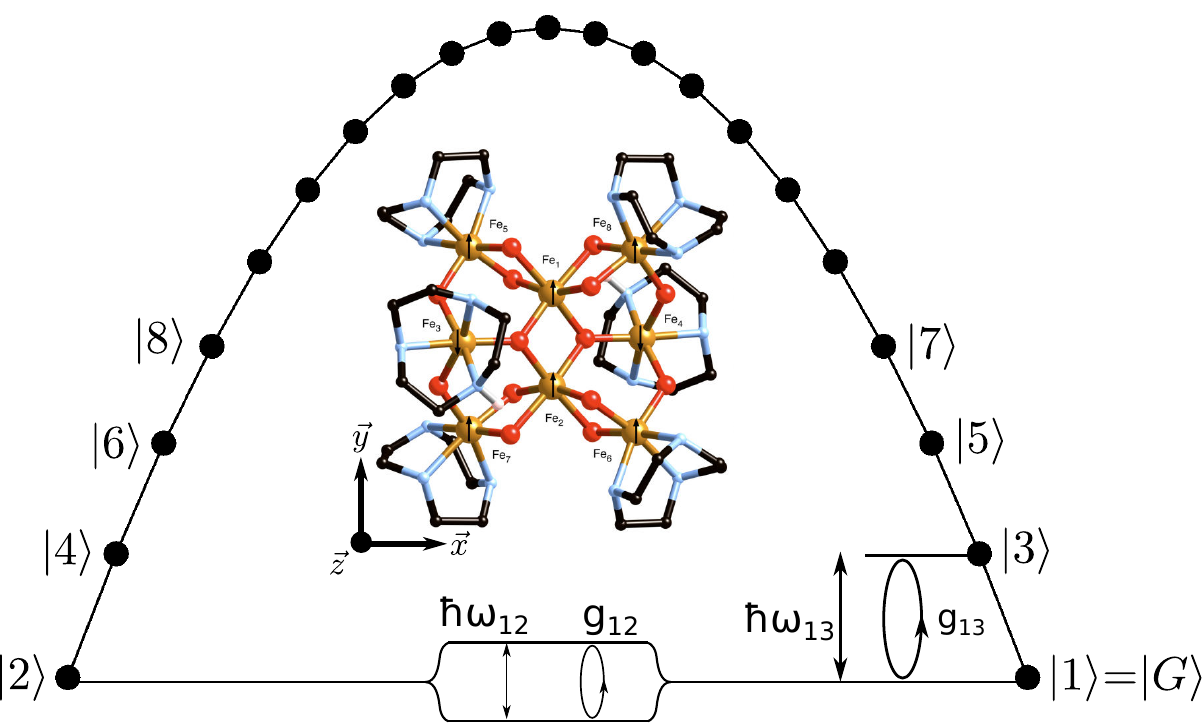}
\caption{Energy level scheme of the [\ce{(C6H15N3)6Fe8O2(OH)12}] single molecule magnet (neglecting off-diagonal terms), shown in the inset and referred to in shorthand as \ce{Fe8} \cite{Weighardt1984}. Two possible selections of states for the use of this SMM as qubit are schematically shown.  In particular \ce{Fe8} has $S=10$ and the following anisotropy parameters: $B_2^0 = \SI{-9.8e-2}{\kelvin}$, $B_2^2 = \SI{4.63e-2}{\kelvin}$ and $B_4^4 = \SI{-5.84e-5}{\kelvin}$ \cite{Wernsdorfer1999}}
\label{fig:Fe8}
\end{figure}

In general, anisotropic contributions are necessary to describe the case of single molecule magnets (SMMs).  These systems are organometallic molecules with a high spin magnetic core usually formed by one or several magnetic ions held in place by different types of organic (non-magnetic) ligands.  Their magnetic configuration is mainly determined by exchange couplings between the ions in this core and by their interactions with the crystal field.  The former give rise to multiplets with well-defined spin values, while the latter generate a magnetic anisotropy, thus also a zero-field splitting within each multiplet lifting the degeneracy of the states. Here, we consider only the ground state multiplet with spin $S$ and neglect its quantum mixing with excited multiplets.  These multiplets are assumed to be sufficiently far away energetically from the ground multiplet that their contributions can be safely ignored. This approximation, widely used to describe the physics of SMMs, is known as the "giant spin approximation"\cite{Wilson2007}. The effective spin Hamiltonian of a SMM in an adequate molecular reference frame (i.e. with the coordinate axes aligned with the anisotropy axes) reads then as follows
\begin{equation}
{\cal H}_{\rm s} = \sum_{k,l} B_{k}^{l}O_{k}^{l}-g_{S}\mu_{\rm B} \left( B_{X}S_{X} + B_{Y}S_{Y} + B_{Z}S_{Z} \right)
\label{eq:GSHamiltonian}
\end{equation}
where the first term describes the zero-field splitting contribution and the second is a Zeeman interaction term, where $g_{S}$ is the g-factor for the SMM spin and $B_{X}$, $B_{Y}$, and $B_{Z}$ are the components of the external magnetic field along the molecular axes $X$, $Y$, and $Z$.  The zero-field splitting term is written in terms of the Stevens effective spin operators $O_{k}^{l}$ (see table \ref{fig:stevens}) while the $B_{k}^{l}$'s are the corresponding magnetic anisotropy parameters.  The molecular symmetry and structure determine which anisotropy parameters are zero.  In some simple cases, where only the second order parameters $B_2^0$ and $B_2^2$ are non-zero, this Hamiltonian is commonly rewritten as:
\begin{equation}
{\cal H}_{\rm s} = D\left(S_Z^2-\frac{1}{3}S(S+1)\right) + E (S_X^2-S_Y^2) - g_{S}\mu_{\rm B} \vec{B}\cdot\vec{S},
\label{eq:GSHamiltonian2}
\end{equation}
where $D=3B_2^0$ and $E=B_2^2$ are the ``traditional'' second-order zero-field splitting parameters.  In the large magnetic field limit ($g_S\mu_\textrm{B}B \gg (2S-1)D$) the Zeeman term becomes the dominant contribution and we recover the evenly separated energy levels and spin eigenstates of the isotropic case.

\begin{table}[tbh]
\centering
\begin{small}
\begin{tabular}{ccl}
\hline
\hline
$k$ & $q$ & $O_k^q$ \\
\hline
2 & 0 & $3S_z^2-sI$\\
 & $\pm 1$ & $c_\pm\left[S_z,S_+\pm S_-\right]_+$\\
 & $\pm 2$ & $c_\pm(S_+^2\pm S_-^2)$\\
\hline
 4 & 0 & $35 S_z^4-(30s-25)S_z^2+(3s^2-6s)I$\\
 & $\pm 1$ & $c_\pm\left[7S_z^3-(3s+1)S_z,S_+\pm S_-\right]_+$\\
 & $\pm 2$ & $c_\pm\left[7S_z^2-(s+5)I,S_+^2\pm S_-^2\right]_+$\\
 & $\pm 3$ & $c_\pm\left[S_z,S_+^3\pm S_-^3\right]_+$\\
 & $\pm 4$ & $c_\pm(S_+^4\pm S_-^4)$\\
\hline
 6 & 0 & $231S_z^6-(315s-735)S_z^4+(105s^2-525s+294)S_z^2-(5s^3-40s^2+60s)I$\\
 & $\pm 1$ & $c_\pm\left[33S_z^5-(30s-15)S_z^3+(5s^2-10s+12)S_z,S_+\pm S_-\right]_+$\\
 & $\pm 2$ & $c_\pm\left[33S_z^4-(18s+123)S_z^2+(s^2+10s+102)I,S_+^2\pm S_-^2\right]_+$\\
 & $\pm 3$ & $c_\pm\left[11S_z^3-(3s+59)S_z,S_+^3\pm S_-^3\right]_+$\\
 & $\pm 4$ & $c_\pm\left[11S_z^2-(s+38)I,S_+^4\pm S_-^4\right]_+$\\
 & $\pm 5$ & $c_\pm\left[S_z,S_+^5\pm S_-^5\right]_+$\\
 & $\pm 6$ & $c_\pm(S_+^6\pm S_-^6)$\\
\hline
\hline
\end{tabular}
Here $[A,B]_+ = (AB+BA)/2$, and $s=S(S+1)$, $c_+=1/2$, $c_-=1/2i$.
\end{small}
\caption{Extended Stevens Operators $O_k^q$ \cite{Stevens1952,Rudowicz2004}}\label{fig:stevens}
\end{table}

One of the simplest situations corresponds to a spin with Ising-like second order anisotropy, which has $B_{2}^{0} < 0$ with all other parameters equal to zero, i.e. to a spin Hamiltonian
\begin{equation}
{\cal H}_{\rm s} = B_{2}^{0} \left[ 3S_{Z}^{2} - S(S+1) \right] - g_{S}\mu_{\rm B} \left( B_{X}S_{X} + B_{Y}S_{Y} + B_{Z}S_{Z} \right).
\label{eq:GSHamiltonian_Ising}
\end{equation}
Such a \emph{diagonal} anisotropy splits the $S$ multiplet into a series of doublets, associated with eigenstates $|\pm m \rangle$ of $S_{z}$. As a function of $m$, the energy then shows a characteristic double-well potential landscape shown in figure \ref{fig:Fe8}.  Off-diagonal anisotropy terms (i.e. those having $l \neq 0$ in equation (\ref{eq:GSHamiltonian}) that connect states with different $m$) can induce quantum tunneling across the magnetic anisotropy barrier, between states $|+m \rangle$ and $|-m \rangle$.  The energy eigenstates become symmetric and antisymmetric superpositions of $\ket{\pm m}$ and their initial degeneracy is removed by a quantum tunnel splitting $\Delta_{m}(0)$.  For energy levels where the tunnel splitting is very small or zero, the degeneracy can also be removed by the application of an external magnetic field $\vec{B}$. Energy splittings can be tuned, to some extent, by varying the intensity and orientation of $\vec{B}$ (see figures \ref{fig:fig_13} and \ref{fig:fig_12} below). In particular, close to $B_{Z}=0$, the splitting between the first excited and ground states $\hbar \omega_{12} \simeq \sqrt{ \left[\Delta_{S}(\vec{B}) \right]^{2} + \xi_{S}^{2} }$, where $\Delta_{S}(\vec{B})$ is the ground doublet field-dependent quantum tunnel splitting and $\xi_{S} = 2g_{S} \mu_{\rm B} B_{Z} S$ is the magnetic bias.  The magnetic field also allows the initialization of the SMM state when the temperature is sufficiently low.  In particular, for $S=10$, and at $T=0.1$ K, the thermal population of the ground state becomes $\gtrsim 99.99$ \% for $B_{Z} \gtrsim $ 34 mT.

It is worth mentioning here that equation (\ref{eq:GSHamiltonian}) applies also to, e.g., NV centres in diamond, which have $S = 1$ and a zero-field splitting determined by second-order anisotropy terms with $B_{2}^{0} \simeq 2.88$ GHz ($0.144$ K) and $B_{2}^{2}/B_{2}^{0} \lesssim 3.5 \times 10^{-3}$ \cite{Amsuss2011}. Therefore, the theoretical framework that follows will enable us to compare both situations.

\subsection{Coupling of a SMM to a quantum electromagnetic radiation field}

The coupling between a spin, described by the Hamiltonian ${\cal H}_{\rm s}$, and a superconducting quantum circuit, described by ${\cal H}_{\rm q}$, is governed by the Zeeman interaction,
\begin{equation}
{\cal H} = {\cal H}_{\rm q} + {\cal H}_{\rm s} - \left( \vec{W}^{{\rm (q)}} V_{\rm q}\right)\vec{S}
\label{eq:zeemancoup}
\end{equation}
\noindent where $\vec{W}^{\rm (q)} = g_{S}\mu_{B}\vec{B}^{\rm (q)}$ is proportional to the magnetic field $\vec{B}^{\rm (q)}$ generated by the superconducting circuit at the spin position and $V_{\rm q}$ is an operator acting on the circuit's variables.  This operator depends on the actual nature of the circuit but will basically involve raising (or lowering) the \emph{circuit} state at the expense of lowering (or raising) the qubit state.

In the case of a spin $1/2$ system, there is no ambiguity in the choice of the quantum computational basis.  Spin states are split by an external field.  The low energy state can then be labeled as $\ket{G}$ (ground or $\ket{0}$) and the high energy state can be labeled as $\ket{E}$ (excited or $\ket{1}$).  However, when working with higher spin systems there are several possible choices for the computational basis.  These higher spins can be treated as two-level systems by focusing only on those two spin levels whose energy difference is in (near) resonance with the circuit's transition frequency $\hbar \omega$. More specifically, we choose the spin ground state $|G\rangle$ and one excited state $|E \rangle$. Two possible choices, relevant to real SMMs, are shown in figure \ref{fig:Fe8}.  For either choice we define the spin transition frequency as the energy difference between the two levels.
\begin{equation}
\hbar\omega_{\rm{G,E}} \equiv \langle E|{\cal H}_{\rm s}|E \rangle - \langle G|{\cal H}_{\rm s}|G\rangle
\label{eq:resonantfrequency}
\end{equation}
\noindent and the transition matrix element or interaction strength is defined as:
\begin{eqnarray}
\hbar g &\equiv& \left| \langle G|\vec{W}^{(q)}\vec{S} |E \rangle \right| \nonumber \\
&=& \left| W^{(q)}_{X} \langle G|S_{X} |E \rangle + W^{(q)}_{Y} \langle G|S_{Y} |E \rangle+ W^{(q)}_{Z} \langle G|S_{Z} |E \rangle\right| \label{eq:Melements}
\end{eqnarray}
\noindent Achieving strong coupling requires that the SMMs can be tuned to resonance with the circuit, i.e. that $\hbar \omega_{\rm{G,E}} \simeq \hbar\omega$ for a given $|E\rangle$, and that the relevant matrix element of the Zeeman interaction is sufficiently large. In the remainder of this section, we discuss how matrix elements $\langle G| S_{I} |E \rangle$, with $I = X, Y, Z$, thus also $g$, depend on the choice of state $|E \rangle$ as well as on the magnetic anisotropies and experimental conditions that can be met with real SMMs. The actual coupling $g$ depends also on the magnetic field generated by a given circuit, thus on its design and geometry. These aspects will be considered in section \ref{sec:resonators} below.

\subsection{Calculation of transition matrix elements}\label{sec:matelem}

\subsubsection{Isotropic spin $1/2$ system}\label{sec:spin1o2}

In the case of an isotropic spin $1/2$ system, analogous to a single electron spin, the situation is very simple.  The Hamiltonian contains only the Zeeman interaction given in equation (\ref{eq:zeeman}).  By applying an external DC magnetic in the Z direction, the spin transition frequency (\ref{eq:resonantfrequency}) is tuned as (taking $g_S=2$):
\begin{eqnarray}
\hbar\omega_{\rm{G,E}} & = & \bra{\downarrow}{\cal H}_{\rm s}\ket{\downarrow} - \bra{\uparrow}{\cal H}_{\rm s}\ket{\uparrow} = \nonumber \\ &=& g_S\mu_\textrm{B} B_Z = (\SI{1.342}{\kelvin\per\tesla})B_Z = (\SI{28}{\giga\hertz\per\tesla})B_Z,
\end{eqnarray}
where \SI{}{\kelvin} and \SI{}{\giga\hertz} have been used as units of energy in the \emph{tuning rate} values.  The spin matrices $S_{X,Y,Z}$ are directly the spin $1/2$ Pauli matrices and their matrix elements $\bra{G}S_{X,Y,Z}\ket{E}$ can be easily calculated:
\begin{equation}
\bra{\uparrow}S_X\ket{\downarrow} = \frac{1}{2} \quad \bra{\uparrow}S_Y\ket{\downarrow} = \frac{i}{2} \quad \bra{\uparrow}S_Z\ket{\downarrow} = 0 \label{eq:matelem1o2}
\end{equation}
In this case, the transition matrix elements are constant and independent of the applied DC field.  Since the $S_z$ matrix element is zero, the quantum circuit's magnetic field must be applied in the plane perpendicular to the tuning magnetic field.  The value of the coupling strength from (\ref{eq:Melements}) will then be simply:
\begin{equation}
\hbar g = g_S\mu_\textrm{B} B^\textrm{(q)} \frac{1}{2} = \mu_\textrm{B} B^\textrm{(q)}
\end{equation}

Higher spin isotropic systems may not be suitable as quantum computing systems.  This is due to the fact that all spin states will be equally spaced along the applied field axis.  This makes addressing a single transition with an external excitation impossible.  If, for example, we choose the two lowest lying states as the computational basis ($\ket{S}$ and $\ket{S-1}$) and populate only the lowest energy level, we would be able to rotate this system into the excited state using an external rf field tuned to the desired frequency.  However, if it were necessary to rotate it back to the ground state, the applied rf field would rotate from the excited state $\ket{S-1}$ into both $\ket{S-2}$ and $\ket{S}$ since they would both be separated from $\ket{S-1}$ by the same energy splitting, namely $g_S\mu_\textrm{B}B_z$.  Therefore, the presence if finite non-linear anisotropy terms is necessary to single out the pair of desired energy levels and to allow the use of systems with $S>1/2$ as qubits.

\subsubsection{Transitions between zero-field split levels}

In this and the next subsection, we consider the case of anisotropic high spin systems.  We will use a generic $S=10$ SMM with ${\cal H}_{\rm s}$ described by equation (\ref{eq:GSHamiltonian}) with only second order anisotropy terms as a prototype for this type of systems. This situation applies to some of the best known SMMs, such as Fe$_{8}$, shown in the inset of figure \ref{fig:Fe8}, or even the archetypal Mn$_{12}$ and allows some first order estimates to be made even for more complex anisotropy configurations.

\begin{figure}
\includegraphics[width=\textwidth]{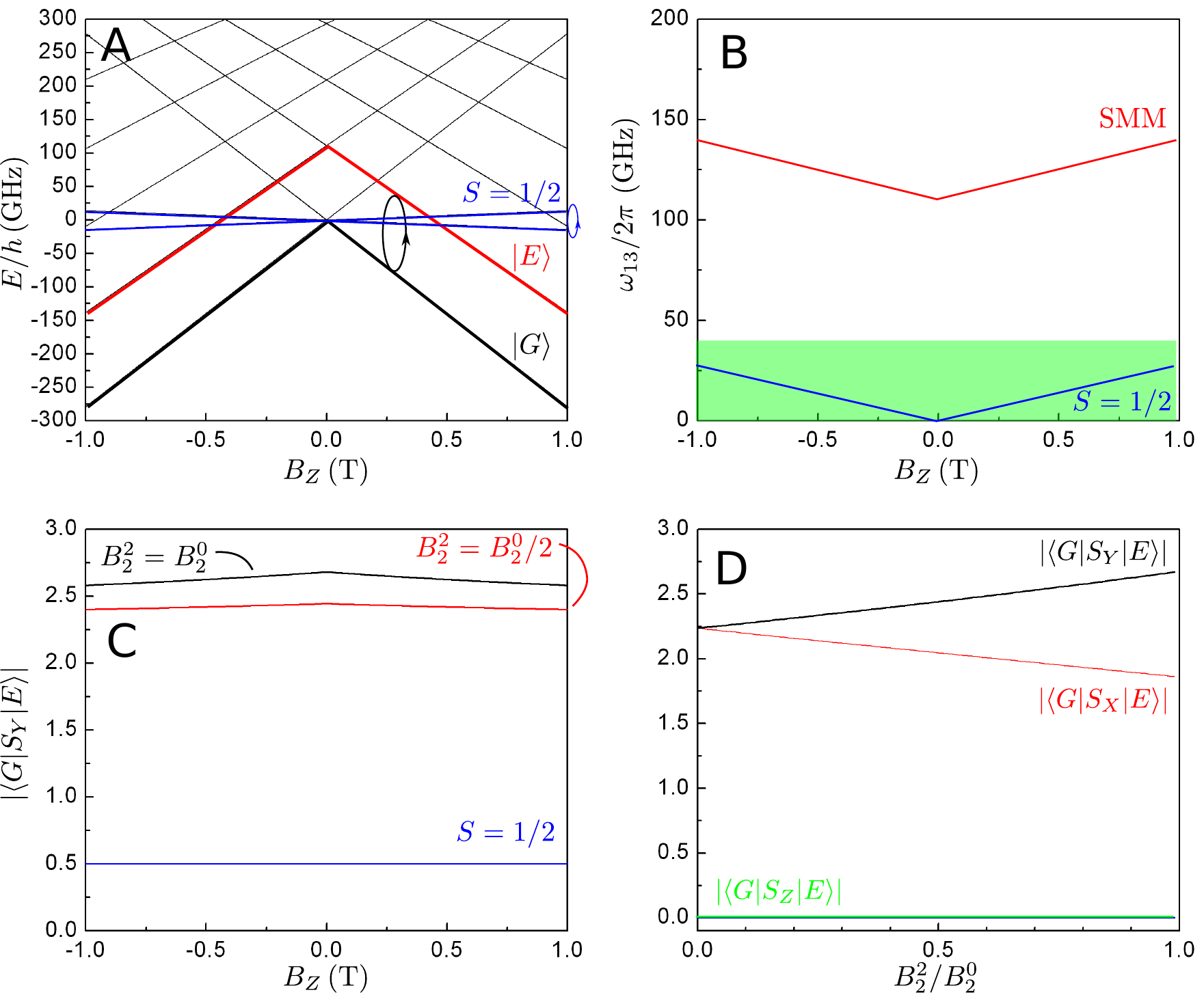}
\caption{Energy levels and transition matrix elements of a generic SMM.  Graph A shows the energy levels (for $S=10$, $B_{2}^{0}/k_{\rm B} = -0.1$ K, and $B^{2}_{2}/k_{\rm B} = 0.1$ K) as a function of the external field parallel to the easy axis ($B_{Z}$).  The thick black and red lines highlight the zero-field split energy levels (i.e., the states $|G\rangle = |1 \rangle$ and $|E\rangle = |3 \rangle$ of figure \ref{fig:Fe8}), the chosen computational basis for which the transition matrix elements are calculated.  The two blue lines show the energy levels for an isotropic spin $1/2$ with $g_S=2$ for comparison.  Graph B shows the level separation for the chosen energy levels from graph A and the separation for the spin $1/2$ case.  The green shaded area corresponds to the 1-\SI{40}{\giga\hertz} band where superconducting resonators and gap-tunable flux qubits can operate comfortably.  Graph C shows the transition matrix elements associated with transitions between the levels highlighted in graph A for two values of $B_2^2$ compared to the value for a spin $1/2$ system.  Graph D shows the transition matrix elements for all three components of $\vec{S}$ as a function $B_2^{2}$.}
\label{fig:fig_13}
\end{figure}

A possible choice of the computational basis, reminiscent of the situation met with NV centres in diamond, is to identify $|E \rangle$ with state $|3 \rangle$, as shown in figure \ref{fig:Fe8}.  Out of the two second order terms in (\ref{eq:GSHamiltonian}), we can neglect $B_{2}^{2}$ in this case since it usually plays a minor role in the determination of the zero field splitting $\hbar\omega_{13}$ unless it is comparable to $B_{2}^{0}$.  With $B_{Z} \simeq 0$ (see figure \ref{fig:fig_13}), this situation corresponds to $|G \rangle \simeq |+S \rangle$ and $|E \rangle \simeq |+S-1 \rangle$ for $B_{2}^{0} < 0$ and to $|G \rangle \simeq |0 \rangle$ and $|E \rangle \simeq |+1 \rangle$ for $B_{2}^{0} >0$. For larger fields applied along the Z direction, the energy splitting changes to $\hbar \omega_{13} = \hbar \omega_{13}(0) + g\mu_{\rm B}B_{Z}$, where $\hbar \omega_{13}(0) = 3(2S-1)B_{2}^{0}$ in the former case and $\hbar \omega_{13}(0) = 3B_{2}^{0}$ in the latter case.  The field is chosen to be applied in the direction of the anisotropy axis Z since it will produce the largest changes to the energy levels (approximately eigenstates of the $S_z$ operator if $B_2^2\sim 0$).  The relevant transition matrix elements will then correspond to transverse spin components $S_{X}$ and $S_{Y}$ and can be calculated analytically using the identities $S_X = (S^++S^-)/2$ and $S_Y=(S^+-S^-)/(2i)$:
\begin{equation}
g \propto |\bra{G}S_{X,Y}\ket{E}| = \frac{1}{2} \sqrt{\left(S-m_{\rm gs}\right)\left( S + m_{\rm gs} + 1 \right)},
\label{eq:g13_for easy spin}
\end{equation}
where $m_{\rm gs}$ is the $S_{Z}$ eigenvalue of the ground state. Depending on the sign of the anisotropy, equation (\ref{eq:g13_for easy spin}) gives
\begin{eqnarray*}
B_{2}^{0} > 0 & \Rightarrow & m_{\rm gs} = 0 \Rightarrow g \propto \frac{1}{2}\sqrt{S (S+1)}\\
B_{2}^{0} < 0  & \Rightarrow & m_{\rm gs} = +S \Rightarrow g \propto \frac{1}{2}\sqrt{2S}.
\label{eq:13_for easy spin2}
\end{eqnarray*}
Enhancements of the order of $S$ or $\sqrt{S}$ (depending on the sign of $B_2^0$ can therefore achieved for high-spin $S$ materials compared to the spin $1/2$ case where the matrix element has a fixed value of $1/2$ (see equation (\ref{eq:matelem1o2})).  Furthermore, $g$ is but weakly affected by external magnetic fields (see figure \ref{fig:fig_13}C). A difficulty associated with this choice of basis is that the zero-field splittings of high-spin SMMs, such as Fe$_{8}$ or Mn$_{12}$, are often very large (e.g. $\hbar \omega_{13} (0) \simeq 114$ GHz for Fe$_{8}$, see figure \ref{fig:fig_13}B) as compared with the typical resonance frequencies of either superconducting resonators \cite{Blais2004,Wallraff2004} ($\omega/2\pi \simeq 1$ to $40$ GHz) or gap-tunable flux qubits \cite{Zhu2010} (for which $\omega/2\pi \simeq 1-10$ GHz).  As we mentioned previously, the effect of $B_2^2$ on the transition matrix elements is expected to be small.  This is confirmed in figures \ref{fig:fig_13}C and \ref{fig:fig_13}D where we see that the addition of a $B_2^2$ term only slightly changes the value of these matrix elements.  According to \ref{fig:fig_13}D its effect is to slightly raise the $\bra{G}S_Y\ket{E}$ value while slightly lowering the $\bra{G}S_X\ket{E}$ value.

From these calculations we can conclude that if we chose the pair $\ket{1},\ket{3}$ as the qubit basis, SMM systems should engineered to preferably have low values for the $B_2^0$ parameter in order to have the zero field splitting $\hbar\omega_{13}$ within the operating range of the desired quantum circuits ($1-\SI{40}{\giga\hertz}$ in our case).  Ideally we would desire a low anisotropy and a high spin since, as we have seen, the spin controls the value of the matrix elements (higher $S$ gives larger matrix elements, equation \ref{eq:g13_for easy spin}) while the anisotropy controls the zero field splitting.  This means that high spin multi-ion SMMs may not be the ideal systems for this approach since, although they have high spins, they tend to have very high $B_2^0$ anisotropy values \cite{Barra1997,Caciuffo1998}.  In chapter \ref{chap:SIMs} we will investigate an alternate type of SMM that is more convenient for this application.

\subsubsection{Transitions between 'spin-up' and 'spin-down' states: photon induced quantum tunneling}

\begin{figure}
\includegraphics[width=\textwidth]{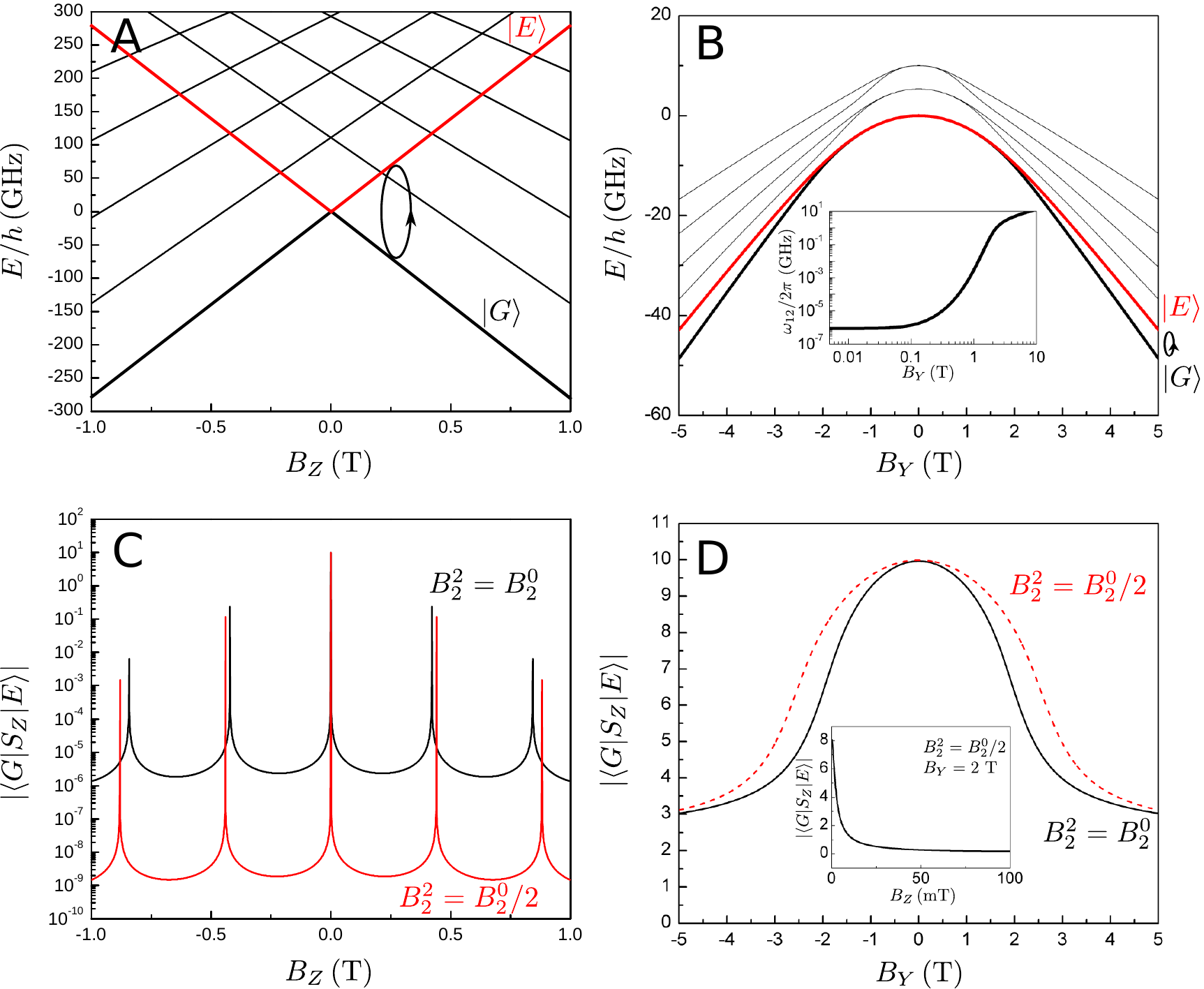}
\caption{Energy levels and matrix elements of a generic SMM (with the same parameters as in figure \ref{fig:fig_13}).  Graph A shows the energy levels as a function of the external field parallel to the easy axis ($B_{Z}$).  The thick black and red lines highlight the tunnel split energy levels (i.e., the states $|G\rangle = |1 \rangle$ and $|E\rangle = |2 \rangle$ of figure \ref{fig:Fe8}), the chosen computational basis for which the transition matrix elements are calculated.  Graph B shows the energy levels as a function of the external field transverse to the easy axis ($B_{Y}$).  The thick lines again highlight the same basis as in graph A.  The inset shows the energy separation between $\ket{1}$ and $\ket{2}$.  Graph C shows the transition matrix elements for this basis for two different values of $B_2^2$ as a function of $B_Z$.  Graph D shows the transition matrix elements for two values of $B_2^2$ as a function of $B_Y$.  The inset shows the same matrix element as a function of $B_Z$ with a fixed $B_Y = \SI{2}{\tesla}$.}
\label{fig:fig_12}
\end{figure}

A second natural choice is to use, as ``computational'' basis for the spin qubit, the two lowest-lying eigenstates of ${\cal H}_{\rm s}$ at zero field, which we denote here (see figure \ref{fig:Fe8}) by $|1 \rangle$ and $|2 \rangle$. For $B_{2}^{2} = 0$, these states correspond to degenerate 'up' and 'down' spin orientations, thus all matrix elements vanish unless $S=1/2$. Off-diagonal anisotropy terms give rise to a finite $\hbar \omega_{12} = \Delta_{S}(0)$, analogous to the tunnel splitting from the example in equation (\ref{eq:tunnel}) and figure \ref{fig:splitting}.  As we saw in that example, at zero-field (or $\xi=0$ in equation (\ref{eq:tunnel})), $|G \rangle \simeq (1/\sqrt{2}) ( |+S \rangle + |-S \rangle )$ and $|E \rangle \simeq (1/\sqrt{2}) ( |+S \rangle - |-S \rangle )$, thus $\langle G|S_{Z} |E \rangle \simeq S$, the other elements being close to zero. This is confirmed by numerical results shown in figure \ref{fig:fig_12}. Considering the high spin of SMMs, this transition can therefore give rise to potentially strong couplings. However, $\Delta_{S}(0)$ often lies in the region of micro-Kelvins or even smaller. For instance, $\Delta_{S} \simeq 10^{-7}$ K ($10^{-11}$ K) or barely $2.1$ kHz ($0.2$ Hz) for Fe$_{8}$ (Mn$_{12}$).  Therefore, a magnetic field needs then to be applied in order to tune $\omega_{12}$ to the circuit frequencies.

Maximum energy changes are obtained when $\vec{B}$ is oriented along the easy magnetization axis $Z$ (figure \ref{fig:fig_12}AC). However, any bias $\xi_{S} \gtrsim \Delta_{S}(0)$ effectively suppresses the overlap between the wave functions of $|1 \rangle$ and $|2 \rangle$ states (that effectively become $|+S \rangle$ and $|-S \rangle$ states) resulting in a dramatic decrease of $g$ with increasing $B_{Z}$.  In our simple example from equation (\ref{eq:tunnel}), this applied field would lead to a bias $\xi\sim 2g_SS\mu_\textrm{B} B_Z$.  Since the splitting $\Delta$ is on the order of \SI{}{\micro\kelvin} energies, fields larger than $\Delta k_B/\mu_\textrm{B} \sim \SI{}{\micro\tesla}$ will eventually suppress the mixing effect.  Also, none of the spin matrix elements can connect the $\ket{\pm S}$ states so all the matrix elements become negligible.  Indeed, in figure \ref{fig:fig_12}C, the matrix elements show narrow peaks at those values of $B_{Z}$ that induce level anti-crossings.  These resonances can be interpreted as photon induced tunneling processes between the quasi-degenerate spin states.  Resonances occur only at every even numbered level crossings (i.e. for $B_{Z} \simeq nB_{1}$, with $B_{1} = 3B_{2}^{0}/g_{S}\mu_{\rm B}$ and $n = 0, 2, \ldots$) because $B_{2}^{2}O_{2}^{2}$ only mixes states $|m \rangle$ and $|m^{\prime} \rangle$ such that $m-m^{\prime}$ is even ($B_2^2$ only contains second powers of $S^\pm$ according to table \ref{fig:stevens}). The width of each resonance (thus also the field region of potential interest for coupling to a circuit) can be increased by enhancing the off-diagonal parameter $B_{2}^{2}$, although it nevertheless remains very narrow even for the maximum $B_{2}^{2} = B_{2}^{0}$.

Alternatively, $\hbar \omega_{12}$ can also be tuned, while retaining a strong overlap between $|1 \rangle$ and $|2 \rangle$, thus a high $\langle G|S_{Z} |E \rangle$ (see figure \ref{fig:fig_12}BD), by a transverse magnetic field $B_{Y}$. This is a highly nonlinear effect (see the inset of figure \ref{fig:fig_12}B), meaning that strong magnetic fields are required to make $\omega_{12}$ close to $\omega$.  The use of stronger magnetic fields also imposes stringent conditions to the alignment of $\vec{B}$ to avoid the presence of a sizable $B_Z$ component.  As follows from the data shown in the inset of figure \ref{fig:fig_12}D, $\vec{B}$ cannot deviate more than about $0.5$ deg (with $B_Y = \SI{2}{\tesla}$) from the $XY$ plane, which can be an experimentally difficult task.

In order to use the tunnel split states of a SMM as the basis for a quantum bit, it is therefore desirable to have a large (as compared to the magnetic bias) tunnel splitting $\Delta$ between these levels.  This would make the matrix elements more robust when applying the necessary magnetic fields to tune the system into resonance with the quantum circuit.  However, from the calculations presented here, archetypal examples like \ce{Fe8} do not have strong enough tunnel splittings to easily allow these levels to be used as a quantum basis since even very small errors in the applied magnetic field direction can suppress the tunneling effect.  However, as we shall see in chapter \ref{chap:SIMs}, different types of SMMs have sizable higher order anisotropy terms that can greatly enhance the zero-field tunnel splitting making them interesting candidates as qubits.

\section{Coupling to superconducting coplanar waveguide resonators}
\label{sec:resonators}

\subsection{Device description and parameters}

\begin{figure}[tbh]
\centering
\includegraphics[width=0.65\textwidth]{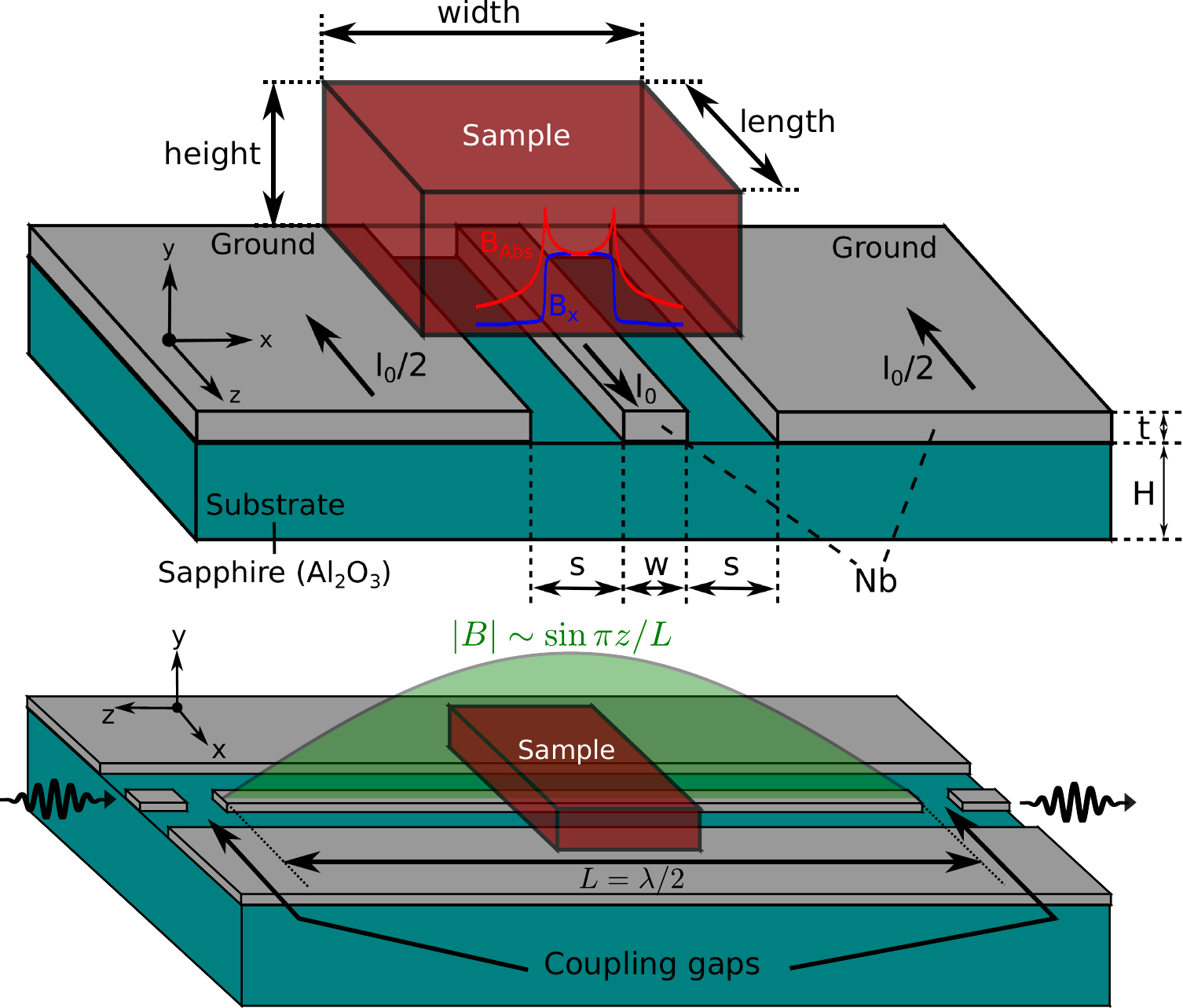}\\[5mm]
\begin{tabular}{|c|c|}
\hline
\multicolumn{2}{|c|}{Resonator dimensions} \\
\hline
s & 25 nm to \SI{7}{\micro\meter} \\
\hline
w & 50 nm to \um{14} \\
\hline
L & $\sim 0.5-\SI{65}{\milli\meter}$ with $f_0=100-\SI{1}{\giga\hertz}$ \\
\hline
t & 150 nm \\
\hline
H & \SI{75}{\micro\meter} \\
\hline
\end{tabular}
\begin{tabular}{|c|c|}
\hline
\multicolumn{2}{|c|}{Sample dimensions} \\
\hline
Width & \um{40} \\
\hline
Length & \um{40} \\
\hline
Height & 0.1 to \um{75} \\
\hline
\multicolumn{2}{|c|}{} \\
\multicolumn{2}{|c|}{} \\
\hline
\end{tabular}
\caption{Basic geometry and dimensions of the CPW resonator and the magnetic samples used for the calculations in this chapter.  The larger values for $s$ and $w$ are used as the starting values in simulations and are scales easily accessible by standard photolithography techniques.  The top graph shows a cross section while the bottom graph shows the length of the resonator and the fundamental mode magnetic field.}
\label{fig:CPWG_diagram}
\end{figure}

Coplanar waveguide resonators (see chapter \ref{chap:CPWG} for more details) are microwave devices that consist of a $\lambda/2$ section of a coplanar waveguide (CPW) that is coupled to external feed lines via gap capacitors. A schematic diagram of such a device is shown in figure \ref{fig:CPWG_diagram}. The fundamental mode resonant frequency is determined by the length of the resonator through the equation $f_0 = \frac{c}{\sqrt{\epsilon_{\rm eff}}}\frac{1}{2l}$. Here $\epsilon_{\rm eff}$ is the effective dielectric constant of the CPW and depends on the waveguide geometry and the dielectric constants of the surrounding media \cite{Simons2004}. As with transmission lines, the electromagnetic mode is described as a voltage and current wave where the current in the centre line is equal and opposite to the current in the ground plates.

Making the resonator out of superconducting materials, such as Nb or NbTi, and using low loss dielectric substrates, such as sapphire, helps to reduce the losses in the system and allows the reduction of the resonator cross section down to the micrometer level while maintaining quality factors of up to $10^{5}-10^{6}$ \cite{Frunzio2005,Barends2007,Goppl2008}.

\subsection{Coherent coupling to individual SMMs and SMM ensembles}\label{sec:coupCPWG_SMM}

We now must introduce the Hamiltonian for a resonator as the superconducting circuit contribution $\mathcal{H}_q$ into equation (\ref{eq:zeemancoup}) and set the corresponding interaction term.  The Hamiltonian for the resonator corresponds to that of a quantum harmonic oscillator and can be written in terms of creation and annihilation operators, $a$ and $a^\dagger$.  The interaction term is given by the Zeeman interaction term $\hat{\mu}\hat{B_q}$, i.e., the product of the dipolar magnetic moment operator of the spin qubit times the field operators for the resonator.  The spin system is reduced to a two level system and projected onto the chosen $\ket{G}$, $\ket{E}$ basis.  Using the rotating wave approximation to cancel anti-rotating terms $a^\dag\sigma^+$ and $a\sigma^-$ \cite{Bina2012}, equation (\ref{eq:zeemancoup}) reduces to the Jaynes-Cummings equation \refeq{eq:jc} \cite{Verdu2009}:
\begin{eqnarray}
\nonumber
{\cal H} & = & {\cal H}_{\rm s} + {\cal H}_{\rm q} - \vec{\mu}\cdot \vec{B}^{\rm (q)} \\
& = & \frac{\hbar\Omega}{2}\sigma^z +
\hbar\omega_r\left(a^\dag a+\frac{1}{2}\right) +
\hbar g(\vec r_j)(a^\dag\sigma^-+\sigma^+a)\label{eq:8}
\end{eqnarray}
\noindent
where we have projected onto the basis formed by the two relevant SMM states $|G\rangle$ and $|E\rangle$, $\hbar\Omega$ is the energy separation between these states and $\omega_r = 2\pi f_0$ is the resonator frequency. $\sigma_{x,z}$ are the (spin $1/2$) Pauli matrices acting on the spin basis and the coupling strength:
\begin{equation}
g(\vec r_j) = \frac{g_S \mu_{\rm B}}{\sqrt{2}}\left| \langle {\rm G}| \vec{b} (\vec r_j)\vec{S} | {\rm E}\rangle \right|
\label{eq:interaction}
\end{equation}
with $ \vec{b} (\vec r_j)$ the value of the field $\vec{b}$ generated by the vacuum current (see below).  The matrix element is calculated for the full spin system (as in section \ref{sec:matelem}).  The position $\vec r_j$ matches the spin location.  Through this section we will assume that the magnetic sample is centered at the position of maximum magnetic field generated by the resonator.  For the fundamental mode this position corresponds to the midpoint of the resonator as seen in figure \ref{fig:CPWG_diagram}.

In order to calculate the collective coupling of SMMs to a single photon, one needs to evaluate the magnetic field generated by the vacuum current fluctuations, $I_0$.  Considering the resonator as an electrical RLC oscillator (see chapter \ref{chap:CPWG}), this current can be found considering that the zero point energy of the resonator is equal to the peak energy stored in the magnetic (or electric) fields:
\begin{equation}
\frac{\hbar\omega}{2} = \frac{1}{2}LI^{2}_0 \quad \Rightarrow \quad
I_{\rm 0}=\omega \sqrt{\frac{\hbar\pi}{2Z_0}} \label{eq:vacuumcurrent}
\end{equation}
where $L=2Z_0/(\pi\omega)$ is the lumped inductance of the resonator \cite{Pozar2011} and $Z_0$ is the characteristic impedance of the transmission line segment that forms the resonator.  Then, taking a standard value of $Z_0\simeq 50\,\Omega$, we find that the vacuum current fluctuations $I_0\simeq(\SI{11.4}{\nano\ampere\per\giga\hertz})(\omega/2\pi)$.  With this value for the vacuum current we can now make quick estimate of the average value for the coupling to a single spin.  Assuming that the field can be approximated by that of a straight wire, we would have $B_{\rm wire} = \frac{\mu_0 I_0}{2\pi r}\sim\SI{0.28}{\nano\tesla\per\giga\hertz}$ for $r\sim\SI{10}{\micro\meter}$ (using the dimensions from figure \ref{fig:CPWG_diagram}).  Then using equation (\ref{eq:interaction}) with matrix elements $\sim 1$, we find that the coupling per spin will usually be small (of the order of 100 Hz depending on the operating frequency).  With this coupling value, the losses both from the qubit and the resonator can easily prevent the attainment of the the coherent coupling limit.  

To explore the possibility of getting higher couplings we will first consider the coupling of a resonator to an ensemble, e.g. a crystal, of $N$ spins. For this, we sum (\ref{eq:interaction}) over each spin at position $\vec r_{\rm j}$.  It is convenient to introduce the collective spin operator
\begin{equation}
b^{\dagger}=\frac{1}{\sqrt{N}\bar{g}}\sum_{j}^{N}g_{j}^{*}\sigma_{j}^{+} \label{eq:collop}
\end{equation}
where $\bar{g}$ is the average coupling, defined as $\bar{g}^{2} \equiv \sum_{j} \left| g_{j} \right|^{2}/N$. In the low polarization level $\langle \sum \sigma_j^\dagger \sigma^-_j \rangle \ll N$ these operators approximately fulfill bosonic commutation relations, $[b,b^{\dagger}] \approx 1$ \cite{Hummer2012}. Equation (\ref{eq:8}) then becomes approximately equal to the Hamiltonian of two coupled resonators,
\begin{equation}
{\cal H}  =  \hbar\Omega b^\dagger b + \hbar\omega_r a^\dag a +
\hbar g_N(a^\dag b + b^\dag a)
\label{eq:collective}
\end{equation}
with an effective coupling given by,
\begin{equation}
g_N = g_s\frac{\mu_\textrm{B}}{\sqrt{2}h}\sqrt{n\int_V \left| \langle G|\vec{b}\cdot\vec{S}|E\rangle \right|^2dV} \label{eq:coupling}
\end{equation}
where we have replaced the sums by integrals and assumed a uniform density $n$. Let us emphasize that equation (\ref{eq:coupling}) leads to a $\sqrt{N}$ enhancement of the effective coupling with respect to that of a single spin (as can be seen by taking an average field and integrating over the interacting sample volume).

To get more accurate estimations of the coupling strength, we use the Comsol Multiphysics AC/DC module to calculate the field distribution given the resonator geometry and currents (see section \ref{sec:comsol}).  This module numerically solves for the magnetic vector potential and electric potential using Ampere's law and current conservation.  We use a 2D geometry where we model only a cross section of the resonator.  This means that the the calculated fields are approximated by those generated by an infinite conductor length.  This approximation holds as long as the SMM crystals are placed close to the resonator center and the crystal length is much shorter than that of the resonator itself, which ranges from around $0.5$ to $65$ mm for the frequencies of relevance here.  More accurate results for larger crystals would require taking into account the variation of the current amplitude along the resonator length (approximately a sine wave).  The conductor cross sections are modeled by ``single-turn coil'' domains with a fixed current perpendicular to the cross section of $I_0$ for the center line and $-I_0/2$ for each of the ground planes.

Even for DC currents, the current density distribution in a superconductor is not uniform and different from that of a normal conductor.  In real superconductors, the superconducting current density decays exponentially with the distance to its surface and the decay constant is the London penetration depth $\lambda_{\rm L}$ \cite{London1935}. For Nb, $\lambda_{\rm L} \simeq 80$ nm at $4$ K and increases as temperature increases toward the critical temperature ($T_{\rm c} \simeq 9$ K) \cite{Kim2003}. Since the thickness of the superconducting lines we consider (see figure \ref{fig:CPWG_diagram}) is of this order or smaller, we need to simulate the current distribution carefully to take this effect into account. As a simple approximation and one that is already implemented in the Comsol package, we use the skin effect of standard conductors to produce the current profiles in the superconducting regions.  This means we use alternating currents in our simulation and tune the frequency $\omega_{\rm ac}$ and material parameters (conductivity $\sigma$) in the simulation to make the skin depth of the conductor
\begin{equation}
\lambda_{\rm skin} = \sqrt{\frac{2}{\sigma\omega_{\rm ac}\mu}}
\label{eq:skineffect}
\end{equation}
equal to $\lambda_{\rm L} = 80$ nm.  This procedure will generate current profiles similar to those expected to be seen in a superconducting wire.

\begin{figure}
\centering
\includegraphics[width=\textwidth]{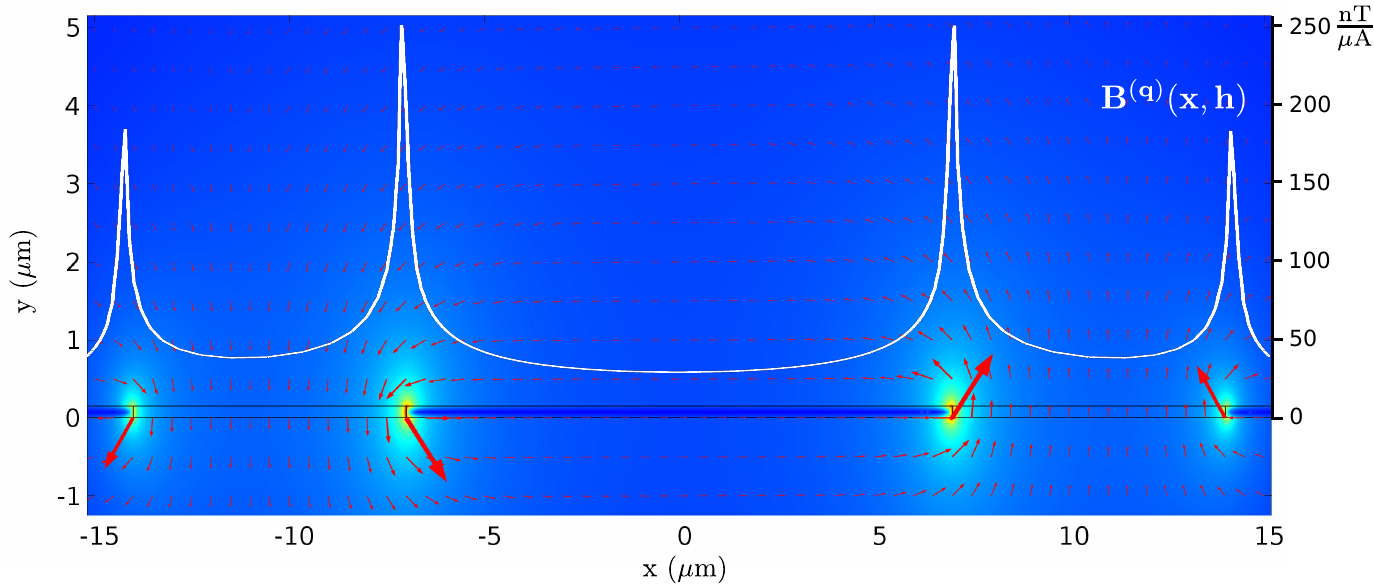}
\caption{Simulated field distribution on a CPW cross-section.  The white profile is the field value calculated at a constant distance from the substrate $y=t$, that is, right at the surface of the superconducting regions.  The actual simulation volume and the ground planes extend far beyond the figure limits.}
\label{fig:Bprofile}
\end{figure}

Taking all this into account, we simulate the magnetic field distribution for the geometry shown in figure \ref{fig:CPWG_diagram}.  A typical magnetic field distribution is shown in figure \ref{fig:Bprofile} (current profiles can be seen in figure \ref{fig:profiles}A below). As expected, the superconducting current and magnetic field concentrate near the edges of the centre line and the inner edges of the ground planes. Using these magnetic field distributions and the matrix element values calculated for different samples, it is possible to obtain the coupling strength from equation (\ref{eq:coupling}) for crystals of varying dimensions.  Writing (\ref{eq:coupling}) explicitly and using the resonator coordinate system ($x,y,z$) (see figure \ref{fig:CPWG_diagram}) we obtain\footnote{When expanding the squared absolute value, a crossed term proportional to $\int_S b_xb_ydS$ also appears.  However, symmetry considerations will force $b_y(x,y) = -b_y(-x,y)$ and $b_x(x,y) = b_x(-x,y)$ making the integrand antisymmetric with respect to the resonator $y$ axis.  Since we assume that our integration domain is centered on this axis we will have $\int_S b_xb_ydS = 0$ and we can thus ignore this term.}
\begin{equation}
g_N = g_s\frac{\mu_\textrm{B}\sqrt{nl}}{\sqrt{2}h}\sqrt{\left|\bra{G}S_x\ket{E}\right|^2\int_S \left|b_x\right|^2 dS+\left|\bra{G}S_y\ket{E}\right|^2\int_S \left|b_y\right|^2 dS},\label{eq:coupling_explicit}
\end{equation}
where $l$ is the crystal length along the resonator direction.  $\bra{G}S_{x,y}\ket{E}$ will correspond to a certain combination of $\bra{G}S_{X,Y,Z}\ket{E}$ given by the rotation between the molecular coordinate system ($X,Y,Z$) and the  resonator coordinate system ($x,y,z$).  As we noted before, we simulate only a cross section so this dependence is only valid for small $l$ compared with the resonator wavelength (see figure \ref{fig:CPWG_diagram}).  Larger $l$ will of course give larger couplings, but it will always increase slower than $\sqrt{l}$.

An appealing aspect of many SMMs crystals (including those considered in this work) is that the magnetic anisotropy axes of all molecules can be aligned with respect to each other.  This enables orienting the crystal so that the fields from the resonator induce the desired transitions for all the spins.  Comparing to the case of NV centers, for example, there are 4 different orientations for the spin centers which means that each of these four subgroups couples differently to the magnetic field.  Each sample and each choice of computational basis can potentially have different optimal orientations of the magnetic anisotropy axes ($X,Y,Z$) with respect to the resonator coordinate system ($x,y,z$).  In our simulations, the axis with the largest absolute value of the transition matrix element points along the $x$-axis of the resonator (i.e. horizontal, see figures \ref{fig:CPWG_diagram} and \ref{fig:Bprofile}) while the second largest is placed along the $y$-axis (i.e. perpendicular to the resonator).  The $b_X^{2}$ and $b_Y^{2}$ integrals entering in equation (\ref{eq:coupling_explicit}) are almost equal for the given geometry\footnote{If the centerline is made wider, the contribution from the $b_X^{2}$ term is somewhat larger than the one from the $b_Y^2$ term.  This can be intuitively understood by noting that the flatter geometry will leave the vertical ($b_Y$) fields around the edges mostly unchanged while they will produce \emph{horizontal} fields ($b_X$) in more of the sample volume)} so rotations that keep the axes with large transition matrix elements in the $x,y$ resonator plane produce very small changes to the overall coupling.  We also calculate the collective coupling of NV centres in diamond crystals, averaging over the four different orientations of their magnetic anisotropy axes. In all these calculations, we consider crystals of fixed length and width (both equal to \um{40}, see figure \ref{fig:CPWG_diagram}) and study how $g_{N}$ depends on the crystal thickness (thus also the number of spins), from $100$ nm up to \SI{100}{\micro\meter}.

\begin{figure}
\centering
\includegraphics[width=\textwidth]{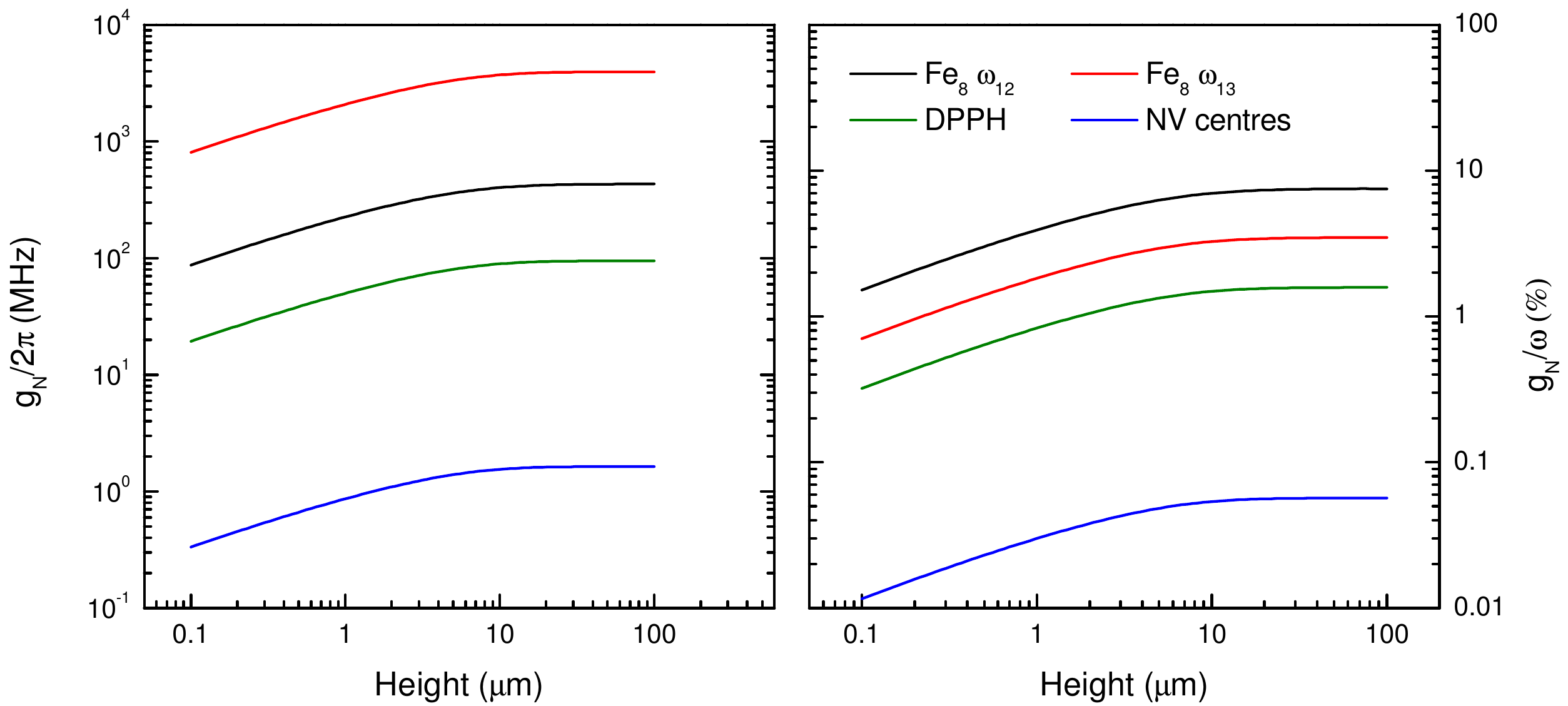}\\[5mm]
\begin{tabular}{|c|c|}
\hline
& Spin density (spins/\SI{}{\centi\meter\cubed)}\\
\hline
\ce{Fe8}  & \SI{5.11e20}{} \\
\hline
DPPH & \SI{2.14e21}{} \\
\hline
NV centers & \SI{1.1e18}{} \\
\hline
\end{tabular}\\[2mm]
\begin{tabular}{|c|c|c|c|}
\hline
Fe$_8\;\omega_{12}$ & Fe$_8\;\omega_{13}$ & DPPH & NV centres \\
\hline
5.75 GHz & 114.6 GHz & 6 GHz & 2.88 GHz \\
($B_Y = 2.325$ T) & & ($B_Z = 0.43$ T) & \\
\hline
\end{tabular}
\caption{Coupling of $\SI{40}{\micro\meter}\times \SI{40}{\micro\meter}\times {\rm height}$ SMM and diamond crystals to a CPW resonator as a function of crystal thickness. On the left we show the total coupling strength and on the right we show the coupling strength normalized by the resonator frequency.  For each sample, $\omega_{ij}$ denotes the transition used and the operating frequencies are detailed in the table above.  The spin densities for each sample are also shown \cite{Wernsdorfer1999,Kiers1976,Amsuss2011}.}
\label{fig:gvsh}
\end{figure}

The results are shown in figure \ref{fig:gvsh} for \ce{Fe8} and different choices of its computational basis $|G \rangle$ and $|E \rangle$.  We also show the values for the DPPH (2,2-diphenyl-1-picrylhydrazyl) free radical \cite{Kolaczkowski1999} as well as for NV centers in diamond \cite{Balasubramanian2009}.  The DPPH system is commonly used in EPR as a calibration sample since it is a very pure isotropic spin $1/2$ system (analogous to the spin $1/2$ in figure \ref{fig:fig_13}).  As would be expected, the dependence on the crystal thickness is essentially the same for all samples.  We see that the coupling first increases with crystal thickness and then saturates once the crystal is thicker than about $10-\SI{15}{\micro\meter}$. This behavior reflects the decay of $\vec{b}$ with the distance $y$ from the resonator surface and shows that the fields are confined to a volume around the center line up to a height similar to the gap size.  Only the spins within this volume significantly contribute to $g_{N}$ since the fields are negligible further away.  This emphasizes the importance of carefully placing the sample on top of the device within the active volume.

When increasing the resonator size, although the field strengths will be lower since the fields are spread over a larger volume, a larger sample volume will feel the magnetic field and more spins will contribute.  For larger crystals or with samples with an appreciable roughness, it is therefore generally better to use larger resonators.  The reason for this can be illustrated by remembering that, for a given frequency $\omega_r$, the energy per photon is constant and independent of the mode volume.  This energy $E_{\rm em}$ is contained in the electromagnetic energy density given by either the magnetic or electric fields:
\begin{equation}
E_{\rm em} = \frac{1}{2\mu_0}\int_V B^2 = \frac{1}{2\mu_0}\langle B^2\rangle_V V \quad \textrm{(constant for different V)}.
\end{equation}
From (\ref{eq:coupling}) we see that $g_N\propto\sqrt{\int_V B^2} = \sqrt{\langle B^2\rangle_V V}$, the same integral appearing in $E_{\rm em}$, and will hence also be constant if we change the active volume $V$ as long as this volume is filled with spins.  This means that, for many spins, there should be no coupling gain when using smaller resonators since the gain in field intensity is offset by the reduction in the number of interacting spins.  However, using smaller resonators has the added difficulty of placing and fixing the sample within the active volume.  This may be inefficient or impossible if the sample has high roughness (compared to the circuit dimensions) or if it can not be adequately attached, since there will be areas with high field that are not adequately filled by the sample.  In contrast, for small samples (again compared to the waveguide gaps) or single spins, the coupling will always be better for smaller devices since the peak field values are much more important than the field spread.

As can be seen in figure \ref{fig:gvsh}, because of their specific characteristics, the coupling to SMM crystals can be very large, much larger indeed than the coupling to NV centres in diamond crystals of equivalent size. The largest couplings $g_{N} \simeq 2-3$ GHz are found for transitions between states $1$ and $3$ of Fe$_{8}$. Yet, this transition is characterized by a very high resonance frequency $\omega_{13} \simeq 114$ GHz. Very large couplings ($g_N \simeq 0.5$ GHz) are also found for transitions between tunnel split states of e.g. Fe$_{8}$, for which  $\omega_{12}$ can be tuned by applying transverse magnetic fields (see figure \ref{fig:fig_12}). However, one then has to deal with rather strong ($\gtrsim 2$ T) and very accurately aligned (typically within less than $0.5$ deg.) magnetic fields (see figure \ref{fig:fig_12}).  We also note that much of this gain is due to the higher achievable densities of molecular systems as compared with NV centers (see table in \ref{fig:gvsh}), the higher matrix elements (see section \ref{sec:matelem}) as well as the higher vacuum currents associated with higher operating frequencies (see \ref{eq:vacuumcurrent}).

The couplings need to be compared with spin decoherence frequencies $\sim 1/T_{2}$, where $T_{2}$ is the phase coherence time. Experiments performed on crystals of Fe$_{8}$ \cite{Takahashi2011} show that $T_{2} \lesssim 500$ ns at liquid helium temperatures and under the best conditions, thus much shorter than $T_{2} \sim 1-2$ ms of NV centres \cite{Balasubramanian2009} at room temperature. Still, the strong coupling limit $g_{N}T_{2}/2\pi \gg 1$ should also be accessible for these molecular materials. Furthermore, for some of the examples given in figure \ref{fig:gvsh}, $g_{N}$ can in fact become a sizable fraction of the resonator frequency, thus opening the possibility to reach and explore the ultra-strong coupling limit with a spin ensemble where the rotating wave approximation no longer holds.

\subsection{Nanoscale resonators}

\begin{figure}[tbh]
\centering
\includegraphics[width=0.8\textwidth]{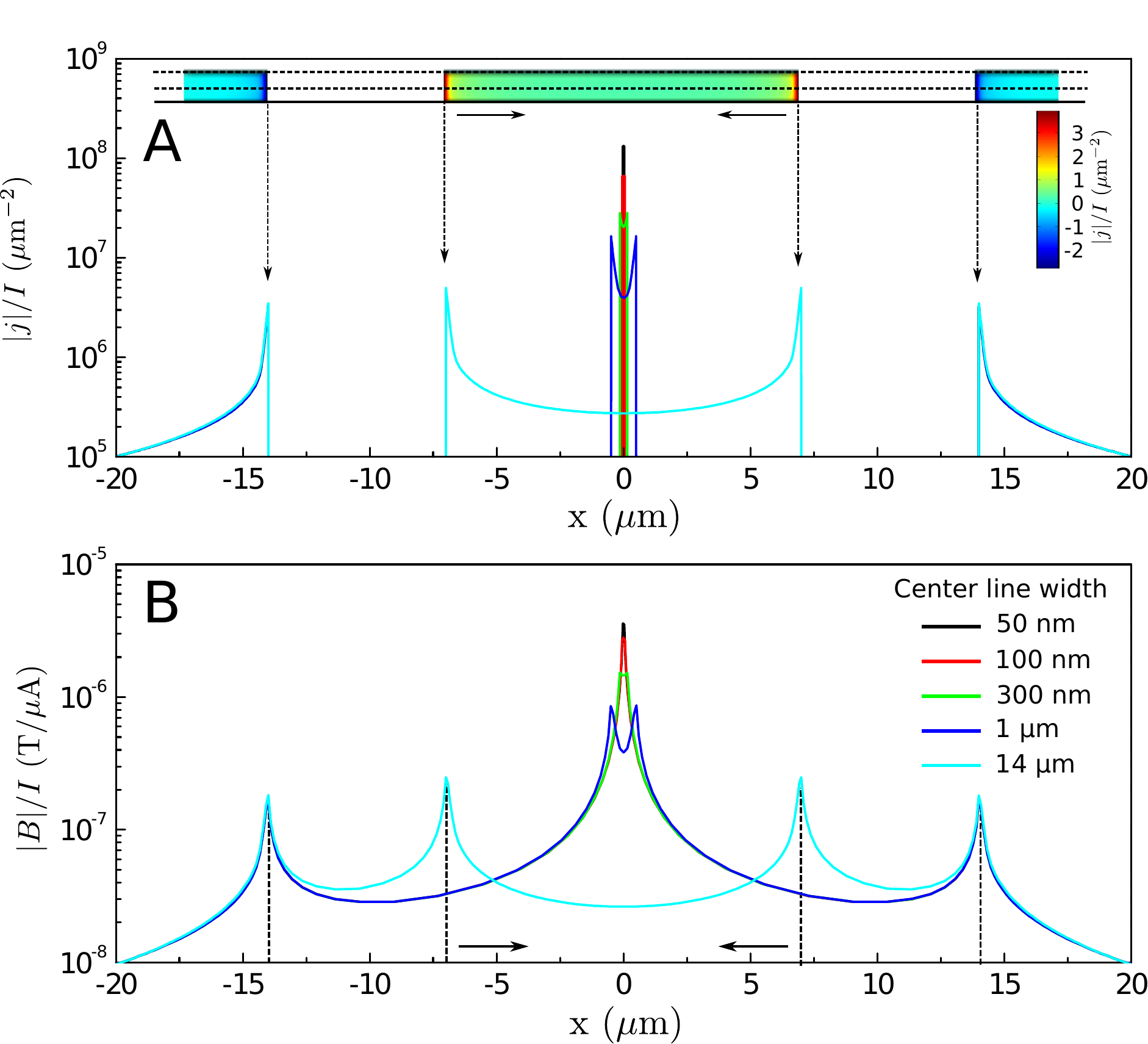}
\caption{Graph A shows the absolute value of the current density for a niobium superconducting CPW for different center line widths.  The  geometry is that given in figure \ref{fig:CPWG_diagram}.  The ground planes have been kept fixed and the center line width $w$ has been given several values.  The profile has been taken at a height of half the thickness of the superconducting layer (marked by the lower dotted line in the inset).  The inset shows a 2D map of the current density for the largest center line width.  The arrows show the dimension that is changed for each profile.  Figure B shows profiles of the absolute value of the magnetic field $B$ at the upper surface of the superconductor layer (shown by the upper dotted line in the inset in graph A) for the same center line widths.}\label{fig:profiles}
\end{figure}

The simulations described in the previous section also allow one to estimate the coupling to a single SMM at any location with respect to the device.  For a molecule placed in between the ground and central lines, we find that $g$ ranges between $100$ Hz and a few kHz, depending on the particular sample. Notice, however, that the magnetic field is enhanced, up to a factor $5$ or so, in narrow regions close to the edges of these lines (figure \ref{fig:Bprofile} and \ref{fig:profiles}). Two distinctive aspects of SMMs, which are not easily found in other qubit realizations, is that they are sufficiently small, with lateral dimensions of the order of $1$ nm, to fit inside these regions and that they can be delivered from a solution with very high spatial accuracy by, e.g. using the tip of an atomic force microscope \cite{Martinez-Perez2011}. The magnetic field generated near the central line edges, thus also the coupling to molecules or molecular ensembles located near them, can be further enhanced by fabricating narrow constrictions. Superconducting circuits with dimensions well below $100$ nm can be fabricated, and even repaired, by either etching with a focused ion beam or by using the same ion beam to induce the growth of a superconducting material from a gas precursor \cite{Martinez-Perez2009}. Provided that these constrictions are much shorter than the photon wave length, they are expected to have very little effect on the general resonator characteristics as we will investigate in chapter \ref{chap:CPWG}.

\begin{figure}[tbh]
\centering
\includegraphics[width=0.8\textwidth]{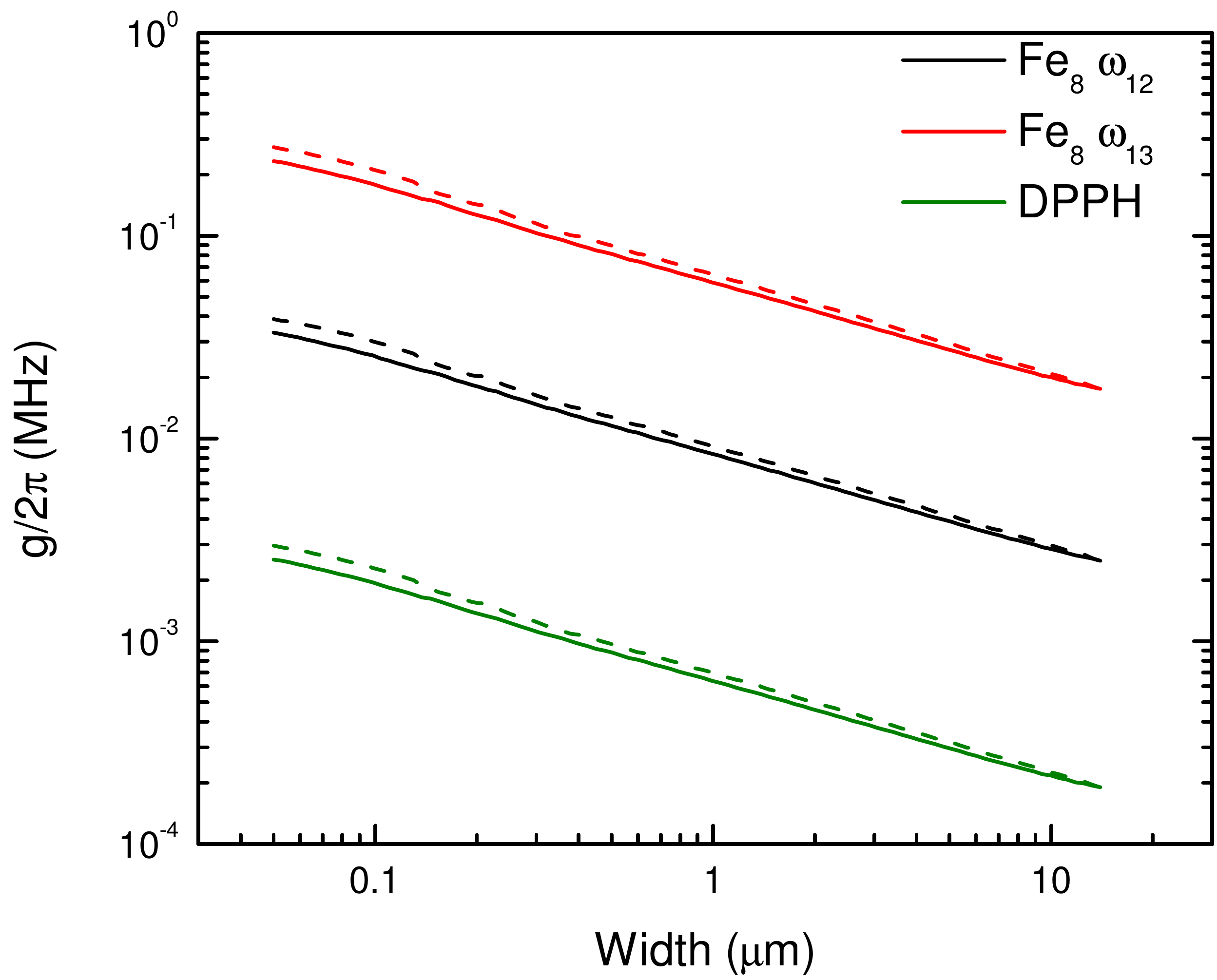}
\caption{Coupling of a single SMM to a CPW resonator as a function of centre line width. The SMM is located at the point of maximum field on the surface of the centre line.  For each sample, $\omega_{ij}$ denotes the transition used and the operating frequencies are detailed in figure \ref{fig:gvsh}.  The dashed lines correspond to scaling both the center line and the gaps while for the solid line only the center line width was scaled (and the gaps increased accordingly).}
\label{fig:gvsw}
\end{figure}

In order to explore this possibility, we have repeated the above simulations for varying center line widths, down to $50$ nm, while keeping the current constant.  Field and current profiles can be seen in figure \ref{fig:profiles}.  We then evaluate the coupling to a single SMM located at the point of maximum field on the surface of the centre line and oriented in such a way as to maximize the transition matrix element. The results are shown in figure \ref{fig:gvsw}.  The two possibilities shown correspond to the values obtained by reducing the centerline width but keeping the total width between the ground planes constant or by scaling both the center line width and gaps.  The latter case would correspond to shrinking the line while keeping the characteristic line impedance constant (see chapter \ref{chap:CPWG} for details).  However, we see only very small differences between the two cases since, although scaling the whole geometry further confines the field, the actual peak field value does not increase substantially as compared to the former case.  We see that reducing the width from \SI{14}{\micro\meter} to $50$ nm can lead to enhancements of an order of magnitude in the coupling strength. Again, the dependence on the geometry is the same for all samples.  NV centers have not been included since it is technically inviable to isolate a single NV center and place it on a transmission line since it is an entity that only exists within a diamond crystal.

The actual location of the maximum field is close to the edges of the centerline where the highest currents can be found.  However, the exact location is less important when the centerline cross section becomes smaller than the London penetration depth for niobium.  The currents in this limit are essentially uniform and the field distribution is similar to that seen from a thin wire.  Figure \ref{fig:profiles} shows the different field intensities for the different geometries considered.

The conclusion here is that achieving strong coherent coupling of a single SMM to such nanoresonators requires that that the decoherence time $T_{2}$ of an individual molecule grafted to a superconducting device can be made significantly longer than $10\,\mu$s. Despite the lack of $T_{2}$ data for truly isolated molecules, it seems that such coherence times can be reached under adequate conditions, i.e. for molecules having a very low concentration of nuclear spins \cite{Wedge2012} and/or adequate molecular structures \cite{Bader2014}.

It is also worth mentioning here that the potential applications of superconducting resonators or transmission wave guides that maximize the magnetic coupling to very small spin ensembles, or eventually enable detecting single spins, extends well beyond the quantum information research field. For instance, these designs might contribute to the optimization of on-chip electron paramagnetic resonance spectrometers for the characterization of magnetic materials \cite{Clauss2013}.

\section{Conclusions: why SMMs?}
\label{sec:conclusions}

The results described in previous sections confirm that, because of their high spins and spin densities, SMMs have the potential to attain very high couplings with superconducting circuits. In addition, the great variety of magnetic molecules enables a vast choice of resonance frequencies. However, for many of the best-known SMMs, such as Fe$_{8}$ or Mn$_{12}$, there are significant obstacles that could prevent their use as quantum bits.  Although their transition matrix elements have large values, these SMMs have a very large magnetic anisotropy due to the fact that their magnetic core consists of multiple magnetic ions.  This introduces several technical difficulties for their coupling to quantum circuits by making the zero field splitting very large ($\sim \SI{100}{\giga\hertz}$) and the tunnel splitting $\Delta$ very small.  The high zero field splitting makes the design of the quantum circuits much more difficult since the typical operating frequencies ($\lesssim\SI{40}{\giga\hertz}$ for CPW resonators and $\lesssim \SI{10}{\giga\hertz}$ for flux-qubits) are much lower than the the zero field split energy level separation.  The low tunnel splitting also makes it difficult to use the tunnel split energy levels as a quantum basis since their use would require the application of strong (above $2$ T) and very accurately aligned (within $0.5$ deg.) magnetic fields.  This leads us to the conclusion that we need to search for SMMs that have lower anisotropy values or stronger tunnel splitting.  For this reason, it will probably be more adequate to work with other species of SMMs such as single ion magnets (molecules with just one magnetic ion) which we will discuss and study in chapter \ref{chap:SIMs}.

Yet, it seems natural to inquire whether SMMs might bring some new possibilities, not easily achievable with other spin systems. A first, {\em quantitative} answer to this question is given by the couplings of SMMs crystals to superconducting resonators.  The collective coupling attains significant fractions, $\sim 10$ \%, of the natural circuit frequency, much larger than those observed so far for, e.g., NV centres in diamond. Under these conditions, the combined system enters the "ultra-strong" coupling limit, meaning that perturbative treatments are no longer applicable to describe the underlying physics.

From a more practical point of view, the attainment of strong coupling conditions might also confer to these systems interest as quantum memories \cite{Imamoglu2009,Wesenberg2009,Marcos2010}. A major difficulty arises though from the short lived spin coherence of these molecular systems. Decoherence times measured on SMMs crystals \cite{Takahashi2011} are still orders of magnitude shorter than those found for, e.g., NV centres \cite{Balasubramanian2009}. Therefore, SMMs cannot be considered for such applications unless coherence times are enhanced significantly. However, chemistry also provides suitable means to minimize the main sources of decoherence. For instance, isotopically purified molecules can be synthesized, in order to decrease the number of environmental nuclear spins \cite{Ardavan2007}. Also, decoherence caused by nuclear spin diffusion can be reduced by using sufficiently rigid ligand molecules \cite{Wedge2012}. Pairwise decoherence caused by dipolar interactions \cite{Morello2006} can be reduced by either dissolving the molecules in appropriate solvents \cite{Wedge2012,Ardavan2007,Schlegel2008,Bertaina2008} or by growing crystals in which a fraction of molecules is replaced by nonmagnetic ones \cite{Martinez-Perez2012}. Working with magnetically diluted samples has, however, a cost in terms of coupling. Therefore, a gain in performance (i.e. a net enhancement of $g_{N}T_{2}/2\pi$) can only be achieved provided that $T_{2}$ grows faster than $1/\sqrt(N)$, a condition that seems to hold in the very low temperature limit $k_{\rm B}T \ll \hbar \omega$, when magnon-mediated decoherence is expected to dominate \cite{Morello2006}. For a given spin density, the strength of dipolar interactions also decreases with $S$, thus it can be reduced by working with low-spin molecules, e.g. single ion magnets containing lighter lanthanide ions (Ce$^{3+}$,Sm$^{3+}$, or Gd$^{3+}$) or $S=1/2$ molecules, e.g. paramagnetic radicals \cite{Chiorescu2010,Abe2011}, Cr$_{7}$Ni molecular rings \cite{Troiani2005,Wedge2012,Ardavan2007} or \ce{Cu} complexes \cite{Bader2014}. The material of choice will therefore largely depend upon the attainment of an optimum tradeoff between maximizing $g_{N}$ and $T_{2}$.

But probably the main interest of SMMs is that they are also {\em qualitatively} different to most other spin systems in that they can be chemically engineered to fulfill very diverse functionalities. Restricting ourselves to the field of quantum information, magnetic molecules can be much more than single spin qubits \cite{Meier2003,Troiani2005}. Some molecular structures \cite{Timco2009,Candini2010,Aromi2012} embody several weakly coupled, or entangled, qubits which can provide realizations of elementary quantum gates  \cite{Luis2011} or act as quantum simulators \cite{Santini2011}. In addition, their multilevel magnetic energy structure can be used to encode multiple qubit states or even to perform quantum algorithms \cite{Leuenberger2001}. Coupling to quantum circuits can provide a method to experimentally realize these ambitious expectations and build architectures similar to the one proposed in figure \ref{fig:fantasy}, provided that one is able to strongly couple, and thus coherently manipulate and read-out, individual molecules.  In this respect, the fact that most SMMs are stable in solution opens the possibility to deposit them, in the form of monolayers or even individually, onto solid substrates \cite{Mannini2010} or at specific locations of a given device that maximize $g$ \cite{Martinez-Perez2011,Urdampilleta2011}. Our simulations show also that it is then possible to reach significantly larger couplings $g$, which can be further enhanced (up to $g/2\pi \sim 100-200$ kHz, see figure \ref{fig:gvsw}) by the fabrication of narrow constrictions in the centre line of superconducting nanoresonators. These results suggest that the strong coupling limit is attainable for individual molecules, using state-of-the art technologies, provided that decoherence times can be made longer than $10-20\,\mu$s. Considering the available experimental evidences \cite{Wedge2012,Ardavan2007}, this limit, and thus the realization of quantum technologies based on SMMs coupled to quantum circuits, seems definitely within reach.

In the following chapters we will further explore if these objectives are feasible by searching for better qubits (chapter \ref{chap:SIMs}) and investigating optimized circuit architectures (chapter \ref{chap:CPWG}).

\bibliographystyle{h-physrev3}
\bibliography{mybiblio}

\chapter{Single Ion Magnets as Qubits}\label{chap:SIMs}

\section{Introduction}
In chapter \ref{chap:Theo1} we found that \emph{classical} SMMs (such as \ce{Fe8} \cite{Wernsdorfer1999}) could potentially couple strongly to circuit QED systems and have applications in the context of quantum computation.  We also found that, depending on the choice of the  qubit basis, there are certain properties that are desirable in a spin system.  If the zero field split energy levels are used, an SMM with a relatively low anisotropy is desired to keep the operating frequencies in a comfortable range (section \ref{sec:matelem}).  If, on the other hand, the tunnel split energy levels are used, the tunnel splitting must be large enough to make the matrix elements resistant to perturbations such as dipolar and hyperfine interactions and to the unavoidable magnetic fields required to tune the frequencies of the qubit.

In order to address these two points, we now turn our attention to a specific family of SMMs known as single ion magnets (SIMs) \cite{Aldamen2008,AlDamen2009,Cardona-Serra2012,Martinez-Perez2012}.  These types of molecules are polyoxometalates (POMs) consisting of a single lanthanoid ion encased in a metal oxide structure that form molecular crystals.  This 4f-magnetic ion is subjected to a crystal field given by the interaction with surrounding ligands that modifies the free-ion anisotropy.  By choosing an adequate ligand structure and ion, these complexes allow a rational design of the spin Hamiltonian of the system.  This allows the optimization of the system to fulfill either of the two previously mentioned requirements for their application as spin qubits.  Also, a crystal or powder sample can be magnetically diluted by replacing the lanthanide ion at the center of the molecule with a non-magnetic \ce{Y} ion.  This does not change the overall crystal structure and allows a reduction in the dipolar interactions among neighboring molecules, thus enhancing the quantum spin coherence.

\begin{figure}[!tb]
\centering
\includegraphics[width=0.75\columnwidth]{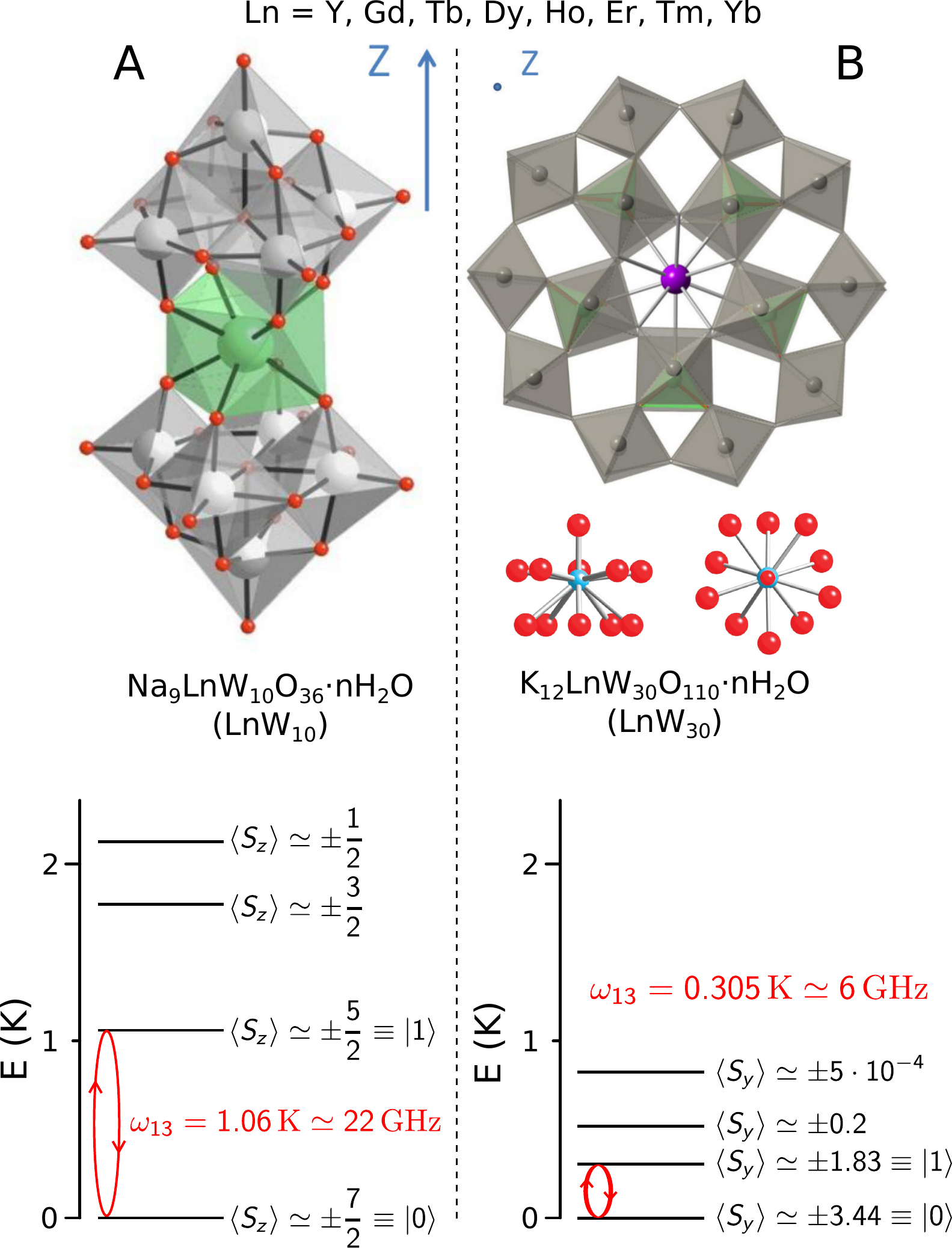}
\caption{Molecular structure of the generic \ce{LnW30} (A) and \ce{LnW10} (B) compounds.  Only the central lanthanide ion and the \ce{W} shell atoms are shown for visibility.  The \ce{Ln} ion can be any lanthanide.  The \ce{LnW30} case also shows the five-fold coordination symmetry structure of the ligands.  Below, the energy level scheme and the quantum basis for the case of $\ce{Ln}=\ce{Gd}$ is shown.}\label{fig:W10W30}
\end{figure}

If an intrinsically isotropic ion such as, \ce{Gd^{3+}}, is chosen, the magnetic anisotropy, and in particular the parameter $B_2^0$, will be entirely determined by the local coordination.  Figure \ref{fig:W10W30} shows two possible structures and their corresponding generic formulas.  The Hamiltonian parameters for both molecules were previously determined from powder EPR experiments and reported in \cite{Martinez-Perez2012}.  The axial \ce{GdW10} structure induces a stronger magnetic anisotropy while the \ce{GdW30} structure is more planar and results in a lower $B_2^0$ value.  In the latter molecule, all the energy levels are found to be within \SI{1}{\kelvin}, thus making it an interesting spin qubit candidate since all allowed transitions are accessible with relatively low microwave frequencies.  In section \ref{sec:GdW30} we will study this molecule both in powder and in crystalline form.  The aims of our work are, on the one hand, to refine the Hamiltonian parameter values, find the magnetic system anisotropy axes relative to the crystal structure and, on the other, to apply pulsed EPR techniques to directly measure the intrinsic decoherence times of each transition and to coherently manipulate the molecular spin, thus showing its viability as a spin qubit.

Adequate symmetries in the ligand structure can also lead to very high tunnel splitting terms in the spin Hamiltonian.  In particular, the \ce{LnW30} structure has an unusual 5-fold symmetry structure that gives rise to $B_6^5O_6^5$ terms in the crystal field Hamiltonian \cite{Cardona-Serra2012}.  The presence of these terms can introduce a very large tunnel splitting if the ground state of the ion is $\ket{\pm 5}$, as is the case with \ce{TbW30}, since $O_6^5$ contains terms proportional to $(S^\pm)^5$ that can connect these states in only two steps.  We therefore expect the \ce{TbW30} molecule to have clearly separated tunnel split energy levels.  Section \ref{sec:TbW30} will be dedicated to the measurements performed on this sample in order to confirm the presence of this large tunnel splitting and, therefore, its potential as a robust spin qubit.

In remainder of the chapter (section \ref{sec:simsCPW}) we will evaluate the couplings expected for SIMs to CPW resonators (with and without constrictions) and compare them to those found other spin systems and for more complex SMMs using the same procedures described in section \ref{sec:coupCPWG_SMM}.  Additionally, in section \ref{sec:fluxqubits} we introduce superconducting flux qubits and explore the possibility of coupling them to SIMs.  Finally, section \ref{sec:conclSIMs} summarizes our conclusions for this chapter.

\section{\ce{K12GdW30}}\label{sec:GdW30}

In this section we will characterize the SIM sample \ce{K12GdP5W30O110} (referred to as \ce{GdW30}).  Its basic molecular structure corresponds to that shown in figure \ref{fig:W10W30}B.  The \ce{Gd^{3+}} ion has spin $7/2$ and an isotropic gyromagnetic g-factor $g_S=1.99$.  Its molecular weight is $M_{\ce{GdW30}}=\SI{8058.25}{\gram\per\mole}$.  Crystals of this compound are obtained by making an over-saturated solution in deionized water and allowing the mixture to set overnight.  The crystals grow up to a few millimeters in size.

\subsection{Powder cw-EPR experiments: Temperature dependence of the magnetic anisotropies}\label{sec:powderEPR_GdW30}

\begin{figure}[!tb]
\centering
\includegraphics[width=0.6\columnwidth]{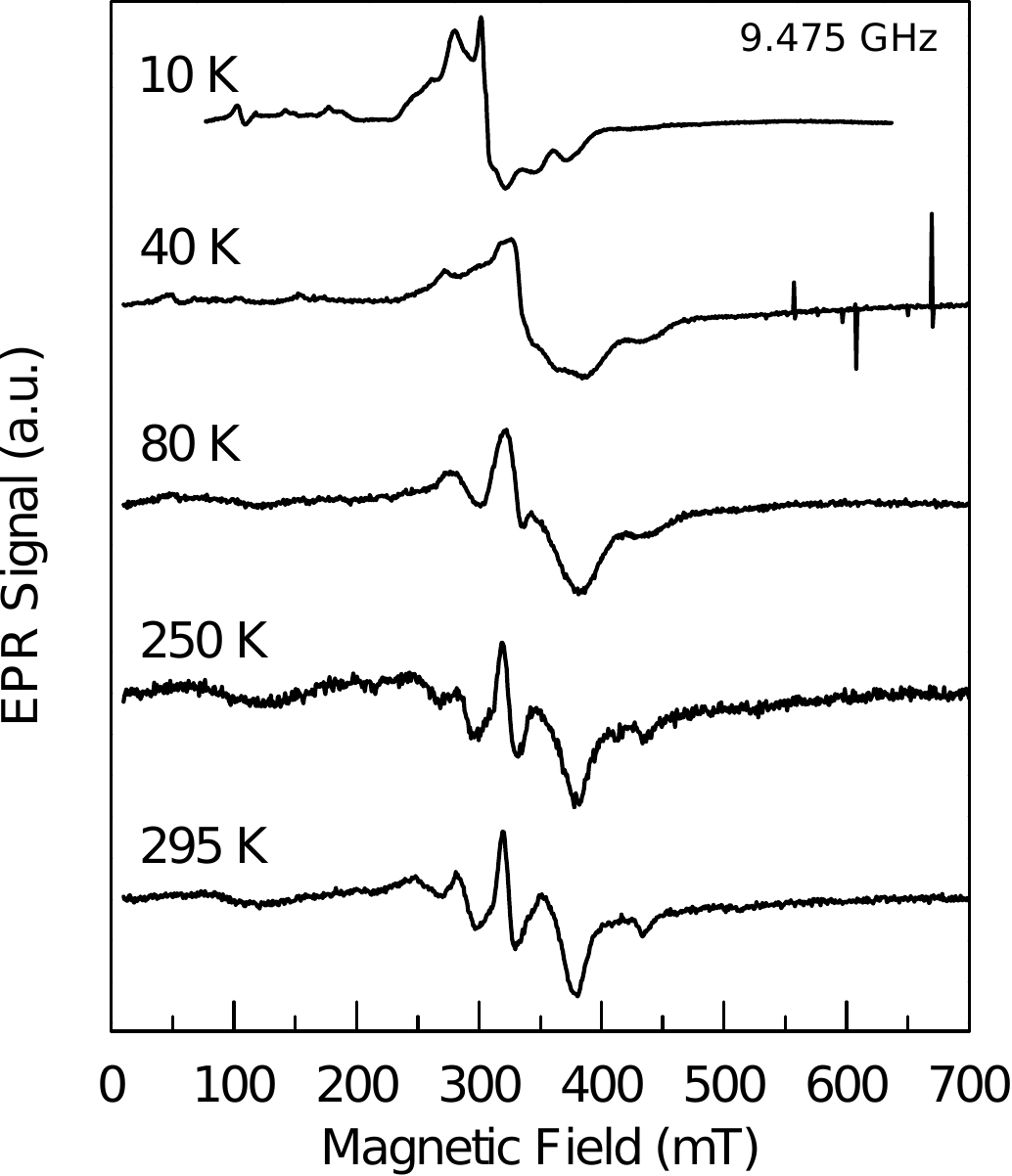}
\caption{\ce{GdW30} Powder X-band EPR spectra (\SI{9.475}{\giga\hertz}) for different temperatures.  Changes in the spin Hamiltonian for different temperatures are evidenced by the spectrum changes.}\label{fig:epr_polvo}
\end{figure}

For an initial determination of the crystal field Hamiltonian parameters, X-band cw-EPR (see section \ref{sec:EPRtech}) experiments were performed on powder samples.  Diluted \ce{Y_{1-x}Gd_xW30} crystals, with $x=0.01$, obtained from an over-saturated solution are ground to powder and then placed into a quartz EPR tube for the experiment.  The spectra obtained are shown in figure \ref{fig:epr_polvo} where we see changes in the spectrum for different experimental temperatures.

\begin{figure}[!tb]
\centering
\includegraphics[width=0.95\columnwidth]{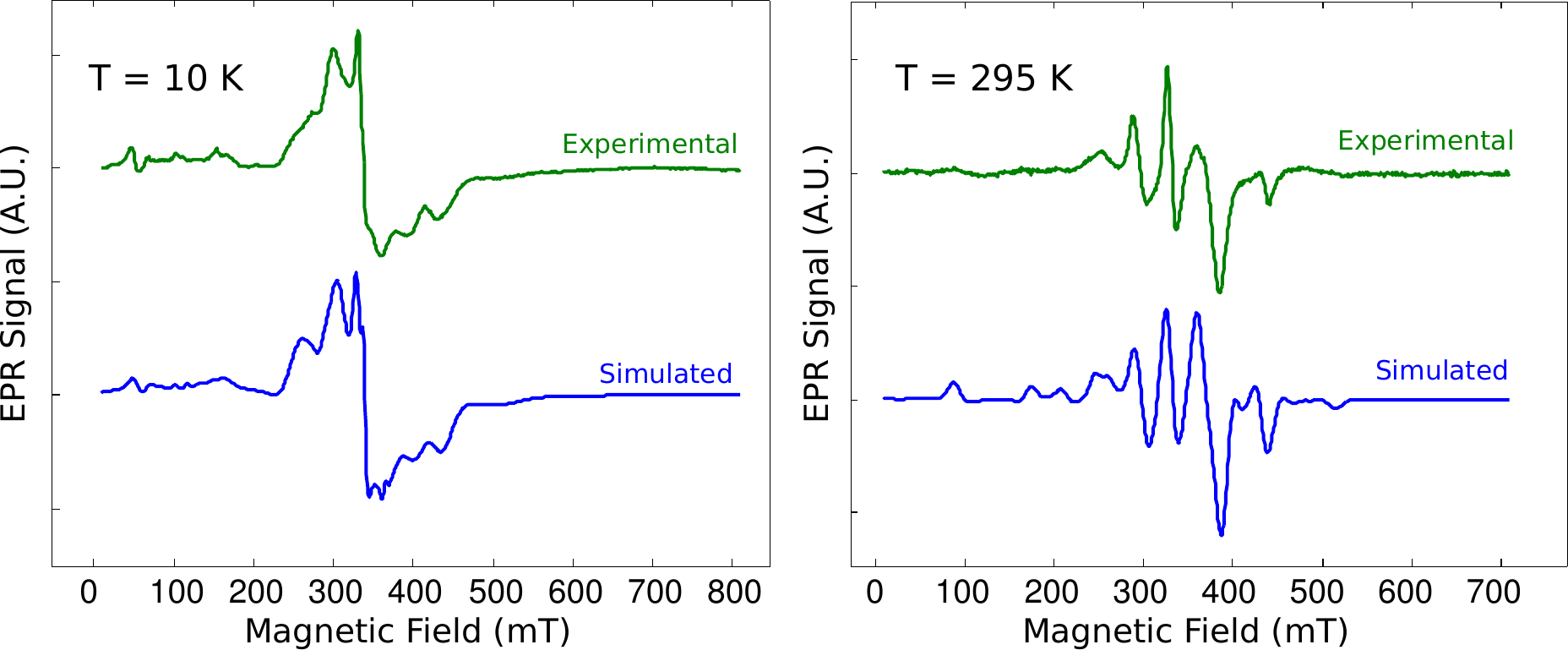}\\[3mm]
\begin{tabular}{rccc}
& $T=\SI{10}{\kelvin}$ & $T=\SI{295}{\kelvin}$ & $T=\SI{10}{\kelvin}$ (from \cite{Martinez-Perez2012})\\
\hline
$B_2^0$ & \SI{427}{\mega\hertz} & \SI{400}{\mega\hertz} & \SI{396}{\mega\hertz} \\
$B_2^2$ & \SI{294}{\mega\hertz} & \SI{86}{\mega\hertz} & \SI{396}{\mega\hertz} \\
$\Delta B_2^0$ & \SI{110}{\mega\hertz} & - & -\\
$\Delta B_2^2$ & \SI{224}{\mega\hertz} & - & -\\
$w$ & \SI{3.0}{\milli\tesla} & \SI{12.1}{\milli\tesla} & -
\end{tabular}
\caption{Measured and EasySpin simulated spectra for \SI{10}{\kelvin} and room temperature (\SI{295}{\kelvin}) X-band cw-EPR experiments on \ce{GdW30} powder.  The given parameters provide reasonably good fits to the experimental data.  The low temperature spectrum requiered strains in $B_2^0$ and $B_2^2$ ($\Delta B_i^j$) as well as an equal peak to peak linewidth $w$.  Considerable changes in the parameters are observed from room temperature to low temperature.}\label{fig:epr_polvo_fitted}
\end{figure}

From previous studies \cite{Martinez-Perez2012} we expect this sample to have contributions to $B_2^0 O_2^0$ and $B_2^2 O_2^2$ terms from the crystal field Hamiltonian \refeq{eq:GSHamiltonian} with no other higher order terms playing an important role.  The values reported previously in \cite{Martinez-Perez2012} were $B_2^0 = B_2^2=\SI{0.019}{\kelvin}=\SI{396}{\mega\hertz}$ at $T=\SI{10}{\kelvin}$.  Using EasySpin \cite{Stoll2006}, we fit the measured spectra at room temperature and \SI{10}{\kelvin} to this type of Hamiltonian containing only these two terms and a common transition line width ($w$, peak to peak).  These three parameters are sufficient to provide a reasonable fit for the room temperature spectrum.  However, reproducing the low temperature spectrum also requires the inclusion of strains in $B_2^0$ and $B_2^2$ to reproduce the spectrum (referred to as $\Delta B_i^j$).  The fits and the values obtained for each set of parameters are shown in figure \ref{fig:epr_polvo_fitted}.  For our subsequent simulations we will use our fitted values for \SI{10}{\kelvin}.  These values differ from those found at room temperature but also from the values reported in \cite{Martinez-Perez2012} although not radically.  The somewhat large temperature dependence of the parameters and the high strain values could be due to the large size of the molecule (over 150 atoms).  It may be the case that, when cooling, each molecule is \emph{strained} in slightly different ways as the lattice vibrations are suppressed giving rise to the parameter distribution measured.  On the other hand, at high temperatures fast molecular vibrations could average out these local differences thus leading to a uniform linewidth, sufficient to reproduce the data.

\subsection{Single crystal X-ray diffraction}\label{sec:xrayGdW30}

\begin{figure}[!tb]
\centering
\includegraphics[width=0.7\columnwidth]{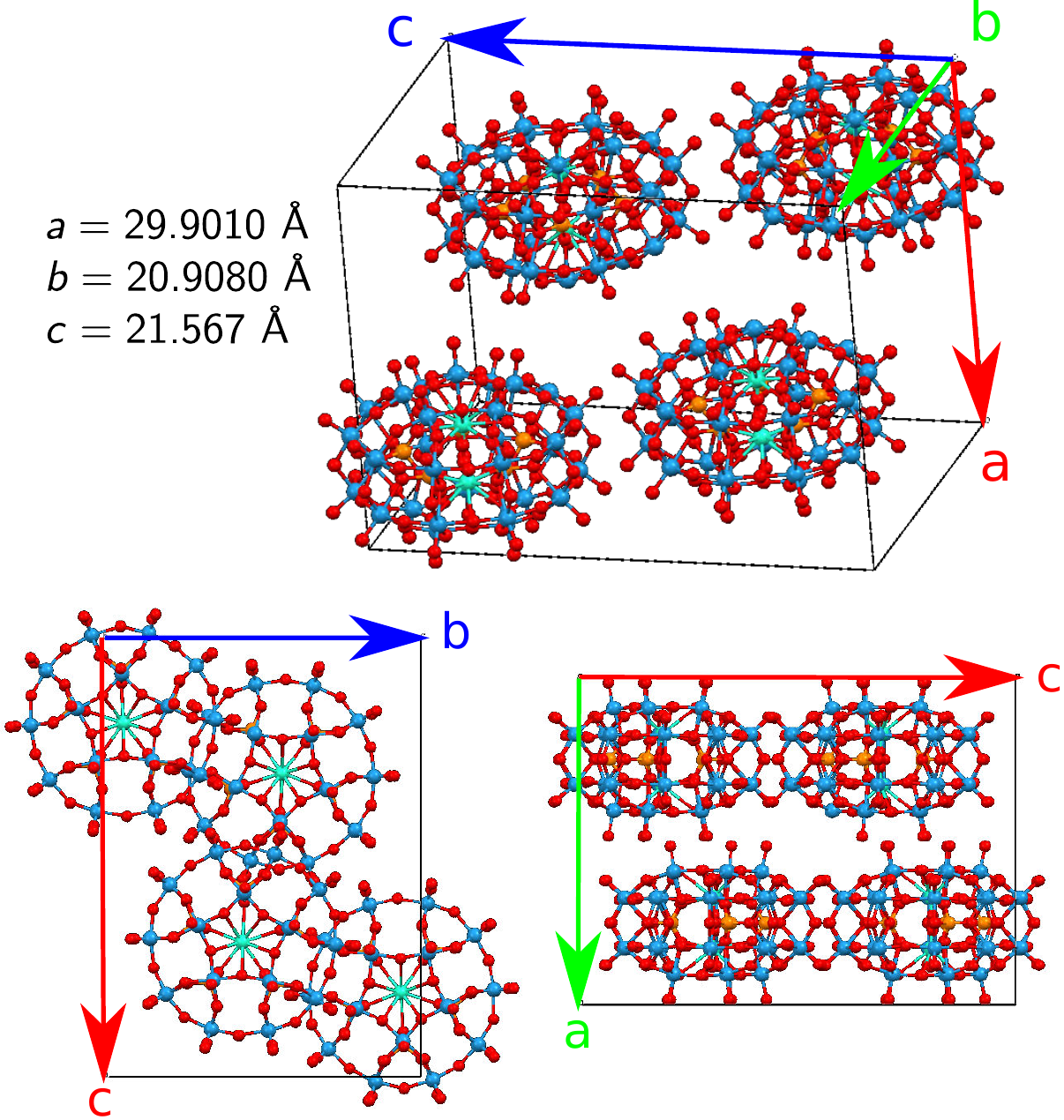}
\caption{Unit cell structure of \ce{LnW30} \cite{Kim1999}.  The chosen unit cell contains 4 \ce{LnW30} molecules.  Many surrounding atoms (including the 12 \ce{K} atoms) have been hidden for visibility.}\label{fig:crystalstructure}
\end{figure}

Previous studies on \ce{K12LnW30} compounds \cite{Kim1999,Aldamen2008,AlDamen2009} reported their crystalline structure.  A unit cell is shown in figure \ref{fig:crystalstructure} along with the cell parameters determined by x-ray diffraction.  The crystal structure is orthorhombic with one long axis (c, \SI{28.9}{\angstrom}) and two shorter axes of almost equal length (a,b of \SI{21.567}{\angstrom} and \SI{20.9080}{\angstrom} respectively) and is the same for all the possible lanthanide ions.  This allows different species of lanthanide to be present in a single crystal and, as mentioned before, gives the possibility of growing magnetically dilute crystals by replacing the lanthanide species by a non-magnetic ion (such as \ce{Y}) in a certain fraction of sites.  The chosen unit cell contains 4 molecules as can be seen in figure \ref{fig:crystalstructure} each with a flattened sphere shape.

In our case, we perform x-ray diffraction experiments to check the integrity and confirm crystal structure of our sample.  Additionally, we are interested in finding the crystal axes relative to the physical crystal faces and edges.  A diluted crystal of \ce{Y_{1-x}Gd_xW30} with $x=0.01$ was taken from a solution of the compound in deionized water and then encased in an epoxy glue.  The use of this glue is necessary to maintain the crystal integrity since these crystals contain interstitial water molecules that sustain the structure.  These water molecules are rapidly lost when the crystal is exposed to normal atmospheric conditions leading to a rapid loss of crystallinity.  The crystals that form in the saturated solution are generally elongated with a roughly rectangular cross section that we will approximate to a cuboid with one long edge.  The chosen crystal (of about $1.1\times 0.31 \times \SI{0.27}{\milli\meter}$) is mounted in a commercial x-ray difractometer.

\begin{figure}[!tb]
\centering
\includegraphics[width=0.8\columnwidth]{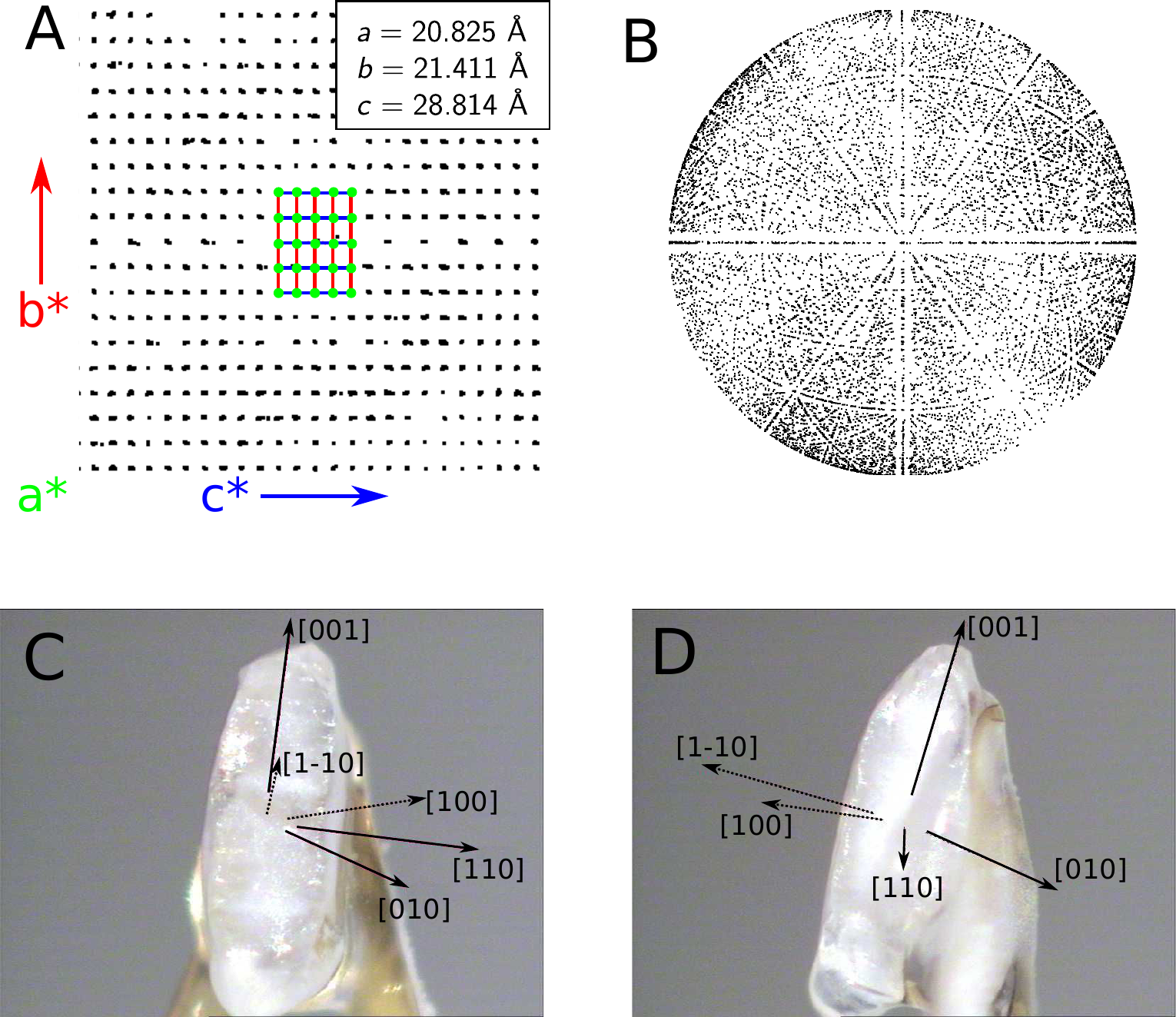}
\caption{Results of x-ray difractometry on \ce{Y_{0.99}Gd_{0.01}W30}.  Graph A shows the measured reciprocal lattice and the lattice parameters measured.  We find good agreement with the orthorhombic structure described in figure \ref{fig:crystalstructure}.  Graph B shows the measured Ewald sphere.  Graph C and D show several crystal axes determined from the measurements relative to the physical crystal.}\label{fig:xray}
\end{figure}

The main difractometry results are shown in figure \ref{fig:xray} where we find that the crystal structure is clearly resolved and is compatible with the expected values from figure \ref{fig:crystalstructure}.  The crystal is of good quality and an almost complete Ewald sphere can be measured (figure \ref{fig:xray}B).  With regards to the orientation of the crystal axes (a,b,c) relative to the crystal edges,  we find that the long edge of the crystal corresponds to the long crystal axis (c or $[001]$), while the lateral faces approximately correspond to the (1-10) and (110) crystal planes.  From the full structure in figure \ref{fig:crystalstructure}, we also conclude that the long edge is in the \emph{molecular plane} defined by the flattened sphere shape of the molecule.  The measurements were repeated for several crystals and this assignment was found to be consistent in all crystals checked.  This information forms the basis to determine, through magnetic measurements, the position of the magnetic axes relative to the crystal structure and to the molecule.

\subsection{Angle dependent magnetic susceptibility}\label{sec:rotGdW30}

\begin{figure}[!tb]
\centering
\includegraphics[width=0.95\columnwidth]{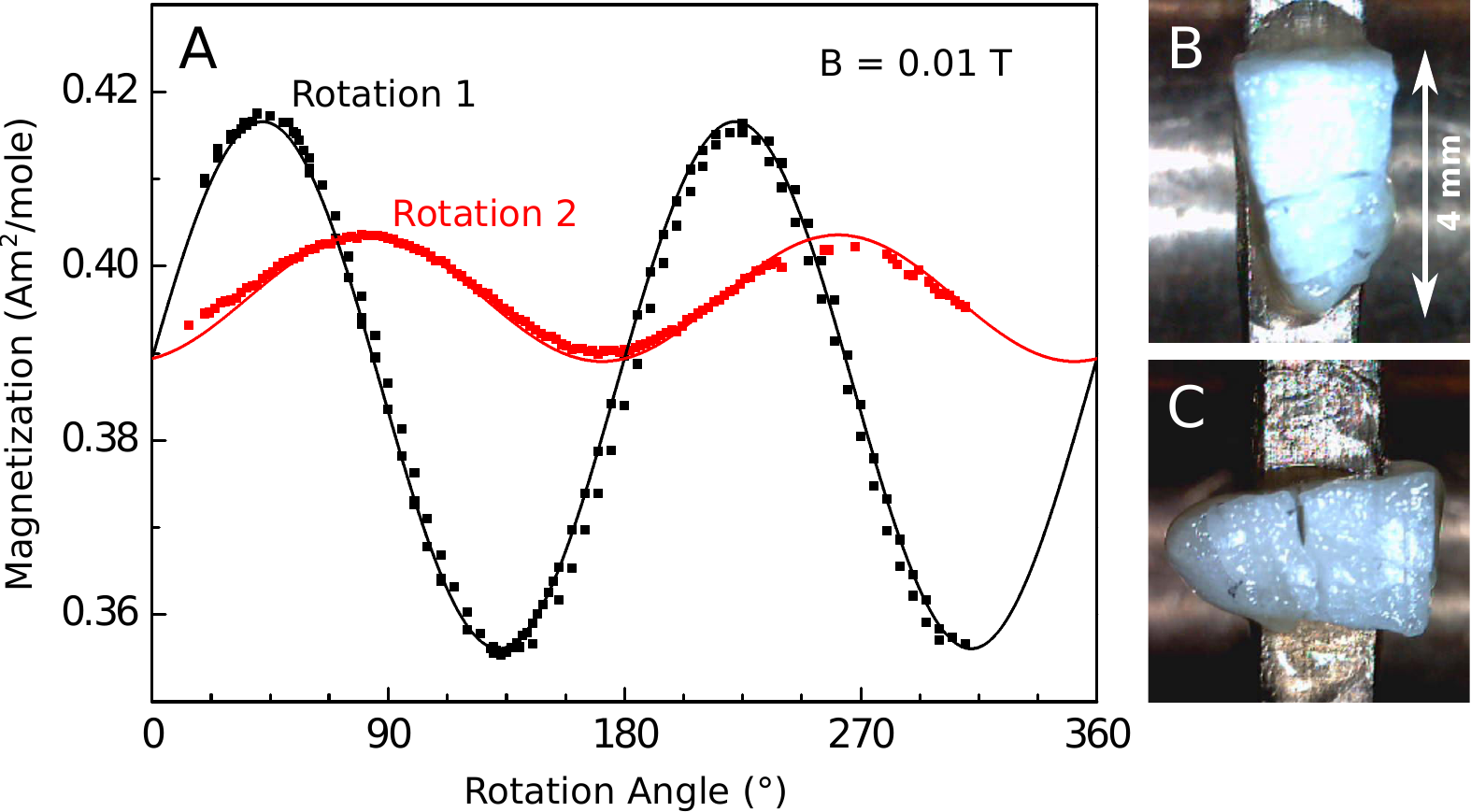}\\[3mm]
\begin{tabular}{|c|c|}
\hline
$\theta = \ang{118.1}$ & $\alpha_0= \ang{55.7}$ \\
\hline
$\phi = \ang{219.0}$ & $\beta = \ang{80.9}$\\
\hline
$C_0 = \SI{-9e-4}{\ampere\meter\squared\per\mole}$ & \\
\hline
\end{tabular}
\caption{Magnetization measured in the MPMS SQUID as a function of rotation angle for \ce{GdW30}.  The magnetization along the direction of the DC magnetic field is measured for a field value of \SI{0.01}{\tesla}.  \ang{0} corresponds to the rotation platform being perpendicular to the DC magnetic field.  Rotation 1 and 2 correspond to the configurations seen in graph B and C respectively.  A small fracture in the crystal can be seen in the photographs.  The continuous lines are fits to the theoretical model with the fitted parameters given in the table above.  The meanings of each parameter are described in the text and summarized in figure \ref{fig:rot_reference}.}\label{fig:rot_GdW30}
\end{figure}

We perform angle dependent magnetization measurements in a commercial MPMS system (see section \ref{sec:mpms}).  A undiluted crystal of \ce{GdW30} was taken from a deionized water solution and placed on the MPMS rotation stage as shown in figures \ref{fig:rot_GdW30}B and \ref{fig:rot_GdW30}C.  Its magnetization was then measured as a function of the rotation angle at a temperature of \SI{2}{\kelvin}.  The crystal was covered by Apiezon N grease to avoid interstitial water loss.  A fracture in the crystal was detected after the measurement.  After a first rotation, the crystal is rotated \ang{90} clockwise on the stage (figure \ref{fig:rot_GdW30}C) and a second rotation measurement is performed.  The results are shown in figure \ref{fig:rot_GdW30}.

With the Hamiltonian parameters derived from EPR experiments at \SI{10}{\kelvin} (figure \ref{fig:epr_polvo_fitted}), the crystal field Hamiltonian can be numerically solved and any observable computed.  Since we are in the linear regime (low field), the following equation for the magnetization holds:
\begin{equation}
\vec{M} = \hat{\chi}\vec{H} = \left(\begin{array}{ccc}
\chi_{xx} & 0 & 0 \\
0 & \chi_{yy} & 0 \\
0 & 0 & \chi_{zz} \\
\end{array}
\right)\vec{H}, \label{eq:susclin}
\end{equation}
where the susceptibility matrix $\hat{\chi}$ is diagonal in the reference system set by the principal anisotropy axes XYZ.  The (molar) values of $\chi_{xx},\chi_{yy},\chi_{zz}$ can be directly obtained from the Hamiltonian \refeq{eq:GSHamiltonian} and, at \SI{2}{\kelvin}, turn out to be:
\begin{eqnarray}
\chi_{xx} &=& 4\pi\cdot\SI{3.95}{\centi\meter\cubed\per\mole} \\
\chi_{yy} &=& 4\pi\cdot\SI{4.31}{\centi\meter\cubed\per\mole} \\
\chi_{zz} &=& 4\pi\cdot\SI{3.44}{\centi\meter\cubed\per\mole}
\end{eqnarray}
With these parameters fixed and using equation \refeq{eq:susclin}, the magnetization for any given direction can be calculated\footnote{The full Hamiltonian was also used to directly calculate the magnetization, but there are no significant differences with the linear response approximation for this experimental regime.}.  It is therefore possible to fit the rotation dependence of the magnetization shown in figure \ref{fig:rot_GdW30}A, and obtain from it the orientation of the rotation axes relative to the magnetic XYZ axes.  Since from x-ray diffraction we have a good idea of where the crystalline axes are relative to the crystal geometry and the rotation axes, we can also deduce the directions of the magnetic axes relative to the the crystal structure.

\begin{figure}[!tb]
\centering
\includegraphics[width=0.6\columnwidth]{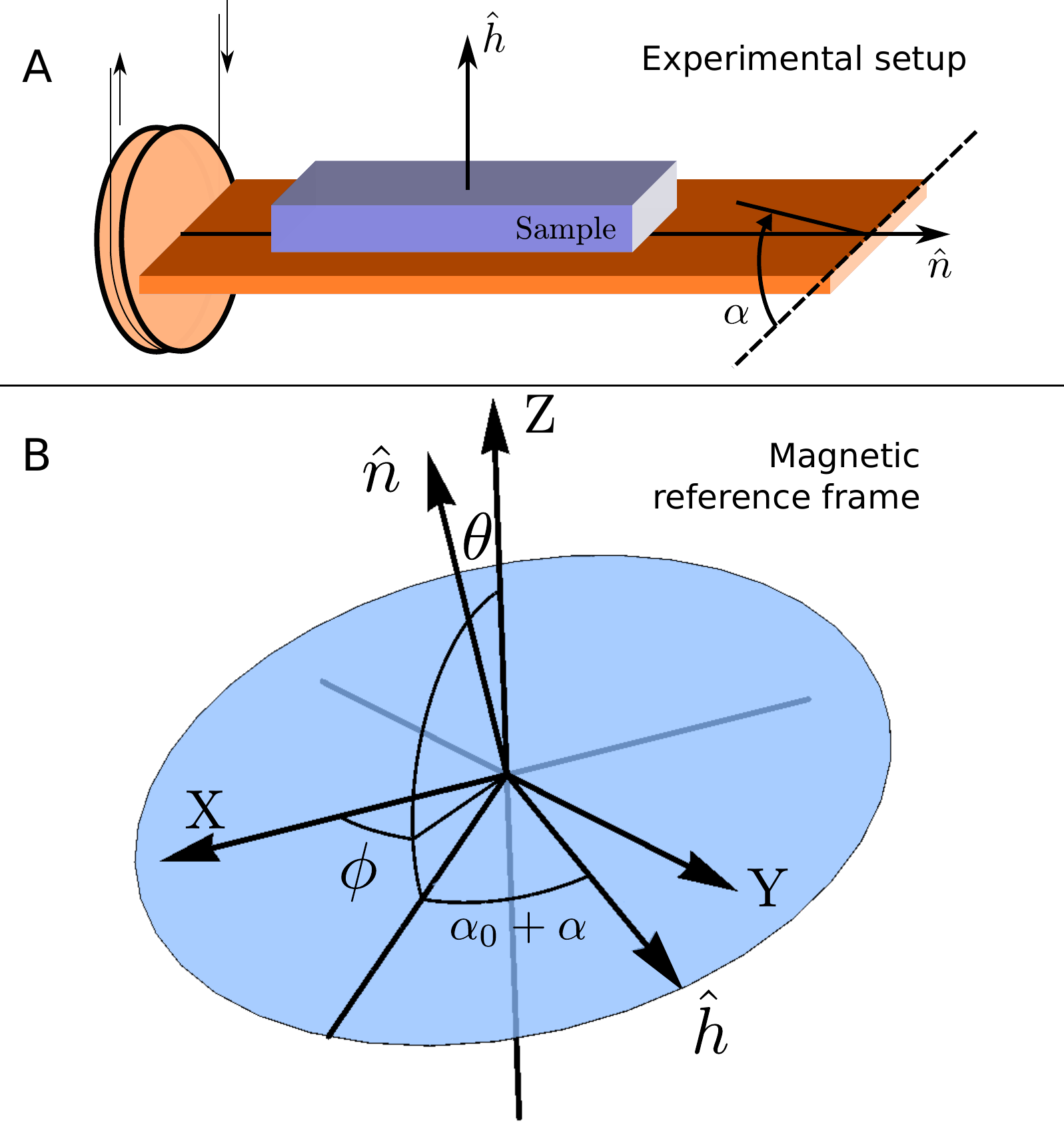}
\begin{tabular}{c|c}
Parameter & Meaning \\
\hline
$\theta$ & Inclination of $\hat{n}$ in the magnetic reference frame \\
$\phi$ & Azimuth of $\hat{n}$ in the magnetic reference frame \\
$\alpha_0$ & Initial rotation of $\hat{h}$ around $\hat{n}$ \\
$\beta$ & Rotation of $\hat{n}$ around $\hat{h}$ for second rotation curve \\
$C_0$ & Constant added background.
\end{tabular}
\caption{Parameters, definitions and configuration for theoretical fits to the magnetic rotation data shown in figure \ref{fig:rot_GdW30}.  Graph A shows the rotation stage and the relevant axes.  Graph B shows these axes in the magnetic reference frame and the angles chosen as parameters to be fitted.}\label{fig:rot_reference}
\end{figure}

In the fitting, we follow the definitions shown in figure \ref{fig:rot_reference}.  We label the rotation axis direction $\hat{n}$ with the spherical coordinates $\theta,\phi$ in the magnetic reference frame.  The DC magnetic field direction, $\hat{h}$, is perpendicular to this rotation axis and rotates around this axis with the angular coordinate $\alpha$.  If these coordinates are set to $\theta= 0$ and $\phi  =0$, $\hat{n}$ is aligned along the magnetic Z axis, while the $\alpha = 0$ position of $\hat{h}$ is then along the magnetic X axis.  For non-zero values of $\theta$ and $\phi$, both $\hat{n}$ and $\hat{h}$ are rotated firstly around the Y axis an angle $\theta$ (inclination) and then around the Z axis an angle $\phi$ (azimuth).  We also note that the initial position of $\hat{h}$ may not correspond to $\alpha=0$ in the experiment, so a angular shift $\alpha_0$ is also introduced as a fitting parameter.  A simulated rotation curve, similar to the experimental ones in figure \ref{fig:rot_GdW30}, is then obtained by sweeping $\alpha$ through the desired angles.  The second rotation curve from figure \ref{fig:rot_GdW30} is obtained by rotating $\hat{n}$ an angle $\beta$ around $\hat{h}$ at its initial angle $\alpha_0$.  With these definitions, we end up with the parameters shown in figure \ref{fig:rot_reference} and the fit results from figure \ref{fig:rot_GdW30}.

Through symmetry considerations we can make some educated guesses at where the magnetic anisotropy axes should lie relative to the molecular structure and compare these predictions to the fit results.  Since \ce{Gd^{3+}} has no intrinsic anisotropy, the magnetic configuration will be determined mostly by the molecular structure (a flattened sphere or disc, figure \ref{fig:crystalstructure}).  Given the Hamiltonian parameters, we know that the two easy axes (higher susceptibility) have similar susceptibility values while the hard axis susceptibility lies further away.  By symmetry then, the magnetic hard axis, i.e. the axis with the lowest magnetization response (Z in this case), is expected to be perpendicular to the molecular plane.  The other two axes (X,Y) should lie somewhere in the molecular plane in which no large differences in susceptibility are expected.  This seems to be compatible with the results obtained from the fit to the magnetization data.  The direction of the axis for the first rotation is close to the [001], a direction that lies in the molecular plane (the long edge of the crystal was aligned with the rotation axis).  If the magnetic Z axis is perpendicular to the molecular plane, this rotation axis should approximately have $\theta =\ang{90}$.  The fitted value is close to this value (\ang{116}) making our original guess plausible.  The deviation is explainable by the fact that the crystal is aligned by hand and, as can be seen in figure \ref{fig:rot_GdW30}B, the wedge shape makes it unclear which crystal edge is parallel to [001].  Also, a rotation around an axis in the molecular plane should have a larger variation in the magnetization since it includes magnetizations along both the hard axis and the \emph{easy} plane.  This is also confirmed by the data where we see a much smaller variation when the axis is placed in a different direction in the second measurement.  We will further refine the magnetic axis orientations using angle dependent cw-EPR experiments in the next section.

\subsection{Crystal EPR experiments}

\begin{figure}[!tb]
\centering
\includegraphics[width=0.9\columnwidth]{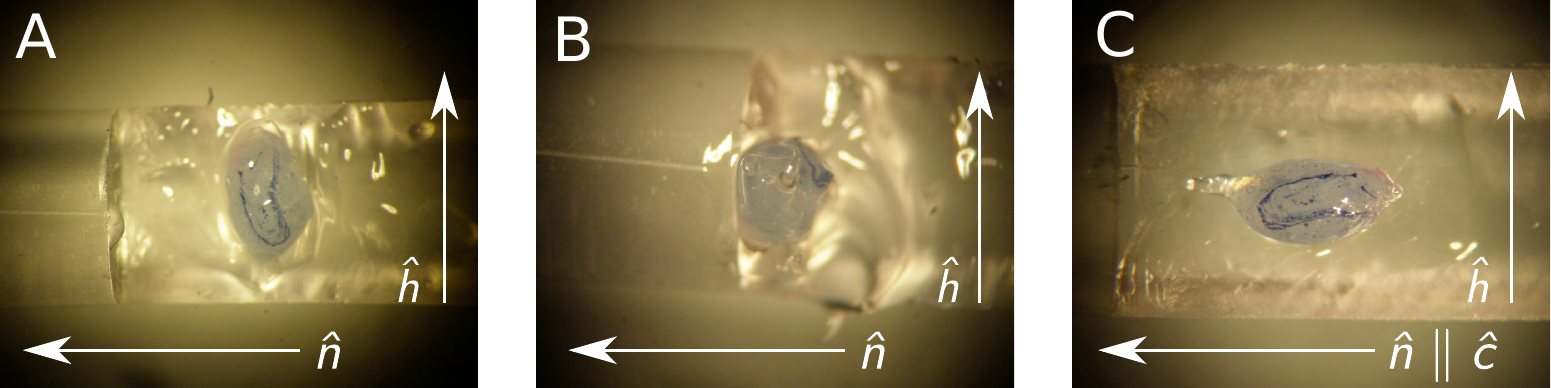}
\caption{\ce{Y_{0.99}Gd_{0.01}W30} crystal on the rotating EPR stage.  cw-EPR spectra were obtained as a function of angle for the three sucessive crystal orientations (A,B,C).  $\hat{n}$ labels the rotation axis while $\hat{h}$ labels the DC magnetic field direction (for a certain angle).  In C, the rotation axis is parallel to the crystal axis $\hat{c}$}\label{fig:rot_epr_photo}
\end{figure}

\begin{figure}[!tb]
\centering
\includegraphics[width=1.\columnwidth]{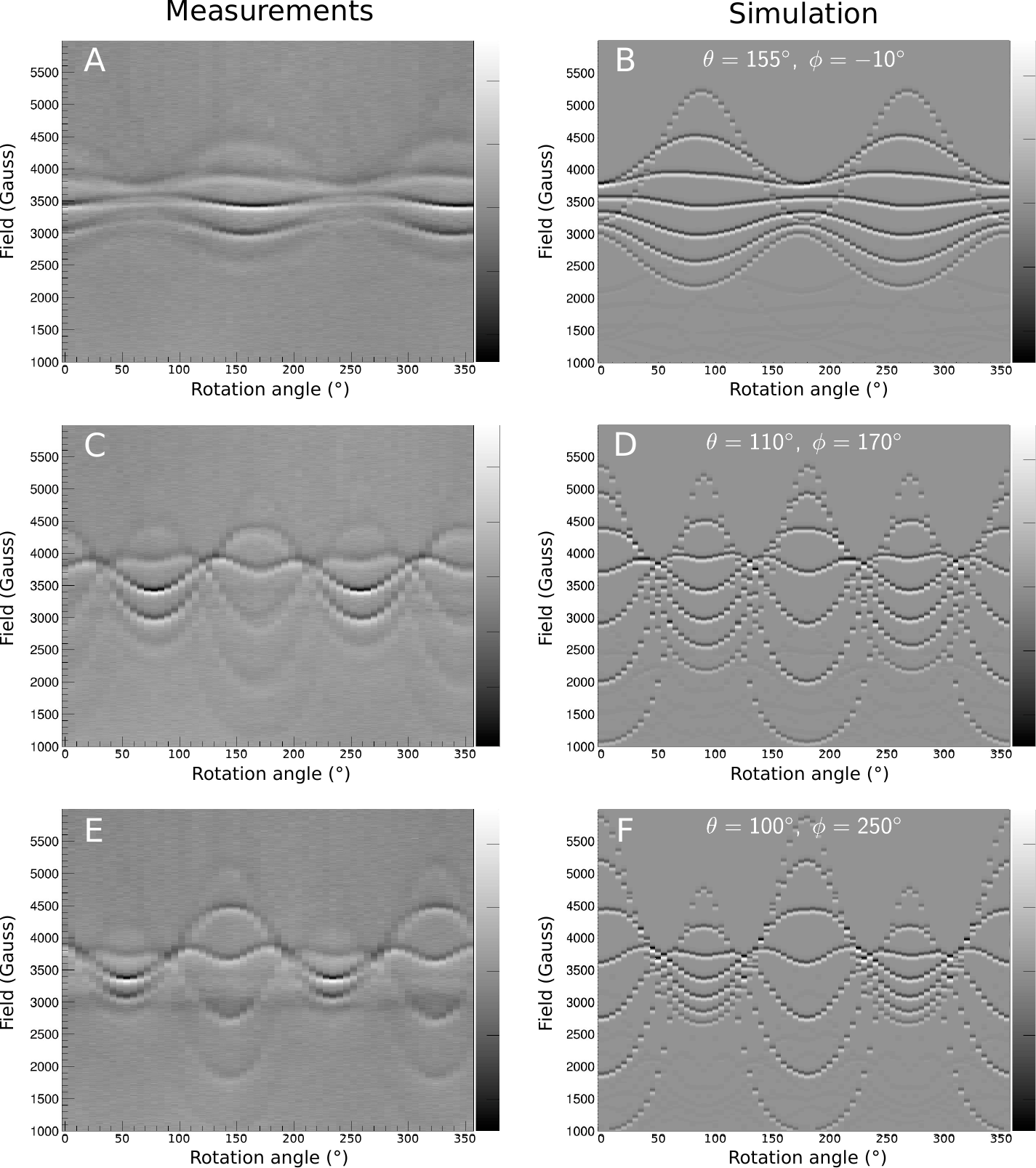}
\caption{Measured (ACE) and simulated (BDF respectively) X-band cw-EPR spectra for a dilute \ce{Y_{0.99}Gd_{0.01}W30} crystal.  Graphs ACE correspond to the configurations shown in figures \ref{fig:rot_epr_photo}ABC respectively.  The simulations are labelled with their corresponding rotation axis according to the conventions from figure \ref{fig:rot_reference}.}\label{fig:rot_epr_GdW30}
\end{figure}

We measured a dilute crystal of \ce{Y_{0.99}Gd_{0.01}W30} using X-band cw-EPR and a rotating stage.  The crystal was encased in an epoxy glue after being removed from its original solution and checked with x-ray diffraction (see section \ref{sec:xrayGdW30}).  Having confirmed its crystal structure, a room temperature cw-EPR measurement for three different orientations of the crystal on the rotating stage was done (figure \ref{fig:rot_epr_photo}).  The measured spectra can be seen in figures \ref{fig:rot_epr_GdW30}A, \ref{fig:rot_epr_GdW30}C, and \ref{fig:rot_epr_GdW30}E.

To compare these measurements to the theory and to obtain the positions of the magnetic axis (similar to the previous section), we use the room temperature Hamiltonian parameters from figure \ref{fig:epr_polvo_fitted} and the EasySpin package to generate simulated spectra.  Two dimensional representation analogous to those obtained from the measurement (figures \ref{fig:rot_epr_GdW30}A, \ref{fig:rot_epr_GdW30}C and \ref{fig:rot_epr_GdW30}E) were generated for a large number of rotation axis directions.  The angle definitions are the same as in the previous section (figure \ref{fig:rot_reference}) with the exception that $\alpha_0$ has always been set to 0 since the EPR rotating stage does not keep its position once the sample is removed from the cavity.  Among the collection of generated spectra, we have searched for the ones that best reproduce the measured data.  The best fitting simulations are represented in figures \ref{fig:rot_epr_GdW30}B, \ref{fig:rot_epr_GdW30}D, and \ref{fig:rot_epr_GdW30}F next to their measured counterparts and with the simulated axis directions.  We see good agreement of these simulated spectra with the experiment.  Also, the three directions are very close to forming \ang{90} between each other, as they should if the crystal positioning (figure \ref{fig:rot_epr_photo}) were perfect.

\begin{figure}[!tb]
\centering
\includegraphics[width=0.7\columnwidth]{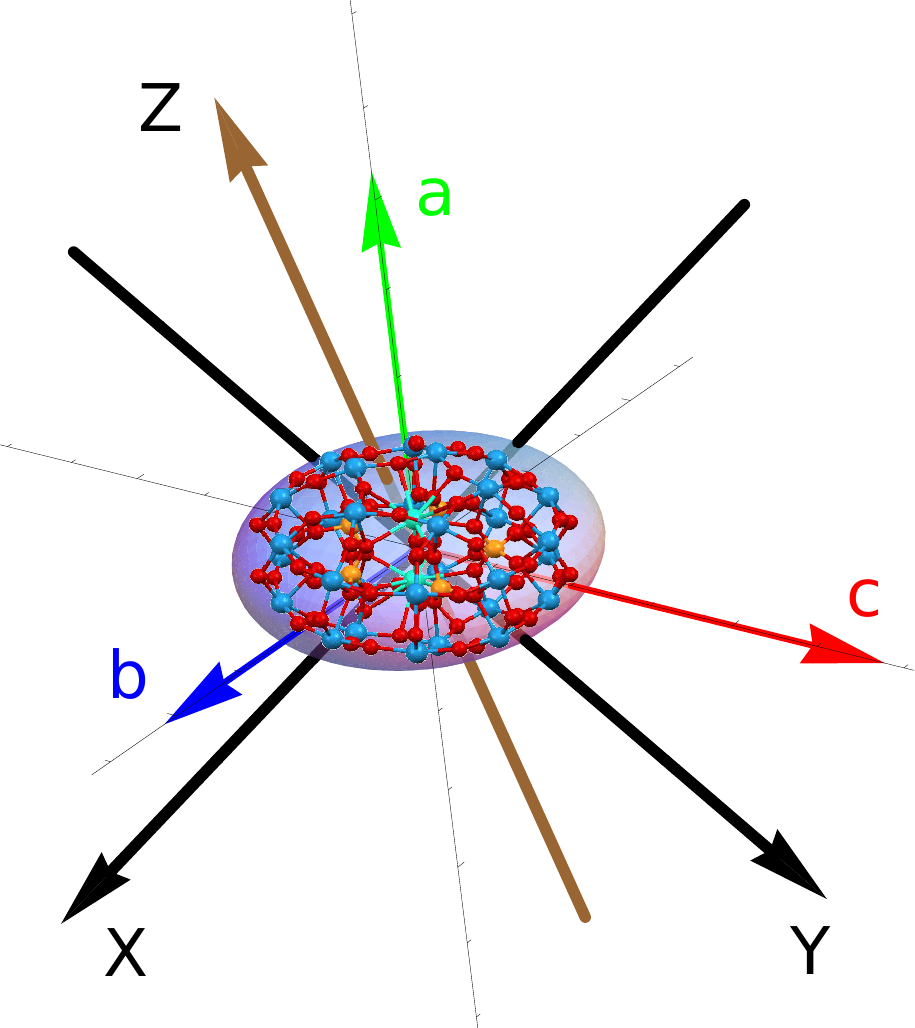}
\caption{Calculated magnetic axes orientation for \ce{GdW30} from angle dependent cw-EPR experiments.  The magnetic hard axis Z is seen to be close to the direction perpendicular to the molecular plane (a).  The two easy axes are close the the molecular plane.}\label{fig:rot_orientation_GdW30}
\end{figure}

Assuming that the rotation axes for figures \ref{fig:rot_epr_GdW30}ACE correspond to the crystal axes [001], [1-10] and [110] respectively (see section \ref{sec:xrayGdW30}), the magnetic axes can be converted to the crystal structure basis and the position of these axis relative to each molecule can be obtained.  With the numbers shown in figure \ref{fig:rot_epr_GdW30} we get the axis orientations shown in figure \ref{fig:rot_orientation_GdW30}.  All three axes are relatively close to the three crystal directions.  Our initial guess is that the magnetic Z axis should be aligned with the a crystal axis and that the other two axes should lie in the molecular plane.  It may be the case that they are indeed tilted but we think it is more likely that at least the Z and a directions are aligned and that the measured deviation is due to experimental error.  The main source of error is probably the fact that the crystal alignment in done by hand in all cases except when the crystal is in the x-ray difractometer.  We note also that the rotation axis for figure \ref{fig:rot_epr_GdW30}EF should be the same as the axis fitted from the magnetization data in figure \ref{fig:rot_GdW30}.  There are however some differences, again probably due to misalignments.  Another factor that could potentially lead to some error is the possible degradation of the crystal structure over time.

\subsection{Pulsed EPR experiments: spin qubit decoherence times}\label{sec:pulsedEPR_GdW30_times}

In this section we investigate the dynamic response of the \ce{GdW30} system using pulsed EPR methods, described in section \ref{sec:EPRtech}.  The same dilute crystal (\ce{Y_{0.99}Gd_{0.01}W30}) from the previous section is mounted in a X-band pulsed EPR cavity and oriented as in figure \ref{fig:rot_epr_photo}A and manually rotated into a position where the cw-spectrum presents the highest field spread between the absorption peaks.  This is necessary to allow us to excite each individual transition independently of the others and to study the relaxation dynamics for each of them.

\begin{figure}[!b]
\centering
\includegraphics[width=0.7\columnwidth]{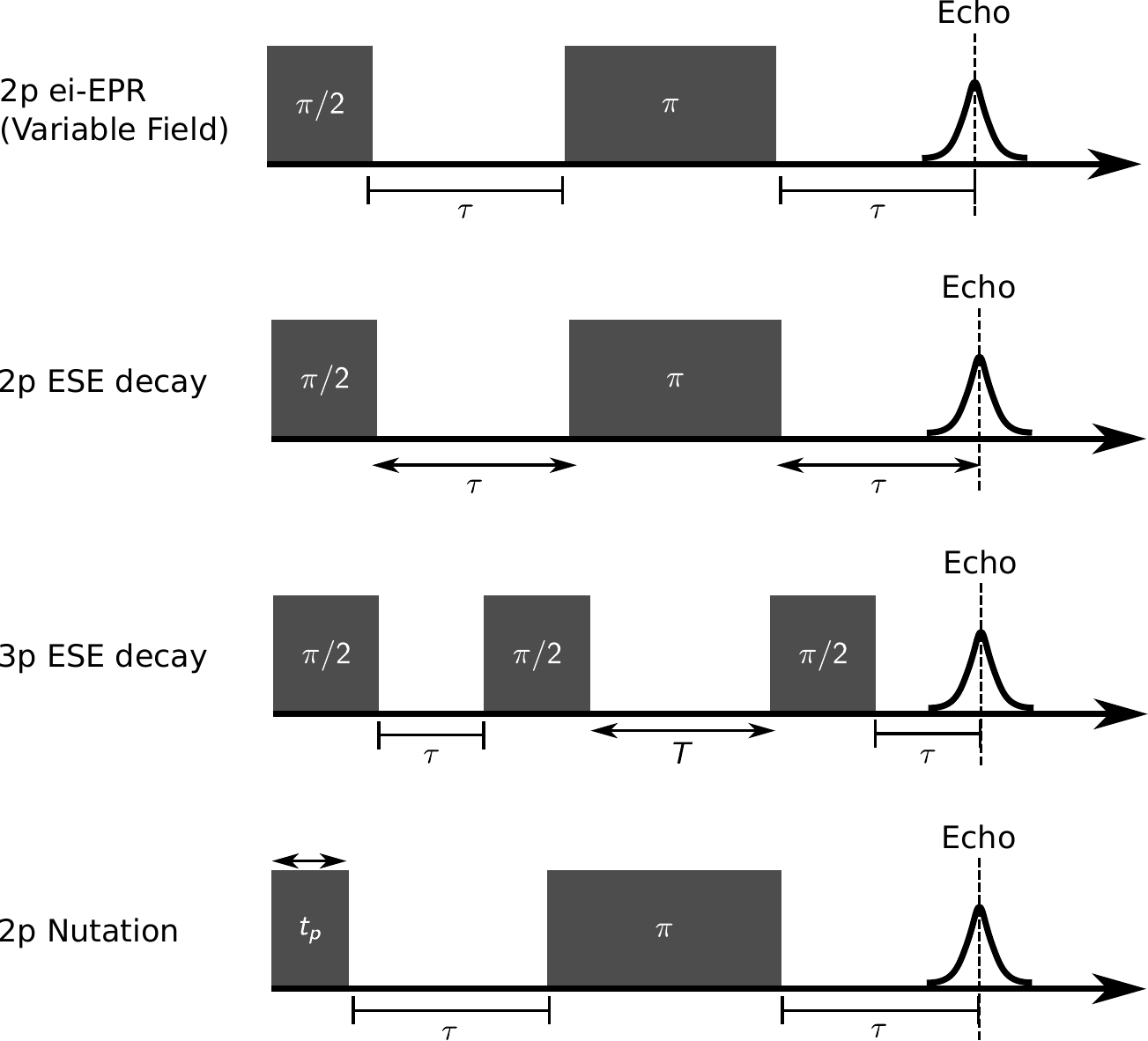}
\caption{Pulsed EPR sequences \cite{Schweiger2001}.  Double ended arrows denote the parameter that is varied for each echo measurement.  Sequence A is an echo-induced EPR (ei-EPR) measurement that produces a signal analogous to a cw-EPR experiment.  The field is swept while the pulse lengths and wait times are kept fixed for each echo.  Sequence B is a two pulse Electronic Spin Echo decay sequence.  Here the field is  tuned to the desired transition and a $\pi/2$--$\pi$ pulse with a variable wait time $\tau$ is applied.  The dependence of the echo intensity on $\tau$ follows an exponential decay whose decay constant is the $\textrm{T}_2$ relaxation time.  Sequence C is a three pulse sequence with a fixed wait time $\tau$ and a variable wait time $T$.  The echo can be shown to decay exponentially as a function of $T$ according to the $\textrm{T}_1$ decay time.  Finally, sequence D consists of two pulses where the first pulse is of variable width and the second is a $\pi$ pulse that produces an echo proportional to the in-plane spin component.  The dependence of the echo on the pulse time $t_p$ is approximately an exponential decay according to $\textrm{T}_2$ modulated by the sample-cavity Rabi frequency.}\label{fig:pulses}
\end{figure}

There are several pulse sequences that will be used throughout this section, each geared towards extracting different information from the sample.  The different sequences used are schematically shown in figure \ref{fig:pulses} along with a description of the parameter each sequence is designed to be sensitive to \cite{Schweiger2001}.  All the measurements shown in this section were performed at $T=\SI{6}{\kelvin}$.

\begin{figure}[!tb]
\centering
\includegraphics[width=0.75\columnwidth]{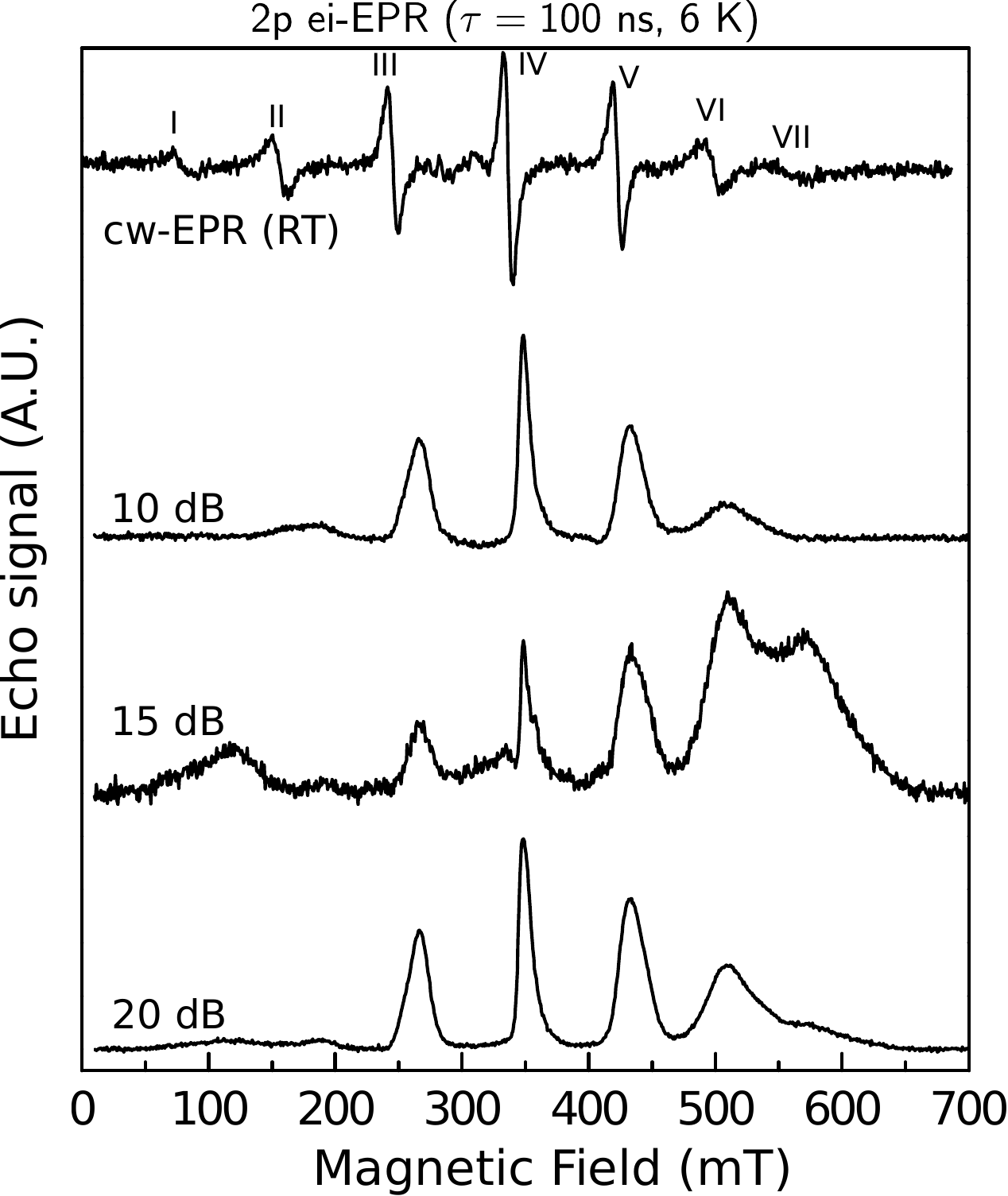}
\caption{X-band 2p ei-EPR of crystalline \ce{GdW30}.  The top graph shows the room temperature cw-EPR spectrum for comparison with the pulsed spectra.}\label{fig:echo1}
\end{figure}

\begin{figure}[!tb]
\centering
\includegraphics[width=0.9\columnwidth]{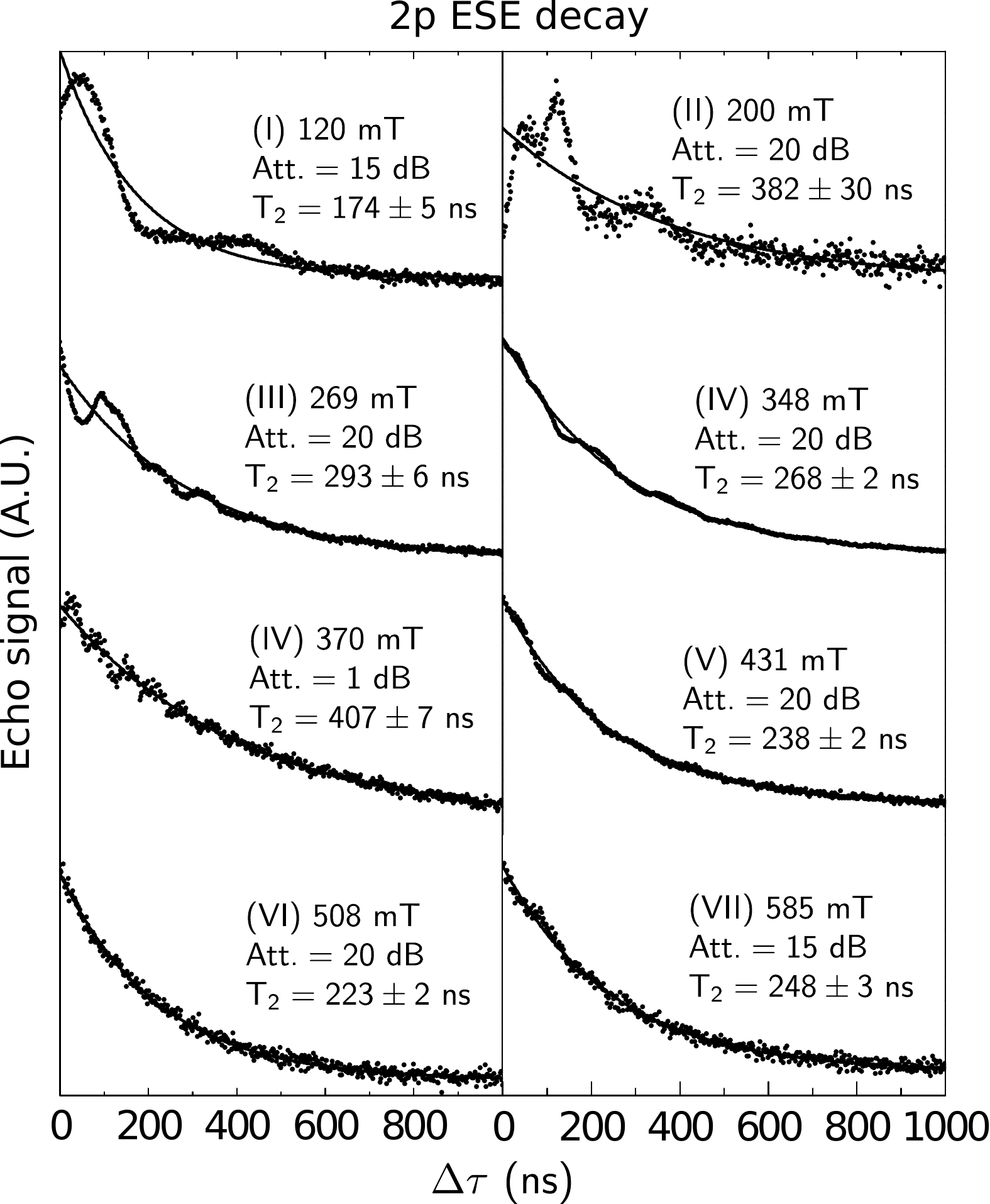}
\caption{X-band 2p ESE decay of crystalline \ce{GdW30}.  Graphs show the echo intensity as a function of $\Delta\tau = \tau -\tau_0$ where $\tau$ is the wait time from figure \ref{fig:pulses}B and $\tau_0 =\SI{100}{\nano\second}$ is a minimum wait time fixed for the experiment.  Each set of data is fitted to an exponential decay function from which the decay time $\textrm{T}_2$ is extracted.}\label{fig:echo2}
\end{figure}

The first measurement performed is an ei-EPR experiment (figure \ref{fig:pulses}A) at \SI{6}{\kelvin}.  This experiment is necessary to determine the resonant magnetic fields for each transition measured in order to perform further measurements on them.  The echo intensity as a function of field and as a function of the pulse power is shown in figure \ref{fig:echo1} along with the room temperature cw-EPR measurement.  The transitions are numbered on the cw-spectrum where the visibility is better.  The ei-spectrum shows that all transitions are still visible at \SI{6}{\kelvin} although their intensity depends on the microwave power used for the pulses.  This is due to fact that for an ei-EPR measurement, the pulse lengths and intensities are optimized for certain transitions (the central one).  The differing matrix elements and Rabi frequencies for each transition do not produce optimal echoes for the remaining transitions and in general get worse the further the transition is from the optimized one.  Also, we find that some transitions have been displaced from their room temperature positions most likely due to the variation in the Hamiltonian parameters from room temperature to low temperature.  We note also that there are two peaks at the central IV position (both are visible when the attenuation is set to \SI{15}{\deci\bel}).  One of the peaks corresponds to the sample while the other is most likely due to copper impurities in the cavity.  We expect that the lower field peak corresponds to the copper impurity while the larger field peak corresponds to the sample.

\begin{figure}[!tb]
\centering
\includegraphics[width=0.9\columnwidth]{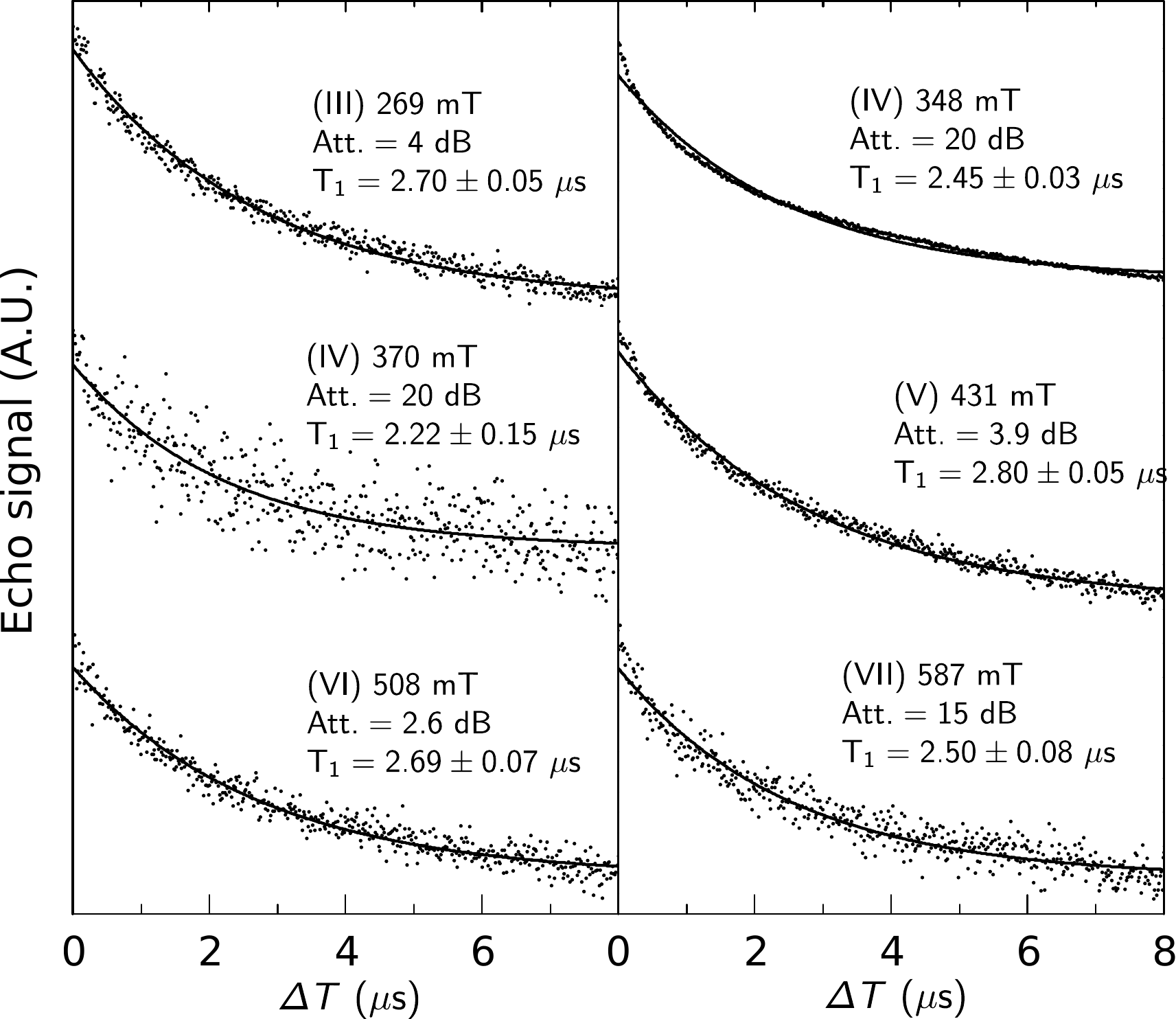}
\caption{X-band 3p ESE decay of crystalline \ce{GdW30}.  Graphs show the echo intensity for $\tau=\SI{100}{\nano\second}$ as a function of $\Delta T = T -T_0$ where $\tau$ and $T$ are the wait times from figure \ref{fig:pulses}C and $T_0 =\SI{100}{\nano\second}$ is a minimum wait time fixed for the experiment.  Each set of data is fitted to an exponential decay function from which the decay time $\textrm{T}_1$ is extracted.}\label{fig:echo3}
\end{figure}

The next step is to extract phase coherence relaxation rate $\textrm{T}_2$ by using a 2p-ESE (figure \ref{fig:pulses}B) sequence.  The results and parameters used for each transition are summarized in figure \ref{fig:echo2}.  The main result here is that all transitions can be coherently manipulated and have similar coherence times, ranging from \SI{130}{\nano\second} to \SI{400}{\nano\second}.  We give the value of $\textrm{T}_2$ for both the IV transitions found in the ei-EPR spectrum (figure \ref{fig:echo1}).

The $\textrm{T}_1$ relaxation time is extracted from a 3p-ESE decay measurement (figure \ref{fig:pulses}C).  The resulting echo intensities for each transition are shown in figure \ref{fig:echo3}.  Only transitions III, IV, V, VI, and VII are shown since no measurable echo signal was achieved for transitions I and II.  Again we find a fairly even value of $\textrm{T}_1$ for all the measured transitions with its value ranging from \SI{2.2}{\micro\second} to \SI{2.8}{\micro\second}, thus much longer that $\textrm{T}_2$ as expected.  The values of $\textrm{T}_1$ and $\textrm{T}_2$ are of the same order of magnitude as those found in previous experiments performed on a powdered sample of the same concentration \cite{Martinez-Perez2012}.

\subsection{Rabi spin oscillations: potential for quantum computation}\label{sec:pulsedEPR_GdW30_rabi}

\begin{figure}[!tb]
\centering
\includegraphics[width=0.9\columnwidth]{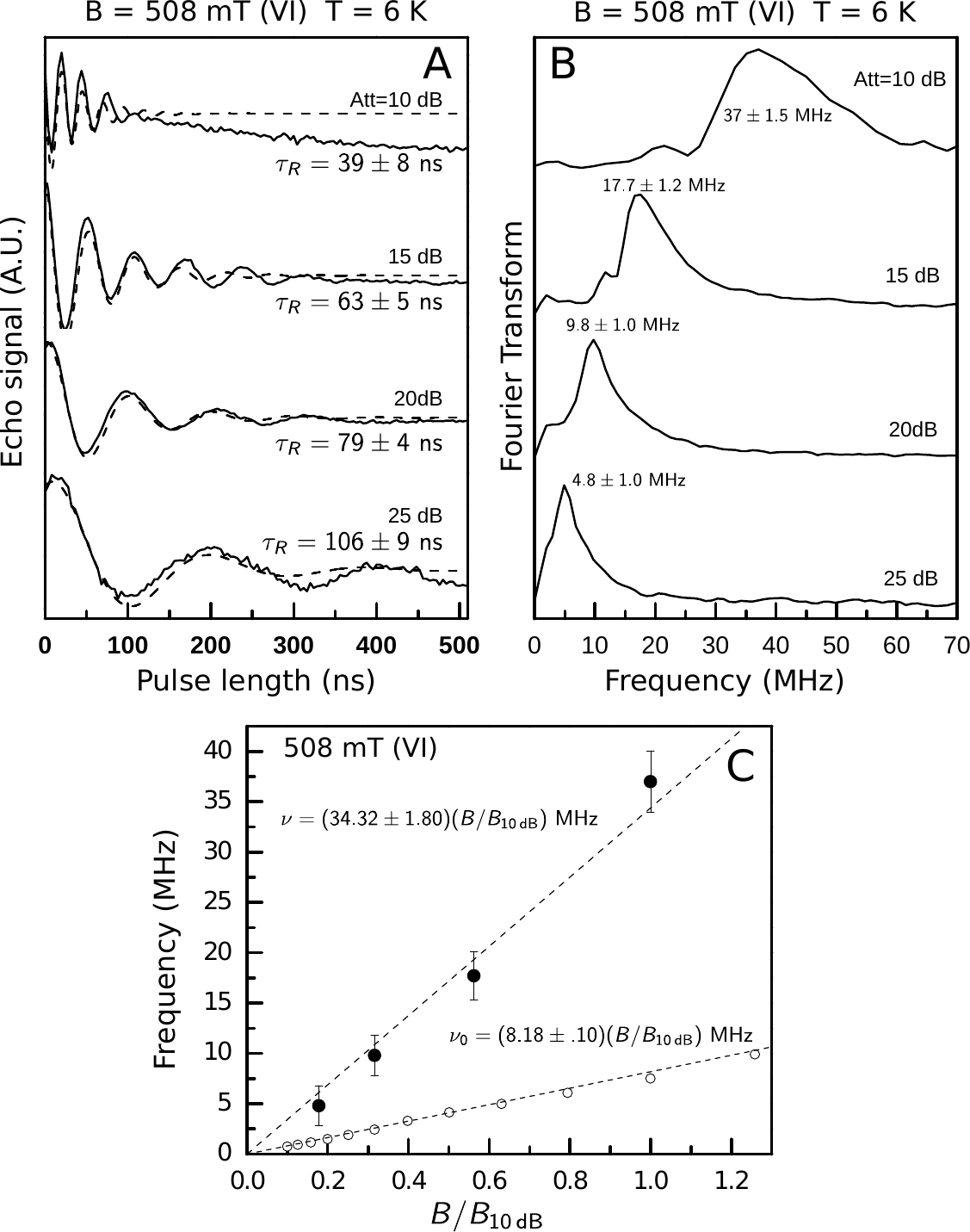}
\caption{Power dependent Rabi oscillations for transition IV of \ce{GdW30}.  The pulse sequence used is shown in figure \ref{fig:pulses}D an a value of $\tau=\SI{100}{\nano\second}$ was used.  Graph A shows the echo intensity as a function of the initial pulse length $t_p$ for different attenuation values.  The dashed lines show fit to equation \refeq{eq:rabifit} and the fitted values of $\tau_B$ are shown along each graph.  Graph B shows the Fourier transform of the measurements from graph A along with the value of the frequency at the maximum (Rabi frequency).  Graph C shows the measured frequency values as a function of the field (normalized by the field at \SI{10}{\deci\bel} attenuation, $B_{10}$.  The filled dots are for \ce{GdW30} while the empty dots are for a sample of natural charcoal and used for calibration.  The Rabi frequency for both samples was fitted using a linear equation.}\label{fig:nutation}
\end{figure}

Nutation measurements were performed according to the pulse sequence shown in figure \ref{fig:pulses}D.  The echo intensity is proportional to the transverse spin component that oscillates when plotted against $t_p$ according to the Rabi frequency of the cavity-spin system (analogous to \refeq{eq:interaction}):
\begin{equation}
\nu_{\rm Rabi} \propto B_{\rm rf}|\bra{n}S_{\hat{b}}\ket{n+1}| \label{eq:rabifrec}
\end{equation}
where $\ket{n}$ are the spin Hamiltonian eigenstates and the matrix element is for the spin component parallel to the rf magnetic field.  The oscillation frequency can directly provide the spin matrix element if a calibration sample is used to solve for the pulse field intensity, $B_{\rm rf}$.  Figure \ref{fig:nutation}A shows the echo signal for transition VI as a function of the pulse length $t_p$ for different microwave powers (i.e. values of $B_{\rm rf}$).  We see the expected Rabi oscillations modulated by the relevant decay time.  The time dependence can be fitted using the following equation:
\begin{equation}
f(t) = Ae^{-t/\tau_R}\cos{(2\pi\nu_{\rm Rabi} t+\theta_0)}+C_0, \label{eq:rabifit}
\end{equation}
where $\tau_R$ is the Rabi decay time, $A$ is a scaling constant, $C_0$ is the background level and $\theta_0$ an initial phase.  Observing the fits of this equation to the data in figure \ref{fig:nutation}A, we see that a single damped Rabi oscillation does not fit the data perfectly.  The Fourier transform of these signals (figure \ref{fig:nutation}B) gives the distribution of Rabi frequencies present for each value of the attenuation.  The existence of a sizable distribution is in line with the fact that we expect a distribution of the Hamiltonian parameters $B_2^0$ and $B_2^2$, but could also be attributed in part to inhomogeneities in the rf magnetic field.

The data also shows that lower attenuations (higher fields) lead to higher frequencies, as expected (see equation \refeq{eq:rabifrec}).  If we plot these frequencies as a function of the field value normalized by the field value at a \SI{10}{\deci\bel} attenuation ($B_{10}$), as shown in figure \ref{fig:nutation}C, we find the expected linear dependence of the Rabi frequency on the rf field strength
\begin{equation}
\nu_R = C \frac{B_{\rm rf}}{B_{10}}\propto MB_{\rm rf},
\end{equation} 
where $M=|\bra{n}S_{{b}}\ket{n+1}|$ is the matrix element for the relevant transition and $C$ is the slope obtained from the linear fit in figure \ref{fig:nutation}C for \ce{GdW30}.  The figure also shows the results of the same experiment performed on a sample of natural charcoal with a linear fit to the data.  Natural charcoal can be used to calibrate the field strengths since it behaves as an isotropic spin $1/2$ system whose matrix element is well defined and equal to $1/2$.  Therefore,
\begin{equation}
\nu_{R,0} = C_0 \frac{B_{\rm rf}}{B_{10}}.
\end{equation}
The matrix element for \ce{GdW30} can be determined as follows:
\begin{eqnarray}
&& \nu = C \frac{B_{\rm rf}}{B_{10}}\propto MB_{\rm rf} \quad  \propto \frac{1}{2}B_{\rm rf} \nonumber \\
&& M = \frac{\nu}{2\nu_0} = \frac{\nu}{2C_0}\frac{B_{10}}{B_{\rm rf}} =  \frac{\nu}{2C_0}10^{10/a}, \label{eq:matconv}
\end{eqnarray}
where $a$ is the attenuation in \si{\deci\bel}.

\begin{figure}[!tb]
\centering
\includegraphics[width=0.65\columnwidth]{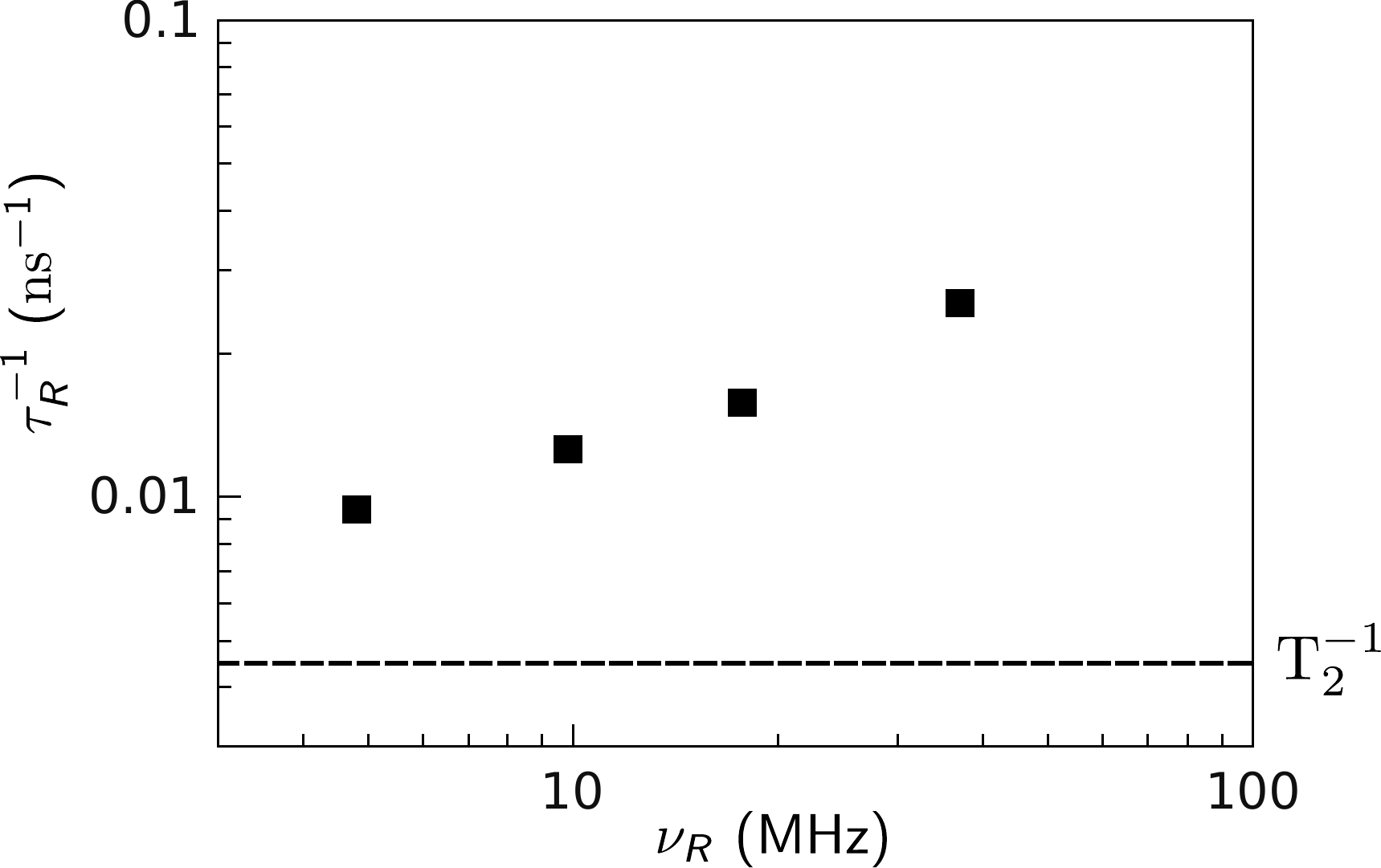}
\caption{$\tau_R$ relaxation times from measurements in figure \ref{fig:nutation} as a function of the Rabi frequency $\nu_R$.  The dashed line represents the $\textrm{T}_2$ value from figure \ref{fig:echo2} for $B=\SI{508}{mT}$.}
\label{fig:tauR}
\end{figure}

For a perfectly homogeneous sample in a fully homogeneous rf magnetic field, the value of $\tau_R$ is limited by the $\textrm{T}_2$ coherence time, intrinsic to each molecule \cite{DeRaedt2012}.  However, besides the intrinsic decoherence, $\tau_R$ is also limited by field inhomogeneities and strains in the Hamiltonian parameters \cite{Bertaina2007,DeRaedt2012} which give rise to a distribution of Rabi frequencies (see figure \ref{fig:nutation}B).  These effects probably account for the shorter $\tau_R$ we find in this experiment when compared to the $\textrm{T}_2$ time obtained from the 2p ESE-decay described in the previous section.  The values of $\tau_R$ obtained from the fits to equation \refeq{eq:rabifit} are shown in figure \ref{fig:tauR} along with the value of $\textrm{T}_2$ obtained for the same transition IV.  For a real sample, $\tau_R$ depends on the Rabi frequency $\nu_R$.  The dependence is expected to be linear \cite{Bertaina2007}:
\begin{equation}
\frac{1}{\tau_R} \simeq c_1\nu_R + \frac{1}{\textrm{T}_2},
\end{equation}
where $c_1$ is a constant that depends on the sample properties.  For low powers, i.e. small $\nu_R$, $\tau_R \rightarrow \textrm{T}_2$.  Although this limit is not reached in this measurement, the dependence of $\tau_R$ on $\nu_R$ appears to be consistent with the previous equation.

\begin{figure}[!t]
\centering
\includegraphics[width=0.85\columnwidth]{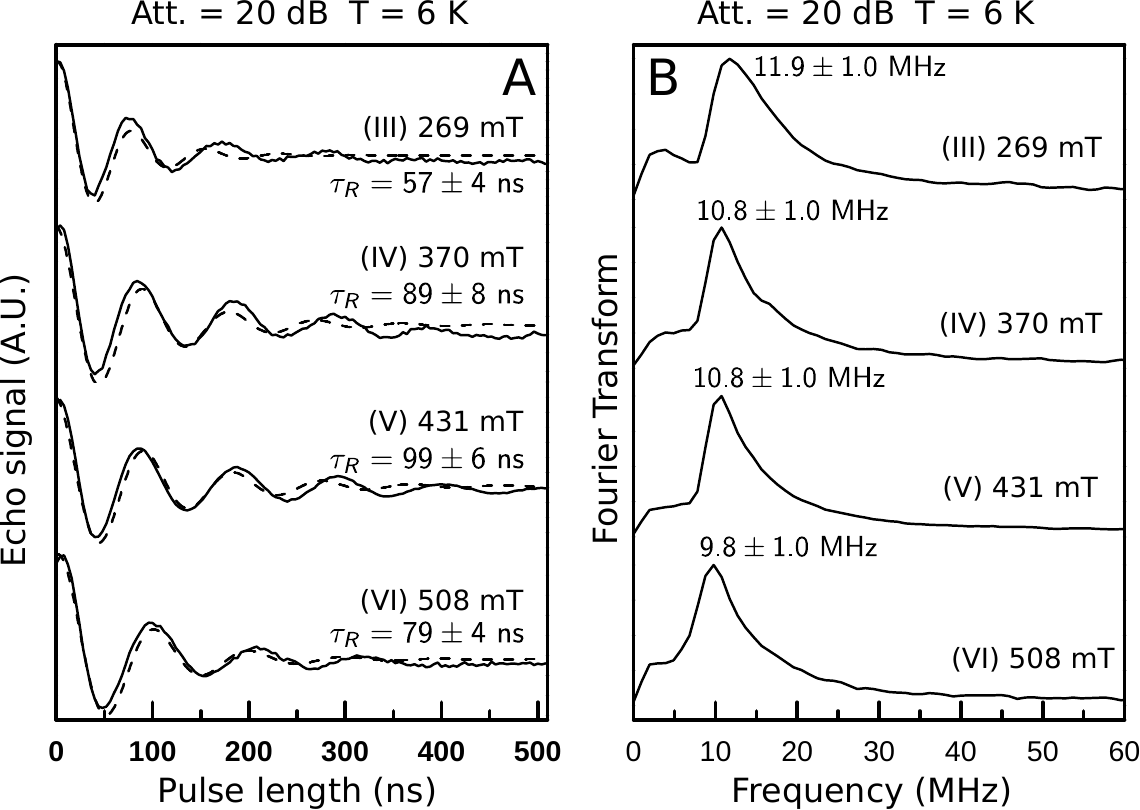}
\caption{Rabi oscillations for transitions III-VI of \ce{GdW30}.  The pulse sequence used is shown in figure \ref{fig:pulses}D an a value of $\tau=\SI{100}{\nano\second}$ was used.  Graph A shows the echo intensity as a function of the initial pulse length $t_p$ for different transitions.}\label{fig:nutation_2}
\end{figure}

\begin{table}[!b]
\centering
\begin{tabular}{c|cc}
Transition & Measured matrix element & Simulated matrix element \\
\hline
I & - & 1.44\\
II & - & 1.85\\
III & 2.3 & 2.03\\
IV & 2.08 & 2.02\\
V & 2.08 & 1.91\\
VI & 1.89 & 1.75\\
VII & - & 1.37
\end{tabular}
\caption{Measured and expected transition matrix elements for \ce{GdW30}}
\label{fig:tabmat}
\end{table}

The same nutation experiments have been performed on different transitions at a fixed attenuation.  The results are shown in figure \ref{fig:nutation_2} along with the Fourier transforms.  We find that for the four transitions measured have very similar Rabi frequencies (around \SI{10}{\mega\hertz}).  Using equation \refeq{eq:matconv} we can obtain the corresponding matrix element values.  These values can be compared to those derived from the Hamiltonian \refeq{eq:GSHamiltonian} and the parameters from figure \ref{fig:epr_polvo_fitted} (\SI{6}{\kelvin}).  The direction chosen for the calculation is one compatible with the orientation of the crystal on the EPR sample stage (the same as in figure \ref{fig:rot_epr_photo}C).  Looking back at the results and simulations from figure \ref{fig:rot_epr_GdW30}EF, we choose a direction that gives the maximum field spread for the absorption features (field along the spherical coordinate direction $\theta = \ang{170}$ and $\phi = \ang{70}$ in the molecular reference frame) and calculate the maximum absolute value of the matrix elements for the spin components transverse to the DC magnetic field.  The results are shown in table \ref{fig:tabmat}.

Considering the uncertainties in orientation and in Hamiltonian parameter values, we find that the transition matrix elements are in good agreement with our measured Rabi frequencies.  This is a good indication that our theoretical model and parameters for this system provide a reasonably good description of its coherent spin dynamics.  As it has been shown for transition IV (figure \ref{fig:nutation}), the decay time $\tau_R<\textrm{T}_2$ for transitions III to VI.  This has important consequences for the achievement of strong coupling to quantum circuits since $\tau_R$ is the \emph{effective} decoherence time relevant for performing quantum operations on ensembles of spins.  When performing quantum operations on spin ensembles, it is not generally possible to cancel out the inhomogeneous broadening (from strains or rf magnetic field inhomogeneities) as is done in a 2p-ESE decay experiment to obtain the intrinsic $\textrm{T}_2$.

\begin{figure}[!tb]
\centering
\includegraphics[width=0.9\columnwidth]{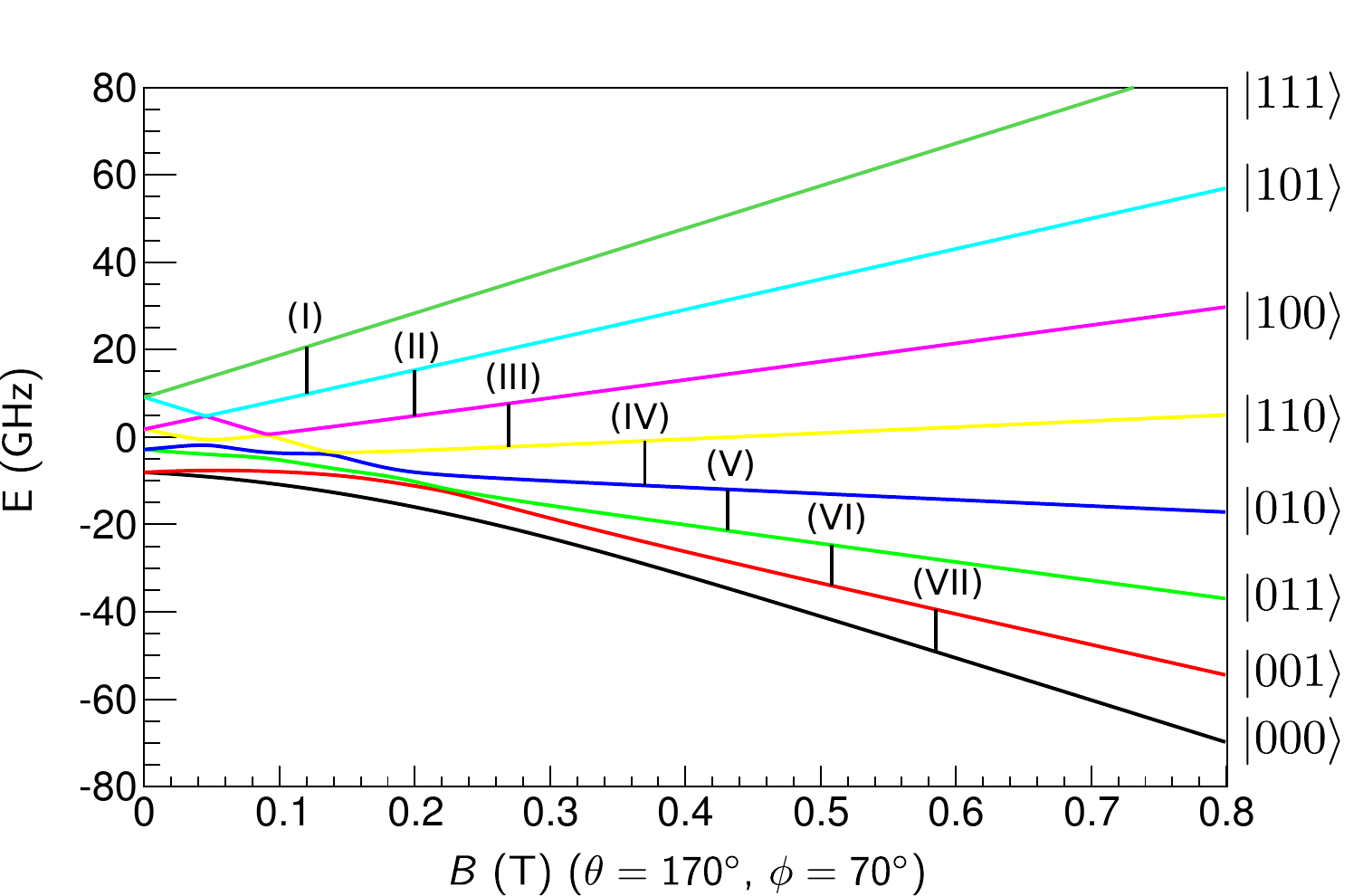}
\caption{Zeeman diagram for \ce{GdW30} and possible level assignment for an equivalent 3-qubit processor.  The vertical lines are the magnetic fields for the seven transitions measured with pulsed EPR methods (see figure \ref{fig:echo2}).}\label{fig:gdw30zeeman}
\end{figure}

In summary, we have found that this system has energy levels easily accessible with relatively low microwave frequencies.  Each transition can used to define a qubit.  Furthermore, the frequencies needed lie in the ranges of a few \si{\giga\hertz} for moderate magnetic fields.  These properties make \ce{GdW30} a very attractive candidate to couple to circuit QED systems.  Another important conclusion that may be drawn from the pulsed EPR experiments is the following.   We have found that the system has seven transitions that are individually addressable by either varying the DC magnetic field and using a constant frequency or by using different frequencies at a fixed field.  All transition frequencies lie within ranges easily accessible with standard microwave electronics (below \SI{10}{\giga\hertz}).  Although the $\textrm{T}_1$ and $\textrm{T}_2$ coherence times are not extraordinary, these times are relatively consistent for all transitions.  Also, at least for the transitions measured, these have matrix elements in line with our theoretical model.  This means it is hypothetically possible to encode 3 full qubits on a single \ce{GdW30} molecule.  The scheme would be similar to what is shown in figure \ref{fig:gdw30zeeman}.  There, we show a Zeeman diagram of the \ce{GdW30} energy levels for a given field direction and the position of the transitions if a frequency of \SI{9.7}{\giga\hertz} is used.  We have labeled each of the spin levels with a three qubit state (from $\ket{000}$ to $\ket{111}$).  Since all the transitions can be individually addressed, a rotation from the ground state into any other combination of states should be possible.  Therefore, \ce{GdW30} can be used as a three-qubit universal quantum processor.  The possibility of using this scheme, at least in the context of circuit QED, would of course be dependent on improving the coherence times for this type of molecule and on the attainment of higher coupling to microwave cavities and circuits.

\section{\ce{K12TbW30}}\label{sec:TbW30}

In this section we turn our attention to the compound \ce{TbW30}, another sample from the \ce{LnW30} series.  As mentioned in the introduction, the five-fold symmetry of the molecular structure coupled with the fact that the non-Kramer \ce{Tb^{3+}} ion has a $S_z=\pm 5$ ground state in this coordination, is expected to give this species an extraordinarily high tunnel splitting.  The ground $\ket{\pm 5}$ doublet will split into its symmetric and antisymmetric superpositions with a large tunnel energy gap $\Delta$ between them, thus leading to a situation analogous to that described by a simple two level system (equation \refeq{eq:tunnel}).  Unlike the case of \ce{Fe8} (section \ref{sec:matelem}) for which $\Delta \sim \SI{e-7}{\kelvin}$ at zero magnetic field, the tunnel splitting in this case should be large enough to survive sizable magnetic fields, therefore making the tunnel split states suitable as a quantum computing basis.  Our aim in this section is to experimentally verify the existence of this gap in our samples, previously predicted in \cite{Cardona-Serra2012}, and to determine its magnitude.  Unfortunately, the anisotropy of the sample and the value of $\Delta$ are expected to be so large that X-band EPR can not be used to determine the energy level spectrum and the full Hamiltonian parameters as in the case of \ce{GdW30}.  We will therefore use other methods to characterize the ground state and tunnel gap.

\begin{figure}[!tb]
\centering
\includegraphics[width=0.9\columnwidth]{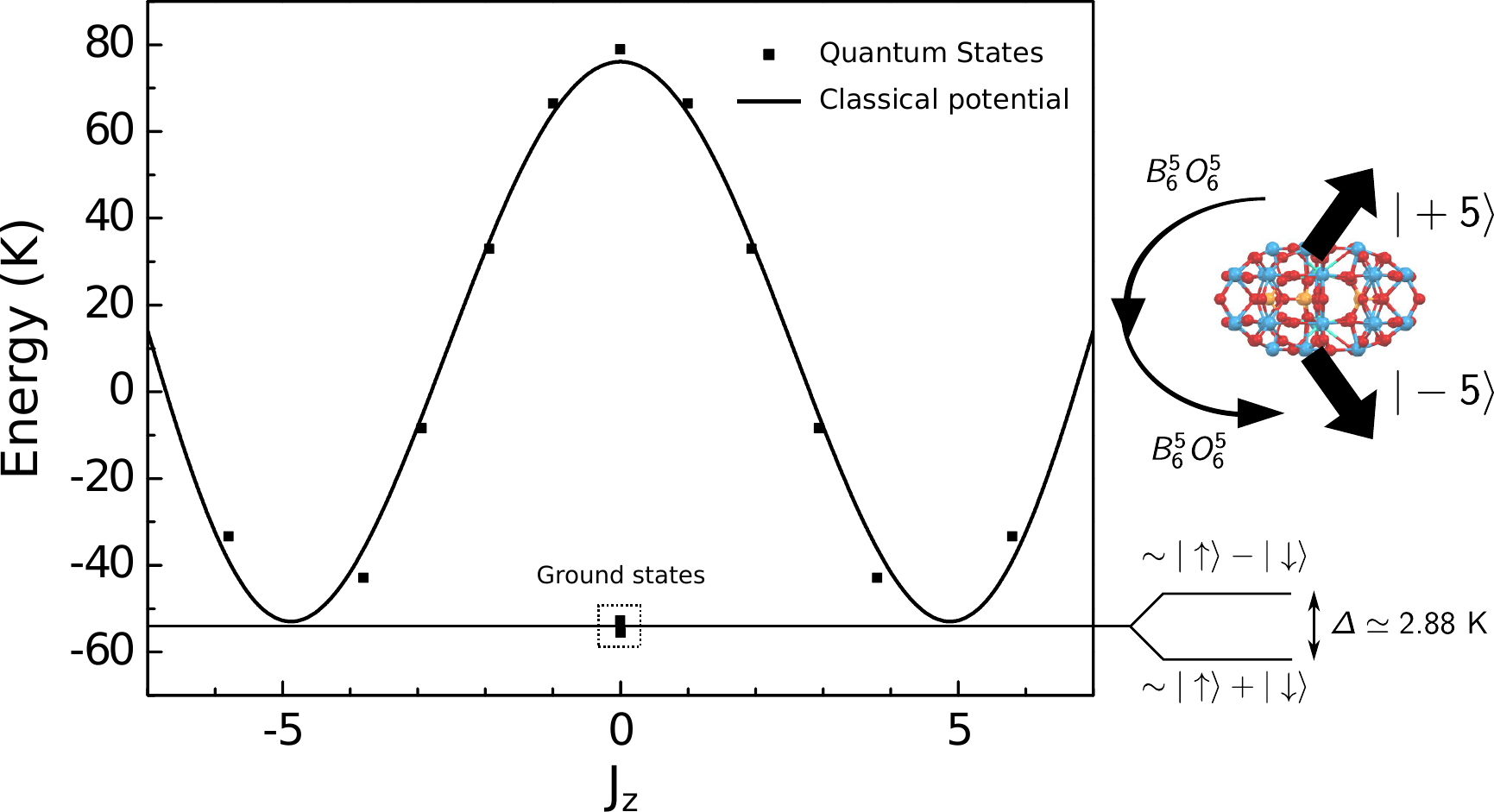}
\caption{Classical potential and quantum energy levels of the \ce{K12TbW30} molecule for the magnetic anisotropy parameters given in \cite{Cardona-Serra2012}.  The tunnel splitting of the ground state $\ket{\pm}$ doublet is schematically shown as well how the $B_6^5O_6^5$ term induces this splitting \cite{Garanin1991}.}\label{fig:potential_TbW30}
\end{figure}

The full crystal field Hamiltonian \refeq{eq:GSHamiltonian} was determined \cite{Cardona-Serra2012} from the simultaneous fit of the powder susceptibilities of \ce{K12LnW30} with $\ce{Ln^{3+}}=\ce{Tb}, \ce{Dy}, \ce{Ho}, \allowbreak \ce{Er}, \ce{Tm}, \ce{Yb}$ using the method suggested by Ishikawa and collaborators \cite{Ishikawa2003}.  The Hamiltonian contains only terms proportional to $B_2^0,B_4^0,B_6^0,B_6^5$.  A diagram showing the classical spin potential and the quantum energy levels at zero field using these parameters is shown in figure \ref{fig:potential_TbW30}.  The critical parameter that induces the tunnel splitting and also the one that is most difficult to determine by the above method is $B_6^5O_6^5$.  It contains terms proportional to $S_+^5$ (see table \ref{fig:stevens}) and is therefore able to connect the $S_z=\pm 5$ states in only two steps \cite{Garanin1991}.  Using the reported values for these parameters, the full Hamiltonian can be seen to have a single magnetic easy axis (Z) and two equal hard axes (X,Y).  Taking into consideration the symmetry of the molecule (see figure \ref{fig:crystalstructure}), we will assume that the easy axis Z lies perpendicular to the molecular plane while the two hard axes are assumed to lie within this plane.  Since, given the temperatures we will be operating at (below \SI{2}{\kelvin}), the population of the higher energy levels will be negligible (the first excited doublet lies about $\sim\SI{8}{\kelvin}$ away from the ground state doublet according to \cite{Cardona-Serra2012}), the thermal and magnetic properties will mostly depend on to the two tunnel split energy states.  In this regime it therefore seems reasonable to describe the system with an effective spin $1/2$ Hamiltonian with a zero field tunneling gap $\Delta$ and a Zeeman interaction term.  It can be shown that the g-factor matrix can only have one non-zero principal value that will correspond to the easy magnetization axis Z \cite{Griffith1963,Abragam1970,Leggett1987}.  The effective Hamiltonian then reduces to:
\begin{equation}
\mathcal{H}_{\rm eff} = \Delta S_x + g_{zz}\mu_BS_ZB_Z  \label{eq:effham}
\end{equation}
where $S=1/2$ and the $\tilde{g}$ matrix only has $g_{zz}\neq 0$.  Using low temperature specific heat and magnetic susceptibility measurements we will determine the values of $g_{zz}$ and $\Delta$ for \ce{TbW30} and use these values to obtain the transition matrix elements and level separations for different applied magnetic fields.

\subsection{Specific heat:  Determination of the tunneling gap}
\begin{figure}[!tb]
\centering
\includegraphics[width=0.9\columnwidth]{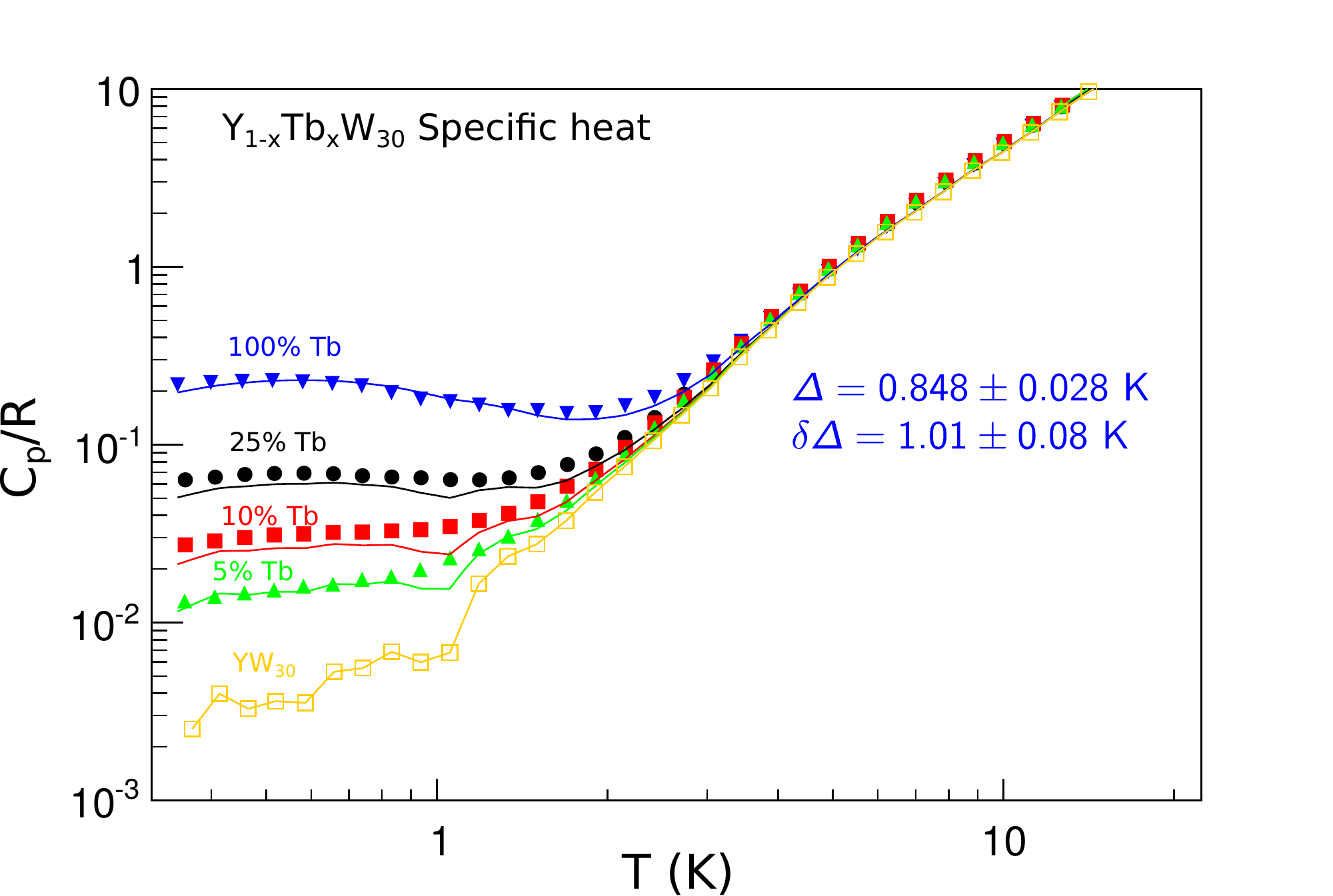}
\caption{Low temperature specific heat for \ce{Y_{1-x}Tb_xW30}.  The solid lines are fits using the Schottky anomaly formula \refeq{eq:schottky} and a (truncated) gaussian distribution of gaps $\Delta$ with $\delta\Delta$ width.}\label{fig:cpTbW30}
\end{figure}

We performed a series of low temperature specific heat measurements using the PPMS \ce{^3He} refrigerator (section \ref{sec:cp}).  Experiments were performed on successive dilutions of the sample, i.e., the \ce{Tb^{3+}} ions are replaced with non magnetic \ce{Y^{3+}} ions in different proportions.  The samples measured had 5\%, 10\%, 25\% and 100\% \ce{Tb} content.  The dilutions are used to more clearly distinguish the magnetic contribution to the specific heat from other possible contributions (lattice, dipole-dipole interactions).  Also, a pure \ce{YW30} sample was measured to get a direct measurement of the lattice contribution.  As mentioned in section \ref{sec:xrayGdW30}, the entire \ce{LnW30} series is expected to have the same crystal structure and, since most of the mass is due to the ligand shell, the lattice vibrations and associated specific heat should be the same for different central ions.  The result of all these measurements is plotted in figure \ref{fig:cpTbW30}.  A broad low temperature anomaly is clearly visible below \SI{2}{K} for all \ce{Tb} concentrations.  Therefore, the anomaly is not associated with from intermolecular interactions, but arises from the energy splitting of each individual molecule.

The magnetic contribution for a two-level system with an energy gap $\Delta$ is the well known Schottky anomaly, given by:
\begin{equation}
\frac{C_p}{R} = \left(\frac{\Delta}{T}\right)^2\frac{e^{\Delta/T}}{(1+e^{\Delta/T})^2}.\label{eq:schottky}
\end{equation}
The experimental data do not follow the simple behavior predicted by equation \refeq{eq:schottky}.  As in the case of \ce{GdW30}, we expect there to be strains on the Hamiltonian parameters for \ce{TbW30}.  In order to take them into account, we fit the low temperature specific heat with a Gaussian (truncated at 0 since $\Delta$ can not be negative) distribution of gap parameters of width $\delta\Delta$.  The magnetic contribution will then be a weighted average of the Schottky anomaly equation \refeq{eq:schottky}.  To this fit we then add the lattice contribution from the \ce{YW30} measurement and obtain the solid lines from figure \ref{fig:cpTbW30}.  We fit only the 100\% \ce{Tb} curve and scale the result by concentration $x$ to obtain the corresponding fits for the magnetically diluted samples.  The central value of the level gap is $\Delta = 0.848\pm\SI{0.028}{\kelvin}$ while the distribution width is $\delta\Delta = 1.01\pm\SI{0.08}{\kelvin}$.  The deviations for lower concentrations are small and most likely due to inaccuracies in the actual concentrations of the samples.  Also, the relatively large $\Delta$ distribution width leads us to the conclusion that, as in the case of \ce{GdW30}, the Hamiltonian parameters may present large strains.

This experiment therefore confirms that a large energy gap $\Delta$, of the order of \SI{1}{\kelvin} separates the two lowest lying energy levels of \ce{TbW30}.  This is in line with what is predicted in \cite{Cardona-Serra2012} although our measured value is somewhat lower than the value reported there, leading us to believe that there may be some differences in the actual Hamiltonian parameters for our sample and the ones derived by fitting the powder magnetic susceptibility.

\subsection{Low temperature magnetic susceptibility}

\begin{figure}[!tb]
\centering
\includegraphics[width=0.9\columnwidth]{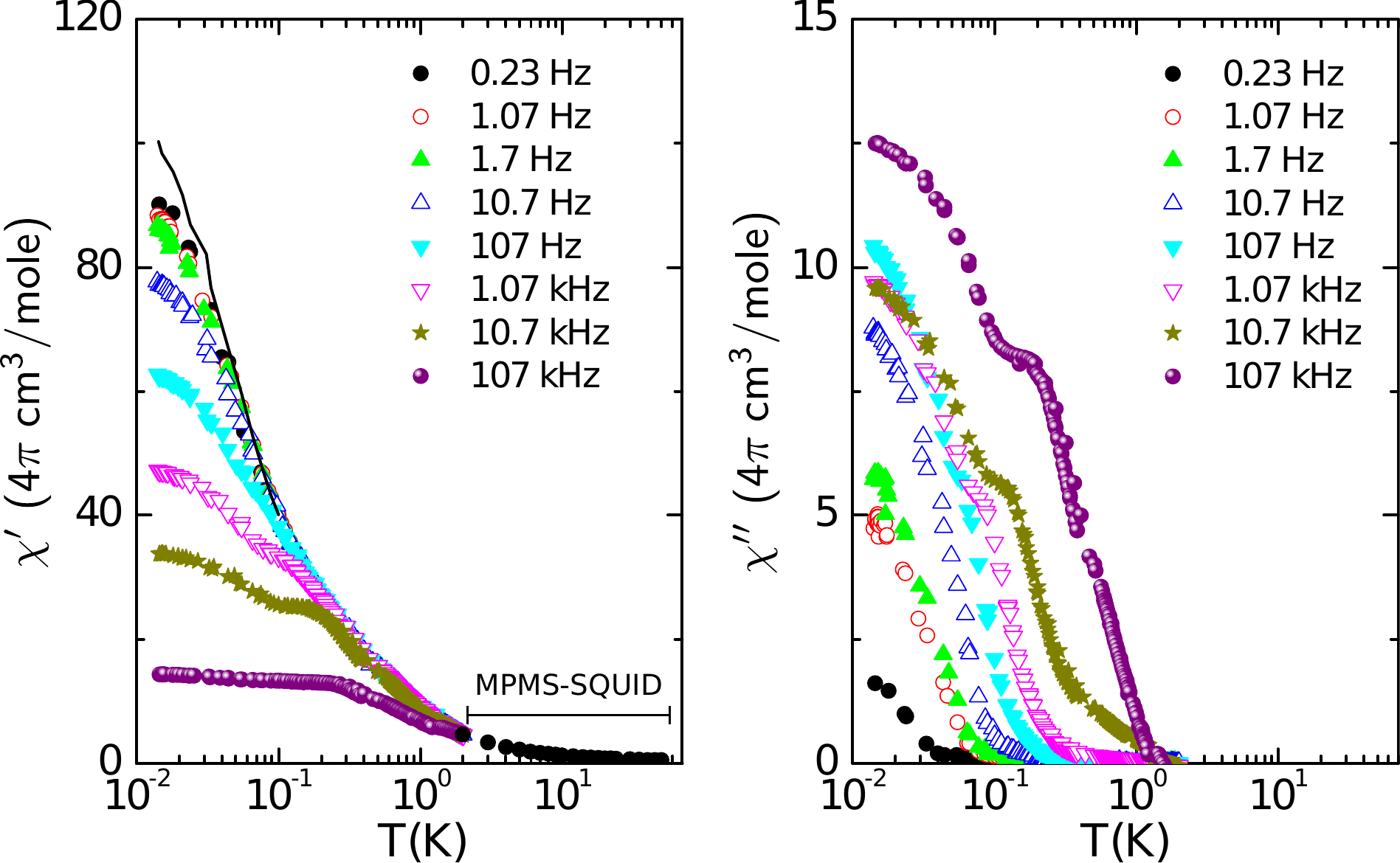}
\caption{Complex ac susceptibility measured along the [1-10] axis of \ce{TbW30} at very low temperatures.  The susceptibility is shown as a function of temperature for different excitation frequencies.  The SQUID output voltage values are normalized to match the susceptibility values obtained from a temperature dependent MPMS measurement from \SI{2}{\kelvin} to \SI{50}{\kelvin} (points marked on the left graph).  The solid black line in the real component graph represents the equilibrium value derived from fits to a double Cole-Cole law (see figure \ref{fig:dilution2}).}\label{fig:dilution1}
\end{figure}

\begin{figure}[!tb]
\centering
\includegraphics[width=0.7\columnwidth]{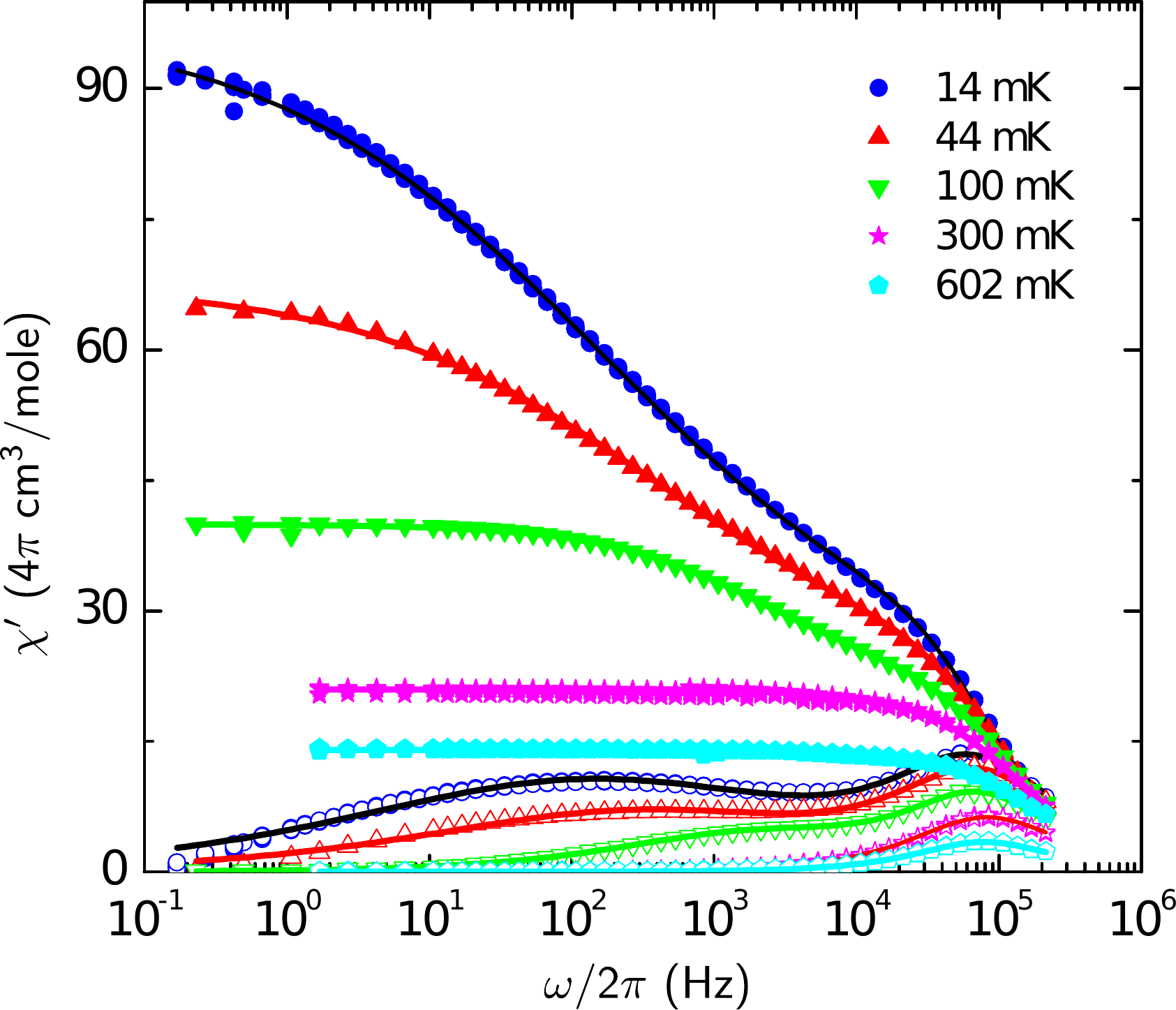}
\caption{Complex ac susceptibility for \ce{TbW30} at very low temperatures.  The susceptibility is shown as a function of excitation frequency for different temperatures.  The SQUID output voltage values are normalized to match the susceptibility values obtained from a temperature dependent MPMS measurement from \SI{2}{\kelvin} to \SI{50}{\kelvin}.  The filled markers correspond to the real component and the empty markers correspond to the imaginary component of the ac susceptibility.}\label{fig:dilution2}
\end{figure}

Very low temperature ac magnetic susceptibility measurements were performed on a single crystal of pure \ce{TbW30} by using a \si{\micro}-SQUID susceptometer \cite{Martinez-Perez2010,Bellido2013}, similar to that shown in figure \ref{fig:dpn2}, placed inside the mixing chamber of a \ce{^{3}He-^{4}He} dilution fridge \cite{Pobell2007}.  The susceptibility was measured for frequencies ranging from \SI{0.23}{\hertz} to \SI{107}{\kilo\hertz} and for temperatures ranging from \SI{14}{\milli\kelvin} to \SI{2}{\kelvin}.  A sample of the measured data points is shown in figure \ref{fig:dilution1} (as a function of temperature) and in figure \ref{fig:dilution2} (as a function of frequency).  The actual susceptibility values are obtained by matching the SQUID output voltages to the magnetization measurements from a commercial MPMS system from \SI{50}{\kelvin} down to \SI{2}{K} using the same crystal orientation as the one used on the \si{\micro}-SQUID.

\begin{figure}[!tb]
\centering
\includegraphics[width=0.7\columnwidth]{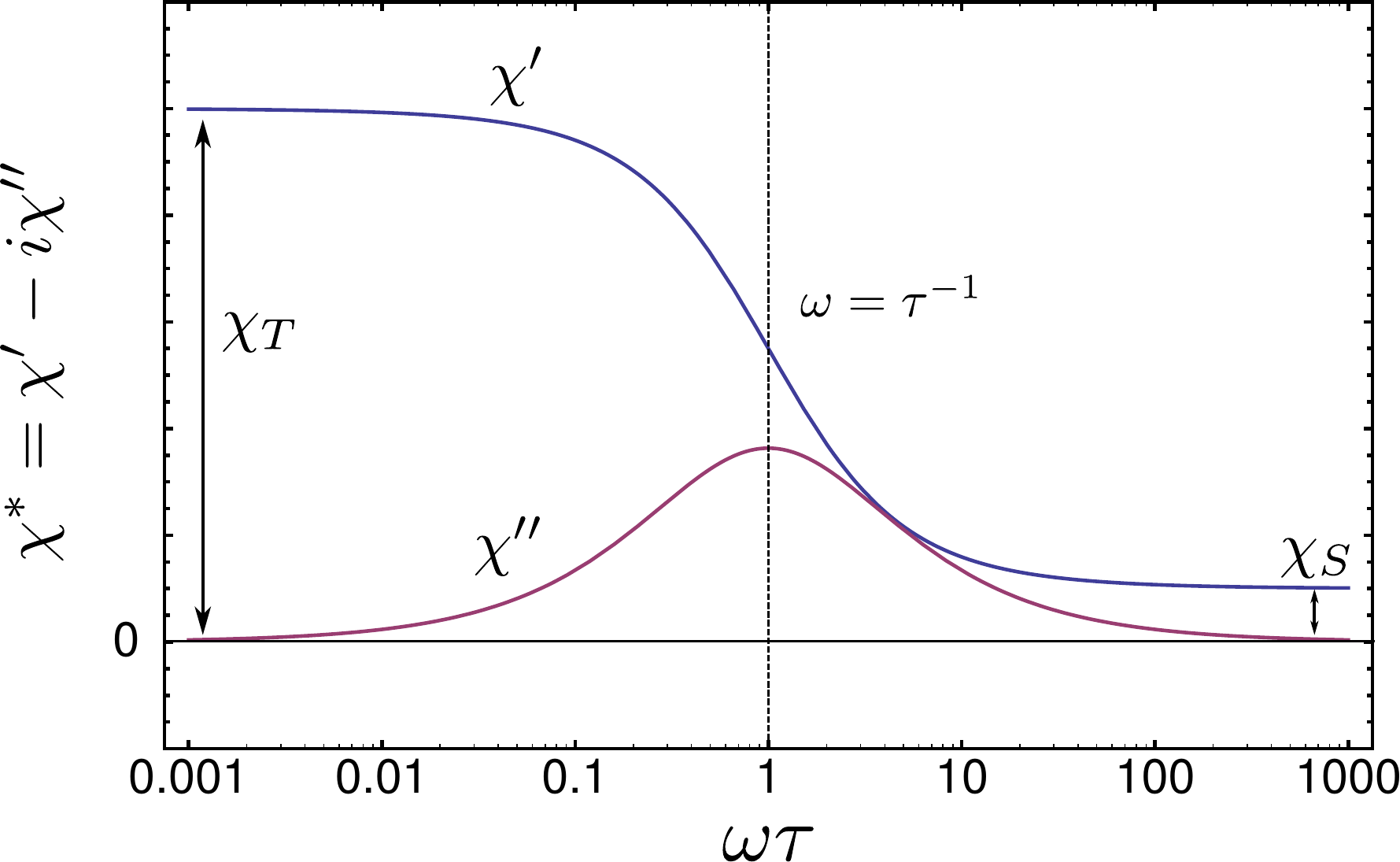}
\caption{Cole-Cole equation for the complex ac susceptibility with $\beta=0.8$ (equation \refeq{eq:cole}).}\label{fig:cole}
\end{figure}

The frequency dependence of the complex ac susceptibility, $\chi^* = \chi' - i\chi''$, can be adequately modeled by a Cole-Cole equation \cite{Cole1941} that describes the dynamics of spins having a distribution of relaxation times $\tau$:
\begin{equation}
\chi^* = \chi_S + \frac{\chi_T-\chi_S}{1+(j\omega\tau)^\beta}\label{eq:cole}
\end{equation}
This model reduces to the Debye equation when $\beta=1$.  $\chi_S$ is the adiabatic or high frequency limit of the susceptibility while $\chi_T$ is the equilibrium low frequency limit.  $\chi_S$ is usually small since at very high frequencies the spins do not have time to relax and do not contribute to the susceptibility while the value of $\chi_T$ typically follows the Curie law where $\chi_T\propto T^{-1}$.  The typical behavior of a system following the Cole-Cole law is shown in figure \ref{fig:cole} where we see that the in phase component has a step when the frequency $\omega\sim 1/\tau$ and the out of phase component has a peak at the same frequency.  Observing our measurements however, it seems evident that there are two characteristic relaxation times and that, therefore, it will be necessary to use a double Cole-Cole equation to fit the frequency dependence at each temperature:
\begin{equation}
\chi^* = \chi_S^{(1)}+ \chi_S^{(2)} + \frac{\chi_T^{(1)}-\chi_S^{(1)}}{1+(j\omega\tau_1)^{\beta_1}} +  \frac{\chi_T^{(2)}-\chi_S^{(2)}}{1+(j\omega\tau_2)^{\beta_2}}
\end{equation}

\begin{figure}[!tb]
\centering
\includegraphics[width=0.6\columnwidth]{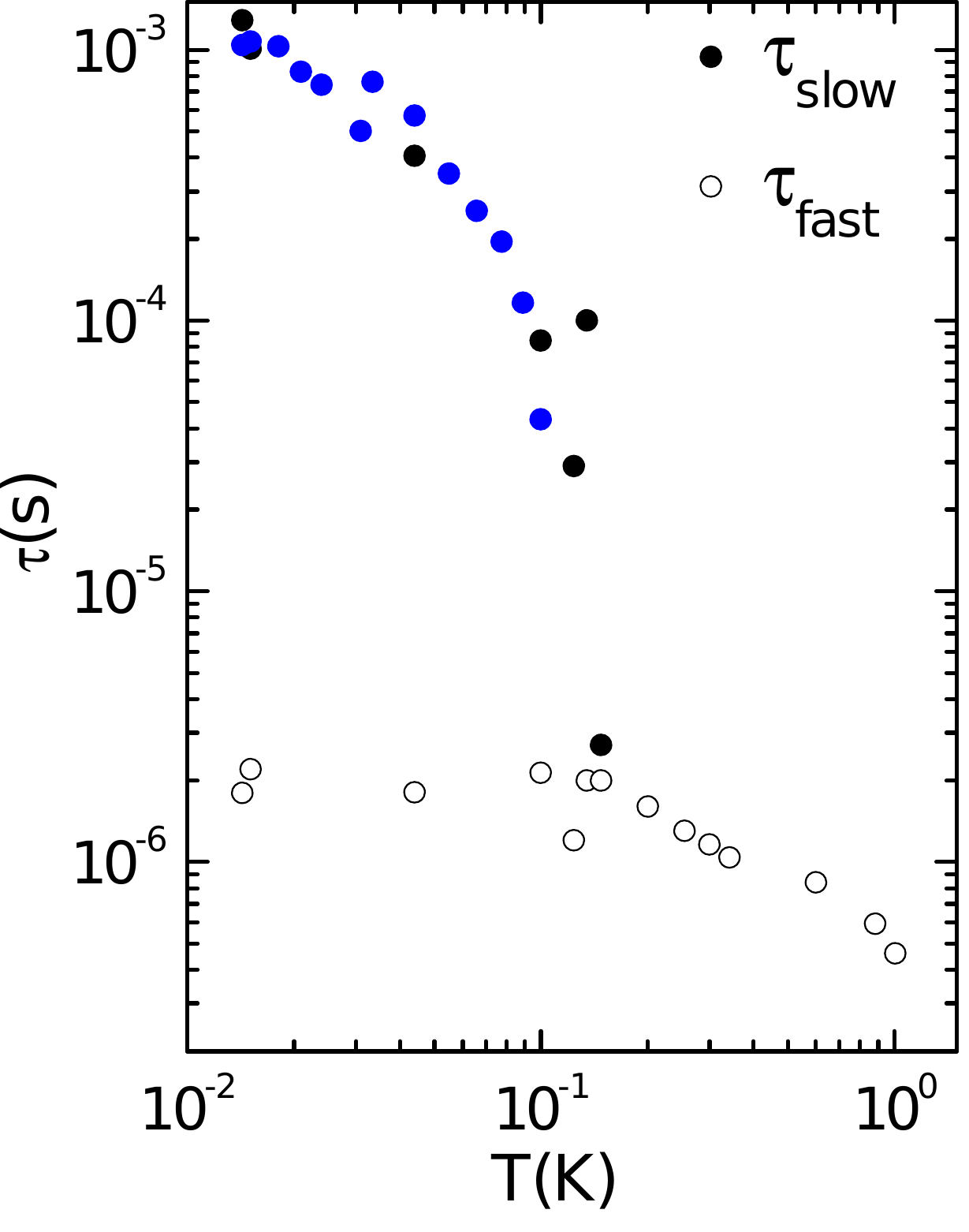}
\caption{Fitted values of the spin relaxation times as a function of temperature for \ce{TbW30}.}\label{fig:dilution3b}
\end{figure}

Fits to this double Cole-Cole equation are shown in figure \ref{fig:dilution2} (solid lines).  The values of $\chi_S$ in the fits have both been taken equal to 0.  From these fits we can obtain the temperature dependence of the two relaxation times ($\tau_1\equiv\tau_{\rm fast}$ and $\tau_2\equiv\tau_{\rm slow}$), shown in figure \ref{fig:dilution3b}, as well as the equilibrium susceptibility (solid line in figure \ref{fig:dilution1}).  At the higher temperatures, the equilibrium susceptibility value can usually be read directly off the graph, but for the lower temperatures the fit is necessary to correctly extrapolate the susceptibility to $\omega\rightarrow 0$.

From the Cole-Cole fits we see that there is a well defined $\tau_{\rm fast}$ that remains constant at lower temperatures and falls off slightly at higher temperatures.  It is of the order of \si{\micro\second} and, we presume, it corresponds to the usual $\textrm{T}_1$ electronic spin relaxation which, at very low temperatures, corresponds to the lifetime of the excited spin state of the doublet.  There is however a second, much slower, relaxation process.  This process becomes observable only at very low temperatures.  Its characteristic time, $\tau_{\rm slow}$, decreases rapidly with increasing temperature and becomes of the order of \si{\milli\second} at $T\rightarrow \SI{14}{\milli\kelvin}$.  The origin of this relaxation is unclear but we speculate that it may be due to the relaxation of nuclear spins through the hyperfine interaction.  \ce{Tb} has a relatively large hyperfine interaction and, although it has been unnecessary to consider up till now, it may be the case that it plays an important role in the relaxation dynamics at low temperatures.  In any case, more work is necessary to fully understand this effect.

\begin{figure}[!tb]
\centering
\includegraphics[width=0.7\columnwidth]{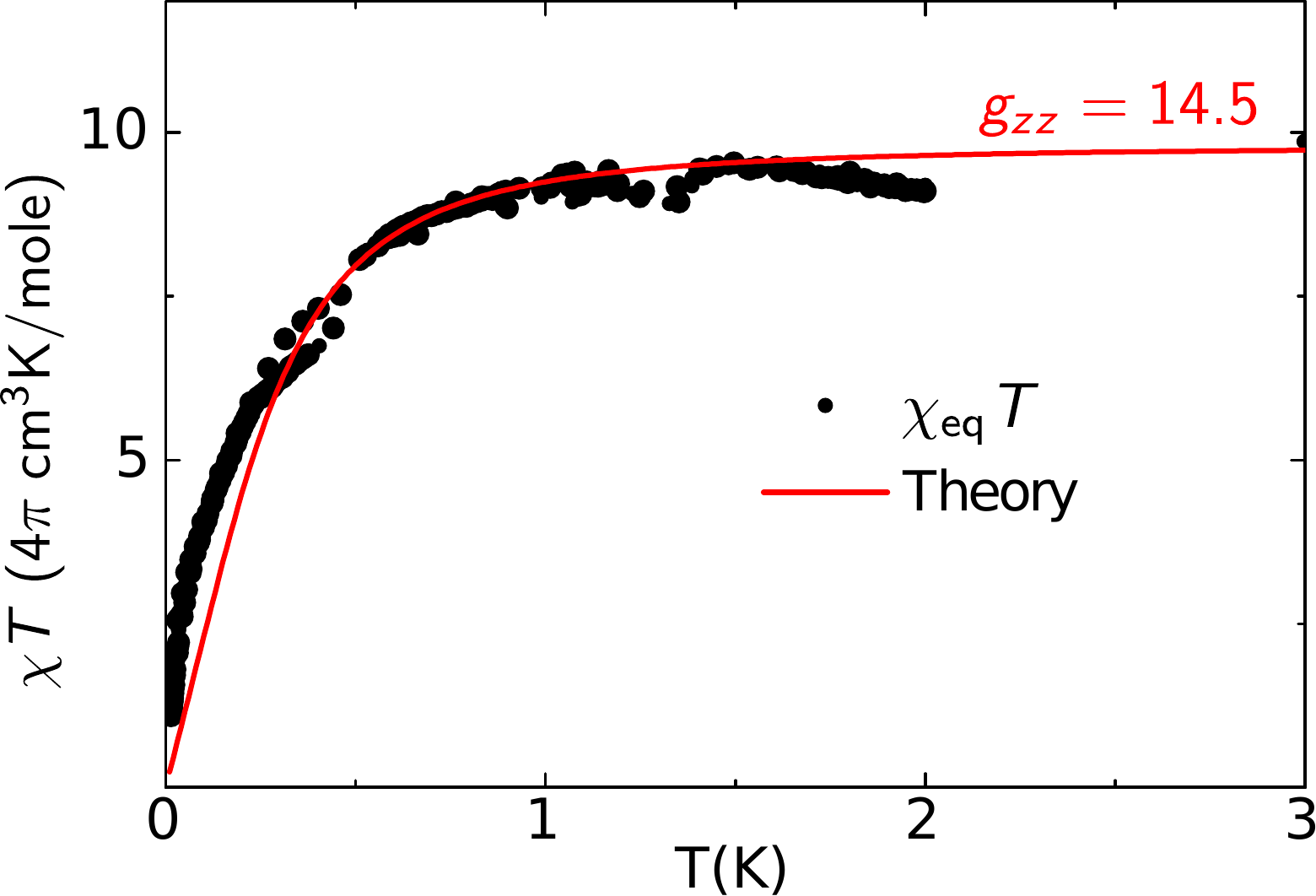}
\caption{Equilibrium $\chi T$ for \ce{TbW30} at very low temperatures.  The solid curve represents the fit using equation \refeq{eq:effham} assuming that the susceptibility is measured along an axis separated \ang{45} from the molecular Z axis.  The gap $\Delta$ in \refeq{eq:effham} was fixed at the measured value from the specific heat measurement (Figure \ref{fig:cpTbW30}, $\Delta = \SI{0.848}{\kelvin}$).  We see Curie law behavior for high temperature but a slower that $T^{-1}$ decay of $\chi$ for low temperatures with $\chi T \rightarrow 0$ as $T\rightarrow 0$.}\label{fig:dilution3}
\end{figure}

We can now obtain the value of $g_{zz}$ for our effective model \ref{eq:effham} by fitting it to the measured equilibrium susceptibility.   Assuming that the crystal structure is similar to that of \ce{GdW30}, that the molecular Z axis is perpendicular to the molecular plane and considering that the ac field was applied along the [1-10] direction, we can deduce that the susceptibility is measured along an axis that forms an approximately \ang{45} angle with the easy Z axis.  Calculating the equilibrium susceptibility in this direction, we obtain the $\chi T$ curve with a fitted value of $g_{zz} = 14.5$ shown in figure \ref{fig:dilution3}.

In order to discuss these results, we find it illustrative to review here the physical origin of the different terms that contribute to the magnetic susceptibility in a generic spin system.   By definition, the expectation value of the magnetization along direction $\hat{b}$ is given by:
\begin{equation}
M_{\hat{b}} \equiv \frac{1}{Z}\sum_n \bra{n}\mu_{\hat{b}}\ket{n}e^{-\beta E_n}, \label{eq:mdef}
\end{equation}
where $\beta = (k_BT)^{-1}$, $\mu_{\hat{b}} = \mu_B\vec{S}\tilde{g}\hat{b}$ is the magnetic moment operator, $\hat{b}$ is the unitary vector in the direction of the magnetic field, and the sum is over the Hamiltonian eigenstates $\ket{n}$ with energies $E_n$.  The susceptibility is defined as the derivative of the magnetization with respect to the magnitude of the field applied in the $\hat{b}$ direction:
\begin{equation}
\chi_{\hat{b}} \equiv \frac{\partial M_{\hat{b}}}{\partial B}.
\end{equation}
Assuming that the field enters in the spin Hamiltonian through a Zeeman term $-\vec{\mu}\vec{B}$, and applying first order perturbation theory we have that the changes in the energy and wave functions are:
\begin{eqnarray}
E_n & \simeq & E_n^{(0)} - \bra{n^{(0)}}\mu_{\hat{b}}\ket{n^{(0)}}\delta B, \\
\ket{n} & \simeq & \ket{n^{(0)}} - \delta B\sum_{n' \neq n} \frac{\bra{n'^{(0)}}\mu_{\hat{b}}\ket{n^{(0)}}}{E_n^{(0)}-E_{n'}^{(0)}}\ket{n^{(0)}},
\end{eqnarray}
where $\delta B$ is a perturbation in the field and the $(0)$ superscript denotes the unperturbed energy levels and states.  With this in mind, it is now straightforward  to compute $\chi_{\hat{b}}$ by explicitly differentiating \refeq{eq:mdef}:
\begin{equation}
\chi_{\hat{b}} = \frac{1}{k_BT} \left[\langle\mu_{\hat{b}}^2\rangle - \langle\mu_{\hat{b}}\rangle^2\right] + \frac{2}{Z^2}\sum_n \sum_{n'\neq n} \frac{|\bra{n}\mu_{\hat{b}}\ket{n'}|^2}{E_{n'}-E_{n}}e^{-\beta E_n}
\end{equation}
where all the $\langle ... \rangle$ mean \emph{thermal} averages.\footnote{$\langle\mu_{\hat{b}}^2\rangle =  \frac{1}{Z}\sum_n \bra{n}\mu_{\hat{b}}\ket{n}^2e^{-\beta E_n}\left(\neq \frac{1}{Z}\sum_n \bra{n}\mu_{\hat{b}}^2\ket{n}e^{-\beta E_n}\right)$}  The first two terms are proportional to $T^{-1}$ and give rise to the Curie law behavior.  They originate from the variations of the energy levels due to changes in the field (proportional to $\partial E_n/\partial B$).  The last term, on the other hand, originates from changes in the eigenstates of the Hamiltonian (proportional to $\partial\ket{n}/\partial B$) and corresponds to the Van Vleck susceptibility.  In a sense, the Curie terms represent the contribution from classical thermal fluctuations while the Van Vleck term represents the changes in the quantum wave function.

\begin{figure}[!tb]
\centering
\includegraphics[width=0.7\columnwidth]{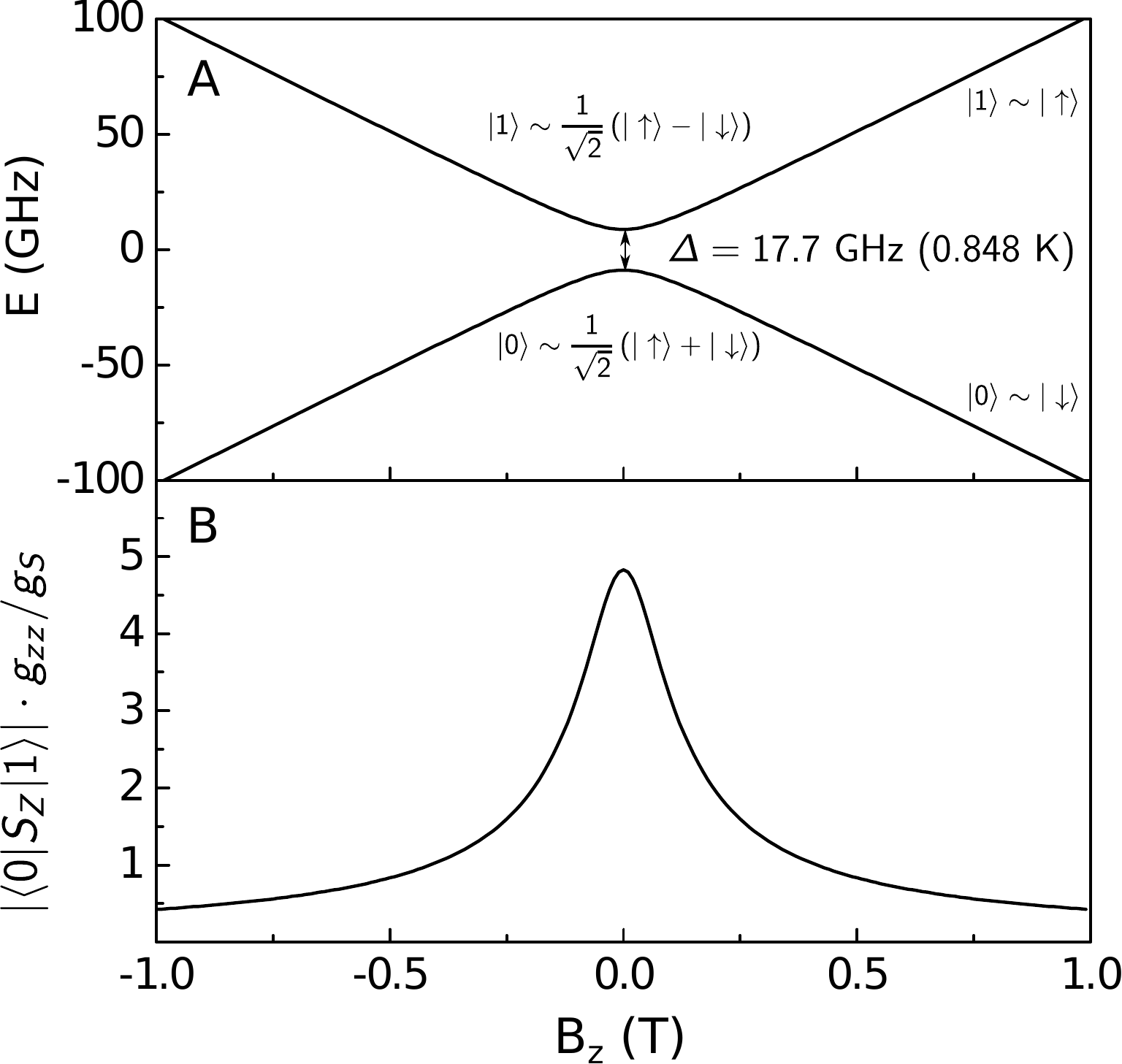}\\[3mm]
\caption{Effective Hamiltonian for the ground state double of \ce{TbW30}.  Graph A shows the energy levels as a function on an applied magnetic field in the Z direction.  Graph B shows the matrix elements of the $S_Z$ matrix normalized by the factor $g_{zz}/g_S$ where is the g-factor from equation \refeq{eq:effham} and $g_S=1.5$ is the g-factor of \ce{Tb}.  This is done so that the effective model matrix elements can be directly compared to matrix elements calculated in full crystal field model.}\label{fig:levels_TbW30}
\end{figure}

In a system with an effective Hamiltonian as in equation \refeq{eq:effham} we find that the terms proportional to $T^{-1}$ are both zero at zero field and temperatures where the population of the excited state is negligible (temperatures smaller than $\Delta$).  This can be seen by simply solving the Hamiltonian and calculating $\langle \mu_{\hat{n}}\rangle$ and $\langle \mu_{\hat{n}}^2\rangle$ or, more intuitively, by looking at the shape of the energy curves in figure \ref{fig:levels_TbW30} whose derivative vanishes at zero field.  This effect directly arises from the large quantum tunnel splitting of this system and suppresses the $T^{-1}$ terms for $k_BT\lesssim \Delta$ leaving the much weaker temperature dependence that characterizes the Van Vleck term.  This agrees with the measured behavior shown in figure \ref{fig:dilution3} where at high temperatures we see a constant value for $\chi T$ corresponding to a Curie law, while at low temperatures the classical contributions are suppressed and $\chi T \rightarrow 0$.  There is good agreement with the effective model predictions and the relatively small deviations are likely due to the presence of the aforementioned strains in the Hamiltonian parameters.

These results serve to further confirm the presence of large tunnel gap and to reaffirm that the \ce{TbW30} system is an interesting spin qubit candidate with the computational basis defined as the two tunnel split states. Its large tunnel gap allows tuning fields to be applied without destroying the matrix elements of photon induced transitions connecting these states (figure \ref{fig:levels_TbW30}B).  Beyond its possible applications as a quantum bit, this sample provides the rather unique opportunity of studying a pure quantum two level spin system, where the ground state and magnetic response is determined fully by tunnel effects.  The ground state of \ce{TbW30} is a robust superposition of up and down spin states with no net magnetization, unlike most other spin systems where the ground state ends up being either $\ket{\pm S}$ as a result of the action of perturbations such as dipole-dipole or hyperfine interactions.

\section{Coupling SIMs to CPW resonators}\label{sec:simsCPW}

After studying two model SIM systems in detail, we now wish to evaluate their coupling to quantum circuits.  The calculation of the coupling of these systems to superconducting coplanar waveguide resonators is completely analogous to the calculation presented for generic SMMs in section \ref{sec:coupCPWG_SMM}.  In fact, the same field simulations and integrals apply and the only change that has to be made is to input the correct operating frequencies and matrix elements in equations \refeq{eq:interaction}, \refeq{eq:coupling} and \refeq{eq:coupling_explicit}.  These parameters are readily derived from the Hamiltonians deduced in the previous sections allowing us to directly compare the couplings and operating conditions to those of polynuclear SMMs studied in section \ref{sec:coupCPWG_SMM}.

\begin{figure}[!tb]
\centering
\includegraphics[width=0.95\columnwidth]{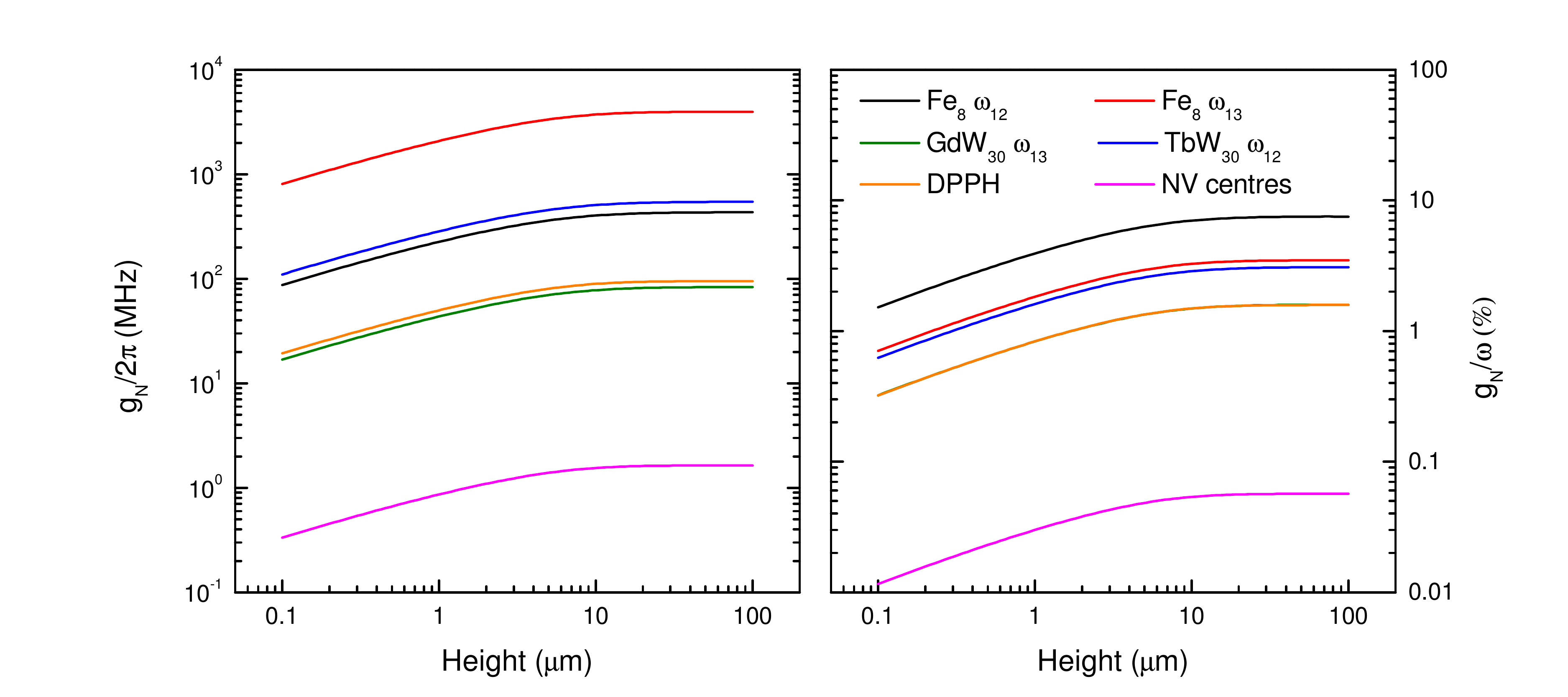}\\[3mm]
\begin{small}
\begin{tabular}{|c|c|}
\hline
& Spin density (spins/\SI{}{\centi\meter\cubed)}\\
\hline
\ce{Fe8}  & \SI{5.11e20}{} \\
\hline
DPPH & \SI{2.14e21}{} \\
\hline
NV centers & \SI{1.1e18}{} \\
\hline
\ce{GdW30}, \ce{TbW30} & \num{3.08e20} \\
\hline
\end{tabular}\\[2mm]
\begin{tabular}{|c|c|c|}
\hline
Fe$_8\;\omega_{12}$ & Fe$_8\;\omega_{13}$ & \ce{GdW30} $\omega_{13}$ \\
\hline
5.75 GHz & 114.6 GHz & 5.23 GHz \\
($B_Y = 2.325$ T) & & \\
\hline
\hline
\ce{TbW30} $\omega_{12}$ & DPPH & NV centres \\
\hline
17.7 GHz & 6 GHz & 2.88 GHz \\
 & ($B_Z = 0.43$ T) & \\
 \hline
\end{tabular}
\end{small}
\caption{Coupling of $\SI{40}{\micro\meter}\times \SI{40}{\micro\meter}\times {\rm height}$ SIM, SMM and diamond crystals to a CPW resonator as a function of crystal thickness. On the left we show the total coupling strength and on the right we show the coupling strength normalized by the resonator frequency.  For each sample, $\omega_{ij}$ denotes the transition used ($\omega_{12}$ denotes tunnel split energy levels while $\omega_{13}$ refers to zero-field split energy levels) and the operating frequencies are detailed in the table above.  The spin densities for each sample are also shown \cite{Wernsdorfer1999,Kiers1976,Amsuss2011,Kim1999}. The \ce{GdW30} and DPPH lines overlap in the right hand graph.}\label{fig:vsH_sims}
\end{figure}

Figure \ref{fig:vsH_sims} shows the calculated collective coupling of a resonator to a crystal with a base size of $\SI{40}{\micro\meter}\times \SI{40}{\micro\meter}$ placed at the center of the resonator for different crystal heights.  We see again that the coupling saturates when the height of the crystal is larger than the gap sizes (\SI{14}{\micro\meter} center line and \SI{7}{\micro\meter} gaps in this case).  In all cases we see that the coupling is much higher that the values found for NV centres in diamond and, in many cases, can amount to a sizable fraction of the operating frequency.

\begin{figure}[!tb]
\centering
\includegraphics[width=0.7\columnwidth]{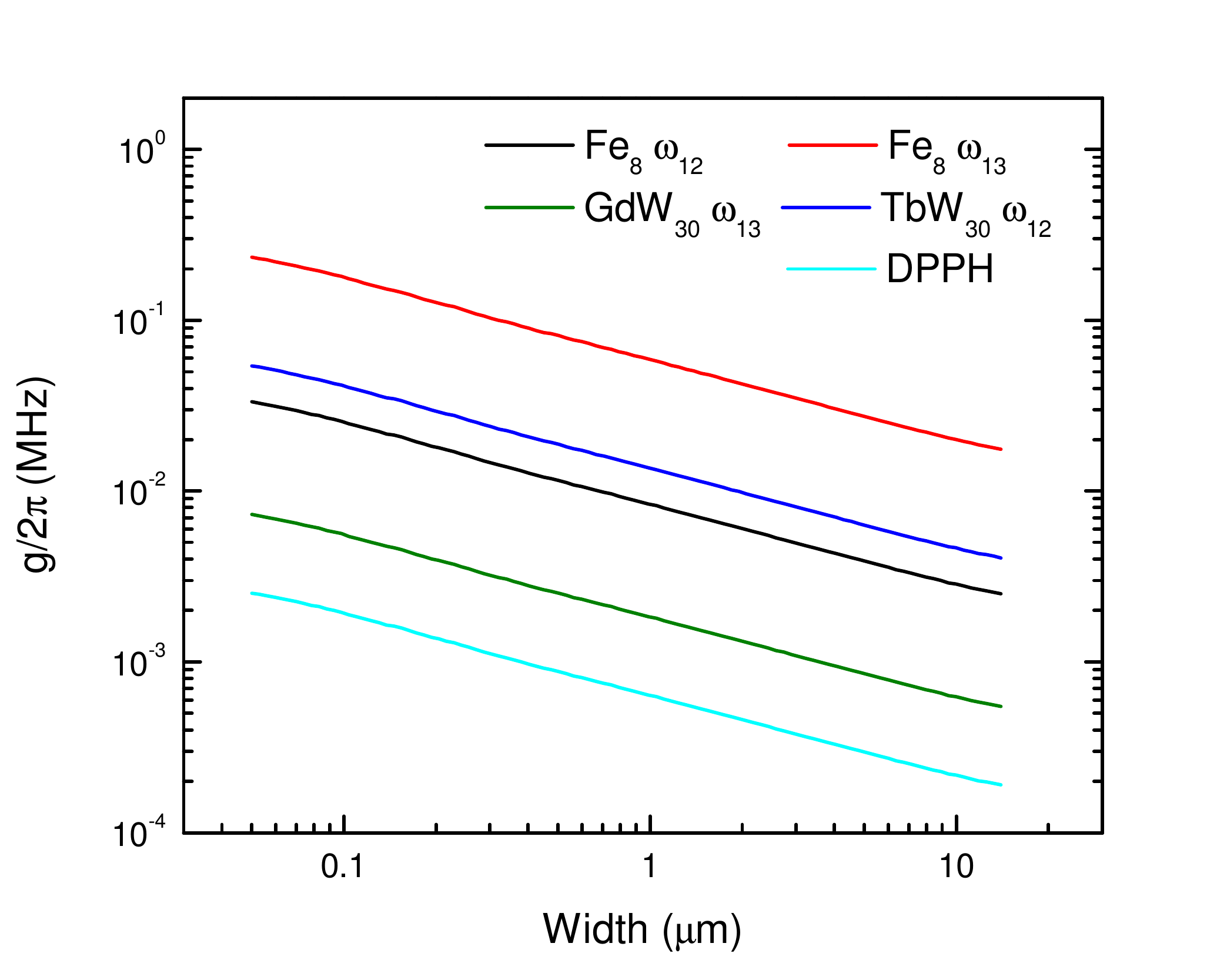}
\caption{Coupling of a single spin to a CPW resonator as a function of centre line width. The spin is located at the point of maximum field on the surface of the centre line.  For each sample, $\omega_{ij}$ denotes the transition used and the operating frequencies are detailed in figure \ref{fig:vsH_sims}.}
\label{fig:vsW_sims}
\end{figure}

In the case of \ce{GdW30}, the values obtained are similar to those for DPPH.  This may seem somewhat disappointing but we must keep in mind that DPPH requires the application of a relatively large magnetic field (almost \SI{0.5}{\tesla}) in order to operate at a frequency similar to \ce{GdW30} which can be a technical issue when working with superconducting circuits (sensitive to background fields).  \ce{GdW30} can operate in this regime without the need of high tuning fields which generally is favorable for the resonator performance.

For \ce{TbW30} we find that the coupling values are close to the values obtained when using the tunnel split energy levels of \ce{Fe8}.  As in that case, the coupling values are very high (hundreds of \si{\mega\hertz}) but the operating conditions for \ce{TbW30} are much more favorable.  As discussed in section \ref{sec:coupCPWG_SMM}, tuning the tunnel splitting $\Delta$ of \ce{Fe8} to the operating regime a strong and very precisely aligned magnetic field, while \ce{TbW30} does not require a tuning field to obtain the same performance.  Even higher values of the coupling are in principle achievable for \ce{Fe8} when using its zero field split energy levels.  However, this requires working at extremely high frequencies of the order of \SI{200}{\giga\hertz}.  In comparison, \ce{TbW30} offers an attractive option because of the much more manageable operating frequency although its coupling is somewhat smaller.

Figure \ref{fig:vsW_sims} shows the coupling calculated for a single spin placed at the maximum rf field location as a function of the center line width.  As discussed in section \ref{sec:coupCPWG_SMM}, we observe an order of magnitude enhancement in the coupling when narrowing the center line down from \SI{10}{\micro\meter} to \SI{100}{nm}.  The relative coupling intensities of the different samples are in line with what has been found for crystal samples and the same considerations about the SIMs favorable operating conditions apply.  It seems therefore possible that single or small ensembles of SIMs could potentially present strong coupling to this type of circuits.


\section{Coupling SIMs to flux qubits}
\label{sec:fluxqubits}

The low anisotropy of \ce{GdW30} leads to a relatively low transition frequency between the ground state and the first excited state.  This property makes this sample an interesting candidate for coupling to another type of quantum circuit, the superconducting flux qubit \cite{Clarke2008}.  Flux qubits are technically limited to operating at up to a few \si{\giga\hertz} while only very small external magnetic fields may be applied (fields such that only a few flux quanta flow through the SQUID loop).  \ce{GdW30} has an operating frequency at zero field of about 5-\SI{6}{\giga\hertz} and it would be well suited to interact coherently with a flux qubit at zero magnetic field.  In this section, we briefly describe the superconducting flux qubit and its basic Hamiltonian and calculate the coupling strengths of \ce{GdW30} as compared to NV-centres.

\subsection{Device description and parameters}

Flux qubits (FQs) are superconducting loops interrupted by, almost always, three Josephson junctions \cite{Orlando1999}.
When half of a flux quanta passes through the loop, the two lowest eigenstates of the qubit Hamiltonian are symmetric and antisymmetric superpositions of counter propagating persistent currents.  Those states define the qubit states, and will be denoted as $\left\{ |\circlearrowleft\rangle,|\circlearrowright\rangle\right\}$. By changing the flux, the qubit is biased to one of those currents. Furthermore, any influence of higher excited levels can be safely neglected for standard qubit parameters. Therefore, for our purposes the FQ can be modeled with the, now familiar to us, two-level system:
\begin{equation}
\label{FQ}
H_{FQ}
= \frac{\epsilon}{2} \sigma_z
+
\frac{\Delta}{2} \sigma_x
\end{equation}
with $\Delta$ the qubit gap, which lies in the GHz regime, and $\epsilon = 2 I_p (\phi_{\rm ext} - \Phi_0/2)$ the  bias term associated with the external flux $\phi_{\rm ext}$ and $I_p$ the persistent current in the loop.  Here, we have chosen the {\it physical} basis where the eigenstates are the clockwise and anticlockwise supercurrents: $\left\{ |\circlearrowleft\rangle,|\circlearrowright\rangle\right\} $.

FQs provide a  platform for hybrid structures because their ability to couple to magnetic moments through the field induced by the supercurrents in the loop. Previous studies have focused on considering the coupling to NV-centres \cite{Marcos2010}, with recent experimental realizations showing promising results \cite{Zhu2010}. Some applications of these hybrid structures to quantum information processing have been recently pointed out \cite{Hummer2012, Xiang2013, Lu2013}.

\subsection{Coherent Coupling to SIMs}
\label{fluxqubitsgeometry}

\begin{figure}
\centering
\includegraphics[width=0.7\textwidth]{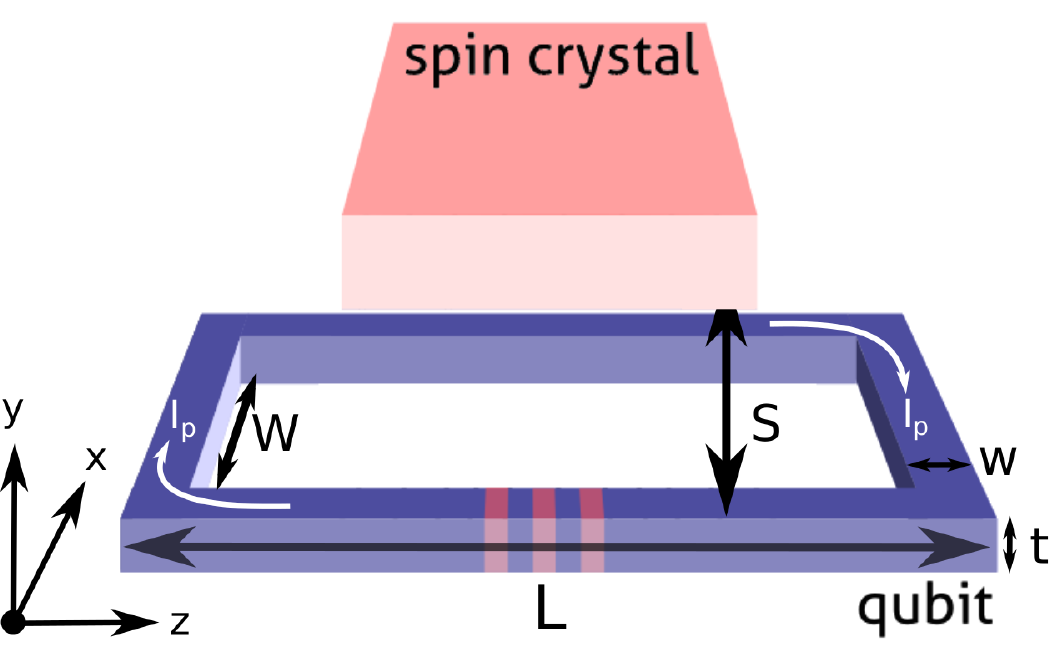}\\[2mm]
\begin{tabular}{|c|c|}
\hline
\multicolumn{2}{|c|}{Flux-Qubit dimensions} \\
\hline
L & 43 $\mu$m \\
\hline
W & 3.25 $\mu$m \\
\hline
w & 1.2 $\mu$m \\
\hline
t & 150 nm \\
\hline
\end{tabular}
\begin{tabular}{|c|c|}
\hline
\multicolumn{2}{|c|}{Sample dimensions} \\
\hline
Height & 0 to 40 $\mu$m \\
\hline
Width & 20 $\mu$m \\
\hline
Depth & 20 $\mu$m \\
\hline
S & 0 to 15 $\mu$m \\
\hline
\end{tabular}
\caption{Basic geometry and dimensions of the flux-qubit and the magnetic samples used for the calculations described in this work.  The flux-qubit dimensions resemble those from \cite{Zhu2011}}
\label{FQ-SMM}
\end{figure}

Within the state of the art of both qubit geometry and parameters, the coupling to single spins is too weak to overcome the losses.
Therefore, in this section we will discuss the coupling between a flux qubit and a spin ensemble. A schematic representation of a possible layout is depicted in figure \ref{FQ-SMM}.

The Hamiltonian describing the combined spin ensemble and the flux qubit can be shown to correspond to a Rabi like model as in the case of the CPW resonator (equation \refeq{eq:8}):
\begin{equation}
\label{rabi}
H = \frac{\Delta}{2} \sigma_x
+
\omega b^\dagger b
+
g \sigma_z (b^\dagger + b)
\end{equation}
where, for simplicity, we have chosen to be at the degeneracy point $\epsilon = 0$ (see equation \refeq{FQ}).
The spin collective modes, $b, \; b^\dagger$,  were defined in equation (\ref{eq:collop}) and the coupling in (\ref{eq:coupling}).
In this case, the magnetic field, $b(\vec r_j)$ is generated by the circulating currents in the qubit (see equation \refeq{eq:coupling}).
The current operator can be written as,
\begin{equation}
I=\sum_{m,n=\circlearrowleft,\circlearrowright}|n\rangle\langle n|I|m\rangle\langle m|=I_{p}|\circlearrowleft\rangle\langle\circlearrowleft|-I_{p}|\circlearrowright\rangle\langle\circlearrowright|=I_{p}\sigma_{z}\, ,
\end{equation}
that justifies the coupling through $\sigma_z$. The magnetic field strength, $b(\vec r)$ entering in the formula for $g$, equation (\ref{eq:coupling}), corresponds to the field generated in a loop with a circulating current $I_{\rm p}$.

\begin{figure}[!t]
\centering
\includegraphics[width=\textwidth]{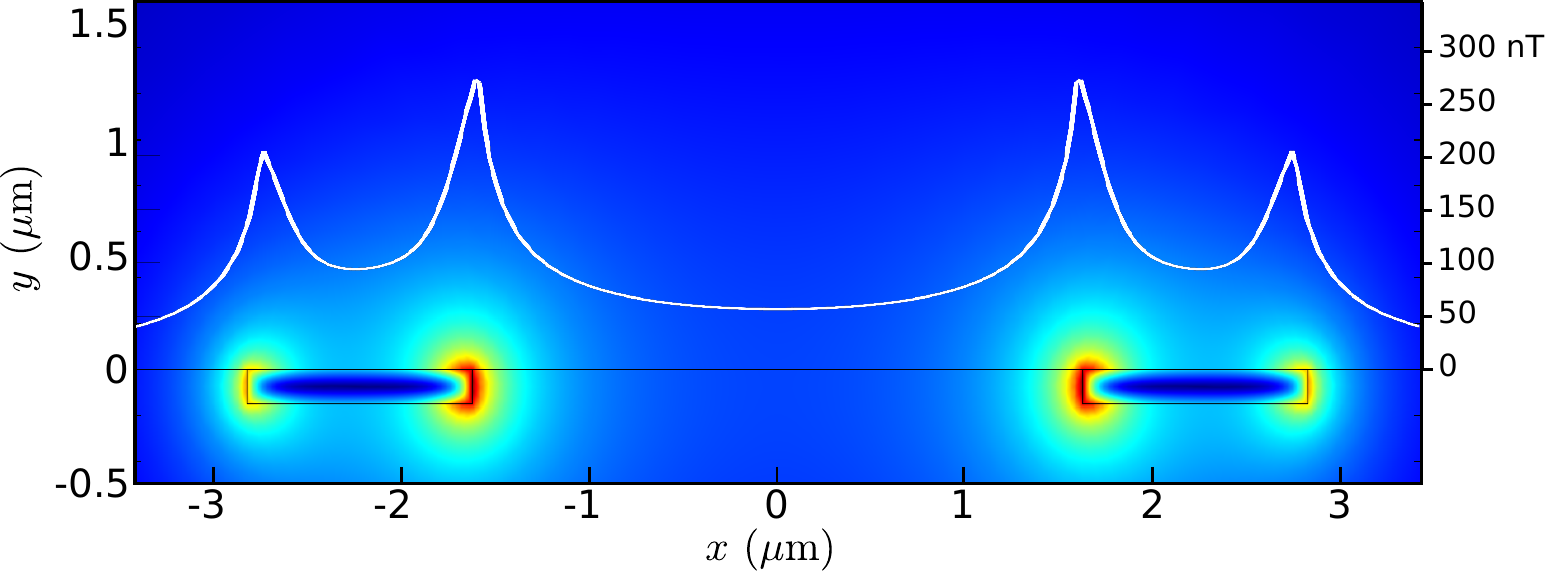}
\caption{Simulated field distribution on a flux qubit cross-section at the centre of the device.  The white profile is the field value calculated at a constant distance right at the surface of the superconducting regions.}
\label{Bprofile_FQ}
\end{figure}

To estimate this coupling, we again perform numerical simulations in Comsol Multiphysics (see section \ref{sec:coupCPWG_SMM}) assuming a superconducting loop with current $I_{\rm p}\simeq \SI{300}{\nano\ampere}$ \cite{Zhu2011}.  As in the previous section, we simulate the field at a cross section at the centre of the flux qubit and choose a crystal size of about half the length of the flux qubit (see figure \ref{FQ-SMM}) to avoid edge effects. We also use the skin effect, as before, to simulate the superconducting current distribution (see equation (\ref{eq:skineffect})). An example of the field distributions found is shown in figure \ref{Bprofile_FQ}.
We complement our numerical studies with an analytical approach.  In order to get a tractable and closed formula for the magnetic field generated, we approximate the qubit by two parallel counter currents, $I_p$.  This yields for the magnetic field:
\begin{equation}
\mathbf{b} =\frac{\mu_{0}\, I_{p}}{2\,\pi}\left(\frac{1}{\left(x+\nicefrac{w}{2}\right)^{2}+y^{2}}\left(\begin{array}{c}
-y\\
x+\nicefrac{w}{2} \\
0
\end{array}\right)\right.
\left.-\frac{1}{\left(x-\nicefrac{w}{2}\right)^{2}+y^{2}}\left(\begin{array}{c}
-y\\
x-\nicefrac{w}{2}\\
0
\end{array}\right)\right)\,.
\label{analytical}
\end{equation}

\begin{figure}[!t]
\centering
\includegraphics[width=\textwidth]{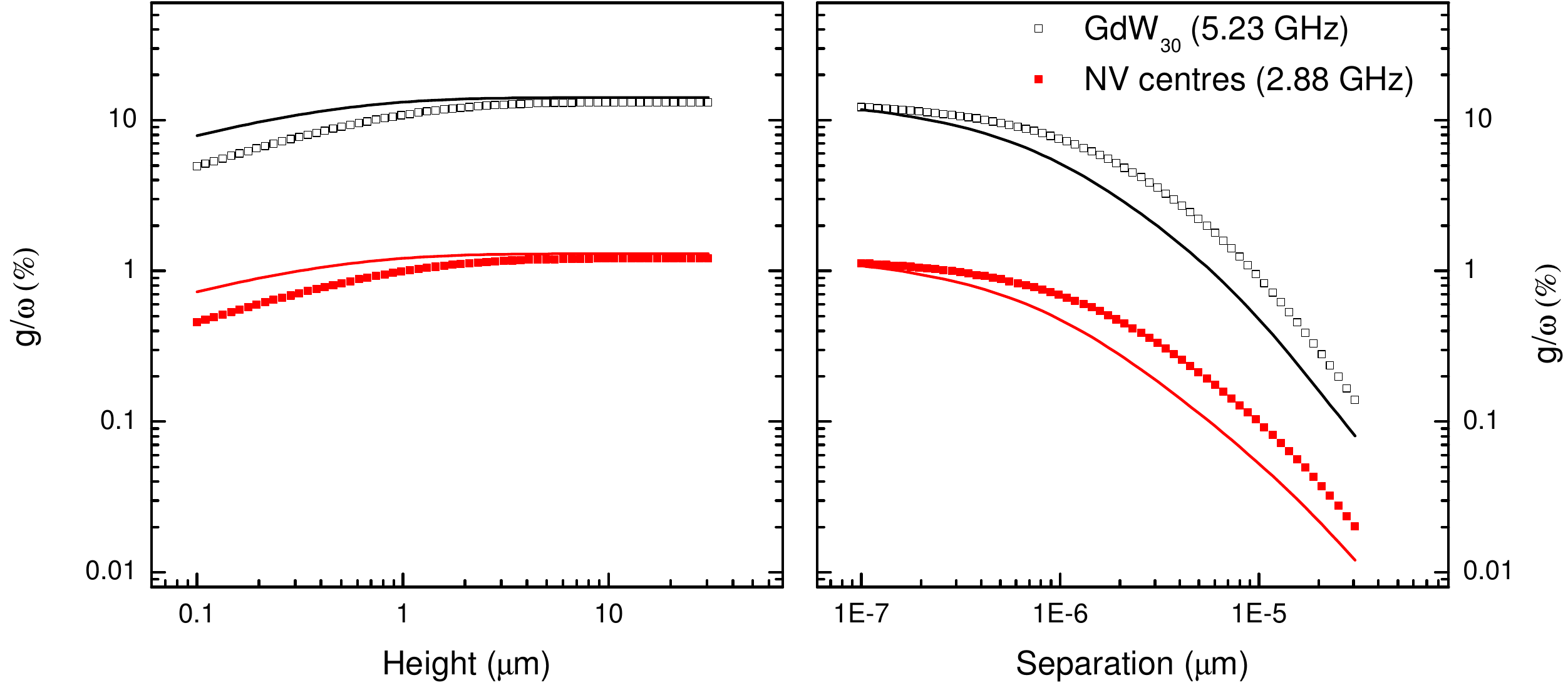}
\caption{Coupling of SMM crystals to a flux qubit as a function of crystal height (left) and as a function of the vertical (i.e. along $y$) separation between the crystal and the device, normalized by the qubit frequency $\omega$. Solid lines represent the analytical estimations that follow from equation (\ref{analytical}) while the dots are from the numerical simulation.}
\label{gvsh_FQ}
\end{figure}

Both numerical and analytical estimates for $g$ are shown in figure \ref{gvsh_FQ}. We plot our results as a function of crystal height and as a function of the vertical separation between the crystal and the flux qubit (S in figure \ref{FQ-SMM}). As with the resonator, we observe a saturation of $g$ beyond a certain height.  This can also be understood by looking at the dependence on separation S. The field saturation occurs between $1$ and $10 \; \mu$m, i.e. when the magnetic field generated by the flux qubit becomes negligible. We also observe that our simple analytical estimation is reasonably close to the numerical results.

In the case of FQs we have compared two species: GdW$_{30}$ and NV-centres, with zero field level splittings of \num{5.23} and \SI{2.8}{\giga\hertz}, respectively. These transition frequencies lie in the range of  available qubit tunneling gaps. The achievement of strong coupling between NV-centres and a flux qubit has been recently reported in \cite{Zhu2010}. From the present results we conclude that spin ensembles tend to couple more strongly to flux qubits than to resonators, see figures \ref{fig:vsH_sims} and \ref{gvsh_FQ} (for \ce{GdW30}, CPW resonators and flux qubits have $g/\omega\sim 1\%$ and $g/\omega\sim 10\%$ respectively).  Also, as in the case of the resonators, the coupling to SMMs is stronger than that to NV-centres. This point is interesting since the qubit-ensemble coupling can reach up to $10 \% $ of the qubit natural frequency. For such a coupling strength, the qubit-spin ensemble system enters the so-called ultrastrong coupling limit \cite{Niemczyk2010,Forn-Diaz2010}. This means that the Rotating Wave Approximation no longer holds (see section \ref{sec:CQED}) and the terms neglected for this approximation have to be included to understand the physics. In the usual case of weaker coupling, one can rotate the qubit basis $\sigma_x \to \sigma_z$ and write the interaction within the Rotating Wave Approximation $g (\sigma^+ a + \sigma^- a^\dagger)$, with $\sigma_\pm = \sigma_x \pm i \sigma_y$. The latter approximation allows a perturbative treatment. Therefore, SMMs are candidates to observe analogues of light-matter interaction beyond perturbative treatments.  We finish by noting that the flux qubit parameters used here were taken from the experimental paper \cite{Zhu2010}.  Further optimization of the parameters and of the flux qubit shape could yield even stronger couplings.

\section{Conclusions}\label{sec:conclSIMs}

In the field of quantum computation, Single Ion Magnets provide many interesting possibilities since they can allow a rational design of their spin Hamiltonian.  The provide a good way to fulfill the two criteria outlined in chapter \ref{chap:Theo1} regarding the desirable qualities of spin systems as qubits:  a relatively low anisotropy or a strong tunnel splitting.  The fact the molecular structure can be tuned and that the ion chosen can be any lanthanide, gives a broad range of possible Hamiltonian parameters.  In this chapter, we have studied two examples in detail each one fulfilling one of the two desired qubit qualities.

Because of its low magnetic anisotropy, \ce{GdW30} is found to have all its energy levels within a \SI{1}{\kelvin} energy range, meaning that the energy any of the spin transitions lie  within a comfortable range for their coupling to quantum circuits.  The Hamiltonian parameters have been found to depend on temperature and to have considerable strains in their specific values probably due to the large molecule size and to the relative fragility of the molecular crystals.  Through angle dependent EPR and susceptibility experiments, the magnetic hard axis has been found to be approximately perpendicular to the molecular plane defined by the flattened sphere shape of the molecule, while the easy axes are found to be close to this plane.  We have also found that all the possible spin transitions can be excited independently while the spin $\textrm{T}_2$ coherence time is found to be of the order of \SI{300}{\nano\second} for all of them.  This not only makes \ce{GdW30} interesting as a qubit, but also could allow for up to three qubits to be encoded on a single molecule potentially making it a universal three qubit processor.

The case of \ce{TbW30} has been found to have a very large tunnel splitting of the order of \SI{1}{\kelvin} thanks to the characteristics of the \ce{Tb^{3+}} ion and the five-fold symmetry of the \ce{TbW30} structure.  Contrary to the case \ce{Fe8} studied in chapter \ref{chap:Theo1}, this makes the transition matrix elements between these two levels resistant to applied magnetic fields which retain high values for high tuning fields.  The existence of the the tunnel gap has been confirmed with low temperature specific heat measurements and an equivalent model for the two tunnel split energy levels has been deduced.  The model correctly describes the low temperature susceptibility data that further confirm the existence of the large tunnel gap.  Beyond the quantum computing applications, \ce{TbW30} presents interesting physics in its own right.  At very low temperatures ($T\ll \Delta/k_B$), its larger tunneling gap, as compared to both dipolar and hyperfine interactions, makes it a model realization of a simple quantum two level system.  This system gives a unique opportunity to study quantum effects on the magnetic properties.  Under these conditions the Van Vleck contributions to the zero field susceptibility dominate over the classical thermal fluctuations (Curie law) making it purely quantum in nature.  Further study of this system could also provide insight into a new kind of spin liquids, induced by zero point fluctuations of each spin and not, as is the case for other realizations, by the topology of spin-spin interactions \cite{Wills2002,Robert2006}.

Finally, we compare the coupling of SIMs to coplanar waveguide resonators to the couplings obtained for other SMM samples in chapter \ref{chap:Theo1}.  In both in crystal form and as individual molecules, the couplings for SIMs are found to be in line with those found for typical SMM systems but with the advantage of having much more convenient and technologically accessible operating conditions.  These conditions allow systems such as flux qubits to also be interesting candidates for coupling to spin crystals.  The operating frequencies lie well within the range of current technologies while the necessary tuning fields can be kept within the necessary limits for the adequate operation of superconducting circuits.  With these results it seems clear that strong coupling to this type of systems should be achievable in the near future.

\bibliographystyle{h-physrev3}
\bibliography{mybiblio}

\chapter{Superconducting Coplanar Waveguide Resonators and Constrictions}\label{chap:CPWG}
\chaptermark{Superconducting CPW resonators and Constrictions}

\section{Introduction}
In chapter \ref{chap:Theo1} we introduced spin systems and single molecule magnets as possible candidates for quantum bits and discussed the problem of coupling single spins to quantum circuits.  Although strong coupling of quantum circuits to spin ensembles should be readily achievable, the coupling to single spins presents a bigger challenge.  We discussed that the coupling could be increased by either improving the spin transition matrix elements or by increasing the magnetic field generated by the quantum circuit.  We have explored the former approach in chapter \ref{chap:SIMs}.  In this chapter, we discuss the latter approach and explore the possibility of narrowing down the wires in a coplanar waveguide (CPW) resonator to concentrate the current and therefore getting enhanced field strengths in the vicinity of this constriction.  As detailed in chapter \ref{chap:Theo1}, the coupling to individual molecules can be greatly enhanced by making the center conductor width $w$ of the order of nm.  In some cases, the coupling for a single molecule can be as high as $g\sim\SI{1}{\mega\hertz}$. In section \ref{sec:basicCPWG} we present the basic concepts involved in the design of CPW resonators.  Section \ref{sec:testCPWG} has details on the fabrication and testing of our own resonators while section \ref{sec:constrictionsCPWG} explores the effects of introducing nanoconstrictions into these systems.  Finally, section \ref{sec:conclusionsCPWG} summarizes the conclusions.

\section{Coplanar Waveguide Resonators}\label{sec:basicCPWG}

\subsection{Basic CPW properties}

A coplanar waveguide consists of a dielectric substrate with conductors on the upper surface (figure \ref{fig:CPWGcross}).  The conductors form a center strip of width $w$ surrounded by two narrow gaps of width $s$ and two ground planes on either side which are assumed to be infinite.  As transmission lines, they carry electromagnetic radiation in current and voltage waves while the magnetic and electric fields are concentrated in the gaps  between the conductors.  The basic parameters describing any transmission line are its capacitance and inductance per unit length ($C',L'$), its characteristic impedance ($Z_0$), and the wave velocity ($v$).  These parameters are related by \cite{Pozar2011}:
\begin{equation}
Z_0=\sqrt{\frac{L'}{C'}}\quad v=\frac{1}{\sqrt{L'C'}} = \frac{c}{\sqrt{\epsilon_\textrm{eff}}} \\
\end{equation}
where $\epsilon_\textrm{eff}$ is defined as the effective relative dielectric constant of the structure and is used when there are different dielectric domains in the structure.  The parameters $L'$ and $C'$ depend on the electric and magnetic material properties and on the waveguide geometry.  For CPW, using quasi-static conformal mapping techniques \cite{Simons2004}, we can arrive at expressions for the transmission line parameters which in the case of a single layer infinitely thick substrate surrounded by air and for thin metal layers ($t\ll s$) reduce to:
\begin{eqnarray}
&\epsilon_\textrm{eff} = \frac{1+\epsilon_r}{2} \quad C'=2\epsilon_0(\epsilon_r+1)\frac{K(k_0)}{K(k'_0)} \quad L'=\frac{\mu_0}{4}\frac{K(k'_0)}{K(k_0)} & \nonumber\\
&Z_0=\frac{c\mu_0}{\sqrt{8(\epsilon_r+1)}}\frac{K(k_0')}{K(k_0)}\simeq \frac{30\pi}{\sqrt{(\epsilon_r+1)/2}}\frac{K(k_0')}{K(k_0)}\;\textrm{(Ohm)} \quad v=\frac{c}{\sqrt{(\epsilon_r+1)/2}}& \label{eq:Z0}
\end{eqnarray}
where $K(x)$ denotes the complete elliptic integral of the first kind, $\epsilon_r$ is the substrate relative electrical permittivity and:
\begin{equation}
k_0 = \frac{w}{w+2s} \qquad k'_0=\sqrt{1-k_0^2}.
\end{equation}
Observing these equations, it is clear that all transmission parameters depend on the geometry only through $k_0$.  Therefore the absolute value of the gaps and center line width can in principle be made arbitrarily small as long as this ratio is kept constant.  For our fabrication method, it is convenient to take these values of the order of \SI{10}{\micro\meter} since it is a comfortable size for photolithography techniques.

\subsection{Distributed circuit model}
\begin{figure}[htb]
\centering
\includegraphics[width=0.7\textwidth]{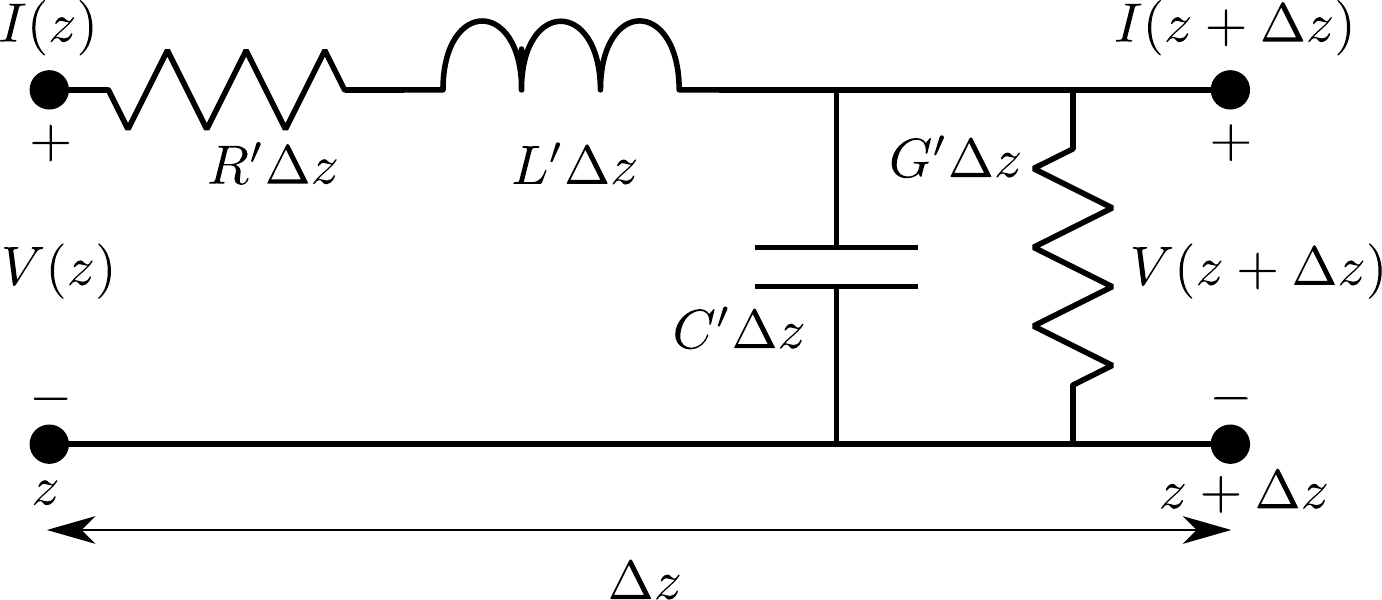}
\caption{Distributed circuit model of a general transmission line}
\label{fig:distrcircuit}
\end{figure}
From the point of view of electrical circuit theory, a transmission line can be described using voltage and intensity waves propagating according to a distributed circuit model \cite{Pozar2011}.  In this model, a waveguide is modeled as a series impedance and shunt admittance per unit length to which Kirchoff's circuit laws are then applied (figure \ref{fig:distrcircuit}).  The values of these impedances and admittances can be derived using Maxwell's equations and the geometric and electric properties of the waveguide and are different for each type of transmission line.  This analysis results in the well known \emph{Telegrapher's equations} \cite{Pozar2011}:
\begin{eqnarray}
\frac{dV}{dz}(z) &=& -Z'I(z) = -(R'+j\omega L')I(z) \\
\frac{dI}{dz}(z)&=& -Y'V(z) = -(G'+j\omega C')V(z)
\end{eqnarray}
Where $L'$ and $C'$ were defined in the previous section as the inductance and capacitance per unit length and we have also introduced $R'$ and $G'$ as the series resistance and shunt admittance per unit length (see figure \ref{fig:distrcircuit}).
This pair of coupled differential equations can be readily solved.  The most general solution has the form of forward and backward traveling current and voltage waves:
\begin{eqnarray}
V(z) &=& V_+e^{-j\beta z} + V_- e^{j\beta z} \\
I(z) &=& \frac{1}{Z_0}(V_+e^{-j\beta z} - V_- e^{j\beta z})
\end{eqnarray}
Where
\begin{eqnarray}
\beta &=& -j\sqrt{(R'+j\omega L')(G'+j\omega C')} \\
Z_0 &=& \sqrt{\frac{R'+j\omega L'}{G'+j\omega C'}}
\end{eqnarray}
In the case of a lossless line, $R' = 0$ and $G' = 0$, so the previous equations reduce to:
\begin{eqnarray}
\beta &=& \omega\sqrt{L'C'} \\
Z_0 &=& \sqrt{\frac{L'}{C'}}
\end{eqnarray}
The wave impedance $Z$ and refection response $\Gamma$ of a waveguide at position $z$ are defined by
\begin{equation}
Z(z)=\frac{V(z)}{I(z)} = Z_0\frac{V_+(z)+V_-(z)}{V_+(z)-V_-(z))},\quad \Gamma(z)=\frac{V_-(z)}{V_+(z)}
\end{equation}
It can easily be seen that these two properties are related by
\begin{equation}
Z(z) = Z_0\frac{1+\Gamma(z)}{1-\Gamma(z)},\quad \Gamma(z)=\frac{Z(z)-Z_0}{Z(z)+Z_0}
\end{equation}
Using these definitions and the solutions to the Telegrapher's equations, we can find expressions for Z and $\Gamma$ at different points ($l=z_2-z_1$) along the waveguide, as well as the voltage at the second point in terms of the voltage at the first point:
\begin{eqnarray}
Z_1&=&Z_0\frac{Z_2+jZ_0\tan(\beta l)}{Z_0+jZ_2\tan(\beta l)} \nonumber\\
\Gamma_1 & = &\Gamma_2e^{2j\beta l}  \label{eq:transfer}\\
V_2 &=& V_1e^{-j\beta l}\frac{1+\Gamma_2}{1+\Gamma_1} \nonumber
\end{eqnarray}
If we want to include small losses ($R'\ll \omega L'$ and $Y'\ll\omega C'$), it is sufficient to consider a lossless line and to replace in $\beta$ in (\ref{eq:transfer}) with a complex propagation constant $\beta_c = \beta - j\alpha$ where, in a first order approximation, $\alpha$ is given by:
\begin{equation}
\alpha = \frac{1}{2}\left(\frac{R'}{Z_0}+Y'Z_0\right) = \alpha_c + \alpha_d
\end{equation}
and $\alpha_{c}$ and $\alpha_{d}$ represent the resistivity and dielectric losses respectively.  In broad terms, the resistivity losses are proportional to the surface resistance $R_s=\sqrt{\frac{\omega\mu_0}{2\sigma}}\propto \sqrt{\omega}$ and the dielectric losses are proportional to $\beta\tan\delta\propto \omega$ where $\tan\delta$ is the loss tangent of the insulator.  Expressions of $\alpha_{c,d}$ as a function of the material properties and geometry for CPW can be found in the literature \cite{Simons2004,Pozar2011,Booth1999}.  However, if we are working with superconducting resonators, radiative losses are expected to be small \cite{Browne1987} and resistive losses should be negligible at temperatures well below the critical temperature for the superconductor \cite{Frunzio2005} leaving dielectric losses as the main contribution.  According to \cite{Simons2004}, dielectric losses in a CPW with an infinite dielectric substrate can be expressed as:
\begin{equation}
\alpha_d = \frac{\omega}{2c}\frac{\epsilon_r}{\sqrt{(\epsilon_r+1)/2}}\tan{\delta}
\end{equation}

\begin{figure}[htb]
\centering
\includegraphics[width=\textwidth]{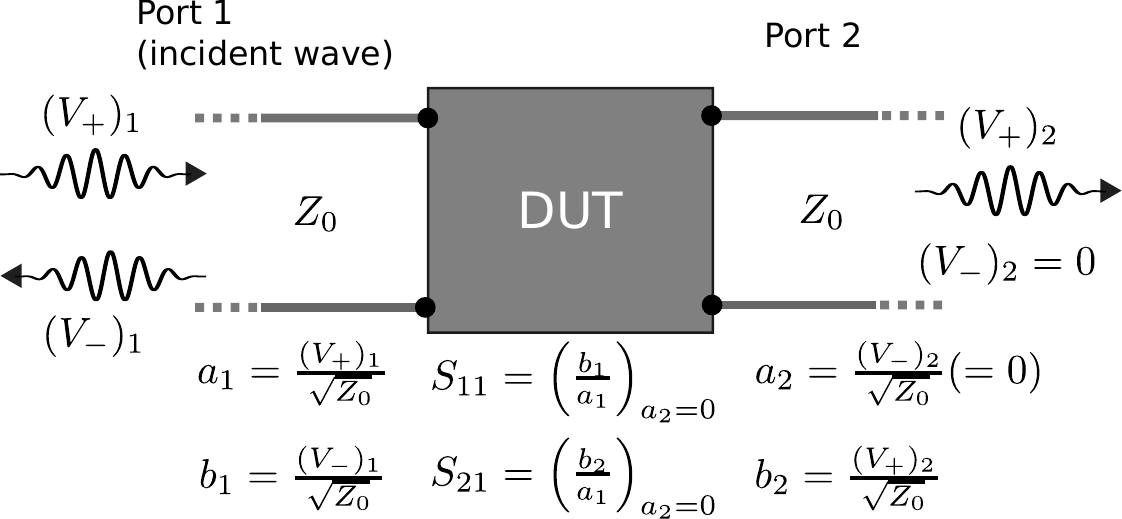}
\caption{Scattering parameter definitions for signal applied to port 1}
\label{fig:sparam}
\end{figure}

Using equations (\ref{eq:transfer}) it is now straightforward to obtain circuit properties such as S-parameters for circuits involving waveguides.  S-parameters are called scattering parameters to convey the idea that they describe how a signal scatters off a device under test.  The usual setup where S-parameters are applied is shown in figure \ref{fig:sparam}.  Here, two identical transmission lines are connected to the device under test and a signal is applied to one of the ports.  The two-port S-parameters are then defined as:
\begin{eqnarray}
&S_{11} = \left(\frac{b_1}{a_1}\right)_{a_2=0} \;
S_{21} = \left(\frac{b_2}{a_1}\right)_{a_2=0} &\; \textrm{(Input on port 1)} \\
&S_{12} = \left(\frac{b_1}{a_2}\right)_{a_1=0} \;
S_{22} = \left(\frac{b_2}{a_2}\right)_{a_1=0} &\; \textrm{(Input on port 2)}
\end{eqnarray}

\begin{figure}[htb]
\centering
\includegraphics[width=\textwidth]{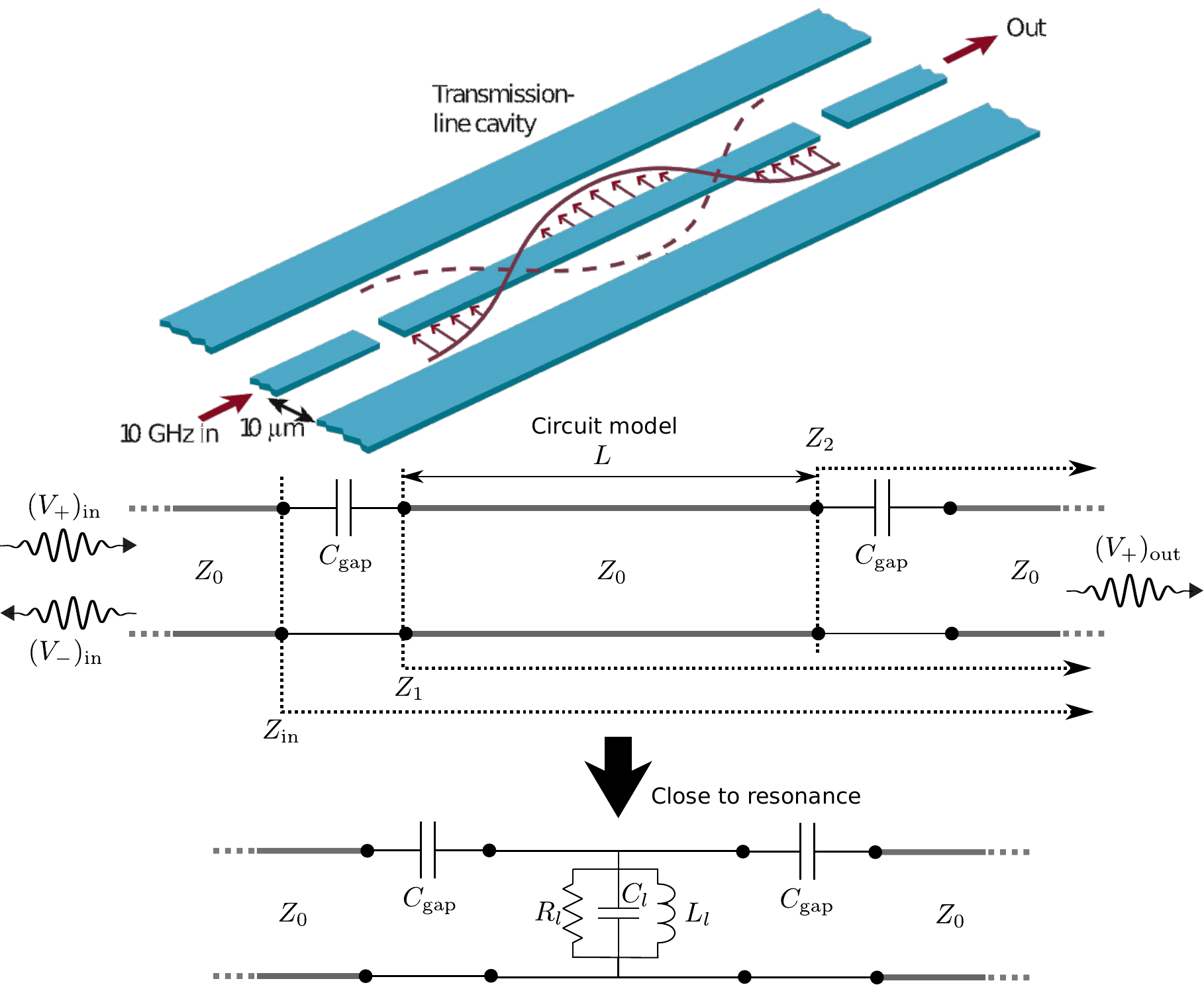}
\caption{Circuit model for a CPW resonator.  The image shows the general structure of a CPW resonator while the lower diagrams show the circuit equivalents.  Close to the resonance condition, the waveguide structure can be exchanged for a lumped element RLC circuit as long as $C_{\rm gap}$ is such that $R\ll 1/(\omega C_{\rm gap})$.}
\label{fig:ResCircuit}
\end{figure}

To illustrate an S-parameter calculation, we analyze the case of our coplanar waveguide resonator capacitively coupled to the external feed lines (figure \ref{fig:ResCircuit}).  This circuit can be fabricated by simply making two narrow cuts in the center line of a CPW leaving a waveguide segment of length $L$.  The equivalent impedance seen at the output $Z_2$ is a series combination of $Z_0$ and $C_{\rm gap}$.  We can then use equation (\ref{eq:transfer}) to obtain the impedance seen at the start of the waveguide segment $Z_1$ in terms of $Z_2$.  Then we can obtain the impedance at the circuit input $Z_{\rm in}$ and the refection response at the input $\Gamma_{\rm in}$:
\begin{eqnarray}
Z_2 &=& \frac{1}{j\omega C_{\rm gap}} + Z_0 \\
Z_1 &=& Z_0\frac{Z_2+jZ_0\tan(\beta L)}{Z_0+jZ_2\tan(\beta L)} \label{eq:Z1}\\
Z_{\rm in} &=& \frac{1}{j\omega C_{\rm gap}} + Z_1 \quad \Rightarrow \quad \Gamma_{\rm in} = \frac{Z_{\rm in} - Z_0}{Z_{\rm in} + Z_0}
\end{eqnarray}

The definition of $S_{11} = (V_-/V_+)_{\rm in}$ implies that $S_{11} = \Gamma_{\rm in}$.  Using the above equations and defining $s=C_{\rm gap}Z_0\omega$ gives
\begin{equation}
S_{11} = \frac{2s+\tan(\beta L)}
{2s(1+js) + (1+2js-2s^2)\tan(\beta L)} \label{eq:s11res}
\end{equation}

\begin{figure}[htb]
\centering
\includegraphics[width=\textwidth]{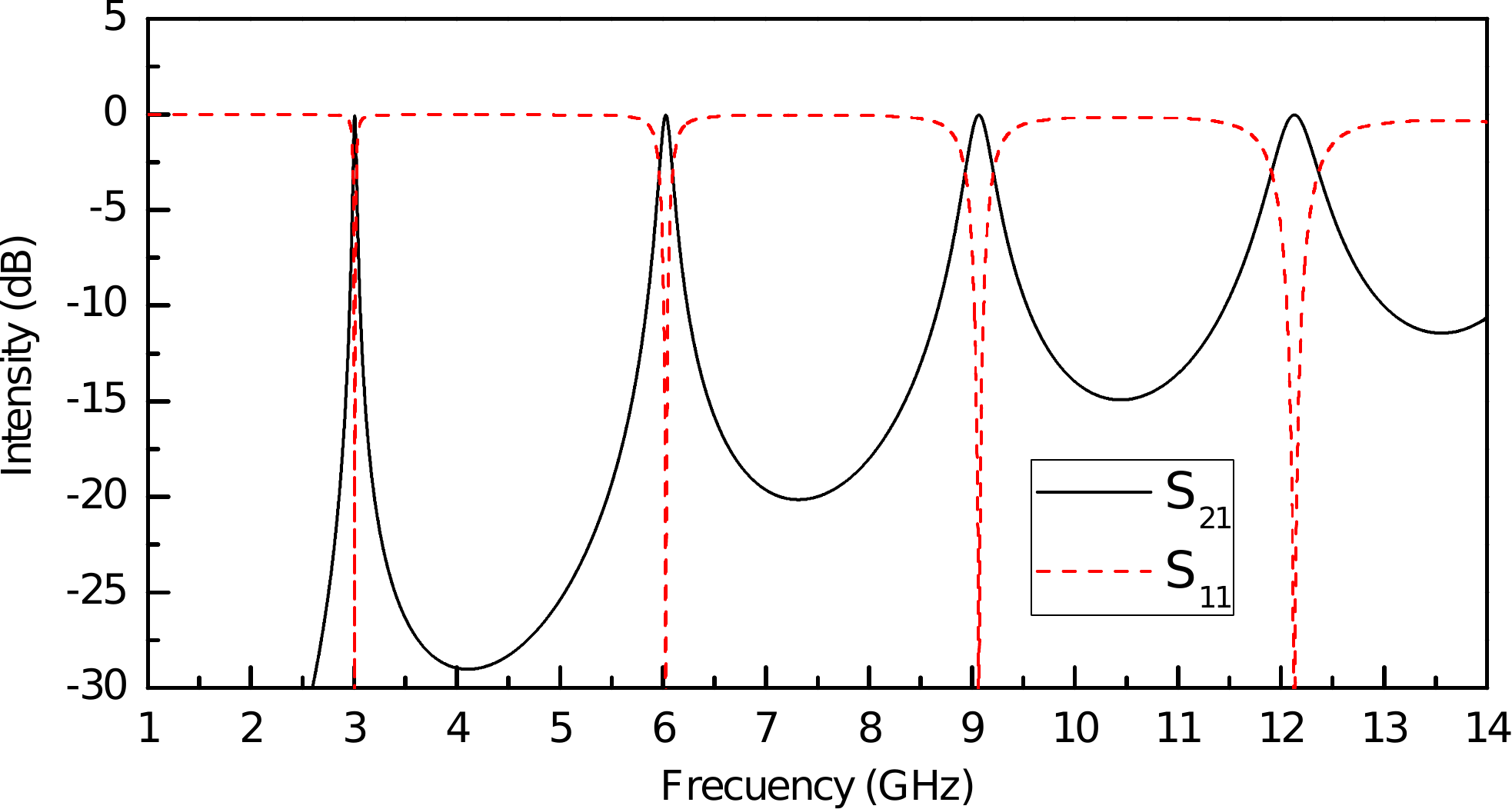}
\\[0.5cm]
\begin{tabular}{|c|c|c|c|c|c|}
\hline
$C_{\rm gap}$ & $Z_0$ & $L$ & $\beta$ & $\epsilon_{\rm eff}$ & $\alpha$ \\
\hline
\SI{100}{\femto\farad} & \SI{50}{\ohm} & \SI{20}{\milli\meter} & $\frac{\omega}{c}\sqrt{\epsilon_{\rm eff}} - j \alpha$ & 5.5 & $(\SI{6d-9}{\per\radian\second\per\metre})\omega$ \\
\hline
\end{tabular}
\caption{Transmission ($S_{21}$) and reflection ($S_{11}$) for an ideal CPW resonator.  The table shows the parameters used in equations (\ref{eq:s21res}) and (\ref{eq:s11res}).}
\label{fig:restheory}
\end{figure}

To obtain the $S_{21} = (V_+)_{\rm out}/(V_+)_{\rm in}$ transmission coefficient we again use (\ref{eq:transfer}) to obtain the voltage $V_{\rm out}$ at the output:
\begin{eqnarray}
V_{\rm in} &=& (V_+)_{\rm in}(1+\Gamma_{\rm in}) \\
V_1 &=& V_{\rm in} \frac{Z_1}{Z_{\rm in}} \\
V_2 &=& V_1e^{-j\beta L}\frac{1+\Gamma_2}{1+\Gamma_1} \\
V_{\rm out} &=& V_2\frac{Z_0}{Z_2} = (V_+)_{\rm out}
\end{eqnarray}
Substituting all these expressions into the definition of $S_{21}$ we finally get:
\begin{equation}
S_{21} = \frac{4s^2e^{j\beta L}}{1+e^{2j\beta L}(j-2s)^2} \label{eq:s21res}
\end{equation}
Setting some typical parameters and plotting $|S_{11}|$ and $|S_{21}|$ as functions of the wave frequency, we see the typical resonance pattern with harmonics at regular intervals (figure \ref{fig:restheory}).  The transmission shows peaks while the reflection is suppressed when the resonance condition is met:
\begin{equation}
L \simeq (n+1)\lambda_n/2
\end{equation}
or in terms of frequency $f = \omega/2\pi$:
\begin{equation}
f_n = \frac{(n+1)v}{2L},
\end{equation}
where $n$ labels the harmonic and $n=0$ corresponds to the fundamental mode.  The losses are higher and hence the peaks are wider and have lower quality factors at higher harmonics.  These losses are due mainly to the fact that the coupling impedance decreases with increasing frequency ($ Z_{\rm gap} \propto \omega^{-1}$), thus making the system more strongly coupled.

When working close to resonance it is sometimes convenient to replace the first circuit shown in figure \ref{fig:ResCircuit} with an equivalent lumped RLC element.  The values of the lumped capacitance ($C_l$), resistance ($R_l$) and inductance ($L_l$) are given by \cite{Goppl2008}
\begin{eqnarray}
L_l &=& \frac{2L'L}{n^2\pi^2} \label{eq:lumpedL}\\
C_l &=& \frac{C'L}{2} \\
R_l &=& \frac{Z_0}{\alpha L}
\end{eqnarray}
These expressions can be obtained taking the limit $\omega \simeq \frac{\pi v}{L}$ and $\alpha L \ll 1$ and assuming large coupling impedances, i.e., very small coupling capacitances.  With this equivalence we can obtain the loaded quality factor $Q_L$ in terms of its internal $Q_{\rm int}$ and external $Q_{\rm ext}$ contributions \cite{Pozar2011}:
\begin{equation}
\frac{1}{Q_L} = \frac{1}{Q_{\rm int}} + \frac{1}{Q_{\rm ext}}. \label{eq:Qsep}
\end{equation}
$Q_{\rm int}$ takes into account only losses that occur inside the resonator such as the conductor resistivity and dielectric losses.  On the other hand, $Q_{\rm ext}$ contains losses through the coupling capacitances $C_{\rm gap}$ to the external feed lines.  It can be shown that they have the following expressions:
\begin{eqnarray}
Q_{\rm int} & = & \omega R_l C_l \\
Q_{\rm ext} & = & \frac{\omega C_l}{2} \frac{1+\omega^2 C_{\rm gap}^2 Z_0^2}{\omega^2 C_{\rm gap}^2 Z_0}
\end{eqnarray}
From a measurement point of view, only $Q_L$ is directly accessible while the $Q_{\rm int}$ contribution can not be immediately separated.  It can however be estimated by also measuring the insertion loss $L_0$, defined as the deviation from unity of the peak transmission value:
\begin{equation}
L_0 = -\left|S_{21}(f_n)\right| \; \textrm{(in dB)}
\end{equation}
Defining the coupling factor as $g=Q_{\rm int}/ Q_{\rm ext}$, the insertion loss $L_0$ and $g$ are then related by:
\begin{equation}
L_0 = -20 \log\left(\frac{g}{g+1}\right)\,\textrm{dB}.\label{eq:L0g}
\end{equation}
This then allows us to separate the quality factor contributions $Q_{\rm int}$ and $Q_{\rm ext}$ using (\ref{eq:Qsep}),(\ref{eq:L0g}) and measurements of $L_0$ and $Q_L$:
\begin{eqnarray}
g & = & \frac{10^{\frac{-L_0}{20}}}{1-10^{\frac{-L_0}{20}}} \\
Q_{\rm int} & = & (1+g)Q_L \label{eq:Qintg}
\end{eqnarray}

\subsection{Design and fabrication} \label{sec:defects}

\begin{figure}[htb]
\centering
\includegraphics[width=\textwidth]{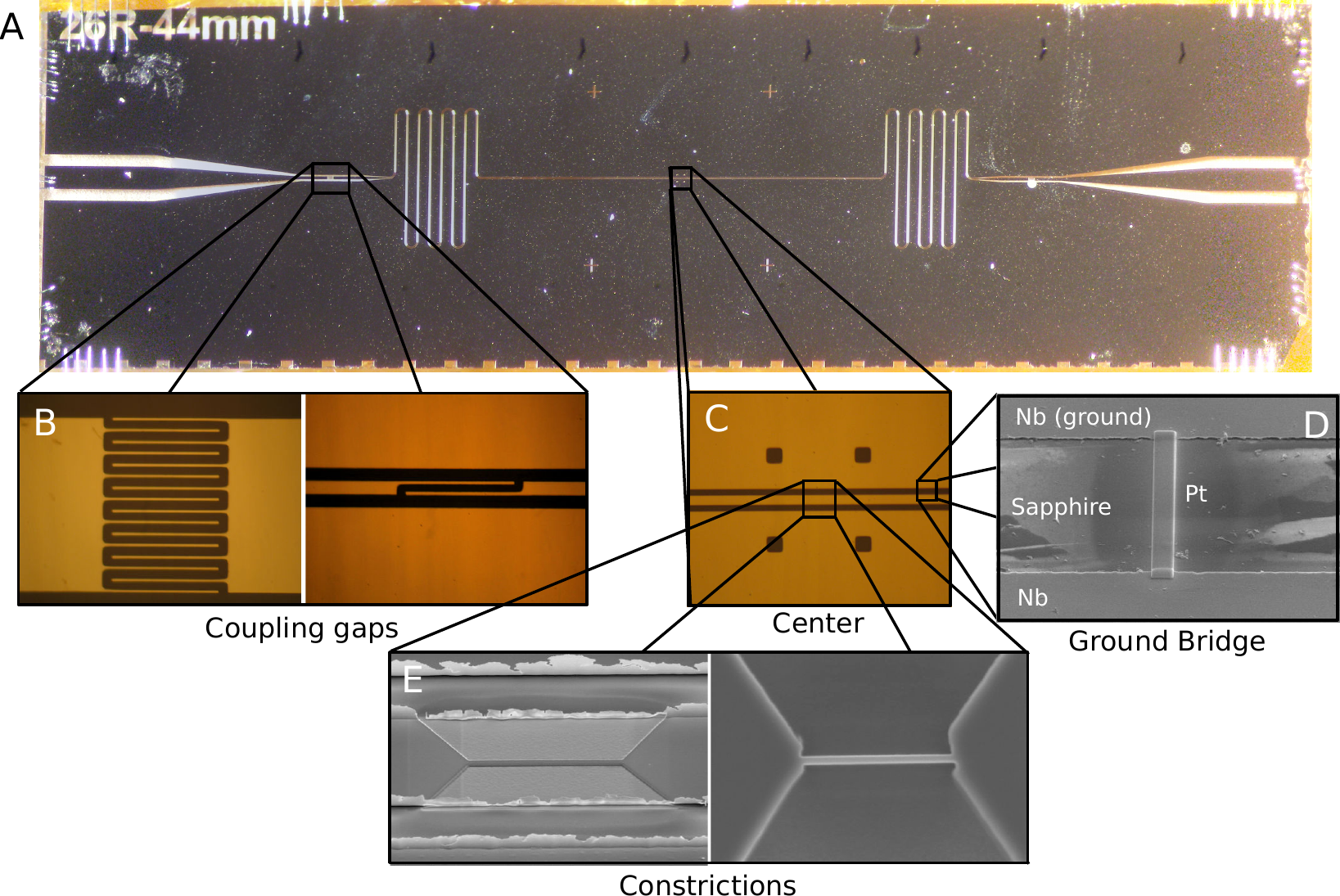}
\caption{Coplanar waveguide resonator.  A) Optical microscope image of a CPW resonator.  The chip is 2 cm in length, by 0.5 cm in width. B) Different coupling gaps with different numbers of fingers.  Larger numbers of fingers give over-coupled resonators.  C) Center region of CPW resonator.  D) Platinum bridge deposited by focused ion beam induced deposition.  This bridge is removed after FIB processing and allows the center line to be grounded adequately during nanoconstriction fabrication. E) Different nanoconstrictions fabricated at the center of the resonator}
\label{fig:litho}
\end{figure}

Using equations (\ref{eq:Z0}) we can set the design dimensions and parameters for our resonators.  We design them to have a resonance frequency around 1.5 GHz and lateral dimensions in the \SI{10}{\micro\meter} range.  The devices are fabricated by optical lithography on sapphire wafers following the procedures detailed in chapter \ref{chap:tech} (section \ref{sec:liftoff}).  Sapphire is chosen as the substrate because of its excellent insulating properties and low dielectric losses ($\tan{\delta}\sim 10^{-5}-10^{-6}$ \cite{Hartnett2006} at room temperature and $\tan{\delta}\sim 10^{-8}-10^{-10}$ at liquid helium temperature \cite{Buckley1994}).  The mask designs consist of large feed lines with \SI{400}{\micro\metre} center lines separated from the ground by \SI{200}{\micro\metre} gaps that narrow down to a \SI{14}{\micro\metre} centerline and \SI{7}{\micro\metre} gaps after going through the gap capacitors.  Several types of gap capacitors with a finger design (see figure \ref{fig:litho}B) were fabricated to allow for over and under-coupled systems depending on the number of fingers.  The finger separation and width are \SI{4}{\micro\meter} and they are \SI{96}{\micro\meter} long.  The length of the cavity is chosen to be 44 mm by making the waveguide meander across the surface.  Even though sapphire is anisotropic, for our purposes it is sufficient to take an average dielectric constant of $\epsilon_r = 10$ \cite{Harman1994}.  Assuming no losses and using (\ref{eq:Z0}) we get the following values for the resonator parameters:
\begin{eqnarray*}
\epsilon_\textrm{eff} &= & 11/2 \\
Z_0 & = & \SI{52.9}{\ohm}\\
L' & = & \SI{0.414}{\micro\henry\per\meter}\\
C' & = & \SI{0.148}{\nano\farad\per\meter}\\
f_0 & = & \SI{1.454}{\giga\hertz}  \quad \textrm{(Unloaded)}
\end{eqnarray*}

Our devices are fabricated on \SI{500}{\micro\metre} thick C-plane sapphire wafers by optical lithography followed by either reactive ion etching (RIE) or by lift-off.  The details of the lithography techniques are covered in section \ref{sec:liftoff}.  An image of a finished resonator is shown in figure \ref{fig:litho}A.

\begin{figure}[phtb]
\centering
\includegraphics[width=0.8\textwidth]{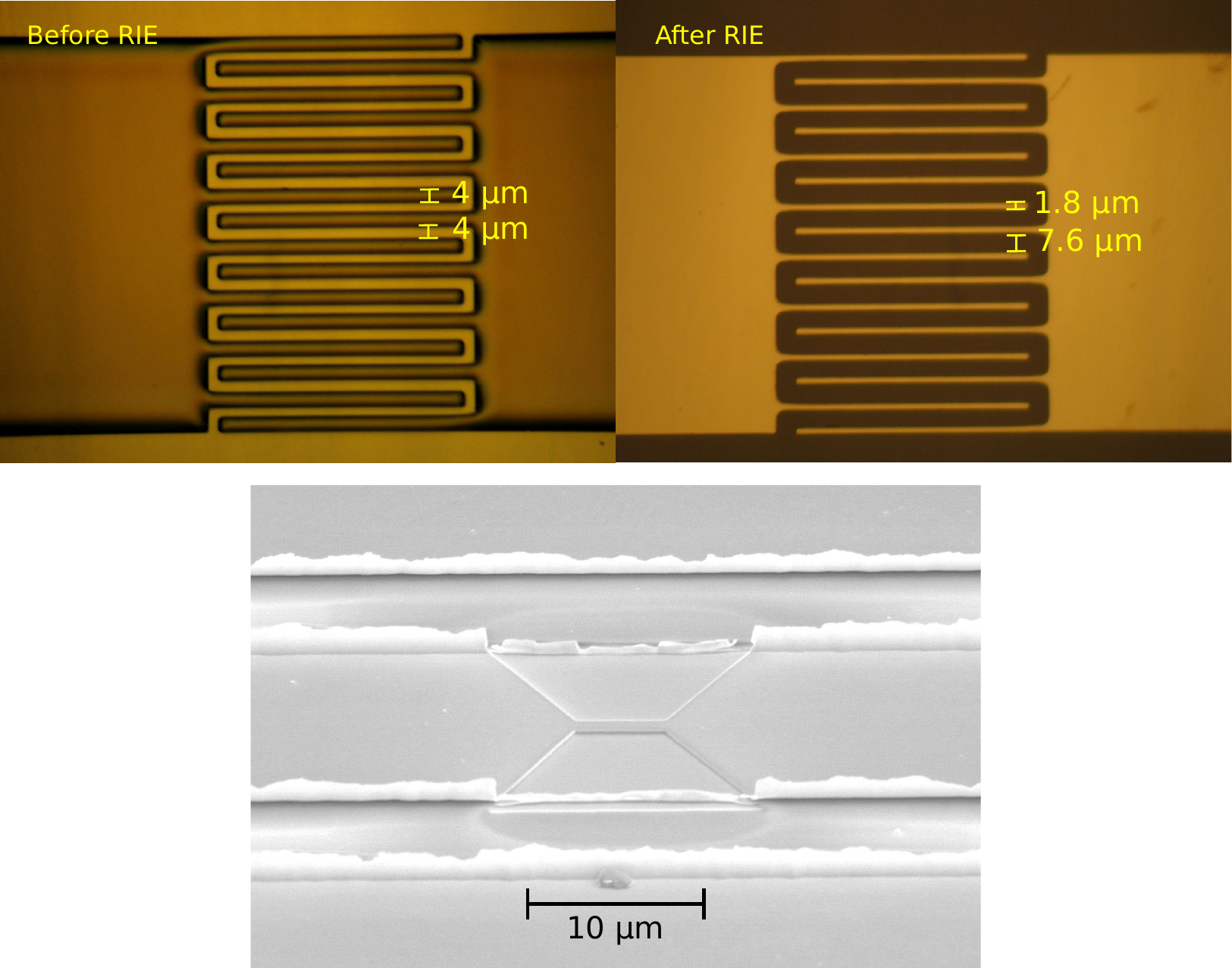}
\caption{Lithography defects.  The top images show microscope images of the differences between the resin pattern and the final etched Nb pattern when RIE was used for fabrication.  The bottom image shows an SEM image of the resin remains on Nb feature edges when liftoff was used for fabrication.}
\label{fig:defects}
\end{figure}

The finished circuit dimensions differ somewhat from the mask specifications especially in the cases that RIE is used (figure \ref{fig:defects} top).  The RIE procedure usually over etches the niobium layer leading to lateral size errors of about 1-\SI{2}{\micro\meter}.  Although these differences are fairly large, the devices generally work acceptably well.  The reproducibility of the mask is better using lift-off.  However, when using lift off, there are often remains of the resin forming high walls on the edges of structures (figure \ref{fig:defects} bottom).  These resonators are also still usable but it is as yet unclear whether and how these structures could affect the transmission properties.

\section{Tests at $T=\SI{4.2}{K}$}\label{sec:testCPWG}

\subsection{Frequency dependence: resonant behavior}
\begin{figure}[phtb]
\centering
\includegraphics[width=0.8\textwidth]{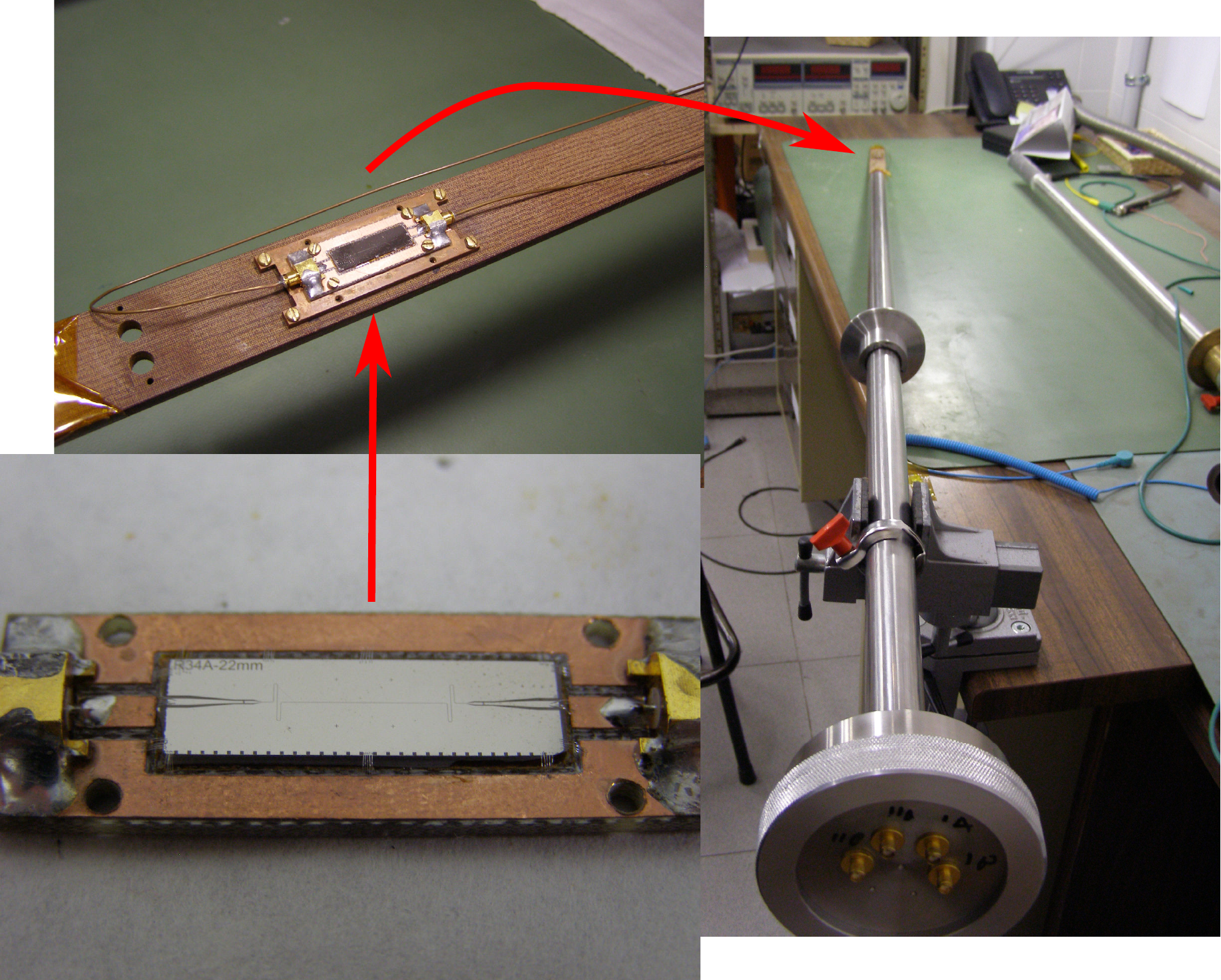}
\caption{Coplanar waveguide resonator mounted on 4 K probe.  The resonator is glued to a PCB and wire bonded to copper pads.  The PCB is then screwed into a copper box and wired with semi-rigid coaxial cables to the 4 K probe.}
\label{fig:probe}
\end{figure}

The resonators are first glued to a fiberglass printed circuit board (FR4 or similar \cite{Association2012}) shown in figure \ref{fig:probe} and wire bonded with aluminum thread to copper pads.  Typically there are between 6 and 10 wire bonds for the center line while the ground planes have between 40 and 60 wire bonds.  The circuit board has SMP type connectors \cite{Technologies} to connect to external semi-rigid cryogenic coaxial cables.  The wires and PCB are inserted into a simple 4 K probe that consists of essentially a steel tube with SMA connectors at one end, to connect to our measurement electronics, and the PCB with the resonator at the other (figure \ref{fig:probe}).

\begin{figure}[tbh]
\centering
\includegraphics[width=\textwidth]{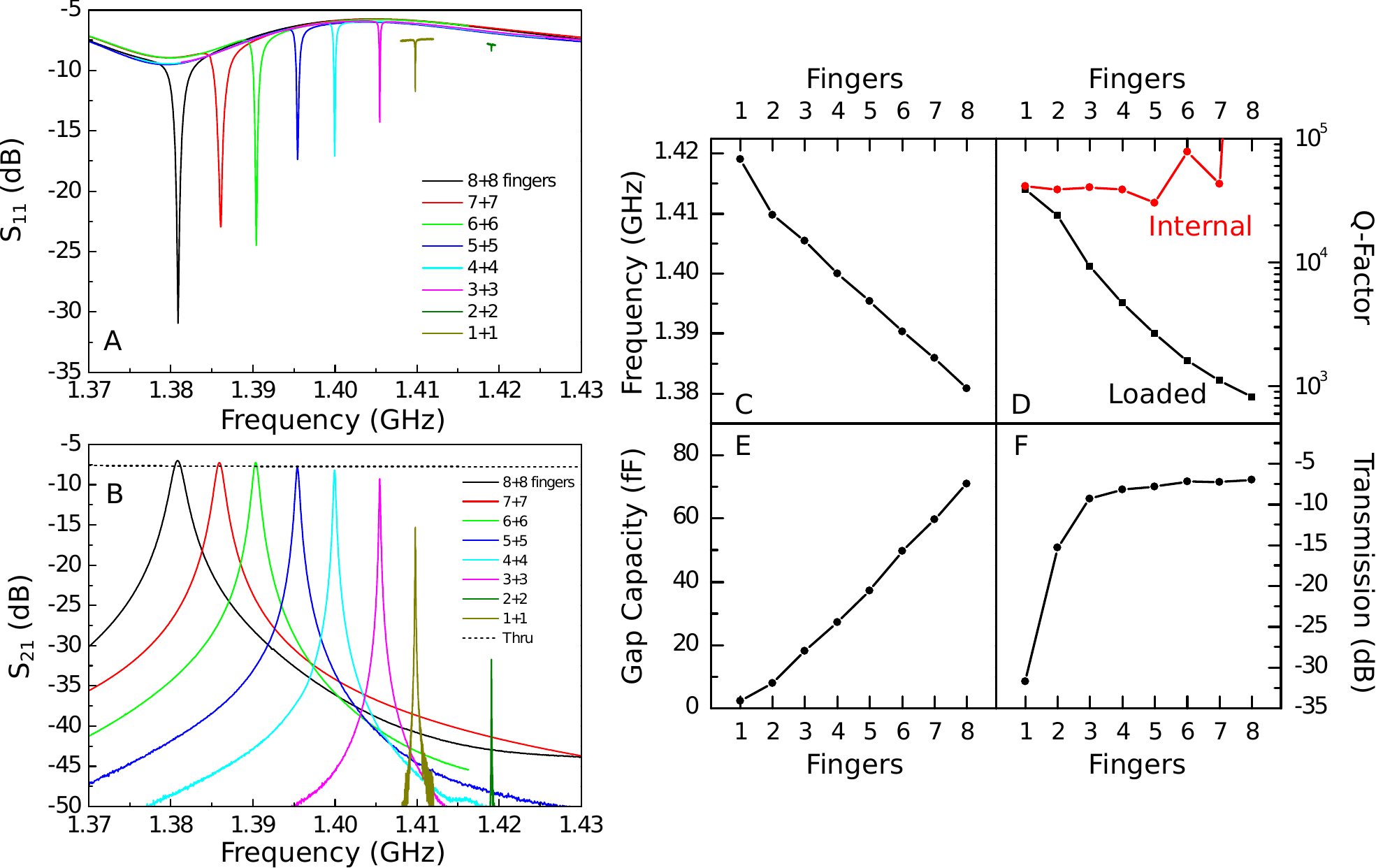}
\caption{Fundamental mode resonances for differently coupled resonators.  The graphs on the left shows the transmission $S_{21}$ and reflection $S_{11}$ as functions of the microwave frequency.  The graphs on the right show the changes in resonant frequency, loaded Q-factor, fitted gap capacity and transmission peak value for different numbers of fingers.}
\label{fig:1stharmonic}
\end{figure}

Once mounted on our 4 K probe, the probe is inserted into a helium bath and the devices are then characterized using a programmable network analyzer \cite{Schwarz} (see also section \ref{sec:pna}).  One batch of resonators fabricated by optical lithography and RIE were found to have the fundamental mode resonances shown in figure \ref{fig:1stharmonic}.  In general for all our transmission measurements, the excitation power is set to the minimum value of -45 dBm at the signal source to avoid any peak distortions.  We see that the resonances move to higher frequencies and have higher Q factors the less coupled they are to the feed lines.  Each peak can be fitted to a Lorentzian line shape (in dB scale):
\begin{equation}
F(f) = \frac{A_0}{(f-f_0)^2+\delta\! f^2/4},
\end{equation}
where $f_0$ is the peak position, $\delta\! f$ is the full width at half maximum and $A_0$ is a multiplicative constant that sets the peak height.  From these fits it is simple to obtain the loaded Q-factor $Q_L=f_0/\delta\! f$ for each resonator.  It is interesting to isolate the internal quality factor $Q_{\rm int}$ to evaluate the internal losses of this series of resonators.  In principle, the internal losses should be the same for the entire series and they all should have similar $Q_{\rm int}$.  To obtain this value we need to use equation (\ref{eq:Qintg}) and (\ref{eq:L0g}) to obtain $Q_{\rm int}$ from the insertion loss $L_0$.  This method is not adequate to obtain this value in very over-coupled cases as the formula is sensitive to very small changes in $L_0$ and will have large errors in these instances.  In any case, $L_0$ is evaluated by comparing the peak transmission with that of a \emph{thru} connection (see \ref{fig:1stharmonic}B).  As can be seen, the transmission peak surpasses the \emph{thru} value and can not give a value for $Q_{\rm int}$.  To partially correct this error, the entire \emph{thru} curve is shifted so that the 8 finger case has $L_0 = 0$ (and an indeterminate $Q_{\rm int}$).  This new curve is then used as the reference for the remaining cases to estimate $Q_{\rm int}$ and plotted as a function of the number of fingers (figure \ref{fig:1stharmonic}D).  The error is smaller for lower couplings, i.e., fewer fingers and should have very small effects in the low coupling cases.  We see the calculated values converge on a value of around $Q_{\rm int}\simeq 4\cdot 10^4$ which we will take to be our standard value for the series.
Figure \ref{fig:1stharmonic} also shows the peak transmission values and resonance frequencies (\ref{fig:1stharmonic}C and \ref{fig:1stharmonic}F).  The figure \ref{fig:1stharmonic}E shows an approximation of the gap capacitance (which determines the external losses) assuming the internal losses are given by our calculated value of $Q_{\rm int}\simeq 4\cdot 10^4$.  This is done taking the absolute value of equation (\ref{eq:s21res}) and using it to fit the peak data via the $C_{\rm gap}$, $L$ and a multiplicative constant to account for the losses in our connecting wires.  The resulting values of $C_{\rm gap}$ are similar to those reported in \cite{Goppl2008}.    It is worth noting that although most of the results shown are for transmission measurements ($S_{21},S_{12}$), much of the same behavior can be seen in the reflection signals ($S_{11},S_{22}$).

\subsection{Magnetic field dependence}\label{sec:magfielddep}

\subsubsection{Niobium thin film properties}
As well as being perfect electrical conductors, superconductors present perfect diamagnetism meaning they expel magnetic field lines that would ordinarily pass through them \cite{Rose-Innes1978, Tinkham2012}.  However, if the applied fields are strong enough the superconducting properties end up being suppressed and the material becomes a normal conductor.  There are in general two types of superconductor that show different field dependence near the critical field.
\begin{itemize}
\item Type I superconductors are generally pure metals (Aluminum, Lead, Mercury) and abruptly lose their superconducting properties when the critical field is reached.  These fields are generally much smaller ($B_c\lesssim \SI{50}{\milli\tesla}$) than for type II superconductors and also have lower critical temperatures ($T_c \lesssim \SI{1}{\kelvin}$).

\item Type II superconductors are characterized by two critical fields and in general can withstand much higher magnetic fields before becoming resistive.  The first critical field is the minimum field capable of creating vortices in the superconducting domain.  These vortices are microscopic areas that allow the magnetic field to pass through them while the rest of the material remains superconducting.  When higher fields are applied up to the second critical field, the density of vortices increases until all the material becomes a normal conductor.
\end{itemize}

Here we work only with niobium which is a type II superconductor \cite{Finnemore1966}.  Using the MPMS and PPMS systems (see sections \ref{sec:mpms} and \ref{sec:PPMS} respectively) we have measured some of the magnetic and DC conduction properties of our niobium thin films.  These properties can give us approximations of the values for the different critical fields and currents and their dependence on the direction of the applied field.

\begin{figure}[htb]
\centering
\includegraphics[width=\textwidth]{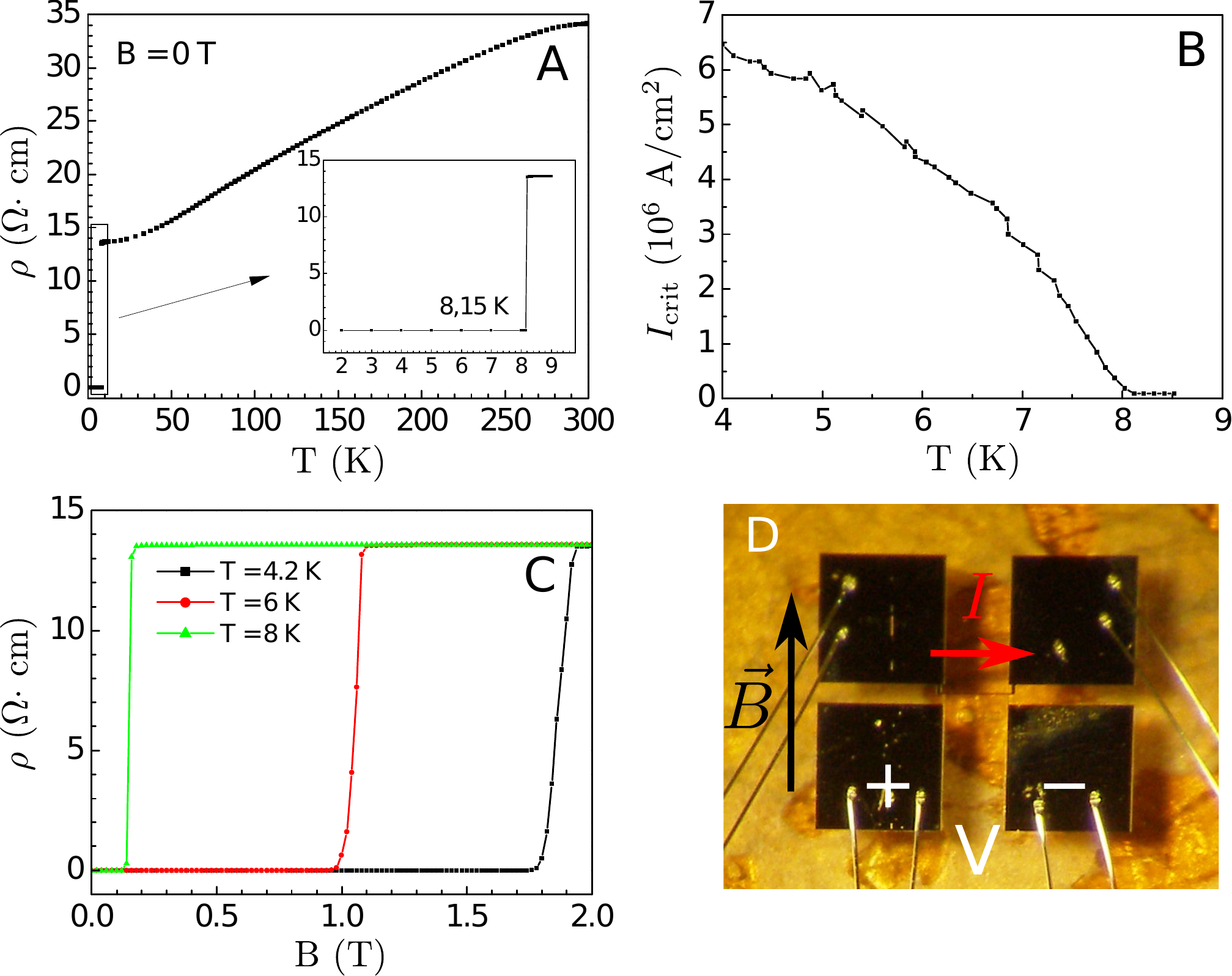}
\caption{DC resistivity performance of Nb resistivity patterns.  The wire measured was \SI{400}{\micro\meter} long and had a cross section of approximately \SI{10}{\micro\meter} by 150 nm (\SI{1.5}{\micro\metre\squared}).  Graph A shows the temperature dependence of the resistivity at zero field with a critical temperature of about \SI{8.15}{\kelvin}.  Graph B shows the critical current density as a function of the temperature obtained from measuring I-V curves at each temperature.  Graph C shows the dependence of the resistivity on the in plane magnetic field at several temperatures.  Graph D shows the geometry of the circuit}
\label{fig:Nbresist}
\end{figure}

Firstly, using a simple 4 point resistance measurement pattern, we have obtained the DC behavior of our niobium circuits including the critical fields, critical temperatures and currents.  The resistivity pattern is fabricated in the same process as the resonators (in the wafer margins) and has a separation of \SI{400}{\micro\meter} between the voltage leads while the wire has a cross section of 150 nm by \SI{10}{\micro\meter}.  This system is mounted on a vertical stage (see section \ref{sec:PPMS}) so that the magnetic field is applied in the wafer plane and perpendicular to the wire.  The results are shown in figure \ref{fig:Nbresist}.  We see from the resistivity measurement that the critical temperature is \SI{8.15}{\kelvin}, close to values found in the literature for niobium thin films \cite{VanHuffelen1993} and for bulk niobium ($T_c=\SI{9.15}{\kelvin}$ \cite{Finnemore1966}).  The critical field $H_{c2}$ (approximately when the zero resistivity state is lost) depends strongly on temperature and can be higher than \SI{1.5}{\tesla} at \SI{4.2}{\kelvin} and even higher (in excess of 2 T) at \SI{2}{\kelvin}.  However, these values of $H_c$ are only for the in plane field case and we expect the critical fiends to be very small if the field were applied perpendicular to the film.

\begin{figure}[htb]
\centering
\includegraphics[width=\textwidth]{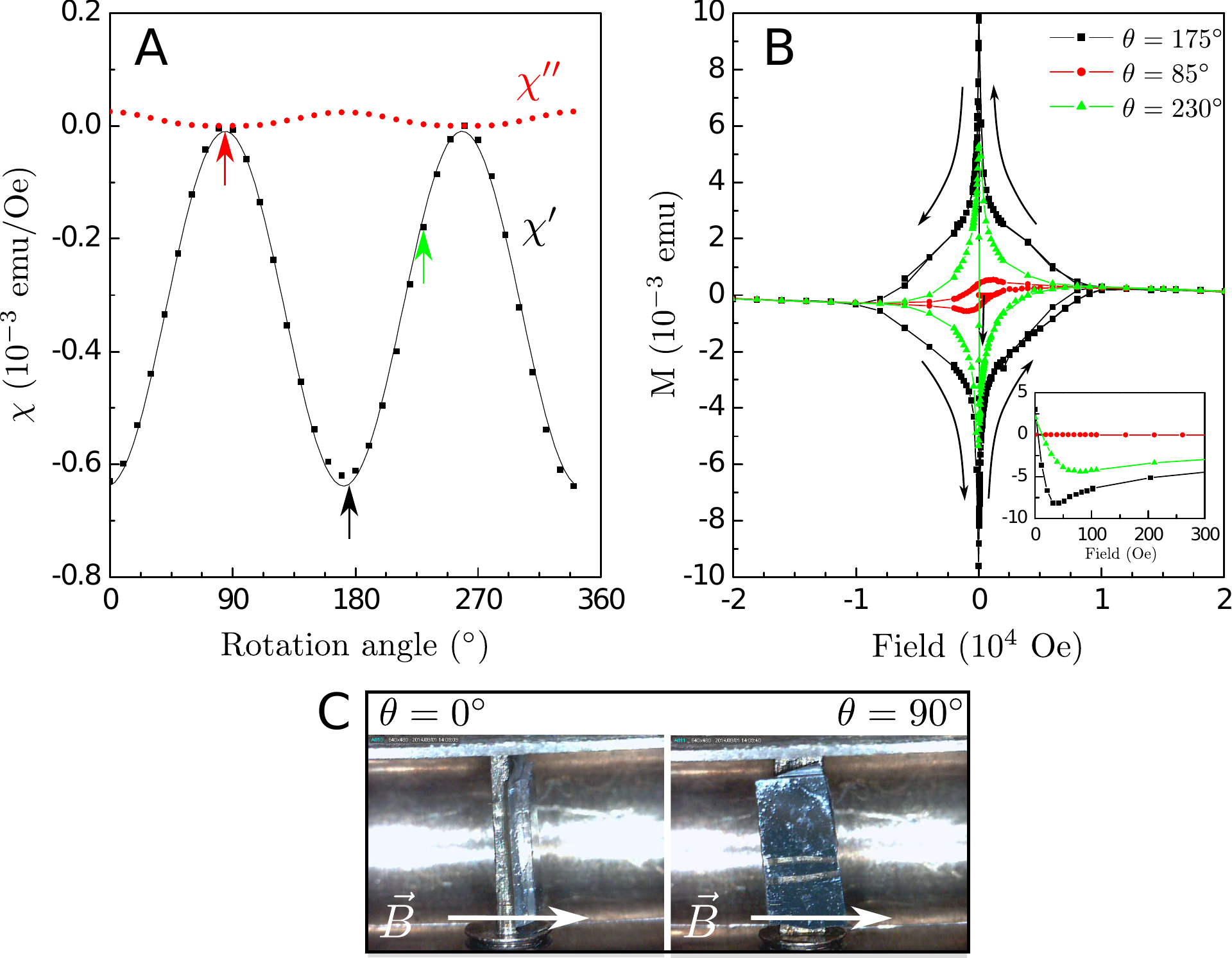}
\caption{Niobium susceptibility and magnetization measurements at $T=\SI{4.5}{\kelvin}$.  Figure A shows the angle dependence of the AC susceptibility for a 4 Oe driving field and 15 Hz frequency.  The colored arrows correspond to the three orientations shown in figure B.  Figure B shows hysteresis cycles at different orientations.  Starting from a non-magnetized state the field was swept successively to 2 Tesla, -2 Tesla and 2 Tesla.  The arrows label the up-sweep and down-sweep curves and the inset shows the initial response when applying the field.  Figure C shows the fragment of niobium on sapphire measured on the MPMS rotation stage and the direction of the applied field (AC and DC).}
\label{fig:nbsusc}
\end{figure}

To check the magnetic properties of the niobium films, we placed a small piece of sapphire wafer with a 150 nm Nb layer on the rotating stage of the MPMS system (section \ref{sec:mpms}).  We then measure the AC susceptibility and DC magnetization as functions of the film orientation and magnetic field.  The results are shown in figure \ref{fig:nbsusc}.  From figure \ref{fig:nbsusc}A, we see that, as expected, the diamagnetic response is much stronger when the magnetic field is applied perpendicular to the niobium film (close to \ang{0} and \ang{180}) and very small when the field is in plane (around \ang{90} and \ang{270}).  Also, the reactive component of the susceptibility $\chi''$, although small, is also larger in the perpendicular orientation.  Figure \ref{fig:nbsusc} shows hysteresis cycles for different field orientations.  In the $\theta = \ang{175}$ case and starting from an unmagnetized state, applying the field in the positive direction gives a strong diamagnetic response that starts to decay at about $H_{c1} \sim 35$ Oe (inset).  After that, the magnetization continues to decay until the material no longer has a magnetic response.  The magnetic flux through the film is trapped when the applied field is removed and the superconducting state is restored leaving a positive magnetization at zero field.  The cycle then repeats in the opposite field direction.  It is interesting to note that the original state is never restored during the cycle.  The cycles are smaller at intermediate angles and in the parallel direction while the effective value of $H_{c1}$ increases ($\sim 75$ Oe at \ang{230} and not visible at \ang{85}).

The hysteresis and the nonzero $\chi''$ susceptibility indicates the presence of vortices at relatively low magnetic fields, whose effect becomes especially important when the field has a non-zero component perpendicular to the film.  On the other hand, these results also show that we should be able to apply moderate magnetic fields without breaking the superconductivity of our circuits.  This is important because, as we can see in chapter \ref{chap:Theo1} and \ref{chap:SIMs}, most candidates for magnetic qubits require magnetic fields fields to tune the level separation into resonance with our circuits.  Applications in micro and nano-EPR experiments would also require applied fields.  The next step is therefore to study the behavior of our resonators at their operating frequency in the presence of magnetic fields to analyze the possible limitations.

\subsubsection{Effects of magnetic fields on resonators characteristics}\label{subsec:resmag}
The appearance of trapped Abrikosov vortices in our wires can alter their transmission properties and hence, the properties of our resonances as well as, according to \cite{Bothner2012} and our susceptibility measurements, introduce hysteresis effects.  We use our setup to insert a 1.409 GHz resonator into our vector magnet (see chapter \ref{chap:tech}) and study its fundamental mode for magnetic fields applied in different directions.  We measure the transmission for each field, fit a Lorentzian line shape to the resonance in each case, and extract the resonance characteristics (resonant frequency, $Q$ and transmission value).

\begin{figure}[htb]
\centering
\includegraphics[width=0.95\textwidth]{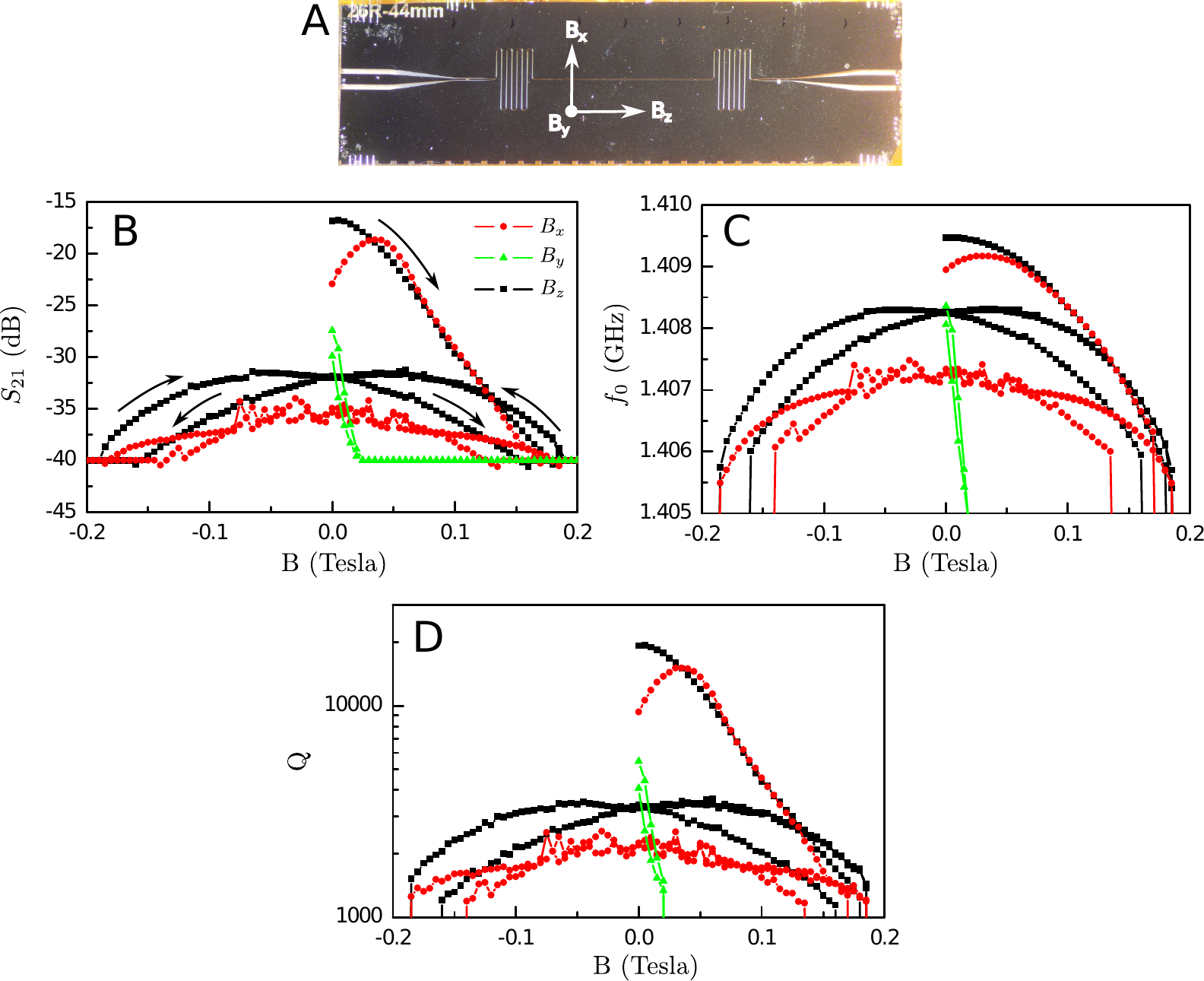}
\caption{Resonator performance in the presence of magnetic fields.  In the absence of field, the original resonant frequency of the resonator is 1.409 GHz with a Q-factor of about 19000 and a transmission value of -16 dB.  Graph A shows the orientation of the magnetic field directions.  Graphs B,C and D show the transmission, resonant frequency and quality factor as a function of the applied magnetic field in the three axis directions. For the in plane directions $B_x$ and $B_z$ a full hysteresis cycle was performed (0 T, 0.5 T, -0.5 T, 0.5 T, 0 T) while the out of plane field $B_y$ was only swept from 0 T to 0.5 T and back to 0 T.  The arrows in graph B show the typical route for the hysteresis cycles for all the graphs.}
\label{fig:hist27R}
\end{figure}

\begin{figure}[htb]
\centering
\includegraphics[width=0.95\textwidth]{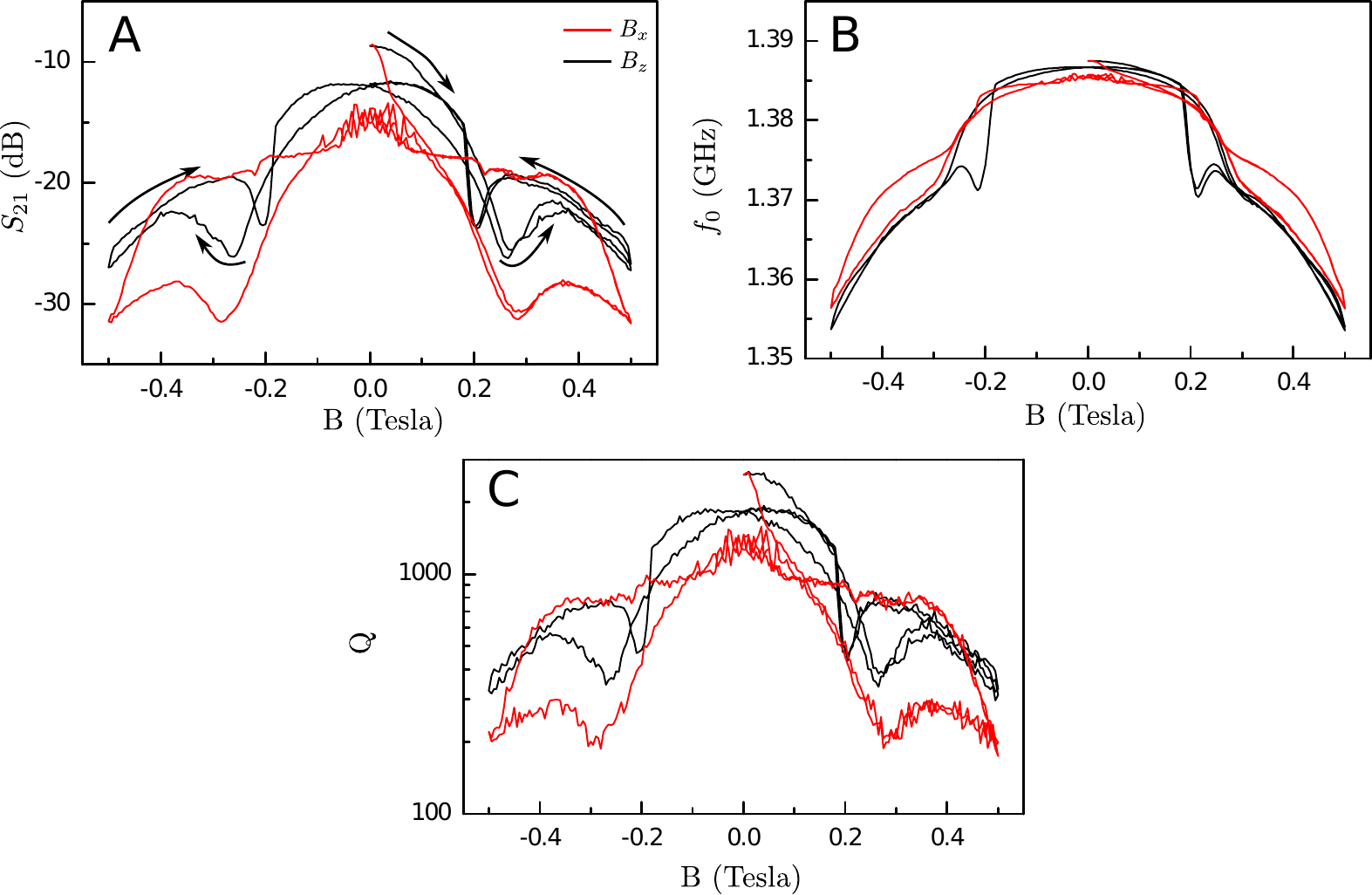}
\caption{Resonator performance in the presence of magnetic fields.  At zero field of field, the original resonance frequency of the resonator is 1.387 GHz with a Q-factor of about 3500 and a transmission value of -8.7 dB.  Graphs A,B and C show the transmission, resonance frequency and quality factor as a function of the applied in plane magnetic field in the $B_x$ and $B_z$ axis directions (see figure \ref{fig:hist27R}A). A full hysteresis cycle was performed for each orientation (0 T, 0.5 T, -0.5 T, 0.5 T, 0 T) The arrows in graph A show the typical route for the hysteresis cycles for all the graphs.}
\label{fig:hist23R}
\end{figure}

The measurements shown in figure \ref{fig:hist27R} reveal clear signs of hysteresis in the resonance characteristics.  According to \cite{Bothner2012}, these effects are due to the appearance of Abrikosov vortices in the niobium film that alter the conductions characteristics.  Also, we find that the original state before the application of any magnetic field is never recovered in the cycle, an effect we already expected from the results of previous susceptibility measurements (figure \ref{fig:nbsusc}).  After each cycle, the magnetic field is oscillated to 0 to reduce the trapped field and to avoid remanent fields in the superconducting magnets.  However, the effect is never entirely removed and causes each cycle to start at a different point.  In any case, we see that this specific resonator ceases to have a usable resonance at about 0.2 T and 0.02 T for the in plane and out of plane field directions respectively.  We also note that when applying the field in the $B_x$ direction we get some noise in the low field region, possibly due to some complex vortex dynamics.  It is important to note that this noise is not time dependent in the sense that it only appears when changing the field.  If the field is kept fixed the transmission signal is stable but its evolution is noisy when the field is moved to different values.  In any case, the impact of these hysteresis effects can be minimized performing measurements preferably on the first upsweep.  It is also worth noting that resonators with lower Q factors are somewhat less susceptible to these effects and their original characteristics can be approximately recovered without having to heat the resonator above $T_c$.  Also, lower Q-factors are usually accompanied with a higher transmission value meaning that the resonance can be tracked up to higher fields without being lost.  As an example, figure \ref{fig:hist23R} shows the hysteresis cycles for a 2500 Q-factor and 1.387 GHz resonator.  We again see noise in the $B_x$ curve at low fields.

\section[Nanometric constrictions in superconducting resonators]{Nanometric constrictions in superconducting co\-pla\-nar waveguide resonators}
\label{sec:constrictionsCPWG}

\subsection{Fabrication}

After the basic characteristics of the devices were measured, the center line is narrowed down from around \SI{10}{\micro\metre} to minimum widths of 50 nm along distances of up to \SI{15}{\micro\metre} at the center of the resonator length (see figure \ref{fig:litho}E).  The constriction was made using a focused ion beam (FIB) system like those described in section \ref{sec:semfib}.  A narrow beam of gallium ions (down to a few nm in diameter) is focused onto the niobium layer and swept over the area to be etched away leaving only a small nanowire.  Although it is possible to extend the constriction length up to \SI{50}{\micro\meter}, they are harder to make and would likely induce greater changes in the resonator properties.  Once the constrictions are made, special care must be taken both with the device manipulation and scanning electron microscope imaging since electrostatic buildup and discharge can easily vaporize the nanowires.  To avoid this during FIB processing and SEM imaging, a short to ground is made by depositing a small platinum bridge by FIB induced deposition (figure \ref{fig:litho}D).  This allows the beam current used to etch the center line or to image the area to be dissipated continually from the resonator center conductor instead of as a sudden electrostatic discharge.  After the processing is complete, the bridge is removed with the ion beam.

\begin{figure}[htb]
\centering
\includegraphics[width=0.8\textwidth]{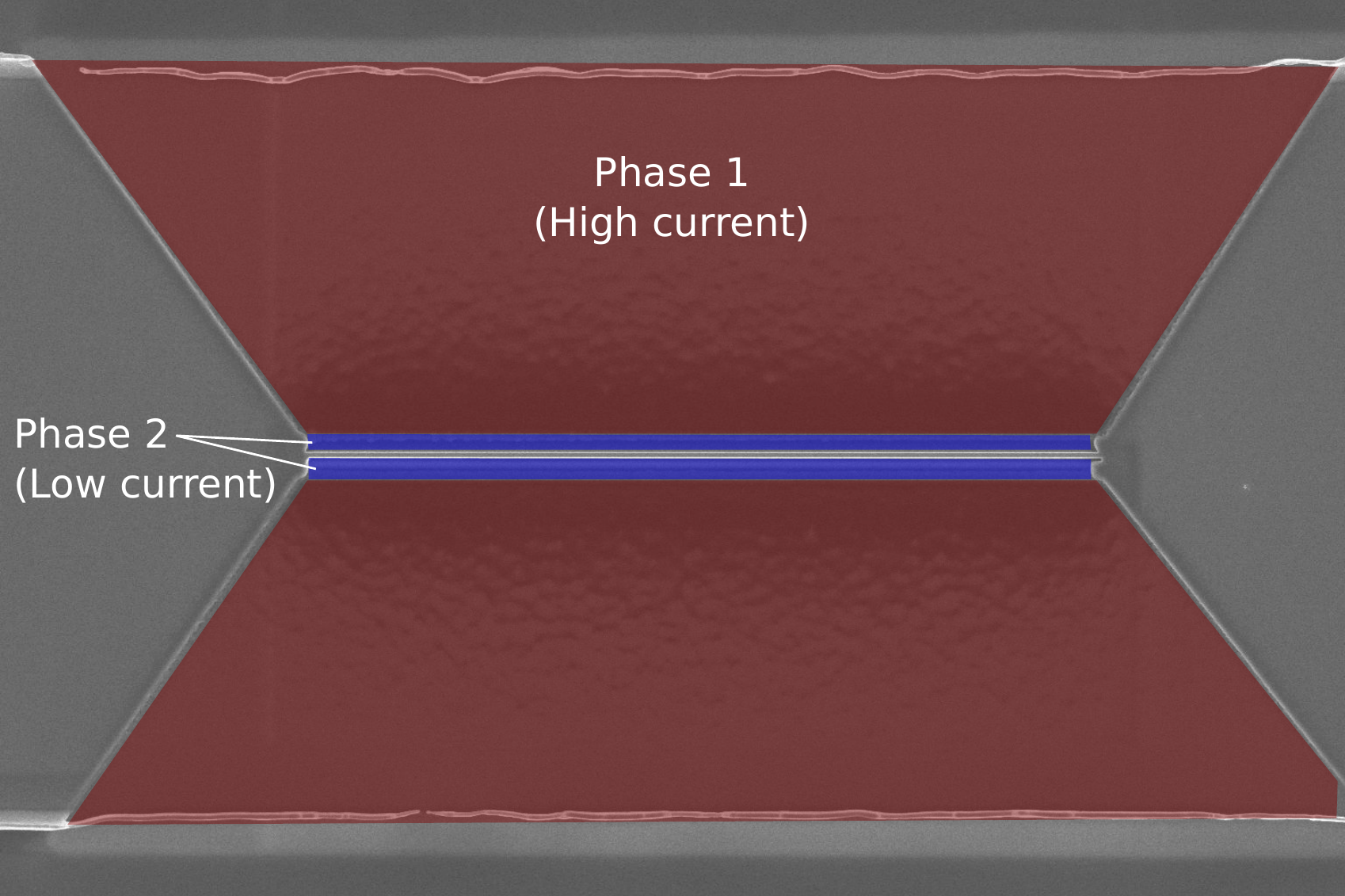}
\caption{Two-phase FIB etching process.  Firstly the red areas are etched using a high beam current.  Finally the blue areas are etched using a small beam current resolve fine features.}
\label{fig:fibphases}
\end{figure}

In most cases, the etching process is done in two phases (figure \ref{fig:fibphases}).  Firstly, the larger areas are etched with a large beam current to save time.  This allows the center line to be narrowed down to about \SI{1}{\micro\meter} in a couple of minutes.  In the second phase we remove the remaining material and create the finished geometry using small beam currents (around 20 pA beam current) that allows for finer details.  The entire procedure, once the sample is loaded and the chamber prepared, takes no more than 10-15 minutes.

\subsection{Effects of constrictions on the resonator properties}

\begin{figure}[phtb]
\centering
\includegraphics[width=0.8\textwidth]{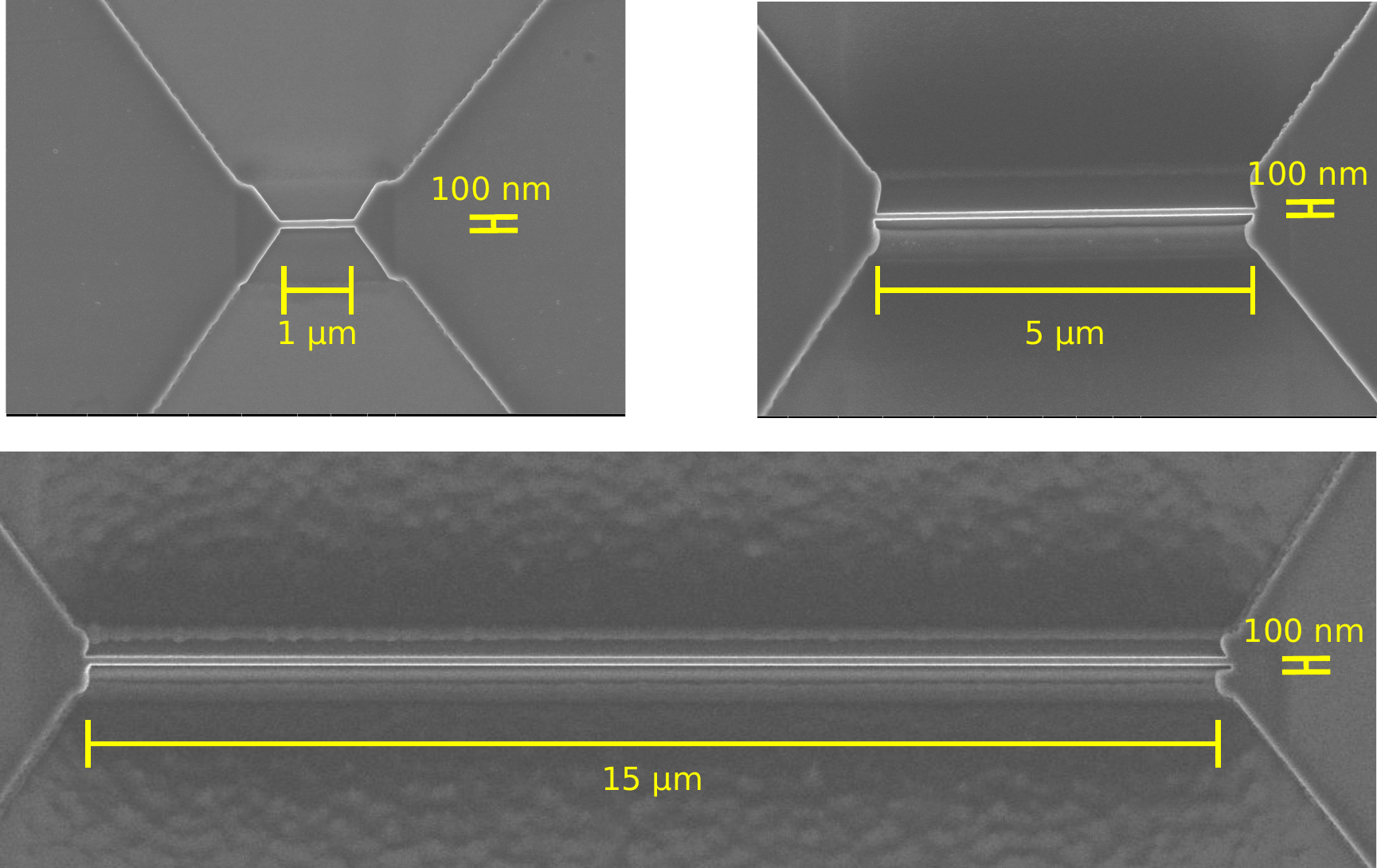}
\caption{Different length constrictions.  The microwave performance of this set of constrictions is compared in figure \ref{fig:constriction_comp}B}
\label{fig:constlength}
\end{figure}

\begin{figure}[phtb]
\centering
\includegraphics[width=0.8\textwidth]{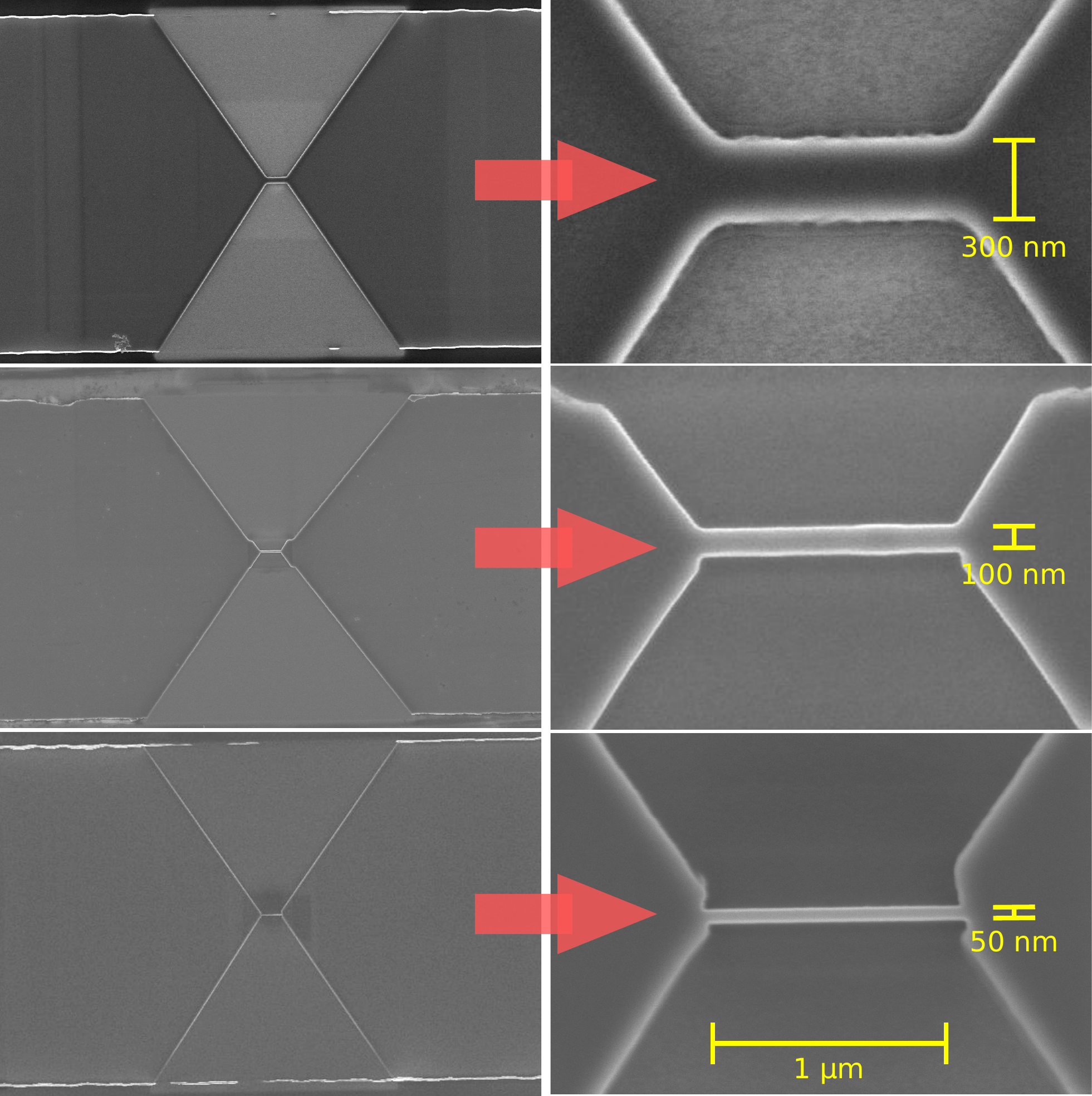}
\caption{Different width constrictions.  The microwave performance of this set of constrictions is compared in figure \ref{fig:constriction_comp}C}
\label{fig:constwidth}
\end{figure}

\begin{figure}[phtb]
\centering
\includegraphics[width=0.45\textwidth]{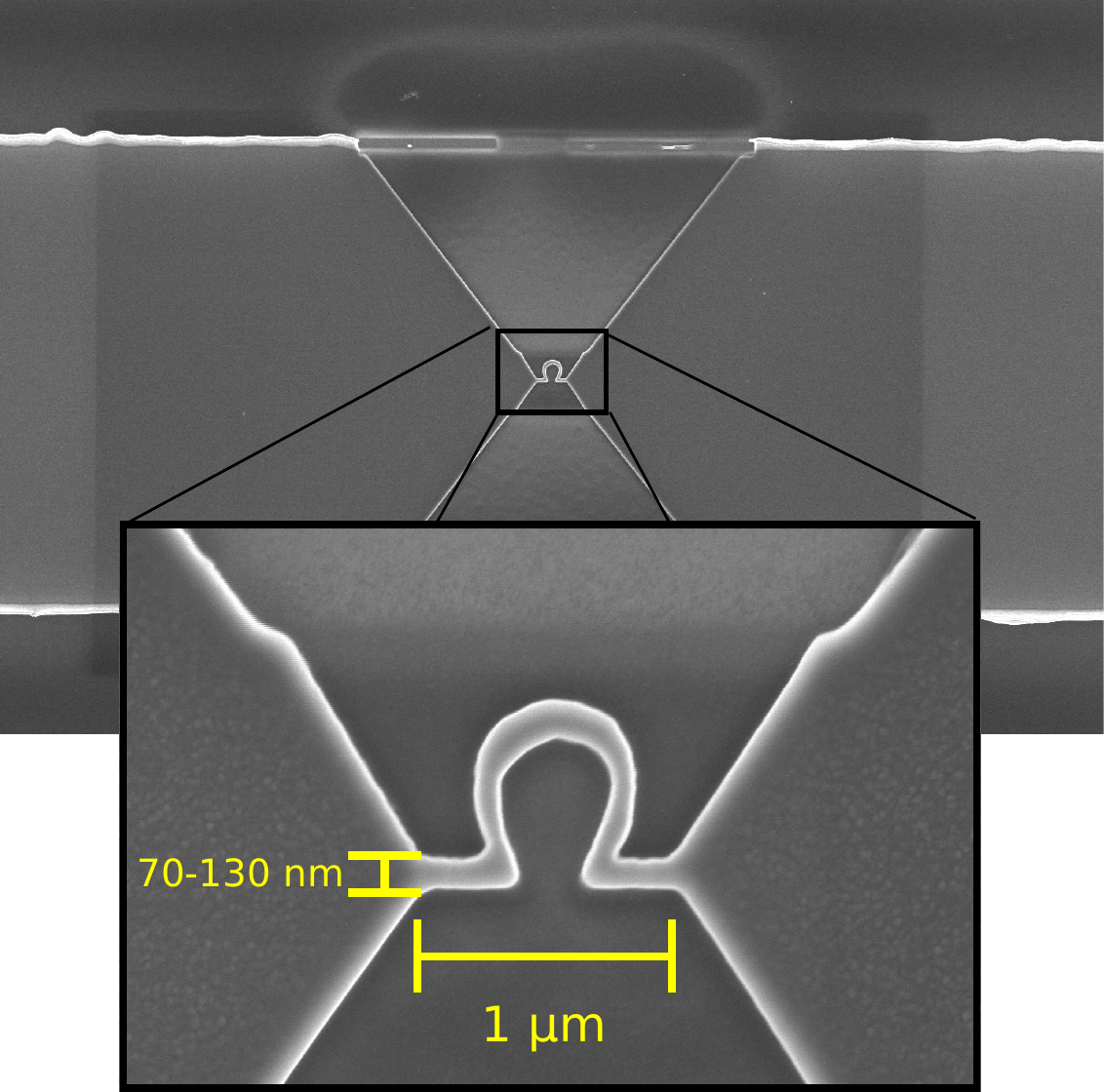}
\caption{Loop constriction  The microwave performance of this set of constriction is shown figure \ref{fig:constriction_comp}A}
\label{fig:constloop}
\end{figure}

\begin{figure}[htb]
\centering
\includegraphics[width=0.75\textwidth]{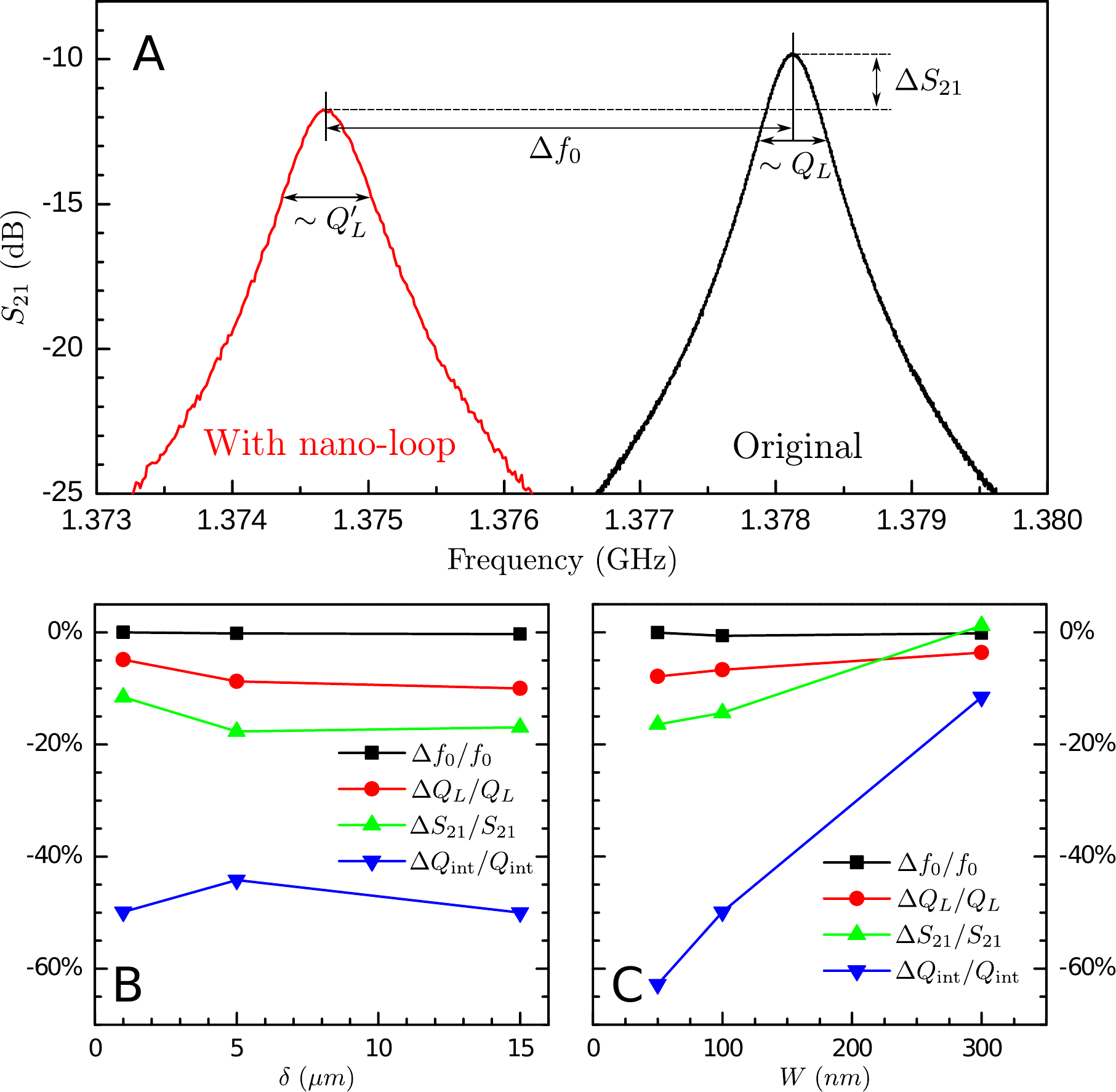}
\caption{Resonator performance for different constriction geometries. A) Resonance comparison parameters for the specific case of the nanoloop constriction (see figure \ref{fig:constloop}).  $\Delta f_0$, $\Delta Q_L$, $\Delta Q_{\rm int}$ and $\Delta S_{21}$ are the variations in resonance frequency, loaded quality factors, internal quality factors and transmission value (in linear scale) respectively.  In particular, the case of the nanoloop has $\Delta f_0/f_0 = -0.25\%$, $\Delta Q_L/Q_L = -22.7\%$, $\Delta Q_{\rm int}/Q_{\rm int} = -54\%$ and $\Delta S_{21}/S_{21} = -36\%$.  B) Variation of resonance parameters for a \SI{100}{\nano\metre} wide constriction and different lengths (see figure \ref{fig:constlength}).  Longer constrictions induce larger variations in the parameters.  C) Variation of resonance parameters for a \SI{1}{\micro\metre} long constriction and different widths (see figure \ref{fig:constwidth}).  Narrower constrictions induce larger variations.  $Q_{\rm int}$ values are calculated from insertion losses as in figure \ref{fig:1stharmonic}.}
\label{fig:constriction_comp}
\end{figure}

To check the changes to the resonance when making a constriction, we study two series of three identical resonators.  In each series we make constrictions varying either the length $\delta$ or the width $w$ and keeping the other constant.  The first series has constrictions of \SI{100}{\nano\metre} width and lengths of \SI{1}{\micro\metre}, \SI{5}{\micro\metre} and \SI{15}{\micro\metre} to explore the length dependence (figure \ref{fig:constlength}) with original values for $f_0\simeq \SI{1.35}{\giga\hertz}$ and $Q\simeq 800$.  Similarly, the second series keeps a constant length of \SI{1}{\micro\metre} while the width is varied through \SI{300}{\nano\metre}, \SI{100}{\nano\metre} and \SI{50}{\nano\metre} (figure \ref{fig:constwidth}) with original values for $f_0\simeq \SI{1.31}{\giga\hertz}$ and $Q_L\simeq 250$.  In order to show that other geometries can also be attained, we show the results for a single case of a loop constriction (see \ref{fig:constloop} and \ref{fig:constriction_comp}A) instead of a straight nanowire (original parameters, $f_0=\SI{1.378}{\giga\hertz}$ and $Q_L=2700$).  The results are analyzed by comparing the resonant frequencies, quality factors and transmission values measured before and after the constrictions were made.  The changes found are shown in figure \ref{fig:constriction_comp}.  As we can see, although there are some changes in the resonator properties, they remain broadly the same after having the constriction made.  The resonant frequency is altered only very slightly while the loaded quality factor changes less than 10\% in all cases giving us an indication that there are some extra losses induced in the system by the constriction.  These are possibly due to the fact that the constriction constitutes a defect for the radiation propagation and reflects some extra radiation from the source.  Since these additional losses must be internal, the effect on $Q_{\rm int}$ is more dramatic (up to 60\%).  These losses also seem to only depend on the width for the studied geometries.

The decrease of $f_{0}$ likely results from the enhancement of the inductance at the constriction. A simple approach is to model the constriction by a region of length $\delta$ and effective inductance per unit length, $l^{\prime}$, larger than its value $l$ outside this region. As it is described in \cite{Jenkins2013}, this effect makes the center line effectively longer for the propagation of RF currents and leads to a close to linear decrease of $f_{0}$ with increasing $\delta$:
\begin{equation}
\frac{\Delta f_{0}}{f_{0}} \simeq -\frac{\delta}{L}\left( \frac{l^{\prime}}{l} - 1 \right)
\label{eq:fvsdelta}
\end{equation}
where L is the resonator length.  The experimental data shown in figure \ref{fig:constriction_comp}B are compatible with $l^\prime/l = 10.9$.

\begin{figure}[htb]
\centering
\includegraphics[width=0.55\textwidth]{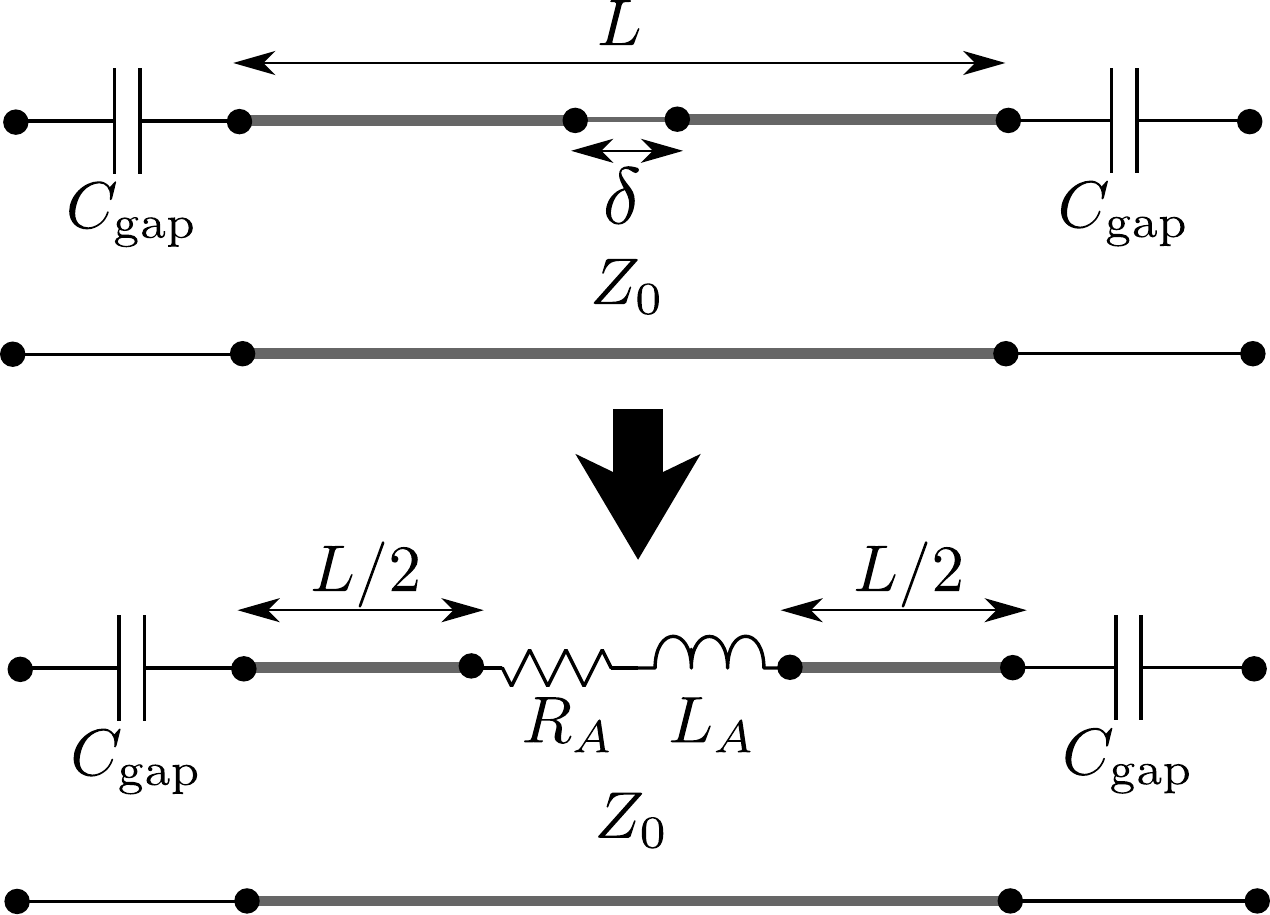}
\caption{Lumped circuit model for constriction}
\label{fig:constriction_circuit}
\end{figure}

The experimental results can also be modeled using an equivalent circuit with two additional lumped elements at the center of the resonator (figure \ref{fig:constriction_circuit}), an inductance $L_{\rm A}$ that accounts for changes of $f_{0}$, and a resistance $R_{\rm A}$ that accounts for $\Delta Q$.  The fitting procedure consists of first measuring the unconstricted resonator and fitting its length $L$ and gap capacity $C_{\rm gap}$ as was done for figure \ref{fig:1stharmonic}.  Keeping these values fixed, we then fit the transmission data after the constriction is made using only the additional lumped elements $L_A$ and $R_A$.  Again for the series of resonators in figure \ref{fig:constriction_comp}, we find that $L_{\rm A}$ lies in the range of a few tens of pH, increasing with $\delta$ as we expected from the above considerations, whereas the effective resistance $R_{\rm A}$ is of the order of a few m$\Omega$'s and fairly independent of $\delta$.

\begin{figure}[htb]
\centering
\includegraphics[width=0.65\textwidth]{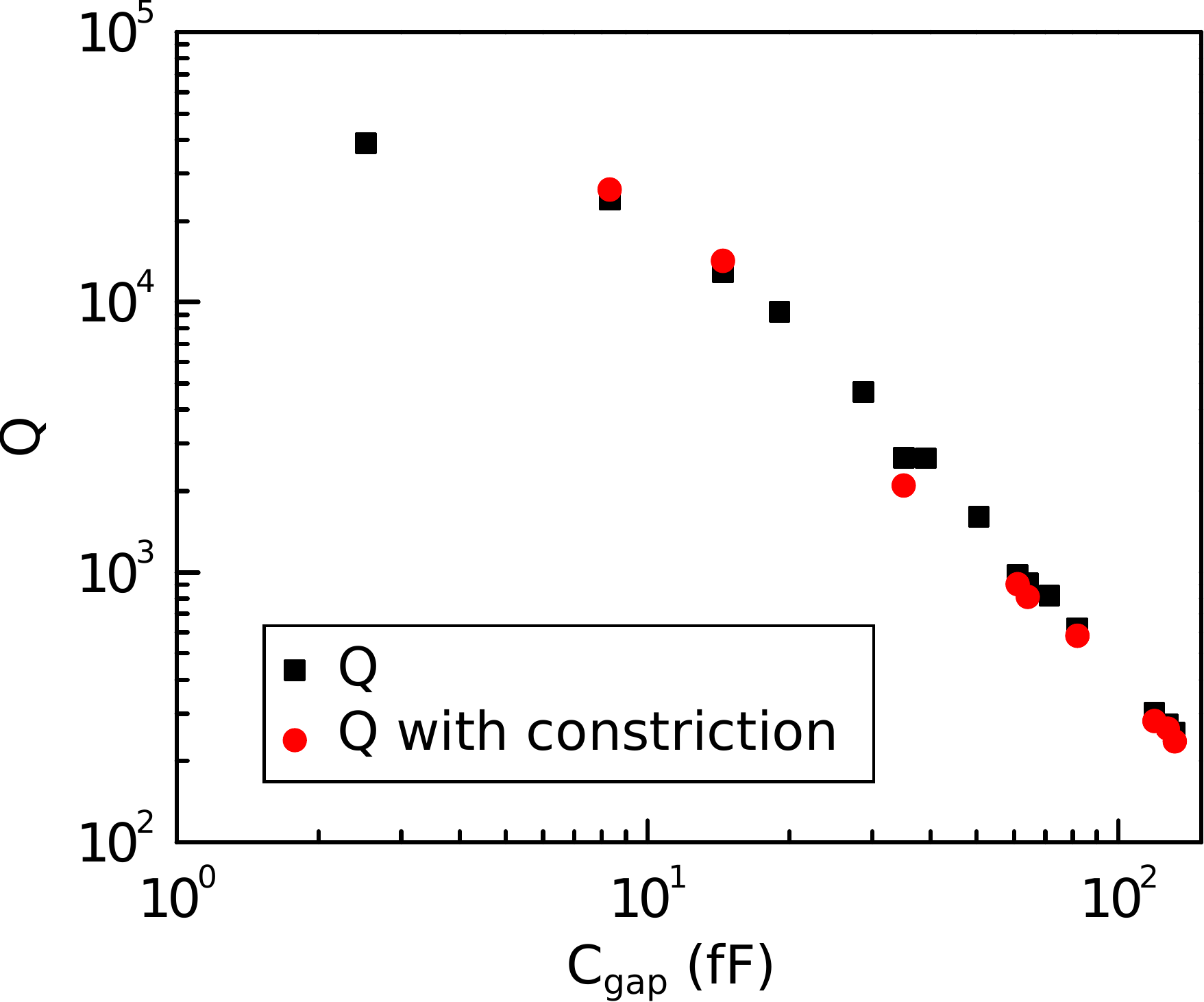}
\caption{Loaded Q-Factors for different resonators with and without constrictions.  The series studied in figure \ref{fig:constriction_comp} are on the right end of the graph while the additional higher Q resonators are on the left end.  Additional resonators without constrictions have also been added to the graph (squares with no circles).  The coupling capacitance is obtained using the same procedure as in figure \ref{fig:1stharmonic} and assumed to be equal before and after a constriction is made.}
\label{fig:QvsCin}
\end{figure}

To see what limits the internal losses may place on fabricating higher $Q$ resonators with constrictions we prepared some \SI{100}{\nano\metre} wide by \SI{15}{\micro\metre} long constrictions in higher $Q$ resonators and checked the change in the $Q_L$.  The effects on $Q$ are shown in figure \ref{fig:QvsCin} where we show the values of $Q$ as a function of the coupling capacitances for all our studied constrictions.  The two high Q cases measured are on the left end of the graph and actually present higher $Q_L$ after the constriction is made.  It is unclear why this happened but we assume that it is probably may be due to the resonator being cleaner during the constriction measurement.  This removes some of the internal losses which play a much more important role in this case since they a much larger influence on loaded quality factor than in the strongly coupled cases.  In any case this shows that the addition of constrictions need not dramatically influence the internal losses.  We also note that the high Q resonators were fabricated using the RIE etching while the low Q are from a batch made by liftoff (see section \ref{sec:liftoff} for details on these procedures).    Even before the constrictions are made, we usually get lower Q factors in the cases where liftoff is used than when RIE is used.  It may be the case that the lithography defects in the liftoff case (section \ref{sec:defects}), as well as introducing extra losses, make the resonators more sensitive to the addition of constrictions.

\subsection{Power effects and critical current density}
\begin{figure}[thb]
\centering
\includegraphics[width=\columnwidth]{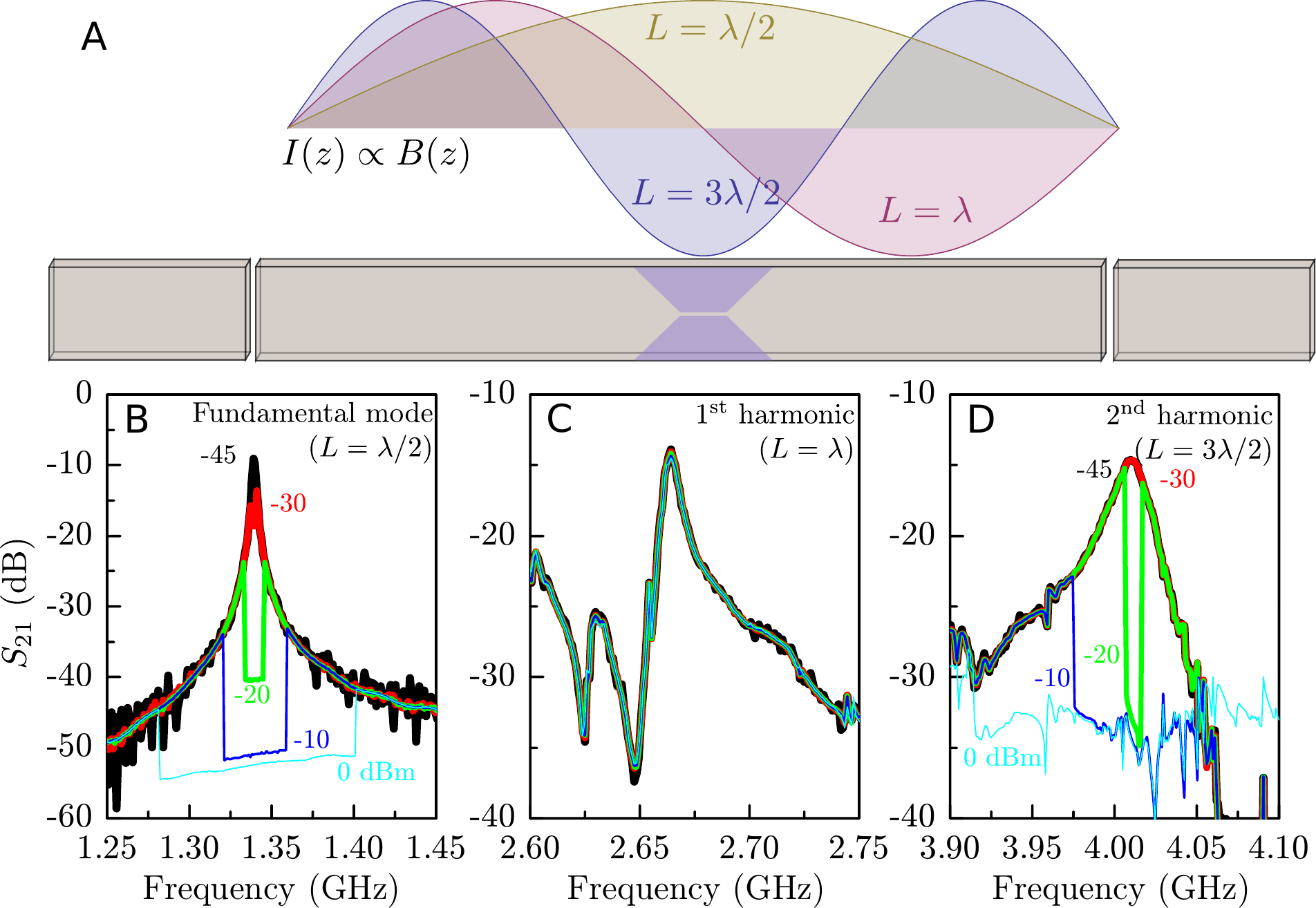}
\caption{Power effects on different resonator modes.  Diagram A schematically shows the centerline of the resonator and the different standing current waves for the first three resonant modes.  Graphs B,C and D graphs show the transmission spectra in a constricted resonator (\SI{1}{\micro\meter} long and \SI{100}{\nano\meter} wide) for these three resonant modes and for increasing excitation power (-45,-30,-20,-10 and 0 dBm).  Modes with a current maximum at the constriction position (The fundamental and second harmonic) show a loss of resonance when the power is increased while modes with no current at the constriction (first harmonic) show very small changes in the resonance.}
\label{fig:power}
\end{figure}

One important difference that we observe when adding the constrictions is the power dependence of the resonances.  Figure \ref{fig:power} shows the performance of a \SI{100}{\nano\metre} wide and \SI{1}{\micro\metre} long constriction at different excitation powers and for the first three cavity modes.  The fundamental mode (\ref{fig:power}B) and second harmonic (\ref{fig:power}D) break down when the microwave power is increased.  This can be explained by noting that, since both the fundamental mode and second harmonic have a standing wave with a current maximum at the center (as shown by figure \ref{fig:power}A) where the cross section has been drastically reduced, it is possible that the current density at the constriction exceeds the critical current for niobium and hence it is becoming resistive.  We find that there is no such effect in the case of the first harmonic. This is understandable along the same lines mentioned above, since there is then almost zero current at the constriction (for this mode, the current has a node at the center).  These experiments directly show evidence for the current flow through the nanowire and for the corresponding enhancement of the current density (thus also the magnetic field).

We can make an estimate of the critical current density in the nanowire by applying the definition of the Q-factor and the formulas (\ref{eq:lumpedL}) for the equivalent RLC circuit for the resonator.
\begin{equation}
Q=\omega\frac{E_{\rm stored}}{P_{\rm diss}}  \label{eq:Qdef}
\end{equation}
The stored energy can be related to the circulating current considering the lumped RLC model for the resonator and remembering that the energy stored in an RLC circuit on resonance is equal to the maximum magnetic energy stored in the inductor:
\begin{equation}
E_{\rm stored} = \frac{1}{2}L_lI_c^2
\end{equation}
Solving for $I_c$, substituting $L_l$ from equation (\ref{eq:lumpedL}) and using the definition of Q from (\ref{eq:Qdef}) we get an expression for the critical current in the nanowire:
\begin{equation}
I_c=\sqrt{\frac{\pi Q P_{\rm loss}}{Z_0}},
\end{equation}
where $P_{\rm loss}$ is the power loss from the resonator (dissipated or lost to the feed lines).  Since our measurement is stationary, we assume that the power arriving at the resonator is equal to the power being dissipated.  Since, on resonance, most of the power is dissipated in 2 equal length wires connecting the network analyzer to the resonator system (input and output), we assume that the dissipated power is half (in dB) that arriving at the output port of the network analyzer.

Using the data shown in figure \ref{fig:power}, we find that the resonance breaks down at an excitation power of $P_{\rm input} = -35$ dBm and an on resonance transmission of $S_{21} = -8.71$ dB.  Therefore $P_{\rm diss} = S_{21}/2 + P_{\rm input} = \SI{-38.4}\;\textrm{dBm} = \SI{0.11}{\micro\watt}$.  Taking the quality factor value of $Q=324$ and designed characteristic impedance of $Z_0\sim \SI{50}{\ohm}$ as well as the cross section of the nanowire (approximately \SI{150}{\nano\meter} thick by \SI{100}{\nano\meter} wide), we get a critical current density of approximately $j_c\simeq\SI{10e6}{\ampere\per\centi\metre\squared}$ which agrees with the order of magnitude of the critical current found for niobium thin films \cite{Huebener1975} and is slightly larger than the value found in our own measurements (see figure \ref{fig:Nbresist}).

\begin{figure}[thb]
\centering
\includegraphics[width=\columnwidth]{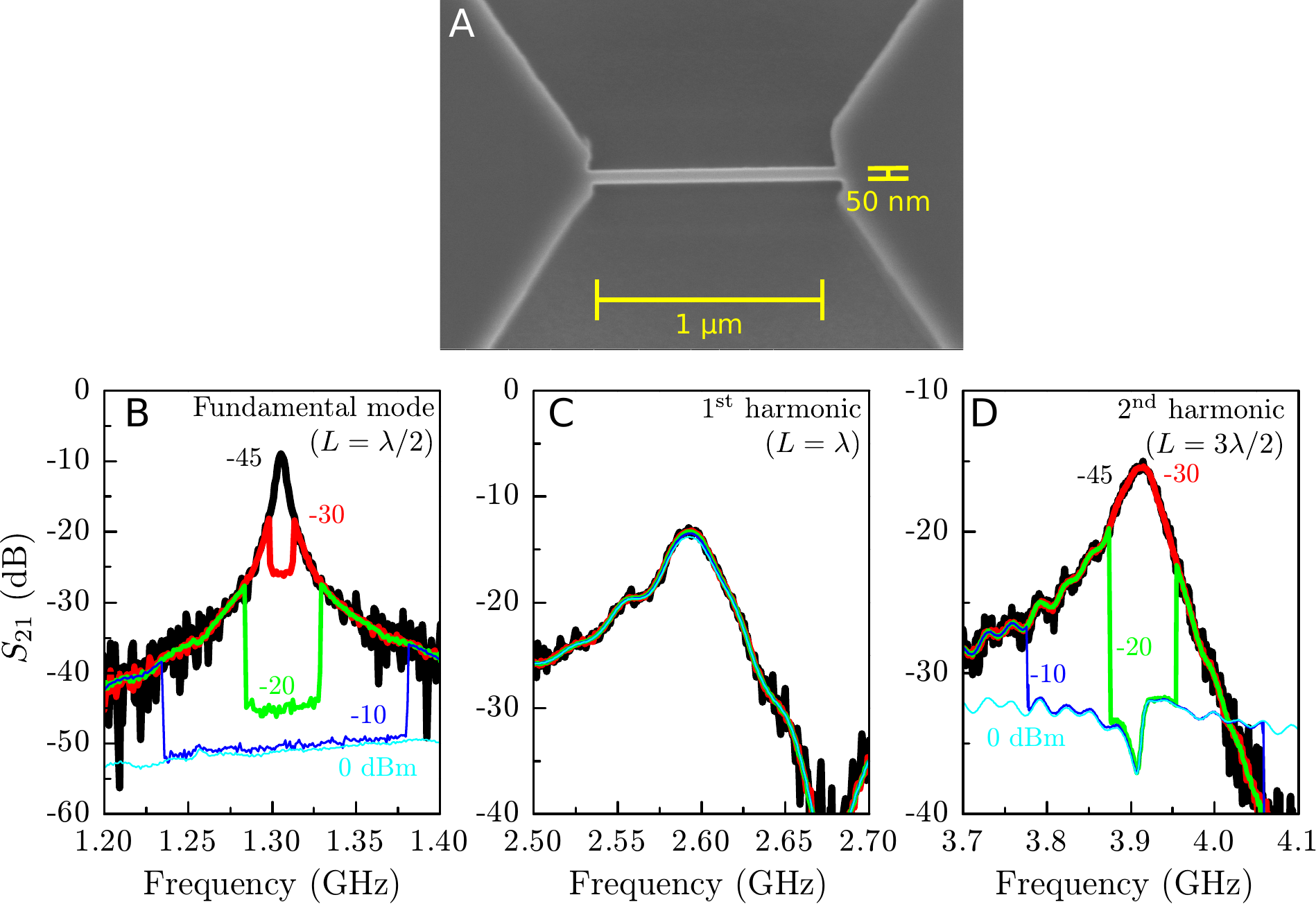}
\caption{Power effects on different resonator modes.  Diagram A shows the 50 nm wide constriction used in this case.  Graphs B,C and D graphs show the transmission spectra in a constricted resonator for these three resonant modes and for increasing excitation power (-45,-30,-20,-10 and 0 dBm).  Modes with a current maximum at the constriction position (The fundamental and second harmonic) show a loss of resonance when the power is increased while modes with no current at the constriction (first harmonic) show very small changes in the resonance.}
\label{fig:power50nm}
\end{figure}

The use of FIB in the fabrication of these constrictions could potentially implant Ga ions in a thin layer near the constriction edges.  This could potentially affect the superconducting properties of our constrictions by making these layers non-superconducting \cite{Hao2009}.  However, in our case the ionic currents used were kept low (around 20 pA) to achieve the maximum resolution in our structures.  According to \cite{Castan-Guerrero2014}, the regions where the implantation of Ga ions takes place are about 10-15 nm thick under these conditions.  This could reduce the effective width of our wires by at most 30 nm in the case that the Ga ions actually make the Nb a normal conductor.  However, in figure \ref{fig:power50nm} we show power sweep data for a 50 nm constriction and find that the calculated critical current density in this case is essentially the same as in the 100 nm case (we have in this case $P_{\rm diss} = -44.34\; \textrm{dBm}$, $Q=236$ and a cross section of \SI{150}{\nano\meter} thick by \SI{50}{\nano\meter} wide), with a value of $j_c\simeq\SI{9.8e6}{\ampere\per\centi\metre\squared}$.  This means that any implanted Ga ions can not be having much of an effect on conduction properties since the effective cross section should have been reduced by up to 50\% and should have reduced the critical current density.

\subsection{Performance in the presence of magnetic fields}

\begin{figure}[htb]
\centering
\includegraphics[width=\textwidth]{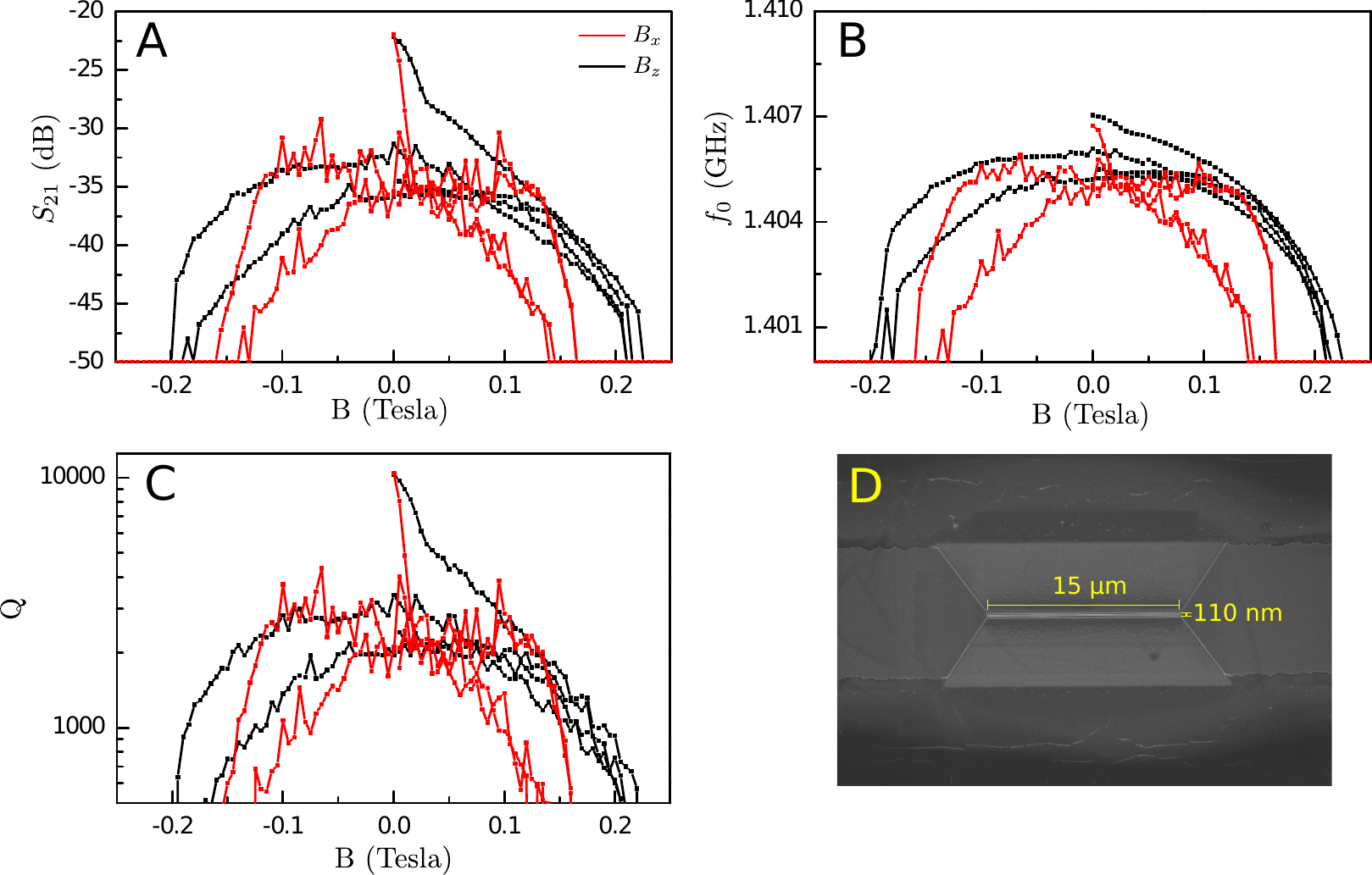}
\caption{Constricted resonator performance in the presence of magnetic fields.  The constriction is \SI{15}{\micro\meter} long and 110 nm wide.  In the absence of field, the original resonant frequency of the resonator is 1.407 GHz with a Q-factor of about 10000 and a transmission value of -22 dB.  Graphs A,B and C show the transmission, resonant frequency and quality factor as a function of the applied in plane magnetic field in the $B_x$ and $B_z$ axis directions (see figure \ref{fig:hist27R}A). A full hysteresis cycle was performed for each orientation (0 T, 0.5 T, -0.5 T, 0.5 T, 0 T).  Graph D shows an SEM image of the constriction.}
\label{fig:hist27R_constr}
\end{figure}

As we did in section \ref{subsec:resmag} for normal resonators, we now check the field dependence of the new resonators with constrictions.  A \SI{15}{\micro\meter} long and 110 nm wide constriction was made in the centerline (figure \ref{fig:hist27R_constr}D) of a resonator studied previously (figure \ref{fig:hist27R}).  Hysteresis cycles were performed for the two in plane axes using the superconducting vector magnet described in section \ref{sec:vecmag} to correctly align the magnetic field.  The results measured at $T=\SI{4.2}{\kelvin}$ are shown in figure \ref{fig:hist27R_constr}.

We see that the cycles in this case are somewhat more erratic than in the non-constricted case.  The curves are noisier, have asymmetries when inverting the polarity of the field and, in the $B_x$ case, the low field noise in still present.  Also the reproducibility of these curves is poor and there are significant differences between successive field sweeps.

We believe the reason for these effects is the vortices introduced in or near the constrictions have large effects on the circulating currents.  If they are found in this narrow area, the currents are forced to pass through them and suffer additional dissipation.  Also, the exact number of vortices, how they are pinned in the structure and their possible rearrangement when changing fields could have large effects in the overall transmission leading to the asymmetries and noise observed in our measurements.  These effects need to be taken into account when performing sample measurements and steps should be taken to avoid them.  The main precaution is to measure preferably starting from a non-magnetized state and to reset the system by heating it beyond $T_c$ after each measurement.  Also, oscillating the magnetic field to zero after each measurement is also helpful in restoring the resonator properties.  Sweeping the magnetic field at lower rates and in narrower ranges during measurements may also be helpful in avoiding vortex movement and rearrangement and make the response smoother.

\subsection{Local MFM measurement of the magnetic field enhancement near a nanoconstriction}

As an additional indication that there is an enhancement of the current density, and hence of the magnetic field, we locally probe the magnetic field generated by a current flowing through one of these constrictions using a magnetic force microscope (MFM) described in section \ref{sec:afmmfm}.  MFM is typically done with a vibrating AFM tip whose amplitude and phase with respect to the tip excitation are continuously monitored.  The tip is coated with a magnetic compound usually containing cobalt to make it sensitive to stray magnetic fields from the sample.  The actual measurement is done in two passes.  During the first pass, the tip is brought into close contact with the surface and is operated in tapping mode.  In this regime, van der Waals interactions between the tip and the substrate dominate.  The tip will then be only sensitive to the topography of the sample, which is recorded.  The second pass is done by lifting the tip a certain distance from the surface (usually of the order of 100 nm) and then following the recorded topography profile.  On account of the strong dependence of van der Waals forces on distance, this suppresses the topography signal from the measurement.  The magnetic signal can be seen as phase displacements in the second pass.  To a first approximation, these phase differences are proportional to the force gradient felt by the tip in the vibration direction (vertical).  The tip-sample interaction energy is approximately proportional to the tip magnetic moment times the stray field.  The force felt by the tip can then be obtained as minus the gradient of this energy:
\begin{equation}
E_{\rm int} = -\vec{\mu}_{\rm tip} \vec{B} \quad \Rightarrow \quad \vec{F} = \nabla(\vec{\mu}_{\rm tip} \vec{B})
\end{equation}
Assuming that the vibration direction is $y$, we get that the phase signal in MFM is proportional to second derivatives of the magnetic field:
\begin{equation}
\Delta\phi \propto = \frac{\partial F_z}{\partial y} = \frac{\partial^2}{\partial y^2}(\vec{\mu_{\rm tip}} \vec{B}) = \mu_x \frac{\partial^2B_x}{\partial y^2}+\mu_y \frac{\partial^2B_y}{\partial y^2}+\mu_z \frac{\partial^2B_z}{\partial y^2}
\end{equation}

\begin{figure}[htb]
\centering
\includegraphics[width=0.9\textwidth]{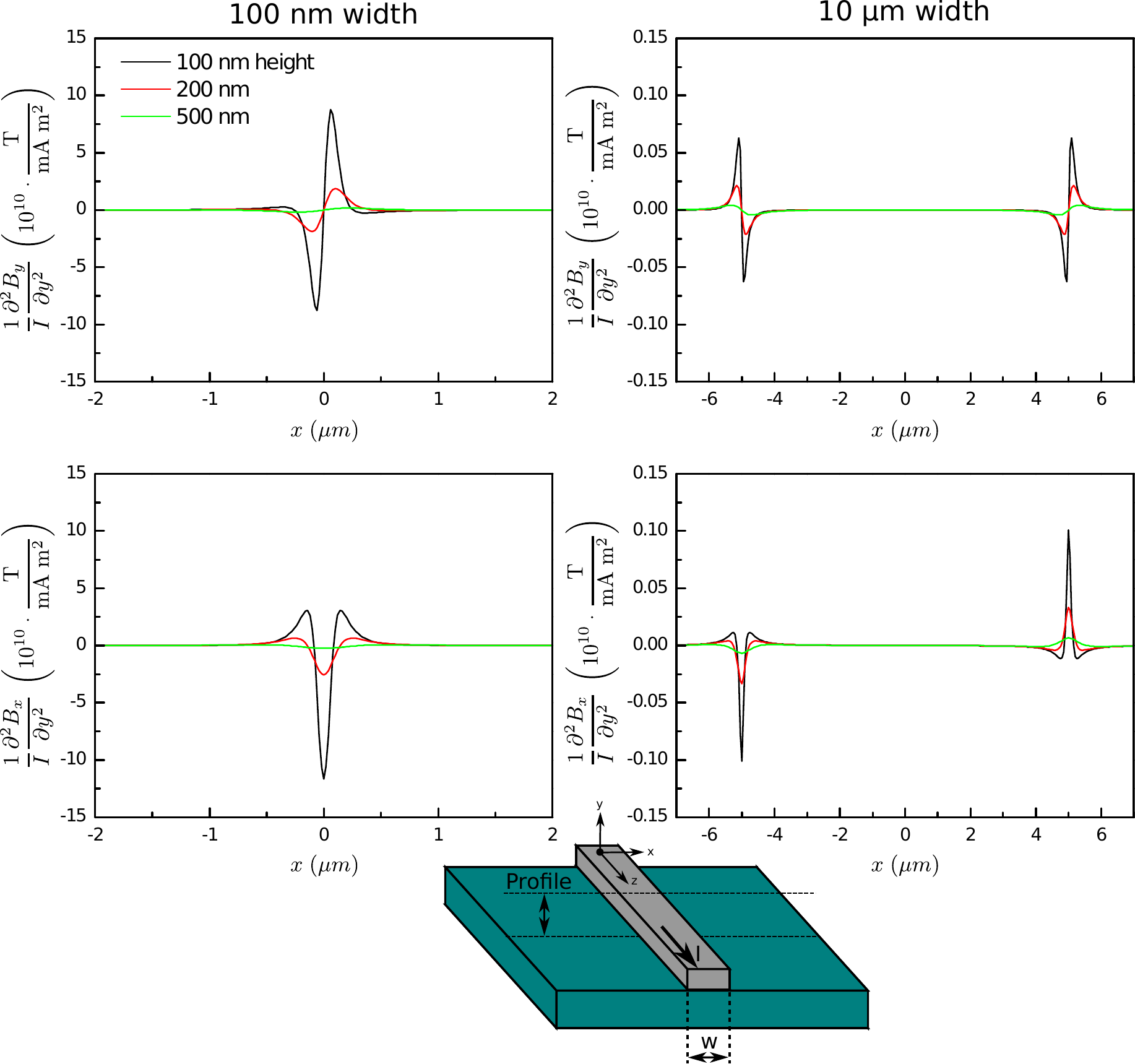}
\caption{Profiles for second derivatives of the magnetic field in the proximity of a rectangular wire.  The graphs show the derivatives of $B_y$ (top) and $B_x$ (bottom) while the graphs on the left are for a 100 nm width wire and the graphs on the right are for a \SI{10}{\micro\meter} width wire.  The lift separations shown are distances from the profile height to the top of the wire surface (100 nm further from the surface).}
\label{fig:MFMtheory}
\end{figure}

Values for these derivatives can be obtained numerically for simple geometries so we can compare the results to experimental MFM signals.  We calculate second derivatives of the magnetic field components generated by 100 nm high wires.  We consider two different wire widths: \SI{10}{\micro\meter}, typical of ``normal'' resonators, and 100 nm, thus of the order of the constrictions made by ion-beam lithography.  Cross sections of the calculated profiles are shown in figure \ref{fig:MFMtheory}. It shows the general behavior of the derivatives of the two non-zero components $B_x$ and $B_y$.  MFM tips are typically magnetized in the y direction, so the signals should qualitatively be similar to the $\frac{\partial^2 B_y}{\partial y^2}$ profiles, with a signal featuring positive and negative edges.  Also, narrowing the wires increases the simulated signal by a factor of about 100 as can be seen observing the scales in figure \ref{fig:MFMtheory}

\begin{figure}[p]
\centering
\includegraphics[width=0.7\textwidth]{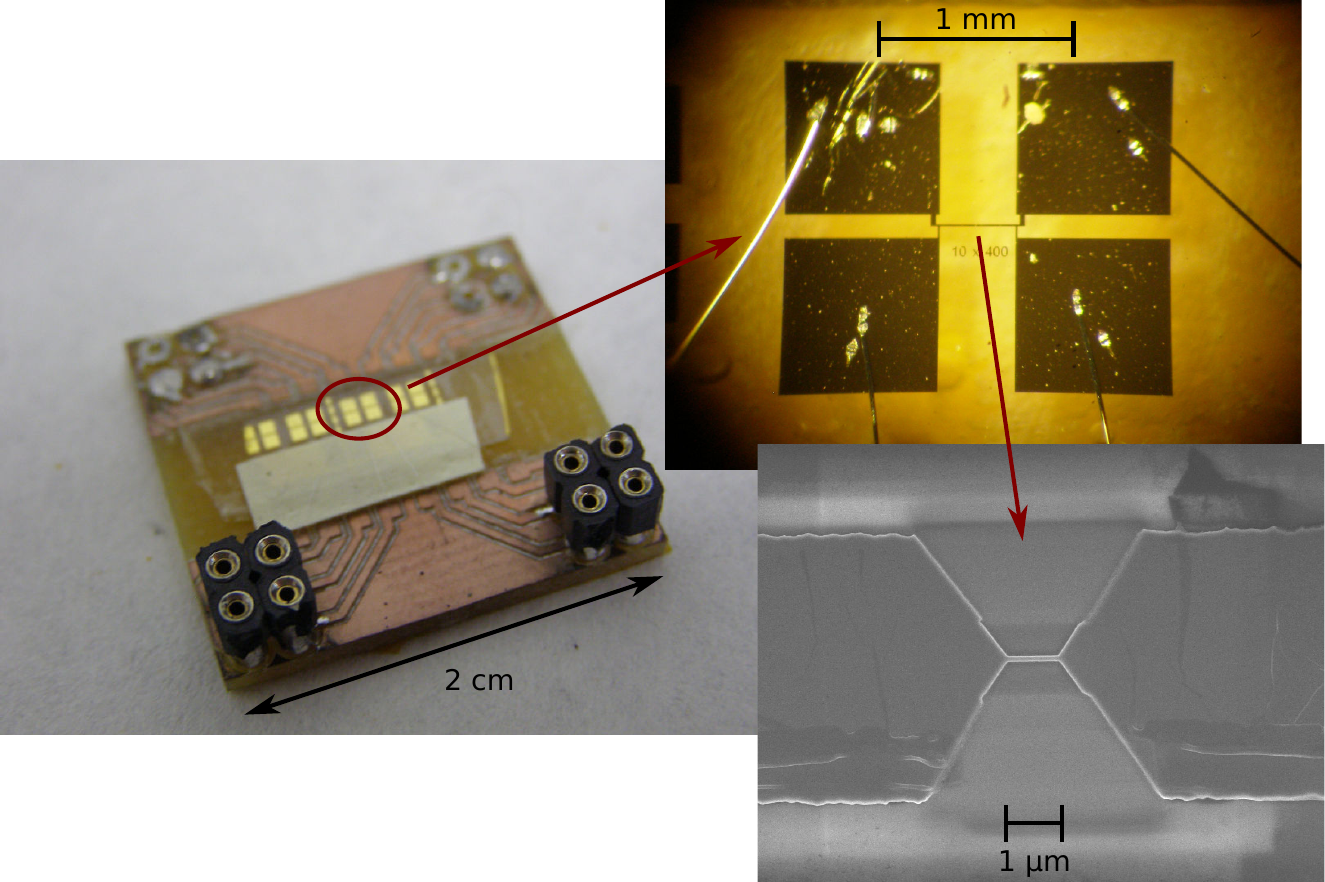}
\caption{Experimental setup for MFM measurement of gold nanowire.  The circuit is mounted on a fiberglass sample holder with copper contacts that are wire bonded to the gold circuit.  The gold circuit (100 nm layer thickness) is constricted at its center using FIB in the same way as the CPW resonator case.  The dimensions of the constriction are \SI{1}{\micro\meter} long by 100 nm wide.}
\label{fig:MFMsetup}
\end{figure}

\begin{figure}[p]
\centering
\includegraphics[width=\columnwidth]{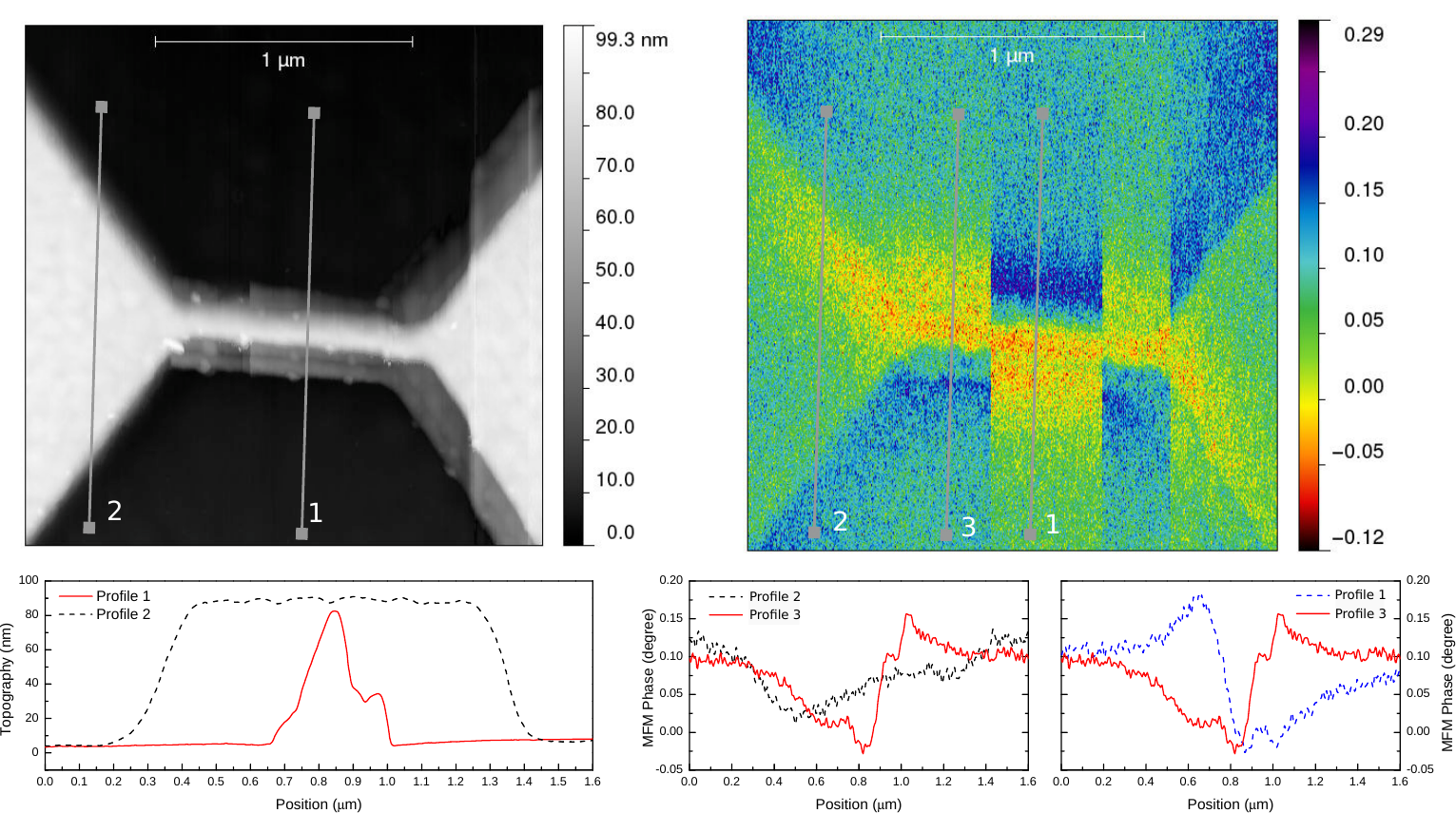}
\caption{Magnetic force microscopy image of a \SI{1}{\micro\metre} by \SI{100}{\nano\meter} constriction.  The circuit is made from a  \SI{100}{\nano\metre} thick gold layer on a sapphire substrate and has a \SI{2.1}{\milli\ampere} DC current.  The graphs on the left show the wire topography while the images on the right show the MFM signal and MFM profiles for different locations.  We see that there is a clear magnetic signal in the wire that is not visible in the wider area.  Also, the changes in contrast in the image correspond to flipping the current direction during the image acquisition.  We also note that there is a appreciable electrostatic background signal.}
\label{fig:afm}
\end{figure}

For the MFM measurements we fabricate a \SI{1}{\micro\metre} long by \SI{100}{\nano\meter} wide constriction out of a \SI{100}{\nano\metre} thick gold layer on sapphire.  This constriction is made on a simple gold circuit with 4 pads, 2 to inject the current and 2 to measure the voltage drop in the wire.  The sapphire is glued to a sample holder with larger copper pads that are wire bonded to the gold circuit.  This whole ensemble is then placed on the AFM stage and connected to a current source and a voltmeter.  A \SI{2.1}{\milli\ampere} DC current is made to circulate in either direction along the nanowire while experiments are performed at room temperature.  Topography and magnetic phase images are shown in \ref{fig:afm}.  These images qualitatively show an enhancement of the magnetic signal (a magnetic phase contrast of $\Delta \phi = 0.26^\circ$ from maximum to minimum) in the constriction area and a negligible signal (indistinguishable from background) in the wider areas of the circuit.  The current was reversed three times during the image acquisition.  These correspond to the contrast changes in the phase image, thus showing that the signal must be magnetic in origin and due to the circulating current.  We also note a constant background signal probably due to an electrostatic potential difference between the metalized and non-metalized areas.  If greater precision were needed, this background could possibly be filtered out using schemes similar to those detailed in \cite{Yongsunthon2001} where electric potential nulling is used.  However, our measurements here are sufficient to conclude that there is a strong field enhancement near the nanowire and that the behavior is qualitatively similar that shown in figure \ref{fig:MFMtheory} for a vertically magnetized tip.

\section{Conclusions}\label{sec:conclusionsCPWG}

In this chapter we have introduced superconducting coplanar waveguide resonators as interesting devices for quantum information processing.  We firstly summarized the distributed circuit model to describe their basic properties required for their design.  We then designed, fabricated and tested resonators with operating frequencies in the 1.4 GHz range with quality factors ranging from 100 to 40000.  We also studied their performance at 4.2 K with and without applied magnetic fields.  We find that magnetic fields can have large effects on the resonators by introducing extra losses in the resonators and generating hysteresis effects related to the appearance of Abrikosov vortices.  We also see these hysteresis effects measuring magnetic susceptibility in Nb thin films.

Once this basic characterization is done, we use FIB to create nanometric constrictions in the centerline with the objective of concentrating the current and enhancing the magnetic field.  After testing these designs we conclude that they do in fact concentrate the magnetic field close to the nanoscale constrictions in the center line.  These devices could potentially be used to achieve strong coupling to magnetic qubits and for nano-EPR experiments.  The resonance characteristics remain largely unchanged although some additional internal losses are added.  Although the resonator frequency is relatively low (\SI{1.3} to \SI{1.4}{\giga\hertz}), we see no reason why these designs should not also work well at higher microwave frequencies.  Due to the reduced wavelength at higher frequencies it is possible that there are somewhat larger changes for longer constrictions since they would be larger compared to the shorter wavelength.  However, the wavelength $\lambda$ remains much larger than our typical constriction length for frequencies $f<\SI{50}{\giga\hertz}$.   Direct evidence of the current concentration is seen by the power effects in the different harmonics of the resonator where we see that for high powers and modes that have current maximums at the constriction position, there is a suppression of the superconductivity.  Also, local measurements of DC currents in nanowires show a magnetic field enhancement directly.  We propose that these systems are promising candidates for further applications in quantum information control and in the characterization of nanoscopic magnetic samples.

\Urlmuskip=0mu plus 1mu\relax
\bibliographystyle{h-physrev3}
\bibliography{mybiblio}

\chapter{Experiments on the coupling of magnetic samples to Coplanar Waveguides and Resonators}\label{chap:samp}
\chaptermark{Coupling samples to Coplanar Waveguides and Resonators}

\section{Introduction}
In the previous chapters we have established that a possible method of building a quantum processor is to use Single Molecule Magnets (SMMs) and, in particular, Single Ion Magnets (SIMs) (chapter \ref{chap:SIMs}) in conjunction with circuit QED systems, namely superconducting CPW resonators.  The strong coupling limit is expected to be achievable using SIMs with large enough samples or using nanometric constrictions (section \ref{sec:resonators}).

The next obvious step would the be to measure this strong coupling.  However, technical limitations at our laboratory currently make it difficult to measure samples in resonators at frequencies beyond about \SI{1.5}{\giga\hertz}.  Therefore, the interesting cases like \ce{GdW30} (section \ref{sec:GdW30}) and \ce{TbW30} (section \ref{sec:TbW30}) that have higher operating frequencies are not currently accessible with our hardware (some transitions between excited states of \ce{GdW30} are accessible), although efforts are being made to improve our systems and such experiments will be possible in the near future.

However, we can test the spectroscopic capabilities of these devices with other samples allowing us to also test whether the nanometric constrictions described in section \ref{sec:constrictionsCPWG} improve the coupling for very small samples.  Additionally, broadband measurements up to \SI{14}{\giga\hertz} can be performed with our systems using Coplanar Waveguide (CPW) transmission lines instead of resonators.  Although the coupling is much lower than in the resonator case, absorption can be detected in a much wider range of frequencies.  This allows the direct mapping of the energy level separation as the magnetic field is swept \cite{Clauss2013}.

We will present results for three different samples:
\begin{itemize}
\item DPPH (2,2-diphenyl-1-picrylhydrazyl)\cite{Kolaczkowski1999}, an EPR calibration sample previously introduced in section \ref{sec:coupCPWG_SMM}.
\item A \ce{Gd} fluoride, \ce{Ca_{1-y}Gd_{y}F_{2+y}}.  Rare earth ions doped into fluoride crystals are one of the basic systems originally used for study of rare earth paramagnetic properties \cite{Abragam1970} in crystal environments.  They exhibit a well known crystal structure and have good stability.
\item A \ce{GdW30} crystal similar to those described in section \ref{sec:xrayGdW30}.  Although the transitions that are most interesting for quantum computation applications lie beyond our resonator operating frequency, some transitions could potentially be visible.
\end{itemize}

The chapter is organized as follows.  In section \ref{sec:wg} we will measure absorption for the different samples on a superconducting CPW transmission line.  Next, we review the theoretical framework describing a system of coupled spins to a resonator system in section \ref{sec:seiji}.  In section \ref{sec:res} we investigate the coupling to standard CPW resonators of different sizes to crystals and droplets.  Later, in section \ref{sec:DPPHconstriction} we will evaluate the performance of a nanometric constriction in improving the coupling to a small sample.  Finally in section \ref{sec:conclusionsMeas} we give the conclusions for this chapter.

\section{Broadband spectroscopy using superconducting CPW}\label{sec:wg}
One disadvantage of standard EPR measurements is that the use of a cavity restricts the working frequencies.  As described in chapter \ref{sec:EPRtech}, EPR setups are built to work in a specific frequency band given by the cavity geometry and its microwave source.  The cavity allows electromagnetic energy to be accumulated within and higher rf fields to be attained allowing the measurement of smaller samples and signals.

It is also possible to measure a sample directly on a transmission line such as those based on CPW designs (chapter \ref{chap:CPWG}).  An open waveguide allows transmission through it with, in principle, no frequency limitation\footnote{At high enough frequencies, the losses would eventually be too high to allow small signals to be detected.} allowing broadband spectroscopy in a single device.  This however comes at the cost of a much reduced energy density in the sample volume as the photons only \emph{see} the sample once as they pass through the device.  In the cavity setup they are stored in a certain volume where they have a higher chance of interacting with the sample before they are dissipated.  Hence, all else being equal, the signal changes associated with absorption of rf photons by a magnetic sample are much weaker when using a transmission line than when a cavity is used.  However, if the coupling is strong enough or the samples large enough, monitoring the transmission through such a system while sweeping the source frequency would allow us to detect an absorption line when the frequency coincides with an energy transition in the sample.  Also, if the transitions depend on an applied magnetic field for instance, it would be possible to map the changes in the energy level spectrum.

As an initial test of our spin systems and as a test of our setup, measurements were performed using transmission lines fabricated and mounted using the same procedures as those used for resonators in chapter \ref{chap:CPWG}.  These measurements also allow us to work with higher frequencies than our current resonators can achieve (up to \SI{14}{\giga\hertz} on a waveguide instead of $\sim\SI{1.5}{\giga\hertz}$ for resonators).  All experiments are performed at liquid helium temperatures (\SI{4.2}{\kelvin}).

\subsection{DPPH on a superconducting CPW}\label{sec:DPPHwg}

\begin{figure}[!tb]
\centering
\includegraphics[width=1.\textwidth]{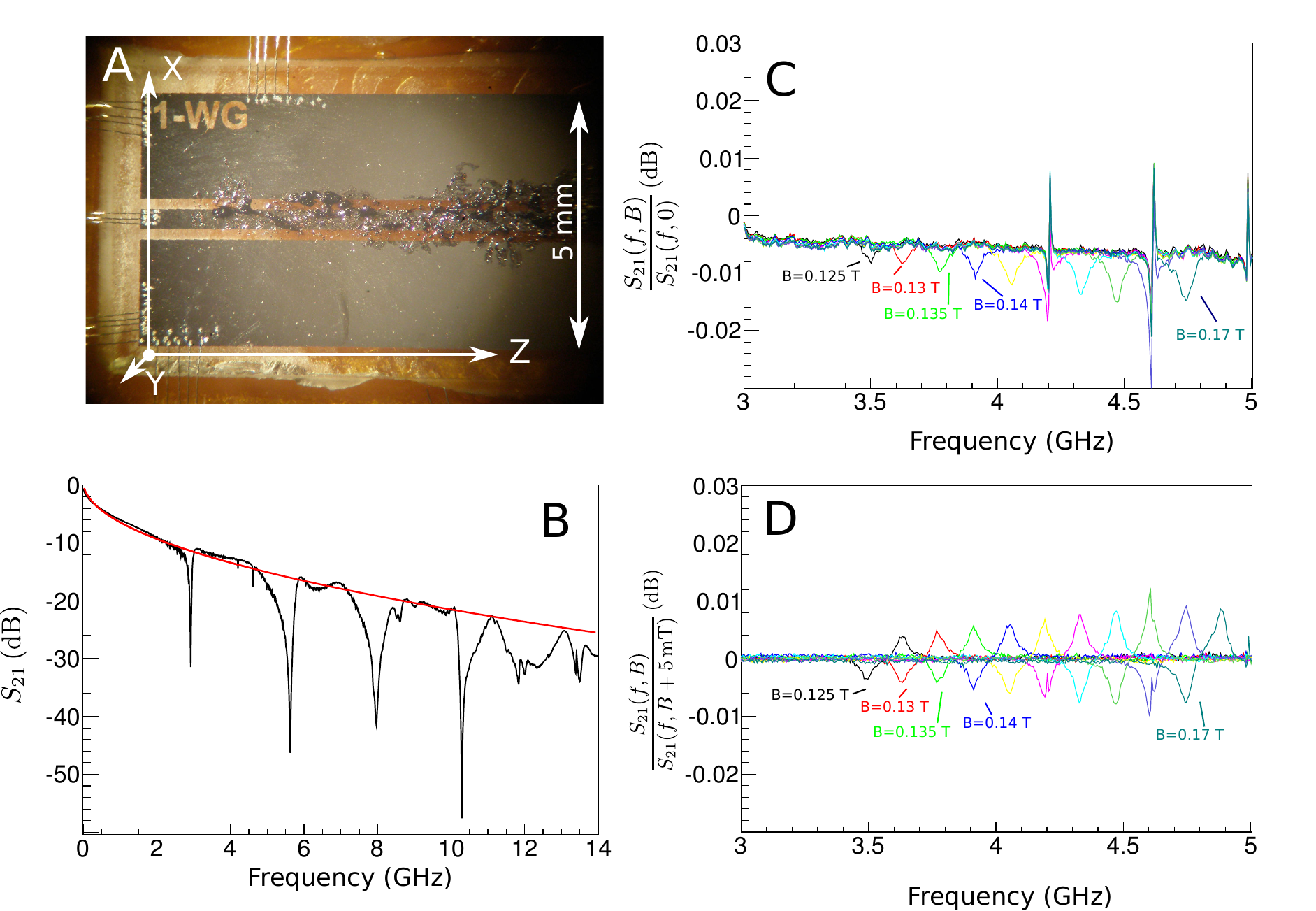}
\caption{DPPH on a niobium CPW transmission line.  Graph A shows a microscope image of the device and sample.  Graph B shows the background transmission spectrum including connecting wires.  The red line shows the expected transmission from a waveguide with only resistive losses (from connecting wires).  Graph C shows a close up of the transmission $S_{21}$ normalized by the transmission at zero field.  Each color was taken at fields \SIlist{0.125;0.13;0.135;0.14;0.145;0.15;0.155;0.16;0.165;0.17}{\tesla} respectively.  Graph D shows the same spectra but each of them normalized by the spectrum at a \SI{5}{\milli\tesla} higher field.}\label{fig:wg1photo}
\end{figure}

\begin{figure}[!tb]
\centering
\includegraphics[width=1.\textwidth]{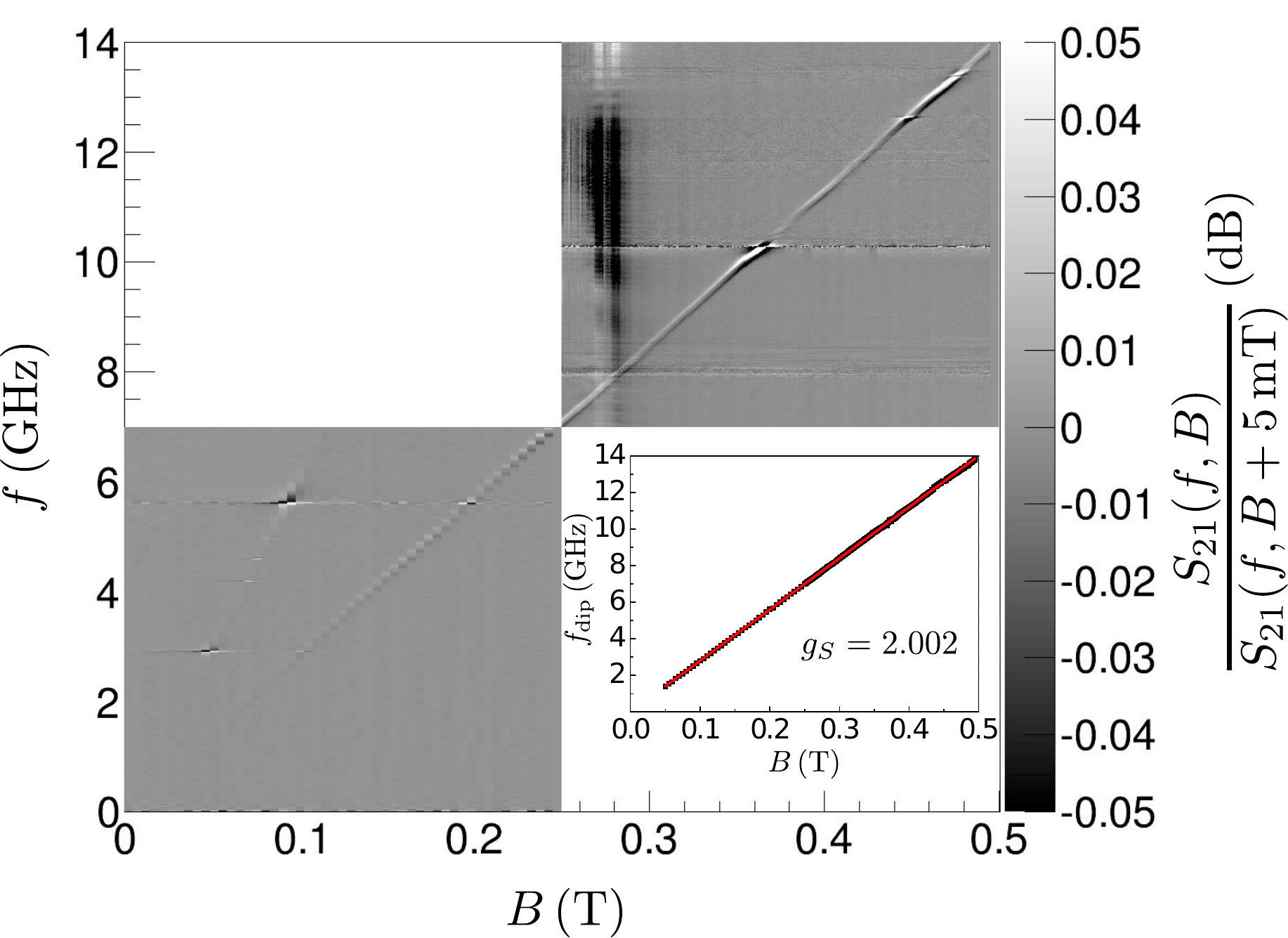}
\caption{The main graph shows the normalized $S_{21}$ transmission as a function of both frequency and magnetic field for DPPH an a niobium CPW transmission line.  An image of the transmission line and the sample is shown in figure \ref{fig:wg1photo}.  The spectrum at each field is normalized by the spectrum at a field \SI{5}{\milli\tesla} higher.  The peak-dip frequency was extracted and is shown in the inset.  The slope gives the g-factor for DPPH.  The measurement was done in two separate sweeps and the blank areas were not measured.}\label{fig:DPPH1}
\end{figure}

As a first test, a Nb on sapphire CPW with a \SI{200}{\micro\metre} gap and \um{400} centerline was used.  A large amount of DPPH powder mixed with Apiezon N grease was spread over the surface (figure \ref{fig:wg1photo}A).  The DPPH spin density can be found using its density ($\rho = \SI{1.4}{\gram\per\centi\meter\cubed}$) and molecular weight ($M_{\rm mol} = \SI{394.32}{\gram\per\mole}$) \cite{Kiers1976} to be \SI{3.55e-3}{\mole\per\centi\meter\cubed}.  The device was then mounted on the \SI{4}{\kelvin} probe and inserted into a superconducting magnet.  The transmission is measured from \SI{10}{\mega\hertz} to \SI{14}{\giga\hertz} while the field B is swept from \SI{0}{\tesla} to \SI{0.5}{\tesla} applied in the Z direction marked in figure \ref{fig:wg1photo}A.  The rf field will be perpendicular to the DC field which, as discussed in section \ref{sec:coupCPWG_SMM}, is the optimal configuration to induce transitions in the sample.  Since DPPH is an spin $1/2$ system with $g\simeq 2$, there should be absorption at frequency:
\begin{equation}
f_0 = \frac{\Omega}{2\pi} = \frac{g_S\mu_{\rm B}}{h}B = (\SI{28}{\giga\hertz\per\tesla})B  \label{eq:dpphlevel}
\end{equation}

The transmission $S_{21}(f,B)$ is a function of the drive frequency and of the DC magnetic field.  The background including the connecting wires at \SI{0}{\tesla} (equivalent to the empty transmission line case) is shown in figure \ref{fig:wg1photo}B.  When compared to the red line representing an ideal waveguide with resistive losses, we see that the profile is very irregular.  However, this \emph{noise} is stable and reproducible from one frequency sweep to the next indicating that these irregularities probably originate from the connections to the device (wire bonds, intermediate connectors, PCB design and materials, etc.).  The sample absorption is too small to be directly distinguished from this noise at a fixed field.  However, if the field is changed, the background irregularities vary slowly and do not in general move from their position.  On the other hand, the sample absorption peak does move when the field is changed according to equation (\ref{eq:dpphlevel}).  Therefore, if we divide each spectrum by, for instance, the signal at \SI{0}{\tesla}, and represent $S_{21}(f,B)/S_{21}(f,\SI{0}{\tesla})$ for different fields we see an absorption peak that changes its frequency for each field (figure \ref{fig:wg1photo}C).  The sample absorption signal is visible in this case but changes in background signal when the magnetic field is increased produce artifacts and distortions in what should be a flat graph with a dip at the corresponding frequency.  Better results can be obtained if we normalize instead by a spectrum with a field value close to the current field.  For example, figure \ref{fig:wg1photo}D shows $S_{21}(f,B)/S_{21}(f,B+\SI{5}{\milli\tesla})$, i.e, the transmission normalized by the signal at a field \SI{5}{\milli\tesla} higher.  We see a much flatter background and both a peak and a dip.  The dip corresponds to the current field, while the peak corresponds to the spectrum used to normalize.  We will use this last method of normalization in most cases.

\begin{figure}[!p]
\centering
\includegraphics[width=1.\textwidth]{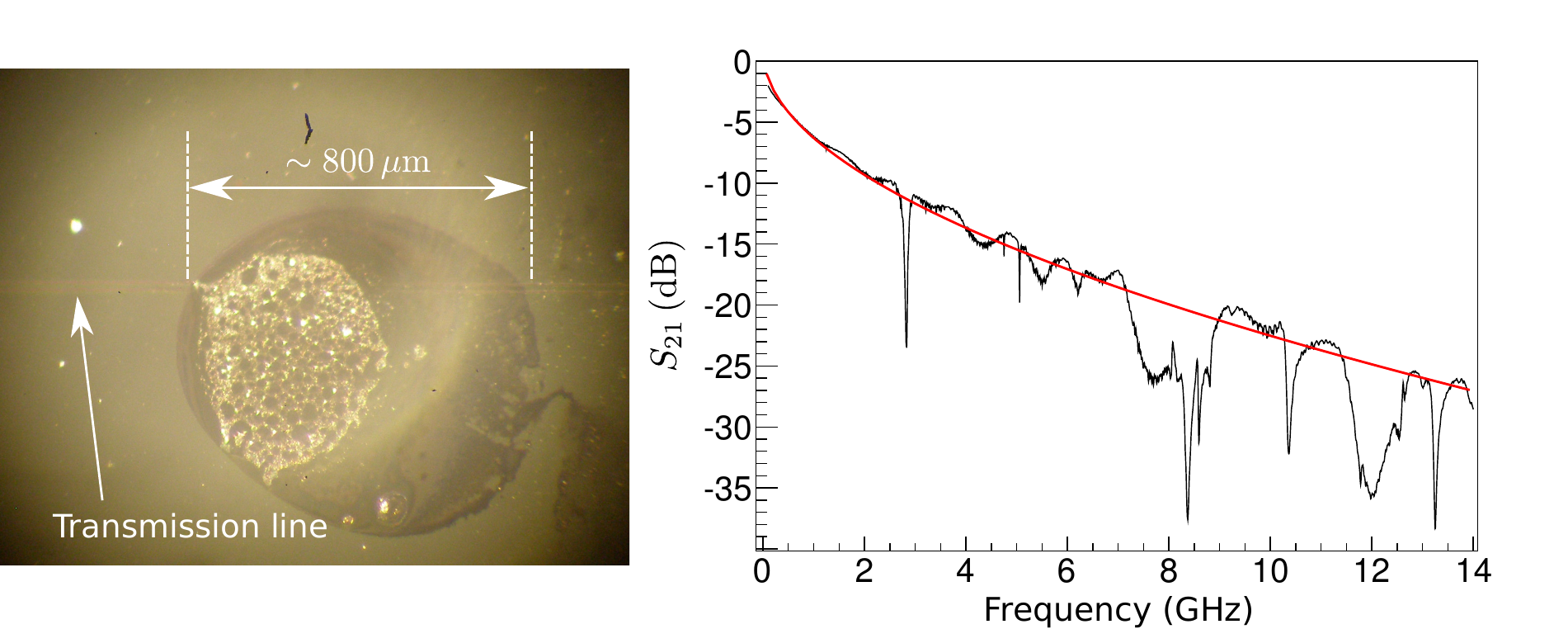}
\caption{Thin transmission line (\um{7} gaps and \um{14} centerline) and a drop of DPPH placed on top.  Only the parts of the drop close (within about 10-\um{20}) to the transmission line contribute to the signal.  The graph shows the background transmission ($S_{21}$).  The red line is the behavior of an ideal waveguide with dielectric and resistive losses (from connecting wires).  The dielectric losses are included in this case to better follow the measured dependence.}\label{fig:wg2photo}
\end{figure}

\begin{figure}[!p]
\centering
\includegraphics[width=1.\textwidth]{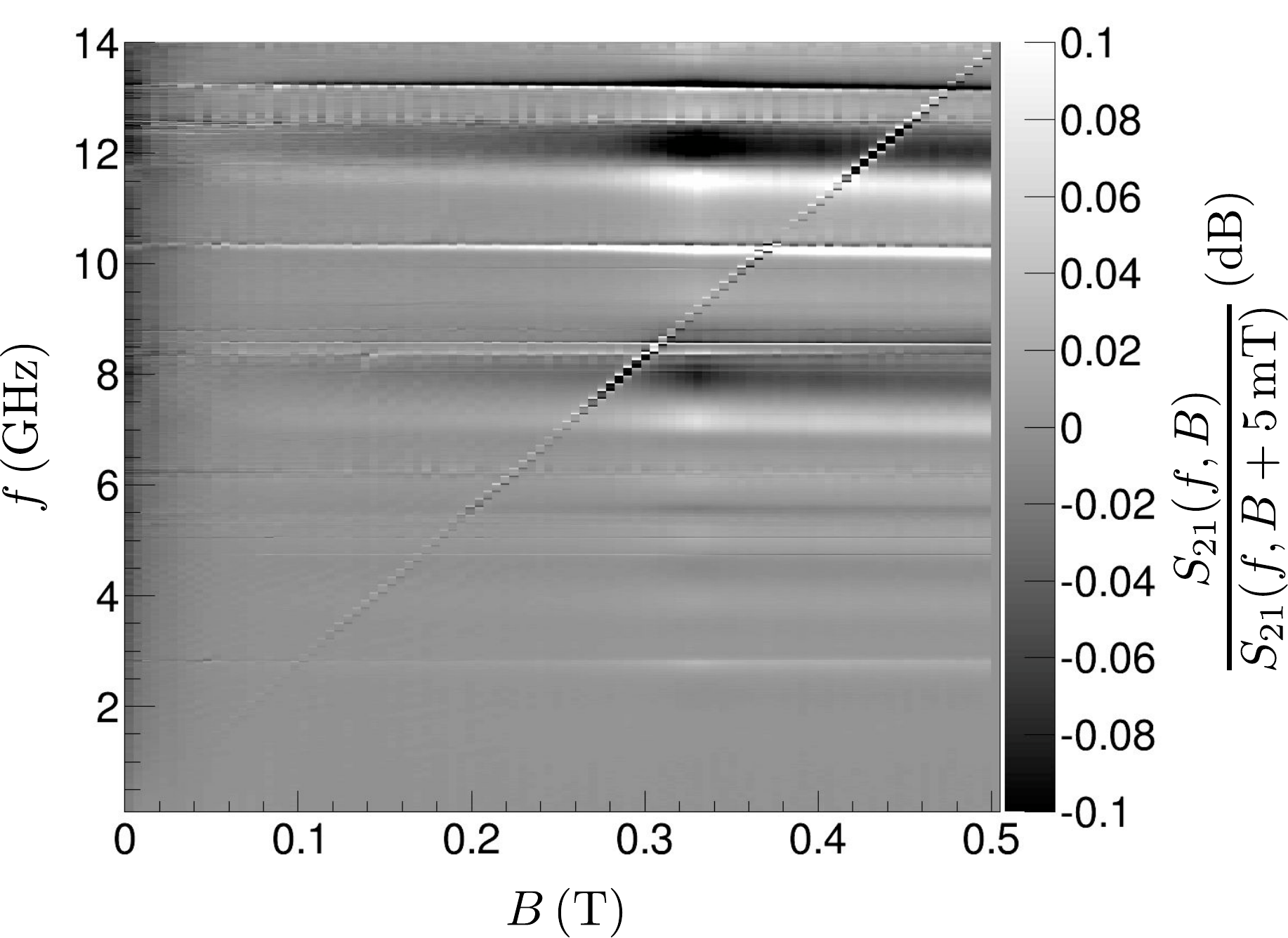}
\caption{The graph shows the normalized $S_{21}$ transmission as a function of both frequency and magnetic field.  The transmission line and the sample are shown in figure \ref{fig:wg2photo}.  The spectrum at each field is normalized by the spectrum at a field \SI{5}{\milli\tesla} higher.  The absoption feature is clearly distinguished from the background noise and distorsions (horizontal bands).}\label{fig:DPPH2}
\end{figure}

Figure \ref{fig:DPPH1} shows the normalized transmission spectra at different fields.  Using the previously detailed normalization, We clearly see a dip-peak in the transmission intensity that changes position according to a $g_S\simeq 2$ slope.  Each spectrum is fitted to a dip-peak function:
\begin{equation}
S_{21} = \left|1 - \frac{v}{\Gamma - i (f-f_0)} + \frac{v}{\Gamma - i(f-f_0-\delta)}\right| \label{eq:dippeak}
\end{equation}
Where $v$ is a coupling strength, $f_0$ is the energy level separation and $\delta$ is the separation between the peak and dip.  The fitted $f_0$ is shown in the inset of figure \ref{fig:DPPH1}.  The fit gives the g-factor of the DPPH sample ($g=2.002$).  We also find that the actual signal difference to the background is $\lesssim \SI{0.05}{\deci\bel}$, corresponding to power absorptions of less than about 1\% relative to the background.

Another experiment using a superconducting CPW is done with a more controlled sample size and a thinner transmission line.  The device is a transmission line with \um{7} gaps and a \um{14} wide center line.  The DPPH powder is used to make a saturated solution in dimethylformamide (DMF) with 5\% glycerol.  A drop is placed onto the transmission line using a micro-pipette and left to dry (figure \ref{fig:wg2photo}A).  The solution is used to more effectively fill the transmission cross section and get the sample close enough to the currents.  As discussed in section \ref{sec:coupCPWG_SMM}, the fields from the transmission line do not reach far beyond the gap and center line width and the sample needs to be placed well within this volume.  The background is shown in figure \ref{fig:wg2photo}B and is again very irregular.  The red line represents the behavior of an ideal waveguide with only resistive and dielectric losses.  In contrast with the case in figure \ref{fig:wg1photo}B where only resistive losses were necessary, here some small dielectric losses need to be included so that the ideal behavior follows the measured spectrum.  This is probably due to the fact that the CPW is much thinner than in the previous example and will therefore have much stronger electric fields causing larger dielectric losses.

Figure \ref{fig:DPPH2} shows the 2 dimensional normalized transmission spectrum as a function of frequency and magnetic field.  The field is again applied in the resonator plane, parallel to the transmission line.  The absorption feature has the same dependence as in figure \ref{fig:DPPH1} and can be fitted in the same way using equation \refeq{eq:dippeak}.  The extracted $g$-factor is $g_S=2.000$.  The horizontal bands correspond to background features that change in intensity with the magnetic field but do not shift their position.  The differences with the background signal were slightly larger in general $\lesssim\SI{0.1}{\deci\bel}$.  We see that even with the sample being much smaller than in figure \ref{fig:wg1photo}, we can get similar signals by choosing a transmission line well adapted to the sample size and by filling the sensitive areas effectively.

\subsection{\ce{CaGdF} on a superconducting CPW}\label{sec:CaGdFline}

\begin{figure}[!tb]
\centering
\includegraphics[width=1\textwidth]{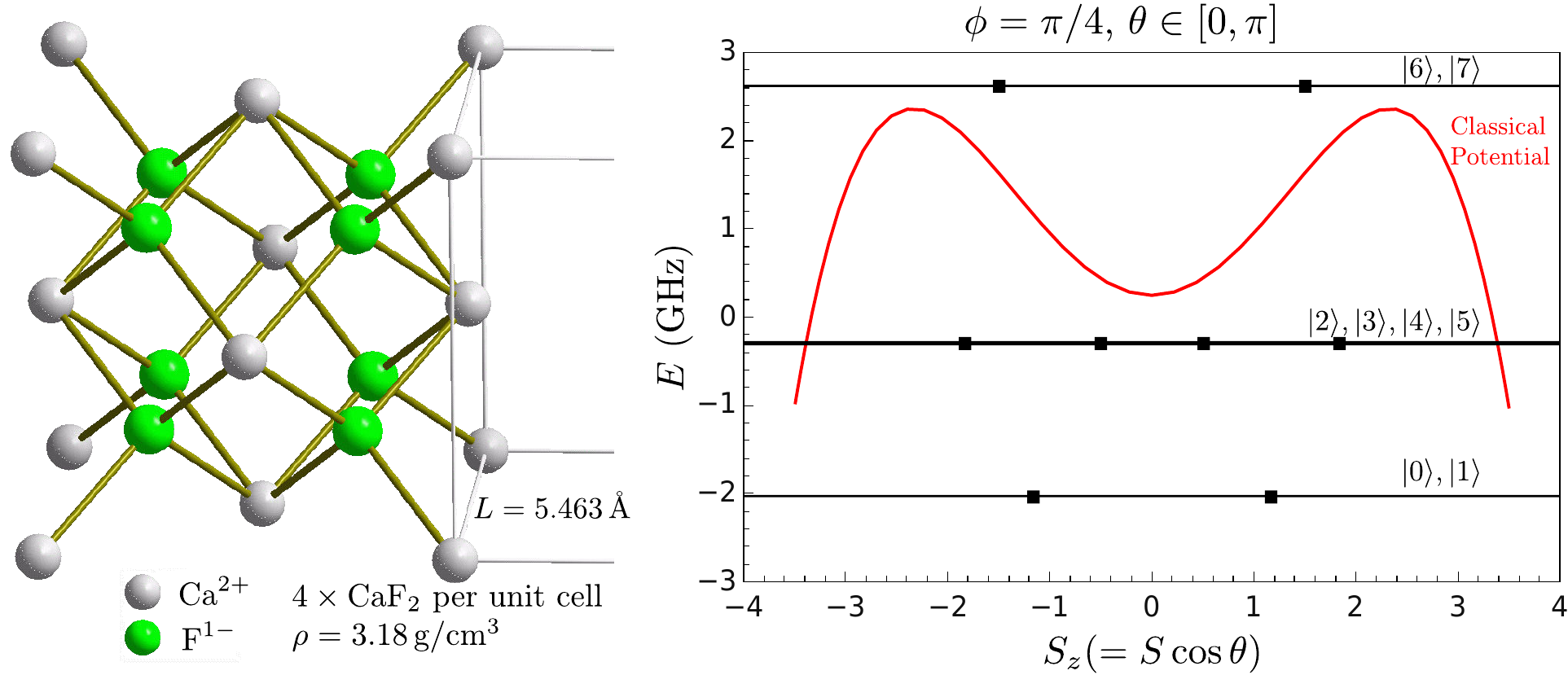}
\caption{Left: Unit cell of a \ce{CaF2} crystal.  When doped with \ce{Gd}, the \ce{Gd^{3}+} ion replaces a \ce{Ca^{2}+} ion.  Only \ce{Gd^{3}+} ions in cubic symmetry sites contribute to EPR signals \cite{Arauzo1997a}.  Right: Energy levels of the zero field states for the \ce{Gd} spin in a \ce{CaF2} crystal.  Horizontal lines represent the quantum energy levels.  The dots represent the energy levels and the $\langle S_Z \rangle$ expectation value for each.  The line at the center has four degenerate energy levels while the other two have two degenerate levels.  The red curve represents the classical potential for the Hamiltonian in equation (\ref{eq:GdFH}) for the polar coordinates $\phi=\pi/4$ and $\theta\in [0,\pi]$ represented as a function of $S_Z$.}\label{fig:CaF}
\end{figure}

Calcium Fluoride \ce{CaF2} crystals have been used extensively for basic studies of rare earth ions in crystalline environments \cite{Abragam1970} due to their good crystal qualities (cubic symmetry) and stability for a broad range of temperatures \cite{Perry2011}.  The crystal unit cell is shown in figure \ref{fig:CaF}.  These crystals are doped with a rare earth ion with a fraction of the \ce{Ca^{2}+} sites being replaced with the dopant.

\begin{figure}[!tb]
\centering
\includegraphics[width=1\textwidth]{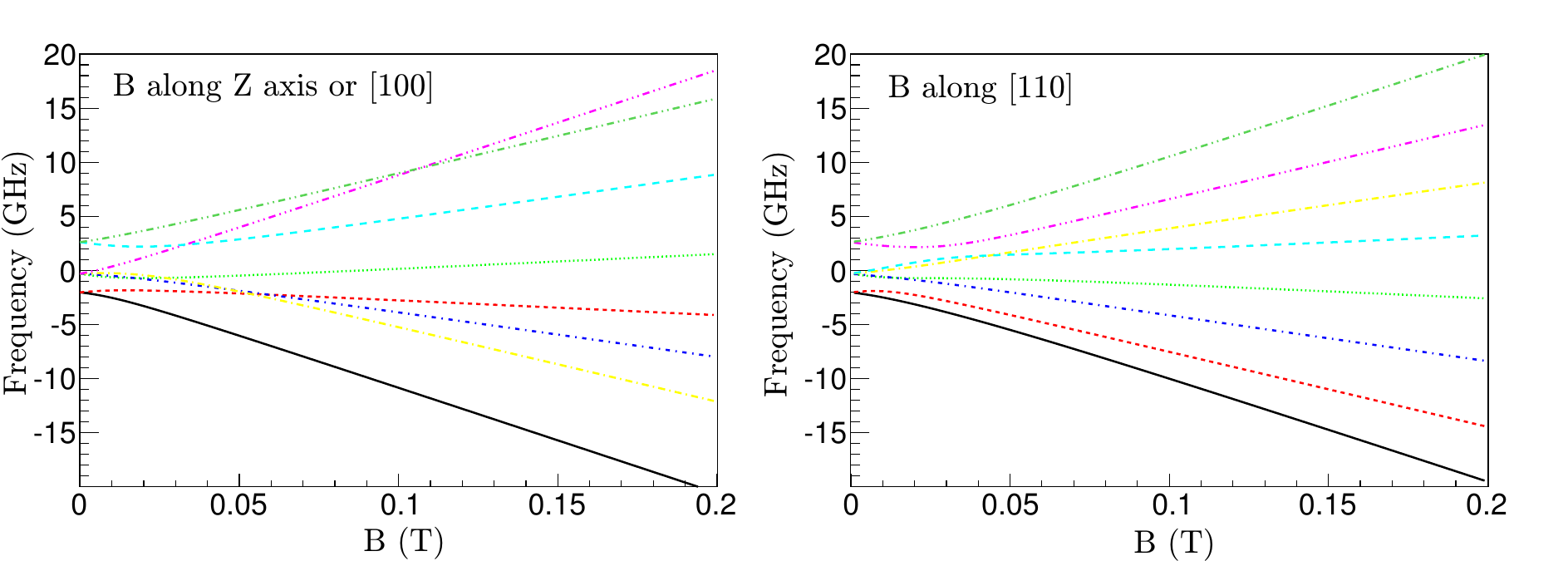}
\caption{Energy levels for magnetic fields applied along two different crystal axes of a \ce{GdCaF} crystal calculated from Hamiltonian \refeq{eq:GdFH}. Direction [110] (right) corresponds to a crystal edge}\label{fig:GdFlevels2}
\end{figure}

\begin{figure}[p]
\centering
\includegraphics[width=0.9\columnwidth]{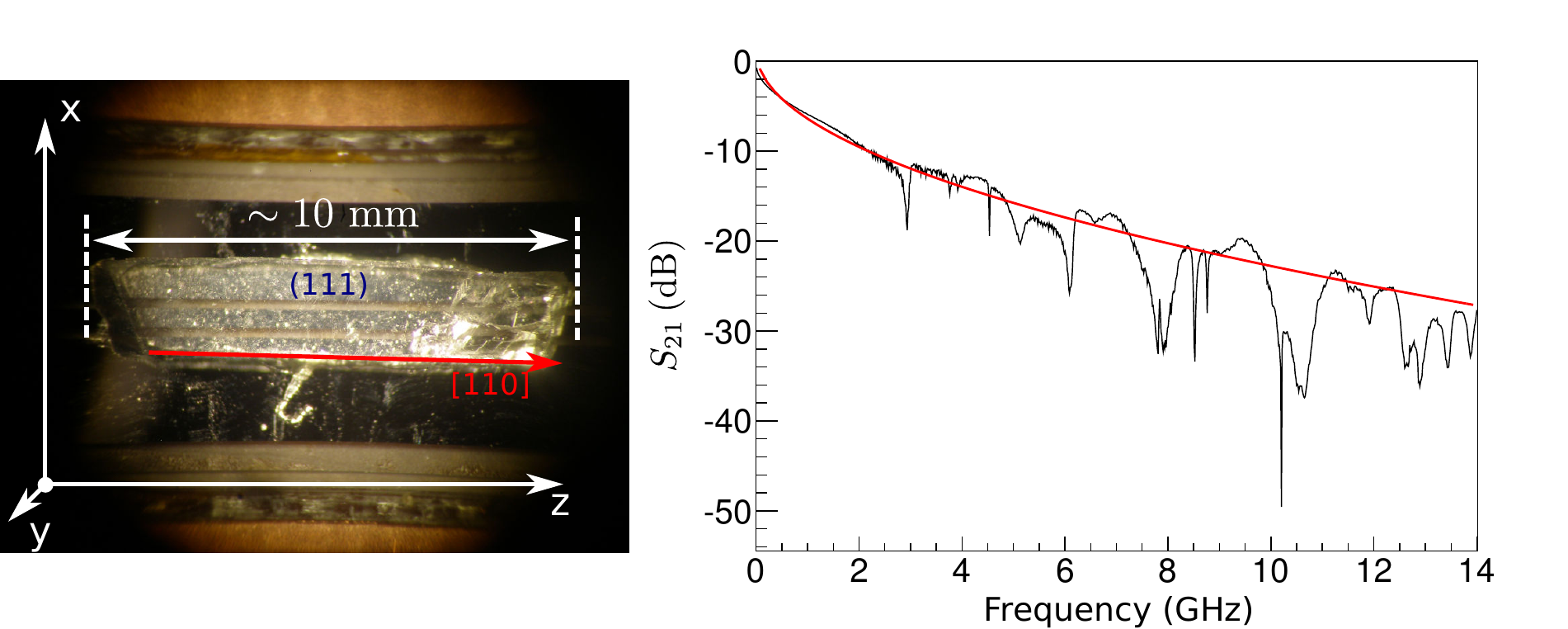}
\caption{The photo shows a \ce{CaGdF} crystal on a niobium CPW transmission line (\um{200} gaps and \um{400} centerline).  The axes approximately show the laboratory field directions.  The red arrow is aligned with the crystal edge corresponding to the $[110]$ crystal direction.  The upper face correspondes to the $(111)$ crystal plane.  The graph shows the background transmission ($S_{21}$).  The red line is the behavior of an ideal waveguide with resistive losses (from connecting wires).}\label{fig:wgGdFphoto}
\end{figure}

In the case of our Gd fluoride crystals, The \ce{Ca^{2}+} ions are substituted with \ce{Gd^{3}+} at a concentration of about 0.2\% per \ce{CaF2} unit.  Only 13\% of these are actually in cubic symmetry sites and contribute to EPR signals \cite{Arauzo1997a}.  Using the density of \ce{CaF2}, $\rho = \SI{3.18}{\gram\per\centi\meter\cubed}$ and the concentration \num{0.026}\%, this works out to a \ce{Gd^{3}+} ion concentration of \SI{1.08e-5}{\mole\per\centi\meter\cubed}.  We note that this concentration is about a factor 300 smaller than the spin concentration in DPPH.  The crystals are usually rectangular and the lateral sides correspond to (111) crystal planes and the edges correspond to [110] crystal directions (figure \ref{fig:wgGdFphoto}).

\begin{figure}[p]
\centering
\includegraphics[width=0.9\columnwidth]{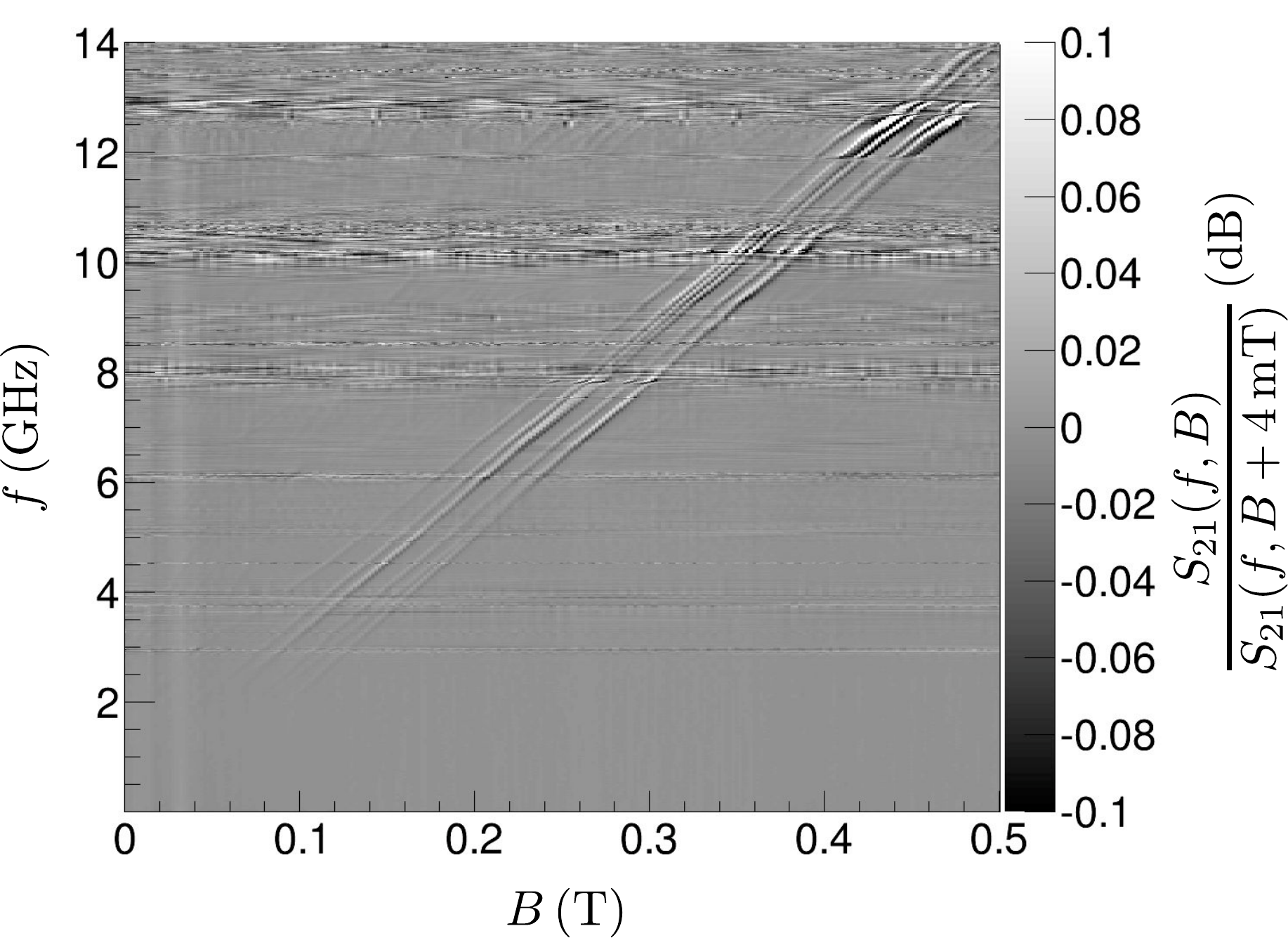}
\caption{The graph shows the normalized $S_{21}$ transmission of a \ce{CaGdF} crystal as a function of both frequency and magnetic field.  The field is applied along the Z laboratory axis which is approximately aligned with the $[110]$ crystal direction.  The transmission line and the sample are shown in figure \ref{fig:wgGdFphoto}.  The spectrum at each field is normalized by the spectrum at a field \SI{4}{\milli\tesla} higher. Several absorption lines can be tracked.  The horizontal breaks and lines are due to poor and noisy transmission in specific frequency regions (mostly field independent).}\label{fig:wgGdF_1}
\end{figure}

\begin{figure}[!tb]
\centering
\includegraphics[width=0.75\columnwidth]{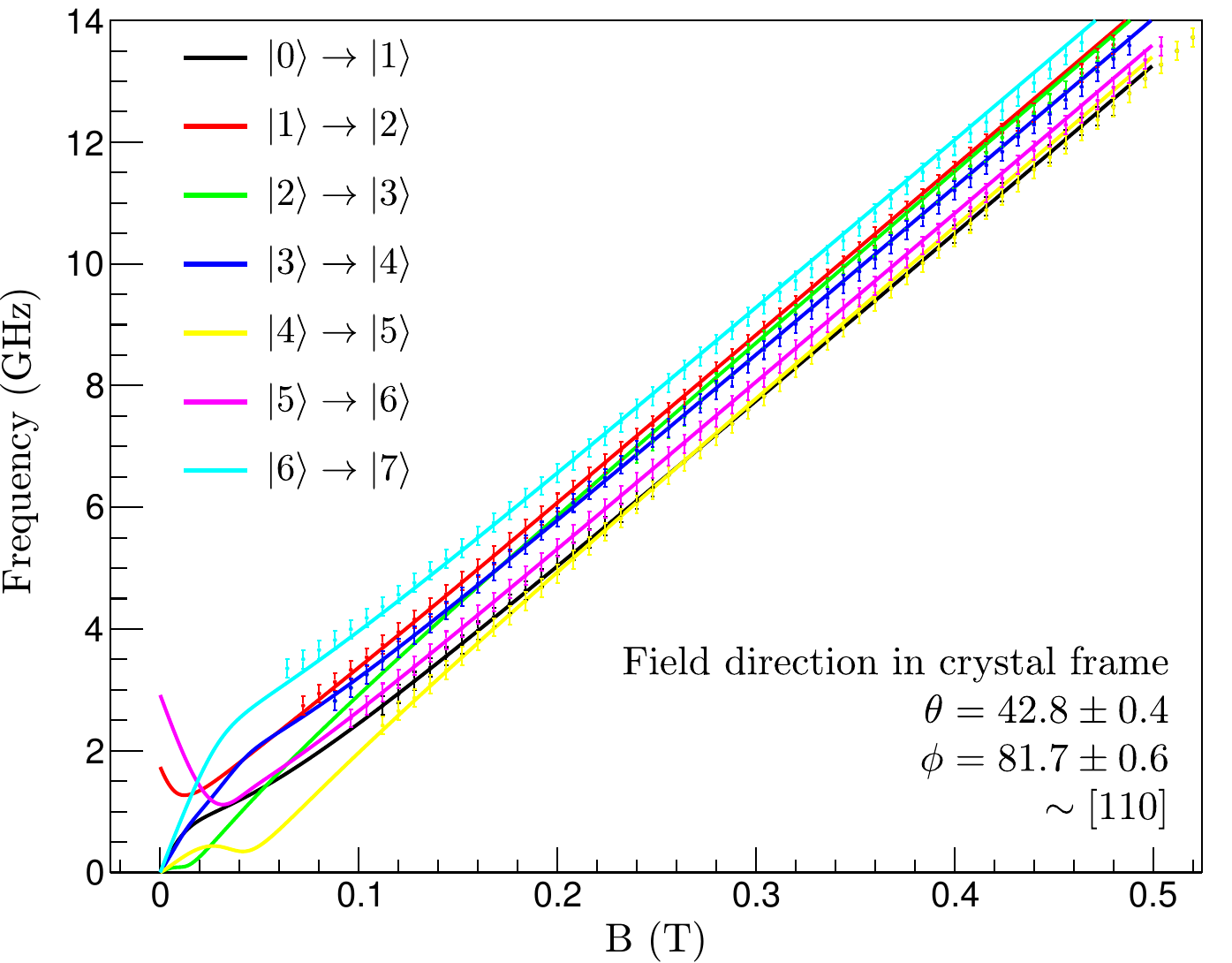}
\caption{Transmission signals extracted from figure \ref{fig:wgGdF_1}.  The points are extracted from the absorption peaks (not all are shown for visibility).  Solid lines are least squares fits of transitions derived from the spin Hamiltonian \refeq{eq:GdFH}.  This allows an estimation of the applied field direction relative to the crystal reference frame (\ref{eq:GdFH}).  In this case it approximately corresponds to the $[110]$ or equivalent (cubic symmetry) crystal direction.  The calculated absorption lines are labelled according to the pair of energy eigenstates connected by the transition.  $\ket{0}$ is the ground state and $\ket{n}$ are the successive excited states.  At high fields they approximately correspond to spin eigenstates in the direction of the applied field, i.e., $\ket{0}\sim\ket{m=-7/2}$, $\ket{1}\sim\ket{m=-5/2}$, ..., $\ket{7}\sim\ket{m=+7/2}$}\label{fig:wgGdF_fit1}
\end{figure}

\begin{figure}[!tb]
\centering
\includegraphics[width=1.\columnwidth]{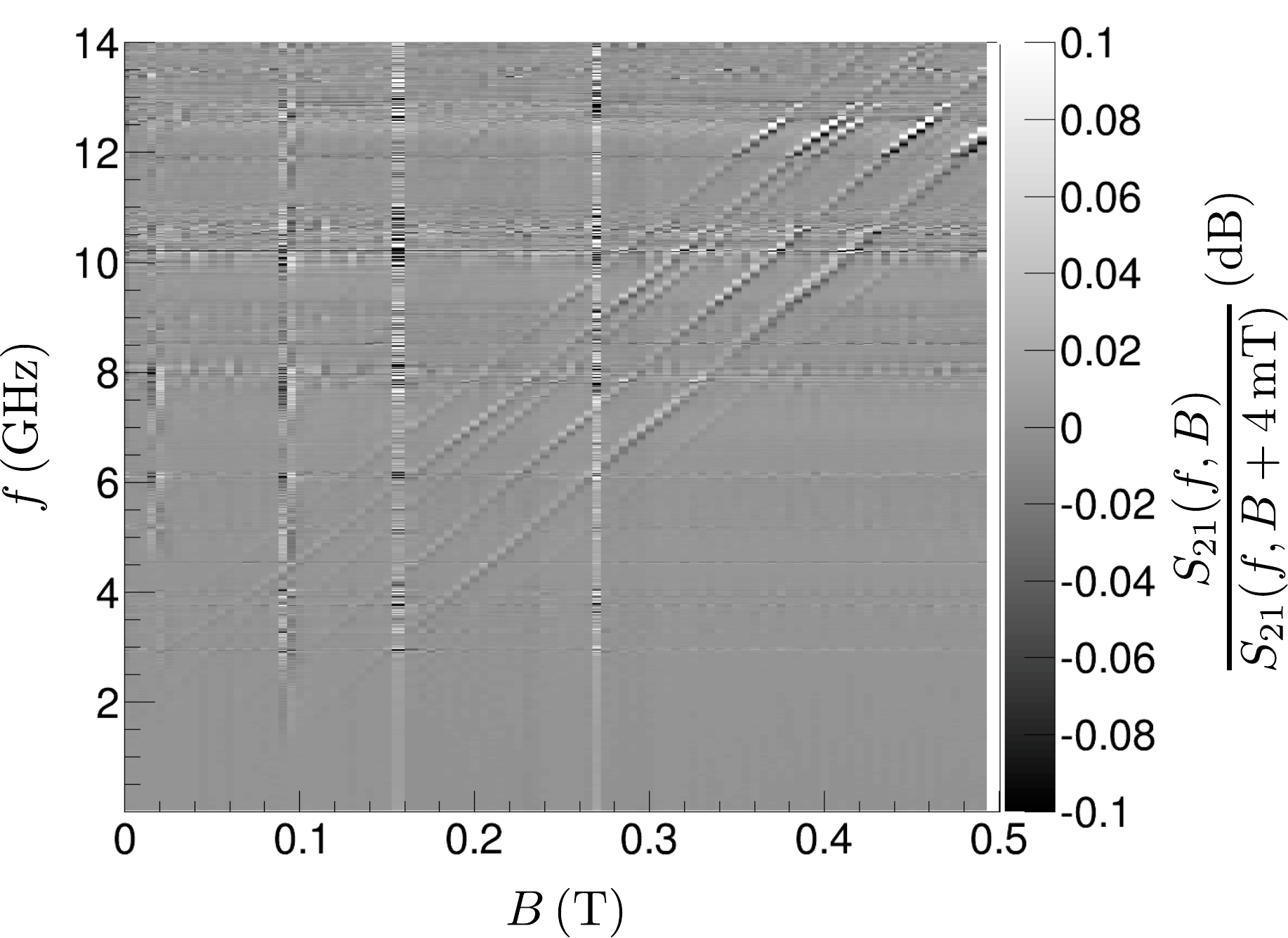}
\caption{The graph shows the normalized $S_{21}$ transmission of a \ce{CaGdF} as a function of both frequency and magnetic field.  The field is applied along the $\theta=45\degree$, $\phi=\arctan{\sqrt{2}}\simeq 55\degree$ laboratory direction which is approximately aligned with the $[100]$ crystal direction.  The transmission line and the sample are shown in figure \ref{fig:wgGdFphoto}.  The spectrum at each field is normalized by the spectrum at a field \SI{4}{\milli\tesla} higher. Several absorption lines can be tracked, which are clearly shifted from those seen in figure \ref{fig:wgGdF_1}.}\label{fig:wgGdF_2}
\end{figure}

\begin{figure}[!tb]
\centering
\includegraphics[width=0.75\columnwidth]{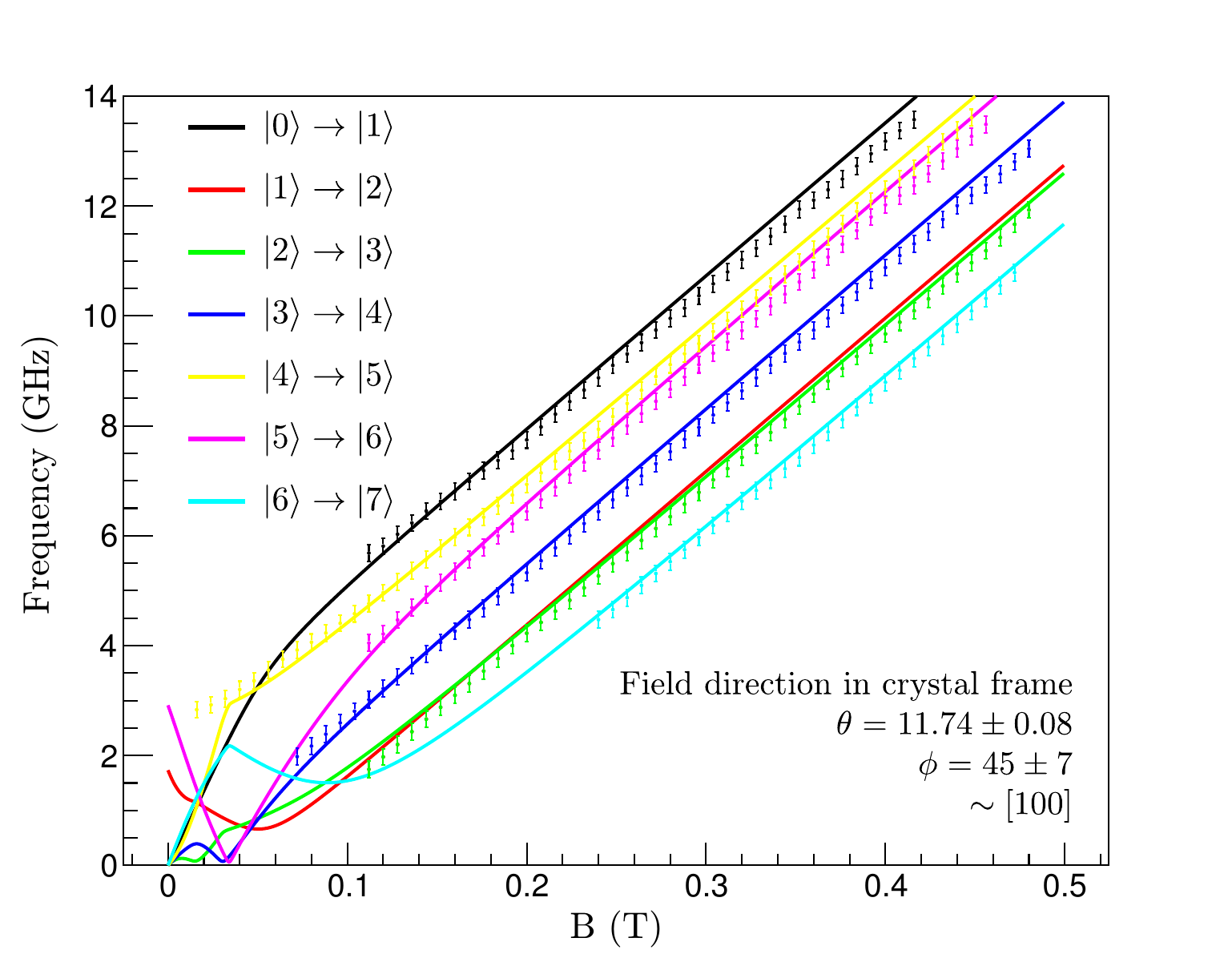}
\caption{Transmission signals extracted from figure \ref{fig:wgGdF_2}.  The points are extracted from the absorption peaks (not all are shown for visibility).  Solid lines are least square fits of the transitions derived from the spin Hamiltonian \refeq{eq:GdFH}.  This allows an estimation of the applied field direction relative to the crystal reference frame (\ref{eq:GdFH}).  In this case it approximately corresponds to the $[100]$ or equivalent (cubic symmetry) crystal direction.  The green and red experimental points are assumed to overlap and only the green are visible in the graph.    The calculated absorption lines are labelled according to the pair of energy eigenstates connected by the transition.  $\ket{0}$ is the ground state and $\ket{n}$ are the successive excited states.  At high fields they approximately correspond to the spin eigenstates in the direction of the applied field, i.e., $\ket{0}\sim\ket{m=-7/2}$, $\ket{1}\sim\ket{m=-5/2}$, ..., $\ket{7}\sim\ket{m=+7/2}$} \label{fig:wgGdF_fit2}
\end{figure}

The Hamiltonian for the rare earth ion spin is given by a crystal field Hamiltonian (equation \ref{eq:GSHamiltonian}) similar to those described for SMMs and SIMs in chapters \ref{chap:Theo1} and \ref{chap:SIMs}.  In this case we have the following parameters \cite{Arauzo1997,Arauzo1997a}:
\begin{eqnarray}
&\mathcal{H}_{\rm s} = \frac{1}{60}b_4\left(O_4^0+5\,O_4^4\right)+\frac{1}{1260}b_6\left(O_6^0-21\,O_6^4\right)-g_{S}\mu_{\rm B} \vec{B}\cdot\vec{S} &\label{eq:GdFH}\\
& S=\frac{7}{2},\quad g_S=1.992,\quad b_4 = \SI{-145.3}{\giga\hertz}, \quad b_6=\SI{-0.3}{\mega\hertz},& \nonumber
\end{eqnarray}
where $O_i^j$ are extended Stevens operators \cite{Stevens1952,Rudowicz2004} (see table \ref{fig:stevens}).  This Hamiltonian can be readily diagonalized for any given field and we find that the zero field split energy levels are all within a \SI{5}{\giga\hertz} energy band (figure \ref{fig:CaF}).  The strong transverse anisotropy terms induce large mixings between the $S_Z$ states that form the energy eigenstates.  For example, the ground state doublet is a mixture of the $\ket{\pm\frac{1}{2}}$ and $\ket{\mp\frac{7}{2}}$ respectively.  The zero field level diagram is shown in figure \ref{fig:CaF} along with the classical potential.  Figure \ref{fig:GdFlevels2} shows the effect of the magnetic field on the energy levels for two different orientations.

A crystal is placed on a wide transmission line (\um{200} gaps and \um{400} center line) and fixed using a thin layer of Apiezon-N grease with its $[110]$ direction aligned with the vector magnet (laboratory) Z axis and its $(111)$ face perpendicular to the Y axis (figure \ref{fig:wgGdFphoto}).  The transmission is then measured following the same procedure used for DPPH in section \ref{sec:DPPHwg}.  The same normalization as in figures \ref{fig:DPPH1} and \ref{fig:DPPH2} is again used.  The separation between the normalizing spectra in this case is \SI{4}{\milli\tesla}.  We firstly apply fields in the Z laboratory axis ($[110]$ crystal direction) and obtain the transmission shown in figure \ref{fig:wgGdF_1}.  We see multiple transmission lines that change position as the field is increased.  These positions can be extracted and compared to the expected absorption lines from the Hamiltonian (\ref{eq:GdFH}).  With the given Hamiltonian parameters, the field direction in the crystal reference frame can then be fitted and compared to the expected direction.  The extracted lines and fits are seen in figure \ref{fig:wgGdF_fit1}.  The fit in this case gives $\theta = 42.8\pm 0.4$ and $\phi = 81.7\pm 0.6$ in spherical coordinates.  This is compatible with the expected $[110]$ result considering that the crystal was aligned by hand.

Using the other components of the vector magnet we can also apply the field in the $[100]$ crystal direction, equivalent to the $S_Z$ direction in the Hamiltonian (cubic symmetry).  Using the orientation of the crystal seen in figure \ref{fig:wgGdFphoto}, the magnetic field must be applied in the direction of the unitary vector $\hat{b} = (1/\sqrt{6},1/\sqrt{3},1/\sqrt{2})$, i.e., $\theta=45\degree$ and $\phi=\arctan{\sqrt{2}}\simeq 55\degree$ in the laboratory frame.  The absorption lines in this case, seen in figure \ref{fig:wgGdF_2}, are clearly different from those in figure \ref{fig:wgGdF_1}.  We again extract the curves and fit them to the expected transitions from equation (\ref{eq:GdFH}) (figure \ref{fig:wgGdF_fit2}).  The fit gives a field direction $\theta = \ang{11.74}\pm \ang{0.08}$ and $\phi = \ang{45}\pm \ang{7}$ in spherical coordinates in the crystal frame.  This is also compatible with the $[001]\equiv [100]$ direction and the deviation is attributable again to a small misalignment of the crystal.

For this sample we only present results obtained with a wide transmission line.  A thin transmission line (\um{7} gaps and \um{14} centerline) was also tested but the signal from the \ce{GdCaF} crystal was too weak and barely visible.  This was due to the inefficient filling of the sensitive area of the transmission line and is a demonstration of the issues discussed in section \ref{sec:coupCPWG_SMM} about the filling of a CPW resonator by a crystal sample.  The same issues apply in the case of an open waveguide.  The layer of grease used to fix the crystal and the surface roughness both contribute to place the spins too far away from the currents as to have a sizable interaction.

\subsection{\ce{GdW30} on a superconducting CPW}\label{sec:GdW30wg}

\begin{figure}[!tb]
\centering
\includegraphics[width=0.95\columnwidth]{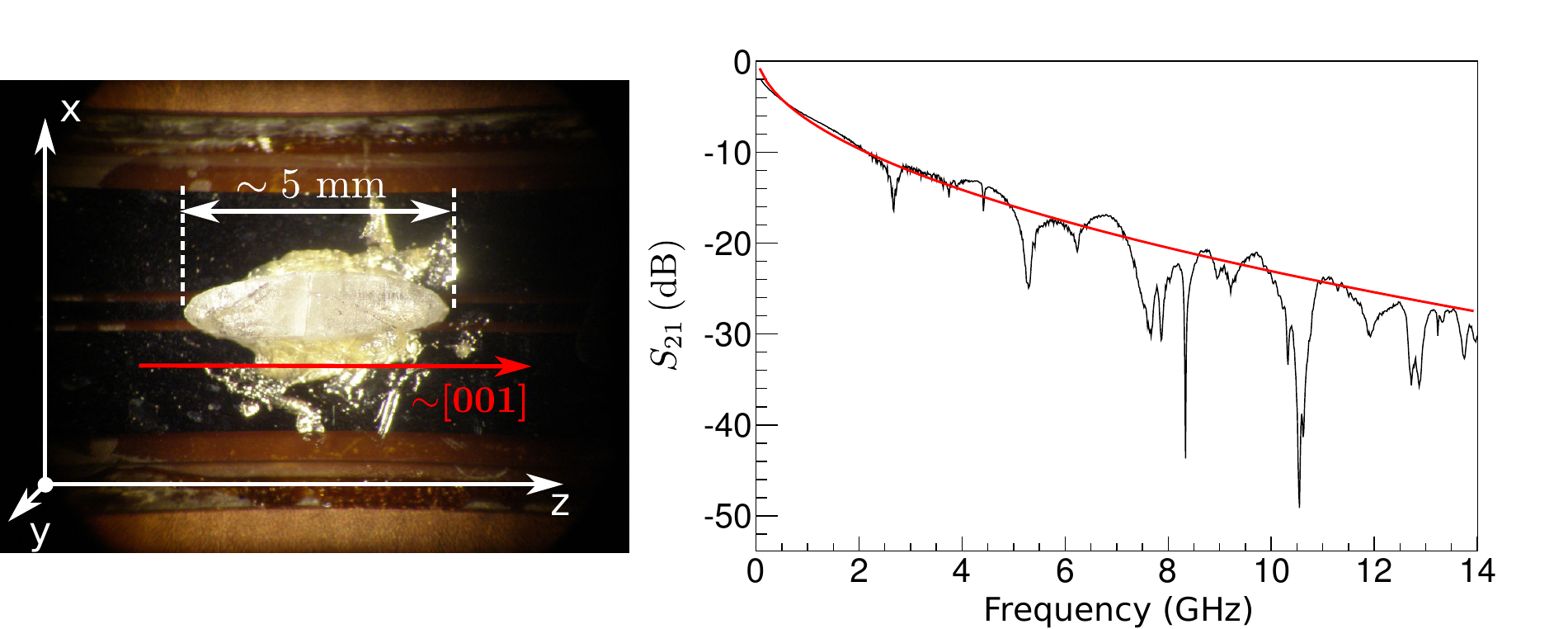}
\caption{The photo shows a \ce{GdW30} crystal on a niobium CPW transmission (\um{200} gaps and \um{400} centerline).  The axes approximately show the laboratory field directions.  The red arrow is aligned along the crystal edge corresponding approximately to the $[001]$ crystal direction.  The graph shows the background transmission ($S_{21}$).  The red line is the behavior of an ideal waveguide with resistive losses.}\label{fig:wgGdW30photo}
\end{figure}

The last sample tested using open waveguides was a \ce{GdW30} crystal.  The Hamiltonian for this sample is described in section \ref{sec:GdW30}.  A several mm long crystal was placed on the previously mentioned wide waveguide (\SI{200}{\micro\metre} gap and \um{400} centerline) as shown in figure \ref{fig:wgGdW30photo}.  The long edge of the crystal was aligned with the Z field and with the waveguide direction.  As described in section \ref{sec:xrayGdW30}, these crystals are produced from an over-saturated solution in deionized water.  Once the crystals are removed from the solution, they rapidly lose interstitial water molecules and the crystal loses its integrity in a few minutes.  In an attempt to avoid this, the crystal was covered in Apiezon N grease to both conserve the crystal and to fix it to the resonator.  The device was then inserted into the vector magnet and the transmission measured, as in the previous sections, as a function of frequency and magnetic field.

\begin{figure}[!tb]
\centering
\includegraphics[width=1.\columnwidth]{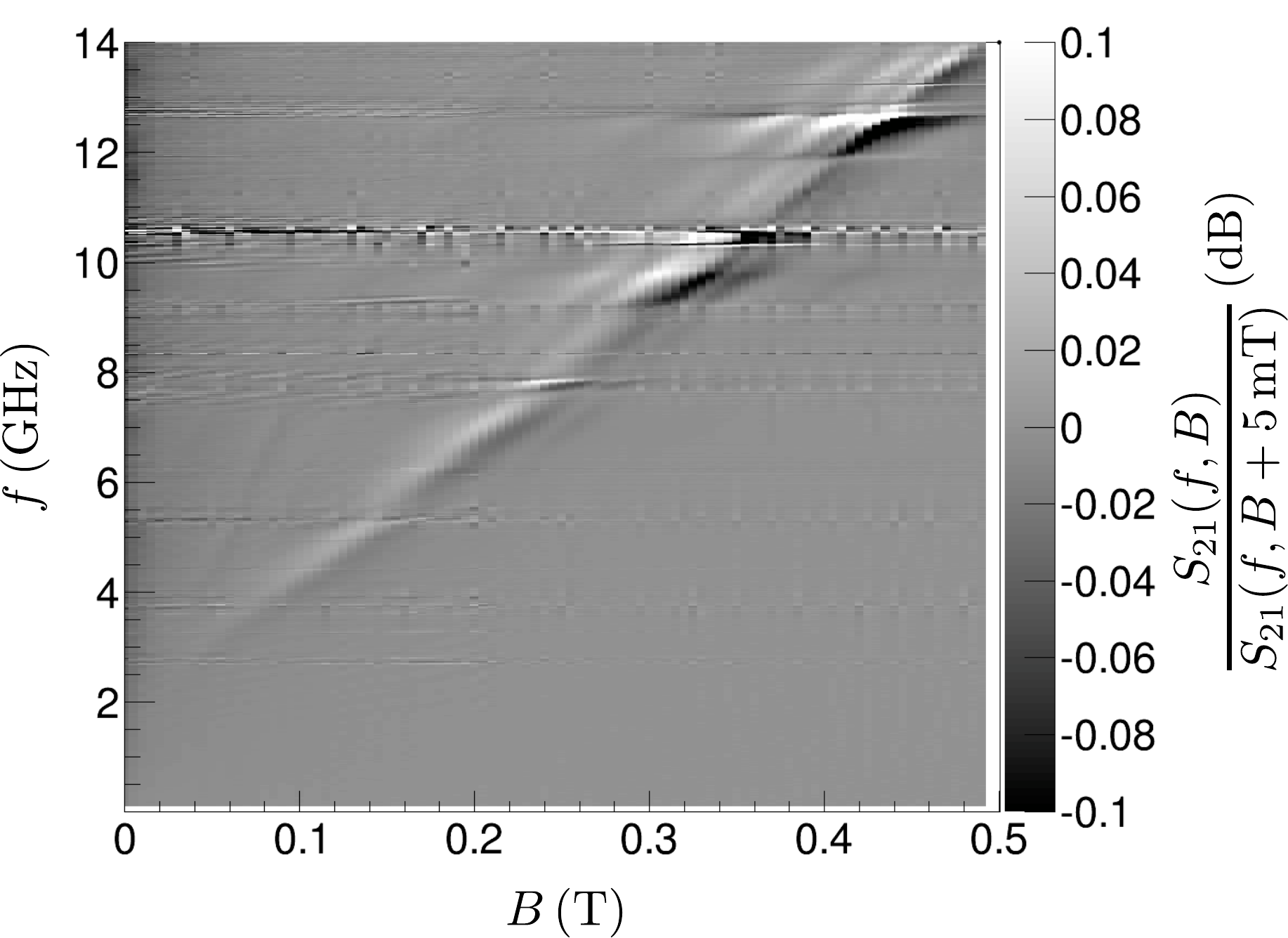}
\caption{\ce{GdW30} crystal on niobium CPW transmission line.  The graph shows the normalized $S_{21}$ transmission as a function of both frequency and magnetic field.  The field is applied along the Z laboratory direction.  The transmission line and the sample are shown in figure \ref{fig:wgGdW30photo}.  The spectrum at each field is normalized by the spectrum at a field \SI{5}{\milli\tesla} higher.  A very wide absorption band is visible probably due to loss of sample integrity.  The horizontal breaks and lines are due to poor and noisy background transmission in specific frequency regions (mostly field independent).}\label{fig:wgGdW30_1}
\end{figure}

\begin{figure}[!tb]
\centering
\includegraphics[width=1.\columnwidth]{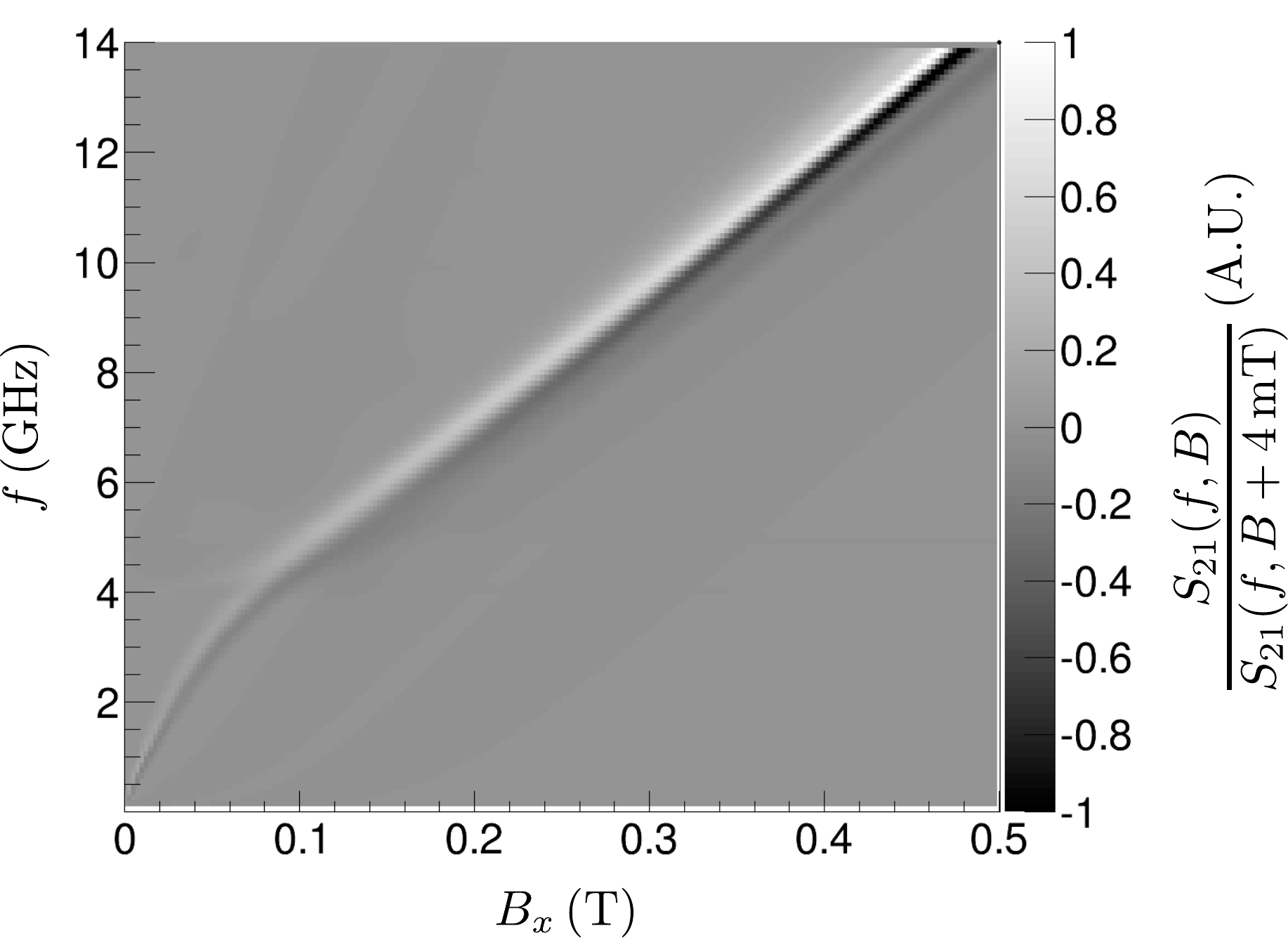}
\caption{EasySpin simulation of a \ce{GdW30} crystal on a transmission line.  The transmission line is emulated by simulating EPR spectra at many different cavity frequencies.  The simulation parameters were set to $T=\SI{4.2}{\kelvin}$, $\textrm{Sample line width} = \SI{16}{\milli\tesla}$ and the field was set along the $x$ molecular axis.  The graph shows the normalized signal as a function of both frequency and magnetic field.  The spectrum at each field is normalized by the spectrum at a field \SI{5}{\milli\tesla} higher to match our experimental protocol.}\label{fig:wgGdW30_1sim2}
\end{figure}

\begin{figure}[!tb]
\centering
\includegraphics[width=1.\columnwidth]{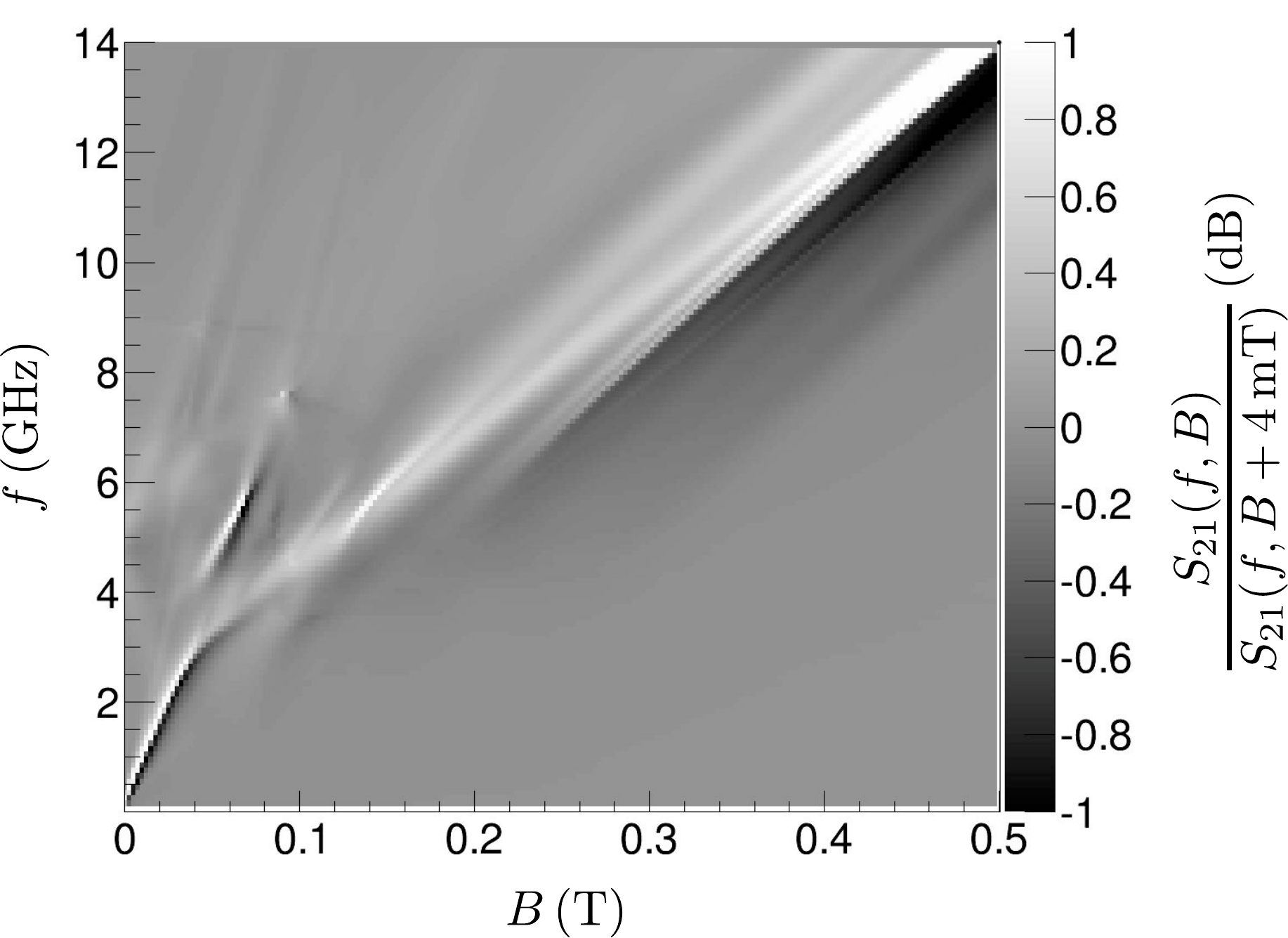}
\caption{EasySpin simulation of a \ce{GdW30} powder on a transmission line.  The transmission line is emulated by simulating EPR spectra at many different cavity frequencies.  The simulation parameters were set to $T=\SI{4.2}{\kelvin}$, $\textrm{Sample line width} = \SI{16}{\milli\tesla}$.  The graph shows the normalized signal as a function of both frequency and magnetic field.  The spectrum at each field is normalized by the spectrum at a field \SI{5}{\milli\tesla} higher.}\label{fig:wgGdW30_1sim}
\end{figure}

The normalized spectrum is shown in figure \ref{fig:wgGdW30_1} again using a \SI{5}{\milli\tesla} field separation for the normalization.  Unlike the results for the \ce{GdCaF} case (figures \ref{fig:wgGdF_1} and \ref{fig:wgGdF_2}), here we see a very wide band instead of well defined peaks.  We interpret this to mean that the crystal lost its integrity in the mounting and cooling process and that, during the measurement, it behaves close to a powder.  To support this assumption, we use EasySpin \cite{Stoll2006} to emulate a similar measurement\footnote{EasySpin is a Matlab \cite{matlab} package that simulates EPR spectra given the sample properties and experimental variables.}.  In our case, we use the Hamiltonian parameters and the $B_2^0$ and $B_2^2$ strains for \ce{GdW30} given in section \ref{sec:powderEPR_GdW30} and do a series of continuous wave simulations with varying frequencies (from \SI{10}{\mega\hertz} to \SI{14}{\giga\hertz}).  We generate spectra for both crystalline and powder samples with the same sample parameters.  The output is set to direct absorption (instead of first derivative) and we normalize the results using the same scheme as in our measurement.  The results are shown in figure \ref{fig:wgGdW30_1sim2} and \ref{fig:wgGdW30_1sim}.  Both graphs are qualitatively similar to the result of the experiment although our measurement may be in between both situations.  Also, some features are absent form our measurement possibly due to having lower sensitivities when using a waveguide than when using a cavity as in the simulation, as well as to other effects derived from the sample degradation (differences in strains and broadening, etc.) not taken into account in the simulations.

This measurement shows that, although the coupling to this type of SIM can be strong enough to do spectroscopy on a superconducting CPW, more precautions must be taken to ensure the integrity of the sample.

\section{Transmission through a resonator coupled to spins (theory)}\label{sec:seiji}
The Hamiltonian describing a quantum two level system (or qubit) coupled to a single mode of the electromagnetic field in a cavity was described in section \ref{sec:CQED}.  The detuning $\Delta = \Omega - \omega_r$, i.e., the difference between the photon frequency of the cavity mode $\omega_r$ and the energy splitting of the two level system $\hbar\Omega$, determines when the two systems are in resonance ($|\Delta|\simeq 0$) and when the systems are far from resonance ($|\Delta|\gg g$, where $g$ is the coupling strength).  Far from resonance, the systems (qubit and resonator) have their independent behavior while, in resonance, the states of the resonator and the qubit are hybridized being mixtures of both photon and qubit states.

As can be intuitively deduced from figure \ref{fig:CQEDdiag}, the transmission far from resonance will consist of a single peak at the resonator frequency $\omega_r$ with a width given by the cavity decay rate $\kappa$.  When we approach the resonance condition ($\Delta\simeq 0$), the peak will split into two peaks separated by $2g$ and reduce its intensity since part of the energy is now transferred to the qubit and will be lost from the transmission channel (only sensitive to photons).  These peaks can be resolved as long as the coupling $g$ is stronger than both the cavity decay rate $\kappa$ and the qubit dephasing rate $\gamma\sim T_2^{-1}$.  It also requires the driving to be low enough so as to have at most a single photon stored in the cavity, so that the transitions will be from the ground state to the first doublet in figure \ref{fig:CQEDdiag}.  At higher drivings, mostly transitions between excited states will be induced averaging out to a single peak spectrum.

The situation is further complicated if, as is the case in this work, the cavity is instead coupled to an ensemble of qubits.  As discussed in section \ref{sec:coupCPWG_SMM}, this broadly leads to a $\sqrt{N}$ enhancement of the effective coupling, with $N$ being the number of qubits and assuming they all couple with the same strength $g$.  The spectrum will display two peaks separated by $2g\sqrt{N}$ if the number of spins is larger than the cavity photon population.  As in the case of a single qubit, if the drive is high enough, the system evolves into a single peak spectrum.

\begin{figure}[!tb]
\centering
\includegraphics[width=0.7\columnwidth]{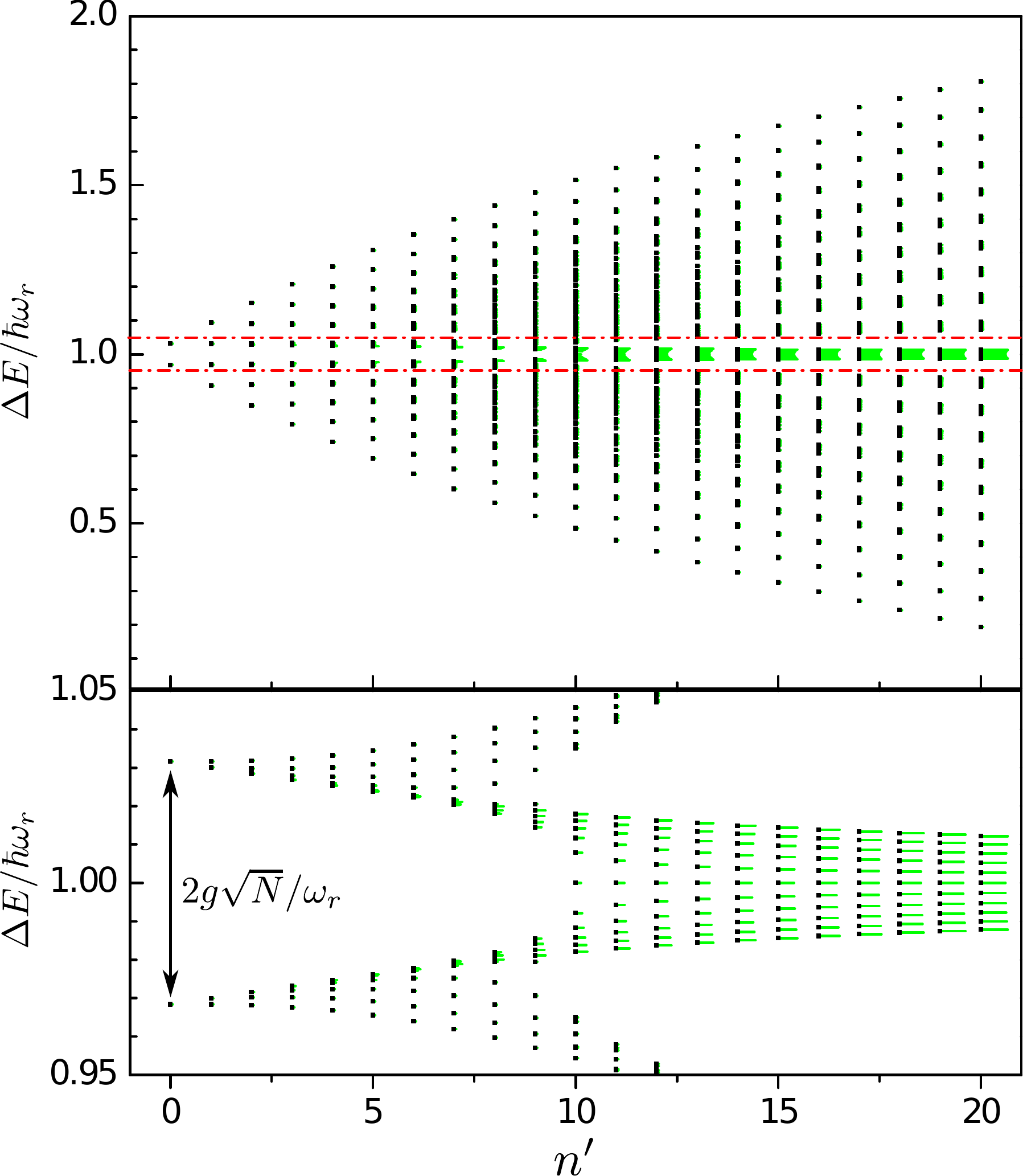}
\caption{Transition frequencies for the Hamiltonian (\ref{eq:ham0}) given by $\Delta E = E_{n'+1,k}-E_{n',k'}$.  The horizontal lines are proportional to the corresponding transition matrix element $\left|\bra{n',k}a\ket{n'+1,k'}\right|^2$.  The lower graph shows a zoom of the central band around the \emph{bare} resonator frequency $\omega_r$.  The parameters chosen for the representation were $\omega_r=1$, $g=0.01$, $\Delta = 0$ ($\Omega = \omega_r$) and $N=10$}\label{fig:seiji1}
\end{figure}

These general remarks on the behavior of the coupled system transmission can be studied in greater detail \cite{Chiorescu2010,Miyashita2012} using the Tavis-Cummings Hamiltonian, a generalized version of the Jaynes-Cummings model, with an additional driving field of intensity $\xi$ \cite{Jaynes1963,Tavis1968,Bishop2008}:
\begin{eqnarray}
\mathcal{H} &=& \mathcal{H}_0+\mathcal{H}_\xi \label{eq:ham1}\\
\mathcal{H}_0 &=& \omega_r a^\dag a+\Omega\left(\sum_i^N S_i^Z + N/2\right) -  g\left(a^\dag\sum_i^N S_{i}^- + a\sum_i^N S_{i}^+\right) \label{eq:ham0}\\
\mathcal{H}_\xi &=& \xi(a^\dag e^{-i\omega_d t}+ae^{i\omega_d t})\label{eq:hamdrv}
\end{eqnarray}
Here, we have set $\hbar=1$ and the qubit or spin ensemble is modeled by $N$ spin $1/2$ systems with spin operators $S_i$ for each spin.  To reduce the spin Hilbert space dimension we consider the approximation where only the total spin couples to the cavity, i.e., we replace:
\begin{equation}
\sum_{i}^N \vec{S}_i = \vec{S}
\end{equation}
where the total spin is $N/2$.  Equation (\ref{eq:ham0}) then reads:\footnote{The coupling $g$ in equation \refeq{eq:confusion} is not to be confused with the g-factor $g_S$ appearing in the Zeeman energy in, for instance, equation \refeq{eq:dpphlevel}.}
\begin{equation}
\mathcal{H}_0 = \omega_r a^\dag a + \Omega(S_Z+N/2) - g(a^\dag S^- + a S^+) \label{eq:confusion}
\end{equation}
Under this Hamiltonian the sum of the total $S_Z$ and the photon number is conserved.  We call this operator the excitation number:
\begin{equation}
N_{\rm ex} = S_Z + a^\dag a \qquad N_{\rm ex}\ket{n,m_z} = (n+m_z)\ket{n,m_z}
\end{equation}
This makes the Hamiltonian block diagonal if the basis $\ket{n, m_z}$ ($n$ is the number of photons and $m_z$ is the $S_z$ quantum number) is ordered according to the excitation number.  Then it can be systematically diagonalized block by block.  Each block is 1 dimension larger than the previous block up to a maximum dimension of $N+1$.  The energy eigenvalues can be labeled $E_{n'k}$ by excitation number $n'$ and a second index $k$ that runs from 1 up to each block's dimension (the minimum of $n'+1$ and $N+1$).  Once the Hamiltonian is diagonalized, we plot the frequencies of the transitions induced induced by the radiation field operators $a$ or $a^\dag$ and the associated matrix elements in figure \ref{fig:seiji1} for the case of zero detuning ($\Delta = 0$).  Note that $a$ and $a^\dag$ connect states with $\pm 1$ excitation, i.e., $\bra{n'_1 k}a^\dag\ket{n'_2 k'}= \bra{n'_1 k}a\ket{n'_2 k'} = 0$ unless $n'_1=n'_2\pm 1$.  We therefore only show transitions between states fulfilling this condition.

Observing this figure, we find that different regimes will be visible in the transmission of microwaves depending on what the number of excitations is.  Firstly we note that the matrix elements are largest in the central band around $\omega_r$ (shown in the lower graph).  If the excitation powers (and temperature) are low enough, only transitions with $n'=0$ will be visible and will be separated by an energy $2g\sqrt{N}$.  If the driving is stronger, the number of excitations will shifted towards the right of the graph and the two peaks slowly approach each other.  Once the the number of excitations is above the number of spins, the larger matrix elements further concentrate in the central band as the two-peak structure is lost and a classical one-peak spectrum is recovered.

The simulation of the actual lineshapes and widths due to the different dephasing mechanisms (associated to the cavity or the qubits) requires simulating the driving field as well as the dissipative effects.  This is done by first converting the Hamiltonian (\ref{eq:ham1}) to a frame rotating at the driving frequency $\omega_d$ through the unitary transformation $U=\exp{\left( i\omega_dt(a^\dag a+S^z)\right)}$.  The dissipation can then be modeled through a Lindblad type master equation \cite{Lindblad1976} for the density matrix $\rho$ of the system:
\begin{eqnarray}
\frac{d}{dt}\rho &=& -i[\mathcal{H},\rho] - \kappa(a^\dag a \rho + \rho a^\dag a - 2a\rho a^\dag) \nonumber\\
&& -\gamma(S^+S^-\rho+\rho S^+S^--2S^-\rho S^+) \nonumber\\
&& -\gamma_z\left( (S_Z)^2\rho+\rho (S_Z)^2-2S_Z\rho S_Z \right) \label{eq:lindblad1}
\end{eqnarray}
The first term in equation \refeq{eq:lindblad1} is the usual time evolution under the Hamiltonian (\ref{eq:ham1}) including the driving term (\ref{eq:hamdrv}).  The term proportional to $\kappa$ gives the cavity decay while the terms proportional to $\gamma$ and $\gamma_z$ describe the qubit dephasing and decay.  For convenience, the matrix equation can be rewritten by taking a vectorization of $\rho$ and defining the \emph{superoperator} $\mathcal{L}$ as follows:
\begin{equation}
\frac{d}{dt}\vec{\rho} =  \mathcal{L}\vec{\rho}. \label{eq:lindblad2}
\end{equation}
We are interested in the steady state transmission and cavity population values so we need to solve the master equation for $d\vec{\rho}/dt  = \mathcal{L}\vec{\rho} = 0$ or equivalently:
\begin{eqnarray}
&& -i[\mathcal{H},\rho] - \kappa(a^\dag a \rho + \rho a^\dag a - 2a\rho a^\dag) \nonumber\\
&& -\gamma(S^+S^-\rho+\rho S^+S^--2S^-\rho S^+) \nonumber\\
&& -\gamma_z\left( (S_Z)^2\rho+\rho (S_Z)^2-2S_Z\rho S_Z \right) = 0. \label{eq:master}
\end{eqnarray}
Solving for $\rho$ allows us to obtain the expectation values of any observable.  In particular:
\begin{eqnarray}
N_{\rm photons} &=& \langle a^\dag a\rangle = \Tr{\rho a^\dag a } \\
Q &=& \textrm{Im}\,\langle a \rangle = \Tr{\rho(ia^\dag - ia)}
\end{eqnarray}
where $Q$ is proportional to the cavity transmission signal \cite{Miyashita2012}.

\begin{figure}[!tb]
\centering
\includegraphics[width=\columnwidth]{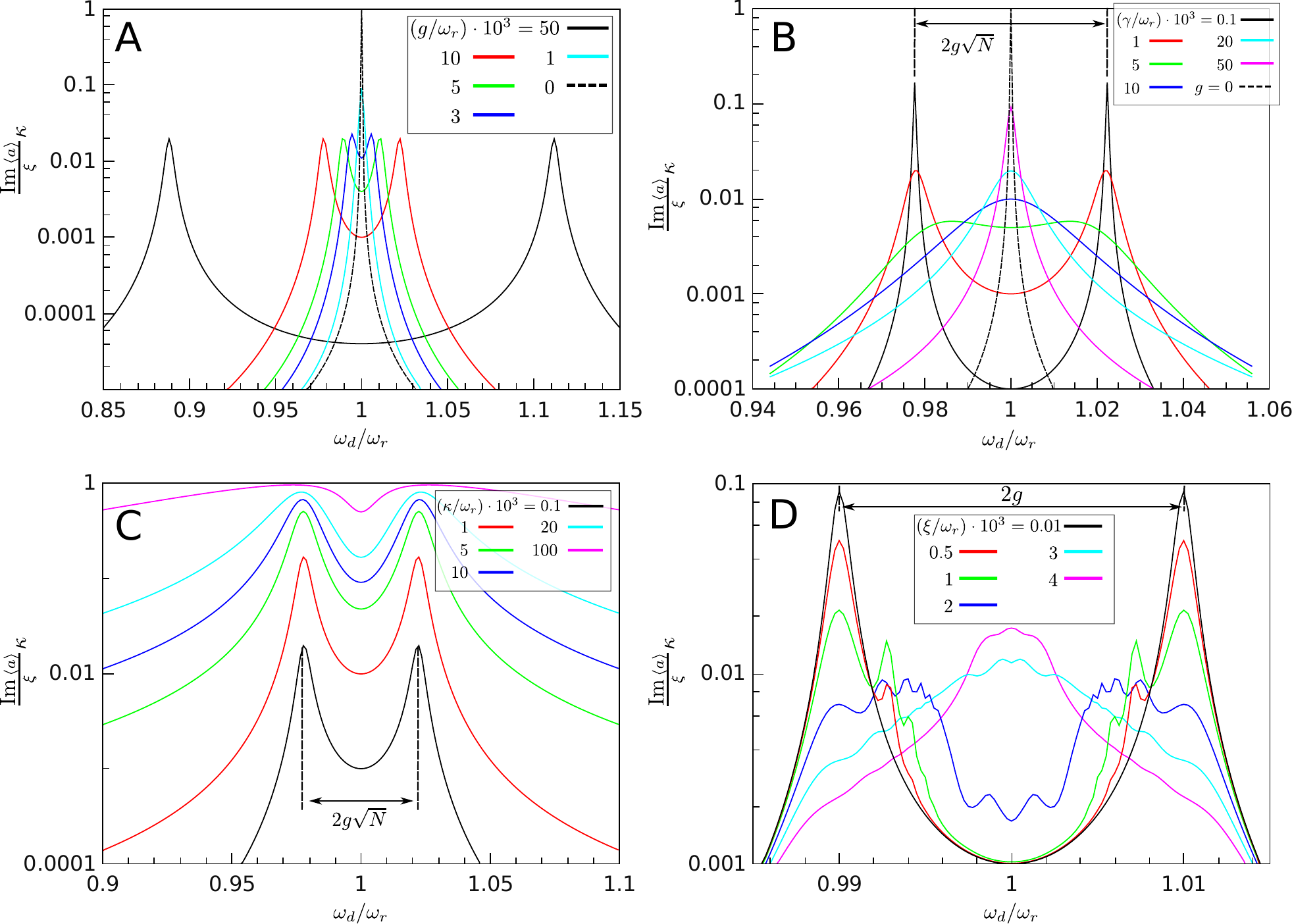}\\[3mm]
\begin{tabular}{|c|c|c|c|c|}
\hline
& Graph A & Graph B & Graph C & Graph D \\
\hline
$g/\omega_r$ & - & 0.01 & 0.01 &  0.01\\
\hline
$\gamma/\omega_r$ & 0.001 & - & 0.001 & 0.001 \\
\hline
$\kappa/\omega_r$ & \num{1e-4} & \num{1e-4} & - & \num{1e-4}\\
\hline
$\xi/\omega_r$ & \num{1e-5} & \num{1e-5} & \num{1e-5} & - \\
\hline
$N_\textrm{spins}$ & 5 & 5 & 5 & 1 \\
\hline
\end{tabular}
\caption{Transmission signal as a function of the driving frequency $\omega_d$ given by the master equation model (\ref{eq:master}) for different parameter combinations.  Each graph shows the dependence for different values of a single parameter and fixed values for the rest.  The parameter values for each graph are shown in the table.  Graph A shows the dependence when changing the spin dephasing $\gamma$, graph B varies the coupling $g$, graph C varies the cavity dissipation $\kappa$ and graph D shows the dependence for different drivings $\xi$.  Graph D only uses a single spin because of computational power limitations.  The normalization used in all cases is such that the cavity transmission value when $g=0$ is 1 when $\omega_d = \omega_r$.}\label{fig:seiji2}
\end{figure}

In resonance ($\Delta = 0$), there are several regimes that can potentially be seen in experiments where the driving frequency $\omega_d$ is swept and the cavity signal is monitored (figure \ref{fig:seiji2}).  For low driving powers, such that the photon population of the cavity is smaller than the number of spins, the cavity resonance splits into two much lower intensity peaks with their centers separated by $2g\sqrt{N}$.  This is in agreement with figure \ref{fig:seiji1} since the low drive only allows the system to explore the low number of excitation states where the dominant transition frequencies are separated by this amount.  The width of the new peaks is a combination of $\gamma$ and $\kappa$ since the states are hybrids of both spin and photon states.  They will be resolved as long as the new widths are smaller than the peak separation (basically $g>\gamma,\kappa$).  A clearly resolved two-peak transmission spectrum is a signature of strong coupling.  If we increase the qubit dephasing rate $\gamma$ (figure \ref{fig:seiji2}A), we see that the two peak structure gradually degrades until we reach the weak coupling limit where both peaks overlap and cannot be resolved.  The two-peak structure also changes when decreasing $g$ (figure \ref{fig:seiji2}B), increasing $\kappa$ (figure \ref{fig:seiji2}C) and also for higher driving intensities where the system is forced to sample states with $n_{\rm ex} \gg N$ (figure \ref{fig:seiji2}D).  This last case is not usually accessible for macroscopic samples and moderate driving, but for systems with low numbers of spins this effect may potentially be observed \cite{Bishop2008}.

\begin{figure}[!tb]
\centering
\includegraphics[width=\columnwidth]{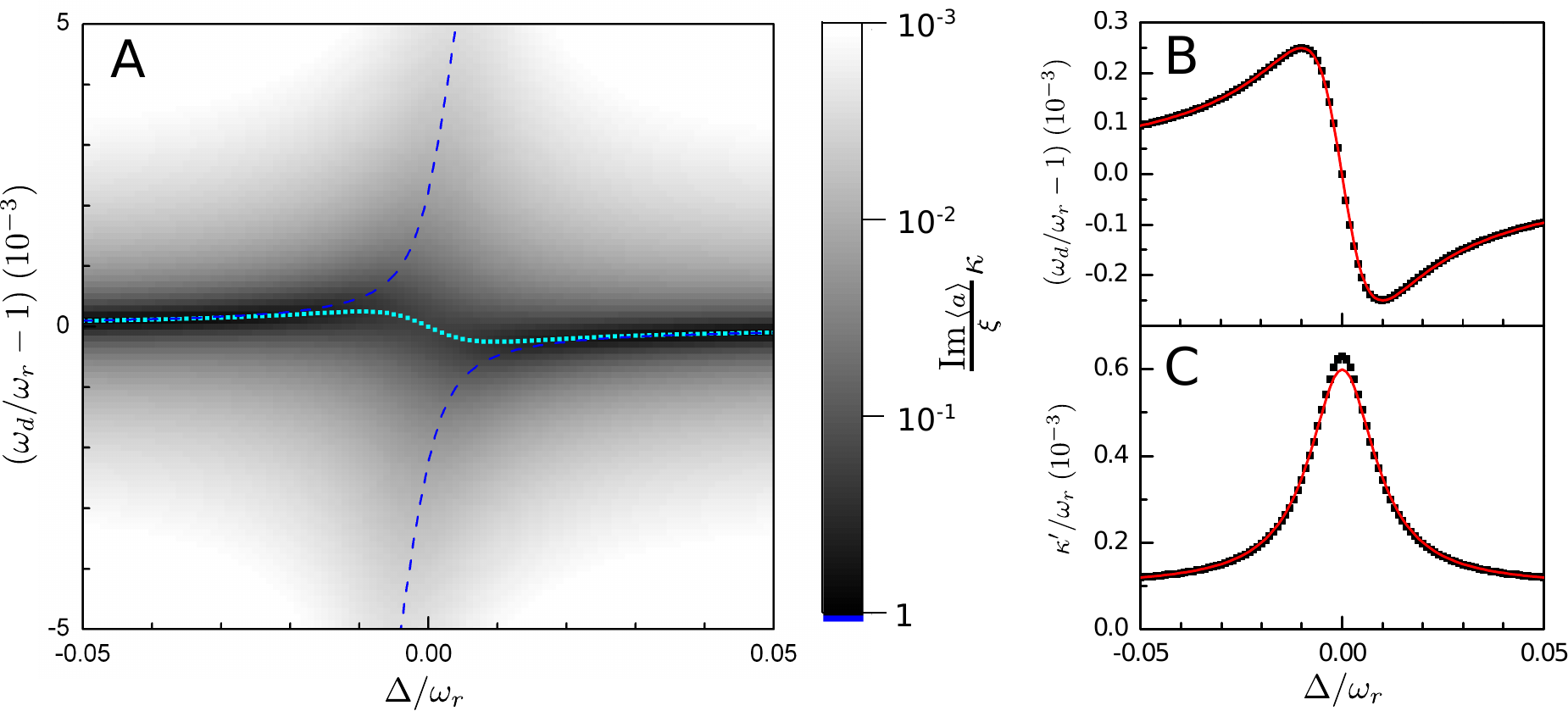}
\caption{Simulated transmission measurement of a weakly coupled resonator-qubit system using the master equation \refeq{eq:lindblad1}.  The parameters used were $N_\textrm{spins}=5$, $\gamma/\omega_r = 0.001$, $\kappa/\omega_r = \num{1e-4}$, $g/\omega_r = 0.001$, $\xi/\omega_r = \num{1e-5}$.  Graph A shows the transmission as a function of both the detuning ($\Delta=\Omega-\omega_r$) and the driving frequency ($\omega_r$).  The two dashed lines track the position of the first two transitions for each detuning value showing an anti-crossing near $\Delta=0$.  The dots mark the peak position determined for each vertical profile ($\Delta = \textrm{constant}$).  Graph B shows the dots from graph A as data points while the curve corresponds to the simple model (\ref{eq:phen1}) with $v=g\sqrt{N}$ and $\Gamma=N\gamma$.  Graph C shows the simulated cavity decay rate $\kappa'$ (or equivalently, the resonance width at half height) for each detuning value as datapoints.  The curve represents the Lorentzian lineshape (\ref{eq:phen2}) with $v=g\sqrt{N}$ and $\Gamma=N\gamma$.}\label{fig:seiji4}
\end{figure}

We note that in cases where $g\ll \gamma$, much of the phenomenology observed when moving in and out of resonance can be described using classical approximations for the field and sample.  In this case, the cavity resonance peak shifts as the detuning $\Delta$ is swept through 0 and the peak width dips around $\Delta=0$.  It can be shown \cite{Bushev2011}, that the functional dependence of these magnitudes is reasonably described by:
\begin{eqnarray}
\omega' = 2\pi f' & = & \omega_r + \frac{v^2(\omega_r - \Omega)}{(\omega_r-\Omega)^2+\Gamma^2},\label{eq:phen1}\\
\kappa' & = & \kappa + \frac{v^2\Gamma}{(\omega_r-\Omega)^2+\Gamma^2},\label{eq:phen2}
\end{eqnarray}
where $\omega'$ and $\kappa'$ are the modified resonance frequency and decay rates and $\Gamma$ and $v$ are equivalent (though not exactly equal) to the qubit decay rate $\gamma$ and qubit-cavity coupling $g$.  The exact correspondence of $v$ and $\Gamma$ to the Hamiltonian parameters can be assessed by simulating a transmission spectrum in the weak coupling case and fitting the results to the phenomenological formula as seen in figure \ref{fig:seiji4}.  If the (now single) resonance peak is fitted to a Lorentzian line shape for each detuning value $\Delta$, and the fitted values are compared to those obtained through the phenomenological model (\ref{eq:phen1}),(\ref{eq:phen2}), the substitution $v=g\sqrt{N}$ and $\Gamma=\gamma N$ exactly reproduces the simulated data.  The scaling of the effective coupling with $\sqrt{N}$ is the expected enhancement obtained by using a spin ensemble.  However, the increase of $\Gamma$ relative to the model's dephasing rate $\gamma$ is an artifact introduced by the approximations of the model.  This scaling of the decay is due to the fact we have replaced the spin ensemble with a giant spin $S=N/2$ and, as such, the master equation \ref{eq:master} includes superradiance of the sample \cite{Dicke1954,Chudnovsky2002}.  This effect does not commonly apply to real crystal samples where each spin decays independently and not as a collective mode.  From these considerations, for weak coupling cases it is reasonable to make the assignment $v=g$ and $\Gamma = \gamma$.

\begin{figure}[!tb]
\centering
\includegraphics[width=0.8\columnwidth]{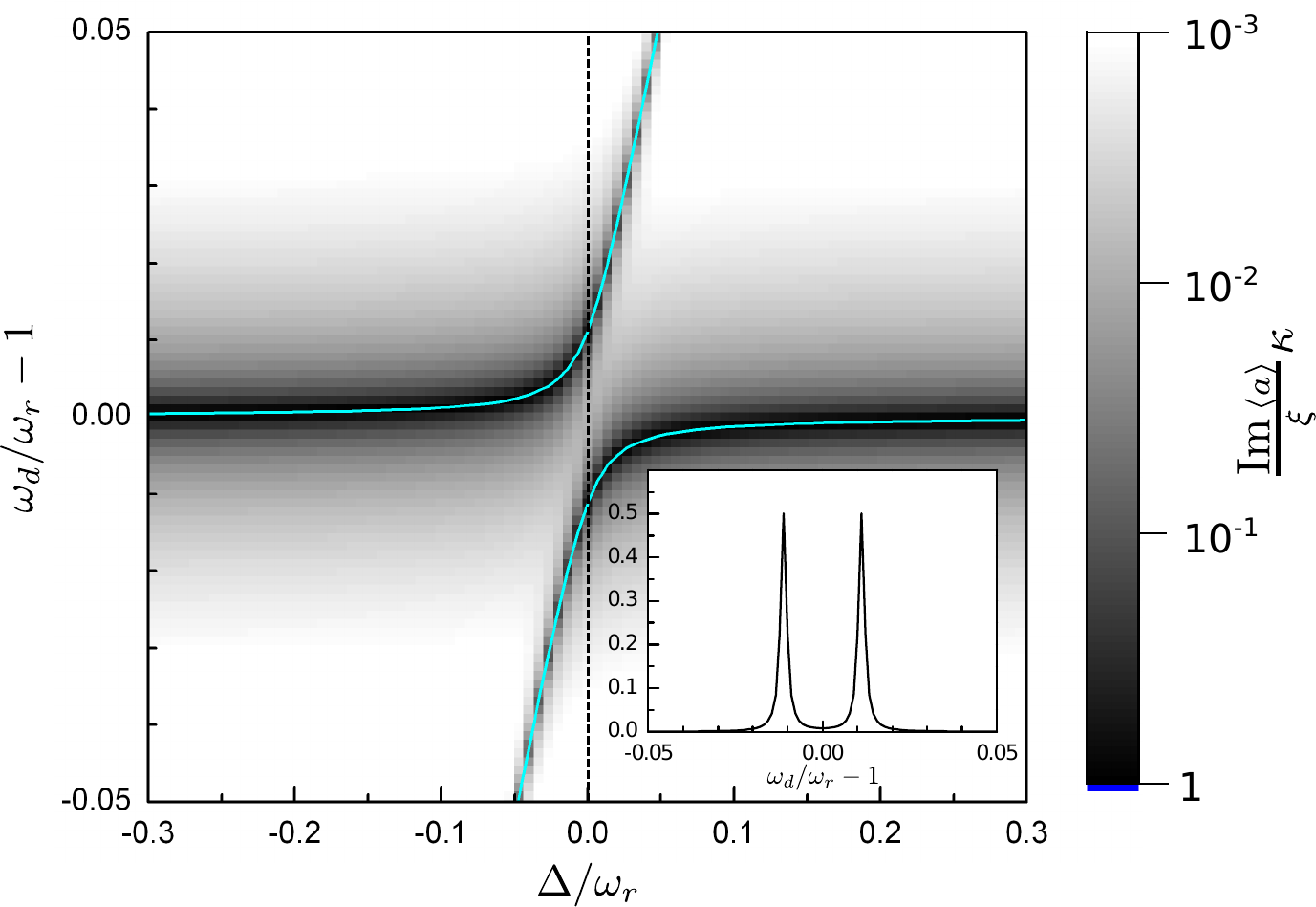}
\caption{Simulated transmission measurement of a strongly coupled resonator-qubit system using the master equation \refeq{eq:lindblad1}.  The parameters used were $N_\textrm{spins}=5$, $\gamma/\omega_r = \num{2e-4}$, $\kappa/\omega_r = 0.001$, $g/\omega_r = 0.005$, $\xi/\omega_r = \num{1e-5}$.  The lines track the position of the first two transitions for each detuning value given by the theory.  The inset shows two peak structure from the profile taken along the vertical dotted line.}\label{fig:seiji3}
\end{figure}

Figure \ref{fig:seiji3} also shows the expected measurement spectrum when the strong coupling conditions are met.  When the detuning approaches $\Delta=0$, the single cavity peak moves away from $\omega_r$ and reduces in intensity while a second peak appears.  Both peaks reach the same transmission value at $\Delta=0$.  As $\Delta$ continues to increase, the first peak gradually disappears and the second peak becomes the original cavity mode.  The position of each of these peaks coincides with the two transition frequencies from the ground state to the first two excited states given by the Hamiltonian (\ref{eq:ham0}) as long as the excitation power is low enough so that $N_\textrm{ex} < N_\textrm{spins}$.

\section{Sample spectroscopy using superconducting CPW resonators}\label{sec:res}
This section discusses the measurements obtained when replacing the CPW waveguides used section \ref{sec:wg} with CPW resonators described in chapter \ref{chap:CPWG}.  The samples used are the same as in \ref{sec:wg}:  DPPH, \ce{CaGdF}, and \ce{GdW30}.  Different sizes of CPW resonators were used for different samples sizes.  All experiments are performed at liquid helium temperatures (\SI{4.2}{\kelvin}).

\subsection{DPPH on a superconducting CPW resonator}\label{sec:DPPHres}

\subsubsection{Droplet of DPPH in DMF solution}\label{sec:DPPHrespippete}
\begin{figure}[!tb]
\centering
\includegraphics[width=\columnwidth]{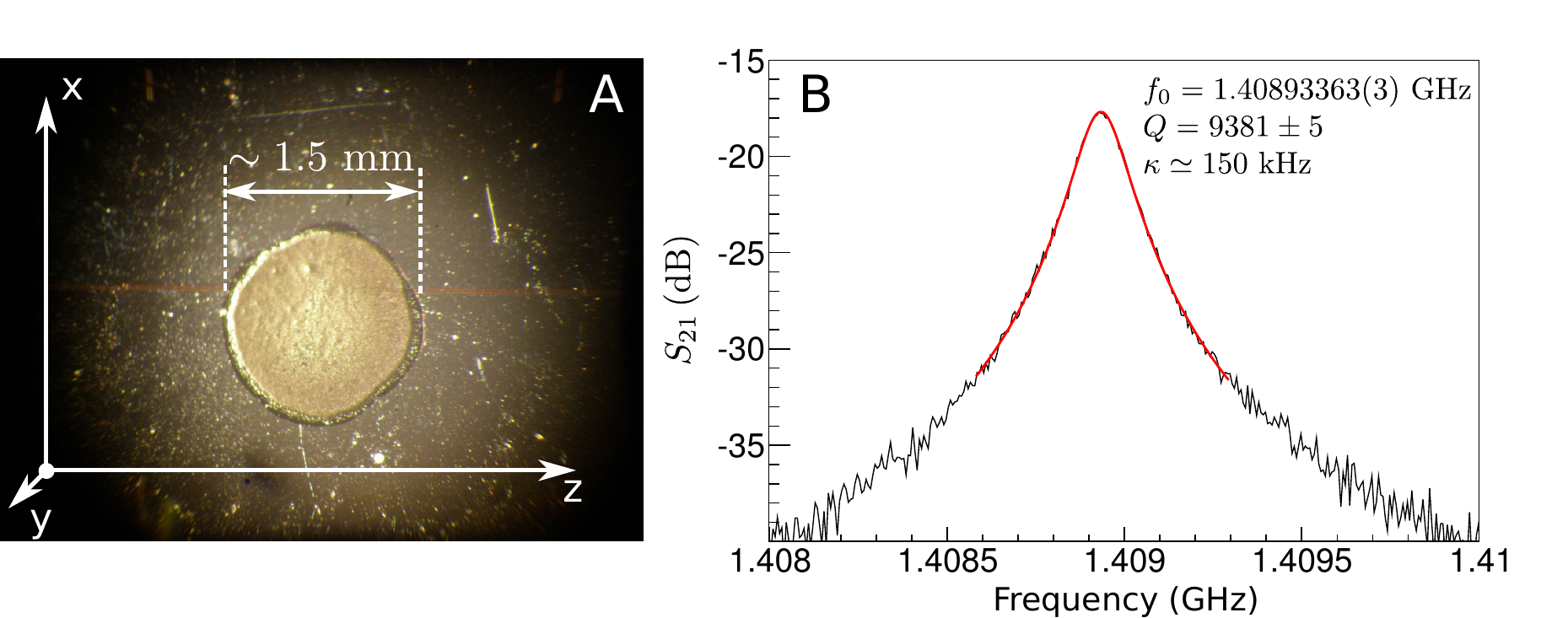}
\caption{Graph A shows a microscope image of a DPPH sample on a niobium CPW resonator and the laboratory reference frame.  Graph B shows the background transmission spectrum of the resonance including connecting wires.  The red line shows a fit of a Lorentzian  line shape to the measured spectrum and the fitted parameters.}\label{fig:resDPPHphoto}
\end{figure}

As in section \ref{sec:DPPHwg}, a saturated solution of DPPH powder in dimethyl sulfoxide (DMF) with and 5\% glycerol is prepared and a drop of the solution is placed on a resonator using a micro-pipette.  The resonator has \SI{7}{\micro\meter} gaps and a \SI{14}{\micro\meter} center line and has a \SI{1.409}{\giga\hertz} resonance frequency and quality factor $Q\sim 10000$, although these values vary when the DC magnetic field is swept.  The resonator and sample along with the background resonance can be seen in figure \ref{fig:resDPPHphoto}.

\begin{figure}[!tb]
\centering
\includegraphics[width=0.8\columnwidth]{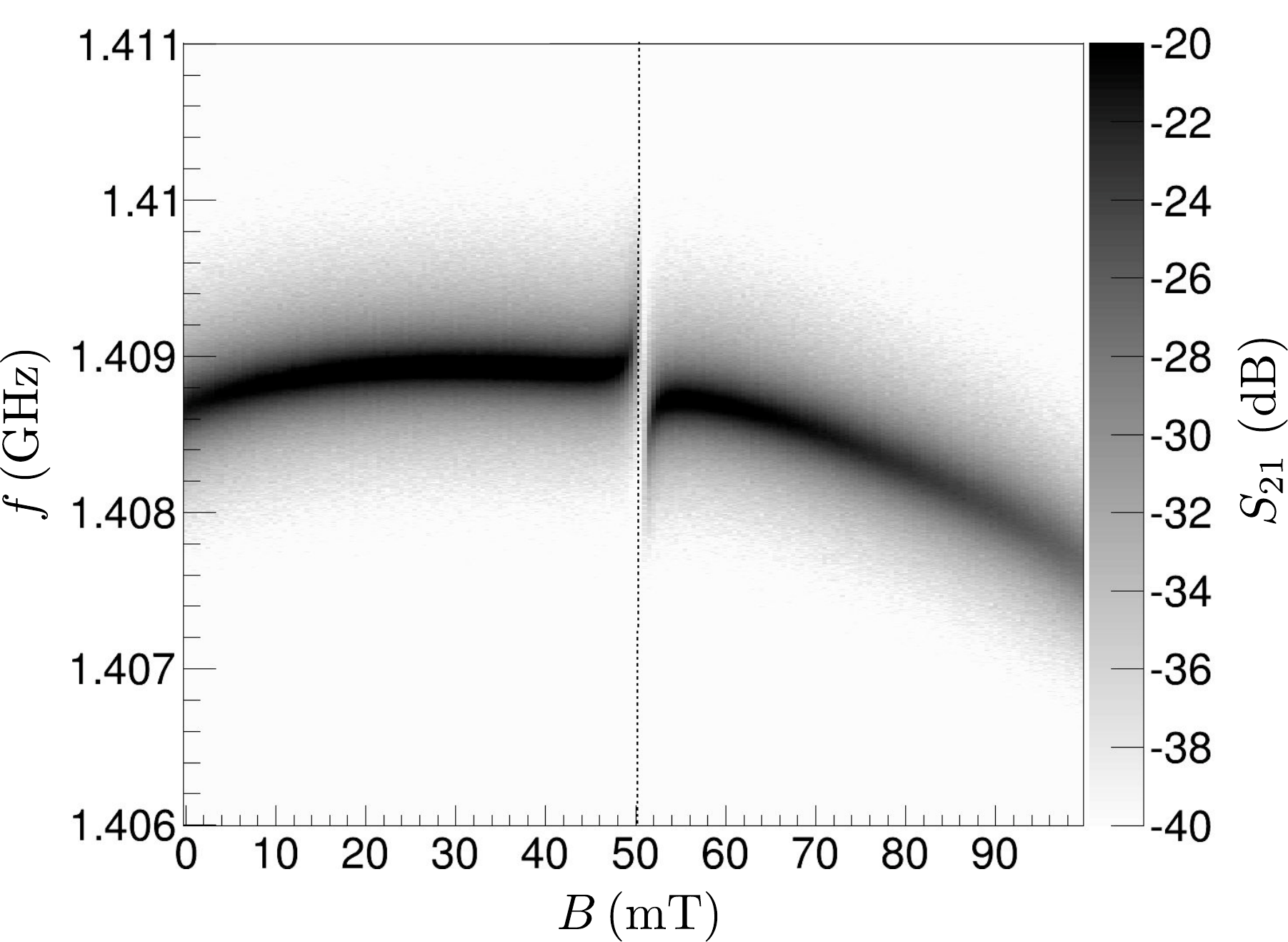}
\caption{Transmission intensity of DPPH on a niobium CPW resonator (greyscale) as a function of the driving frequency and the DC magnetic field applied along the laboratory Z axis (parallel to the transmission line).  The dotted line shows the expected transition frequency for the DPPH sample ($f_s = g_S\mu_\textrm{B}B/h$).}\label{fig:DPPH27R1}
\end{figure}

\begin{figure}[!tb]
\centering
\includegraphics[width=\columnwidth]{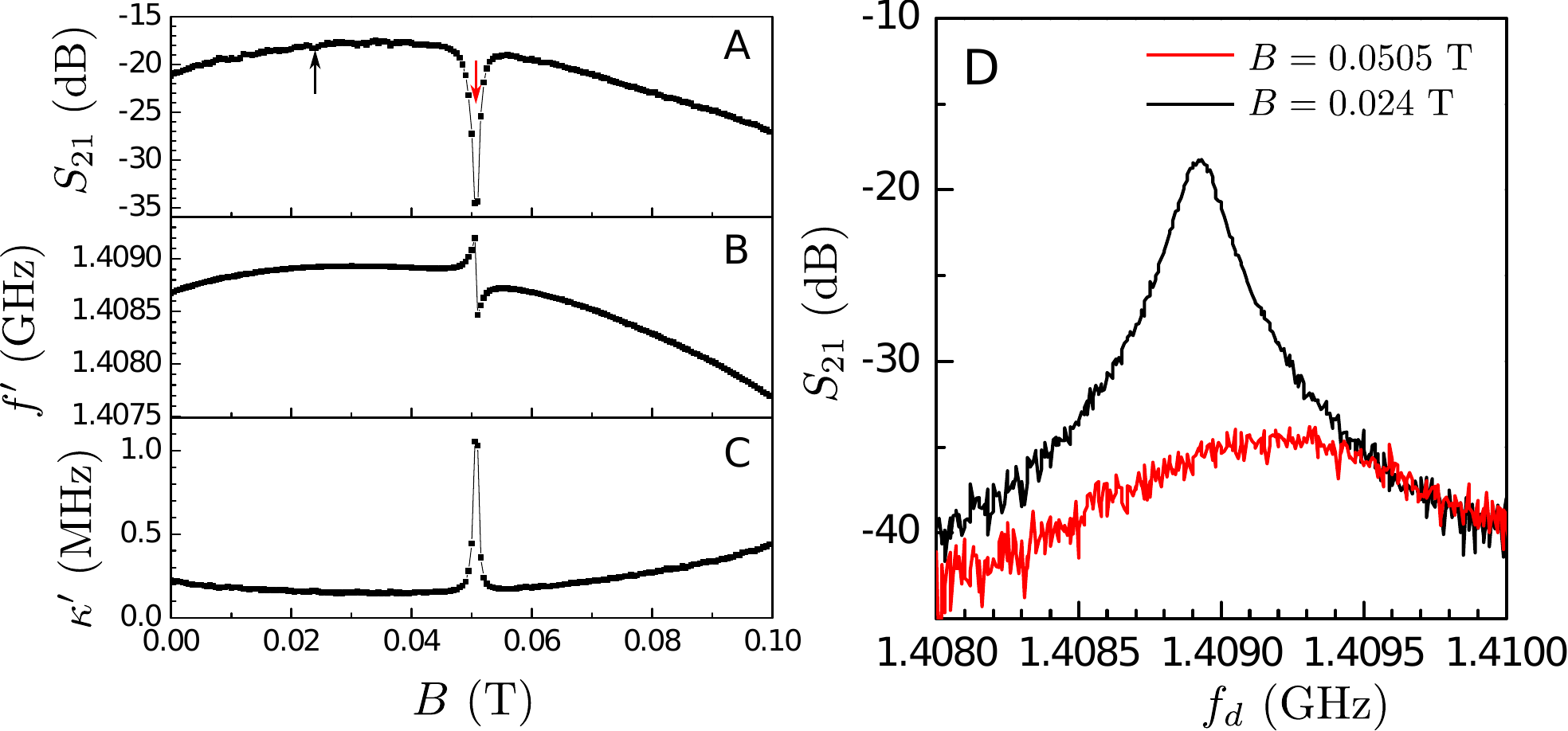}
\caption{Resonance properties as a function of field for the measurement in figure \ref{fig:DPPH27R1}.  Graphs A,B and C show the measured values of peak $S_{21}$, peak frequency ($f'$) and peak width at half maximum (FWHM or $\kappa'$).  All three show the absorption feature at a field value compatible with a sample $g$-factor $g_S\simeq 2$.  Graph D shows the transmission spectrum at two values of the magnetic field, one out of resonance and another in resonance with the sample.  The two chosen fields are marked with arrows in graph A.}\label{fig:DPPH27R2}
\end{figure}

The full transmission measurement with varying field and excitation frequency is shown in figure \ref{fig:DPPH27R1}.  The incident power applied from the network analyzer was $-40$ dBm.  We see the transmission peak of the resonator move across the spectrum as the field is increased due to the influence of this applied field.  When the resonance condition is met, the peak is suppressed and becomes much wider (figure \ref{fig:DPPH27R2}D).  The progression of the resonance characteristics (peak intensity, resonant frequency, peak width) can be extracted by fitting a Lorentzian line shape to the spectrum acquired at each field.  The results of these fits are shown in figure \ref{fig:DPPH27R2}.  Again the resonance feature can be clearly seen in graphs A,B and C.

Only a single peak was detected in this measurement.  We estimate that, for the powers used and losses detected at the output, the number of photons stored in the cavity is approximately $10^{10}$.  For a rough estimate of the number of interacting spins we use the dimensions from the photograph in figure \ref{fig:resDPPHphoto}A (about \SI{1.5}{\milli\meter} by \um{20} by \um{20} since from section \ref{sec:coupCPWG_SMM} we see that only spins within about \um{10} of the center line contribute) and assume a sample density similar to the bulk sample (may be overestimated).  Only the excess of spins in the ground state contribute to the signal so this number must be reduced by a factor 0.008 (see equation (\ref{eq:boltzfact}) below).  This gives us about $\sim 10^{13}$ contributing spins to be compared with the $10^{10}$ photons.  This means that the hybrid system is probably not saturated (i.e., that $n_{\rm ex} < N$) but that the coupling was too weak to overcome the dephasing rates of the system.

\begin{figure}[!tb]
\centering
\includegraphics[width=\columnwidth]{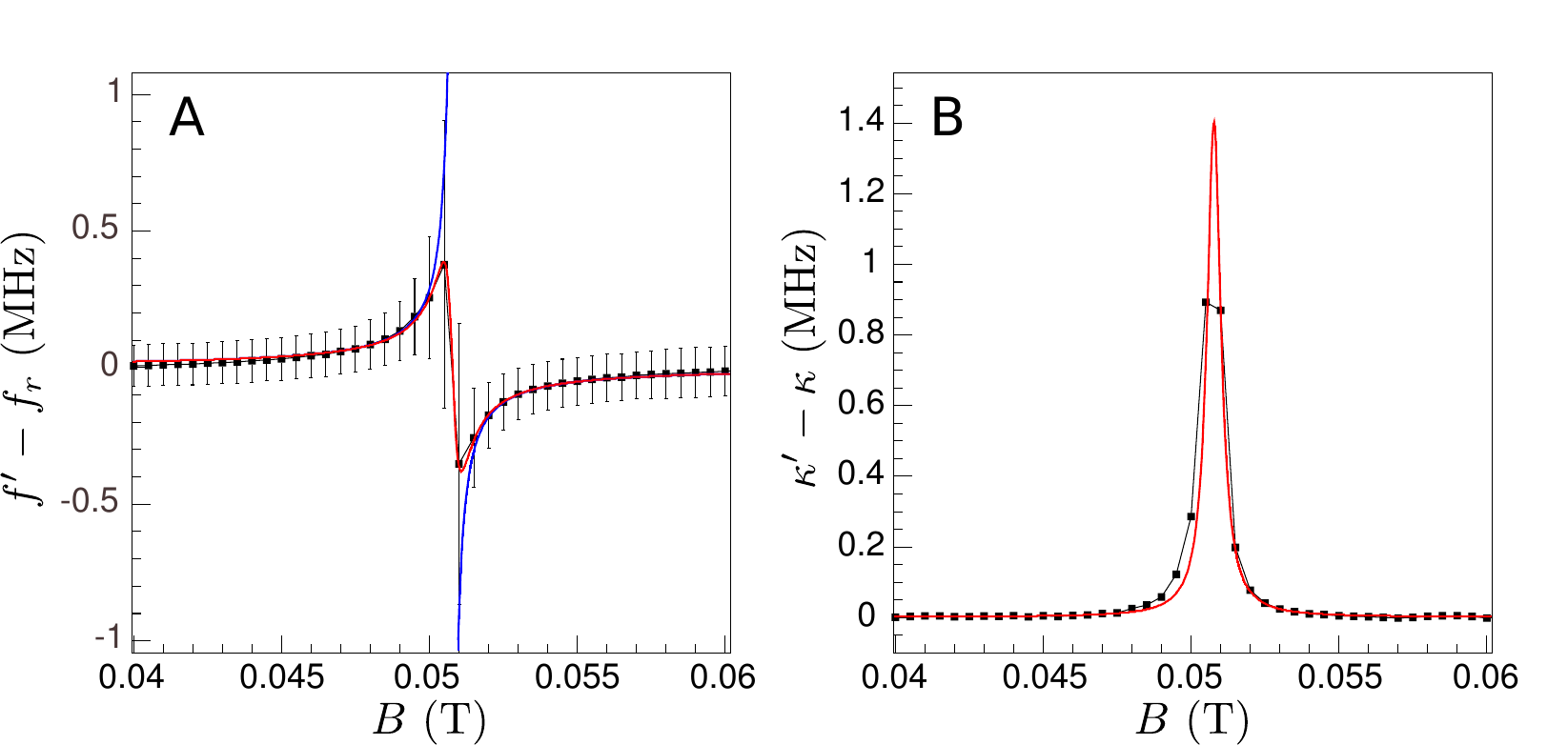}
\begin{tabular}{c|c}
Frequency fit (Graph A) & Linewidth fit (Graph B) \\
\hline
$ g_\textrm{eff} = 2.505 \pm \SI{0.024}{\mega\hertz}$ & $ g_\textrm{eff} = 3.34 \pm \SI{0.04}{\mega\hertz}$ \\
$ \Gamma = 8.20 \pm \SI{0.17}{\mega\hertz}$ & $\Gamma =  7.9 \pm \SI{0.3}{\mega\hertz}$\\
$g_S = 1.9818\pm0.0003$ & $g_S = 1.9822\pm0.0003$
\end{tabular}
\caption{Fit of the phenomenological model from equations (\ref{eq:phen1}) and (\ref{eq:phen2}) to the measured frequency $f'$ and measured linewidth $\kappa'$ extracted from figure \ref{fig:DPPH27R2}.  The backgrounds have been removed and the difference to cavity background is shown.  Graph A shows the frequency fit and graph B shows the linewidth fit (both in red).  The error bars correspond to the value of $\kappa'$ for each field.}\label{fig:DPPH27R3}
\end{figure}

Taking the graphs in figure \ref{fig:DPPH27R2}B and \ref{fig:DPPH27R2}C and fitting them to classical equations (\ref{eq:phen1}) and (\ref{eq:phen2}), we can approximately determine the effective coupling strength ($g_\textrm{eff}=v$) to the sample and the spin linewidth.  The result of this fit is shown in figure \ref{fig:DPPH27R3} with $g_\textrm{eff}\sim \SI{3}{\mega\hertz}$ and $\Gamma\sim\SI{8}{\mega\hertz}$ confirming that we are in the weak coupling regime ($g<\Gamma$).  The amplitude of the frequency change in figure \ref{fig:DPPH27R3}A is equal to $v^2/\Gamma$ while the separation of the upward and downward peaks is equal to $\Gamma$.  We note that the limiting factor here is the spin dephasing since the free cavity linewidth $\kappa\sim\SI{200}{\kilo\hertz}$ is smaller than both $\Gamma$ and $g_\textrm{eff}$.  Also, the fitted values of $\Gamma$ are compatible with those found in the literature for DPPH \cite{Goldsborough1960,Alger1968,Narkowicz2005}.  For comparison, the blue lines in figure \ref{fig:DPPH27R3}A show the positions of the ideal transitions for a strongly coupled system with the same $g_\textrm{eff}$ from the fit.  They are given by $(E_{\overline{\pm,0}}-E_{\uparrow,0})$ from equations (\ref{eq:jcenergy}) and (\ref{eq:Eground}):
\begin{equation}
\hbar\omega_\pm = (E_{\overline{+,0}}-E_{\uparrow,0}) = \frac{\hbar(\omega_r+\Omega)}{2} \pm \frac{\hbar}{2}\sqrt{4g^2+(\Omega-\omega_r)^2} \label{eq:theotrans}
\end{equation}
where $\Omega=g_S\mu_B/\hbar B$ is the spin transition frequency.

Additionally, the expected values for $g_\textrm{eff}$ can be estimated from figure \ref{fig:gvsh} and equation (\ref{eq:coupling_explicit}).  The coupling scales according to $\sqrt{nl}$ and is proportional to the operating frequency (through the rf field values).  From figure \ref{fig:gvsh} we have $g_\textrm{eff}\sim \SI{100}{\mega\hertz}$ for a \um{40} long sample that fills the effective height at a \SI{6}{\giga\hertz} operating frequency and \SI{0}{\kelvin} temperature.  Scaling this value using the measurement conditions we get an estimate of:
\begin{eqnarray}
B &=& \frac{1-e^{-\SI{1.5}{\giga\hertz}/\SI{4}{\kelvin}}}{1+e^{-\SI{1.5}{\giga\hertz}/\SI{4}{\kelvin}}} \simeq 0.008 \quad \textrm{(Boltzmann factor)}\label{eq:boltzfact}\\
g_\textrm{eff} &\sim & \SI{100}{\mega\hertz}\times\sqrt{B}\times\sqrt{\frac{\um{1000}}{\um{40}}}\frac{\SI{1.5}{\giga\hertz}}{\SI{6}{\giga\hertz}}\simeq \SI{10}{\mega\hertz},
\end{eqnarray}
which is reasonably close to the measured value of \SI{3}{\mega\hertz} considering the uncertainties in the estimation of the sample size and spin density inside the cavity.

\subsubsection{DPPH pellets: experiments in the vicinity of strong coupling}

\begin{figure}[!b]
\centering
\includegraphics[width=\columnwidth]{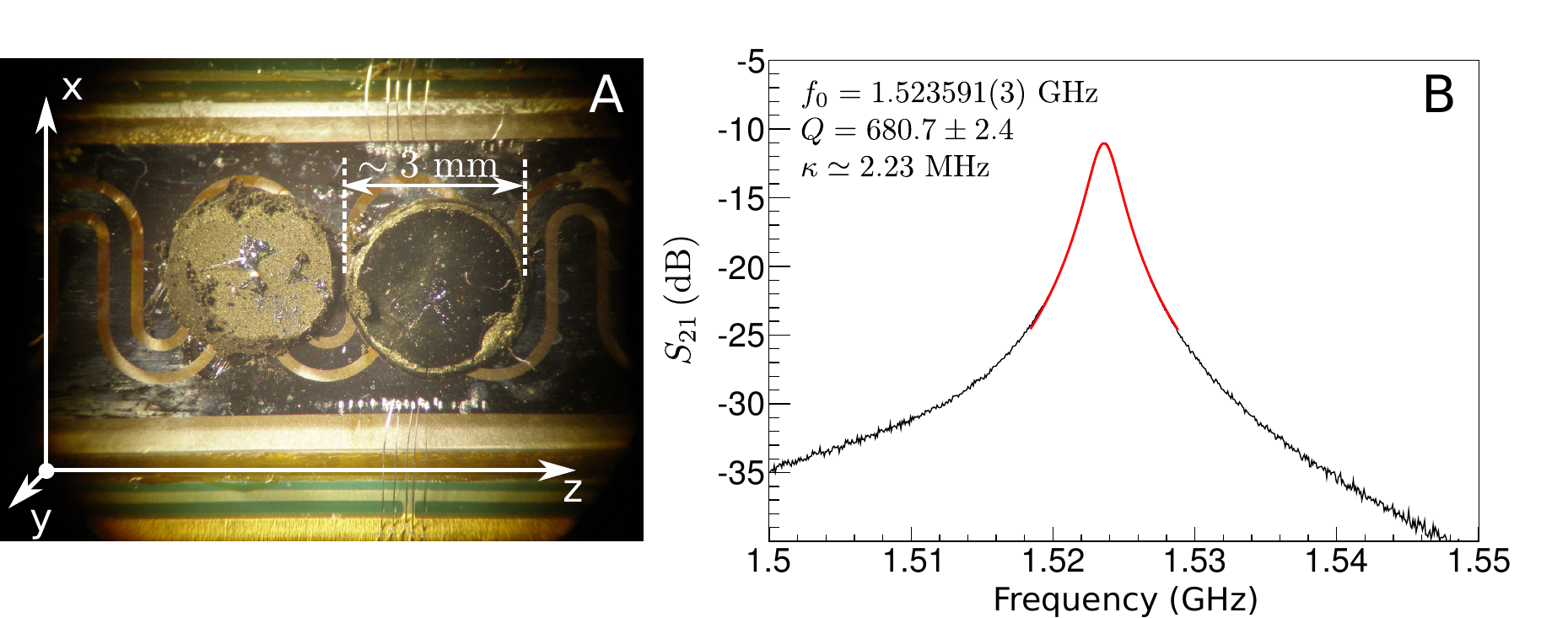}
\caption{Graph A shows a microscope image of DPPH pellets on a niobium CPW resonator and the laboratory reference frame.  Graph B shows the background transmission spectrum of the resonance including connecting wires.  The red line shows a fit of a Lorentzian  line shape to the measured spectrum and the fitted parameters.}\label{fig:DPPH11R1}
\end{figure}

\begin{figure}[!t]
\centering
\includegraphics[width=1\columnwidth]{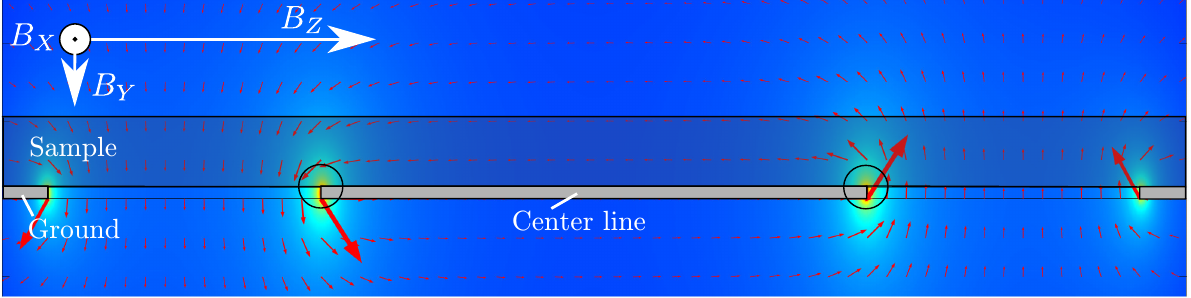}
\caption{Cross-section and simulated field profile for a CPW.  The laboratory (magnet) reference frame corresponding to the experiment is shown along with the sample}\label{fig:bprofile_DPPH}
\end{figure}

To achieve higher couplings, we perform another experiment using a larger sample consisting of compressed pellets made of DPPH powder.  These pellets are approximately \SI{3}{\milli\meter} in diameter and can be seen in figure \ref{fig:DPPH11R1}A on a wide center line resonator (\um{400} line and \um{200} gaps).  As before, we show the background resonance and resonator properties in figure \ref{fig:DPPH11R1}B.  The incident power applied from the network analyzer is chosen to be $-45$ dBm.  The meandering structure of the resonator and the sample placement makes it preferable to apply the tuning field in the X direction since an important contribution to the coupling will come from the rf field parallel to the resonator plane.  Recalling the calculations from section \ref{sec:coupCPWG_SMM}, we see in figure \ref{fig:bprofile_DPPH}, that the most intense rf magnetic field is concentrated around the edges of the centerline and ground planes where there are both $Z$ and $Y$ rf field components.  These are the dominant regions that contributing to the $\int B_{X,Y,Z}^2dS$ integrals appearing in equation \refeq{eq:coupling_explicit}.  Since the field has components in both the $Z$ and $Y$ directions, both contribute approximately equally to the coupling (i.e. $\int B_{Z}^2dS \sim \int B_{Y}^2 dS$) even for very flat samples while the $X$ direction has $\int B_{X}^2dS = 0$.  The tuning fields for the sample must be applied either in the $Z$ or $X$ direction since fields in $Y$ rapidly degrade the resonator performance.  However, any spin matrix element for components parallel to the tuning field are suppressed for a spin $1/2$ system.  Therefore, from \refeq{eq:coupling_explicit} we expect a $\sim\sqrt{2}$ larger coupling when applying the DC field in the laboratory $X$ direction (which suppresses no contribution) than in the $Z$ direction (which suppresses the $\int B_{Z}^2dS$ contribution).

\begin{figure}[!p]
\centering
\includegraphics[width=0.9\columnwidth]{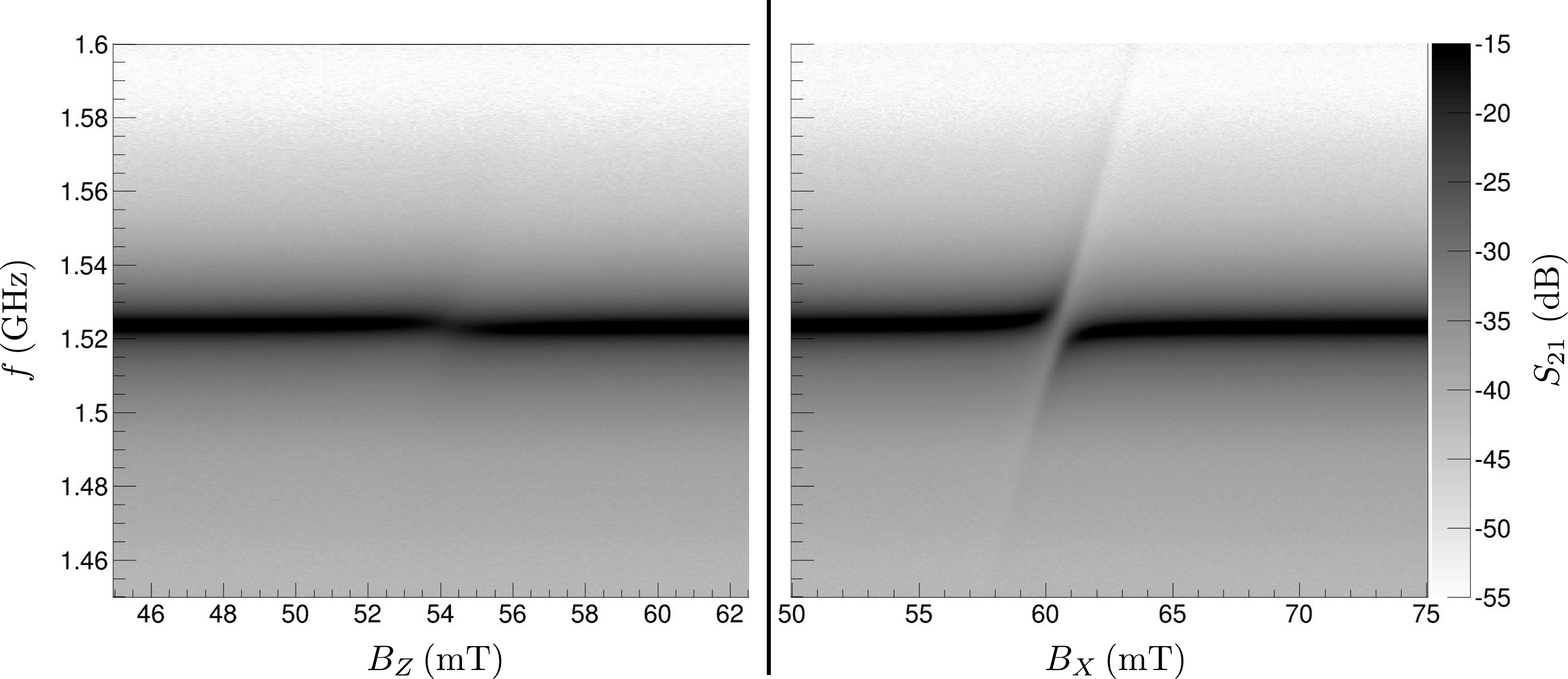}
\caption{Transmission intensity of DPPH pellets on a niobium CPW resonator as a function of the driving frequency and the DC magnetic field applied along the laboratory Z axis (graph A) and X axis (graph B).}\label{fig:DPPH11R2}
\end{figure}

\begin{figure}[!p]
\centering
\includegraphics[width=0.9\columnwidth]{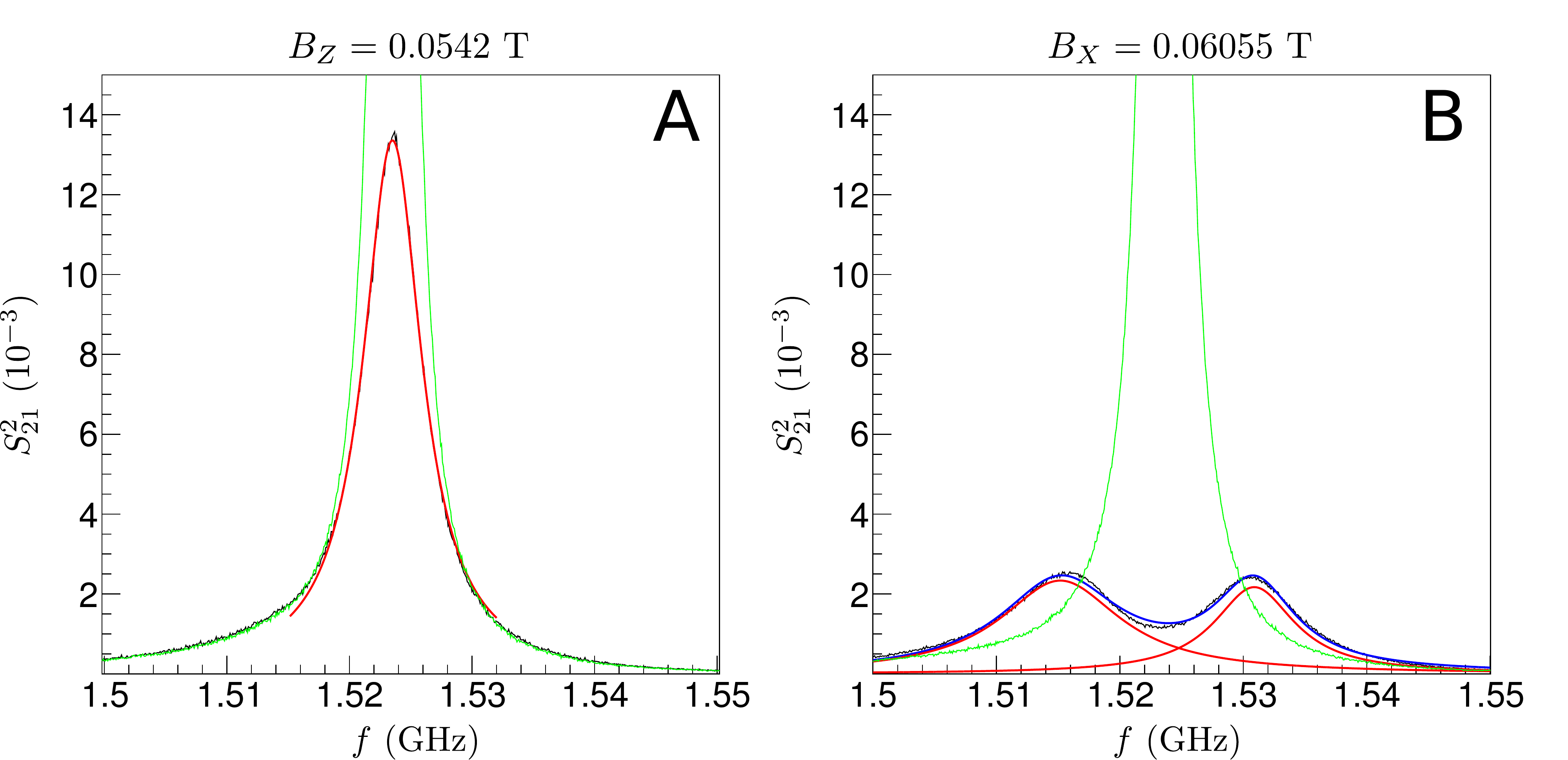}\\[3mm]
\begin{tabular}{c|c}
Graph A & Graph B\\
\hline
$f_{0} = 1.52359907(9)$ GHz & $f_{0} = 1.52361722(8)$ GHz \\
$\kappa = 2.30974(28)$ MHz & $\kappa = 2.21833(26)$ MHz\\
\hline
$f_1 = 1.523508(6)$ GHz & $f_1 = 1.515178(18)$ GHz\\
$\Gamma_1 = 5.819(17)$ MHz &  $\Gamma_1 = 11.86(6)$ MHz\\
& $f_2 = 1.530887(16)$ GHz \\
& $\Gamma_2 = 7.95(5)$ MHz \\
& $g \sim (f_2-f_1)/2 = 7.855(12)$ MHz 
\end{tabular}
\caption{Transmission profile at $B_Z=\SI{0.0542}{\tesla}$ (graph A) and $B_Z=\SI{0.06055}{\tesla}$ (graph B - average of profiles at $B_X=\SI{0.0605}{\tesla}$ and $B_X=\SI{0.0606}{\tesla}$).  In graph A, the red line is a Lorentzian fit to the data.  In graph B, a double Lorentzian peak structure is used to fit the data.  The red lines are the individual peaks while the blue line is the sum of the two peaks.  $f_{1,2}$ and $\Gamma_{1,2}$ are the center frequency and peak widths of each peak respectively.  The measured coupling is $g_\textrm{eff} \simeq \SI{7.9}{\mega\hertz}$.  The green line in both A and B represents the cavity transmission when the sample is out of resonance.} \label{fig:DPPH11R3}
\end{figure}

Similar experiments to those done on the droplet sample (figure \ref{fig:resDPPHphoto}) give the transmission spectrum seen in figure \ref{fig:DPPH11R2} where large differences can be seen depending on the direction of the DC magnetic field ($Z$ or $X$).  In \ref{fig:DPPH11R2}A, the result is similar to that of figure \ref{fig:DPPH27R1} although harder to appreciate with the scales used.  However, in \ref{fig:DPPH11R2}B, we are much closer to a strong coupling regime.  The spin absorption line can be seen to approach the resonator frequency where the levels become hybrids of photon and spin states and are split by $2g_\textrm{eff}$.  If we take the profile at the resonance point (figure \ref{fig:DPPH11R3}B), a two peak structure is clearly distinguishable while in the $Z$ case (figure \ref{fig:DPPH11R3}A) we see only a single peak structure.  The profiles can be fitted to single and and two Lorentzian lines from which we can get an estimate of the coupling (table in figure \ref{fig:DPPH11R3}).  We note also that the resonance in the X direction is displaced from the $g_S\simeq 2$ and appears to be at $g_S\simeq 1.8$.  Since the resonance appears to be at the expected field when the field is applied in the Z direction, we assume this is due to an error in the magnet calibration.  We will assume that the sample does in fact have $g_S=2$ and correct this error with a field shift as this error is most likely not an issue with the sample (DPPH has been commonly used as a marker and for calibration in standard EPR experiments).

\begin{figure}[!tb]
\centering
\includegraphics[width=1\columnwidth]{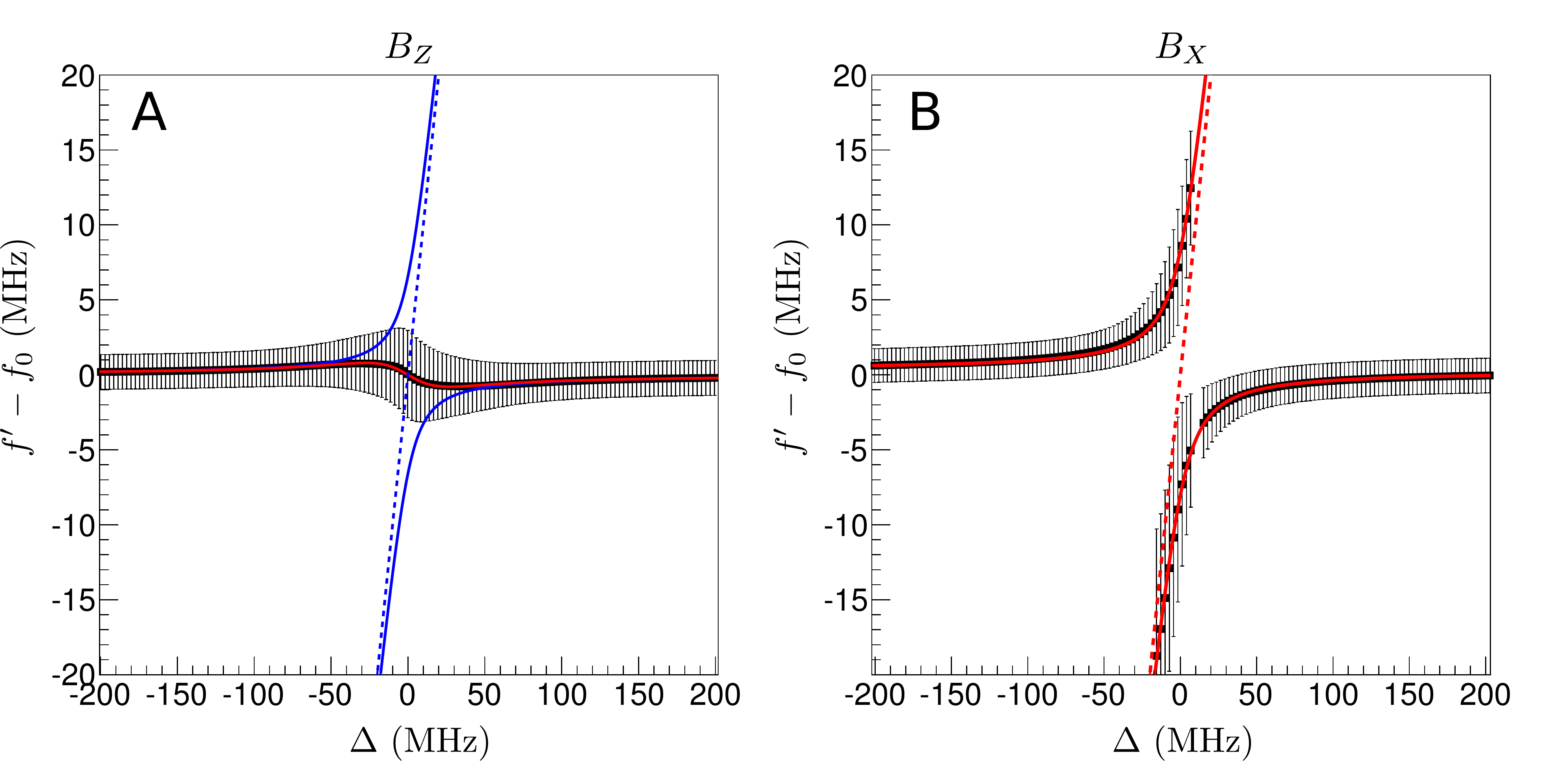}
\begin{tabular}{c|c}
Graph A fit & Graph B fit \\
\hline
$ g_\textrm{eff} = 5.99 \pm \SI{0.05}{\mega\hertz}$ & $ g_\textrm{eff} = 8.172 \pm \SI{0.021}{\mega\hertz}$ \\
$ g_S = 2.00802(26)$ & $g_S =  2$ (fixed)\\
$ f_r=\omega_r/2\pi = \SI{1.52345(3)}{\giga\hertz}$ & $f_r = \omega_r/2\pi =\SI{1.523446(7)}{\giga\hertz}$ \\
$ \Gamma = \SI{27.193(27)}{\mega\hertz}$ & $\Gamma \sim 5-\SI{7}{\mega\hertz}$ (figure \ref{fig:DPPH11R5})\\
\end{tabular}
\caption{Peak positions as a function of the detuning $\Delta = g_S\mu_B/h B_{X,Z} - f_r$.  The positions are fitted to equation \refeq{eq:phen1} in graph A and to equation \refeq{eq:theotrans} in graph B.  The blue lines in graph A represent the behavior from equation \refeq{eq:theotrans} with the coupling $g$ from the fit (in red).  In both graphs, the dotted line represents the spin transition energy as a function of the detuning.  The error bars have length $\kappa'$, i.e. the peak width at each detuning value.}\label{fig:DPPH11R4}
\end{figure}

As we did in the droplet case, we track the resonance characteristics as the field is swept in both cases and plot the measured peak positions as a function of the field.  In the Z direction case we fit the behavior to the weak coupling formula \refeq{eq:phen1} while in the X direction case we use the level separations \refeq{eq:theotrans}.  The results are shown in figure \ref{fig:DPPH11R4}A and \ref{fig:DPPH11R4}B respectively.  From the fit results we find a value of the coupling $g_Z\simeq \SI{6}{\mega\hertz}$ for the Z direction and $g_X\simeq \SI{8.2}{\mega\hertz}$ for the X direction that approximately fulfill the expected relation $g_X/g_Z\sim\sqrt{2}$.  The differences are probably due to the fact that the resonator is not perfectly aligned with the fields and that its meandering structure has areas that are not oriented exactly as in figure \ref{fig:bprofile_DPPH}.

\begin{figure}[!tb]
\centering
\includegraphics[width=0.6\columnwidth]{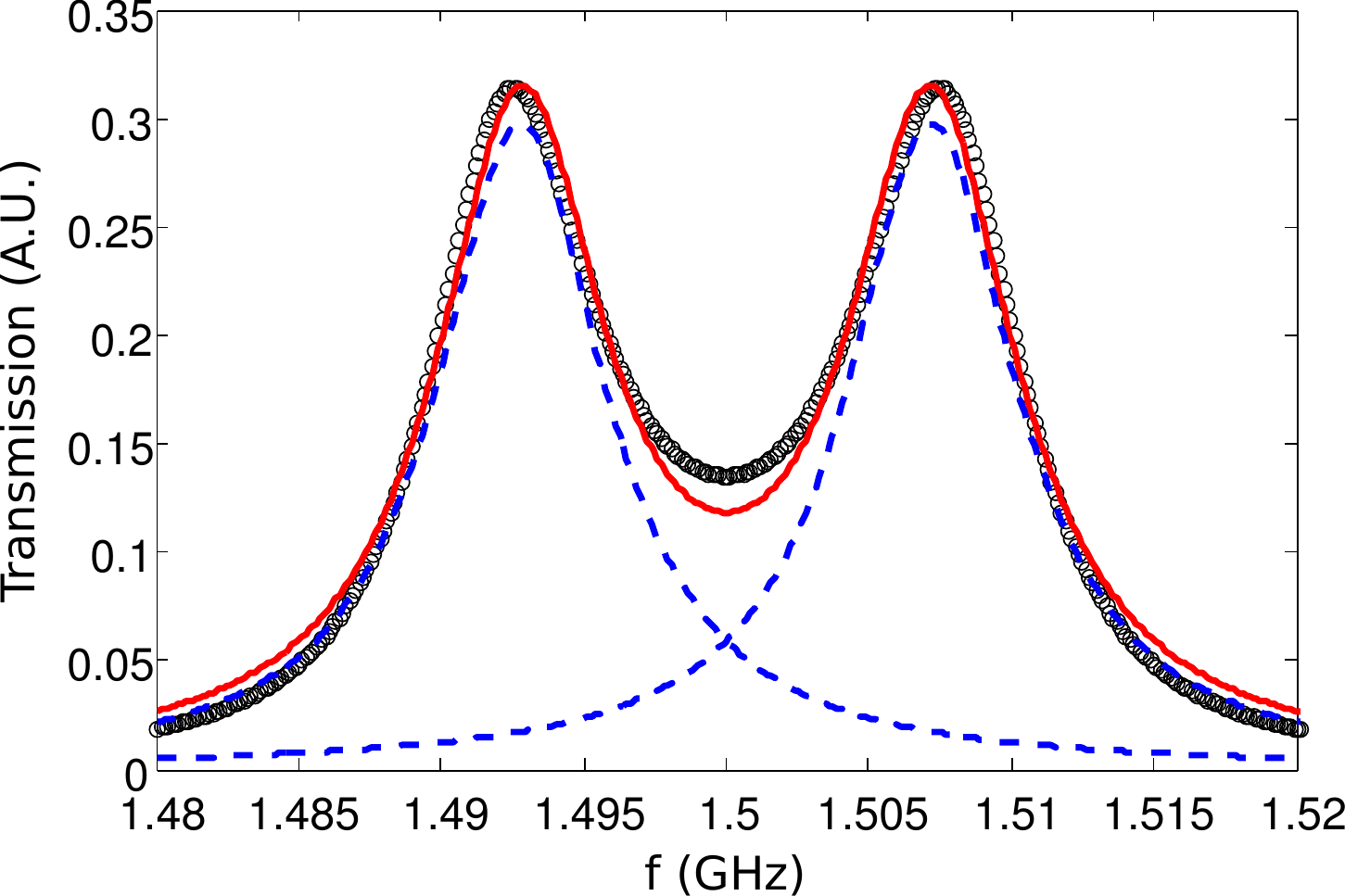}
\begin{tabular}{c|c}
Simulation & Double Lorentzian fit \\
\hline
$\omega_r/2\pi = \SI{1.5}{\giga\hertz}$ & $\omega_r/2\pi = \SI{1.5}{\giga\hertz}$ \\
$g = \SI{8}{\mega\hertz}$ & $g = \SI{7.2}{\mega\hertz}$\\
$\kappa = \SI{2}{\mega\hertz}$ & $\Gamma = \SI{7.2}{\mega\hertz}$ \\
$\gamma = \SI{5}{\mega\hertz}$ & 
\end{tabular}
\caption{Simulation using parameters similar to those measured (see section \ref{sec:seiji}).  A single spin and low excitation power was used.  From comparison with the data measured in figure \ref{fig:DPPH11R4}B, we can estimate that the spin linewidth in the experiment was around 5-\SI{7}{\mega\hertz}.  Also, the double Lorentzian fit to the simulation underestimates the value of $g$, which is also observed comparing the results from the fits in figure \ref{fig:DPPH11R3}B and \ref{fig:DPPH11R4}B.}\label{fig:DPPH11R5}
\end{figure}

The phenomenological model \refeq{eq:phen1} provides an estimate of the spin linewidth for the Z field case while the fit for the X case does not provide the spin linewidth directly.  To estimate this linewidth value we return to the theoretical model from section \ref{sec:seiji}.  If we introduce the measured values of $g$, $\kappa$, $\omega_r$ and try different values of $\gamma$, we can obtain profiles similar to those seen in figure \ref{fig:DPPH11R3}B.  From here we find that the spin linewidth is about $\gamma\sim 5-\SI{7}{\mega\hertz}$ (figure \ref{fig:DPPH11R5}).  We note also that, if we fit the simulated data to a double Lorentzian lineshape, this fit underestimates the coupling value.  This is the same as what we see in our fits from figure \ref{fig:DPPH11R3}B and \ref{fig:DPPH11R4}B.  We also see from the two linewidth values, $\gamma_Z\sim \SI{27}{\mega\hertz} > \gamma_X \sim 5-\SI{7}{\mega\hertz}$.  This is unexpected as the linewidth should be isotropic and, therefore, probably indicates some field homogeneity issue present in the Z axis but not in the X axis.  This could lead to a larger inhomogeneous broadening when the magnetic field is applied along Z, possibly linked to the large sample size.  We have therefore achieved conditions close to strong coupling for DC fields parallel to the X axis, whereas the combination of a larger linewidth and a weaker coupling lead to a classical behavior for fields parallel to the Z axis.

\subsection{\ce{CaGdF} on a superconducting CPW resonator}

\begin{figure}[p]
\centering
\includegraphics[width=0.5\columnwidth]{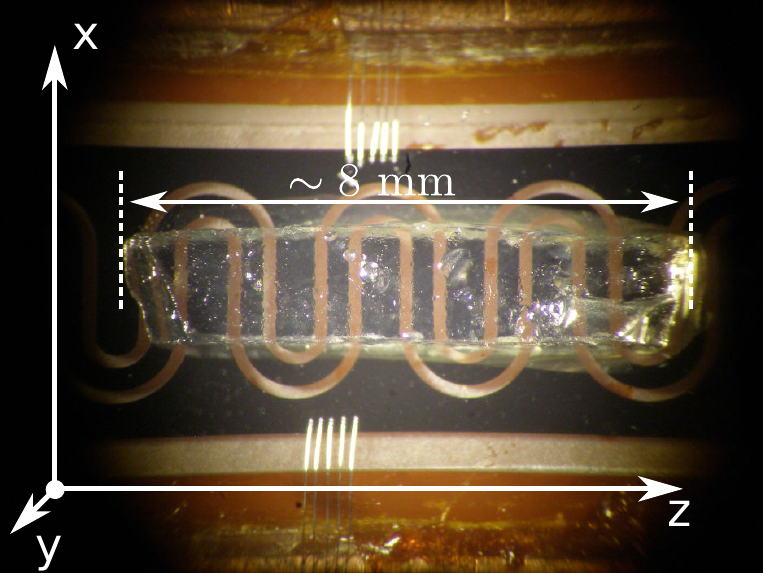}
\caption{Microscope image of \ce{CaGdF} crystal on a niobium CPW resonator.  The crystal has the [110] axis paralell to the long crystal edges, while the top face is the $(111)$ plane.  The laboratory (magnet) reference frame is also shown.  }\label{fig:GdF11R1}
\end{figure}

\begin{figure}[p]
\centering
\includegraphics[width=1\columnwidth]{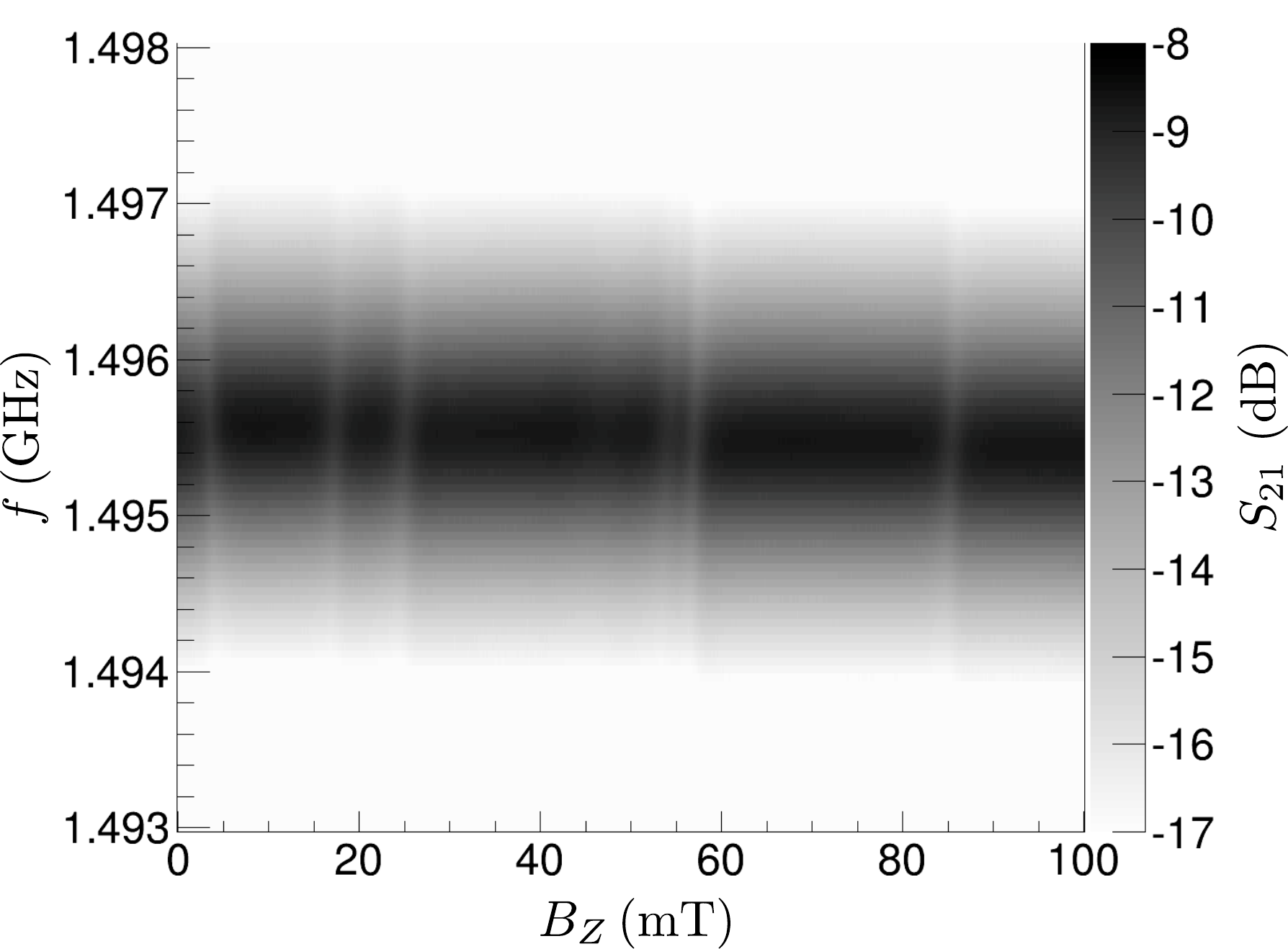}
\caption{Transmission intensity of a \ce{CaGdF} crystal on a niobium CPW resonator as a function of the driving frequency and the DC magnetic field applied along the laboratory Z axis (perpendicular to the transmission line direction and paralell to the surface).}\label{fig:GdF11R2}
\end{figure}

We now test the same high quality \ce{CaGdF} crystal used in section \ref{sec:CaGdFline} on a wide resonator (\um{200} gaps and \um{400} center line) like the one from figure \ref{fig:DPPH11R1}.  The sample and resonator are shown in figure \ref{fig:GdF11R1}.  The experiment is similar to the previous one on DPPH pellets (\ref{sec:DPPHres}) using appropriate field and frequency ranges with the magnetic field applied along the Z laboratory axis.  The incident power applied from the network analyzer was $-30$ dBm.  The full transmission spectrum is shown in figure \ref{fig:GdF11R2}.

\begin{figure}[!tb]
\centering
\includegraphics[width=0.75\columnwidth]{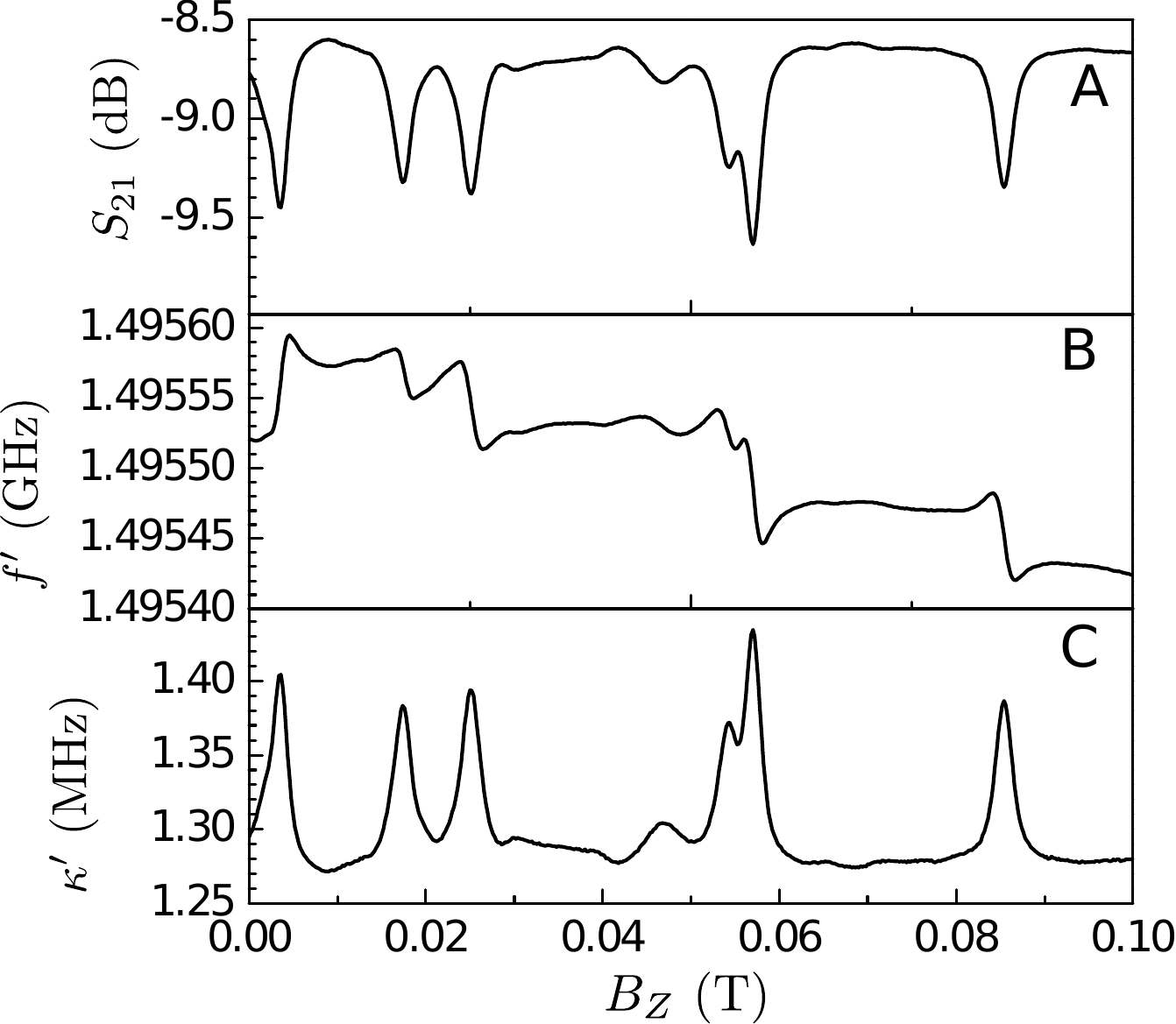}
\caption{Peak transmission (A), resonant frequency ($f'$, B) and peak width ($\kappa'$, C) for the experiment in figure  \ref{fig:GdF11R2}.  Up to 7 absorption lines can be distinguished.}\label{fig:GdF11R3}
\end{figure}

\begin{figure}[!tb]
\centering
\includegraphics[width=0.8\columnwidth]{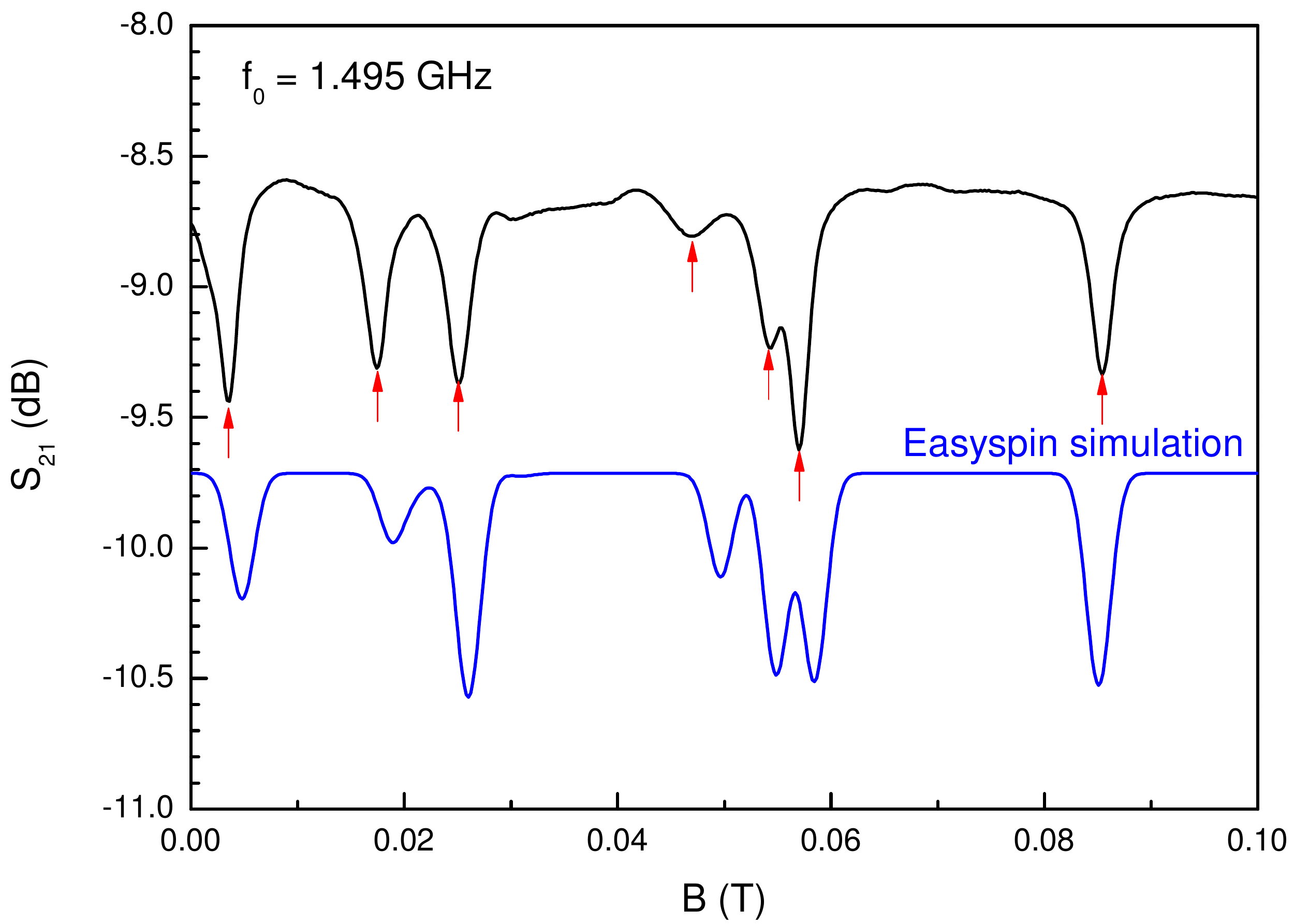}
\caption{Cavity transmission from figure \ref{fig:GdF11R3}A compared to an EasySpin simulated EPR absorption spectrum for Hamiltonian \refeq{eq:GdFH}.  The simulation has a cavity frequency of \SI{1.4956}{\giga\hertz} and a gaussian broadening of \SI{2.3}{\milli\tesla} for each transition.  The angular orientation of the field is $\theta = \ang{42.25}$ and $\phi = 0$ in the crystal frame, approximately the expected orientation given the experimental setup (see figure \ref{fig:GdF11R1}).}\label{fig:GdF11R4}
\end{figure}

Since the sample has an anisotropic Hamiltonian and spin $7/2$ (see equation (\ref{eq:GdFH})), in this case we find multiple features at different transition frequencies.  We fit the lineshape at each field to extract the dependence of the transmission value, resonant frequency and linewidth for each applied field (figure \ref{fig:GdF11R3}).  This experiment is analogous to a continuous wave EPR experiment and, as such, the expected transitions can be simulated using the EasySpin Matlab package \cite{Stoll2006}.  Using the parameters shown in equation (\ref{eq:GdFH}) and taking into account the orientation of the crystal, we obtain the absorption spectrum shown in figure \ref{fig:GdF11R4}, which is in good agreement with the measured spectrum.

\begin{figure}[!tb]
\centering
\includegraphics[width=0.65\columnwidth]{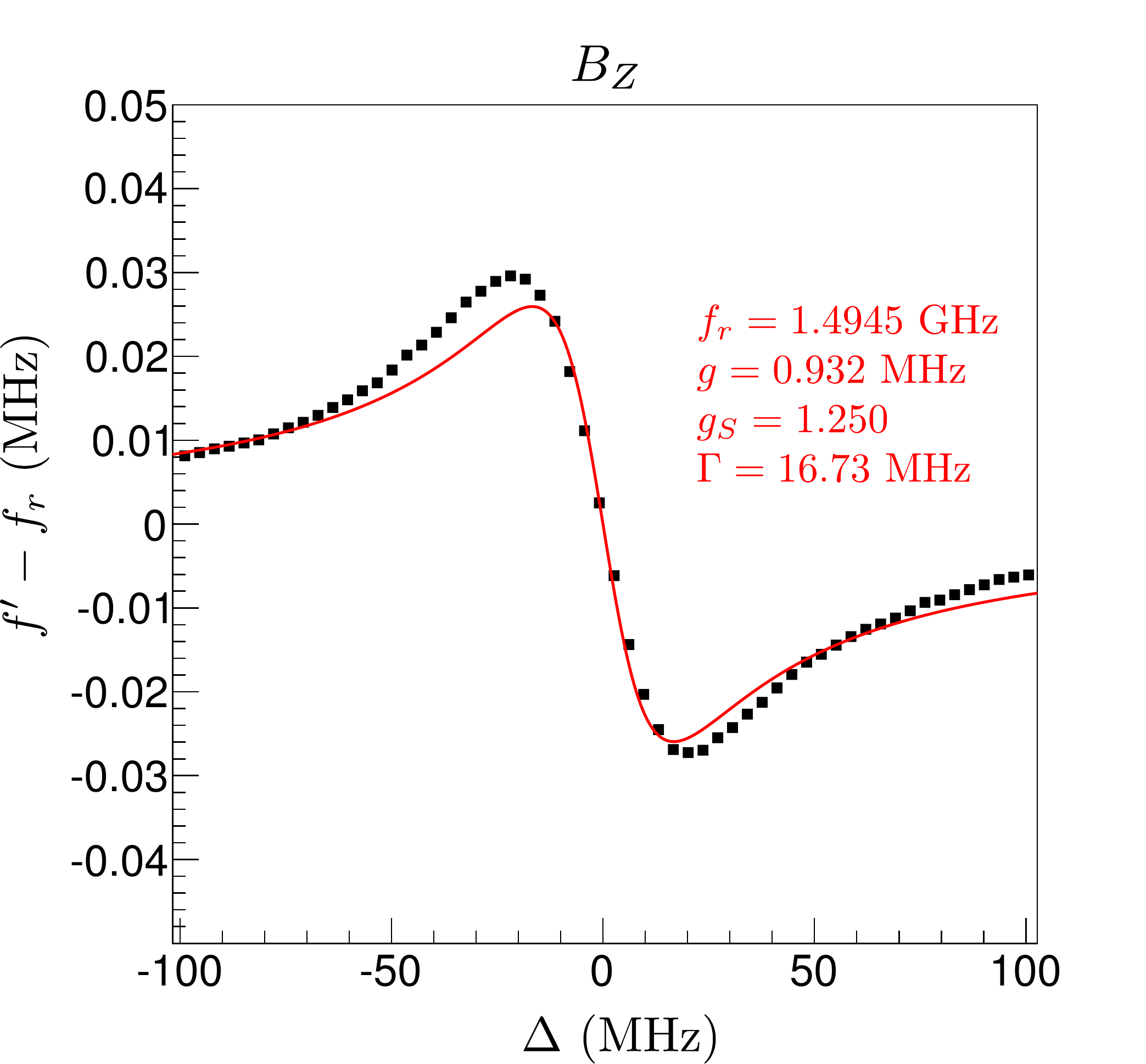}
\caption{Fit to absorption peak at $B_Z\sim \SI{0.085}{\tesla}$ from figure \ref{fig:GdF11R3}B.  The horizontal scale has been rescaled to show detuning $\Delta$ instead of magnetic field ($\Delta = g_S\mu_B/h B - f_r$.  The curve is fitted the equation \refeq{eq:phen1} and the fit results are shown on the graph (red).}\label{fig:GdF11R5}
\end{figure}

Comparing to the DPPH case from the previous section, it is clear that the couplings are rather weak in this case.  If we observe the scale of the frequency dependence of the resonance (figure \ref{fig:DPPH11R3}B), we see that the oscillations associated to each transition (equal to $v^2/\gamma$ according to \refeq{eq:phen1}) are very small compared to the DPPH case (see for example, figure \ref{fig:DPPH27R2} or \ref{fig:DPPH11R4}) where the cavity drift due to the applied field was not readily appreciable compared to the absorption feature.  The values for the coupling are also hard to extract from the frequency dependence since many of the lines overlap or have large background drifts compared to the signal.  In any case, the coupling value of one of the more isolated lines can be estimated from the frequency dependence.  Taking the peak at the highest field, we obtain the fit shown in figure \ref{fig:GdF11R5}, with a coupling of $g\sim \SI{0.9}{\mega\hertz}$, which is indeed small considering the large sample size.

The cause for this relatively low coupling is likely the fact that the field is not applied along the optimal direction to induce coupling to the sample.  As mentioned in the previous section (section \ref{sec:DPPHres}, DPPH pellets), due to the geometry of the wide resonator used, it is preferable to apply the tuning field in the X laboratory direction since the application of the field in the Z direction suppresses one of the components contributing to the coupling strength (rf fields parallel to the DC field do not generally contribute to transitions).  A $\sqrt{2}$ factor enhancement is expected to be attainable if the DC field had been applied the X direction.  Additionally, the spin density of the sample is also low (spin density of about $\rho_s \simeq \SI{6.5e18}{\per\centi\meter\cubed}$ compared to $\SI{2e21}{\per\centi\meter\cubed}$ for DPPH) while the dephasing rate is somewhat high due to the interaction of \ce{Gd^{3+}} magnetic moments with neighboring fluorine nuclear spins \cite{Arauzo1997}.  All these effects contribute to make it even harder to achieve strong coupling regimes with this type of sample.  Our measured linewidth may be overestimated since this sample can also be susceptible to the line broadening effects we saw for the DPPH pellets.

\subsection{\ce{GdW30} on a superconducting CPW resonator}

\begin{figure}[!tb]
\centering
\includegraphics[width=0.5\columnwidth]{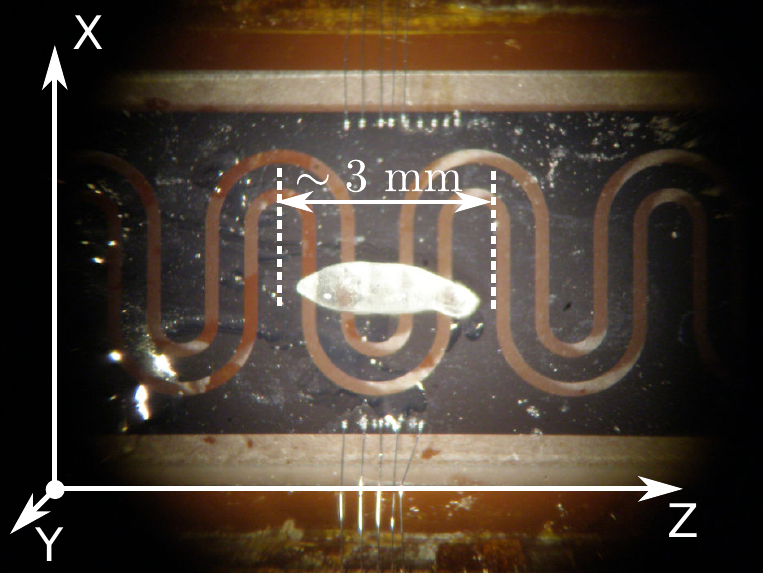}
\caption{Diluted \ce{GdW30} crystal on a niobium CPW resonator.}\label{fig:GdW30_11R_1}
\end{figure}

\begin{figure}[!tb]
\centering
\includegraphics[width=0.7\columnwidth]{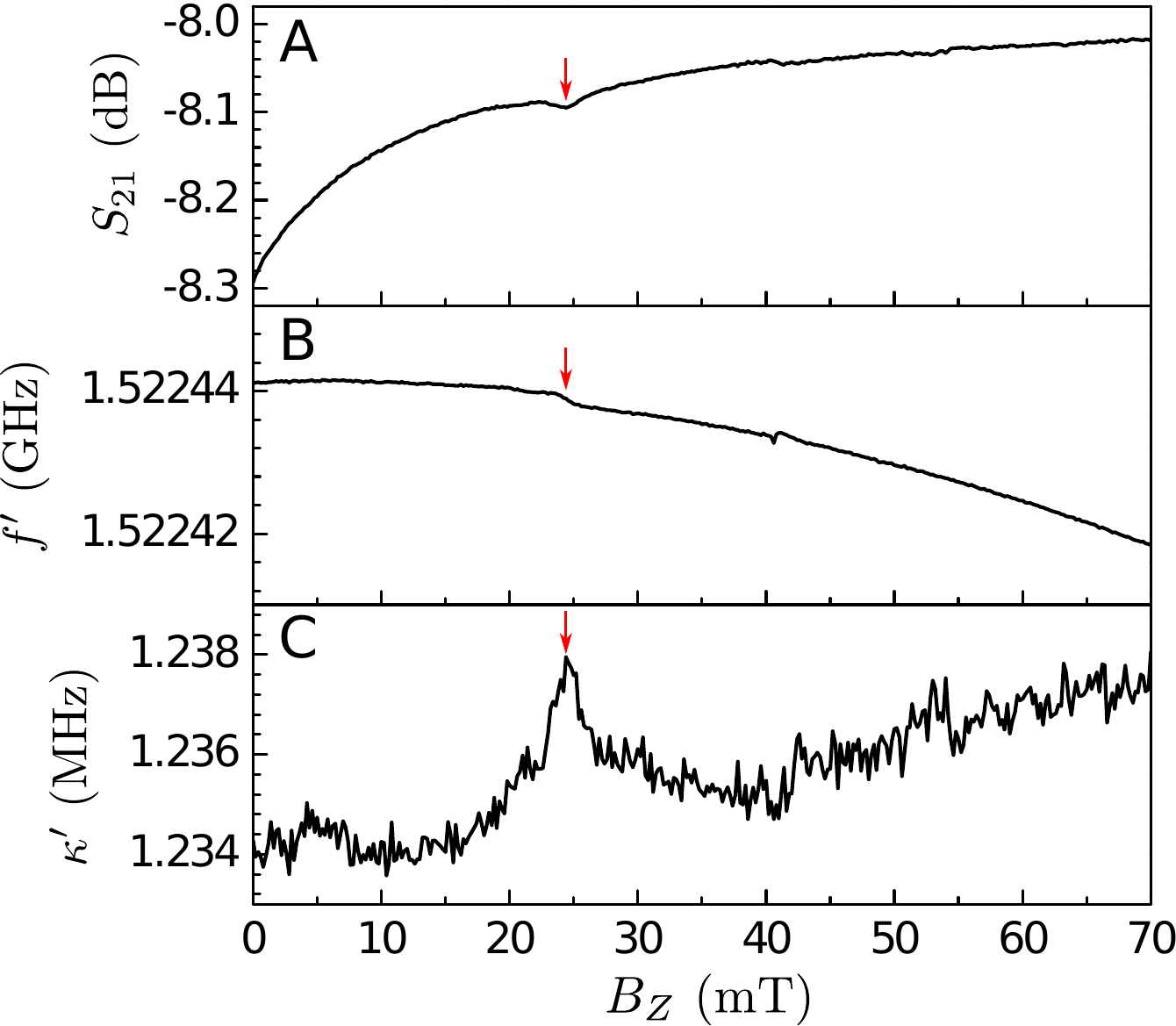}
\caption{Peak transmission properties for the experiment performed on a \ce{GdW30} crystal.  The graphs A,B and C show the transmission value, peak frequency and width respectively.  Only a small absorption feature is visible at about \SI{25}{\milli\tesla}.}\label{fig:GdW30_11R_2}
\end{figure}

Even though or resonators are not at the ideal frequency to access \ce{GdW30} samples, we attempt a measurement with a diluted crystal (\ce{Y_{0.99}Gd_{0.01}W30}) placed on the same resonator used for the DPPH pellets and the \ce{GdCaF} crystal (figure \ref{fig:GdW30_11R_1}).  A similar experiment to that done in the case of \ce{GdCaF} is performed.  The incident power applied from the network analyzer was $-20$ dBm.  The transmission peak properties are extracted for each field by fitting the frequency dependence to a Lorentzian lineshape.  The corresponding values of the peak transmission value, peak frequency and peak width are shown in figure \ref{fig:GdW30_11R_2}.  No features were visible on the spectrum except for a small absorption at about \SI{25}{\milli\tesla}.

The signal is far too weak to be properly fit, but a rough estimate can be made by looking at the separation of the upward and downward peak in the frequency dependence (figure \ref{fig:GdW30_11R_2}B, about $\Gamma\sim\SI{2}{\milli\tesla}\sim \SI{120}{\mega\hertz}$) and the height of the peak in the width dependence (figure \ref{fig:GdW30_11R_2}C, $\kappa'-\kappa\sim \SI{2}{\kilo\hertz}$.  At resonance ($\Delta = \Omega-\omega_r = 0$) from equation \refeq{eq:phen2} we should have $\kappa'-\kappa = v^2/\Gamma$.  This gives us a coupling of at best $v\sim g\sim\SI{490}{\kilo\hertz}$, much lower than in the other cases.  This absorption is most likely associated to the spin $\pm 1/2$ transition of the \ce{Gd^3+} ion that is probably the only transition accessible at these frequencies.  The value seems to be in line with what may be expected considering the low spin density ($\rho_s\simeq \SI{3.07e18}{\per\centi\meter\cubed}$ since the crystal is diluted to 1\%), even when compared to \ce{CaGdF} although differences in sample size and placement makes a direct comparison with the \ce{CaGdF} experiment difficult.  The high linewidth ($\Gamma\sim \SI{120}{\mega\hertz}$) indicates a low $T_2\sim\Gamma^{-1}\sim \SI{10}{\nano\second}$ which, when compared to the EPR measurements from section \ref{sec:pulsedEPR_GdW30_times} ($T_2\sim \SI{300}{\nano\second}$), seems to indicate that additional inhomogeneous broadening effects must be present.  The crystal is likely subject to the same kind of $B_2^0$ and $B_2^2$ strain effects (section \ref{sec:powderEPR_GdW30}) as the sample measured on an open waveguide (section \ref{sec:GdW30wg}) as well as to some degradation effects that make the sample more powder-like than crystalline.

\section{Measurement of the signal enhancement on a nanometric constriction} \label{sec:DPPHconstriction}
\begin{figure}[!tb]
\centering
\includegraphics[width=1\columnwidth]{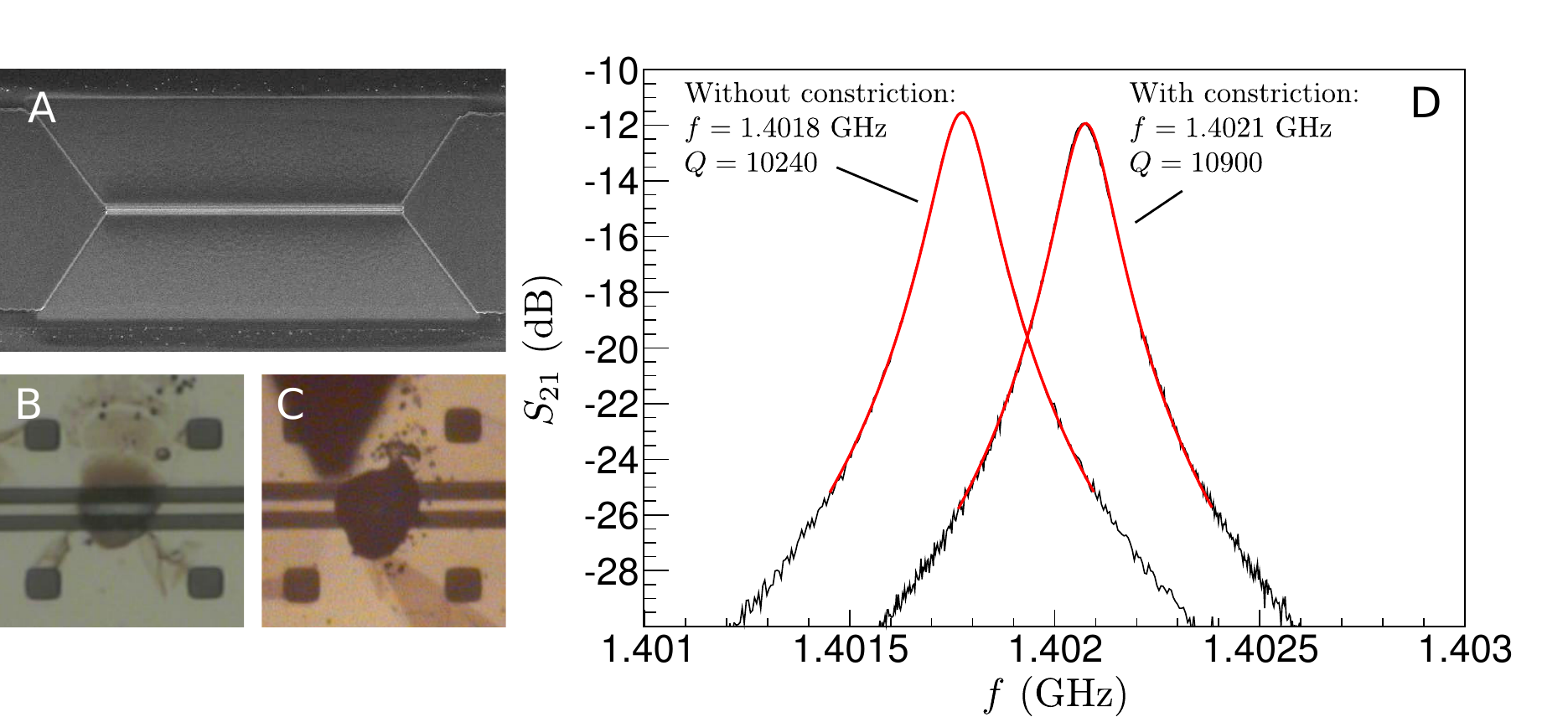}
\caption{DPPH droplets deposited using a dip-pen AFM tip on a niobium CPW resonator with and without a constriction.  Graph A shows a constriction that has a length of \um{15}, a width of \SI{100}{\nano\meter} and \SI{150}{\nano\meter} thickness.  Graphs B and C show microscope images of two droplets deposited onto a resonator without and with a constriction respectively.  Both were prepared following the same procedure and should in principle contain similar sample quantities.  Graph D shows the background transmission spectra for both resonators.  We see that both have very similar characteristics.}\label{fig:constr1}
\end{figure}

In this section we return to the DPPH sample, but in this case we measure a small droplet deposited using a dip-pen tip (see section \ref{sec:DPN}) on a thin CPW resonator (\um{14} center line and \um{7} gaps) with and without a nanometric constriction in its center transmission line.  The different droplets measured are shown in figure \ref{fig:constr1} along with the constriction used and the background transmission spectra.  Both droplets were prepared using the same procedure and taken from a solution of DPPH in DMF and 5\% glycerol with a concentration of \SI{10}{\milli\gram\per\milli\litre}.  The resulting deposits have about \um{60} diameter and a thickness of between 50 and \SI{100}{\nano\meter} as determined by AFM microscopy.

\begin{figure}[!tb]
\centering
\includegraphics[width=1\columnwidth]{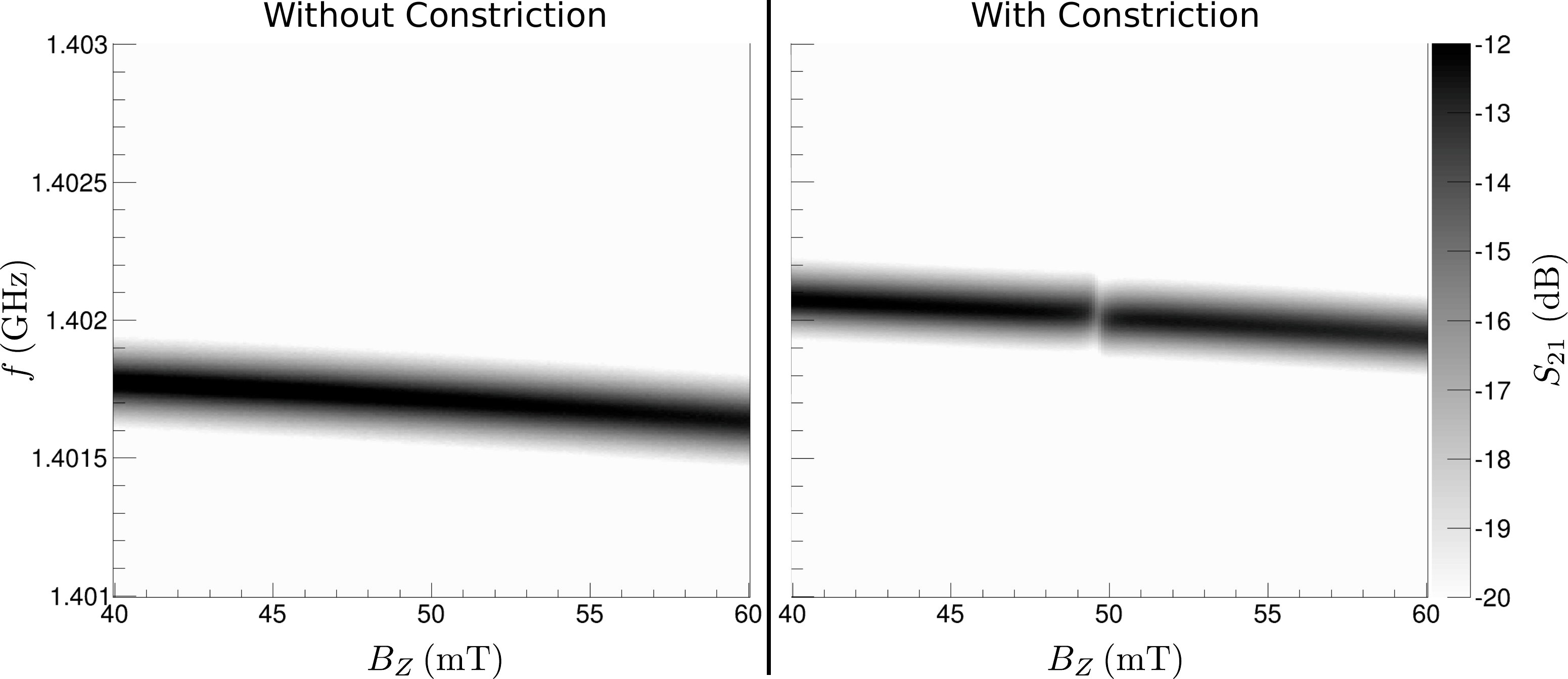}
\caption{Transmission measurement of DPPH droplets on a niobium CPW resonator without (A) and with (B) a nanometric constriction.  The greyscale shows the measured transmission intensity as a function of the driving frequency and the DC magnetic field applied along the laboratory Z axis.}\label{fig:constr2}
\end{figure}

\begin{figure}[!tb]
\centering
\includegraphics[width=0.65\columnwidth]{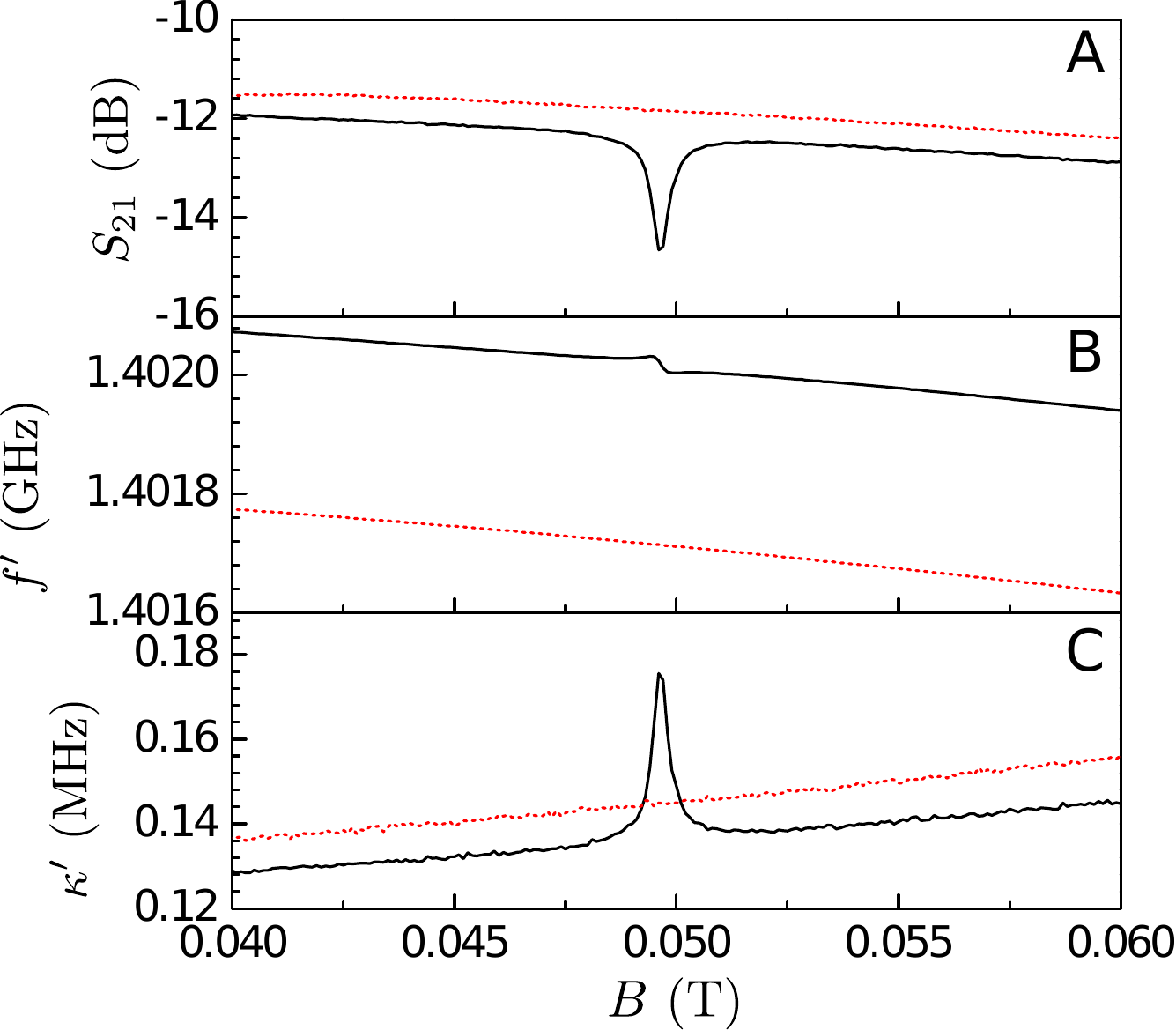}
\caption{Peak transmission (A), resonant frequency ($f'$, B) and peak width ($\kappa'$, C) for the experiments in figure  \ref{fig:constr2}.  An absorption feature is visible in the constriction case that is absent in the normal case.}\label{fig:constr3}
\end{figure}

The experiments are performed in the same manner as in section \ref{sec:DPPHres} with our minimum excitation power of -51 dBm output from the network analyzer.  The field, in the Z direction, is swept over the range 0-\SI{0.1}{\tesla} at the center of which the DPPH resonance should be detected.  The full measurement results are shown in figure \ref{fig:constr2} for both cases.  The absorption of the droplet is invisible for the resonator having no constriction.  In contrast, when the droplet is deposited onto the constriction we see an absorption signal at the expected field.  In figure \ref{fig:constr3} we compare the resonance properties for each field value and clearly see that there is an absorption effect in the constriction case that is absent when a normal resonator is used.

\begin{figure}[!tb]
\centering
\includegraphics[width=0.6\columnwidth]{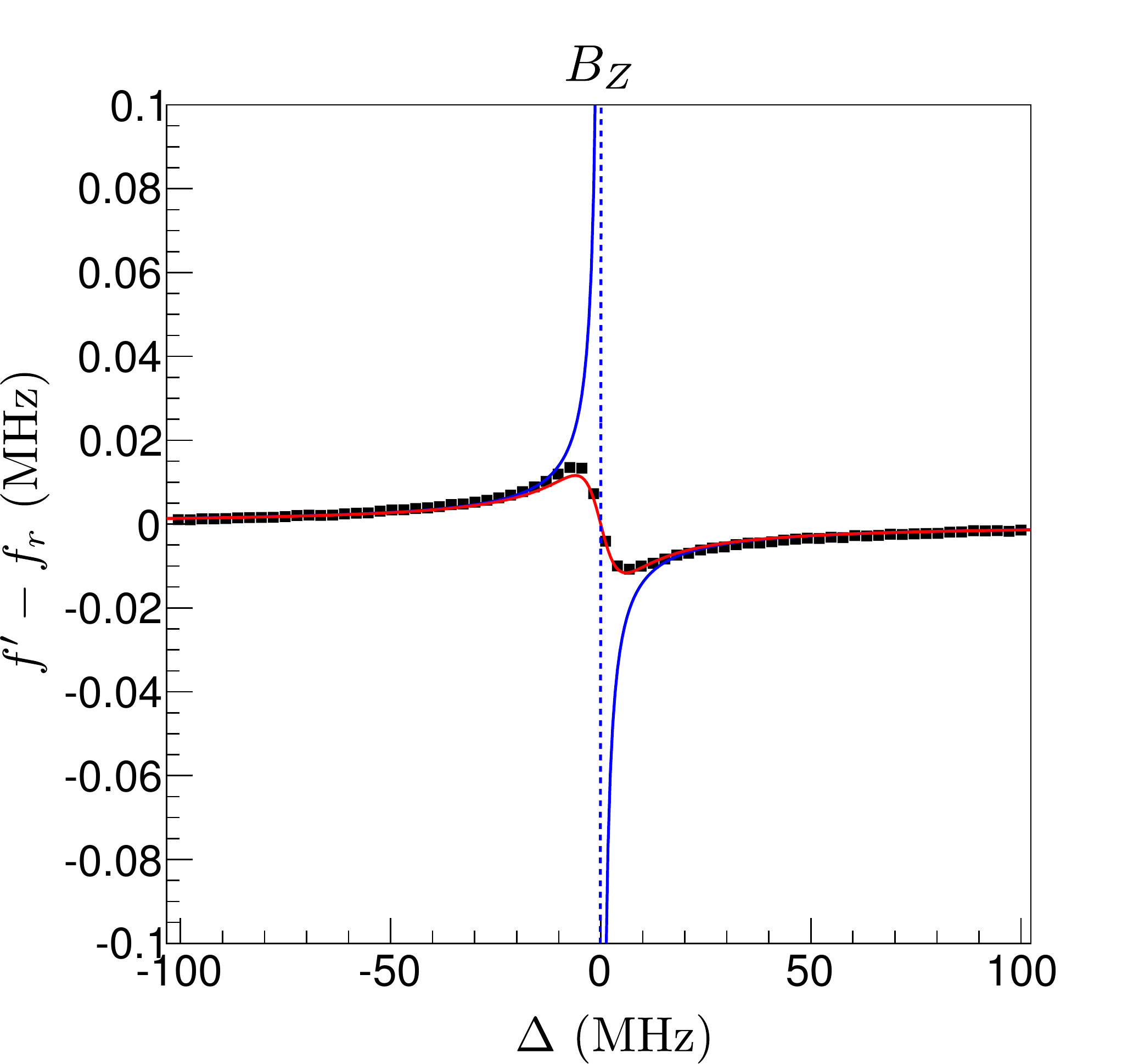}
\caption{Fit of phenomenological model \refeq{eq:phen1} to the absorption of the constricted resonator system.  The frequency deviation from the background value is plotted as a function of the field detuning $\Delta = g_S\mu_B/hB_Z - f_r$.  The red line represents the fit while the blue lines represent the theoretical level positions from equation \refeq{eq:theotrans} with the parameters from the fit.  The dotted line represents the uncoupled spin transition energy.  The values of the fit parameters are $g = \SI{370}{\kilo\hertz}$, $g_S=2.017$ and $\Gamma = \SI{6}{\mega\hertz}$.}\label{fig:constr4}
\end{figure}

As usual, we fit the frequency dependence from figure \ref{fig:constr3}B to the model equation \refeq{eq:phen1} to obtain the measured coupling value and spin linewidth.  This fit is shown in figure \ref{fig:constr4} where we find that the coupling is about $g\simeq \SI{370}{\kilo\hertz}$, about an order of magnitude smaller than in the case of the micropippete drop (section \ref{sec:DPPHres}).  The linewidth is found to be about $\Gamma \simeq \SI{6}{\mega\hertz}$, similar to the previous DPPH measurements.

The low value of the coupling in addition to the large linewidth do not allow us to see strong coupling in this case.  We do however get a significant enhancement when compared to the non-constricted resonator.  The values of the coupling determined from the experiments agree well with simulated predictions for this geometry.  As described in section \ref{sec:coupCPWG_SMM}, we can simulate the magnetic fields for cross sections of the waveguide with the geometries and sample sizes involved in this case.  We then find that, taking into account only the cross section, the ratio between the constriction case and the sample considered in section \ref{sec:DPPHrespippete} should be about equal to 1.  If we then also take into account that the coupling scales with the square root of the sample length (\SI{1.5}{\milli\meter} for the large drop and \um{15} for the current sample) we find approximately the order of magnitude reduction for the constriction sample.  On the other hand, if we do not include the constriction in the simulation we find about two orders of magnitude reduction in the coupling, which is compatible with the fact that we are unable to to see any absorption in our non-constricted resonator.

The minimum number of spin $1/2$ systems detectable is hard to ascertain from this experiment since the drop extends beyond the constriction region making it hard to evaluate what the active volume is.  A drop more confined to the constriction region could allow us a better estimation.  It is however clear that the enhancement must come from the constriction region meaning that we could in principle have sample only in a small volume surrounding the wire.  According to our simulations for this device's geometry, this active region's cross section could be reduced down to \um{1} by \SI{200}{\nano\meter} without a significant effect on the field integrals contributing to the coupling.  Therefore, for our sample we estimate that a volume of $\um{15}\times\um{1}\times\SI{200}{\nano\meter}$ contributes which, assuming the bulk spin density of DPPH, amounts to $\sim 10^9-10^{10}$ spins.  Taking into account the Boltzmann factor for the experimental temperature (4 K), the actual contributing spin number in this experiment could be around $\sim 10^6-10^7$.  We now consider the signal and noise levels in our experiment.  The effect is most easily detectable in the linewidth where the noise level is roughly \SI{4}{\kilo\hertz} compared to a \SI{35}{\kilo\hertz} variation on resonance.  This variation is proportional, according to equation \refeq{eq:phen2}, to $g^2$ and therefore also proportional to the sample length.  This means that the length of the sample or constriction could be reduced about a factor 8 (down to around \um{2} in length) and still be able to detect the sample.  Therefore, for this sample and these experimental conditions, our minimum detectable number of spins $1/2$ should be of the order of $10^5-10^6$.

The results for couplings for this type of resonator can potentially be improved in several ways.  Moving to higher frequencies allows for higher energy densities in the cavity and therefore higher couplings ($g_{\rm eff}$ scales approximately linearly with the radiation frequency, see equation \refeq{eq:vacuumcurrent} and \refeq{eq:coupling_explicit}).  This increase in frequency also improves the Boltzmann factor by reducing the excited state population.  Ideally, the temperature should be lowered and operating frequency increased such that the exited state is not thermally populated.  This would make almost all spins present contribute to the signal.  Narrower constrictions, like those shown in chapter \ref{chap:CPWG} can also be used to further increase the rf magnetic field.  To measure even smaller samples care must also be taken to further reduce the excitation powers and some signal amplification may be necessary.  In this case we estimate that the driven cavity stores about $10^9$ photons for our input powers, similar to the estimated number of spins.  Although it does not seem to be the case here, the sample could be saturated if the number of spins is reduced and the power is not, producing effects like those shown in figure \ref{fig:seiji2}D.

\section{Summary and Conclusions}\label{sec:conclusionsMeas}

In this chapter we have presented results from measurements of several magnetic samples on different superconducting devices based on CPW circuits.  The results show that spectroscopy and EPR-like measurements of different samples can be reliably performed using our setup and devices.

First we present measurements of the sample responses when placed on an open CPW and observing the transmission value through the system for a range of magnetic fields.  Although the signals are relatively small, measuring using an open waveguide allows the use of a broad range of frequencies and fields (in our case frequencies from \SI{10}{\mega\hertz} to \SI{14}{\giga\hertz} and fields of up to \SI{1}{\tesla}).  In this setup, absorptions associated to the different spin transitions can be detected after adequately normalizing the signal.  Both DPPH and \ce{GdCaF} crystals give well defined transitions that can be tracked for a large range of fields.  In both cases, the observed lines match well with the lines expected given the spin Hamiltonians.  The \ce{GdW30} crystals produce broad absorption bands that can be explained by the presence of inhomogeneous broadening effects due to $B_2^0$ and $B_2^2$ strains (section \ref{sec:powderEPR_GdW30}) combined with a possible loss of cystallinity when the sample is not properly protected.

Next we review the theoretical framework used to describe a coupled resonator and spin system \cite{Chiorescu2010,Miyashita2012} and present some examples for the expected transmission signals expected in experiments.  We show that, as long as the number of spins is larger than the number of photons in the cavity, strong coupling ($g_\textrm{eff}>\gamma,\kappa$) can be detected by observing a two peak transmission spectrum when the spin transition is tuned to the cavity frequency.  These peaks will be separated $2g\sqrt{N} = 2g_\textrm{eff}$ under these conditions, where $N$ is the number of spins and $g$ the coupling to each individual spin.  We also show that the phenomenological equations \refeq{eq:phen1} and \refeq{eq:phen2} provide a good measure of the coupling and spin decay rates for weak coupling cases ($g_\textrm{eff}<\gamma,\kappa$).

In the next section, we present measurements of the same three samples on superconducting CPW resonators with resonant frequencies around \SI{1.5}{\giga\hertz}.  Different resonator designs are chosen depending on the sample size.  In the case of DPPH, both droplets from a solution and pellets were measured and the single expected $g\simeq 2$ transition is detected.  Coupling to both samples was detected with results in line with the expected couplings from our simulations (section \ref{sec:coupCPWG_SMM}).  In the case of DPPH pellets, experiments show that the strong coupling regime can be approached and two peaks were distinguishable if the field is applied along the correct direction given the resonator geometry.  The couplings to the different DPPH samples were of the order $g_\textrm{eff}\sim 1-\SI{10}{\mega\hertz}$ while the spin linewidths were $\gamma\sim 5-\SI{20}{\mega\hertz}$.  The cavity linewidths $\kappa$ were always below both of these values (as low as $\sim\SI{300}{\kilo\hertz}$ in some cases).

When measuring the more complex \ce{GdCaF} crystal multiple spin transitions are detected corresponding to the energy levels calculated from the crystal field Hamiltonian.  Also, the simulated EasySpin EPR spectrum is in good agreement with the observed absorption lines.  The coupling and signal in this case is found to be somewhat weaker than in the DPPH case considering the sample size.  This can be understood considering that the spin density of this crystal is lower than that of DPPH and that the spin linewidth is increased by the presence of fluorine atoms.

Although our resonators do not operate at the ideal frequencies for this sample, an attempt was made to measure a \ce{GdW30} crystal.  Only a very weak absorption was detected and associated with the $\pm 1/2$ transition of the \ce{Gd^{3+}} ion.  However the system appears to suffer the same inhomogeneous broadening and loss of crystallinity effects as in the waveguide measurement.

Finally, a performance test of a constricted resonator was carried out.  A \um{60} diameter drop of DPPH solution was placed on a \um{15} long and \SI{100}{\nano\meter} wide constriction and an absorption signal was clearly detected.  A similar drop deposited on a normal resonator did not display any absorption signal.  The measured couplings to the sample are in line with estimations from Comsol simulations like those done in \ref{sec:coupCPWG_SMM}.  This confirmation of the expected couplings is a promising result for further experiments involving better spin systems, with higher transition matrix elements and better coherence times.  Also, experiments with lower temperatures, higher frequencies and more confined samples could soon allow the attainment of strong coupling regimes for single spins or small ensembles.

In conclusion, we have shown that spectroscopy using superconducting circuits is an interesting tool for sample characterization.  They allow EPR-like experiments with different frequencies and field intensities depending on the types of waveguides or resonators designed.  They also allow the characterization of samples at very low temperatures and with very small sample sizes.  As well as for sample characterization, testing on devices with constrictions show promising results for applications in quantum computing architectures.  The expected performance enhancement according to the simulations done in previous chapters (chapter \ref{chap:Theo1}) is confirmed at least in the case of a spin $1/2$ sample.  This result allows us the prospect that, with improvements to the experimental conditions and using spin samples with better properties (SMMs and SIMs for example), an all-spin quantum processor similar to the concept described in chapter \ref{chap:Theo0} may be feasible.

\bibliographystyle{h-physrev3}
\bibliography{mybiblio}

\chapter{Conclusions}\label{chap:concl}

In this final chapter we will give a summary of the main points raised in this thesis work and give our general conclusions.

In the introduction to this thesis (chapter \ref{chap:intro}) we have reviewed some of the different approaches that have been taken in order to achieve the goal of building a working quantum computer.  Although this goal has considerable technical hurdles to be overcome, much progress has been made in the three decades since the original proposal.  The current consensus is that there are no fundamental roadblocks barring the way to this goal and that it is only a matter of investing enough time and money before, at the very least, a basic working quantum computing system is built.  There are many different architectures and approaches being developed, each with their distinct advantages and weaknesses.  It is entirely possible that a better quantum computing architecture could be achieved by combining different systems into a single system.  Therefore there is still much value in the development and study of new qubit systems that may, under certain circumstances, provide better alternatives to the main candidates currently considered.

In this context we have chosen to study the possibility of combining techniques taken from the field of Cavity and Circuit Quantum Electrodynamics with the field of molecular magnetism.  Molecular magnets and Single Molecule Magnets (SMMs) provide interesting alternatives to charge qubits because of their chemical tunability and their potential to serve different roles within a quantum computing system.  They can be designed to act as single qubits or to encode several qubits on a single molecule or even to act as quantum logic gates.  However, many of these interesting applications depend on the possibility of coupling single molecules to photons in a microwave cavity, a challenge that we have endeavored to address and investigate throughout this work.

A first result of this work is that we have identified those steps that are to be taken in order to improve the coupling strength of single magnetic entities (atoms or molecules) to quantum superconducting circuits if they are to be used in quantum computing.  These steps involve both searching for better spin systems and improving the rf magnetic fields attainable in cavities.

Concerning the former, we have found that, because of their high spin and high spin densities, Single Molecule Magnet crystals should be able strongly couple to resonators achieving coupling strengths up to 3 orders of magnitude stronger that Nitrogen Vacancy centers in diamond.  Furthermore, we find that individual molecules could have coupling strengths of up to hundreds of \si{\kilo\hertz} provided that resonators with sufficiently narrow central lines, of the order of a few tens of \si{\nano\meter} can be fabricated.  However, typical SMMs are not without their drawbacks.  Their high anisotropy can make the energy separation between the $\ket{0}$ and $\ket{1}$ states too large (up to \SI{100}{\giga\hertz}) to be comfortably manipulated with conventional Circuit QED systems.  If tunnel split energy levels are used as the quantum basis, the low tunnel gap requires an extraordinarily accurate control of the magnetic fields necessary to tune the SMM into resonance.  We therefore have arrived at the conclusion that SMMs with lower anisotropies or engineered to have strong spin tunneling effects my provide the most suitable candidates.

Single Ion Magnets (SIMs) are a family of SMMs that can present these desired qualities.  These consist of a single lanthanide ion encased in a non-magnetic ligand shell.  The presence of only a single ion makes them conceptually simpler that their multi-ion counterparts.  Also, their anisotropy Hamiltonian is determined by the lanthanide ion species and the shape of the ligand shell.  We have shown two cases, each fulfilling one of the two desired qualities (a low anisotropy or a strong tunnel splitting).
\begin{itemize}
\item In the case of \ce{GdW30}, the \ce{Gd^{3+}} ion has no intrinsic anisotropy making the molecular spin Hamiltonian entirely determined by the ligand shell.  The geometry of the shell is such that the 8 eigenstates of the Hamiltonian all lie within an energy range of \SI{1}{\kelvin}.  Using a combination of electron paramagnetic resonance (continuous wave and pulsed) and magnetic susceptibility measurements we confirm that this is indeed the case and that each transition can be coherently addressed independently using easily manageable rf frequencies ($\sim\SI{10}{\giga\hertz}$).  This means that not only are the energy levels such that it is a good qubit candidate, it also means that it could hypothetically encode three qubits in a single molecule.
\item Due to the spin anisotropy of the \ce{Tb^{3+}} ion and the fivefold symmetry of the ligand shell, \ce{TbW30} presents an extraordinarily high quantum tunnel gap that makes the tunnel split states good candidates to be used as a quantum basis.  Using specific heat and susceptibility measurements we construct an effective Hamiltonian to describe the low energy and low temperature physics of the system and indeed find a large tunnel gap of $\sim\SI{1}{\kelvin}$.  This system is also interesting in its own right as it offers the rather unique chance to study the magnetic behavior of a pure quantum two level system.  In addition, the fact that the tunnel gap is larger than any of the perturbation arising from hyperfine or dipolar interactions should contribute to enhance the quantum spin coherence.
\end{itemize}
Finally the couplings expected to coplanar waveguide resonators (with and without constrictions) are found to be very large for both \ce{GdW30} and \ce{TbW30} thus confirming their potential for quantum computing.

Concerning the quantum circuits and their optimization, we have designed and fabricated \ce{Nb} transmission lines and resonators.  These resonators are tested before and after nanometric constrictions (down to \SI{50}{\nano\meter} in width and up to \SI{15}{\micro\meter} in length) in the centerline are made using focused ion beam etching with no substantial changes in the resonator transmission properties.  These constrictions therefore provide a method to improve the coupling to very small samples both for quantum computing applications or for micro-EPR (electron paramagnetic resonance) type experiments.

We have also carried out the first tests of the coupling of magnetic samples to resonators at $T=\SI{4}{\kelvin}$.  Although the operating frequencies of our resonators are not adequate to properly measure the SMM samples studied previously, many tests with other similar samples can be done.  We have performed broadband spectroscopy using open coplanar waveguides of macroscopic samples.  Broadband spectroscopy measurements were performed on samples of DPPH (a spin $1/2$ free radical), \ce{Gd} doped fluoride crystals (\ce{GdCaF}) and \ce{GdW30} crystals using open superconducting transmission lines.  The measurements provide clear signals in a broad frequency and field range (from 1-\SI{14}{\giga\hertz} and 0-\SI{1}{\tesla}) in line with the given the theoretical models.

Macroscopic samples of DPPH were coupled to superconducting coplanar waveguide resonators approaching the strong coupling regime.  The transmission is seen to have the double peak structure characteristic of strongly coupled systems.  Also, the direction of the applied magnetic field is seen to have an important effect on this coupling.  Although they do not achieve strong coupling, samples of \ce{GdCaF} and \ce{GdW30} were also measured with results that are good agreement with their theoretical models.

Finally, the performance enhancement for small samples when using a constriction was also tested.  Using a micrometric drop of DPPH deposited using an AFM tip on a nanometric constriction, an enhancement in the coupling relative to that of an unmodified resonator was detected.  The enhancement is in agreement with the predictions derived from our numerical field simulations.

In conclusion, although the results presented here are promising, there is still much work to be done.  There is still a long way to go in order to achieve the performance necessary to make proposals of all-spin quantum processors, such as the one given in the introduction, possible.  We are already working on the following steps in this direction such as designing resonators with higher frequencies in line with the requirements of SMM samples as well as preparing our systems to measure at very low temperatures.  In light of these results, we hope to have shown that, although the initial proposal of an all-spin quantum processor (figure \ref{fig:fantasy1}) may have seemed far fetched, quantum processors that combine spin systems and circuit QED systems could be within reach.

\renewcommand{\chaptermark}[1]{\markboth{#1}{}}
\chapter*{Conclusiones}\label{chap:conclesp}
\chaptermark{Conclusiones}

A continuación damos un resumen de los principales puntos tratados en esta tesis y damos nuestras conclusiones generales.

En la introducción a esta tesis (capítulo \ref{chap:intro}) hemos revisado algunas de los diferentes métodos que se han seguido para alcanzar el objetivo de construir un ordenador cuántico.  Aunque este objetivo presenta dificultades técnicas considerables, se ha progresado mucho en las tres décadas que han pasado desde la propuesta original.  El consenso actual es que no existen barreras fundamentales que impidan que se alcance esta meta y que es solo cuestión de invertir suficiente tiempo y dinero para conseguir construir, como mínimo, un sistema de computación cuántica básico.  Se están desarrollando múltiples arquitecturas siguiendo diferentes metodologías, cada una con sus ventajas e inconvenientes específicos.  Es enteramente posible que la mejor arquitectura de computación cuántica sea una combinación de múltiples tipos de sistemas en uno único.  Por tanto, sigue habiendo gran valor en el desarrollo y estudio de nuevos candidatos a qubits que pueden, bajo determinadas condiciones, proporcionar alternativas a los principales qubits que se consideran en la actualidad.

En este contexto hemos elegido estudiar la posibilidad de combinar técnicas que provienen del campo de la electrodinámica cuántica de cavidades y circuitos (CQED y Circuit QED) con las del campo de magnetismo molecular.  Los imanes moleculares (SMMs) proporcionan interesantes alternativas a los qubits de carga debido a la posibilidad de ajustar químicamente sus cualidades y porque potencialmente pueden desarrollar diversos roles dentro de un sistema de computación cuántica.  Pueden ser diseñados para actuar como qubits individuales o para codificar múltiples qubits en una sola molécula o incluso para actuar como puertas lógicas cuánticas.  Sin embargo, muchas de estas interesantes aplicaciones depende de la posibilidad de acoplar moleculas individuales a los fotones de una cavidad de microondas, un reto que hemos intentado abordar a lo largo de este trabajo.

Un primer resultado de este trabajo es que hemos identificado los pasos necesarios que permitirían mejorar la intensidad del acoplo de entidades magnéticas individuales (átomos o moléculas) a circuitos superconductores cuánticos si se han de poder utilizar en computación cuántica.  Estos pasos involucran buscar mejores sistemas de espín y mejorar la intensidad de los campos rf alcanzables en cavidades.

El lo que respecta al primer punto, hemos encontrado que, gracias a su alto espín y alta densidad de espines, cristales de SMMs deben poder acoplarse fuertemente a resonadores alcanzando intensidades de acoplo hasta 3 órdenes de magnitud superior que para vacantes de nitrógeno en diamante.  Además, encontramos que moléculas individuales podrían tener intensidades de acoplo de cientos de \si{\kilo\hertz} si se fabrican resonadores con lineas centrales lo suficientemente delgadas, del orden de decenas de \si{nm}.  Sin embargo, los SMMs típicos tienes algunas desventajas.  Su alta anisotropía puede hacer que la separación energética entre los niveles $\ket{0}$ y $\ket{1}$ sea demasiado grande (hasta \SI{100}{\giga\hertz}) para ser manipulado cómodamente con sistemas de Circuit QED convencionales.  Si se usan niveles desdoblados por efecto túnel como la base cuántica, la pequeña separación asociada al efecto túnel requeriría un control extraordinariamente preciso de los campos magnéticos necesarios para llevar al SMM a la condición de resonancia.  Llegamos por tanto a la conclusión que SMMs con anisotropías menores o diseñados para tener un efecto túnel fuerte pueden proporcionar los candidatos más adecuados para ser qubits de espín.

Los imanes de un solo ion (Single Ion Magnets, SIMs), son una familia de SMMs que pueden presentar estas cualidades deseadas.  Consisten en único ión de lantánido encapsulado en una estructura no magnética de ligandos.  La presencia de un solo ion hace que sean conceptualmente más sencillos que los sistemas con múltiples iones.  Ademas, su Hamiltoniano de anisotropía magnética está determinado por la especie de lantánido utilizada y por la forma de la estructura de ligandos.  Hemos estudiado dos ejemplos, cada uno de los cuales cumple una de las dos cualidades deseadas (baja anisotropía o efecto túnel fuerte).
\begin{itemize}
\item En el caso de \ce{GdW30}, el ion \ce{Gd^{3+}} no tiene anisotropía intrínseca haciendo que el Hamiltoniano de espín esté completamente determinado por los ligandos.  La geometría de la estructura es tal que los 8 autoestados del Hamiltoniano se encuentran todos dentro de un rango de energía de \SI{1}{K}.  Utilizando una combinación de resonancia paramagnética electrónica (onda contínua y pulsado) y susceptibildad magnética, confirmamos que efectivamente es el caso y que cada transición puede excitarse independientemente de forma coherente utilizando frecuencias de rf fácilmente manejables ($\sim\SI{10}{\giga\hertz}$).  Esto significa que, además de tener niveles adecuados para ser un buen qubit, podría hipotéticamente codificar hasta tres qubits en una sola molécula.
\item Dada la anisotropía del ion de \ce{Tb^{3+}} y la symmetria pentagonal de la estructura de ligandos, el \ce{TbW30} presenta un desdoblamiento por efecto túnel extraordinariamente alto haciendo que los estados desdoblados sean buenos candidatos para formar la base cúantica de un qubit.  Utilizando medidas de calor específico y susceptibilidad magnética construimos un Hamiltoniano efectivo que describe el límite de baja energía y bajas temperaturas del sistema confirmado la existencia de un desdoblamiento por efecto túnel del orden de \SI{1}{\kelvin}.  Este sistema también es interesante en si mismo ya que ofrece la oportunidad de estudiar el comportamiento magnético de un sistema cúantico de dos niveles puro.  Adicionalmente, el hecho de que el desdoblamiento por efecto túnel sea mayor que cualquier perturbación que proviene de la interacción hiperfina o dipolar, debe contribuir a mejorar la coherencia del espín.
\end{itemize}
Finalmente, los acoplos esperados a resonadores coplanares superconductors (con y sin constricciones) resultan ser grandes para ambos \ce{GdW30} y para \ce{TbW30}, confirmando así su potencial para computación cuántica.

En cuanto al diseño de circuitos cuánticos y a su optimización, hemos diseñado y fabricado lineas de transmisión y resonadores coplanares de Nb.  Estos resonadores se comprobaron antes y después de practicarles constricciones nanométricas (hasta \SI{50}{\nano\meter} de ancho y \SI{15}{\micro\meter} de largo) en la línea central utilizando haz de iones focalizado,  no se observándose cambios sustanciales en las propiedades de transmisión de los resonadores.  Estas constricciones proporcionan un método para acoplarse a muestras pequeñas tanto para aplicaciones en computación cuántica como para experimentos tipo micro-EPR (Resonancia paramagnética electrónica).

También realizamos las primeras pruebas para acoplar nuestros resonadores a muestras magnéticas a $T=\SI{4}{\kelvin}$.  Aunque las frecuencias de operación de nuestros resonadores no son adecuadas para medir correctamente las muestras de SMMs estudiadas previamente,  se han realizado muchas pruebas con otras muestras similares.  Se ha hecho espectroscopía de banda ancha utilizando líneas de transmisión sobre muestras macroscópicas.  Se realizaron medidas de banda ancha sobre muestras de DPPH (un radical de espín $1/2$), sobre cristales de fluorita dopados con \ce{Gd} (\ce{GdCaF}) y sobre cristales de \ce{GdW30} utilizando líneas de transmisión abiertas.  Las medidas dan señales claras en un amplio rango de frecuencias y campos magnéticos (de 1-\SI{14}{\giga\hertz} y de 0-\SI{1}{\tesla}) que están en linea con los resultados esperados dados los modelos teóricos.

Muestras macroscópicas de DPPH se acoplaron a resonadores coplanares superconductores en régimen cercano al acoplo fuerte.  Se observó la estructura de doble pico en la señal de transmisión característica del régimen de acoplo fuerte.  Además, se ha observado que la dirección del campo magnético aplicado tiene un efecto importante sobre el valor de este acoplo.  Aunque el régimen de acoplo fuerte no se consigue en el caso de \ce{GdCaF} y \ce{GdW30}, las medidas de estas muestras están de acuerdo con sus modelos teóricos.

Finalmente, la mejora de rendimiento para muestras pequeñas utilizando constricciones también se comprobó.  Utilizando una gota micrométrica de DPPH depositada utilizando una punta de AFM, se detectó una mejora del acoplo respecto del de un resonador sin modificar.  La mejora está de acuerdo con las predicciones derivadas de nuestras simulaciones numéricas del campo magnético.

En conclusión, aunque los resultados presentados aquí son prometedores, todavía queda mucho trabajo por hacer.  Todavía queda mucho camino que recorrer para conseguir el rendimiento necesario para hacer que las propuestas de un procesador cuántico de espín, como el presentado en la introducción, sean posibles.  Ya estamos trabajando en los siguientes pasos, diseñando resonadores que operen a frecuencias más altas y preparando nuestros sistemas para operar a muy bajas temperaturas.  A la luz de estos resultados, esperamos haber mostrado que, aunque nuestra propuesta inicial de procesador cuántico de espín pudiera parecer algo descabellada, procesadores cuánticos que combinen sistemas de espín con sistemas de Circuit QED puede que estén al alcance.

\cleardoublepage

\chapter*{List of publications}

\begin{itemize}
\item \emph{Nanoscale constrictions in superconducting coplanar waveguide resonators} \\
M.~D.~Jenkins, U.~Naether, M.~Ciria, J.~Sesé, J.~Atkinson, C.~Sánchez-Azqueta, E.~del~Barco, J.~Majer, D.~Zueco, F.~Luis, \\ 
\textbf{Applied Physics Letters, 105, 16, 162601, 2014} \\
e-print: \texttt{arXiv:1409.1040 [cond-mat]}

\item \emph{Magnetic anisotropy of polycrystalline magnetoferritin investigated by SQUID and electron magnetic resonance} \\
F.~Moro, R.~de~Miguel, M.~D.~Jenkins, C.~Gómez-Moreno, D.~Sells, F.~Tuna, E.~J.~L.~McInnes, A.~Lostao, F.~Luis, J.~van~Slageren, \\
\textbf{Journal of Magnetism and Magnetic Materials, 361, 188-196, 2014}

\item \emph{Coupling single-molecule magnets to quantum circuits} \\
M.~D.~Jenkins, T.~Hummer, M.~J.~Martínez-Pérez, J.~García-Ripoll, D.~Zueco and F.~Luis, \\
\textbf{New Journal of Physics 15 095007, 2013} \\
e-print: \texttt{arXiv:1306.4276 [cond-mat]}

\item \emph{Surface-Confined Molecular Coolers for Cryogenics} \\
G.~Lorusso, M.~D.~Jenkins, P.~González-Monje, A.~Arauzo, J.~Sesé, D.~Ruiz-Molina, O.~Roubeau, M.~Evangelisti\\
\textbf{Advanced Materials, 25, 21, 2984-2988, 2013} (cover) \\
e-print: \texttt{arXiv:1212.1880 [cond-mat]}

\item \emph{Mn$_{12}$ single molecule magnets deposited on $\mu$-SQUID sensors: the role of interphases and structural modifications} \\
E.~Bellido, P.~González-Monje, A.~Repolles, M.~D.~Jenkins, J.~Sesé, D.~Drung, T.~Schurig, K.~Awaga, F.~Luis, D.~Ruiz-Molina, \\
\textbf{Nanoscale, 5, 24, 12565-12573, 2013}
\end{itemize}

\cleardoublepage

\chapter*{Agradecimientos}

Por último, pero no por ello menos importante, solo queda ya agradecer a la gran cantidad de profesionales, compañeros, amigos y familiares cuyo esfuerzo ha contribuido a hacer posible esta tesis.  Muchas gracias a todos por vuestra ayuda y apoyo.  Sin vosotros no habría sido posible.

Como no podía ser de otra forma, tengo que agradecer primero a Fernando Luis, mi director de tesis por darme la oportunidad de venir a Zaragoza y haberme guiado durante estos cuatro años y medio.   Gran parte del mérito de esta tesis es suyo y sin su empeño, dirección y esfuerzo no habría sido posible llegar a este punto.

En segundo lugar quiero agradecer también a todos los compañeros, pasados y presentes, del grupo Molchip.  Muchas gracias a Javier Sesé, a Marco Evangelisti, a Agustín Camón, a Olivier Robeau, a Guilia Lorusso, a Ana Repollés, a Juan José Morales, a Ainhoa, a Enrique Burzuri y a Maria José Martinez-Perez por su ayuda y apoyo durante este tiempo.  Gracias también a Marisa y a Aurora por los ánimos y por conseguir que toda nuestra burocracia saliera adelante.

En cuanto a colaboradores del ICMA y la Universidad de Zaragoza, muchas gracias a Pablo Alonso por las múltiples medidas de EPR y por ayudarnos con el análisis y discusión de los resultados.  Gracias a Santiago Celma por su ayuda con las medidas de microondas y permitirnos usar sus equipos para nuestros experimentos.  Agradecer a Jose Luis García Palacios y a David Zueco por su ayuda con muchos de los aspectos teóricos.  Gracias también a Miguel Ciria por su ayuda con las medidas de MFM.  Al Servicio de Medidas Físicas, en particular a Ana Arauzo, Enrique Guerrero y Ahinoa por la formación en los equipos PPMS y MPMS así como su ayuda en la realización de las medidas con estos equipos (y el Nanoscan...).  Gracias a Batriz Diosdado del Servicio de Diffracción de Rayos X por la ayuda con la caracterización de nuestros crystales y gracias a los miembros del Servicio de Instrumentación Electrónica por la ayuda en la fabricación de PCBs y diferentes soldaduras de cables y conectores.  Gracias al Servicio de Criogenia por haber hecho lo posible por acomodar siempre nuestras solicitudes de helio y nitrógeno (muchas veces con poca antelción).  Gracias al equipo de la sala blanca del INA (Laura Casado, Rubén Valero, Rosa Córdoba e Isabel Rivas y Celia Castán) por su ayuda con los procedimientos de litografía óptica y los procedimientos de Dual Beam (y los cortes de obleas de zafiro...).

Gracias a el grupo de Anabel Gracia en el INA y en especial a Maricarmen Pallarés y Rocío de Miguel por su colaboración y ayuda con las deposiciones por Dip-Pen y por estar siempre dispuestas a hacer intento más de depositar DPPH en la constricción.

En cuanto a nuestros colaboradores en el resto de España, muchas gracias al grupo de Eugenio Coronado en Valencia (en particular a Salvador Cardona-Serra, Alejandro Gaita Ariño y Yan Duan) y al grupo de Guillem Aromi en Barcelona por proporcionarnos la mayoría de las muestras con las que hemos trabajado estos años.  Agradecer también a Jose-Luis García Ripoll por sus años de colaboración con nuestro grupo.

También agradecer a Daniel Ruiz y su grupo mi primera estancia y mi acogida en el CIN2 de Barcelona y, en particular, gracias a Pablo Gonzalez-Monje (cambia er papé...) y a Elena Bellido por su ayuda con los sistemas de AFM y Dip-Pen.

Thank you also to Johannes Majer and his group at TU Wien, for their ongoing participation in our research and providing their expertise in low temperature microwave measurements.

Many thanks go also to Enrique del Barco and his group (Jim, Marta, Alvar, Diego, Hajra, Simran and Asma) at the University of Central Florida for making me feel welcome during my stay and for their ongoing collaboration in our research (y por enseñarme a jugar al mus).

A nivel algo más personal, quiero agradecer la compañía y amistad de los innumerables becarios del ICMA y el INA con los que he pasado estos años.  Soy ya de los últimos que quedan de mi ``generación'' de becarios y últimamente entran menos de los que salen...  A Celia Castán, Clara Guglieri, David Coffey, Mariafer Acosta, Cristina Extremiana, Ana Repollés, Luis Serrano, Adriana Figueroa, Roberto Boada, Montse Perez, Lorena Marco, Luis Alfredo y muchos más...  gracias a todos por los cafés, comidas y fiestas y por los cumpleaños sorpresa y por vuestro apoyo durante todo este tiempo.  Gracias también por su compañía a mis últimos compañeros de despacho, Eduardo Burillo y Fernando Quijandría, y a los compañeros que se toman el café en el despacho.

Fuera ya del ámbito académico, gracias a mis amigos del grupo Reunitas por tantos domingos de juegos y partidas de LoL (ugh...).  Por otra parte, gracias a Pepe-sensei y a los compañeros y amigos de AKZ y del Kajuki por introducirme al kendo y ayudarme a llegar a primer kyu (y con un poco de suerte... ¿quizá shodan?).  No me olvido de mis amigos y compañeros de la Universidad de Granada y de la carrera, muchos de los cuales ya han terminado sus tesis (en particular, un saludo para Rafa Roa y Bruno Zamorano).  Muchas gracias también a Aixa Alarcón que durante tantos años ha estado apoyándome, a Paco Cano por sus sabios consejos y a Miguel Alberto Pelaez por venir a visitarme cuando estuve en Orlando.

Quiero también dar las gracias a mi antiguo director de tesis, Jose Ignacio Illana, que me introdujo por primera vez en el mundo de la investigación y que sin ninguna culpa se quedó sin becario a mitad de tesis, y a los compañeros del Departamento de Física de Altas Energías (Roberto Barceló, Adrian Carmona, Manel Masip, Paco del Águila,...).  Hasta día de hoy todavía me duele haberme ido y haberos dejado colgados.

No podía faltar aquí agradecer el apoyo de mis padres y mis hermanos y toda mi familia, sin los cuales ni esta tesis ni casi cualquier otro logro que he obtenido habrían sido posibles.

Y por último, disculparme de nuevo con aquellos de los que no me he acordado al escribir esto.  Estoy seguro de que tedré que echarme las manos a la cabeza más de una vez cuando vaya dándome cuenta de quienes me he olvidado...

\end{document}